# Super*B*
# Progress Report

# The Collider

# Super*B* Progress Reports

## The Collider

December 27, 2010

**Abstract**

This report details the progress made in by the SuperB Project in the area of the Collider since the publication of the SuperB Conceptual Design Report in 2007 and the Proceedings of SuperB Workshop VI in Valencia in 2008.

M. E. Biagini, R. Boni, M. Boscolo, B. Buonomo, T. Demma, A. Drago, M. Esposito,
S. Guiducci, G. Mazzitelli, L. Pellegrino, M. A. Preger, P. Raimondi, R. Ricci, U. Rotundo,
C. Sanelli, M. Serio, A. Stella, S. Tomassini, M. Zobov
**Laboratori Nazionali di Frascati dell'INFN**, I-00044 Frascati, Italy

K. Bertsche, A. Brachman, Y. Cai, A. Chao, R. Chestnut, M. H. Donald, C. Field, A. Fisher,
D. Kharakh, A. Krasnykh, K. Moffeit, Y. Nosochkov, A. Novokhatski, M. Pivi, C. Rivetta,
J. T. Seeman, M. K. Sullivan, S. Weathersby, A. Weidemann, J. Weisend, U. Wienands,
W. Wittmer, M. Woods, G. Yocky
**SLAC National Accelerator Laboratory,** Stanford, California 94025, USA

A.Bogomiagkov, I. Koop, E. Levichev, S. Nikitin, I. Okunev, P. Piminov, S. Sinyatkin,
D. Shatilov, P. Vobly
**Budker Institute of Nuclear Physics**, Novosibirsk 630090, Russia

F. Bosi, S. Liuzzo, E. Paoloni
**INFN Pisa, Università di Pisa, Dipartimento di Fisica**, I-56127 Pisa, Italy

J. Bonis, R. Chehab, O. Dadoun, G. Le Meur, P. Lepercq, F. Letellier-Cohen, B. Mercier,
F. Poirier, C. Prevost, C. Rimbault, F. Touze, A. Variola
**Laboratoire de l'Accelerateur Lineaire, IN2P3/CNRS**, Universite Paris-Sud 11, F-91898,
Orsay, France

B. Bolzon, L. Brunetti, A. Jeremie
**Laboratoire d'Annecy-le-Vieux de Physique de Particule, IN2P3/CNRS**, F-74941,
Annecy-le-Vieux, France

M. Baylac, O. Bourrion, J.M. De Conto, Y. Gomez, F. Meot, N. Monseu, D. Tourres, C. Vescovi
**Laboratoire de Physique Subatomique et de Cosmologie**, **IN2P3/CNRS**, F- 38026,
Grenoble, France

A. Chancé, O. Napoly
**CEA, IRFU,** Centre de Saclay, F- 91191, Gif sur Yvette, France

D.P. Barber,
**Deutsches Elektronen-Synchrotron (DESY)**, 22607 Hamburg, Germany, and
**Cockcroft Institute, Daresbury Science and Innovation Centre,** Warrington WA4 4AD, UK,
and **University of Liverpool**, Liverpool L69 7ZE, UK

S. Bettoni, D. Quatraro,
**CERN**, CH-1211, Geneva, Switzerland

## Contents

# 1. Super*B* executive summary

With this document we propose a new electron positron colliding beam accelerator to be built in Italy to study flavor physics in the B-meson system at an energy of 10 GeV in the center-of-mass. This facility is called a high luminosity B-factory with a project name "Super*B*". This project builds on a long history of successful e+e- colliders built around the world, as illustrated in Figure 1.1. The key advances in the design of this accelerator come from recent successes at the DAFNE collider at INFN in Frascati, Italy, at PEP-II at SLAC in California, USA, and at KEKB at KEK in Tsukuba Japan, and from new concepts in beam manipulation at the interaction region (IP) called "crab waist". This new collider comprises of two colliding beam rings, one at 4.2 GeV and one at 6.7 GeV, a common interaction region, a new injection system at full beam energies, and one of the two beams longitudinally polarized at the IP. Most of the new accelerator techniques needed for this collider have been achieved at other recently completed accelerators including the new PETRA-3 light source at DESY in Hamburg (Germany) and the upgraded DAΦNE collider at the INFN laboratory at Frascati (Italy), or during design studies of CLIC or the International Linear Collider (ILC).

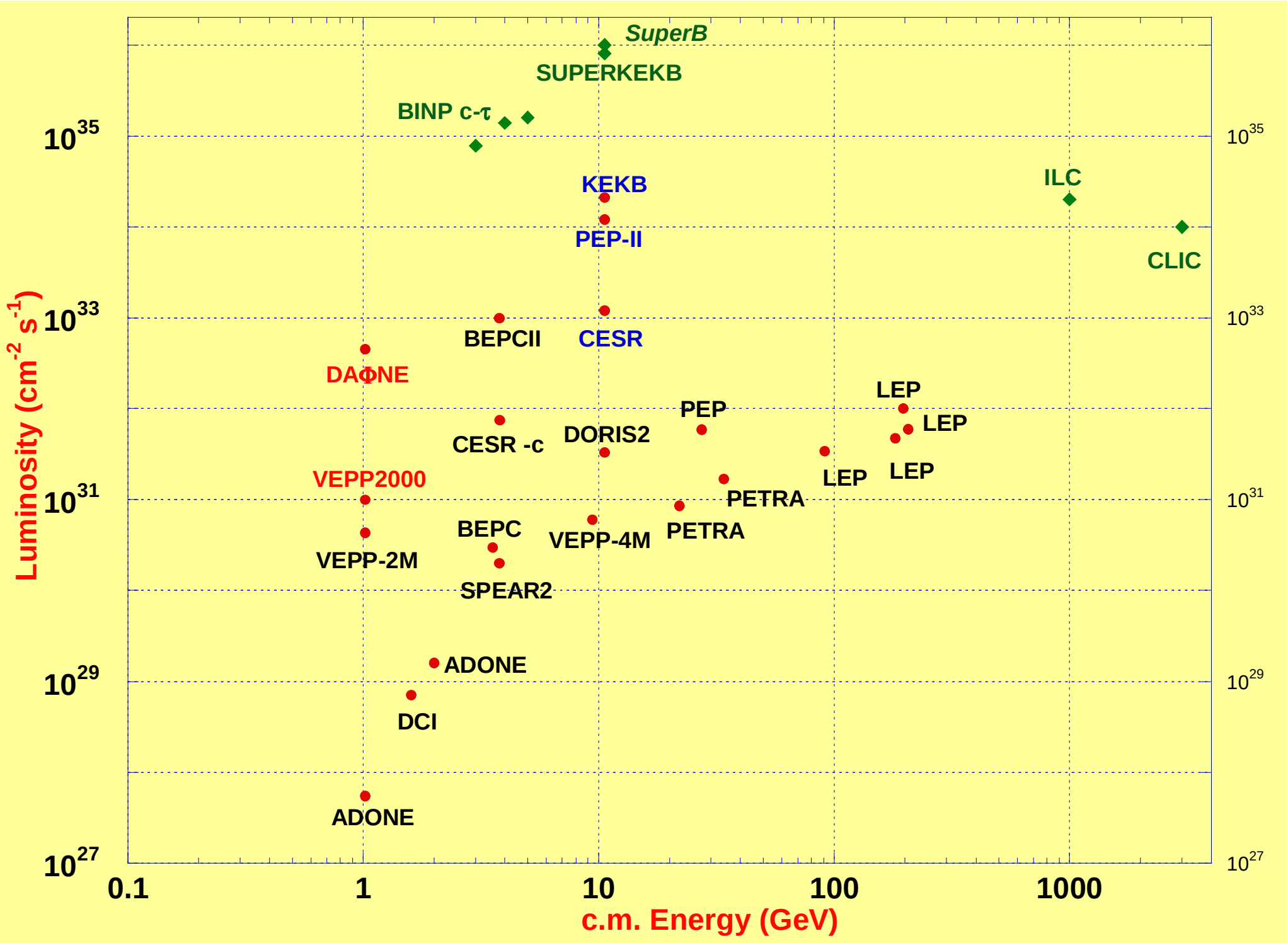

Figure 1.1 Peak luminosity versus e+e- collider center-of-mass energy. Super*B* is shown at the center of the plot at a luminosity of $10^{36}$/cm$^2$/s.

The project is to be designed and constructed by a worldwide collaboration of accelerator and engineering staff along with ties to industry. To save significant construction costs, many components from the PEP-II collider at SLAC will be recycled and used in this new accelerator. The interaction region will be designed in collaboration with the particle physics detector to guarantee successful mutual use.

The accelerator collaboration will consist of several groups at present universities and national laboratories. In Italy these may include INFN Frascati and the University of Pisa, in the United States SLAC, LBNL, BNL and several universities, in France IN2P3, LAPP, and Grenoble, in Russia BINP, in Poland Krakow University, and in the UK the Cockcroft Institute.

The construction time for this collider is a total of about four years. The new tunnel can be bored in about a year. The new accelerator components can be built and installed in about 4 years. The shipping of components from PEP-II at SLAC to Italy will take about a year. A new linac and damping ring complex for the injector for the rings can be built in about three years.

The commissioning of this new accelerator will take about a year including the new electron and positron sources, new linac, new damping ring, new beam transport lines, two new collider rings and the Interaction Region. The new particle physics detector can be commissioned simultaneously with the accelerator. Once beam collisions start for particle physics, the luminosity will increase with time, likely reaching full design specifications after about two to three years of operation.

After construction, the operation of the collider will be the responsibility of the Italian INFN governmental agency. The intent is to run this accelerator about ten months each year with about one month for accelerator turn-on and nine months for colliding beams. The collider will need to operate for about 10 years to provide the required 50 $ab^{-1}$ requested by the detector collaboration.

Both beams as anticipated in this collider will have properties that are excellent for use as sources for synchrotron radiation (SR). The expected photon properties are comparable to those of PETRA-3 or NSLS-II. The beam lines and user facilities needed to carry out this SR program are being investigated.

# 2. Super*B* introduction

## 2.1 A history of B-Factories

A Super B-Factory, an asymmetric energy $e^+e^-$ collider with a luminosity of order $10^{36}$ $cm^{-2}s^{-1}$, can provide a uniquely sensitive probe of New Physics in the flavour sector of the Standard Model.

The PEP-II and KEKB asymmetric colliders [1, 2] have produced unprecedented luminosities, above $10^{34}$ $cm^{-2}s^{-1}$, taking our understanding of the accelerator physics and engineering demands of asymmetric $e^+e^-$ colliders to a new parameter regime.

This very high luminosity, coupled with the innovation of continuous injection and the high efficiency of the accelerators and detectors, has allowed each of these machines to produce 500 to 1000 $fb^{-1}$ of accumulated data. The study of New Physics effects in the heavy quark and heavy lepton sectors, however, requires a data sample two orders of magnitude larger, hence the luminosity target of $10^{36}$ for Super*B*.

Attempts to design a Super B Factory date to 2001. The initial approach at SLAC and KEK had much in common: they were extrapolations of the very successful B Factory designs, with increased bunch charge bunches, somewhat reduced $\beta_y$* values, and crab cavities. These proposed designs reached luminosities of 5 to 7x$10^{35}$, but had wall plug power of the order of 100 MW.

This daunting power consumption motivated us to adapt linear collider concepts from the SLC and ILC to the regime of high luminosity storage ring colliders. The low emittance design presented herein reaches the desired luminosity regime with beam currents and wall plug power comparable to those in the current B Factories.

The parameters for a Flavour Factory based on an asymmetric energy $e^+e^-$ collider operating at a luminosity of order $10^{36}$ $cm^{-2}s^{-1}$ at the Υ(4S) resonance and $10^{35}$ $cm^{-2}s^{-1}$ at τ production threshold are described below. Such a collider could produce an integrated luminosity in excess of 12,000 $fb^{-1}$ (12 $ab^{-1}$) in a running year ($10^7$ s) at the Υ(4S).

The construction and operation of modern multi-bunch $e^+e^-$ colliders have brought about many advances in accelerator physics in the area of high currents, complex interaction regions, high beam-beam tune shifts, high-power RF systems, control of beam instabilities, rapid injection rates, and reliable up-times (90%). The successful operation of the currently operating B Factories has proven the validity of their design concepts:

- Colliders with asymmetric energies work;
- Beam-beam energy transparency conditions provide only weak constraints;
- Interaction regions with two energies can be built for both head-on and small angle collisions
- IR backgrounds can be handled successfully;
- High-current RF systems can be operated with excellent efficiency;
- Beam-beam tune shift parameters up to 0.06 - 0.09;
- Good injection rates can be sustained. Continuous injection is now in routine operation, largely removing the distinction between peak and average luminosity;
- The electron cloud effect (ECI) can be managed; and
- Bunch-by-bunch feedback works well with 4 ns bunch spacing.

Lessons learned from SLC, and more recent ILC studies and experiments (FFTB, ATF, ATF2), have also produced and proven new concepts:

- small horizontal and vertical emittances can be produced in a damping ring having a short damping time.
- very small beam spot sizes and beta functions can be produced at the inter- action region.

The design of the Super*B* $e^+e^-$ collider combines extensions of the design of the current B Factories with new linear collider concepts to produce an extraordinary leap in B Factory luminosity without increasing beam currents or power consumption. The luminosity L of an $e^+e^-$ collider is given by the expression:

$$L = \frac{N^+N^-}{4\pi\sigma_y\sqrt{(\sigma_z\tan\theta/2)^2+\sigma_x^2}}f_c$$

$$\sigma_{x,y} = \sqrt{\beta_{x,y}\varepsilon_{x,y}}$$

where fc is the frequency of collision of each bunch, N is the number of particles in the positron (+) and electron (−) bunches, σ is the beam size in the horizontal(x), vertical (y) and longitudinal (z) directions, ε is the beam emittance, β is the beta function (in cm) at the collision point in each plane and θ is the crossing angle between the beam lines at the interaction point (IP).

In this chapter we will describe the principles of the design of a new asymmetric collider that can reach a peak luminosity of $10^{36}$ $cm^{-2}$ $s^{-1}$, with beam currents and bunch lengths similar to those of the currently operating $e^+e^-$ factories, through the use of smaller emittances and a new scheme of crossing angle collision.

## 2.2 The large crossing angle and crab waist concepts

High luminosity can be achieved in colliders acting on the parameters as in the following formula:

$$L = f_{coll} \frac{N^+N^-}{4\pi\,\sigma_x\sigma_y} R_l$$

where $f_{coll}$ is the collision frequency, $N^+$ and $N^-$ are the number of particles per beam, $\sigma_x$ and $\sigma_y$ are respectively horizontal and vertical rms beam sizes and $R_l$ is a reduction factor which takes into account geometrical and "hourglass" effects.

The first approach chosen by KEKB Super B-Factory for luminosity upgrade was to shorten the bunches to 3 mm (to decrease $\beta_y$* at the IP, without incurring in the "hourglass" effect)., decrease the beam emittances and betatron functions at the IP, so decreasing beam sizes, and increase the beam currents to 9.4 A and 4.1 Amp [3]. However during the past year the design has significantly changed to converge to beam parameters very similar to those of Super*B*.

The option chosen for Super*B* to produce a peak luminosity in excess of $10^{36}$ $cm^{-2}s^{-1}$ is based on the "*crab waist*" (*CW*) scheme [4] for beam-beam collisions which combines several potentially advantageous ideas. This option is now being applied to the upgraded DAΦNE Φ-Factory at LNF, Frascati.

The first ingredient of this scheme is the large Piwinski angle: for collisions under a crossing angle $\theta$ the luminosity $L$ and the horizontal $\xi_x$ and the vertical $\xi_y$ tune shifts scale as (see for example in [5]):

$$L \propto \frac{N\xi_y}{\beta_y} \tag{1}$$

$$\xi_y \propto \frac{N\sqrt{\beta_y}}{\sigma_x\sqrt{1+\phi^2}} \approx \frac{2N\sqrt{\beta_y}}{\sigma_z\theta} \tag{2}$$

$$\xi_x \propto \frac{N}{\sigma_x^2(1+\phi^2)} \approx \frac{4N}{(\sigma_z\theta)^2} \tag{3}$$

Piwinski angle $\phi$ is defined as:

$$\phi = \frac{\sigma_z}{\sigma_x} tg\left(\frac{\theta}{2}\right) \approx \frac{\sigma_z}{\sigma_x}\frac{\theta}{2} \tag{4}$$

$\sigma_x$ being the horizontal rms bunch size, $\sigma_z$ the rms bunch length, $N$ the number of particles per bunch. Here we consider the case of flat beams, small horizontal crossing angle $\theta << 1$ and large Piwinski angle $\phi >>1$. In the *CW* scheme the large Piwinski angle is obtained by decreasing the horizontal beam size and increasing the crossing angle. As a result, both luminosity and horizontal tune shift increase, and the overlap area of the colliding bunches is decreased proportionally to $\sigma_x/\theta$. So, if the vertical beta function $\beta_y$ is comparable to the overlap area size:

$$\beta_y \approx \frac{\sigma_x}{\theta} << \sigma_z \tag{5}$$

several advantages follow:

a) small spot size at the IP, i.e. higher luminosity (see eq. (1)),
b) reduction of the vertical tune shift (see eq. (2))
c) suppression of the vertical synchro-betatron resonances [6].

In addition, in such a collision scheme there is no need of decreasing the bunch length to gain luminosity, then relaxing the problems of HOM heating, coherent synchrotron radiation of short bunches and excessive power consumption.

Long-range beam-beam interactions are expected to limit the maximum achievable luminosity when the bunch distance is short (the so called "parasitic collisions", PC). Thanks to the large crossing angle and small horizontal beam size in the *CW* scheme, the beams separation at the PC is large in terms of $\sigma_x$, automatically solving this problem.

The choice of large Piwinski angle, beneficial to the luminosity, introduces new beam-beam resonances and may strongly limit the maximum tune shifts achievable (see for example in [7]). The *CW* transformation is expected to solve such a problem. It actually contributes to suppress, through the vertical motion modulation by the horizontal oscillations, betatron and synchro-betatron resonances usually arising in collisions without *CW*. The *CW* scheme is realized by a couple of sextupole magnets on the two sides of the IP, as shown in Fig. 2.1. To provide the exact compensation the sextupoles must have a phase advance with respect to the IP of π the horizontal plane and at π/2 in the vertical one.

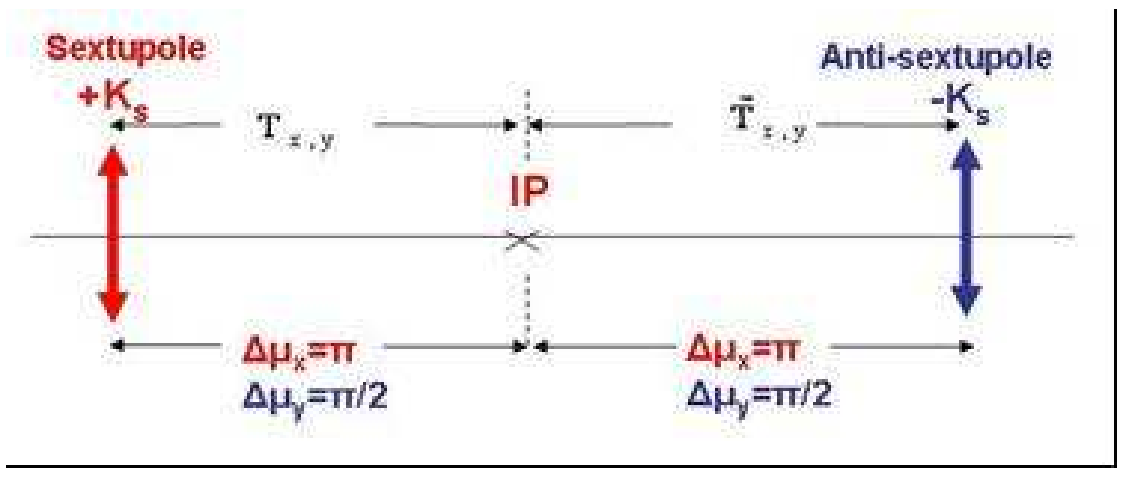

Figure 2.1: *Crab waist* correction by sextupole lenses.

As an example of how the *CW* transformation actually works, Fig.2.2 below shows the Super*B* bunch charge density envelopes at the IP when colliding without (top) and with (bottom) the *CW* sextupoles. In red is the Low Energy Beam, in blue the High Energy one, whom distribution is shown only near the overlap region. For sake of clarity, in the picture the crossing angle has been reduced by a factor of 4, to enhance the *CW* transformation effect.

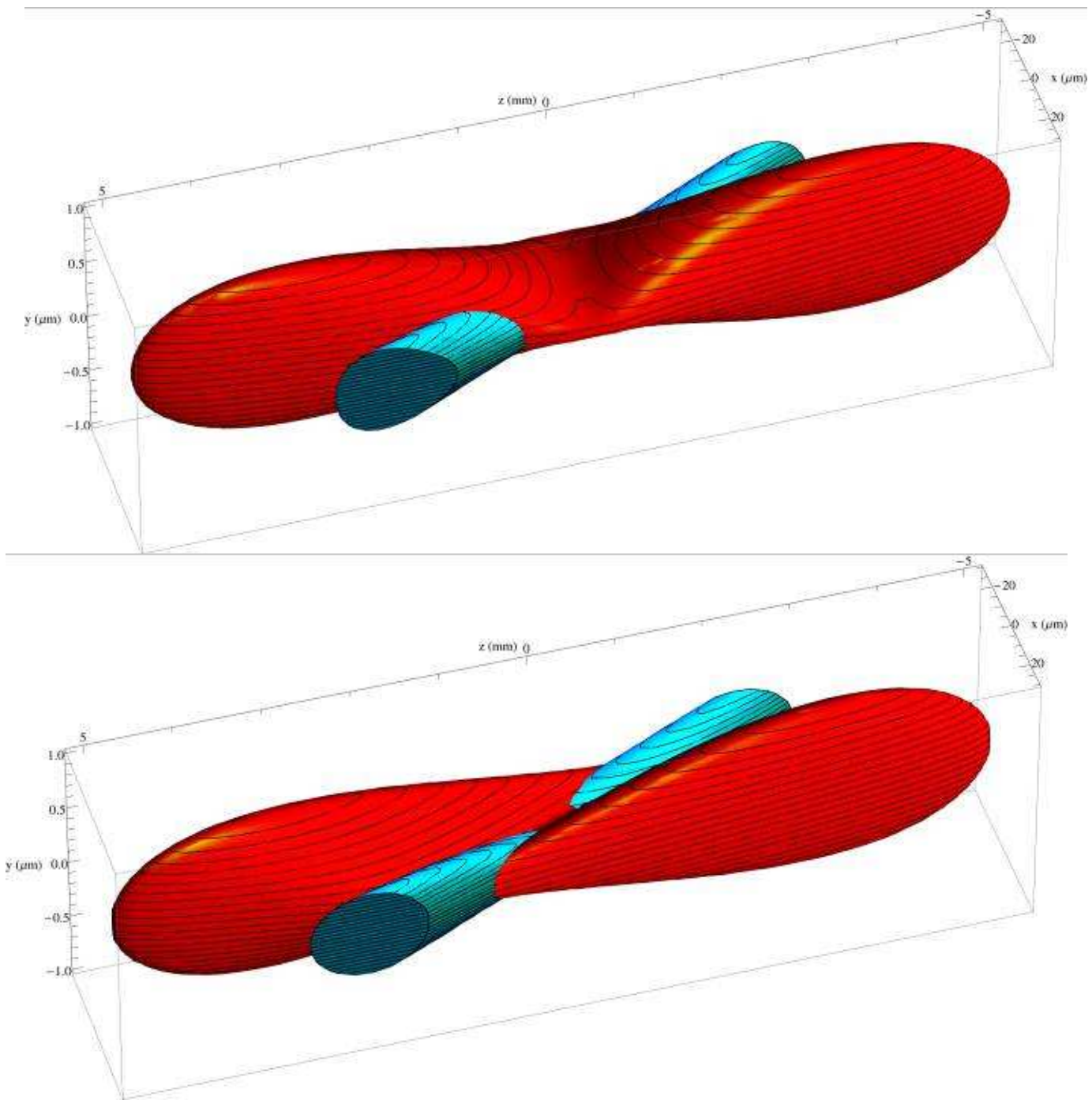

Figure 2.2: Sketch of the large Piwinski angle and crab waist scheme for SuperB.

Top: CW transformation OFF. Bottom: CW transformation ON.

The contour lines are points with the same y coordinate. When there is no *CW* transformation (top plot) the waist line is orthogonal to the axis of one bunch (LEB in this example). Otherwise, when the *CW* transformation is on (bottom plot) the waist moves to the axis of the other beam (HEB here). As a consequence, each beam collides with the other in the minimum $\beta_y$ region, with a net luminosity gain. Actually, besides the geometrical gain just mentioned, because of the *CW* transformation the non linear component of the beam-beam forces decreases, hence reducing the emittance growth due to the collision. The *CW* transformation acts on the y-plane as described by the following formula:

$$y \quad \rightarrow \quad y - \chi x y' / tg(2\theta) \qquad (6)$$

Where $\chi$ is the crab coefficient (of the order of one or less), x(y) is the particle horizontal (vertical) coordinate, y' is the vertical slope.

## 2.3 Super*B* parameters

### Nominal parameters for $10^{36}$ cm$^{-2}$ s$^{-1}$

The IP and ring parameters have been optimized based on several constraints. The most significant are:

- to maintain wall plug power, beam currents, bunch lengths, and RF requirements comparable to present B Factories;
- to plan for the reuse as much as possible of the PEP-II hardware;
- to require ring parameters as close as possible to those already achieved in the BFactories, or under study for the ILC Damping Ring or achieved at the ATF ILC-DR test facility [8];
- to simplify the IR design as much as possible. In particular, reduce the synchrotron radiation in the IR, reduce the HOM power and increase the beam stay-clear;
- to eliminate the effects of the parasitic beam crossing;

- to relax as much as possible the requirements on the beam demagnification at the IP;
- to design the final focus system to follow as closely as possible already tested systems, and integrating the system as much as possible into the ring design.

Column 1 of Table 3-1 shows the baseline parameters set that closely matches these criteria. Further details on beam-beam simulations and lattice design will be presented in the following sections.

Table 3.1: Super*B* parameters for baseline, low emittance and high current options, and for tau/charm running.

| | | Base Line | | Low Emittance | | High Current | | Tau-charm | |
|---|---|---|---|---|---|---|---|---|---|
| **Parameter** | **Units** | **HER (e+)** | **LER (e-)** | **HER (e+)** | **LER (e-)** | **HER (e+)** | **LER (e-)** | **HER (e+)** | **LER (e-)** |
| **LUMINOSITY** | **$cm^{-2}\ s^{-1}$** | **1.00E+36** | | **1.00E+36** | | **1.00E+36** | | **1.00E+35** | |
| **Energy** | **GeV** | 6.7 | 4.18 | 6.7 | 4.18 | 6.7 | 4.18 | 2.58 | 1.61 |
| **Circumference** | **m** | 1258.4 | | 1258.4 | | 1258.4 | | 1258.4 | |
| **X-Angle (full)** | **mrad** | 66 | | 66 | | 66 | | 66 | |
| **$\beta_x$ @ IP** | **cm** | 2.6 | 3.2 | 2.6 | 3.2 | 5.06 | 6.22 | 6.76 | 8.32 |
| **$\beta_y$ @ IP** | **cm** | 0.0253 | 0.0205 | 0.0179 | 0.0145 | 0.0292 | 0.0237 | 0.0658 | 0.0533 |
| **Coupling (full current)** | **%** | 0.25 | 0.25 | 0.25 | 0.25 | 0.5 | 0.5 | 0.25 | 0.25 |
| **Emittance x (with IBS)** | **nm** | 2.00 | 2.46 | 1.00 | 1.23 | 2.00 | 2.46 | 5.20 | 6.4 |
| **Emittance y** | **pm** | 5 | 6.15 | 2.5 | 3.075 | 10 | 12.3 | 13 | 16 |
| **Bunch length (full current)** | **mm** | 5 | 5 | 5 | 5 | 4.4 | 4.4 | 5 | 5 |
| **Beam current** | **mA** | 1892 | 2447 | 1460 | 1888 | 3094 | 4000 | 1365 | 1766 |
| **Buckets distance** | **#** | 2 | | 2 | | 1 | | 1 | |
| **Ion gap** | **%** | 2 | | 2 | | 2 | | 2 | |
| **RF frequency** | **MHz** | 476. | | 476. | | 476. | | 476. | |
| **Revolution frequency** | **MHz** | 0.238 | | 0.238 | | 0.238 | | 0.238 | |
| **Harmonic number** | **#** | 1998 | | 1998 | | 1998 | | 1998 | |
| **Number of bunches** | **#** | 978 | | 978 | | 1956 | | 1956 | |
| **N. Particle/bunch ($10^{10}$)** | **#** | 5.08 | 6.56 | 3.92 | 5.06 | 4.15 | 5.36 | 1.83 | 2.37 |
| **$\sigma_x$ effective** | **µm** | 165.22 | 165.30 | 165.22 | 165.30 | 145.60 | 145.78 | 166.12 | 166.67 |
| **$\sigma_y$ @ IP** | **µm** | 0.036 | 0.036 | 0.021 | 0.021 | 0.054 | 0.0254 | 0.092 | 0.092 |
| **Piwinski angle** | **rad** | 22.88 | 18.60 | 32.36 | 26.30 | 14.43 | 11.74 | 8.80 | 7.15 |
| **$\Sigma_x$ effective** | **µm** | 233.35 | | 233.35 | | 205.34 | | 233.35 | |
| **$\Sigma_y$** | **µm** | 0.050 | | 0.030 | | 0.076 | | 0.131 | |
| **Hourglass reduction factor** | | 0.950 | | 0.950 | | 0.950 | | 0.950 | |
| **Tune shift x** | | 0.0021 | 0.0033 | 0.0017 | 0.0025 | 0.0044 | 0.0067 | 0.0052 | 0.0080 |
| **Tune shift y** | | 0.097 | 0.097 | 0.0891 | 0.0892 | 0.0684 | 0.0687 | 0.0909 | 0.0910 |
| **Longitudinal damping time** | **msec** | 13.4 | 20.3 | 13.4 | 20.3 | 13.4 | 20.3 | 26.8 | 40.6 |
| **Energy Loss/turn** | **MeV** | 2.11 | 0.865 | 2.11 | 0.865 | 2.11 | 0.865 | 0.4 | 0.17 |
| **Momentum compaction ($10^{-4}$)** | | 4.36 | 4.05 | 4.36 | 4.05 | 4.36 | 4.05 | 4.36 | 4.05 |
| **Energy spread ($10^{-4}$) (full current)** | **dE/E** | 6.43 | 7.34 | 6.43 | 7.34 | 6.43 | 7.34 | 6.43 | 7.34 |
| **CM energy spread ($10^{-4}$)** | **dE/E** | 5.0 | | 5.0 | | 5.0 | | 5.0 | |
| **Total lifetime** | **min** | 4.23 | 4.48 | 3.05 | 3 | 7.08 | 7.73 | 11.4 | 6.8 |
| **Total RF Wall Plug Power** | **MW** | 16.38 | | 12.37 | | 28.83 | | 2.81 | |

The machine is also designed to have flexibility for the parameters choice with respect to the baseline. In particular:

- The horizontal emittance can be decreased by about a factor 2 in both rings by changing the partition number (by changing the RF frequency as done in the LEP or the orbit in the Arcs) and the natural Arc emittance by readjusting the lattice functions.
- The Final Focus system as a built-in capability of about a factor 2 in decreasing the IP beta functions.
- The RF system will be able to support higher beam currents (up to a factor 1.6) than the baseline ones, when all the available PEP-II RF units are installed.

Based on these considerations, columns 2 and 3 in Table 3-1 shows different parameters options:

The “Low Emittance” case relaxes the RF requirements and all the problems related to high current operations (including wall-plug power) but put more strain on the optic and the Tuning capabilities.

The “High Current” case has the opposite characteristics. The requirements on vertical emittance and IP beta functions are relaxed but the high currents issues are enhanced (instabilities, HOM, synchrotron radiation, wall-plug power etc…).

The cases shown have several parameters kept as much constant as possible (bunch length, IP stay clear etc…), in order to reduce their impact on other unwanted effects (Detector background, HOM heating etc…).

In overall the collider should be flexible enough to reach the target specifications, superseding the encountered limitations by pushing more the less critical parameters.

## 2.4 Energy scaling for operation at the τ/charm threshold

Super*B* can operate at a lower center-of-mass energy with a somewhat reduced luminosity. In order to operate at τ /charm threshold energies (in the vicinity of 3.8 GeV) with minimal modifications to the machine, beam energies will be scaled, maintaining the nominal energy asymmetry ratio used for operation at the center- of-mass energy of the Υ (4S).

All magnet currents will be rescaled accordingly. In order to provide the necessary damping at low current wigglers will be installed in the straight sections (dispersion free) and in the ARCs, in a relative number matched to achieve the desired beam parameters (emittance etc…). About 15-20m of wigglers will be needed, their total lengths depends from the chosen magnetic field in the wigglers (to be studied). The permanent magnets in the IR will be replaced with weaker versions. The main differences in the ring properties will be:

- Lower energy by a factor of about 2.6-2.8 per ring;
- Longer damping time by a factor of about 2.0 per ring;
- Decreased Touschek lifetime by a factor of 3-6
- Increased sensitivity to collective effects.

Luminosity should scale linearly with energy (see formula in Sec. 3.1.2). However, the damping time and collective effects will result in a further decrease the luminosity. In general, the luminosity dependence is less then linear with respect to the damping time (about $1/\tau$ 0.3−0.5). However, given all factors, we expect that operations at lower energy will require a decrease of the beam current and an increase of the beam emittance. It is thus reasonable to expect a luminosity about 10 times smaller than that at 10.58 GeV. The last column in Table 3-1 shows the parameters for the run at the τ/charm.

## 3. DAΦNE upgrade results

Relevant modifications [1] to the machine have been realized in 2007, aimed at implementing the new "large Piwinski angle (*LPA*) and crab waist (*CW*)" collision scheme. A layout of the upgraded DAΦNE is shown in Fig. 3.1, and the main hardware changes are briefly illustrated in the following.

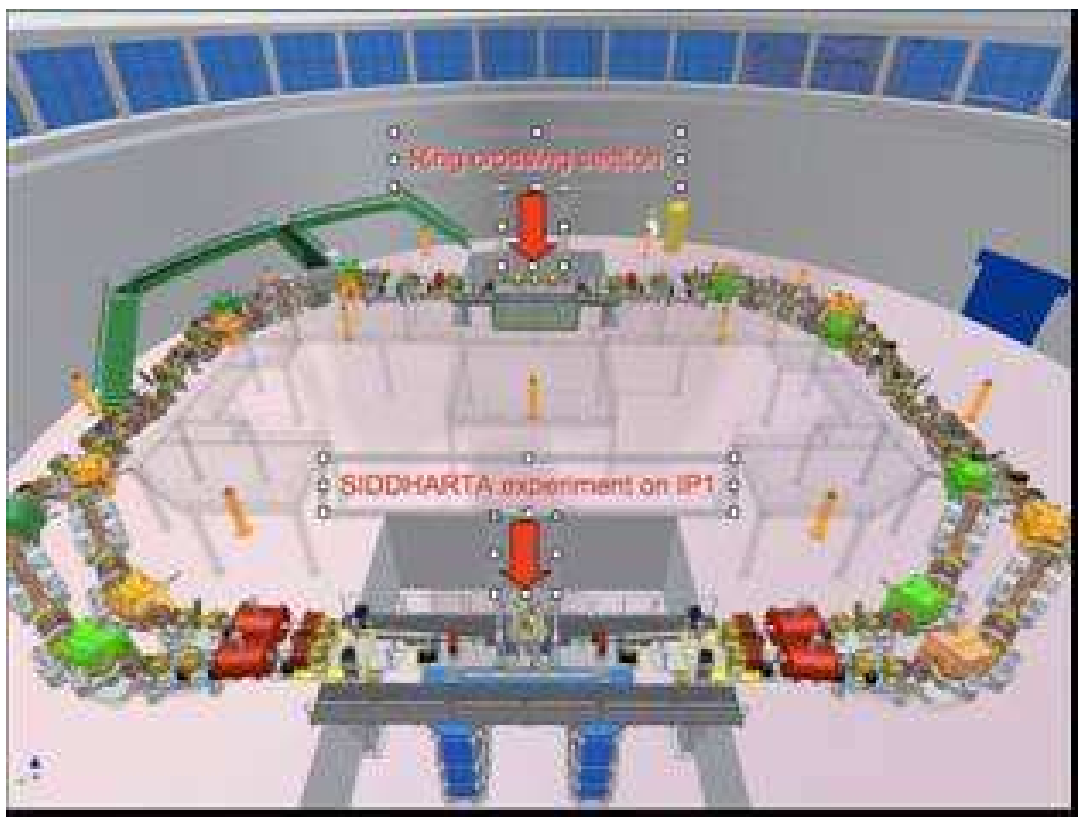

Figure 3.1: Upgraded DAΦNE layout.

### 3.1 Hardware upgrades

The KLOE Interaction Region (IR1) has been modified for the installation of the SIDDHARTA experiment, and equipped with new quadrupoles to be able to lower $\beta^*$ at the IP. The total crossing angle has been increased from 30 mrad to 50 mrad, by removing the splitter magnets and rotating the two sector dipoles in the long and short arcs adjacent to the interaction regions of both rings. New beam pipes have been designed for this scheme. Existing sextupoles are used for the *CW* transformation. Fig. 3.2 shows the comparison between the KLOE IR1 layout (top) and the upgraded one (bottom).

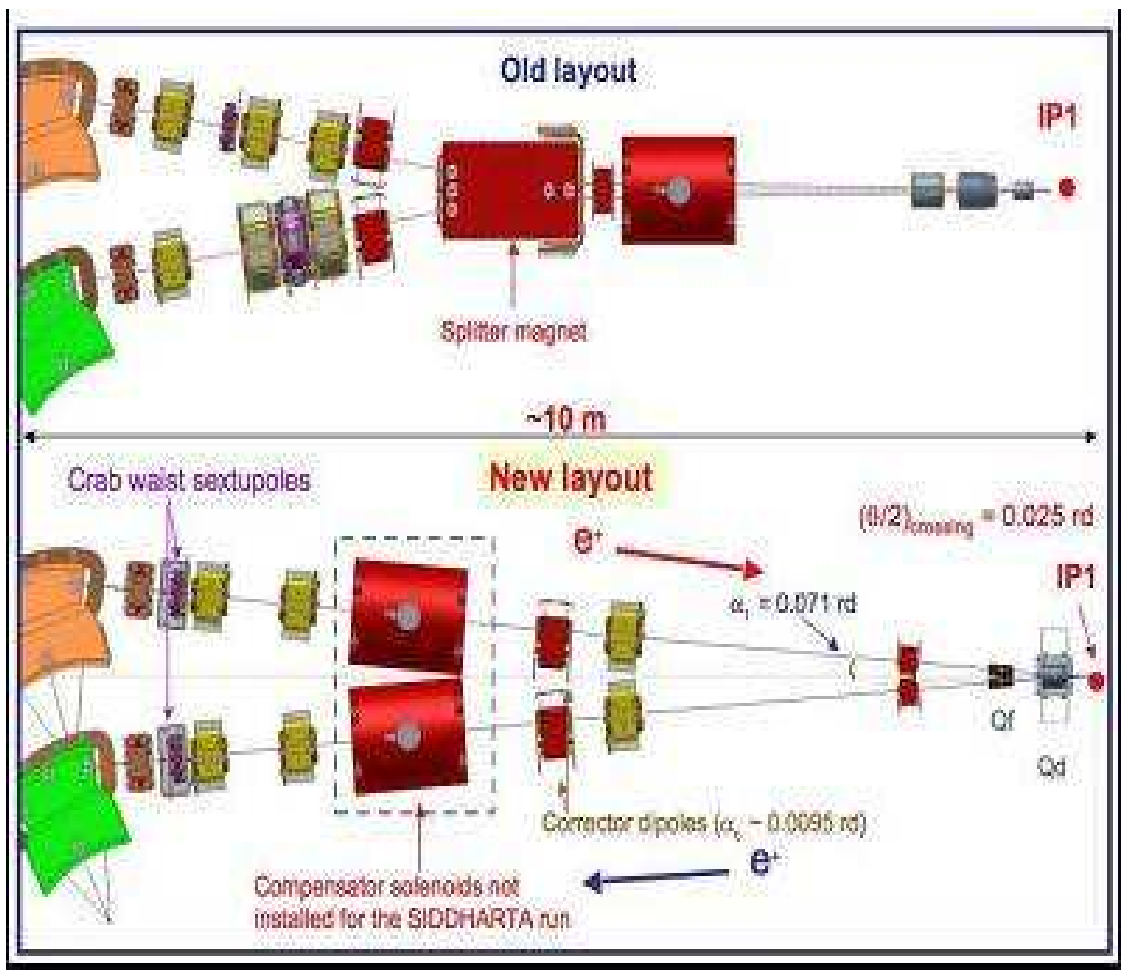

Figure 3.2: Half view of old (top) and new (bottom) IR1 layout.

New permanent magnet quadrupole doublets are needed in order to focus the beams to the smaller $\beta^*$ at the IP. The first quadrupole of the doublet, QD0, is horizontally defocusing, and common to both beams in the same vacuum chamber: it provides a strong separation of the beams. The following horizontally focusing quadrupoles, QF1, are particularly small, in order to fit separated beam pipes for the two beams. The new configuration almost cancels the problems related to beam-beam long range interactions (PC), because the two beams experience only one parasitic crossing inside the defocusing quadrupole where, due to the large horizontal crossing angle, they are very well separated.

The *CW* sextupoles have been installed at both ends of the IR1. Four additional electromagnetic quadrupoles have been installed on both sides of IP1 to get the proper phase advance between the *CW* sextupoles and the IP.

The second IR (IR2) has also been completely rebuilt, in order to provide full beam separation without low-β. A new beam pipe at IP2, providing complete separation between the two beams, has replaced the old one. This is geometrically symmetric to IR1, and its vacuum chamber is based on the same design criteria. Independent beam vacuum chambers are obtained by splitting the original pipe in two *half-moon* shaped sections, providing full vertical beam separation. The problem of the beam-beam long range interaction in this non colliding section is then naturally solved, allowing at the same time to relax the ring optics requirements imposed by beam separation at IP2.

New, fast kickers have been designed and built, based on a tapered strip with rectangular vacuum chamber cross section. Compared to the present DAΦNE injection kickers the new ones have a much shorter pulse (~12 ns instead of ~150 ns), better uniformity of the deflecting field, lower impedance and the possibility of higher injection rate (max 50 Hz). Moreover a smooth beam pipe and tapered transitions reduce the kicker contribution to the total ring coupling impedance. All these features improve the maximum storable currents, colliding beams stability and backgrounds hitting the experimental detector during injection

New bellows have been developed and installed in IR1 and in the ring. The transverse horizontal position of two wigglers in the long arcs has been moved by -2.5 mm in both rings, in order to reduce the non-linear terms in the magnetic field predicted by simulations and affecting the beam dynamics.

### 3.2 Achieved results

The maximum luminosity achieved experimentally with the CW sextupoles ON is about a factor of 2.7 higher than the ideal one predicted numerically for the case of CW sextupoles switched OFF. This is a clear proof that the crab waist concept works. However, in order to complete the studies we have dedicated several hours tuning the collider with the CW sextupoles off. Fig. 3.3 shows a comparison of the luminosity as a function of beam current product obtained with crab sextupoles ON and OFF. The maximum single bunch luminosity reached in the latter case was of the order of 1.6-1.7x$10^{30}$ cm$^{-2}$s$^{-1}$. This result is also consistent with numerical predictions. It should be noted that another limitation becomes very important in collision without crab waist sextupoles: besides much bigger vertical blow up, a sharp lifetime reduction is observed already at single bunch currents of 8-10 mA. That is why the red curve in Fig. 3.3 is interrupted at much lower currents. By including beam-beam interaction in the dynamic aperture simulations, which take into account lattice nonlinearities, it can be seen that their effect is not dramatic for the case of the crab sextupoles on since the beam size blow up is only by about 8% higher with respect to the ideal simulations. No lifetime reduction is indicated by the simulations. In the case of the CW sextupoles off the beam tails are much longer for the nonlinear lattice exceeding an aperture of 80 $\sigma_y$ in the vertical plane, which was estimated to be the dynamic aperture limit. Already at 10 mA per bunch the calculated lifetime sharply drops down in agreement with experimental observations.

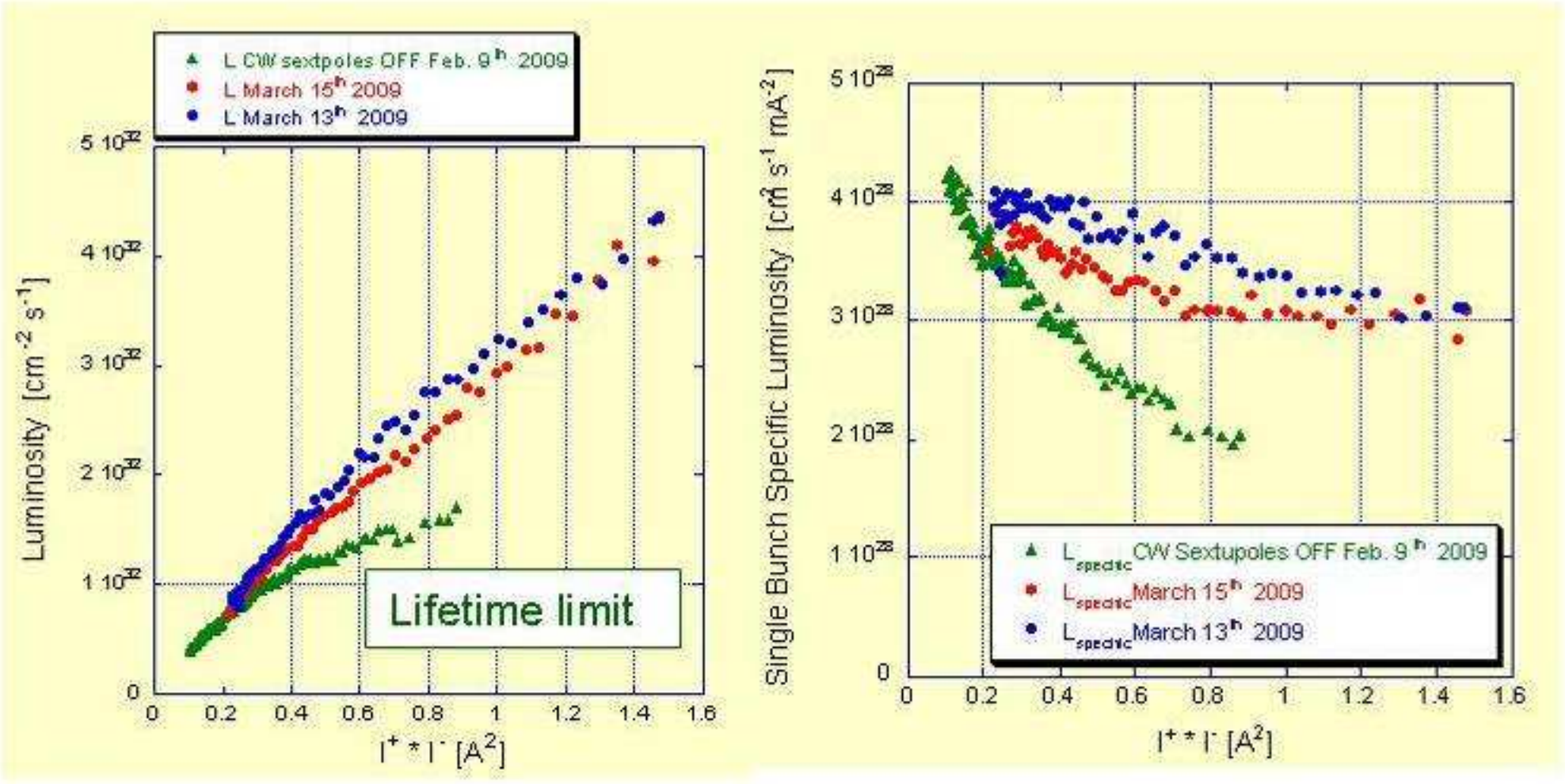

Figure 3.3: Luminosity vs product of beam currents (left) and specific luminosity vs product of beam currents (right), for two record shifts with crab sextupoles ON (read and blue dots) and with crab sextupoles OFF (green).

The convolved IP vertical beam size in collision has been measured by means of a beam-beam scan technique. A measured $\Sigma_y$ of 5.6 μ is compatible with the value obtained by using the coupling value (κ~0.7%) as measured at the Synchrotron Light Monitor (SLM), being the single vertical beam size at the IP1 of the order of 4 μ. Another striking proof of the crab sextupoles effectiveness is shown in Fig. 3.4 where the positrons transverse beam profile measured at the SLM with crab sextupoles OFF (left plot) and with crab sextupoles ON (right plot) is shown. The measurement was taken while colliding is a strong-weak regime: namely 1Amps electrons beam current against 0.1Amps of positrons beam current. It is evident that the transverse beam size is smaller and its shape remains Gaussian during collision with the sextupoles ON.

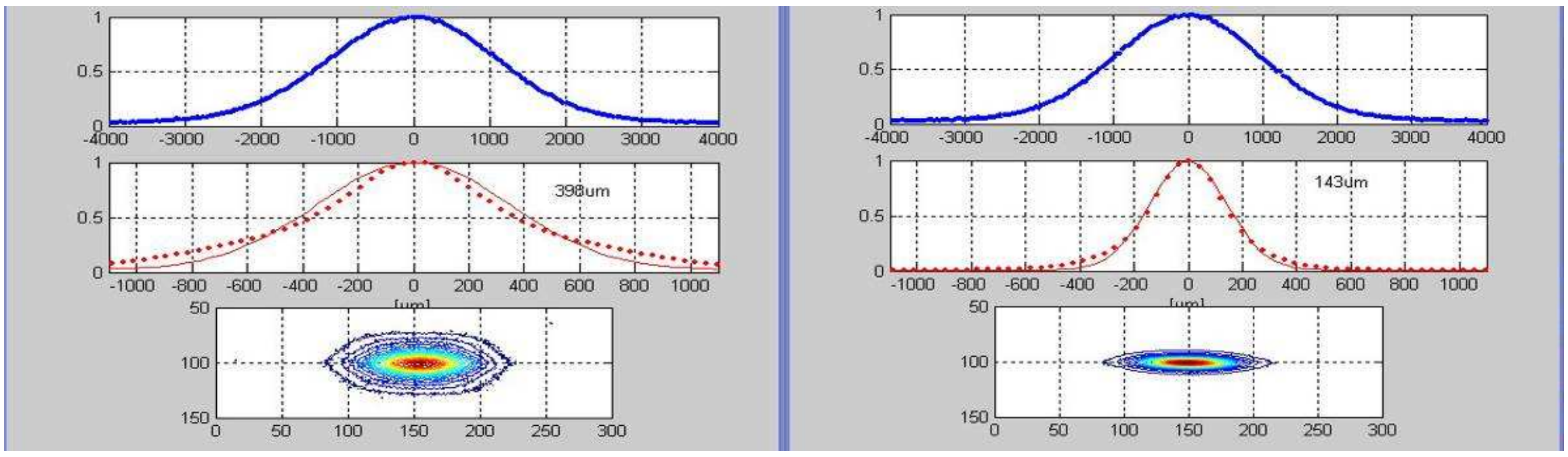

Figure 3.4: Transverse positron beam profile as measured at the SLM with crab sextupoles OFF (left) and with crab sextupoles ON (right) for beams in collision (103 bunches).

Fig. 3.5 summarizes the DAΦNE performances in the running years, showing the improvement due to the new collision scheme. Table 3.1 summarizes the luminosity and corresponding parameters at the interaction point (IP) for the best DAΦNE luminosity runs for the three main experiments carried out on the collider. The first and the second column correspond to the results achieved with the KLOE and FINUDA detectors before the DAΦNE upgrade based on the crab waist concept. The third column shows results obtained during the current run with the SIDDHARTA experiment after the collider upgrade.

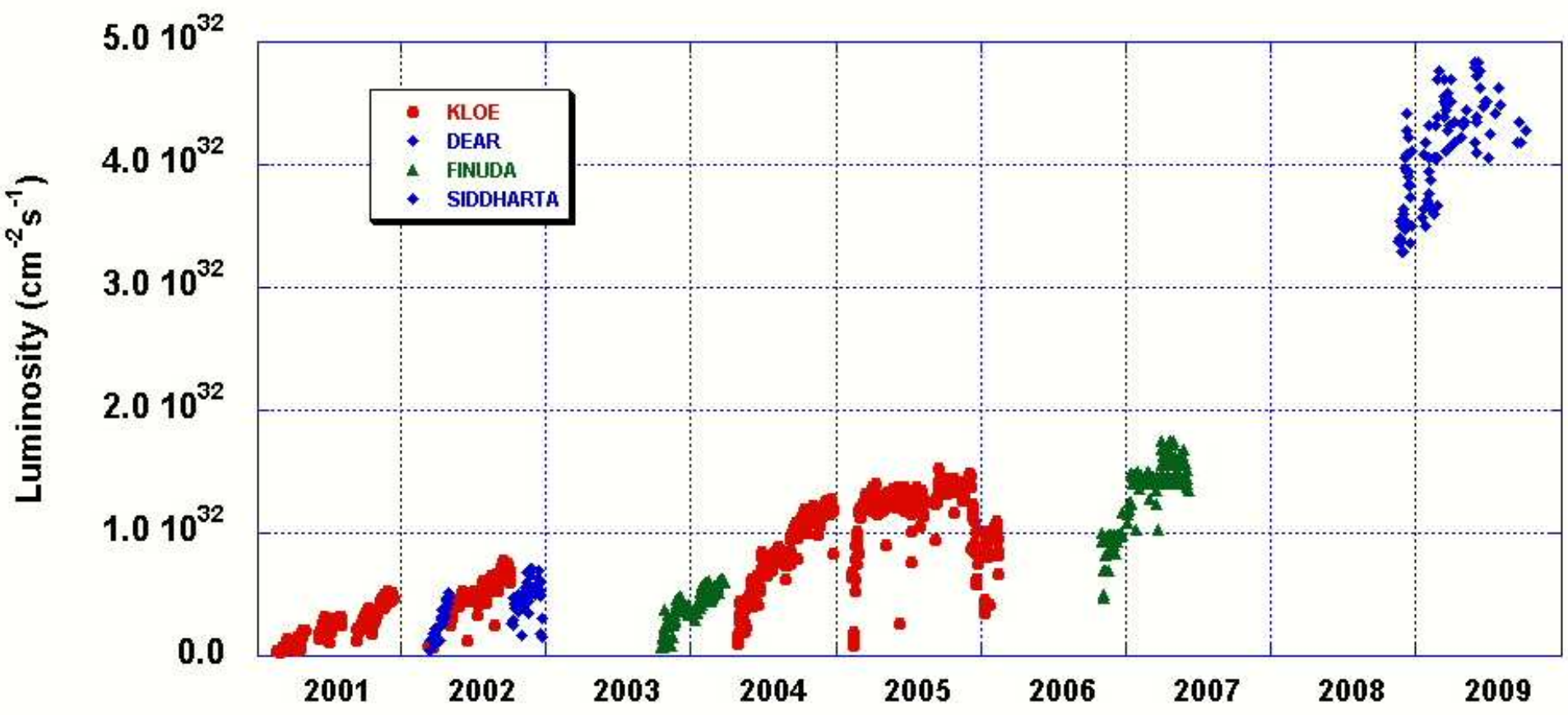

Figure 3.5: Luminosity vs years of running and different detectors (red KLOE, blue DEAR and SIDDHARTA, green FINUDA).

Table 3.1: DAΦNE best luminosity and IP parameters for past (KLOE, FINUDA) and present (SIDDHARTA) experimental runs

| **Parameter** | **KLOE** | **FINUDA** | **SIDDHARTA** |
|---|---|---|---|
| ***Date*** | ***Sept. 05*** | ***April 07*** | ***April 09*** |
| Luminosity, $cm^{-2}s^{-1}$ | $1.5x10^{32}$ | $1.6x10^{32}$ | $4.5x10^{32}$ |
| $e^-$ current, A | 1.38 | 1.5 | 1.43 |
| $e^{+-}$ current, A | 1.18 | 1.1 | 1.1 |
| N. of bunches | 111 | 106 | 107 |
| $\varepsilon_x$, mm mrad | 0.34 | 0.34 | 0.25 |
| $\beta_x$, m | 1.5 | 2 | 0.25 |
| $\beta_y$, m | 1.8 | 1.9 | 0.93 |
| Cros. Angle, mrad | 2x12.5 | 2x12.5 | 2x25 |
| Y-tune shift | 0.0245 | 0.029 | 0.042 |

In summary, DAΦNE has proved that the "large Piwinski angle and crab waist" concept definitely works. Ideal strong-strong simulations agree within 20-25% with experimental results and the agreement is expected to be even better if we include in simulations other luminosity limiting factors.

The "crab waist" sextupoles proved to be of great importance for the collider luminosity increase, since much lower luminosity is achieved with crab sextupoles off, with a larger blow up and a sharp lifetime reduction is observed for bunch currents > 8-10 mA. This is in accordance with beam-beam simulations taking into account the realistic DAΦNE nonlinear lattice.

As a consequence of the very good results, the DAΦNE scientific program has been approved for the next 3 years with an upgraded KLOE detector.

# 4. Super*B* layout at LNF

The Super*B* facility will require a big complex of civil infrastructure. The main construction, which will house the final part of the LINAC, the injection lines, the damping rings, and the storage rings, will be mainly underground. A footprint of the Super*B* layout on the LNF area is shown in Fig. 4.1 All the service buildings are foreseen in the LNF side while in the ENEA area only the underground tunnel is placed. The storage rings will have an elliptical shape with the major axis of about 500 m and the minor axis of about 400 m, for a total length of the circumference of about 1260 m. In the aim to reuse at maximum the civil infrastructure of the LNF and in particular the DAΦNE facility, the machine was located on the site considering the possibility to extend the existing DAΦNE LINAC up to about 350m. Due to the slope of the hill, the end of the LINAC tunnel will be about 20 m below the ground surface from which two injection tunnels housing the injection lines, depart in two opposite directions in order to reach the storage ring tunnel. The storage ring tunnel plane could have a slope of about 1.8 deg versus the horizontal one in order to match firstly the hill slope and secondly to not interfere with the foundations of the existing buildings in the laboratory and ENEA side. Another possibility is to dig the storage ring tunnel horizontally and quite deep in order to pass below the DAΦNE foundations in the north part of LNF. In this second case, as a result, the tunnel is about 40m deep in the area of the strait section (south LNF) and the injection tunnels must have a double curvature, one in horizontal plane and the other in the vertical one. In the north part of the laboratory, the tunnel is located below the DAΦNE hall giving the possibility to reuse all the most important civil infrastructure like, power supply hall, DAΦNE and hopefully the KLOE control room.

In the south side of the laboratory, where the new guest house building has been built, two main service buildings of about 700m$^2$ each of covered area are foreseen to accommodate at least 12 klystrons and modulators. These buildings are designed to have an underground part to locate other devices mainly components for the cooling plants like pumping units, magnet power supplies and control devices. Nevertheless the two roof areas can be used for heat exchanger and air conditioning machines allocation. On the opposite side (north-west) is foreseen the collider hall, a 16x30 m$^2$ building with a shaft able to lower the heavy magnet component in the pit. The surface part is a large covered area with a strong bridge crane. Four safety egresses are foreseen only in the LNF side according to the Italian regulations and law. One of them is located in the collider hall building, another one in connection with the klystrons and modulators Hall, the third one in the south-east side of the tunnel and the fourth in the north-east side of the tunnel. The LINAC will have an access at the beginning and at the end of the surface building; meanwhile the underground part will have two safety egresses in the central and final part (the access at the end of the surface building can be used also as access of the beginning of the underground part of the LINAC). Other buildings are foreseen for offices; assuming about 200 more physicists are resident every day are foreseen in the laboratory. The ENEA side presently is considered only for underground civil constructions but in the agreement between INFN and ENEA could be considered for example the usage of existing civil infrastructure.

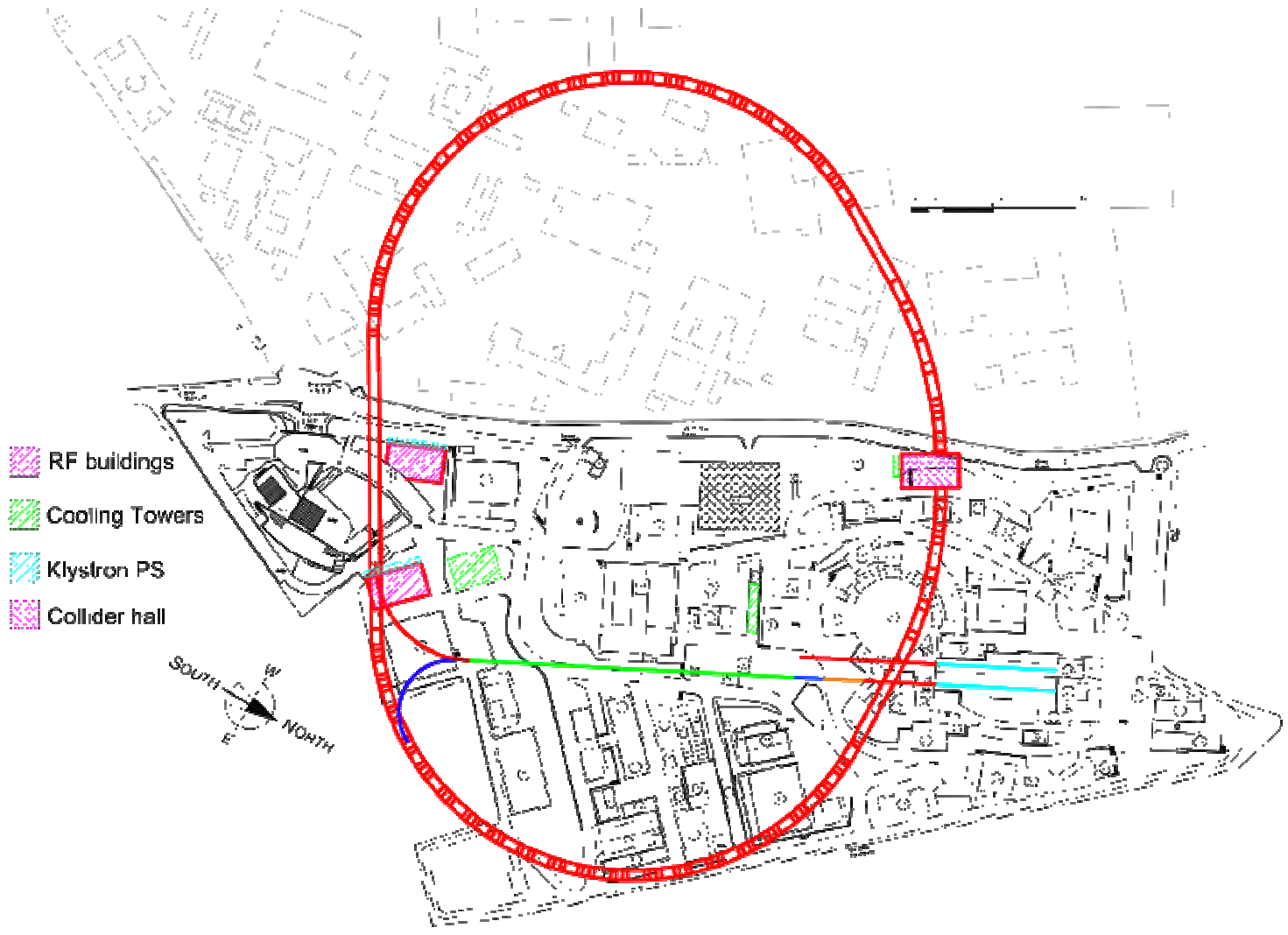

Figure 4.1: Super*B* footprint at LNF.

# 5. Interaction Region

## 5.1 Introduction

The interaction region (IR) design has to bring the two low emittance beams into and out of collision. The high luminosity of the accelerator is achieved primarily with the implementation of very small $\beta_y$* values. However, $\beta_x$* also needs to be small in order to achieve the design luminosity. These conditions are primary driving terms in the design of the IR. The final focus magnets (QD0 and QF1) must be as close as possible to the collision point in order to minimize chromatic and other higher-order aberrations from these magnet fields.

Initial designs of the IR incorporated a shared (both beams are inside) final focus quadrupole in order to get this magnet as close as possible to the interaction point (IP). However, with a non-zero crossing angle, a shared magnet invariably bends one or both of the beams. The bending can produce unwanted additional emittance because the shared magnet is quite strong even when the crossing angle is minimized (~+/-15 mrad). In addition, the bending of the outgoing beams generates significant luminosity based backgrounds for the detector.

These issues have led to an IR design with an increased crossing angle (+/-33 mrad) in order to use separate focusing elements for each beam. The QD0 magnet is now a twin design of side-by-side super-conducting quadrupoles. The magnet windings are designed so that the fringe field of the neighboring magnet can be canceled maintaining high quality quadrupole fields for both beams. Further details about the magnet design can be found in the section on the design of the final focus magnets.

## 5.2 IR design

Table 5.1 lists the accelerator parameters used to design the interaction region. The QD0 magnets are placed as close to the IP as possible while maintaining enough space between the two beams to accommodate the super-conducting cold mass and windings and a space of at least 5 mm between the cold mass and the warm bore beam pipe. The beam pipe (assumed to be 1 mm thick) must be warm because the pipe intercepts significant synchrotron radiation (SR) power from the last bend magnet.

Table 5.1: List of accelerator parameters important for the interaction region design.

| Parameter | HER (e+) | LER (e-) |
|---|---|---|
| Beam Energy (GeV) | 6.70 | 4.18 |
| Beam current (A) | 1.89 | 2.45 |
| $\beta^*_x$ (mm) | 26 | 32 |
| $\beta^*_y$ (mm) | 0.25 | 0.21 |
| Emittance X (nm-rad) | 2.00 | 2.46 |
| Emittance Y (pm-rad) | 5.00 | 6.15 |
| Crossing angle (mrad) | ±33 | |
| Beam-stay-clear | 30σ in x uncoupled and 10σ in y fully coupled | |

As shown in Table 5.1, we are using a definition for the beam-stay-clear (BSC) of 30 uncoupled beam sigmas (all of the beam emittance in the horizontal) in the X plane and 10 fully coupled beam sigmas (50% of the total emittance) in the Y plane. The BSC envelopes, the crossing angle, the space needed for the magnet windings and the space needed for the cryostat dictate how close we can place the face of the final focus magnet (QD0) to the IP. In our design, we have the QD0 face located 0.6 m from the IP. The QD0 magnets are 0.4 m long and are placed parallel to the each other and parallel to the detector magnetic field axis which evenly divides the crossing angle. In order to achieve vertical focusing of each beam as close to the IP as possible we have placed permanent magnet slices around first the LEB and then around the HEB in front of the cryostat for the QD0 and QF1 magnets. Figure 5.1 is a layout of the interaction region and Table 5.2 lists some of the magnet properties and dimensions for QD0 and QF1. Figure 5.2 shows a layout drawing of the interaction region out to 12 m.

Table 5.2: Dimensions and field strengths of QD0, QF1.

| | QD0 | | QF1 | |
|---|---|---|---|---|
| | HER | LER | HER | LER |
| Cold mass inside R (mm) | 22.5 | 32.5 | 50 | |
| Cold mass outside R (mm) | 28.5 | 38.5 | 60 | |
| Length (m) | 0.4 | | 0.3 | |
| Dist. from face to IP (m) | 0.6 | | 1.8 | |
| Gradient (T/cm) | -1.025 | -0.611 | 0.640 | 0.358 |
| Field at inside R (T) | 2.31 | 1.99 | 3.20 | 1.79 |

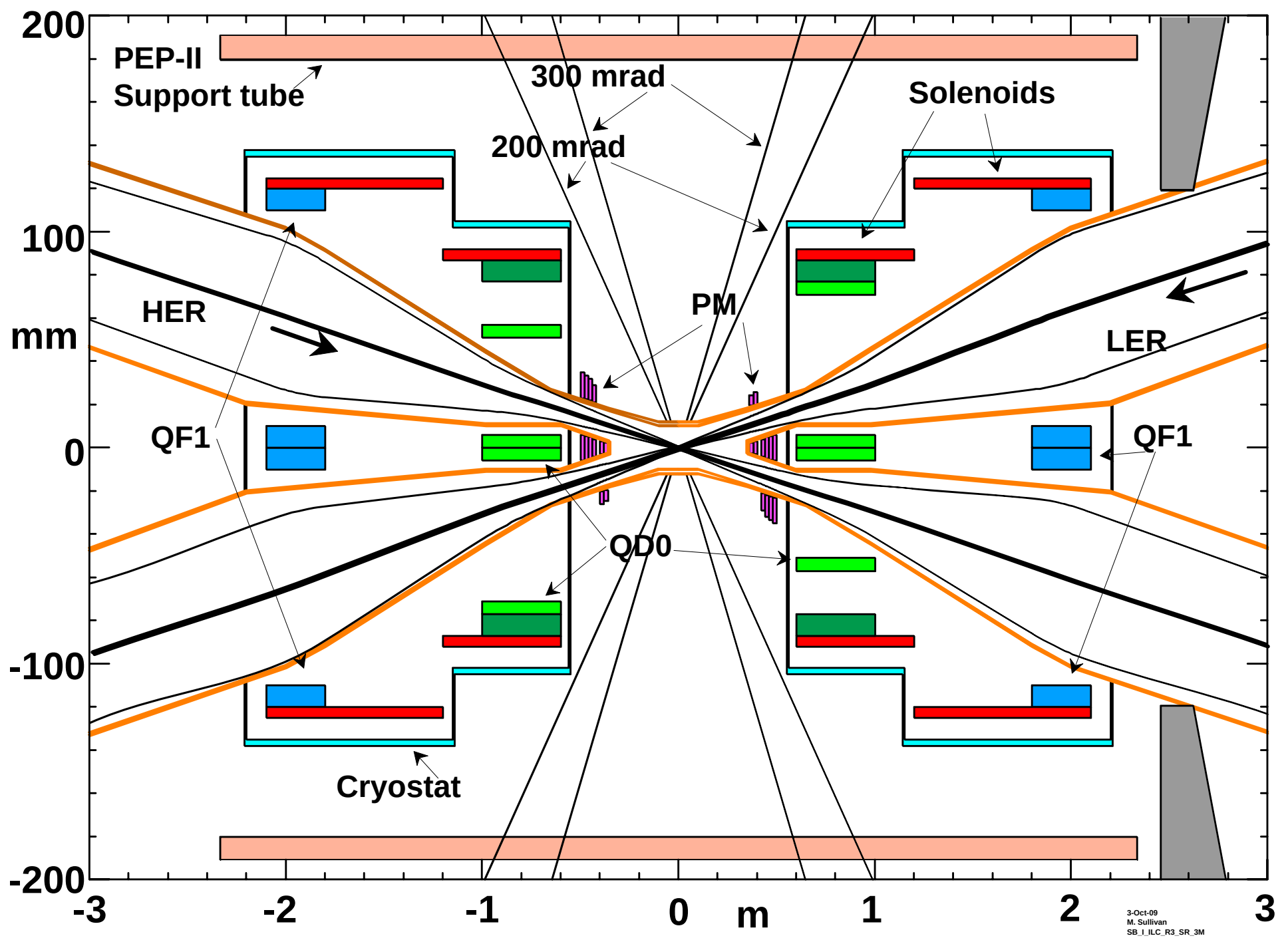

Figure 5.1. Layout of the Interaction Region. Note the change in scale for the x dimension on the left side of the drawing. The outline of the cryostat is only approximate. There is more detail concerning a cryostat design in a later section. For reference, the support tube from PEP-II is drawn in the picture as well as the forward door of the BaBar detector (the gray regions on the right). The forward door has too narrow an opening and this aperture will have to be widened if the door steel is to be reused.

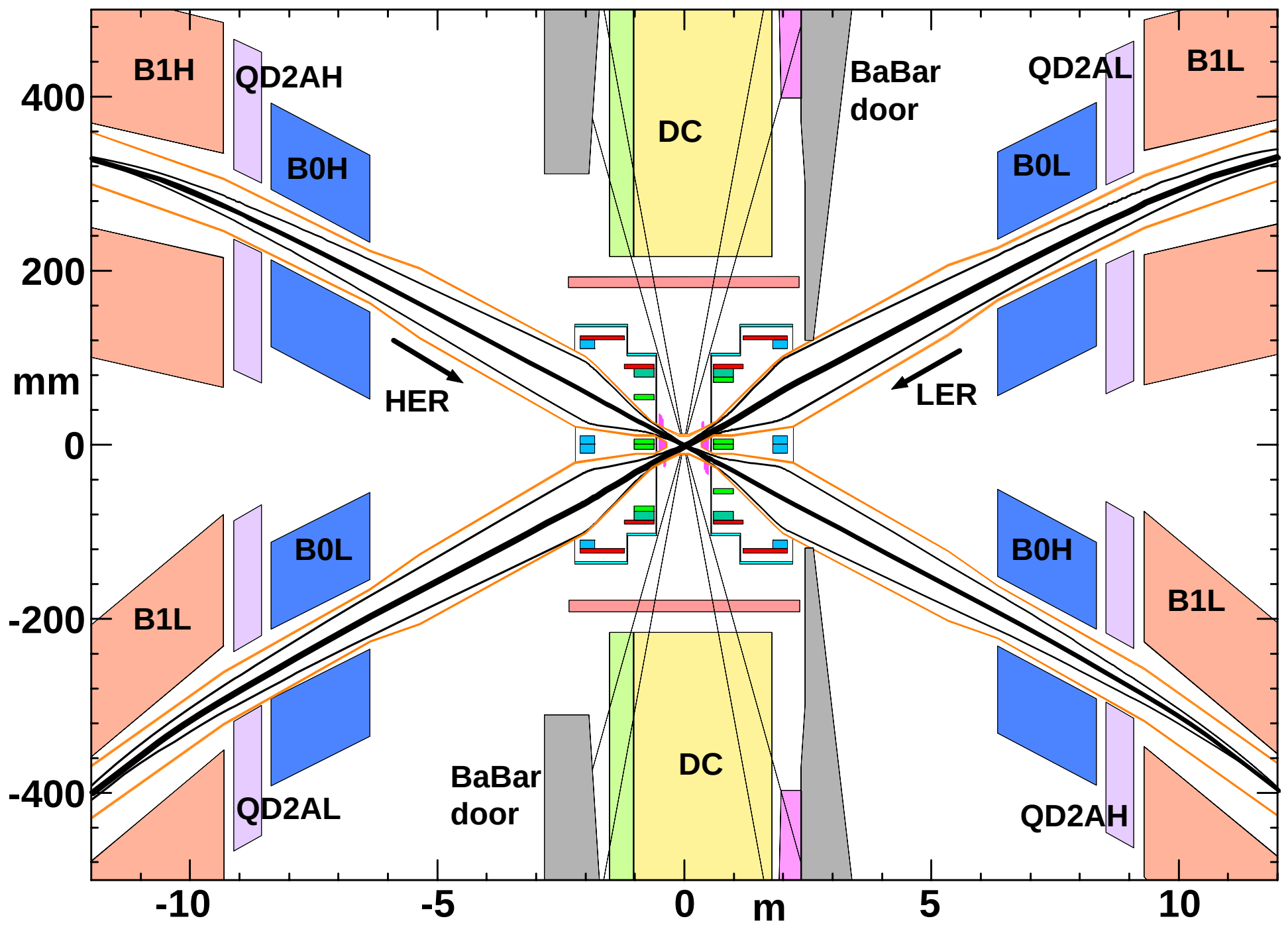

Figure 5.2. Note the expanded scale on the left side of the drawing. The B1L magnets are the last standard bend magnets. The B0L and B0H magnets are softer bending dipole magnets that reduce the synchrotron power from the beam bending inside the cryostats. The BaBar detector is drawn in to indicate the relative size of the new interaction region compared to PEP-II.

## 5.3 Permanent Magnets

In order to get focusing elements as close to the collision point as possible we have incorporated permanent magnet (PM) slices between the cryostat and the IP. These PM slices can generate significant gradient fields and yet are very compact and can fit between the two vacuum beam chambers (see Figs. 5.1 and 5.2). In our design, we are using Neodymium magnets ($Ne_2Fe_{14}B$) with a $B_r$ of 13.8 kG. In this region (0.35-0.45 m from the IP) the BSC is larger vertically than it is horizontally. Therefore we are employing an elliptical design for the magnet slices in order to get as much field strength as possible. The elliptical design [1] is an extension of the standard Halbach design [2] for building quadrupole field magnetic slices. Table 5.3 lists the dimensions and field properties of these magnetic slices.

Table 5.3: Dimensions and parameters of the permanent magnet slices.

| Slice | IP face (m) | Length (m) | Hor. R1 (mm) | Hor. R2 (mm) | Ver. R1 (mm) | Ver. R2 (mm) | Gradient T/cm | Beam |
|---|---|---|---|---|---|---|---|---|
| 1 | 0.36 | 0.2 | 8.50 | 13.50 | 12.50 | 19.85 | 0.8363 | LER |
| 2 | 0.38 | 0.2 | 8.75 | 14.75 | 13.00 | 21.91 | 0.8949 | LER |
| 3 | 0.40 | 0.2 | gap | gap | gap | gap | 0 | |
| 4 | 0.42 | 0.2 | 9.00 | 17.00 | 14.00 | 26.44 | 0.998 | HER |
| 5 | 0.44 | 0.2 | 8.50 | 18.50 | 14.50 | 31.56 | 1.0989 | HER |
| 6 | 0.46 | 0.2 | 8.75 | 19.75 | 15.00 | 33.86 | 1.0946 | HER |
| 7 | 0.48 | 0.2 | 8.75 | 20.75 | 15.50 | 36.76 | 1.0999 | HER |

## 5.4 Solenoid compensation

The cryostats on either side of the IP will contain super-conducting solenoids to compensate the detector solenoidal field. The QD0 and QF1 magnet designs need zero or very small external fields in order to keep the magnetic field intensities in an acceptable range. The overall solenoid compensation will extend out far enough to cancel the fringing field of the detector. This is important because the beams go through the fringe field of the detector significantly off-axis due to the large crossing angle and this large off-axis trajectory can produce vertical emittance [3]. The compensation solenoid windings out at the fringe field of the detector may not be super-conducting. If the field strength is low enough a normal conducting winding may be suitable. The compensation solenoid windings can be tapered to match the fringe field of the detector. This present scheme leaves the central part of the detector uncompensated since we have no room for over-compensating windings in front of the QD0 magnets. We will have to employ either extra compensating solenoids outside of the detector region or extra skew quads in the interaction region or both of these options to fully compensate the detector magnetic field and maintain the low coupling of the beams.

## 5.5 Energy changes

The SuperB accelerator is optimized for running on the upsilon 4S resonance. However, the physics program calls for some running time on the other three upsilon resonances as well as having the capability of performing an energy scan from the upsilon 4S (10.58 GeV) up to over 11 GeV center-of-mass energy ($E_{cm}$). In addition, there is a desire to be able to lower the beam energies down to the Tau-charm threshold (about 4 GeV CM). Changing the energy of either or both beams is complicated in the IR due to the coupled nature of the super-conducting final focusing quadrupoles as well as the inclusion of permanent magnets. However, by changing the beam energies of both beams so as to maintain the same magnetic field ratio in the QD0 and QF1 magnets we can adjust the strength of the coupled QD0 and QF1 magnets without affecting the field quality of these magnets. This allows us to move the CM energy of SuperB down to the other upsilon resonances as well as perform an energy scan above the 4S resonance. Table 5.4 shows solutions for the upsilon resonances. The energy scan above the 4S will have similar values of field strengths. In this case, the permanent magnet strength goes down as the beam energies increase which means we have to increase the strengths of the QD0 and QF1 magnets. We also need to run off resonance (40 MeV $E_{cm}$ below the 4S) about 10% of the running time. Although we do not explicitly show a solution here this should be quite straightforward.

Moving the accelerator $E_{cm}$ down to the Tau-charm threshold (4.07 GeV) will mean we will have to remove most, if not all, of the PM focusing slices we have installed between the cryostats. These magnets would be too strong for the significantly lower beam energies. This will require rapid access to the central part of the detector and we will discuss plans for achieving this in a following section. Once most or all of the permanent magnets are removed we can then just lower the QD0 and QF1 strengths, keeping the field ratios constant until we have the desired $E_{cm}$.

Table 5.4: Beam energy solutions for running on the other upsilon resonances. Note that we want to maintain a constant magnetic field ratio between the two QD0 and QF1 magnets. Note also that the permanent magnet K values change with the beam energy because the field gradient is constant in these magnets. The beam energies are chosen to make the QD0 and QF1 field ratios between the LER and HER constant. We maintain the same β* values at the IP and we keep the same matching conditions (beta and alpha values) at 8.456 m constant. These requirements slightly over-constrain the problem. Hence we don't perfectly keep the ratio of the field gradients constant. However, by slightly adjusting the beta function matching conditions located at 8.546 m from the IP we can easily keep the magnetic field ratios constant.

| Resonance | Upsilon 4S | Upsilon 3S | Upsilon 2S | Upsilon 1S |
|---|---|---|---|---|
| Ecm (GeV) | 10.5794 | 10.3554 | 10.0236 | 9.4609 |
| HER | | | | |
| E (GeV) | 6.694 | 6.553 | 6.343 | 5.988 |
| QD0 (T/cm) | - 0.97584 | - 0.95329 | - 0.91969 | - 0.86285 |
| QF1 (T/cm) | 0.60408 | 0.59132 | 0.57232 | 0.54019 |
| LER | | | | |
| E (GeV) | 4.18 | 4.091 | 3.96 | 3.737 |
| QD0 (T/cm) | - 0.63941 | - 0.62522 | - 0.60435 | - 0.56882 |
| QF1 (T/cm) | 0.37412 | 0.36616 | 0.35445 | 0.33450 |
| QD0 ratio | 1.52617 | 1.52472 | 1.52179 | 1.51693 |
| QF1 ratio | 1.61466 | 1.61491 | 1.61469 | 1.61490 |
| γ | 1.02785 | 1.02787 | 1.02787 | 1.02791 |
| Boost (γβ) | 0.23763 | 0.23773 | 0.23775 | 0.23793 |

## 5.6 IR vacuum chamber

The central vacuum chamber is a circular tube of beryllium with a water-cooled layer. The inside radius of the chamber is 10 mm. The central chamber is ±15 cm long with the window for the physics events defined as ±4cm. A 300 mrad angle of acceptance equals 3.3 cm of z length for a beam pipe of 1 cm radius. A flange pair with a small bellows is attached at each end of the central chamber. Outboard of the flange pair the chamber gradually widens in the x dimension as the beams diverge due to the crossing angle. At 0.35 m from the IP, the beampipe splits into two separate chambers. The chambers now become larger in the y dimension than they are in the x dimension and this is where the PM slices are installed. The beam pipe is made of copper with water cooling channels on the outside. A set of beam position monitors (bpms) are located on each beampipe just before the cryostat. As mentioned earlier, the beam pipes inside the cryostats are at room temperature. Several Watts of synchrotron power strike these chambers from the last upstream bend magnets (see below). We plan to cool these chambers either with water pipes brazed to the inside surface of the vacuum chamber or with water channels machined out of the chamber walls. These water channels would have no beam pipe vacuum to water joints. The outboard end of the cryostat has another set of vacuum flange pairs to separate the cryostat from the rest of the beam pipes connecting the IR vacuum to the rest of the ring vacuum. The chambers just outboard of the cryostats will have as much vacuum pumping as possible. These chambers will be similar to the very low pressure chamber used in the upstream part of the HEB in the PEP-II accelerator. The PEP-II chamber achieved a pressure of less than 1 nTorr at a full beam current of nearly 2A.

## 5.7 Synchrotron Radiation

Backgrounds from synchrotron radiation are an important aspect of the IR design. If not properly controlled, these backgrounds can overwhelm the detector readout system as well as damage the inner layers of the detector. The background rate from this source can jump many orders of magnitude if the masking design does not properly cover all possible beam conditions. Since it is very difficult for any design to cover all possible beam conditions, a necessary part of any design is a background rate detector that can abort the beam if the background rate gets too high. We will not discuss any further details of a background rate monitor except to say a rate monitor similar to the type used in PEP-II is envisioned.

In the SuperB design, the primary synchrotron radiation background comes from the radiation generated by the beams as they travel through the final focus magnets. We call this radiation quadrupole radiation as it is generated in the last quadrupoles before the IP. Another kind of synchrotron radiation comes from dipole magnets where the entire beam is bent. Dipole radiation has about 10 times more power than quadrupole radiation. However, the dipole radiation usually has a lower critical energy photon spectrum making it somewhat easier to control. Quadrupole radiation comes about from the focusing (or bending) of the off-axis beam particles as they travel through the quadrupole and the critical energy of this radiation tends to be significantly higher than dipole radiation.

Typically, the final focus magnets are first an X focusing magnet followed by a Y focusing magnet as the beam approaches the IP. For flat beam designs this is the preferred orientation. Round beam designs usually don't have a preference, however, the focusing trajectories are still very similar. The synchrotron radiation generated by the first magnet (the X focusing magnet) is usually the more difficult radiation to control. The reason is that for all final focusing systems the first magnet must over-focus the beam since the following magnet focuses in the other dimension and hence partially defocuses the beam in the X dimension. The requirement of over-focusing generates steeply angled beam trajectories that make it difficult to place masking that can protect the detector beam pipe without encroaching on the BSC. Figure 5.3 illustrates this issue.

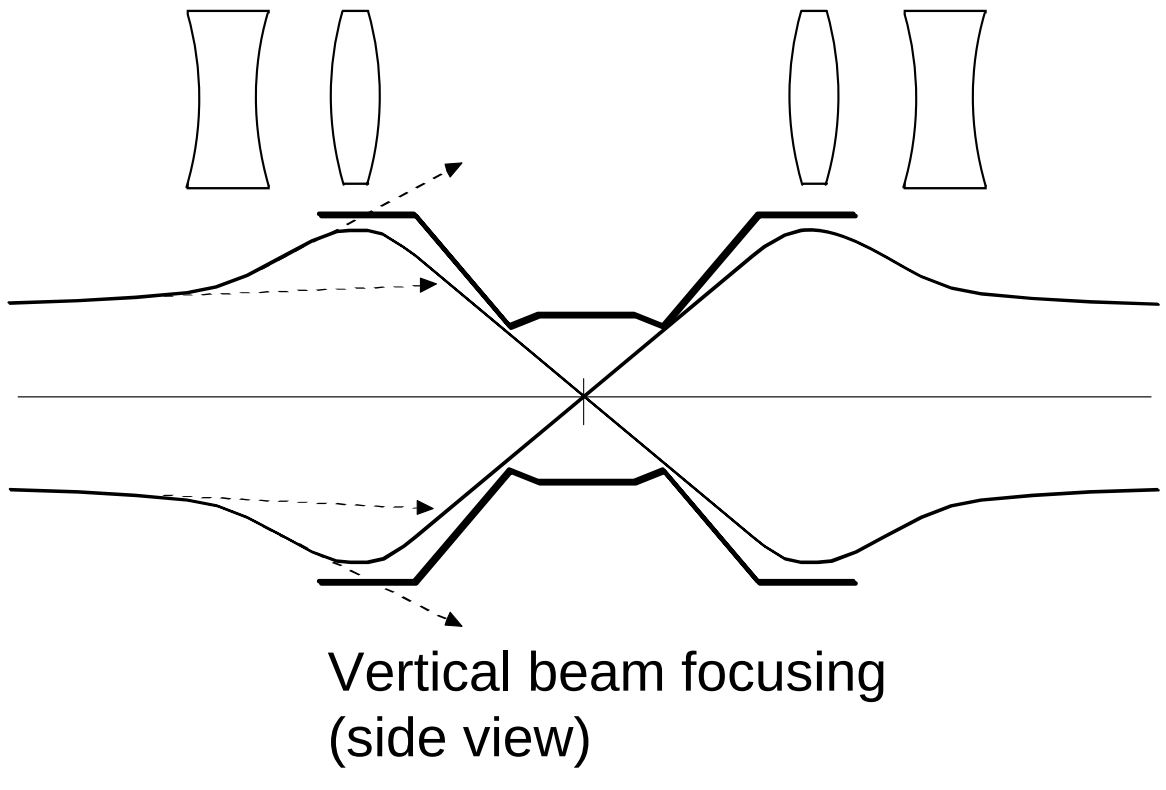

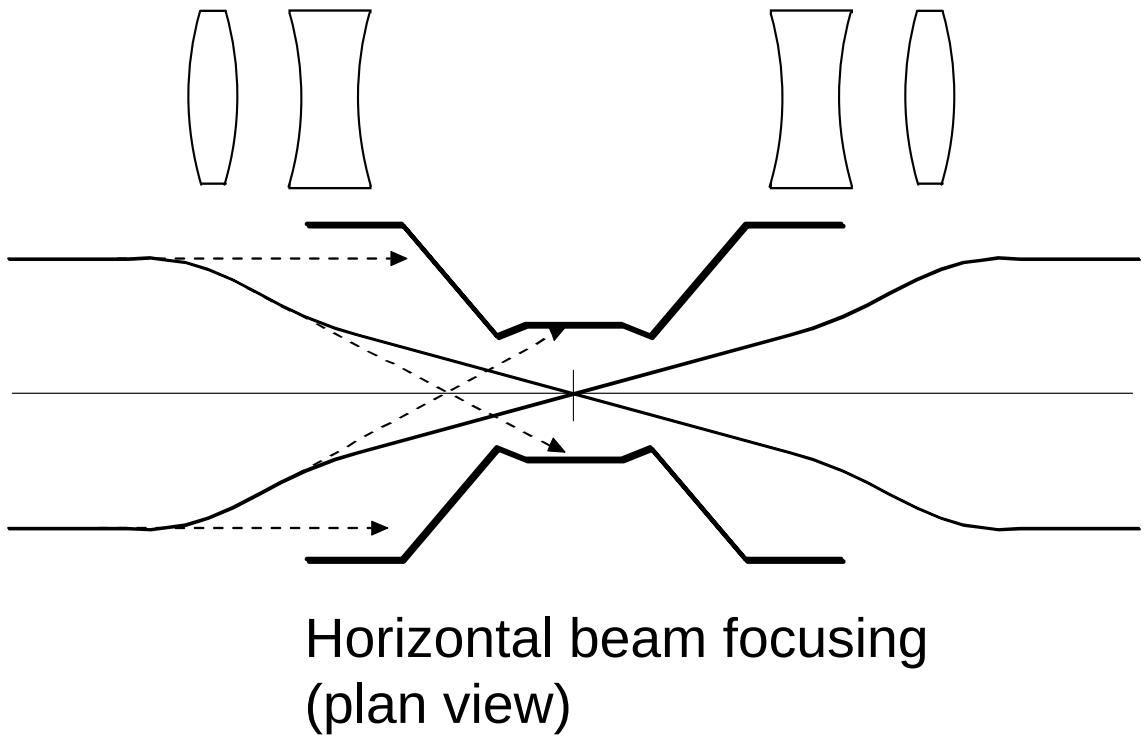

Figure 5.3. Illustration of synchrotron radiation photon trajectories from vertically focusing and horizontally focusing magnets. As seen in the first illustration, the photons produced by the vertically focusing magnet are generally easier to mask. However, the second illustration shows that the over-focused horizontal trajectories produce synchrotron photons that cross through the beam envelope and strike the other side of the detector beam pipe. As depicted in the figure, these photons are the most difficult to prevent from directly hitting the thin central beam pipe.

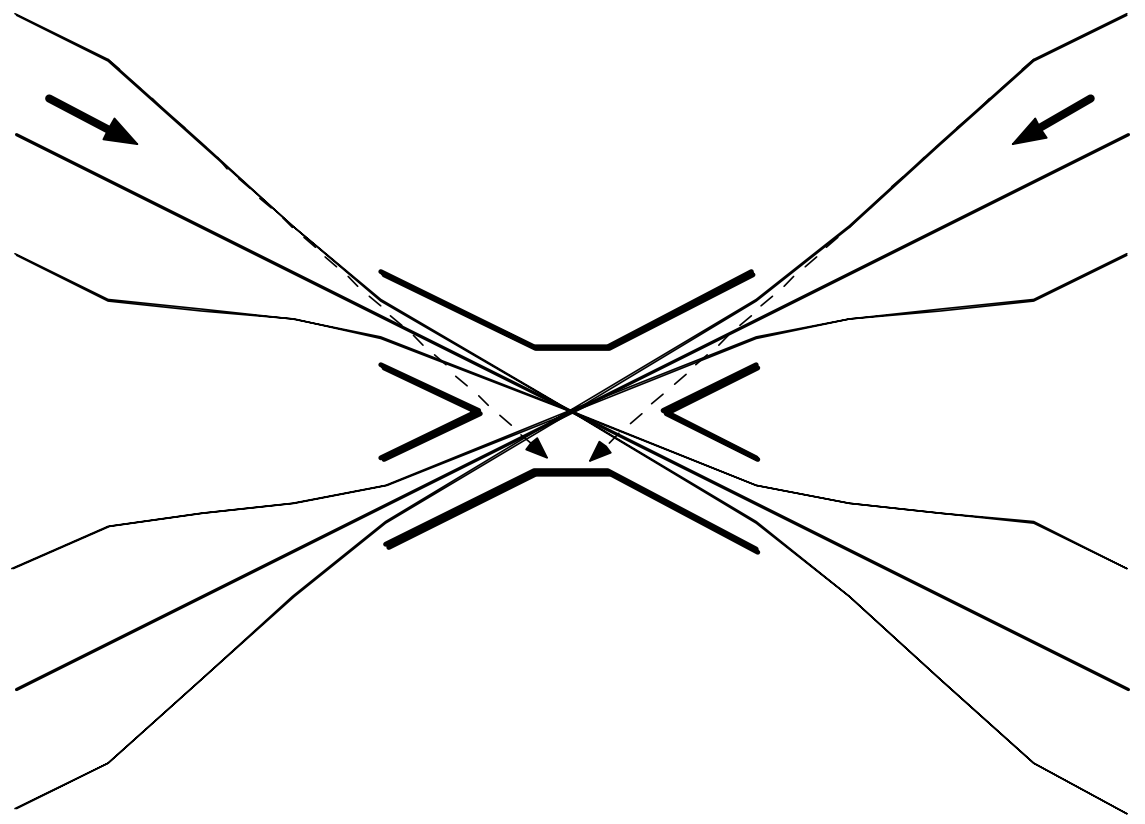

Figure 5.4a. Synchrotron radiation photon directions from a straight crossing angle geometry. The central beam pipe is directly struck by the photons generated by the extreme beam particles.

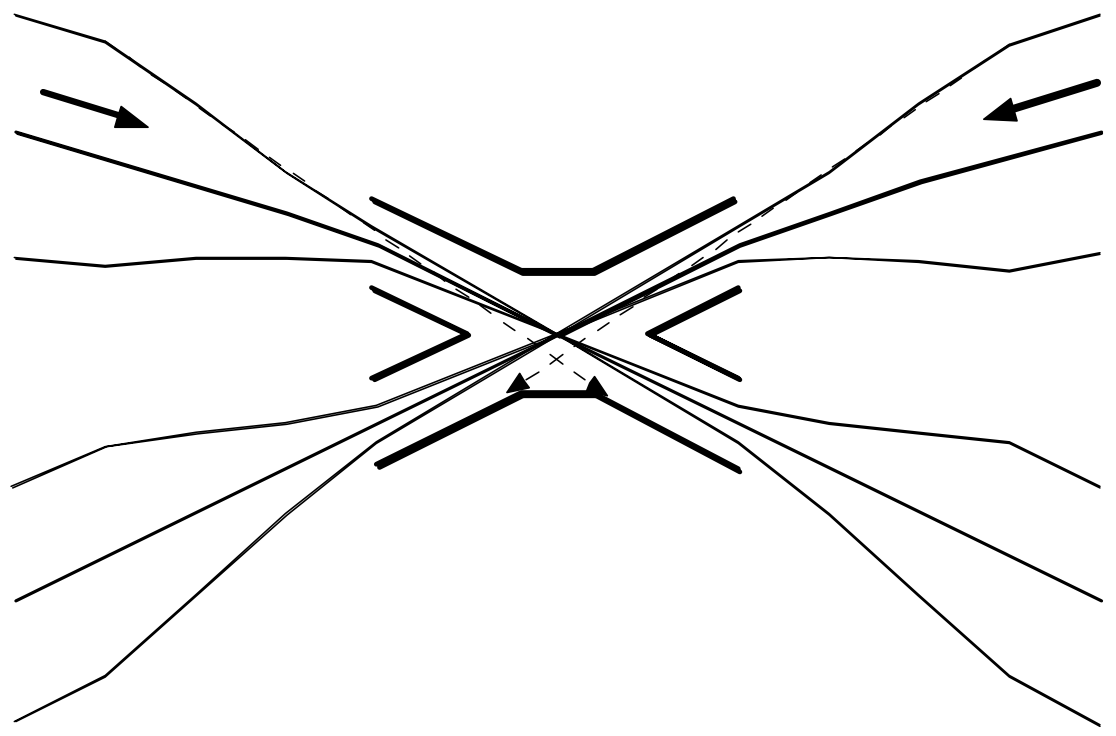

Figure 4b. Synchrotron radiation photon directions from a design where we introduce a small amount of beam bending in the QD0 magnet. Note the photon trajectories from the most extreme beam particles now miss the central beam pipe. The outgoing beams are not bent in this drawing. However, in many cases, in order to preserve symmetry the outgoing beams are also bent.

The crossing angle of the SuperB design tilts the SR from the final focus magnets so that one side of the detector beam pipe is less likely to be hit by SR. However, this geometry makes the other side of the beam pipe easier to hit. The crossing angle asymmetry can be exploited by introducing a slight bend angle in the vertical focusing magnet (in this case QD0). This redirects the SR generated by the horizontally focusing magnet (QF1) toward the side of the detector beam pipe that is easier to shield. Figures 5.4a and 5.4b illustrate this philosophy.

## 5.8 Beam tails distribution

As seen from the above illustrations the beam particle density at high beam sigma values is an important factor in determining the background rate from SR. For the SR background study, we trace the beam particles out to 20σ in X and 45σ in Y. At these high beam sigmas the background rate is dominated by the assumed non-gaussian beam tail distribution. Figure 5.5 shows a plot of the assumed tail distribution used in the SR background studies. The beam lifetime for the SuperB design is 5-10 minutes at the design luminosity. This lifetime is dominated by the luminosity or by the effects of the beams interacting with each other. This is not a beam-beam effect. This means that we should be able to collimate the beam at low beam sigma values (~10-15σ) with little impact on the lifetime. Therefore by designing the interaction region so that detector backgrounds are acceptable for beam particles out to 20σ in X and 45σ in Y we develop some margin in the design.

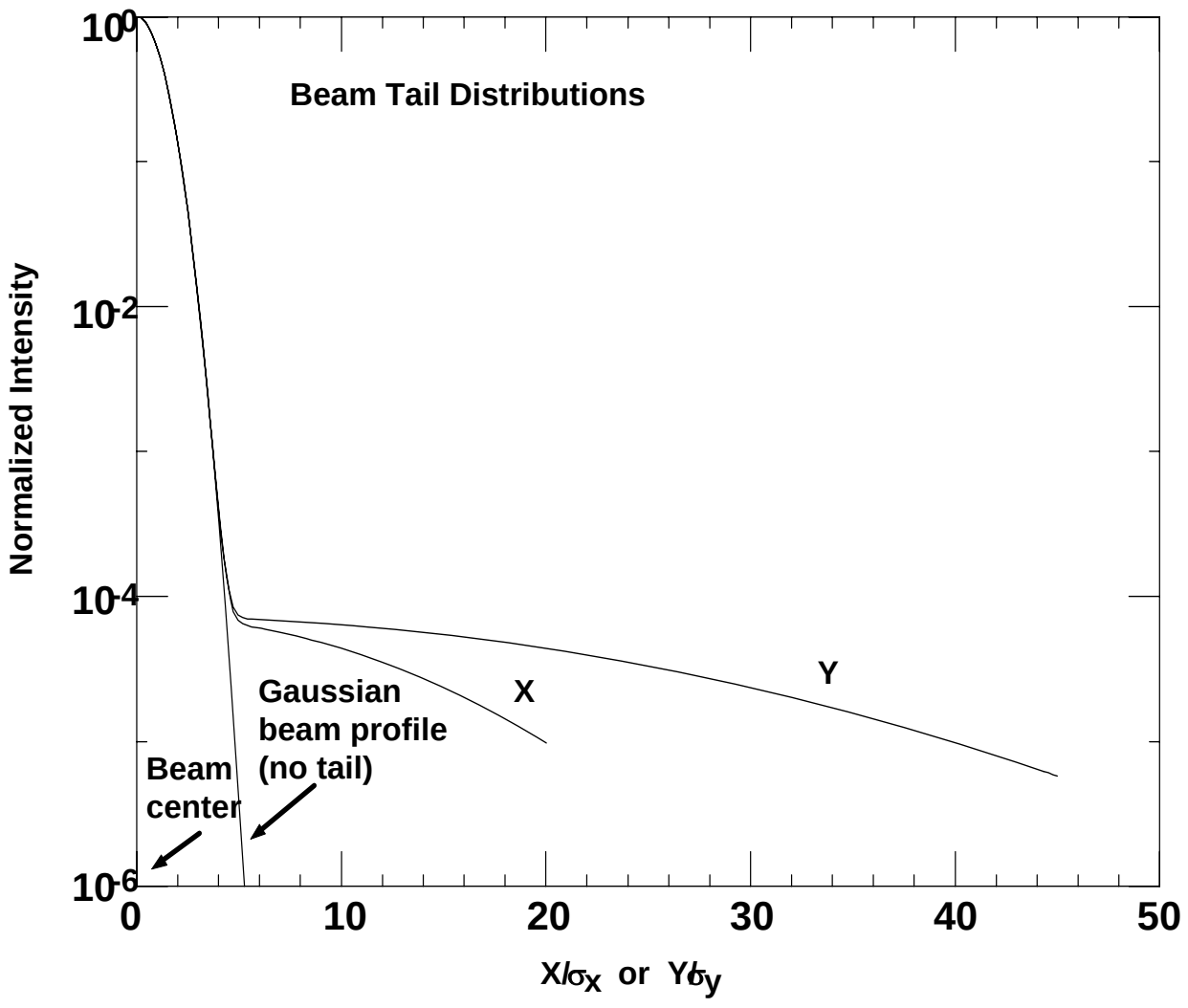

Figure 5.5. Plot of the beam tail distributions used in the SR background calculations. These tails populate beam particles out to 20σ in X and 45σ in Y with an intensity of about $10^{-5}$ of the main core of the beam. This translates to a beam tail lifetime estimate of less than one hour which means that these tail distributions are conservatively high. There is about 1-2% of the total beam bunch population in these tail distributions.

## 5.9 Synchrotron Radiation Backgrounds

Table 5.5 summarizes the synchrotron radiation rates striking various beam pipe surfaces near the central beam pipe. No SR directly strikes the physics window part of the central one cm radius beam pipe. We define the physics window as ±4cm from the collision point. The table also includes the backscatter rate from the surfaces that do intercept SR as well as the calculated solid angle acceptance from these surfaces of the physics window. Assuming the backscattered photon angular distribution is isotropic (this is a conservative estimate) we can estimate the rate of scattered photons hitting the physics window from these nearby surfaces.

Table 5.5: Summary of the photon rates from sources that produce photons that strike nearby beam pipe surfaces. We tally only photons > 10 keV. We are assuming all surfaces are Cu and that the backscatter coefficient is 0.03 of the incident rate. Coating these surfaces with a higher Z metal (Ag or Au) will significantly reduce the backscatter rate from these surfaces and we intend to do this after a more detailed study which will indicate the best coating material and the best coating thickness. The Z locations are with respect to the IP and the +Z direction is in the HEB direction.

| Surface Z loc. (m) | γ > 10 keV γ/xing | Watts | Source | Back-scatter γ/xing | Calc. SA/2π | Inc. on det. beam pipe γ/xing |
|---|---|---|---|---|---|---|
| 0.1 | 748 | 0.0016 | LER Q1,Q0 | 22 | 0.011 | 0.24 |
| -0.06 | 163 | 0.00028 | LER Q1,Q0 | 5 | 0.084 | 0.42 |
| -0.07 | 5298 | 0.0095 | LER Q1,Q0 | 159 | 0.043 | 6.8 |
| -0.1 | 1.00E4 | 0.0185 | LER Q1,Q0 | 300 | 0.011 | 3.3 |
| -0.15 | 3.92E4 | 0.0759 | LER Q1,Q0 | 1176 | 2.7E-3 | 3.2 |
| -0.2 | 1.14E5 | 0.233 | LER Q1,Q0 | 3423 | 1.1E-3 | 3.8 |
| -0.35 | 1.30E6 | 3.00 | LER Q1,Q0 | 3.90E4 | 1.95E-4 | 7.6 |
| | | | | | | |
| -0.1 | 216 | 0.00033 | HER Q1,Q0 | 6 | 0.011 | 0.66 |
| 0.06 | 1614 | 0.0022 | HER Q1,Q0 | 48 | 0.084 | 4.0 |
| 0.07 | 4.36E4 | 0.0603 | HER Q1,Q0 | 1308 | 0.043 | 56 |
| 0.1 | 7.49E5 | 1.02 | HER Q1,Q0 | 2.25E4 | 0.011 | 248 |
| 0.2 | 8.50E5 | 1.13 | HER Q1,Q0 | 2.55E4 | 1.1E-3 | 28 |
| 0.35 | 1.81E7 | 25.1 | HER Q1,Q0 | 5.43E5 | 1.95E-4 | 106 |
| Totals | | | | | | |
| γ/xing | | | | | | 364 |
| γ/sec | | | | | | 8.65E10 |

## 5.10 Final Focus quadrupoles design

The Final Focus doublets, where the beams pass each other with significant horizontal separation due to the crossing angle at the IP, need to provide pure quadrupole fields to each beam in order to minimize the background rate in the detector which would be produced by the bending of off-energy particles if a dipole component were present. Very good field quality is also required to preserve the dynamic aperture of the rings. The beam separation, though significant for a shared magnet, is still small for separate magnets. Because of this and because of the high gradient required by the SuperB final focus, neither a permanent magnet design nor a conventional multi-layer configuration are viable solutions. Therefore a novel design, with two separate super-conducting quadrupoles, one for each beam line, with helical-type windings, had been investigated. The magnet requirements are listed in Table 5.6.

The QD0 will be formed by three windings: two small quadrupoles (qq) each one winded around one of the two beam lines and a large external quadrupole (Q) embracing both of them (see Figg. 5.6 and 5.7). The internal radius of the warm bores of the qq is determined by the beam stay clear envelopes (see Fig. 5.1). The limited space (22 mm) encompassed by the two warm beam pipes is the main source of issues of this magnet since the warm to cold transition, the mechanical support of the windings and the windings of the qq themselves have to be fitted inside this very limited space (see Fig. 5.6).

The warm to cold transition can be made as small as 5mm leaving at the thinner point only 12mm for the cold mass.

Table 5.6: QD0 specifications for HER and LER part.

| Parameter | HER | LER |
|---|---|---|
| Energy (GeV) | 6.7 | 4.18 |
| Gradient (T/cm) | 1.025 | 0.611 |
| Magnetic center (mm) | 22 | -20 |
| Cold mass Internal radius (mm) | 32.5 | 22.5 |
| Front face distance from the IP (m) | 0.58 | |
| Magnetic length (m) | 0.40 | |

The magnetic design of the qq will be based on the double-helix concept that can produce a *theoretically* perfect multipole field [4] inside the whole warm inner bore of the magnet. However, since the distance between the centres of two inner magnets is of the same order of magnitude of their radius, the leaking field of the magnet surrounding the LER produces intolerably high multipolar components on-top to the LER beam line and vice versa. A novel design concept was developed [5] to eliminate this effect. The two qq magnets are designed in such a way that the superposition of the inner field of one magnet and the leaking field of the nearby one produces the desired quadrupole.

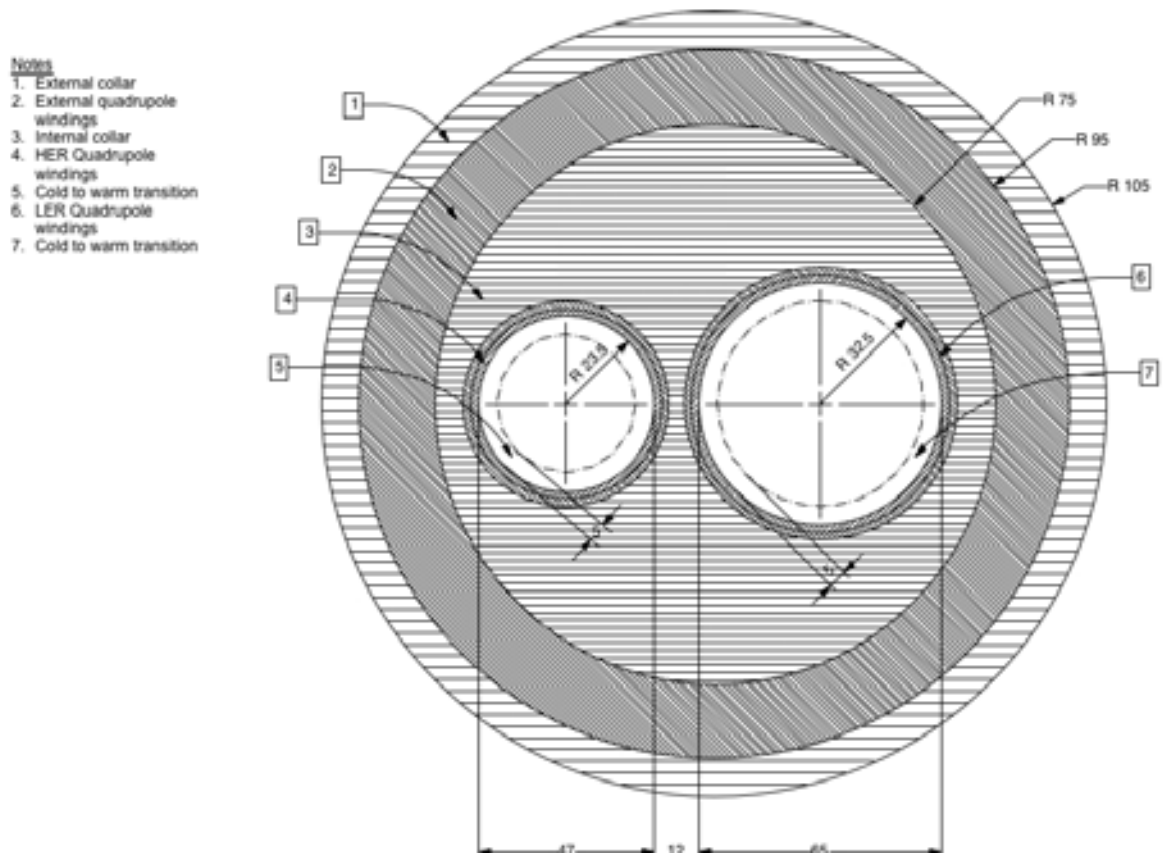

Figure 5.6: Nested quadrupoles mechanical design for the QD0 (Q&qq configuration). The dimensions are expressed in mm.

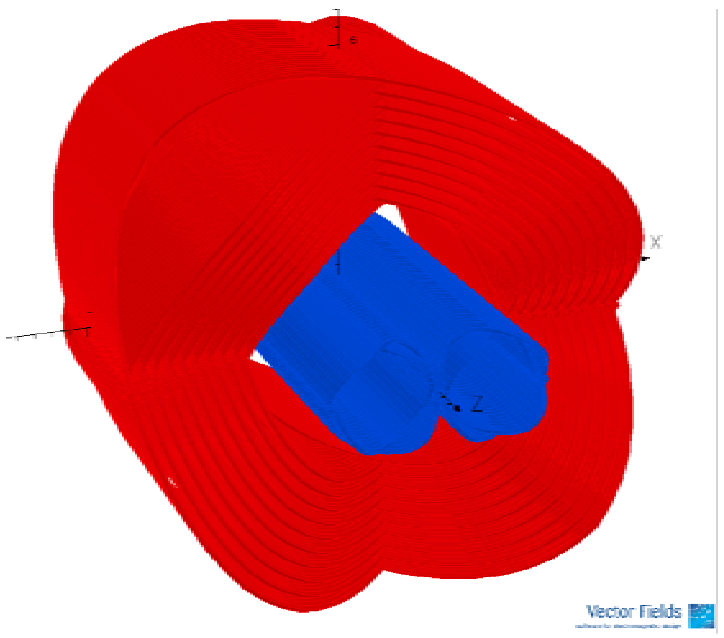

Figure 5.7: Nested quadrupoles magnetic design for the QD0 (Q&qq configuration). The outer quadrupole windings (Q) are represented in red. The *twin* quadrupoles (qq) in blue.

### Simulation and optimization

3D finite element methods [6] have been used to check the validity of the novel compensation scheme. The sextupole and octupole strengths relative to the quadrupole had been determined from the simulations at a reference radius of 5mm from the beam line showing that a field quality of $10^{-5}$ is actually reachable. The margin to quench on the load line was also estimated from the simulations. The maximum B field on the SC surface is 5.5 T at the working point. As a mater of fact, even at 1.9K the maximum current density that the available SC wires can safely carry at 5.5T is insufficient to generate with the qq alone the needed gradients over the needed bore while keeping the thickness of the magnets small enough to be fitted inside our allotted space.

To overcome this problem the external quadrupole Q generates part of the gradient relieving the load from the qq. The gradient and the neutral axis of Q had been determined minimising the current in the qq while keeping fixed to the design value the field gradients and the magnetic neutral axis. To achieve this a dipolar component had been added to the qq. The results of the optimization are reported in Tab. 5.7, the behaviour of the vertical component of the B field as a function of the radial displacement is showed in Fig. 5.8.

This configuration should assures a 20% margin on the load line using an high current grade niobium titanium round wire (Φ=1.3 mm, Cu/SC = 1).

Table 5.7: Dipole and quadrupole field generated by the Q and qq magnets.

| Parameter | HER | LER |
|---|---|---|
| Q gradient (T/cm) | 0.5 | |
| qq gradient (T/cm) | 0.525 | 0.111 |
| qq dipole (T) | 0.95 | -0.40 |

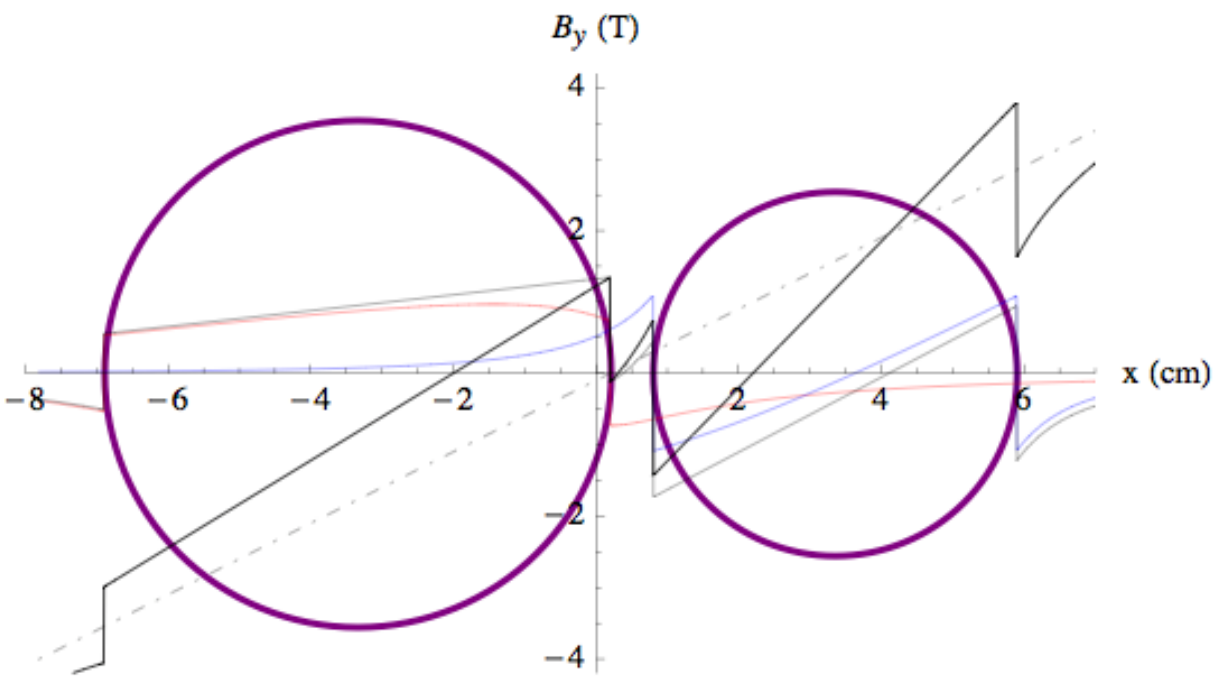

Fig 5.8: The vertical component of the magnetic field as a function of the radial coordinate. The continuos thick line is the total $B_y$ field, the red (blue) line is the $B_y$ generated by the LER (HER) magnet only, the thin black line is their superposition and the dot dashed line is the $B_y$ generated by the outer quadrupole.

## 5.11 Synchrotron radiation from dipoles

The last bending magnets upstream of the collision point will send a fan of synchrotron radiation down the beam line and into the region near the IP. In particular, for SuperB, the beam pipes inside the cryostats will receive significant levels of power from this last bending fan. The beam pipe is smallest under the permanent magnets and therefore only a minimal amount of dipole radiation gets through this restriction and into the central region at the IP. The radiation from this last bending magnet has been softened by adding a low field bend magnet in front of the last regular bend magnet. The lower magnetic field lowers the photon critical energy and reduces the total SR power that comes from the low field bend magnet. Table 5.8 shows the power levels of the SR from the last dipoles as well as the contribution from the final focus magnets that hit beam pipe surfaces within 2 m of the IP.

Table 5.8: Synchrotron radiation power numbers for beam pipe surfaces near the IP.

| Z location m from IP | Location Upstream or down | Power (W) HER | LER |
|---|---|---|---|
| 1.9-1.6 | Upstream | 59.4 | 4.6 |
| 1.6-1.5 | Upstream | 118.6 | 14.9 |
| 1.5-1.4 | Upstream | 194.7 | 20.9 |
| 1.4-13. | Upstream | 131.6 | 14.2 |
| 1.3-1.2 | Upstream | 128.4 | 13.8 |
| 1.2-1.1 | Upstream | 163.0 | 17.5 |
| 1.1-1.0 | Upstream | 87.1 | 9.1 |
| 1.0-0.9 | Upstream | 44.0 | 4.8 |
| 0.9-0.8 | Upstream | 29.7 | 3.3 |
| 0.8-0.7 | Upstream | 43.8 | 4.8 |
| 0.7-0.65 | Upstream | 29.7 | 3.3 |
| 0.65-0.625 | Upstream | 0.14 | 0.021 |
| 0.625-0.6 | Upstream | 0.10 | 0.014 |
| 0.6-0.5 | Upstream | 0.0055 | 5.9x10-4 |
| 0.5-0.35 | Upstream | 2.5 | 0.40 |
| 0.05-0.1 | Downstream | 0.041 | 0.0042 |
| 0.35 | Downstream | 0.41 | 0.18 |
| 0.6 | Downstream | 0.80 | 0.26 |
| > 0.6 | Downstream | 8744 | 691 |

## 5.12 Luminosity Feedback

The PEP-II collider employed a "fast dither" system for luminosity feedback. This system used a set of dedicated air-core Helmholtz coils around a thin stainless steel section of beam pipe to deflect the HEB beam in position (x and y) and angle (y') at the IP. Deflection was simultaneous in all three dimensions, at three separate frequencies near 100 Hz. Lock-in detection of the luminosity signal allowed separation of the three components and calculation of beam steering corrections. The corrections were applied to standard DC correctors and the beam could be corrected at rate of 1 Hz [8].

SuperB has much smaller beam sizes at the IP and thus presents much more stringent requirements on beam alignment. We plan to use a system similar to PEP-II, but dithering the more easily deflected LEB beam and operating with approximately an order of magnitude higher bandwidth. Dither frequencies will be near 1 kHz, which will allow beam correction at about 100 Hz.

We need to dither x, y and y' at the IP. For this, we desire dither coils for both the x and y planes at a location near the IP where $\sqrt{\beta}\sin\psi$ is large, and another set of coils near the IP at a location where $\sqrt{\beta}\cos\psi$ is large. We plan to place an x and a y coil set as close to the IP as reasonable (about 3.5 m from the IP, just outside of the detector solenoid field), and a second coil set between the final two bend magnets (B1), preferably between the quad (QD2) and sextupole (SDM2) which are between these two bend magnets (see Fig. 5.14).

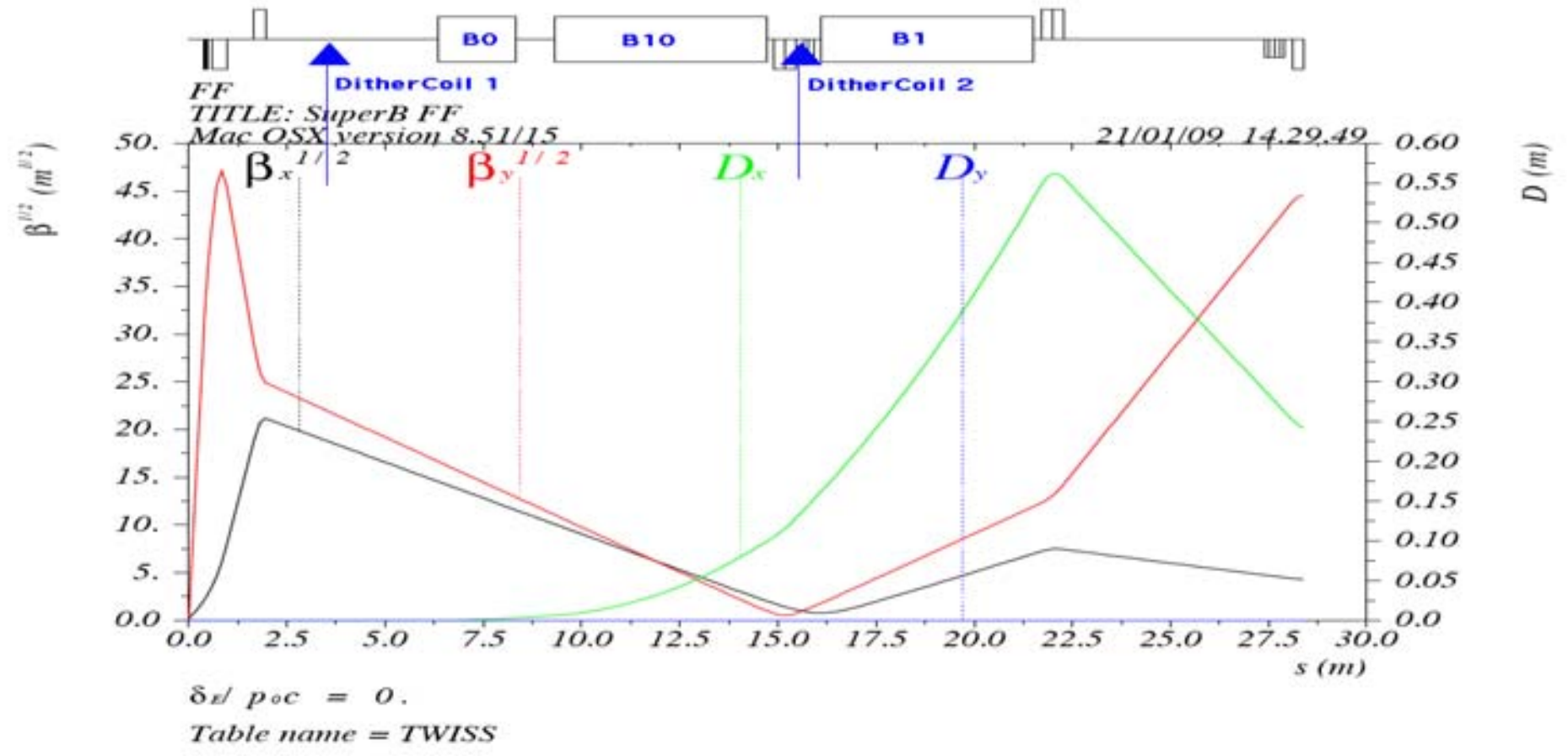

Figure 5.14: Dither coil locations in SuperB LEB lattice

Under normal operation, the dither amplitude should be large enough to be detectable but small enough to have minimal impact on the luminosity. A luminosity modulation of about 1% worked well for PEP-II; this requires shifting the beam by about 0.2 σ. Idealized coil excitations are shown in Table 5.9; coupling will mix these excitations together to some extent. Larger dither amplitude will be helpful during commissioning and for machine studies.

Table 5.9: Dither coil excitations for a shift of 0.2 σ, giving a 1% luminosity reduction

| **Parameter** | **Deflection** | **Coil 1 Excitation** | **Coil 2 Excitation** |
|---|---|---|---|
| X | 2 μm | 7 G-cm | 7.4 G-cm |
| Y | 8 nm | 0.36 G-cm | -0.06 G-cm |
| Y’ | 200 μrad | -2 G-cm | -60 G-cm |

The beam pipe needs to provide good conductivity for beam HOMs, but poor shielding at dither frequencies. Assuming a 5 mm bunch, the inside of the beam pipe should provide multiple skin depths at frequencies above about 7 GHz. The beam pipe conductivity should be poor enough that induced eddy currents at dither frequencies do not induce phase shifts of more than a few degrees.

A ceramic pipe with a 1-2 μm Cu coating fits these requirements well. The skin depth of Cu is about 2 μm at 1 GHz. The phase shift induced in a 5 cm diameter, 1 μm thick Cu pipe at 1 kHz is about 0.2 degrees. The electrical resistance of this pipe is about 0.1 Ω/m, causing about 0.4 W/m power dissipation with a 2 A beam.

Dither coils will be curved “saddle” coils with a $\cos\theta$ current distribution and an outer ferrite cylinder to act as a shield and flux return. This design is similar to CRT deflection coils, and provides much better shielding and efficiency than the open Helmholtz coil design used in PEP-II. Estimated coil parameters are about 1 Ω, 2 mH, and 10 cm length, with either a 9 cm ID (coil 1) or a 5 cm ID (coil 2). Coil sensitivities will be about 50 and 150 G-cm per amp for coils 1 and 2 respectively, with maximum currents of about 2 A and nominal currents (for deflections in Table 5.9) of less than 400 mA. Custom coil sets with these parameters can be purchased from commercial vendors for about $1200 US per coil location.

The corrections will be divided between a slow and a fast component. The slow corrections (slower than about 1 Hz) will be made through normal dipole correctors. Faster corrections will be made through the dither coils themselves, as the dipole correctors will not pass these frequencies. The coil design described above allows enough headroom for these corrections.

The fast dither coils have other applications in addition to fast luminosity feedback. Their high frequency capability will provide a useful diagnostic for identifying sources of electrical noise. They can also be used to scan or raster the beam at larger amplitudes to find collisions. The coil design described above will allow rapid scanning of the collision point by about 25 μm in x and 2 μm in y in just a few milliseconds. A larger search range can be achieved by superposing a slower scan with normal beam correctors.

## 5.13 Fast IP luminosity feedback

The "Dither" feedback is not the only option that can be considered for implementing a luminosity feedback system at the Interaction Point. Another approach, the fast IP feedback, can be implemented in parallel or, eventually in alternative, to the first system. This second project is based on a completely different design and it is inspired to the IntraTrain IP feedback proposed for the ILC (International Linear Collider) by Phil Burrows (Oxford Un.) and presented at PAC07, in the poster: "The FONT4 ILC Intra-Train Beam-Based Digital Feedback". The fast IP luminosity feedback consists basically of an orbit feedback that can work at the IP separately in each of the x, y and y' (or x') planes at much higher response frequencies than the "standard" orbit feedbacks. Furthermore these latter systems should be not operative in the collision area to avoid instability and conflicts between the feedbacks of the two rings.

As well known, SuperB rings are specified to operate at ultra-low vertical emittances and to make stable collisions between beams with vertical sigma* of the order of ~20 nm. Of course problems to a perfect stable match at the IP can easily come from two main reasons:

a) mechanical vibrations from seismic sources or vehicular traffic around the collider buildings;

b) ripples coming from the electromagnet power supplies.

Both cases can produce "slow" shift or drift of the beams decreasing luminosity.

The key components of the Fast IP Feedback system are:

- beam position monitors (BPM) to pickup the beam position, two BPM in case of angle correction;
- signal adapting analog front end circuits with remote gain control;
- analog to digital conversion (ADC) at 12 or 14 bits, using an ADC with differential input;
- DSP's (Digital Signal Processors) inside one FPGA (Field Programmable Gate Array) to produce a position correction output from the raw BPM signals;
- operator interface remotely controlled;
- digital to analog conversion (DAC) at 16 bits;
- amplifiers to provide the required output signal levels;
- kickers for applying position (or angle) correction to the beam;
- fast IP feedback is foreseen to work with a propagation delay < 150ns and should reasonably run at ~ 5-10 MHz depending on the implemented algorithm.

If necessary the use of DSP's will allow the downloading of more sophisticated algorithms that can be optimized for possible beam jitter scenarios at the IP. The Fast IP feedback can be implemented for the SuperB by the same kind of digital processing unit that will be used for the bunch by bunch transverse feedback, but with different programs inside. Using the same hardware will be an advantage from both economical and maintenance points of view.

In case of angle correction, the ADC dual inputs can be used to take input signals from two different pickups and the adapting analog electronic stages. In principle, the fast IP feedback could work in bunch by bunch mode but, if used in this way, it should wait one revolution turn to kick the beam, becoming much slower, and it will not replace the transverse (betatron) bunch by bunch feedback systems, that uses different algorithms and power amplifiers, so this hypothesis is not very attractive. Of course the Fast IP feedback should be installed as close as possible to the IP and at least two implementation schemes should be considered. In the first implementation, the input signal comes from a beam and the output goes to correct the same beam, in the second scheme, maybe more interesting, the feedback works as a "follower", acquiring the signal from both the beams 1 and 2 and correcting only the beam 2 to follow the first one in the horizontal or in the vertical plane. Sophisticated algorithms can be studied to have correction signals compatible with very low vertical beam emittances and dimensions, avoiding to feeding noise to the beams, whilst no direct connection with the luminosity monitor is foreseen. Practical problems could arise from the space needed to allocate the kickers (~20 cm striplines plus the tapering cones) near to the collision area. To save space, kicker could be designed with four striplines in horizontal-vertical or in diagonal position to serve for more planes.

In conclusion, as recommendation, at least one fast IP feedback system, working in the LER vertical plane and taking input signals as difference from both the two beam vertical positions with the goal to make the LEB following the HEB, should be foreseen in the SuperB design to cope with the uncertainty due to the trajectory drifts of so small beams.

## 5.14 Machine Detector Interface

The machine detector interface (MDI) is defined on one side by the goals of the Super*B* Physics program[9] and on the other side by the feasibility and operability of the machine and of the detectors. The angular coverage of the detector must be greater than 95.5% in the laboratory centre of mass frame and the resolution on the proper time of decay of the B mesons must match or exceed the BaBar value in spite of reducing the Lorentz boost of the centre of mass frame. The amount of material present inside the detector acceptance must be kept at a minimum to preserve the performance of the vertex, tracking and calorimetric devices. The MDI design must also take into account the space needed for the ancillary sub detector services (mechanical support, read out electronics, cooling, monitoring, etc.) that necessarily will be placed outside the angular coverage of the Super*B* detector.

A rough sketch of the space left for the machine elements nearby the Interaction Point is presented in Fig. 5.15. The hatched region is the cross section of the volume available for the machine elements and their mechanical support. The requested detector acceptance and the inner support tube of the drift chamber are represented by dashed lines.

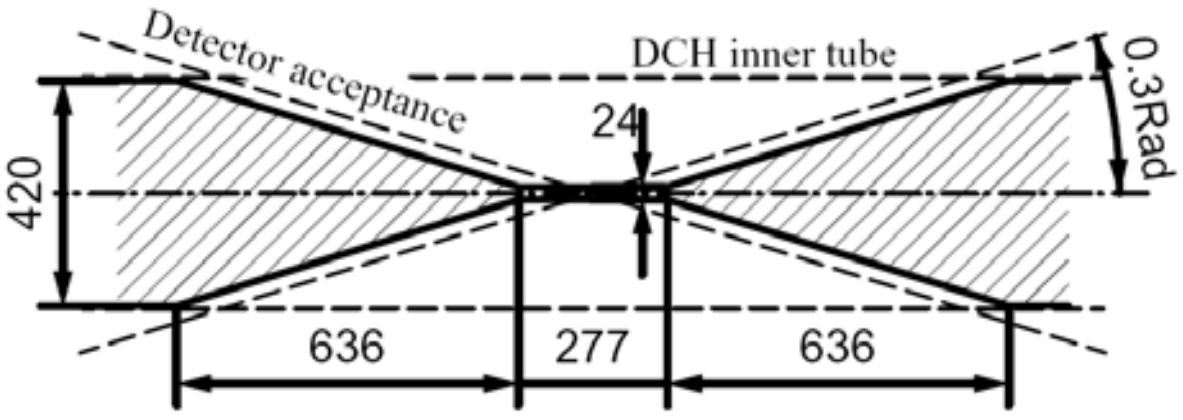

Figure 5.15: Layout of the MDI regions. The hatched region is the cross section of the space available for the machine elements. The detector angular acceptance and the DCH inner tube is also sketched. Linear dimensions are expressed in mm.

**Proper time measurements**

An important part of the Super*B* research program is represented by the measurement of the time dependent CP asymmetries. All these measurements require a very accurate determination of the longitudinal distance $\Delta z$ among the decay vertices of the two short lived *B* mesons from which their decay proper time $\Delta\tau$ is extracted.

Approximately $\Delta\tau \sim \Delta z/\beta\gamma$ , where $\beta\gamma$ is the Lorentz factor of the Y(4S) in the laboratory reference frame.

The resolution on $\Delta\tau$ is worsened by the Super*B* reduced boost, hence, to achieve equal or better performance with respect to BaBar a better resolution on $\Delta z$ is required. At the B factories this resolution is dominated by the multiple Coulomb scattering that affects the decay products traversing the material of the beam pipe and of the tracking devices.

It is, in this respect, of uttermost importance to reduce the amount of material of the beam pipe in the angular acceptance window of the detector and to reduce its radius in order to place the first layer of the Silicon Vertex Detector (SVT L0) as close as possible to the IP in order to minimize both the multiple scattering mean angular deflection and its lever arm.

Monte Carlo simulations of physics events indicate that the BaBar performance level can be reached by a light Silicon Vertex Tracker whose inner layer (L0) thickness is halved with respect to the BaBar vertex layers and by a beryllium beam pipe (outer radius ~1 cm, thickness ~1.5 mm) with an inner 4μm gold coating and a water jacket for cooling purposes (total radiation length order of 0.5% $X_0$).

Preliminary background studies indicate that the dominant source of background for the SVT L0 is two-photon pair production occurring at the IP. The total cross section of this process evaluated by the DIAG36 Monte Carlo generator is 7.3 mb. Only a small fraction of these events are seen by the Super*B* detector: most of the particles produced by this process are soft enough to be confined by the detector solenoidal field inside the beam pipe so to escape unseen in the very forward/backward acceptance hole of the SVT L0.

Assuming a detector solenoidal field of 1.5T and an inner L0 radius of 1.5 cm the production rate of particles impinging on the detector is of the order of 8MHz/cm$^2$. This rate was evaluated using DIAG36 [10] to generate the primary particles and an ad-hoc developed Geant4 based program to simulate their interactions with the machine and the detector material.

The rate of this background quickly increases as the inner vertex detector approaches the IP limiting the smallest L0 radius and hence the minimum machine energy asymmetry tolerable while keeping the time dependent CP asymmetries measurements in the Super*B* research program.

The limit is mainly set by the bandwidth of the read out logic and by the overall time resolution and dead time provided by the detector and its front end electronics.

The baseline configuration is a L0 based on state of the art thin silicon striplet sensors mounted in a 13cm long barrel configuration placed around the beam line.

The simulations suggest limiting to 1.9cm the inner radius of this barrel to keep the background rate at reasonably low level. This limit can be lowered to 1.5 cm using the presently under development thin pixel silicon sensors.

**Radiative Bhabha**

In addition to all of the well known and dangerous background sources that scale with the beam currents and have been seen in the high current colliders such as PEP-II and KEKB, the Super*B* detector will be exposed to major threats arising from the factor hundred increase in the luminosity.

Off energy particles are produced at the IP via the radiative Bhabha scattering reaction with a rate directly proportional to the machine luminosity.

The design of the downstream part of the final focus must guarantee an almost loss-less transport of these off energy particles to the outside of the detector volume in order to prevent debris from the generation of electromagnetic showers from reaching the detector.

The design of the interaction region proposed in the CDR [9] was based on a permanent magnet quadrupole shared among the HER and LER. The magnetic axis of this quadrupole was shifted toward the upstream incoming beam to reduce the radiation dose on the SVT L0. The dispersion generated by the dipole component seen by the downstream outgoing beam over-steered the off-energy particles making most of them collide with the vacuum chamber walls near the IP and inside the detector. Secondary particles produced by the electromagnetic showers were absorbed by a very thick (order of 12 cm) tungsten hollow cylinder containing the beam line to shield the detector against this background.

The problem is eased in the present IR design by the super-conducting double QD0 that provides pure quadrupole fields for each beam line. The dispersion on the downstream doublet, hence the off-energy particle loss rate, is greatly reduced.

The effect of this background is evaluated with a Monte Carlo simulation. Primaries which are off-energy particles from the radiative Bhabha scattering are generated by the BBBrem [11] package. Their transport in the final focus magnets and their interaction with the vacuum chamber wall and with the detector is simulated with a Geant4 based program. Detector occupancies and radiation damage has been evaluated. Even the double QD0 IR design will need a 3 cm thick tungsten shield ( equivalent to 8.5 $X_0$ at normal incidence) to ensure a reasonably low occupancy in the Drift Chamber detector.

## 5.15 Assembly and Rapid Access

The cryostats of the final focus magnets (QD0 and QF1) reside entirely inside the detector magnetic field. The outer dimensions of the cryostats are designed to be outside of the 300 mrad acceptance needed for the detector. However, the detector wants to get as close to the outer radius of the cryostats as possible. This means the inner radius of the drift chamber is as small as possible, minimizing the space between the outer radius of the cryostat and the drift chamber. The anti-solenoids in the cryostat will exert a large expulsion force once they are energized. We also want access to the central region between the two cryostats. A question of initial assembly also arises. In order to solve all of these requirements we have developed a concept of assembly and rapid access as well as machine component support which we describe below.

Each cryostat will have a strong rigid support that comes up from the ground on either side of the detector and reaches in through the detector doors to support the cryostat. These two supports will be tied together underneath the detector with some sort of solid connection. For the backward cryostat the support is cantilevered enough so that this cryostat can be pushed through the entire detector. This gives us access to the central section between the cryostats. The central section then is composed of a central vacuum chamber that contains the Be section for the physics window. The central beam pipe is bolted to the beam pipes coming from the two cryostats using a flange pair. There are also small bellows at each flange pair joint to eliminate any stress on the central Be chamber. Assembly is accomplished by sliding the backward cryostat in through the detector and then bringing up the forward cryostat and bolting the two cryostats together with the central vacuum chamber. Figures 5.16, 5.17 and 5.18 are a sequence of layouts depicting the rapid access scenario.

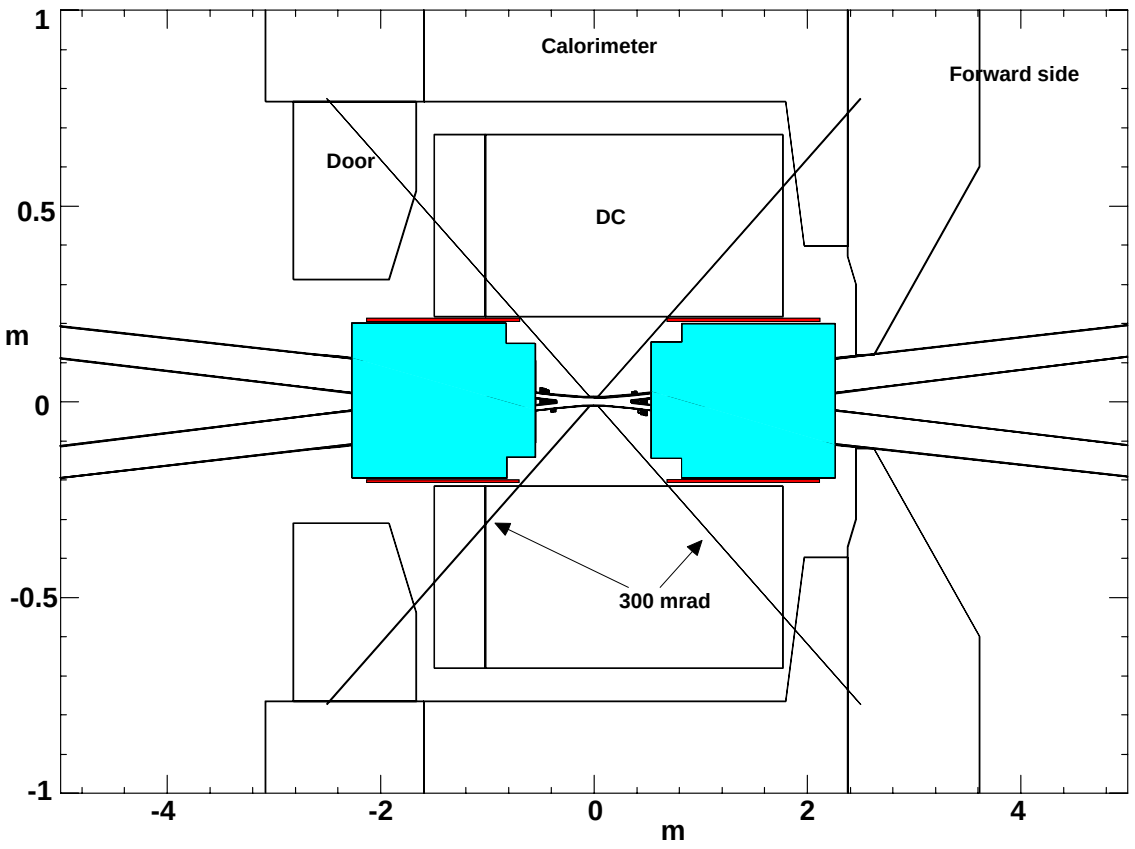

Figure 5.16: Layout of the interaction region Note the exaggerated vertical scale depicting the horizontal plane.

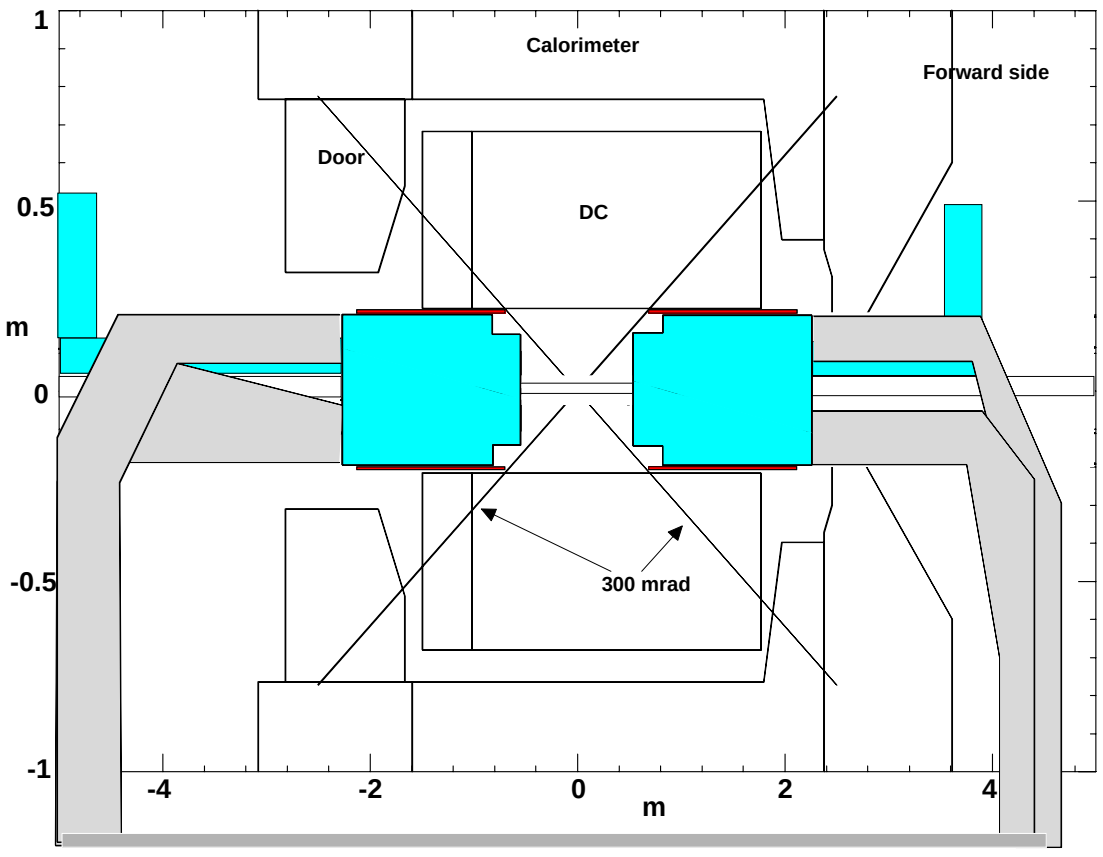

Figure 5.17: Side view with drawings of cryostat supports.

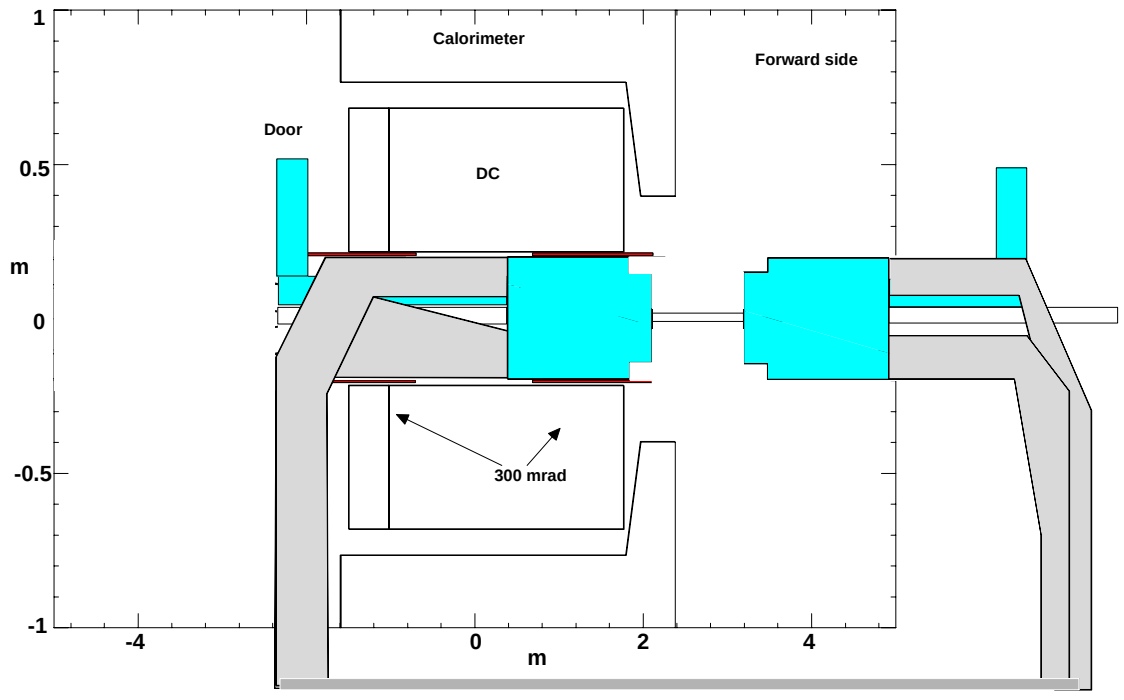

Figure 5.18: Cryostat supports placed on rails and slid to the forward side of the detector for access to the central region. The cryostats can remain cold and connected up. The SVT can also remain connected up.

## 5.16 IR Magnet Cryogenic System

### *Design Goals*

The cryogenic system for the SuperB IR superconducting magnets should meet a number of goals. These are:

1. Reliable and safe operation of the magnets
2. Meet the space requirements of the detector system
3. Allow the use of warm beam tubes
4. Ability to cool down and warm up the magnets independent of the detector solenoid
5. Ability to move the magnet system to allow access to the vertex detector
6. Use of the excess capacity of the Babar refrigerator/liquefier for magnet cooling

### System Description

In order to save space, reduce heat leak and simplify the design, the superconducting IR magnets are grouped into 2 separate helium vessels where they are bath cooled by pressurized He II at 1.9 K. Figure 5.19 is a preliminary Piping and Instrumentation Diagram (P&ID) that shows one side of the IR magnet set. An essentially identical layout would exist on the other side of the IP. The two helium vessels are contained within a single 80 K thermal shield and vacuum vessel. This is in turn attached via a vacuum jacketed transfer line to an interconnect box that sits outside the outer boundaries of the detector.

Bath cooling by pressurized He II was chosen to allow for higher performance magnets and to take advantage of the mechanism of internal convection heat transfer found in He II. This permits the transfer of large amounts of heat with out the need for boiling or forced flow.

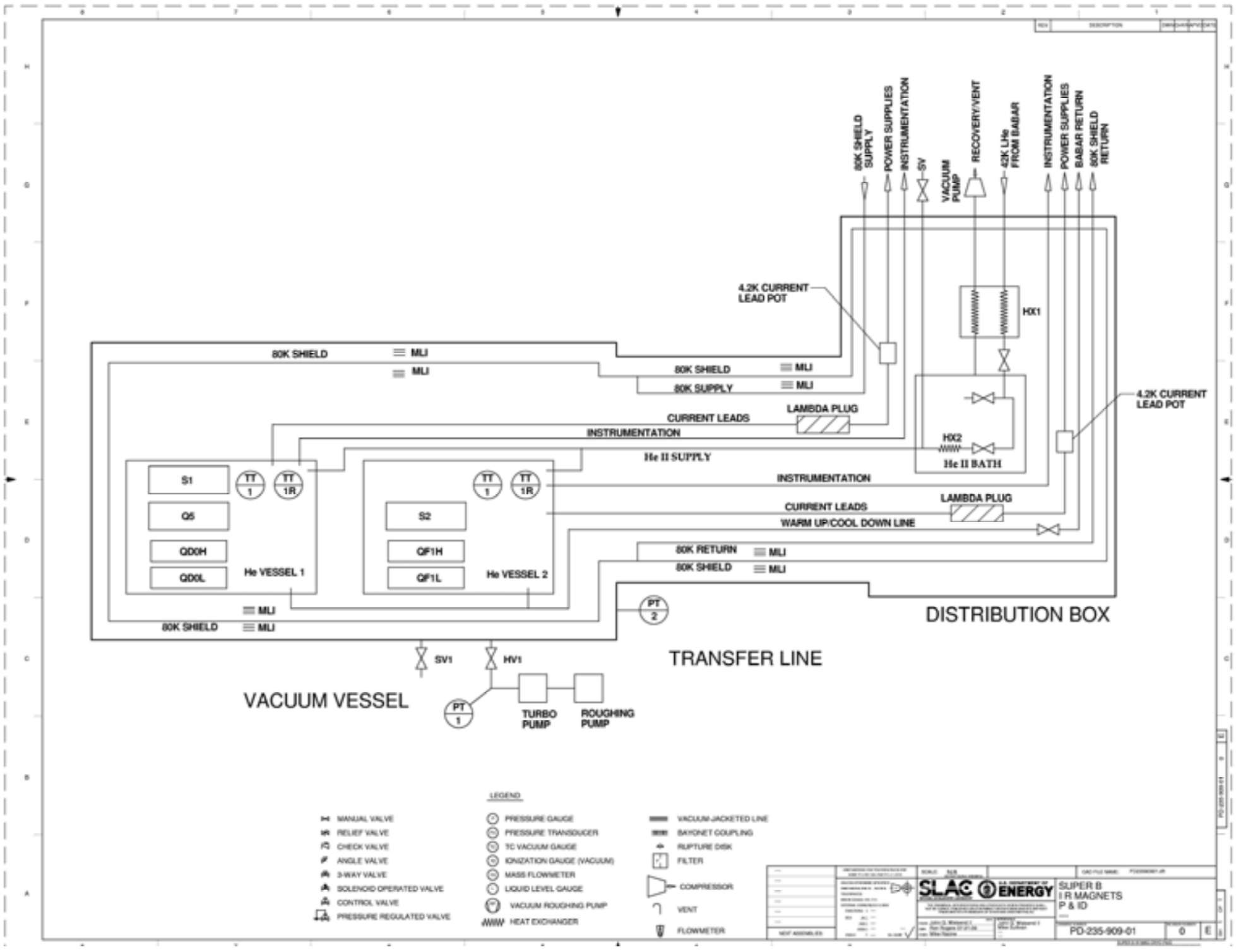

Figure 5.19: IR Magnet Cryogenic System Piping & Instrumentation Diagram

Fig. 5.20 is a schematic that shows how the system would permit access to the vertex detector while the magnets are still cold (though not powered). The concept is that the magnets, transfer line, distribution boxes and vertex detector would be mounted on a sliding support and that the length of the transfer line on one side would be large enough to allow the indicated translation. The distribution boxes are envisioned to be connected to the refrigeration plant by flexible transfer lines.

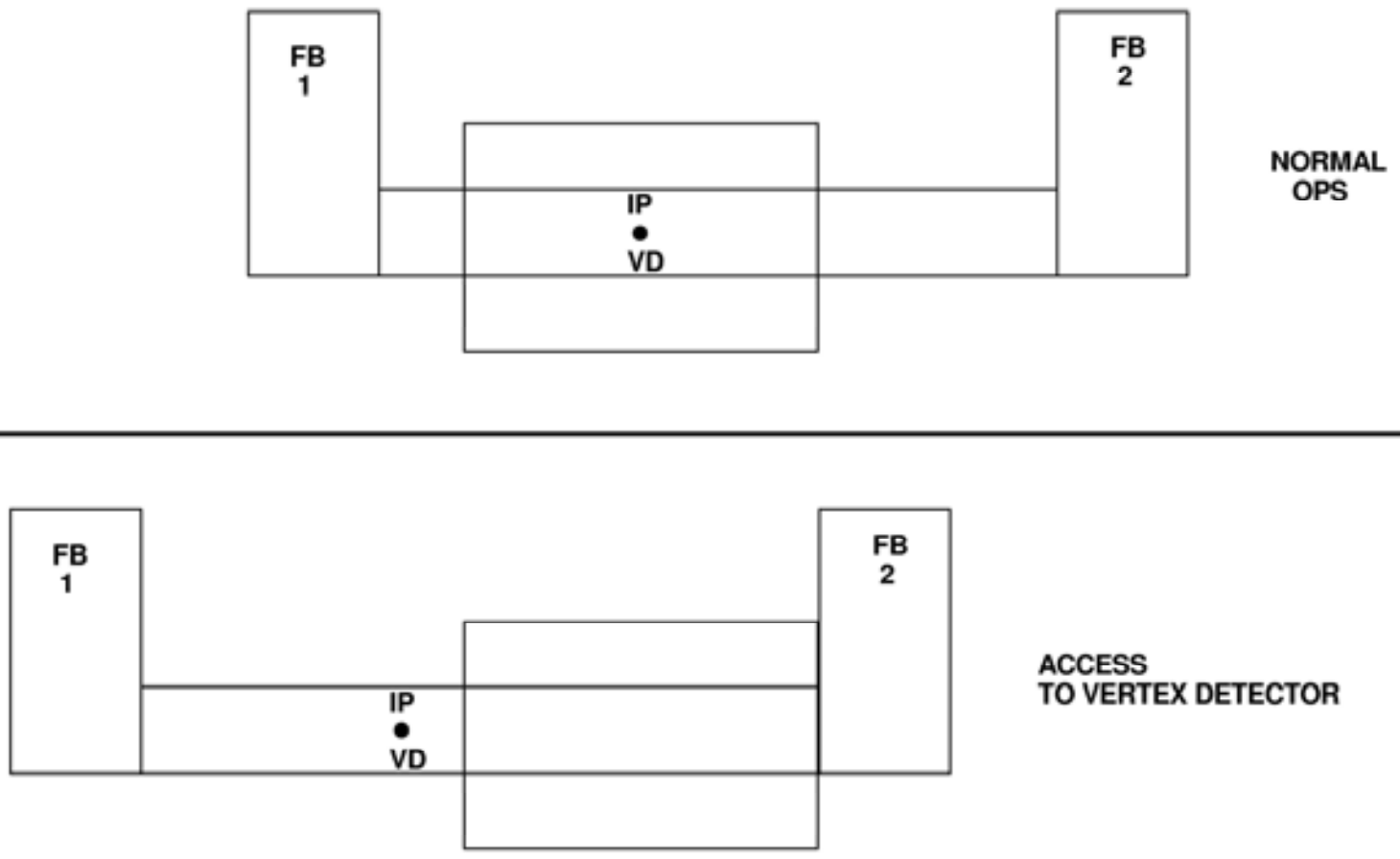

Figure 5.20: Movement of IR Magnet System and IP to Provide Access to Vertex Detector

**Cryostat Description**

Fig. 5.21 shows a solid model based on expected magnet and detector dimensions of the cryostats. This view does not show the 80 K shield and shows the vacuum vessel as see through but does show the helium vessels and the magnets. These cryostats will use standard construction techniques with stainless steel helium and vacuum vessels, copper thermal radiation shield and multilayer insulation in the vacuum space between both the vacuum vessel wall and the 80 K shield and between the 80 K shield and the helium vessel. Stainless steel piping will supply the pressurized He II, allow for cool down and warm up of the magnets and provide access for both instrumentation and magnet current leads. Though not shown in this view the system is designed to allow the beam tubes to operate at room temperature.

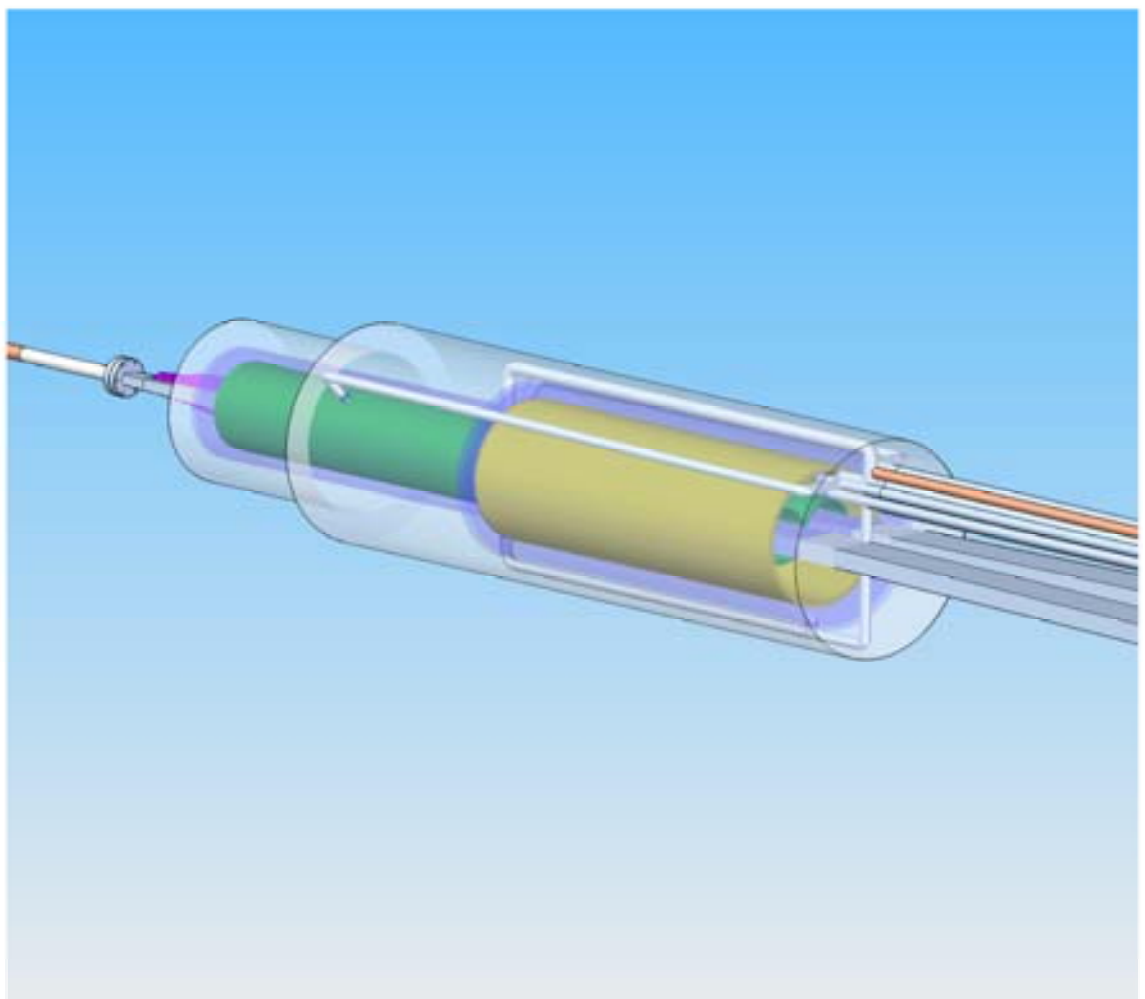

Figure 5.21: Proposed IR Magnet Cryostat System

Fig. 5.22 shows the cryostat and transfer line system (with the outer vacuum vessel and transfer line wall visible) in relation to the space currently allocated to this system, It shows that the system as designed does fit within the space given.

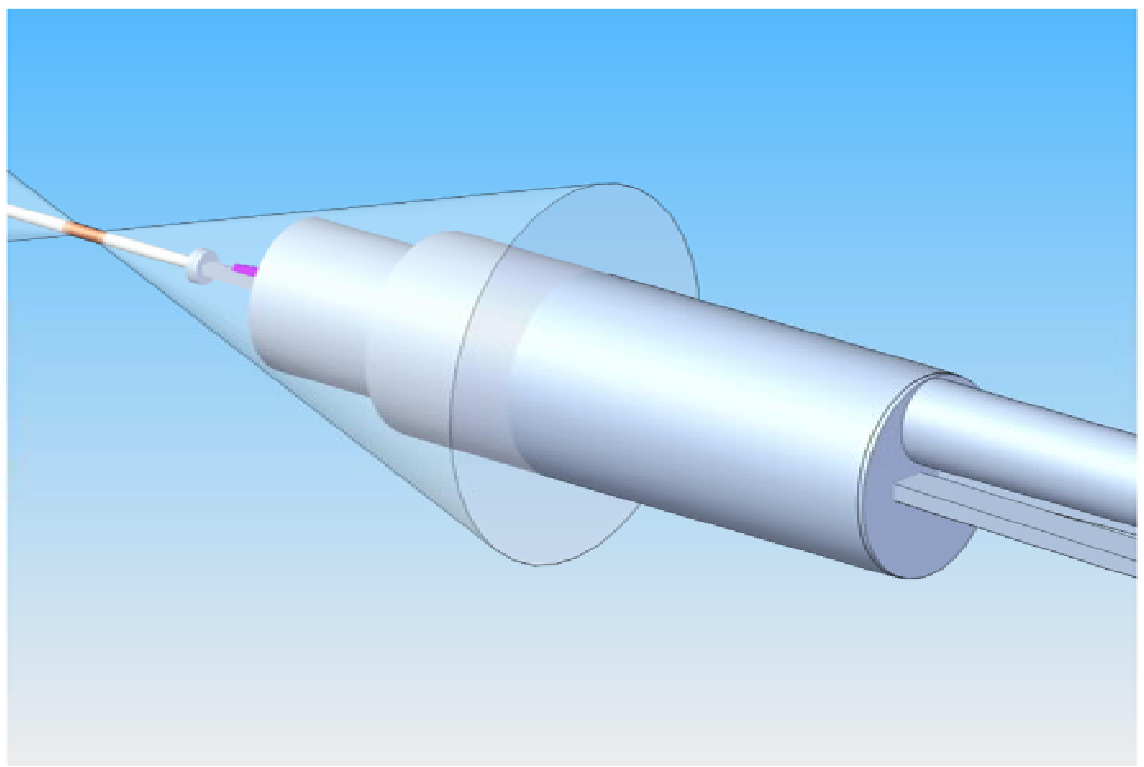

Figure 5.22: Proposed IR Magnet Cryostat System Showing Detector Space Limits

**Interconnect Box & Transfer Line Description**

The interconnect boxes (1 per side) provide both the connection of the magnet system to the cryogenic refrigeration plant and magnet power supplies as well as providing the He II refrigeration for the system. As shown in Figure 5.19, the interconnect box contains the helium vapor/helium liquid heat exchanger, JT valve and saturated He II/pressurized He II heat exchanger. This set of components converts the 4.2 K liquid from the BaBar refrigerator into the He II used to cool the magnets. The heat leak from the magnet system is transferred via internal convection to the heat exchanger in the interconnect box which then boils off the helium in the saturated He II bath. The vapor is pumped off via a warm vacuum pump which maintains the bath at 1.9 K (2299 Pa). Connections to the 80 K shield circuit along with the warmup/cooldown line also pass through these boxes.

The cross section of the transfer line connecting the interconnect box and the magnet cryostats is shown in Figure 5.23. The line contains all the service pipes needed for the magnet system along with an 80 K shield and blankets of multilayer insulation enclosed in a vacuum space. The diameter of the He II supply line is sized to allow up to 24 W of heat to be transferred via internal convection. The other line sizes are approximate based on experience and will be refined as the design develops.

#### Connections to the BaBar Refrigerator

The existing BaBar refrigerator has two sets of connections that supply helium at 4.2 K in addition to those that supply the detector solenoid. These would be connected to the interconnect boxes via flexible transfer lines. The return flows would be pumped off by the warm vacuum pump, cleaned of any impurities (such as pump oil) and returned to the suction side of the main compressors. The 80 K shield flows could either be connected in parallel to the detector solenoid shield flow or be supplied via another 80 K He gas source. It is possible that the temperature of this shield may be lowered somewhat to better match the supply from the refrigerator.

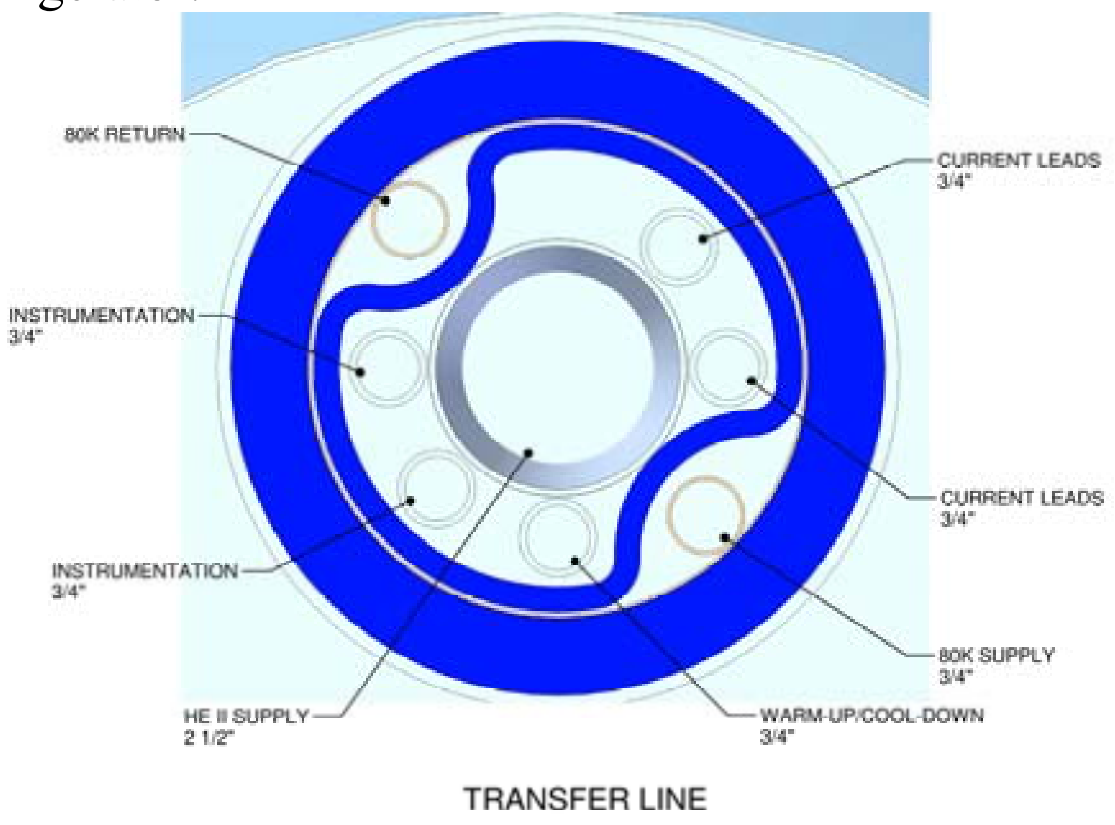

Figure 5.23: Cross Section of IR Magnet Transfer Line

Connecting the IR magnets to the refrigerator in this manner will allow the IR magnets to be cooled down or warmed up independently of the detector solenoid. However, the refrigerator will of course need to be cold for the IR magnets to be at operating temperature.

#### Estimated Cooling Capacity and Heat Leaks

Operating the Babar refrigerator at full compressor flow (150 g/s) will easily result in a production of 3.5 g/s of 1.9 K saturated Helium. This translates into 77 W of total (both sides together) cooling at 1.9 K. The maximum heat transferable by internal convection through the 2.5 inch He II supply line is 24 W per side at 1.9 K.

These limits can be compared to the estimated heat loads from the magnet system. For one side of the IP the estimated magnet cryostat static heat into the 1.9 K fluid is 18 W. This is based on a conservative calculation of thermal radiation heat leak (assuming warm beam tubes as well as an estimate of 2 W conduction heat leak to 1.9 K per each anticipated support in the magnet cryostat and transfer line. The 18 W per side can be compared to the 24 W available capacity per side.

Still to be calculated is the heat load due to LDI/DT losses (dependent on magnet design) and any heat load due to ionizing radiation.

In this conceptual design, the expected heat load appears to be consistent with available cooling capacity. However, this will need to be checked again as a more detailed design evolves. The internal convection heat transfer limit can be raised by increasing the cross sectional area of the He II supply line so there is additional cooling margin available.

#### Utilities

This cooling system requires few utilities in addition to those already supplied for the BaBar refrigerator. Additional electrical power will be required to power the magnets and the vacuum pump for pumping the He vapor from the saturated He II bath. The amount of power needed is still to be determined but it should be consistent with that required for large detector halls. Depending on its design, the vacuum pump may also require water cooling. It is expected that the cryogenic controls for the IR magnet system will be part of the controls for the BaBar refrigerator.

## 5.17 Luminosity Monitor

The PEP-II collider used a zero-angle luminosity monitor that detected the gamma rays from the radiative Bhabha scattering process. The detector was located next to the HER beam pipe where the radiative Bhabha gamma rays from the LEB were intercepted by the vacuum pipe. Because of the synchrotron radiation power, the vacuum chamber wall was at a shallow angle and was water-cooled, thus presenting about 1.5 radiation lengths to the incoming gamma rays. On the other side of the interaction region, the synchrotron radiation from the HEB required a cooled copper beam pipe capable of absorbing 70 kW, and this thickness of material prevented a useful radiative Bhabha signal from being detected.

The beam pipe design for SuperB allows for a luminosity monitor similar to the PEP-II type to be installed at 7-10 m from the IP. With the energy of the HEB lower than at PEP-II, the possibility of also measuring the radiative Bhabha rate on that side will be reexamined. In the HEB downstream case, one possible approach is to make use of muon pair production in the vacuum pipe wall. An estimate of the rate of penetrating muons is 100 kHz at full luminosity. With this signal, luminosity could be monitored, with 1% statistics, at 10 Hz.

The main background signal is BGB generated by the incoming beam. This background signal is integrated over the length of the beam trajectory near the collision point following the last inbound bend. In the PEP-II case this was 42 cm, and led to insignificant backgrounds. Although the straight section in the SuperB design is an order of magnitude longer, the vacuum pumping and conductance will be improved to compensate, and the background situation is expected to be comparable.

An effect that was present at PEP-II, but will be of greater importance at SuperB, is the suppression of the radiative Bhabha signal occurring at small beam dimensions [12].This effect is explained in [13] in terms of the cutoff of transverse integrals of the electromagnetic field to match the oncoming beam size. Using the prescription of Kotkin and Serbo [14], we find that, at SuperB, low energy gammas from the scattering will be suppressed by ~40%. For gammas close to the beam energy, the suppression will be ~20%. For small changes in the spot size the effect does not vary enough to cause confusion in using the radiative Bhabha signal for machine diagnostics or feedback. However, these calculations do not yet include the effects of crab waists, and an investigation is warranted for conditions where misleading signals may appear.

The luminosity counters at PEP-II functioned for various purposes. A signal proportional to luminosity, with an averaging time constant of ~0.5 sec, was sampled frequently by the data collection system, displayed, provided to BaBar and archived. This was also used for luminosity feedback. A faster feedback signal was developed, responding as fast as about 1 kHz, and a signal was buffered for purposes of fast event diagnostics. In addition, a system using the full bandwidth of the counter continuously monitored the relative luminosity of all colliding bunch pairs, completing a survey of all bunches within about 3 seconds. The detector technique that drove these signals used a 4 radiation length converter just outside the beam pipe to rejuvenate the shower from the escaping gamma rays close to the detector. The resulting electrons and positrons produced Cherenkov light in fused silica blocks. Some of the light was extracted through prismatic edges and ducted through air light guides to PMTs capable of resolving pulses at the full 238 MHz of the machine. Output signals were split and processed by different types of electronics systems. The very high rate of the radiative Bhabha signal at PEP-II meant that the detectors behind the shower converter had to withstand in excess of 1 Grad per year. In fact the detector shielding was part of the Personnel Protection System of the machine. At SuperB the radiation level from the radiative Bhabhas will be two orders of magnitude higher. There is some evidence that the fused silica used at PEP-II could withstand at least several times the PEP-II dose. But it would be very difficult to test at 100 Grad per year, and so palliative design work is needed. The most straight-forward change would be to use a much thicker shower converter to filter out most of the radiation dose, and detect only a very penetrating fraction of the gamma rays.

The process of developing an initial design will start with the radiative Bhabha spectrum, and simulate showers through various thicknesses of material with EGS or Geant. The distribution of the surviving tracks will be parameterized, and used to study the propagation of Cherenkov light through a range of possible detector models based on the PEP-II design. For converting the Cherenkov pulse to an electronic signal, it is likely that the fast PMT technique will remain desirable at SuperB, for its ability to monitor the individual bunch pairs. Presently, this restricts the choice of PMTs to a few of the fastest models. As an alternative, and a possible path to future shorter bunch spacing, microchannel plate PMTs would give an improvement in timing, but traditionally they have a limited lifetime in terms of their ability to deliver signal charge. This should be re-evaluated for modern microchannel plate versions, while allowing for the possibility that modern high-gain fast pulse amplifiers might help by reducing the wear on the microchannel plates.

On the other hand, for purposes other than monitoring individual bunch pairs, slower techniques may have advantages, and deserve to be examined. Fast, narrow gap, ion chambers could in principle provide a more stable sensitivity than PMTs, while being intrinsically radiation hard. This will require some research and development to ensure that ion charge saturation does not prevent high rate operation. If this is successful, a preliminary estimate shows that a current of ~1 microamp could be delivered by such a device without amplification. In principle, at full luminosity, this technique would allow measurement with <1% statistical uncertainty at megahertz rates. Suitable readout instrumentation would allow the luminosity signal to be coupled to a fast feedback system. An alternative to direct readout of ionization might be to measure the fluorescence of the gas at optical wavelengths. More speculative is the possibility of using a fast R/F pulse from the excess of electrons over positrons in the gamma ray shower, if ringing after the pulse and background noise can be managed.

Although a careful selection of discriminator thresholds allowed the PEP-II system to operate with satisfactory linearity in counting mode with 238MHz bunch crossings, the use of GHz digitizers synchronized to the machine timing brings the possibility of improved linearity of response over a wider dynamic range of luminosity. Somewhat analogously to bunch by bunch processing at PEP-II, data would be accepted to a deep memory in a burst some tens of milliseconds long, analyzed for luminosity per bucket in a fast processor, and the cycle repeated indefinitely. Following PEP experience, feedback may be needed to stabilize the relative timing between the luminosity signal processing and the collider distributed timing pulses.

Studies on the PEP-II luminosity counter system showed that it was possible to provide a luminosity signal with a response time in the range of 1 to 2 msec and signal resolution of at worst a few percent. A similar detector at SuperB would be positioned behind much more shielding, and so larger fluctuations in the gamma ray showers would decrease the effective statistical weight of each measurement. Part of the design study will be to evaluate this effect, and to achieve a balance between an enhanced response time and the need for shielding. It seems likely that a performance at least as fast as at PEP-II can be achieved. Using either a fused silica Cherenkov detector or possibly a narrow-gap ion chamber system, it appears very likely that a relative luminosity signal, good to 1%, can be provided at a rate approaching 1kHz. Such a signal would be available for a beam position feedback system to maintain the beams in collision within a few percent of maximum luminosity.

# 6. Rings lattice

## 6.1 Introduction

The SuperB HER and LER lattices need to comply with several constraints. These include the extremely low emittances and IP beam sizes needed for the high luminosity, as well as damping times, beam lifetimes and polarization for the electron beam.

The SuperB rings can be basically considered as two Damping Rings (DR), similar to the ILC and CLIC ones, with the constraint to include a Final Focus (FF) section for collisions. So, the challenge is not only how to achieve low emittance beams, but also how to choose the other beam parameters in order to reach a very high luminosity with reasonable lifetimes and small beam degradation. Inspirations from the design of the linear collider DRs, as well as from lattices of the last generation synchrotron light sources, are being very useful to define SuperB lattice characteristics. Nevertheless a new "Arc cell" design has been adopted for SuperB and is now under study for the ILC-DR.

All the SuperB lattice studies so far have been quasi parameters free. After an intense optimization work, the parameters corresponding to asymmetric emittances and beam currents for the two rings seem to be more consistent with other requirements. For instance:

- Larger emittance and lower beam current in the LER is necessary to keep under control the emittance dilution due to Intra-Beam-Scattering (IBS).
- Higher beam current in the LER is necessary to minimize the synchronous phase spread difference between the two rings due to the gap transient.

The SuperB Rings consist of the two main lattice systems:

- *ARCs*, whose main functions are to:
  - bend the beams back into collision;
  - generate the design horizontal emittance.

- *FF System*, which consists of an extremely low-β insertion and a Crab Waist scheme requiring a special optics that:
  - provides the necessary beam demagnification at the IP;
  - corrects its own chromaticity;
  - provides the necessary conditions and constraints for the Crab Waist optics.

The first version of the Arcs lattice (described in the CDR [1]) has been inspired by the low emittance ILC Damping Ring lattice, but has evolved since then towards a more compact and performing design which is described below.

## 6.2 Rings layout

Layout of the HER (positrons) and LER (electrons) is shown in Fig. 6.1, where the rings are in the horizontal plane, and IP is at the top of the Figure. The FF is connected to the two Arcs in two half-rings (one inner, one outer) and a long straight section on the opposite side. The straight section comes naturally to close the ring and readily accommodate the RF system and other necessities. A "parasitic crossing" for the two rings without beam collision will be provided in this utility region.

The LER and HER FF bending systems provide the same total bending angle and the specified 66 mrad crossing angle at IP. The latter requires bending angle asymmetry with respect to IP in one or both rings. The present design uses the same bending configuration in the LER and HER FF, but reversed with respect to each other in order to produce the crossing angle. The latter is obtained by introducing ±33 mrad asymmetry with respect to IP, where the left hand side LER FF is mirror symmetric to the right hand side HER FF and vice versa. This configuration simplifies the FF geometrical and optical match. First, it yields a symmetric overall FF geometry which simplifies the geometric match to the left and right hand side Arcs. Secondly, it allows identical HER and LER FF optics (reversed relative to each other) thus simplifying the optimization. The strengths of the FF dipole magnets are adjusted for ≈2.1 m radial separation between the HER and LER beam lines in the Arcs and in the long straight section. The main difference between the LER and HER FF design is that the LER includes a Spin Rotator (SR) insertion with solenoids at each end of the FF while the HER has a simple FODO section at this location.

The "parasitic crossing" between the two rings is arranged at one end of the Arc near the left hand side of the long straight section as shown in Fig. 6.1. The crossing is obtained by a proper lengthening of a drift space in the HER Arc cell nearest to the straight section in such a way that the two beam lines cross with minimal interference and provide the specified ≈2.1 m separation in the straight section.

The present rings design satisfies all the requirements: luminosity, beam polarization, compatibility with PEP-II hardware, wall-plug power, etc. The ring circumference is ≈1258 m.

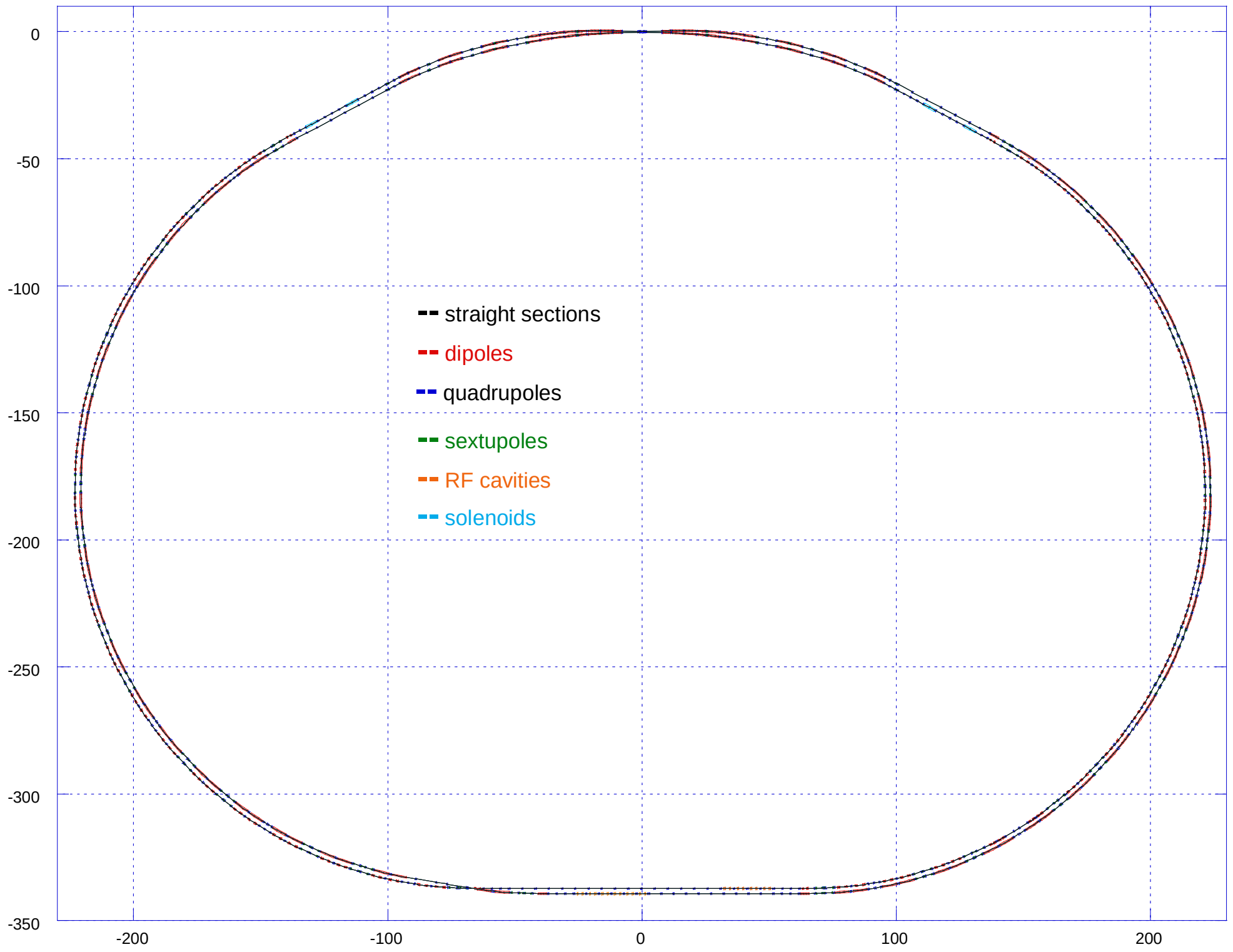

Fig. 6.1 – Layout of the HER and LER rings.

## 6.3 Arc lattice

The Arc design has been constantly ameliorated since the CDR [1]. The design has evolved in order to:

- improve the transverse dynamic aperture;
- improve the energy acceptance;
- improve the flexibility in modifying its parameters (emittance, etc.) during the run;
- decrease its natural chromaticity;
- increase the momentum compaction for a given emittance;
- increase all the instability thresholds;
- increase its tuning ability in order to achieve the target parameters;
- relax the tolerances;
- decrease its complexity.

Fortunately, a lot of these improvements are positively correlated. For instance, a lower chromaticity makes the sextupoles weaker, thus yielding a larger dynamic aperture and weaker head-tail instability.

The design provides safety margins on the required specifications. For instance, presently the Arcs transverse acceptance exceeds 100σ whereas the machine physical aperture is of the order of 40σ.

The SuperB Damping Ring is using a similar Arcs design which provides excellent performance.

The HER and LER Arcs lattices are conceptually the same. The main difference is that the Arc dipoles in the LER are a factor of 3 shorter than in the HER in order to obtain approximately the same emittance values at unequal beam energies. Geometrically, the HER and LER Arcs are parallel to each other in the horizontal plane while separated by ≈2.1 m in the radial direction as shown in Fig. 6.1 and 6.2. The two horizontal crossings (IP and parasitic) result in each ring having one inner and one outer Arc. Both the inner and outer Arcs have the same total bending angle, but the outer Arc is made longer by increasing the drift space around the dipole magnets in order to make the two Arcs concentric with constant separation.

The Arc lattice consists of short and long cells shown in Fig. 6.3 and 6.4. These cells are originally based on the TME type lattice in order to minimize the emittance. However, the standard TME cell optics is modified in this design by splitting the central dipole in two halves and inserting a focusing quadrupole between them as shown in Fig. 6.3 for the short cell. This extra quadrupole increases the cell tuning ability which yields a better horizontal focusing at the dipoles for a lower emittance. Phase advance in the short cell is adjusted to near $\mu_x \approx 3\pi/2$, $\mu_y \approx \pi/2$ for optimal compensation of sextupole non-linear chromaticity.

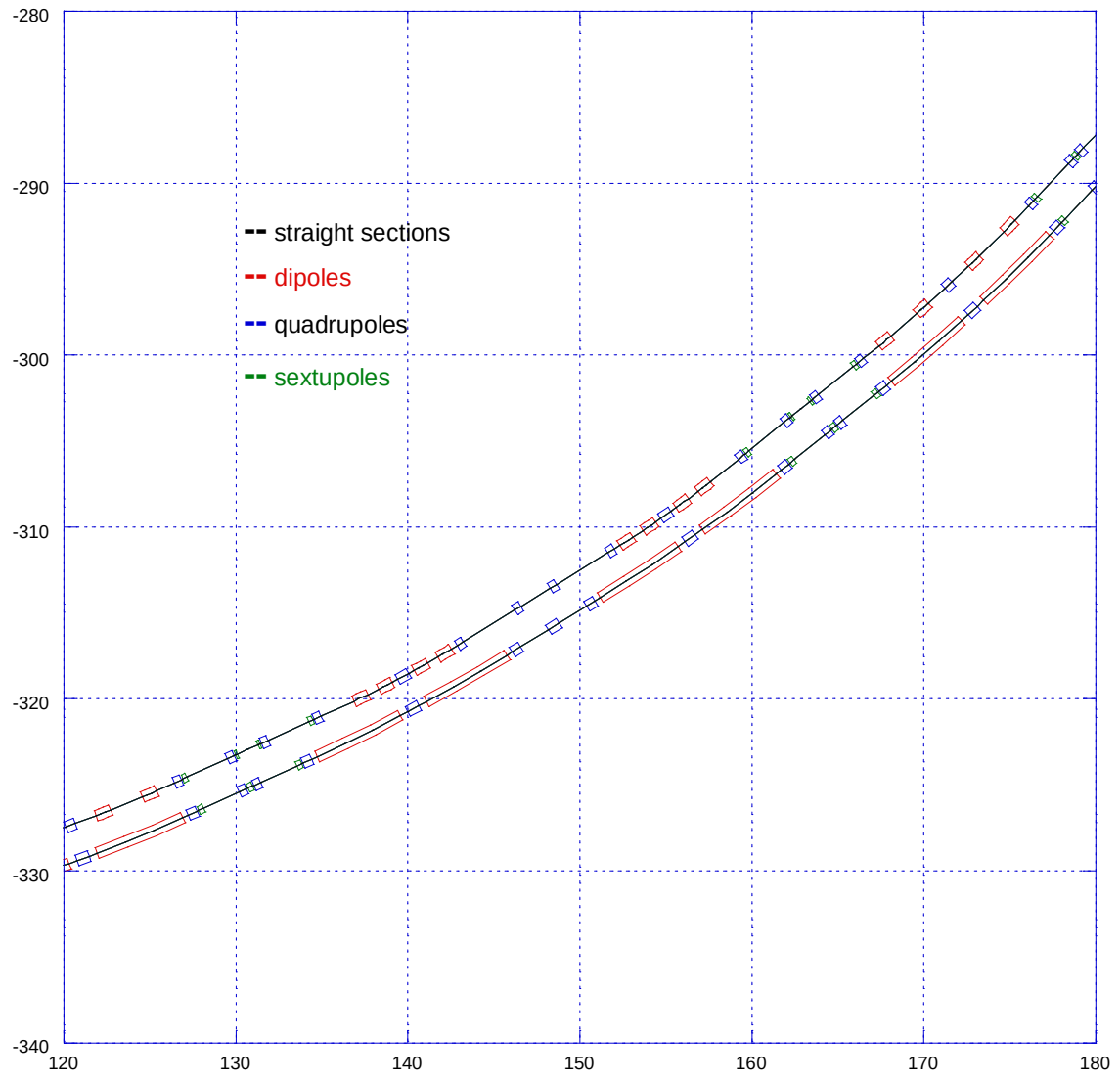

Fig. 6.2 – Layout of the HER (outer) and LER (inner) Arc cells.

The long cell in Fig. 6.4 is approximately a combination of two short cells. It uses 5 independent quadrupole families for maximum flexibility which helps to maximize β-functions and dispersion at sextupole locations in order to decrease their strengths and their non-linear effects on dynamic aperture. In order to minimize emittance and the 2nd order chromatic tune shift, bending angle in the two dipoles near the center of the long cell is reduced by 8 mrad compared to the other Arc cell dipoles. Phase advance in the long cell is matched exactly to $\mu_x = 3\pi$, $\mu_y = \pi$ for –I transformation in both planes.

A high horizontal phase advance in the Arc cells is required for a low emittance. But the vertical phase advance can be made lower for a lower Arc chromaticity. In this design, the Arc vertical phase advance is made 3 times lower than the horizontal one. This allows to maintain –I transformation in the long cells and ~π/2 transformation in the short cells in both planes. The long and short cells are arranged periodically one after the other in each Arc as shown in Fig. 6.5. A dispersion suppressor cell at each Arc end has the dipole and quadrupole strengths adjusted for dispersion cancelation.

Horizontally and vertically correcting chromatic sextupoles are inserted at the beginning and end of the short and long cells, where β-functions and dispersion are at maximum as can be seen in Fig. 6.3, 6.4. In this case, the identical sextupoles form –I pairs which provide local cancellation of the sextupole 2nd order geometric aberrations and the 2nd order dispersion leaving only the higher order terms due to finite sextupole length and partial overlap of the pairs. To minimize the chromatic W-functions and non-linear chromatic tune shift generated by the Arc sextupoles and to maximize dynamic aperture, phase advance in the short cell is optimized.

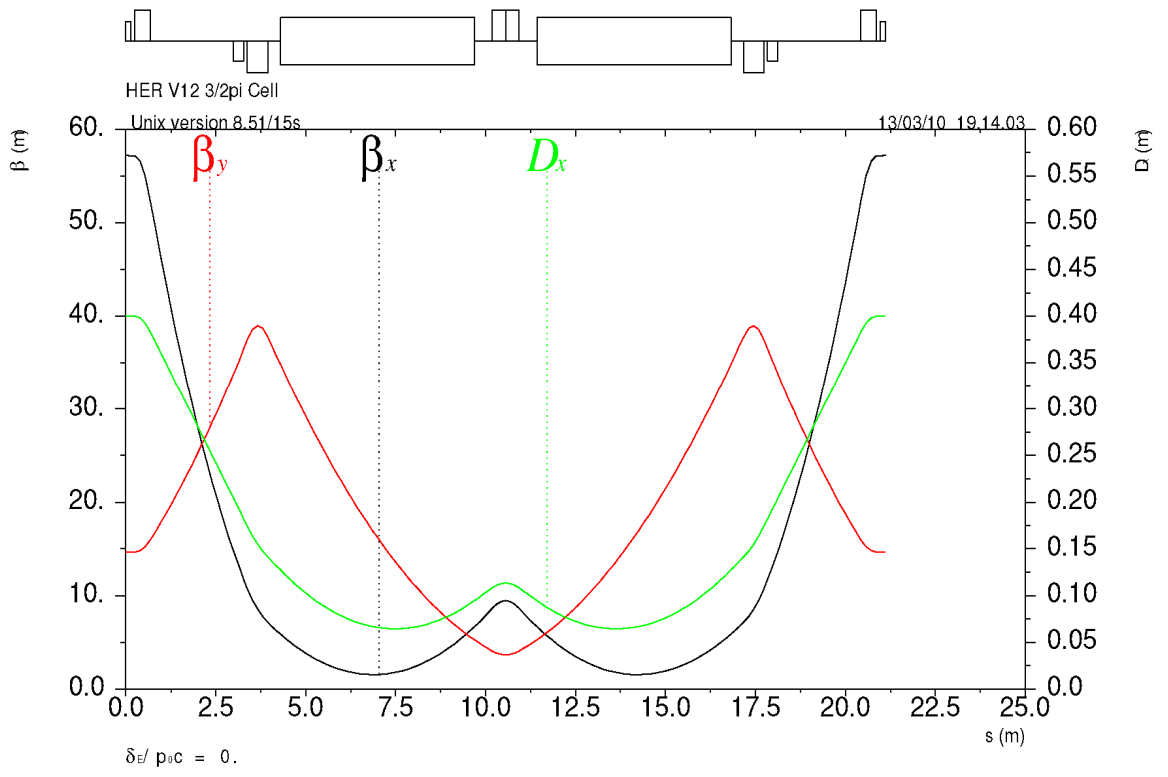

Fig. 6.3 – Lattice functions in the HER short Arc cell.

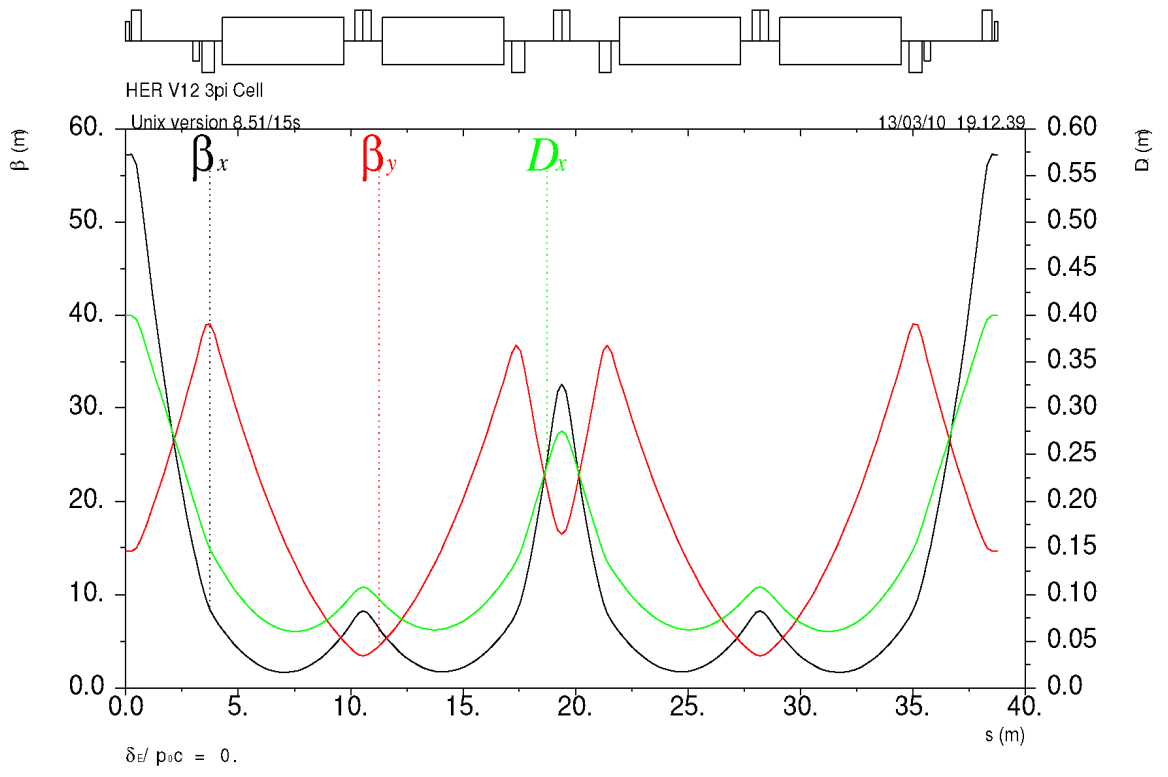

Fig. 6.4 – Lattice functions in the HER long Arc cell with –I transformation.

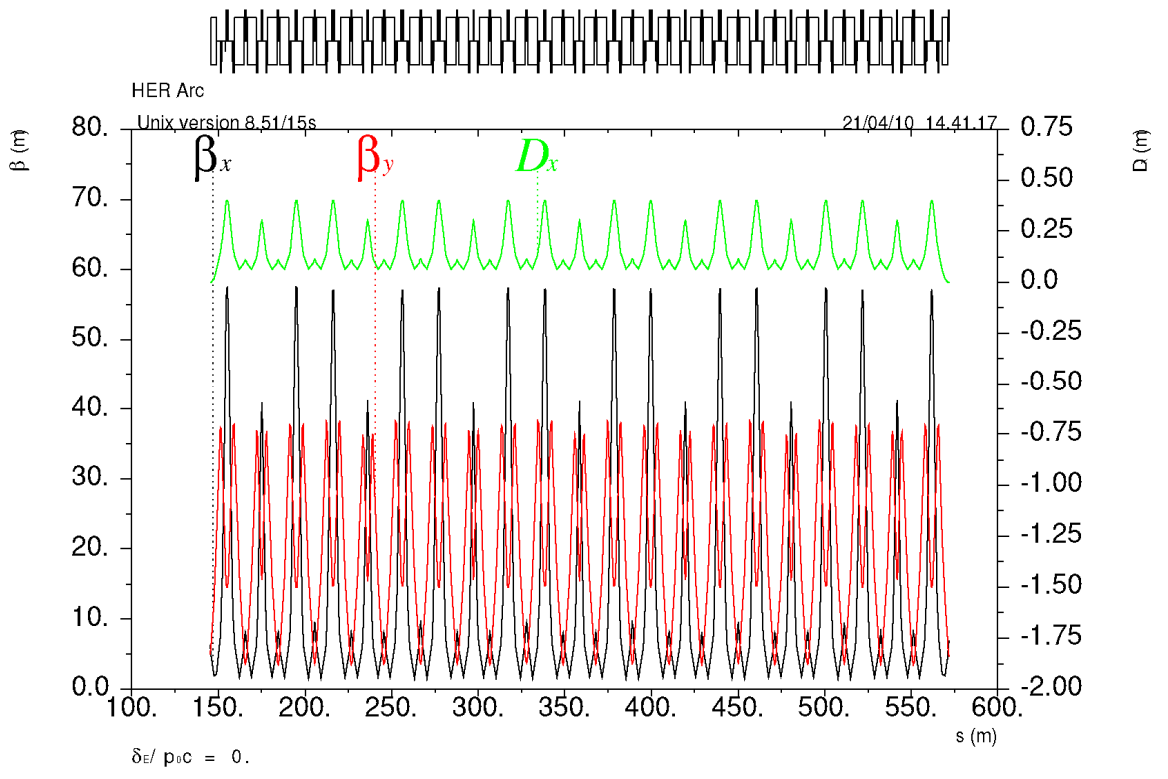

Fig. 6.5 – Lattice functions in one HER Arc.

## 6.4 Final Focus lattice

The Final Focus is the most crucial system for achieving the SuperB performance. The luminosity goal is based on capability of the FF to de-magnify the vertical beam size at the IP down to 35 nm and beyond. In addition, the FF design has to ensure the full functionality of the Crab Waist optics that has been proven fundamental to minimize the beam-beam unwanted non-linearities (see Chapter 3).

This is the biggest challenge we encounter. The FF system is based on the acquired know-how and experience of the systems developed for the Linear Colliders. It is based on the optics developed for the “Next Linear Collider” (NLC) that had been successfully tested on a dedicated single pass beam line, the “Final Focus Test Beam” (FFTB) built at SLAC, where beam sizes down to 70 nm had been measured.

Several modifications have been made in order to adapt such FF optics to a ring operation. In particular:

- All bending angles have the same sign to meet the ring requirements (rather than the Linear Collider ones). For example, they have to generate a specific value of bending angle between the IP and Spin Rotator (see Chapter 16). In addition, their quantity, locations and bending radii have been optimized for minimal ring emittance and maximum dispersion at the Chromatic Correction Section (CCS) sextupoles.
- Two additional sextupoles in phase with the IP, at the beam waist locations upstream the CCS, provide a great increase of the FF demagnification capabilities and bandwidth.
- Crab Waist sextupoles have been added at each end of the FF. It has been proved that this is the only possible location in order to preserve dynamic aperture of the system (CDR [1]). Fig. 6.6 presents lattice functions in the current FF design, and Fig. 6.7 shows a close-up of the IP region layout.
- Dipole lengths and bending angles have been adjusted in order to meet the geometrical constraints. The dipoles in the Y-CCS section are made weaker on one side and stronger on the other side to satisfy the ±33 mrad asymmetry with respect to the IP. The FF dipoles are also made shorter than the Arc (and PEP) magnets (4 m instead of 5.4 m) to obtain the proper separation of the two rings (≈2.1 m average).

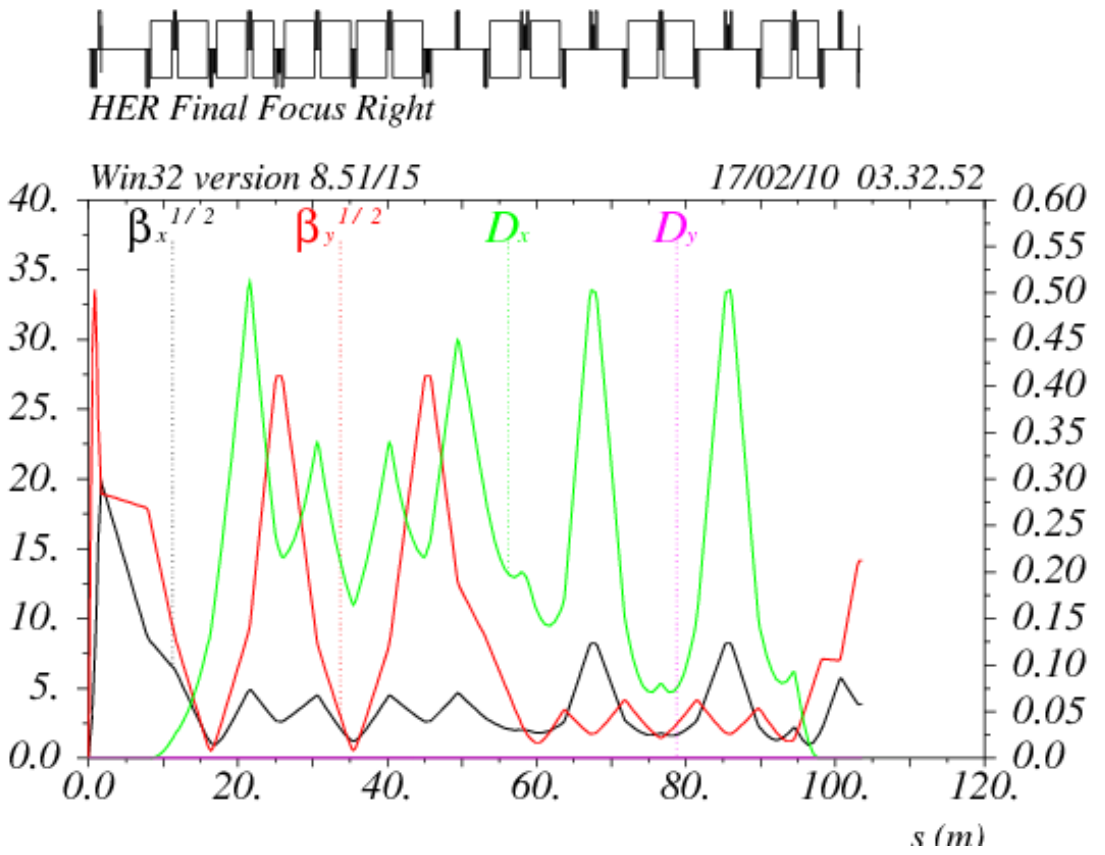

Fig. 6.6 – Lattice functions in the HER Final Focus, where IP is on the left.

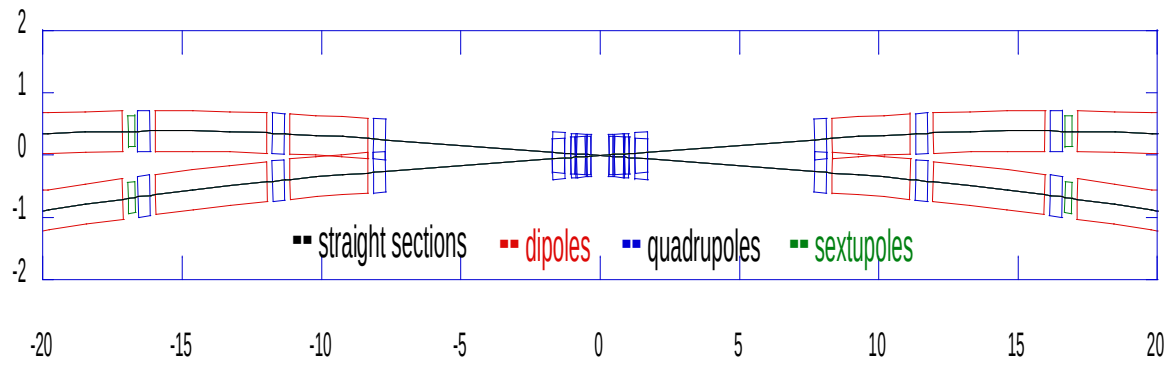

Fig. 6.7 – Close-up of the IP region layout.

## 6.5 Final Focus R&D

SLC at SLAC is the only collider that had operated with a simplified version of the SuperB FF optics. It had two FF sections (for left and right sides of IP) operating in single pass mode, each one equivalent to half of the SuperB FF in terms of complexity. The achieved demagnification ($\beta_y$ = 2 mm) is about 10 times larger than the one required at the SuperB ($\beta_y$ = 0.2 mm). It took 4 years to fully commission the SLC system and make it reliable.

FFTB at SLAC had the most similar FF system to the SuperB FF. It achieved a comparable demagnification and vertical beam size. About 1 year of total beam time was needed to fully commission the FFTB.

ATF2 at KEK is a single pass beam line to test the ILC FF design. It is presently under commissioning. This system is about 30% simpler than the SLC FF, hence it should be easier for commissioning and reliable delivery of the design performance. Its design beam parameters are very similar to the FFTB and SuperB ones.

SuperB has four FFTB-like FF sections that have to simultaneously operate at full specifications. They have the extra requirement that the beam is not dumped after IP, but has to remain virtually unchanged after passing through two of such systems (in each Ring).

## 6.6 Polarization insertion

Several schemes have been studied in order to provide a longitudinal polarization at the IP. All the polarization schemes must satisfy physical constraints and quantized conditions that are intrinsic to the spin dynamics, since the rotation of the spin in a given plane is directly related to:

- beam energy;
- bend angle in a given section;
- integrated field of a solenoid (if used).

### *Polarization with Spin Bumps*

All the polarization schemes that do not use solenoids to rotate the spin require vertical bumps. So far no feasible solutions were found for such schemes that would not have a great impact on the machine design parameters, in particular the vertical IP beam size.

### *Polarization with Spin Rotators*

Bend angle values in the SuperB FF optimized for maximum luminosity differ from the ones required for proper spin dynamics (there are very few quantized choices). Presently, two Spin Rotator (SR) schemes are developed, one for having a net polarization in the HER (6.7 GeV) and the other for the LER (4.18 GeV).

The required integrated solenoid field is ~70 times larger in the HER SR and 40 times in the LER SR compared to the Detector solenoid field. The Spin Rotator tuning properties in terms of coupling compensation, sensitivity to errors, etc. have not been studied yet.

### *Polarization in HER*

Significant modifications of the HER FF were required to provide the proper conditions for the bends between the IP and the SR solenoids. In particular, the Spin Rotator has to be placed in the middle of the FF. In addition, the number of –I cells in the CCS sections has to be increased for optimization of chromatic bandwidth and emittance and preservation of the FF properties. This HER solution presents the following drawbacks:

- tuning ability of the FF optics while maintaining the functionality of the Crab Waist is harder;
- dynamic aperture reduction due to the crab sextupoles is doubled;
- FF bandwidth is halved (range for minimum achievable $\beta_y^*$ is halved);
- energy acceptance is halved and Touschek lifetime is 4 times shorter;
- a Dogleg is required in the long straight section opposite to the IP;
- emittance and energy loss increase by ≈15%;
- circumference increases by ≈1 km relative to the SuperB without SR.

### *Polarization in LER*

Similar efforts have been made in order to modify the LER FF to make it compatible with the solenoid insertions. In this case the impact is minimal since it is possible to satisfy the SR optics requirements by just re-optimizing the FF dipole magnets. The main consequences are:

- HER emittance increases by ≈5%;
- circumference increases by ≈100 m relative to the SuperB without SR;
- polarization lifetime is ≈20 min ( >2 hrs in the HER SR case);
- energy asymmetry is reduced to 1.603 (1.70 in the HER SR case).

Given the above considerations it was chosen to include the Spin Rotators in the LER ring.

The SR solenoids are inserted between the FF and the Arcs. Fig. 6.8 shows the lattice functions and Fig. 6.9 the layout of one SR section. Detailed description of the polarization constraints and rationale can be found in Chapter 16.

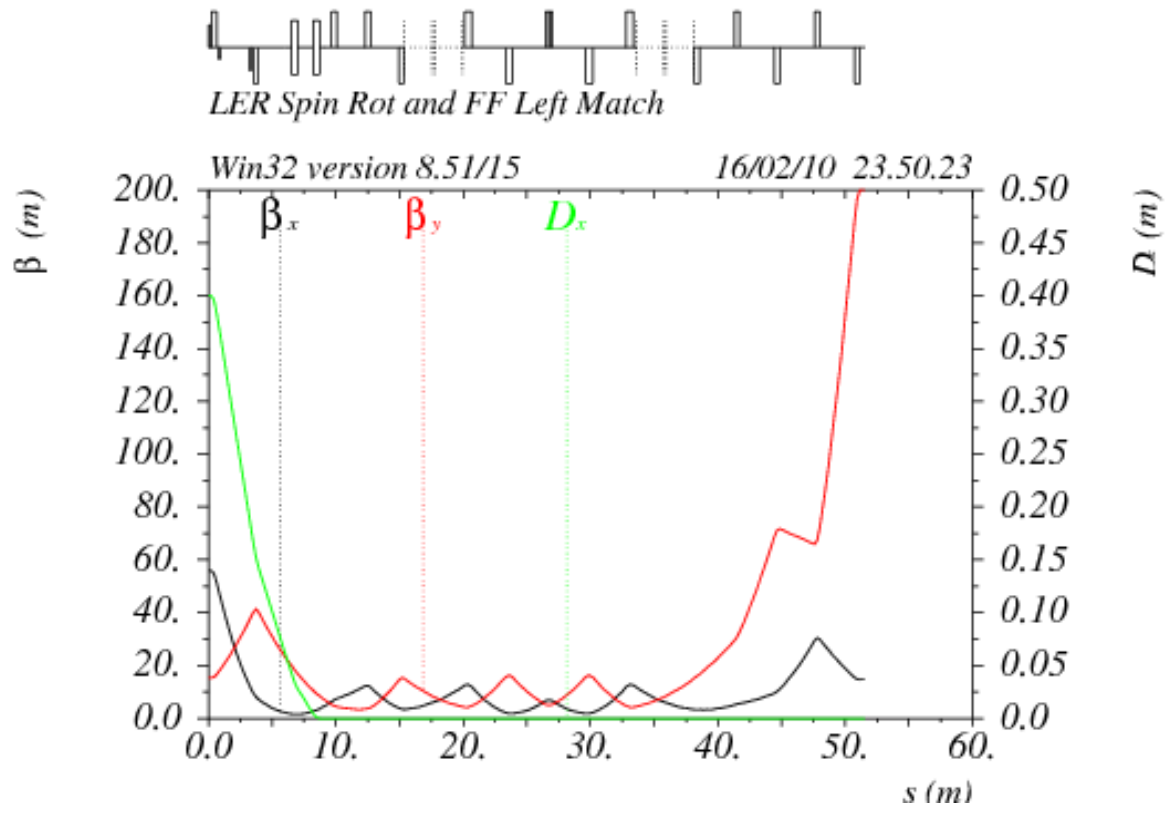

Fig. 6.8 – Lattice functions in the LER Spin Rotator section.

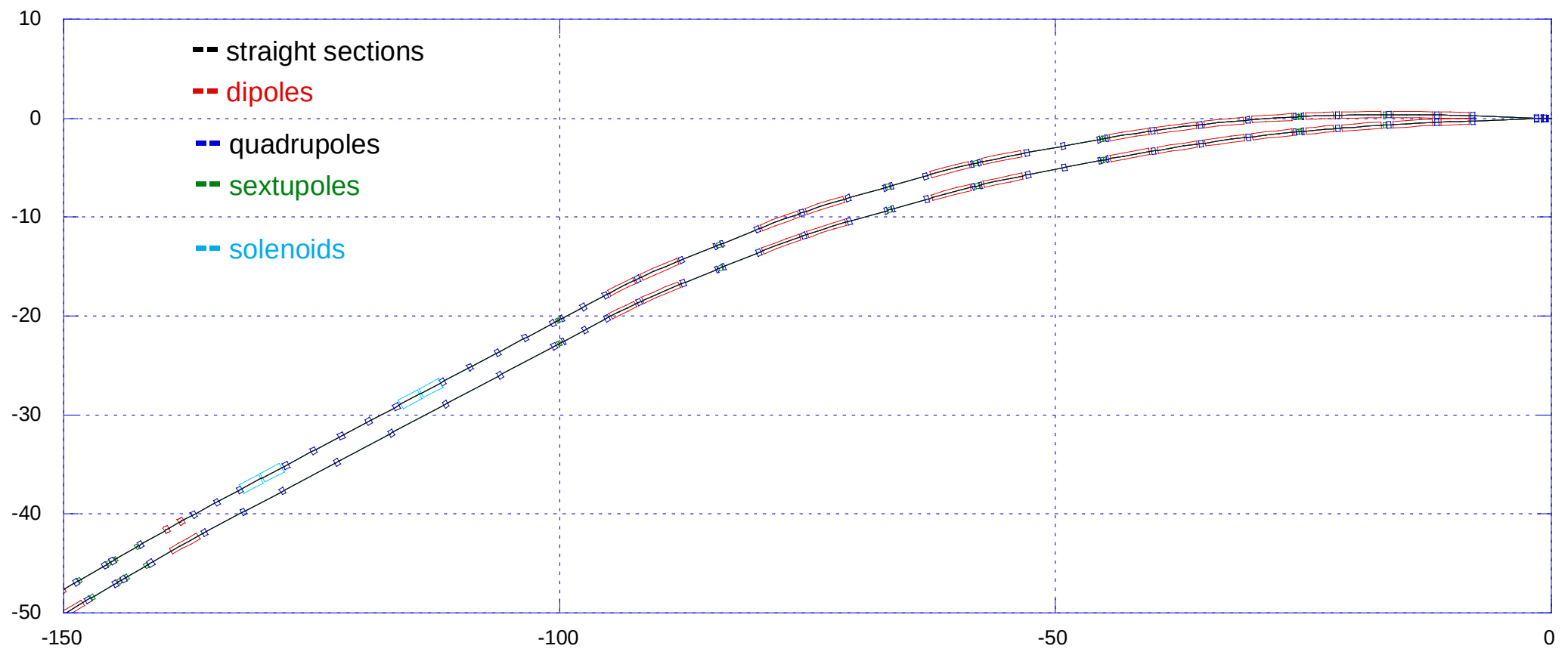

Fig. 6.9 – Layout of the LER Spin Rotator section in the outer ring, where SR solenoids are in blue.

## 6.7 Long straight section

The long straight section located on the opposite side of the rings relative to the FF contains the RF cavities and the tune trombones. More than 50 meters of FODO lattice are left free for additional hardware (more RF cavities, wigglers etc.). The straight section layout along with the parasitic crossing and injection cells in the adjacent Arcs is shown in Fig. 6.10. Lattice functions in the HER straight section and injection section are presented in Fig. 6.11.

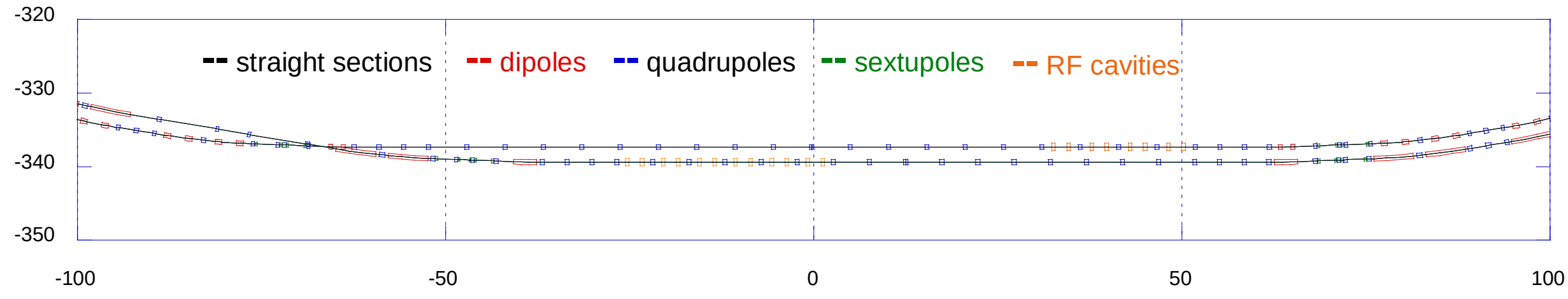

Fig. 6.10 – Layout of the long straight section with RF cavities. The parasitic crossing and the HER injection are in the nearest Arc cell on the left side of the straight, and the LER injection section is in the Arc on the right side of the straight.

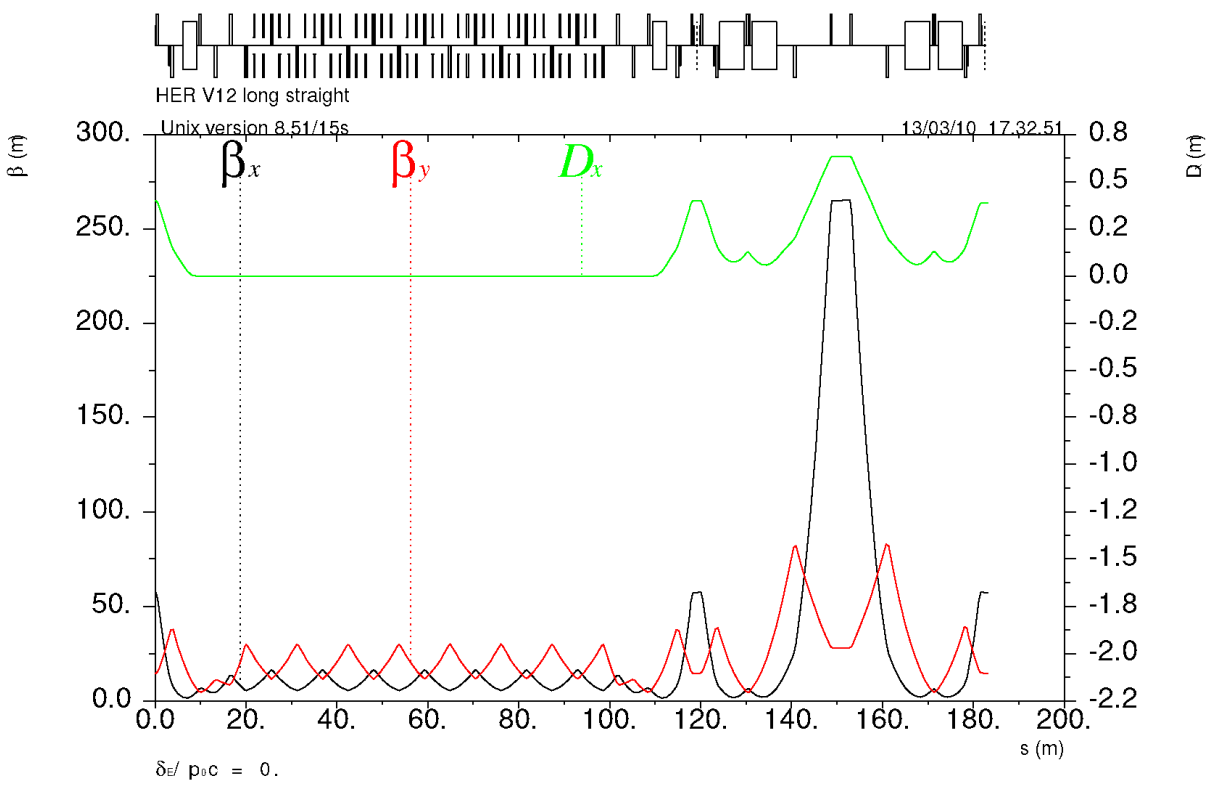

Fig. 6.11 – Lattice functions in the HER straight section and adjacent injection section. Beam direction is from left to right. The LER has a similar optics.

### *RF cavities*

The rings RF accelerating cavities are included in the FODO lattice of the long straight section as shown in Fig. 6.10 and 6.11. The HER and LER cavities are placed far from the nearest Arc bend magnets in order to minimize the synchrotron radiation load on them. The standard RF configuration includes 14 HER cavities and 8 LER cavities, but the ample extra space in the long straight allows installation of additional cavities. Each pair of cavities occupies a drift between quadrupoles in a half FODO cell adjusted to accommodate the PEP-II size cavities.

### *Tune trombone*

FODO cells in the long straight section, including those with the RF cavities, will be used as a tune trombone to control and adjust the HER and LER betatron tunes. This is achieved by variation of the FODO cell phase advance while keeping the long straight matched to the adjacent sections. For tuning in the machine operation, tuning knobs using linear adjustment of the quadrupole strengths can be created.

## 6.8 Injection section

Beam injection will be performed in the horizontal plane in both rings. Both the HER and LER injection sections are created in the Arcs by a proper adjustment of one long cell in each ring.

The HER injection section is obtained by lengthening the central part of the long cell by ≈26 m in order to increase space for injection, attain a large β-function at the septum and provide the "parasitic" rings crossing near the left hand side of the long straight. It includes injection bump kickers at each end and injection septum in the middle.

Layout of the HER injection section is shown in Fig. 6.10 with the "parasitic" crossing on the left hand side. Fig 6.12 and 6.13 show the injection lattice functions and the injection orbit bump at septum, respectively.

The LER injection section has a similar design, but it is shorter than in the HER. Its length is kept the same as in the standard long cell in order to maintain the Arc geometry. The cell magnets have been readjusted for maximum horizontal β-function at the septum and minimal kicker strength (i.e. maximum $R_{12}$ between the kickers and septum). The kickers are placed at the cell ends and the septum is in the middle. The LER injection lattice functions and the orbit bump are shown in Fig. 6.14 and 6.15, respectively.

The two identical injection kickers are separated by a horizontal phase advance of 540° in order to close the injection bump. The required kicker angular deflection is of the order of 0.4 mrad.

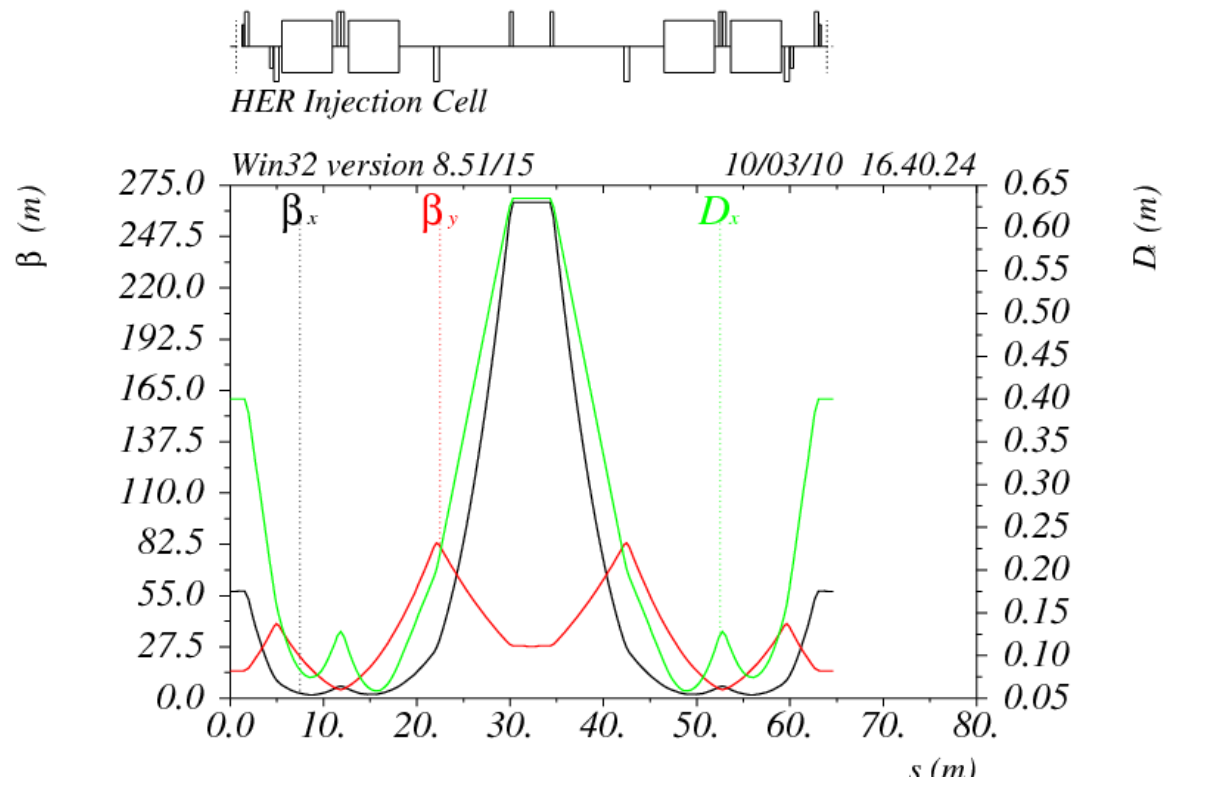

Fig. 6.12 – Lattice functions in the HER injection section.

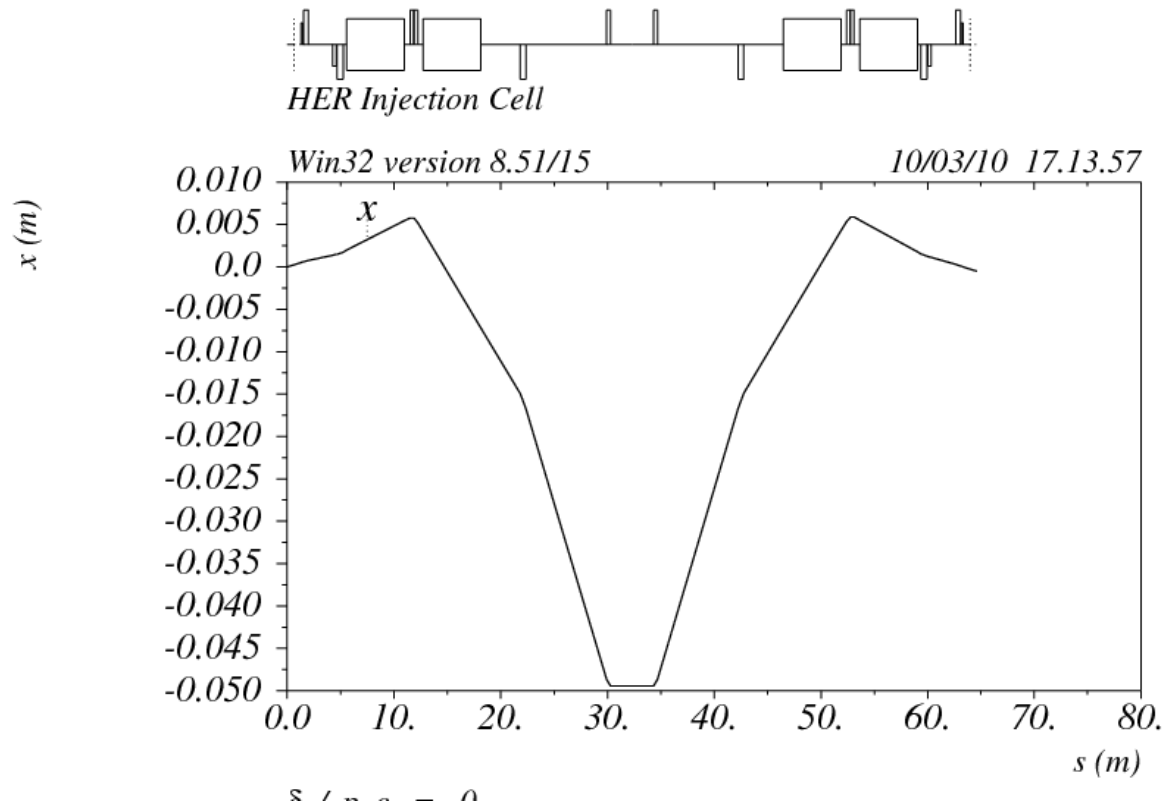

Fig. 6.13 – Horizontal injection bump in the HER.

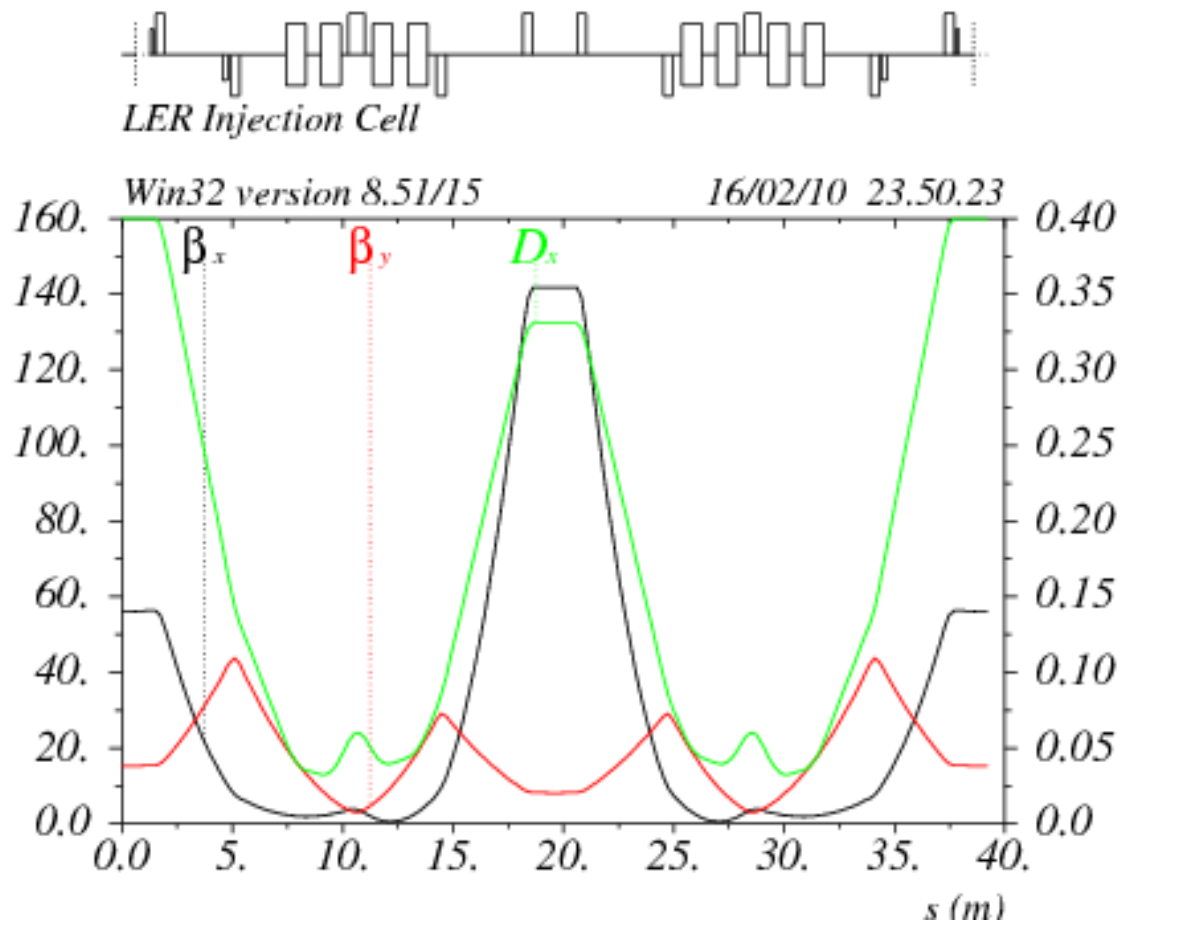

Fig. 6.14 – Lattice functions in the LER injection section.

## 6.9 Complete Ring lattice

Lattice functions in the complete HER and LER are presented in Fig. 6.16 and 6.17, respectively. The lattice direction is clockwise for both rings in these Figures, starting from the middle of the long straight section.

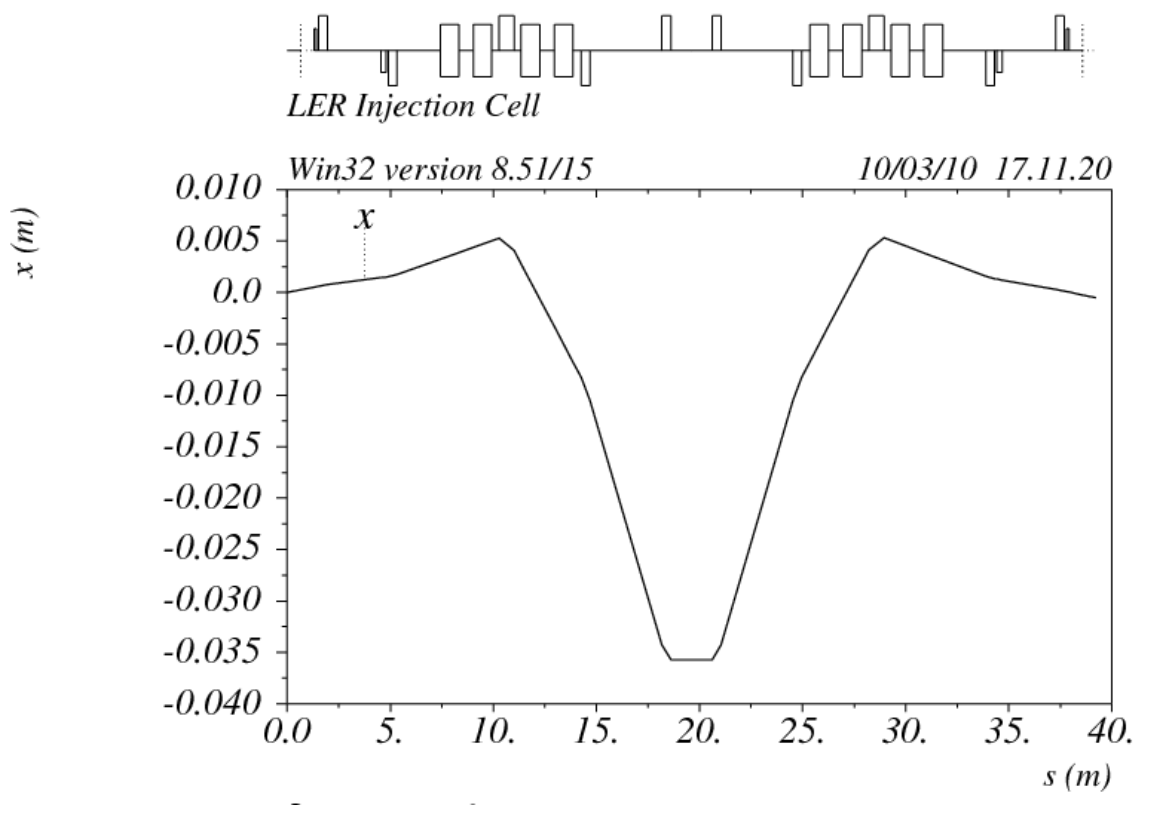

Fig. 6.15 – Horizontal injection bump in the LER.

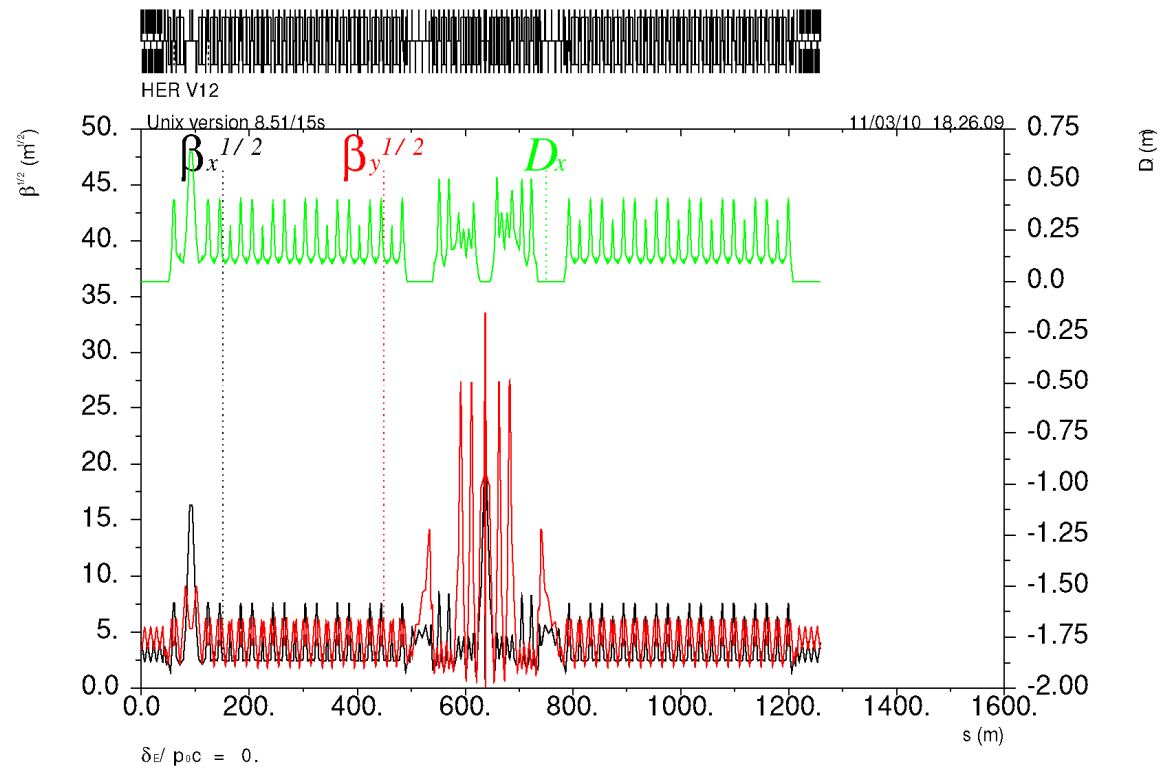

Fig. 6.16 – HER lattice functions, where IP is in the middle.

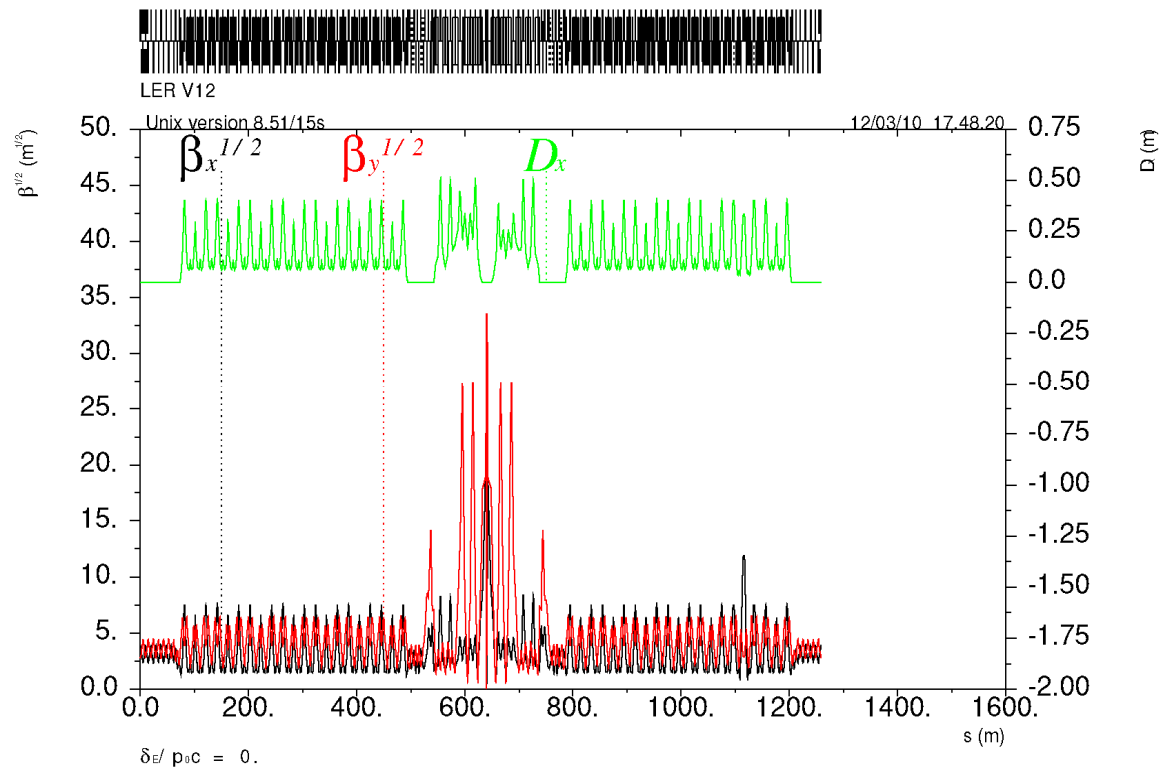

Fig. 6.17 – LER lattice functions, where IP is in the middle.

## 6.10 Rings dynamic aperture

### Introduction

The SuperB lattice has been converted from the *MAD* input file to that of the *Acceleraticum* simulation code; and size of the dynamic aperture was explored under various conditions. Main parameters for HER and LER lattices by *Acceleraticum* are listed in Table 6.1. The fractional tunes (0.54, 0.57) are recommended as optima for reaching high luminosity: beam-beam simulations for bare linear lattice have shown that due to the moderate value of the beam-beam parameter ($\xi_y \approx 0.1$) neither beam blow up nor beam lifetime reduction are observed for the design luminosity (see Chapter 8). However a slightly different fractional tune point (0.575, 0.595) has been chosen which provides relatively large momentum dynamic aperture (see Fig.6.23) increasing when the horizontal tune moves away from the half integer resonance (Fig.6.24).

Table 6.1: Main SuperB parameters.

| **Parameters** | **LER** | **HER** |
|---|---|---|
| Circumference, $L$ (m) | 1258 | |
| Energy, $E$ (GeV) | 4.18 | 6.70 |
| Compaction factor, $\alpha$ | $4.04 \cdot 10^{-4}$ | $4.35 \cdot 10^{-4}$ |
| Emittance coupling factor, $\kappa$ (%) | 0.25 | 0.25 |
| Horizontal emittance, $\varepsilon_x$ (nm) | 1.83 | 1.98 |
| Energy spread, $\sigma_E$ | $6.68 \cdot 10^{-4}$ | $6.31 \cdot 10^{-4}$ |
| Damping times, $\tau_x/\tau_z$ | 40.6/20.3 | 26.7/13.3 |
| Betatron tunes, $\nu_x/\nu_z$ | 42.575/18.595 | 40.575/17.595 |
| Synchrotron tune | 0.0129 | 0.0135 |
| Natural chromaticity, $\xi_x/\xi_z$ | -137/-449 | -134/-447 |
| Beta functions @ IP, $\beta_x/\beta_z$ (cm) | 2.6/0.0274 | 2.6/0.0274 |
| Beam size @ IP, $\sigma_x/\sigma_z$ (μm) | 7.16/0.037 | 6.88/0.035 |
| Bunch length, $\sigma_s$ (cm) | 0.408 | 0.420 |

All damping parameters are calculated for synchrotron radiation only and no IBS is taken into account at this stage.

The dynamic aperture study consists of:

(1) simulation of a stable particle motion area under as full as possible list of perturbations: chromatic sextupoles, crab sextupoles, magnet fringe fields, kinematic terms, synchrotron oscillation, lattice errors, etc.;

(2) optimization of the dynamic aperture by nonlinear correctors and tune point modification if necessary.

However this section will concentrate mainly on the first issue. Should the dynamic aperture require further optimization, the second step will be performed during the TDR stage.

### Perturbation sources

All sextupoles can be classified in three families: (1) crab sextupoles, (2) IR chromatic sextupoles which compensate chromaticity of the FF quadrupoles (both for tunes and betas), and (3) arc chromatic sextupoles which correct chromaticity of the rest part of the ring. The crab sextupoles are placed at the proper betatron phase advance relative to the IP and to each other. This phasing is equal to the $-I$ condition which cancels the second order aberrations outside of the sextupole pair. The IR chromatic sextupoles are arranged in non-interleaved pairs which also provide effective cancellation of the second order terms. The arc $-I$ sextupole pairs are partially overlapped.

Optical functions profile for the IR straight section (where the bulk of nonlinear perturbation concentrates) is shown in Fig. 6.18.

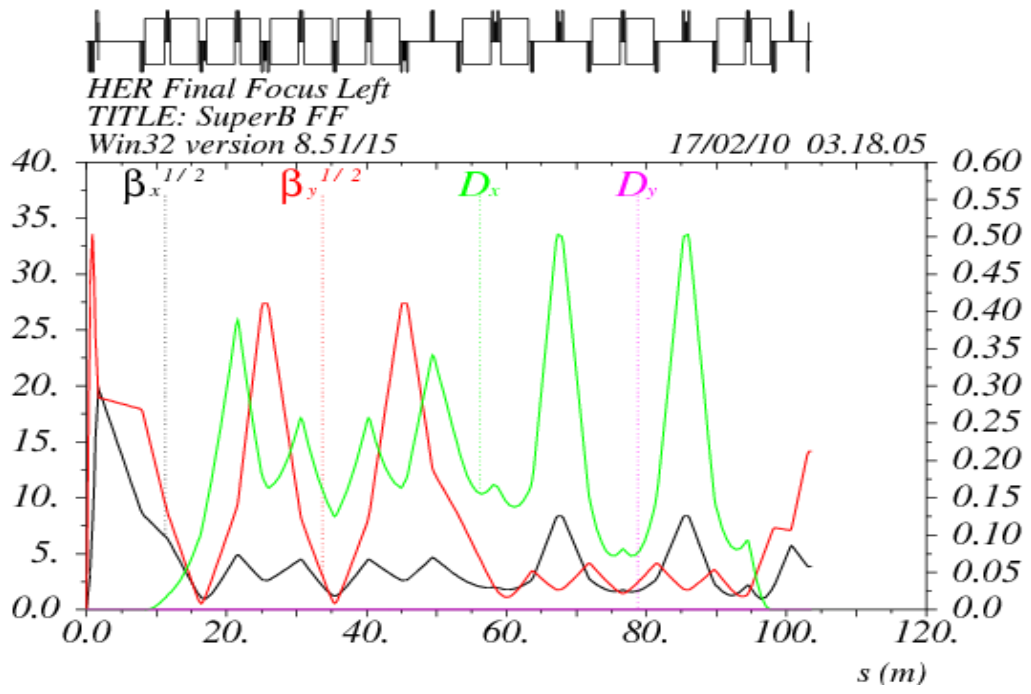

Figure 6.18: IR optical functions: two peaks of $\beta_y$ at S-position of ≈25 m and ≈45 m correspond to the vertical IR chromatic sextupoles while two peaks of $\beta_x$ at S ≈ 65 m and ≈85 m relate to the horizontal ones.

Finite length of sextupole magnets in case of $-I$ separation gives effective cancellation for the second order terms, but generates higher order terms. With accuracy up to third order in initial coordinates, a transformation through such sextupole pair is given by [2]:

$$x_1 = -x_0 - p_{x0}L - \frac{(k_2L)^2}{12}\left(x_0^3 + x_0y_0^2\right)L^2 + O(L^5),$$
$$p_{x1} = -p_{x0} - \frac{(k_2L)^2}{6}\left(x_0^3 + x_0y_0^2\right)L - \frac{(k_2L)^2}{12}\left(3p_{x0}x_0^2 + 2p_{y0}x_0y_0 + p_{x0}y_0^2\right)L^2 + O(L^5),$$
$$y_1 = -y_0 - p_{y0}L - \frac{(k_2L)^2}{12}\left(x_0^2y_0 + y_0^3\right)L^2 + O(L^5), \qquad (1)$$
$$p_{y1} = -p_{y0} - \frac{(k_2L)^2}{6}\left(x_0^2y_0 + y_0^3\right)L - \frac{(k_2L)^2}{12}\left(p_{y0}x_0^2 + 2p_{x0}x_0y_0 + 3p_{y0}y_0^2\right)L^2 + O(L^5),$$

Note, that cubic monomial of the obtained expression differs from the standard octupole (like for example $x_1 = x_0 + p_{x0}L - \frac{k_3}{12}\left(x_0^3 - 3x_0y_0^2\right)L^2$) and thus it is not possible to cancel exactly this monomial by means of octupole lenses. The value of cubic perturbation for a finite length sextupole is the square of the integrated strength $\left(k_2L\right)^2$.

This effect can reduce the dynamic aperture but in our case surprisingly the third order nonlinearity from the quadrupole fringe fields and the kinematic terms (see below) recover partly damage from the sextupole third order aberration and increase the dynamic aperture.

In the simulation the sextupole magnets are treated as a sequence of (many) symplectic kicks spaced along the magnet length, so the higher order effects due to the non-zero sextupole length are included.

Fringe fields exist for all kinds of magnets and they are an important source of nonlinearities in beam dynamics. If the betatron functions reach high values in quadrupole magnets, the quadrupole fringe field nonlinearities should be carefully taken into account.

In the hard edge approximation, a symplectic 6D map through the rising field of the quadrupole is given by the formulae below [3], where $k_1$ and δ are the quadrupole strength and energy deviation respectively:

$$x = x_0 + \frac{k_1}{12(1+\delta)}\left(x_0^3 + 3x_0y_0^2\right),$$
$$p_x = \left[p_{x0} - \frac{k_1}{4(1+\delta)}\left[p_{x0}\left(x_0^2 + y_0^2\right) - 2p_{y0}x_0y_0\right]\right]\cdot\left[1 - \frac{k_1^2}{(1+\delta)^2}\frac{\left(x_0^2 - y_0^2\right)^2}{16}\right],$$
$$y = y_0 - \frac{k_1}{12(1+\delta)}\left(y_0^3 + 3x_0^2y_0\right), \qquad (2)$$
$$p_y = \left[p_{y0} + \frac{k_1}{4(1+\delta)}\left[p_{y0}\left(x_0^2 + y_0^2\right) - 2p_{x0}x_0y_0\right]\right]\cdot\left[1 - \frac{k_1^2}{(1+\delta)^2}\frac{\left(x_0^2 - y_0^2\right)^2}{16}\right].$$
$$z = z_0 - \frac{k_1}{12(1+\delta)^2}\left[p_{x0}x_0\left(x_0^2 + 3y_0^2\right) - p_{y0}y_0\left(y_0^2 + 3x_0^2\right)\right]$$

Since the betatron functions in SuperB lattice reach rather low minima (especially the vertical one at IP), the effect of the kinematic terms on the nonlinear beam dynamics has been considered. Expecting the main contribution from the IP section, we take into account only the leading kinematic term described by:

$$H_2 = \frac{1}{8}\left(p_x^2 + p_y^2\right)^2 \qquad (3)$$

It is worthwhile to note that the kinematic effect does not depend only on β-functions (as magnetic nonlinear terms do) but on this Twiss parameter:

$$\gamma = (1+\alpha^2)/\beta.$$

For instance the detuning coefficients, defined as:

$$\Delta\nu_x = C_{xx}J_x + C_{xy}J_y,$$
$$\Delta\nu_y = C_{yx}J_x + C_{yy}J_y,$$

where $J$ is the action, can be estimated in the first perturbation order as [4]:

$$C_{xx} = \frac{3}{16\pi}\oint\gamma_x^2(s)ds,$$
$$C_{yy} = \frac{3}{16\pi}\oint\gamma_y^2(s)ds,$$
$$C_{xy} = C_{yx} = \frac{1}{8\pi}\oint\gamma_x(s)\gamma_y(s)ds.$$

Fig. 6.19 shows the distribution of the vertical γ-parameter along the beam orbit. The maximum gamma relates to the minimum beta at a waist position; and it is clearly seen that this occurs only close to the IP. This fact allows us to consider the kinematic terms in the FF straight section only and save processor time. In this case Hamiltonian equations corresponding to the Hamiltonian in Eq. 3 can be solved explicitly and the relevant transformation has been implemented in the computer code.

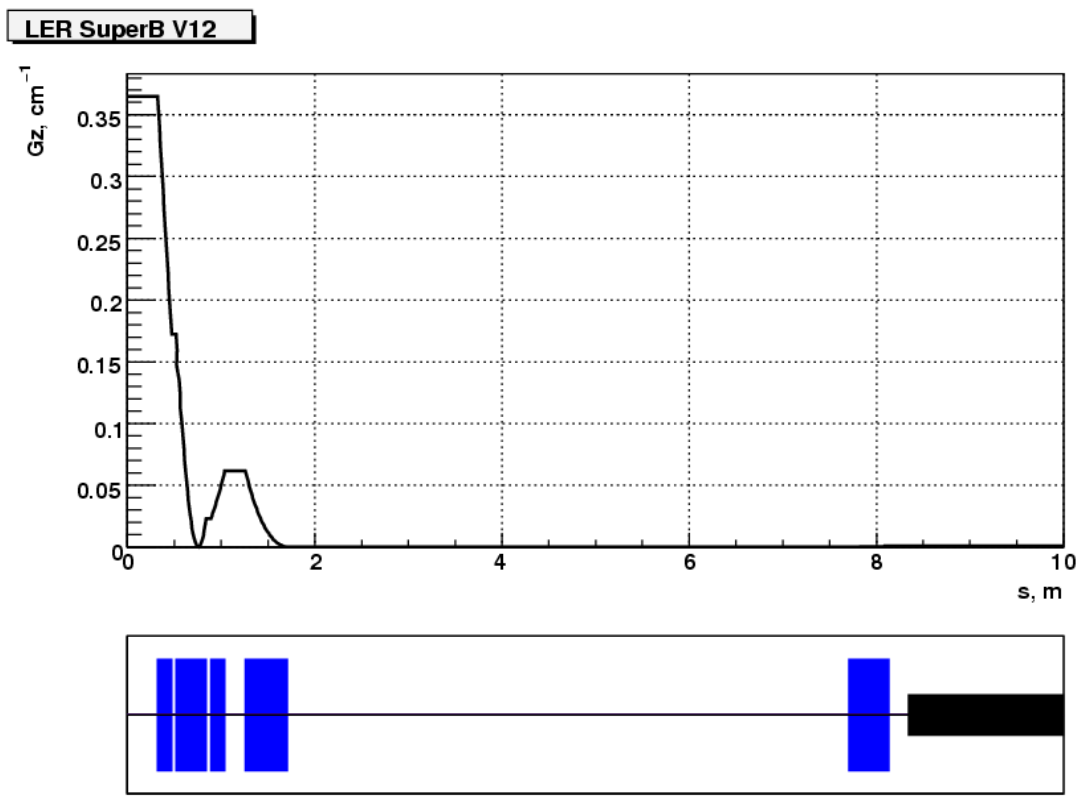

Figure 6.19: Twiss function $\gamma_z(s)$ near IR starting from the IP.

Because of the extremely large β-peaks, the SuperB lattice is sensitive to the magnetic field imperfection placed at the azimuth of high betas. One kind of such imperfection should be mentioned here: small optical detuning of the $-I$ transformation produces imperfect cancellation of the second order terms. If $\delta\mu_{x,y}$ is the phase advance error in the sextupole pair, the residue second order variables transformation has the form:

$$\Delta x = \frac{(k_2 L)}{2}\left(x_0^2 - y_0^2\right)\cdot \beta_x \cdot \delta\mu_x ,$$

$$\Delta p_x = \frac{(k_2 L)}{2}\left[\left(y_0^2 - 3x_0^2\right)\cdot \alpha_x \cdot \delta\mu_x + 2y_0^2 \cdot \alpha_y \cdot \delta\mu_y\right], \quad (4)$$

$$\Delta y = -(k_2 L) x_0 y_0 \cdot \beta_y \cdot \delta\mu_y ,$$

$$\Delta p_y = (k_2 L) x_0 y_0 \left(\alpha_x \cdot \delta\mu_x + 2\alpha_y \cdot \delta\mu_y\right).$$

## Dynamic aperture calculation

As usually, dynamic aperture is defined for 1000 particle revolutions (corresponding to 13 synchrotron oscillations) for both on and off-energy cases. Observation point is at the IP with the beam parameters listed in Table 6.1.

As it was mentioned above, third order effects from the quadrupole fringe fields and the kinematic terms may cure the third order effect from the chromatic sextupoles. However the crab sextupoles reduce effectiveness of this recovering and below we distinguish two cases with and without the crab sextupoles.

Fig. 6.20 shows the on-energy dynamic aperture for LER and HER with and without the crab sextupoles. In spite of the crab sextupoles substantially decreasing the dynamic aperture, this still seems enough to reach the project luminosity with $\xi_y \approx 0.1$.

In principle, this effect can be improved by weak sextupole correctors placed close to the crab sextupoles, but in the present lattice there is lack of space in the vicinity of the crab sextupoles. Main parameters of the crab sextupoles are given in Table 6.2.

Table 6.2: Crab sextupole parameters.

| Parameters | LER | HER | Units |
|---|---|---|---|
| Length, $L$ | 35 | 35 | cm |
| Strength, $K_2$ | 33.34 | 33.34 | $m^{-3}$ |
| Horizontal beta, $\beta_x$ | 14.6 | 14.6 | m |
| Vertical beta, $\beta_z$ | 200 | 200 | m |

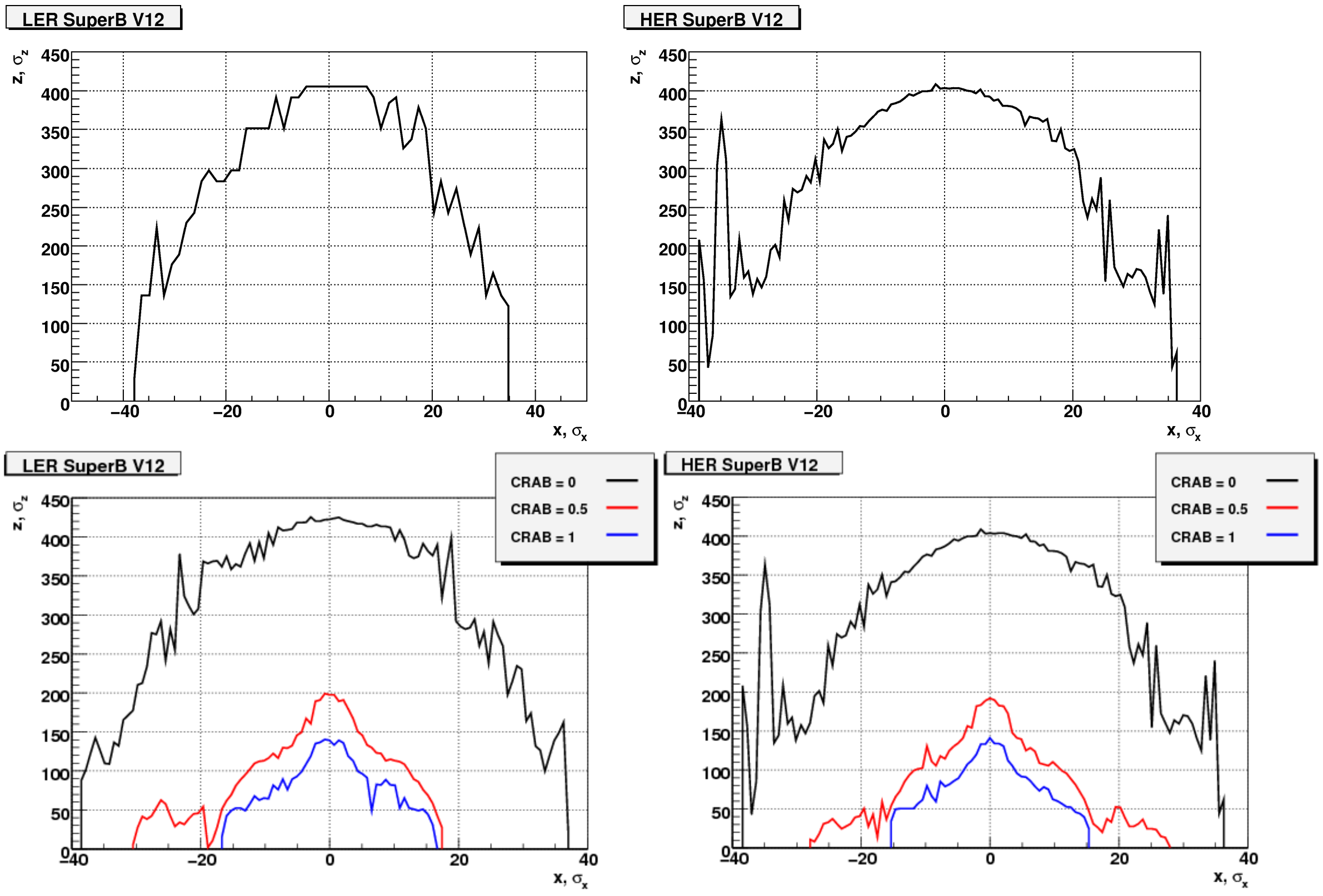

Figure 6.20: On-energy dynamic aperture without (top) and with (bottom) crab sextupoles.

For a better understanding of the crab sextupole effect one can examine the phase space portraits of the particle motion with and without the crab sextupoles (Figs. 6.21 and 6.22).

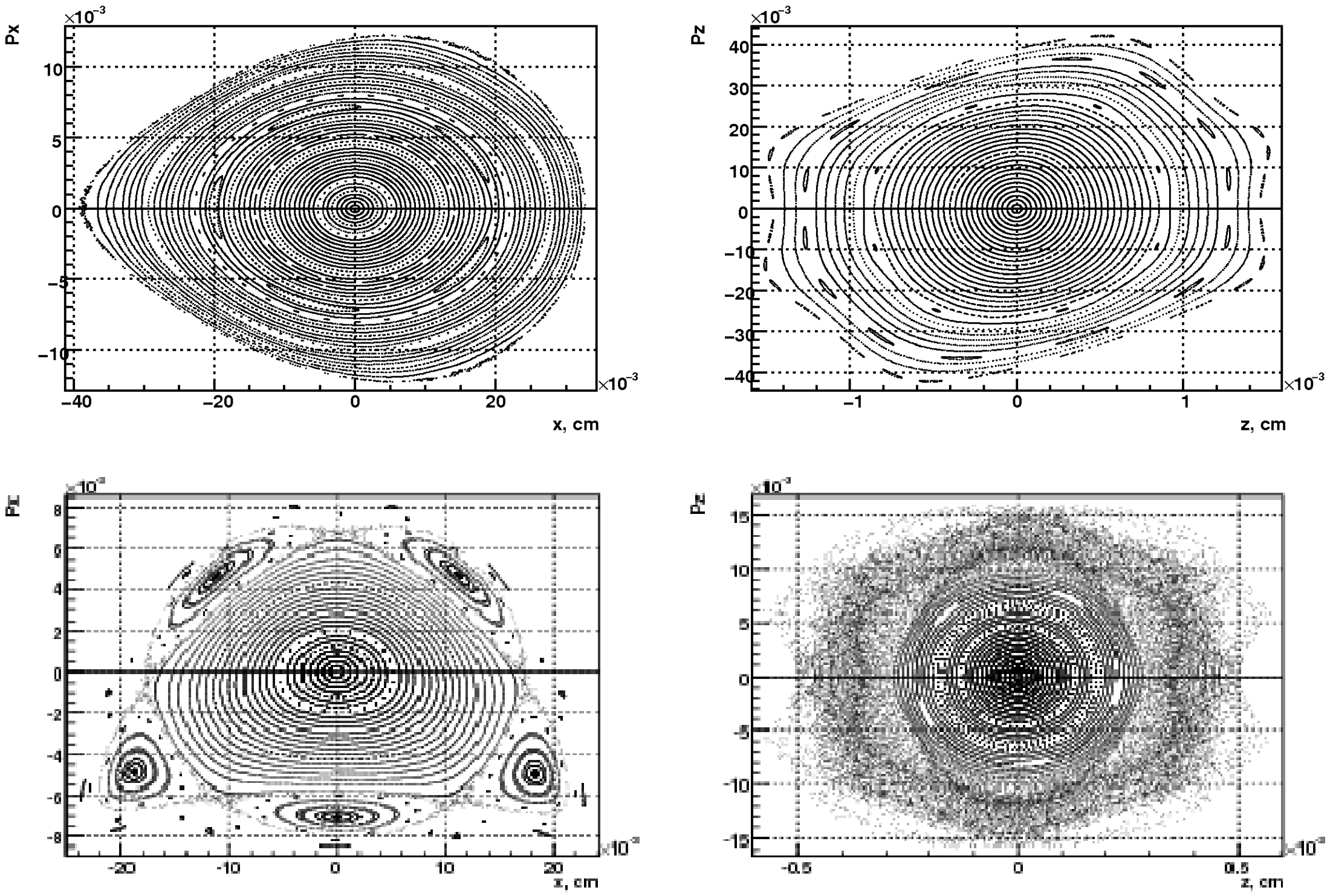

Figure 6.21: Phase trajectories for the LER without (top) and with (bottom) crab sextupoles.

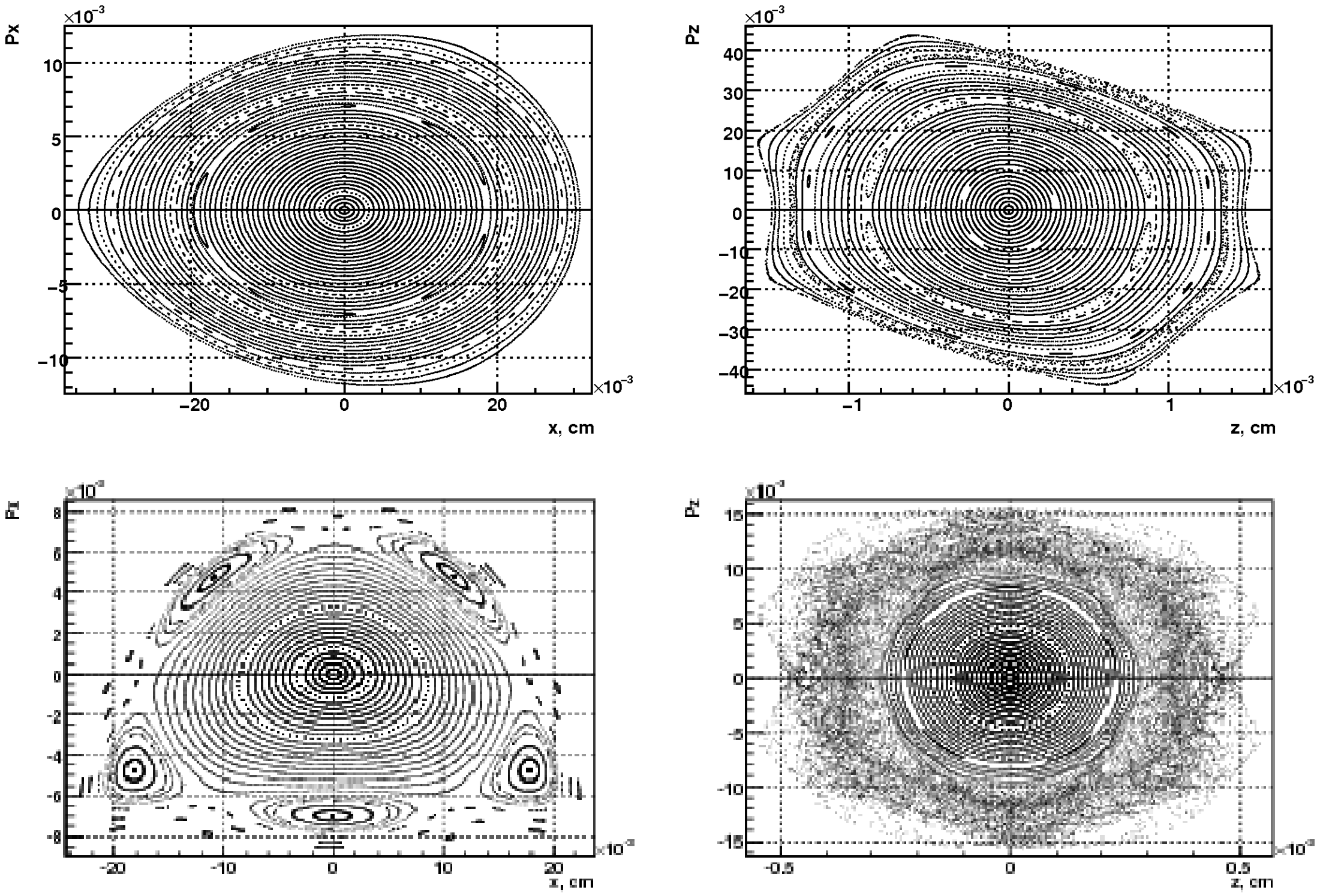

Figure 6.22: Phase trajectories for the HER without (top) and with (bottom) crab sextupoles.

The crab sextupoles introduce strong irregularity in the vertical motion and produce a rather wide 5$^{th}$ order resonance island inside the stable area of the horizontal motion. In spite of the fact that for an "ideal motion" this resonance is stable and the dynamic aperture extends beyond its separatrix, one can expect that lattice errors and imperfections can destroy the regular trajectory above the 5$^{th}$ resonance and the dynamic aperture can shrink.

Besides the phase portraits, useful information on the nonlinear system behaviour can be found from the tune-amplitude dependence (see Figure 6.23).

For the IR sextupoles only, the particle amplitude increases when the tunes approach the half integer resonance. It is well known that close to the half integer resonance the relevant tune-amplitude term demonstrates the resonance behavior:

$$\nu(A) \sim \frac{A^{2k}}{2\nu - n},$$

so the tune at high amplitude very fast reaches the unstable region. When the quadrupole fringe fields and kinematic effects are included, they change the sign of the tune-amplitude dependence, so the tunes move away from the half integer resonance and the dynamic aperture opens up.

The crab sextupoles do not significantly change the tune-amplitude behavior (see Table 6.3) but (a) introduce rather strong nonlinear betatron coupling (it is clearly seen from the vertical phase space smear with the crab sextupoles in Fig. 6.21, 6.22) and (b) generate high order resonances (see horizontal phase space trajectories in Fig. 6.21, 6.22). The first order coefficients for the tune versus action dependence $J \sim A^2$ are summarized in Table 6.3. Tune-amplitude dependence coefficients are a very useful tool to estimate quantitatively the strength of different perturbation sources. Regarding the oscillation modes distortion, one can say that the vertical motion prevails, after that the coupling mode comes and the last is the horizontal motion. Regarding the perturbation sources, the strongest are the IR sextupoles and then the quadrupole fringe fields and the kinematic terms (with the sign opposite to the IR sextupoles).

Off-energy dynamic aperture with and without the crab sextupoles is shown in Fig. 6.24. Both plots include the effect of synchrotron oscillations. RF parameters for this simulation are listed in Table 6.4.

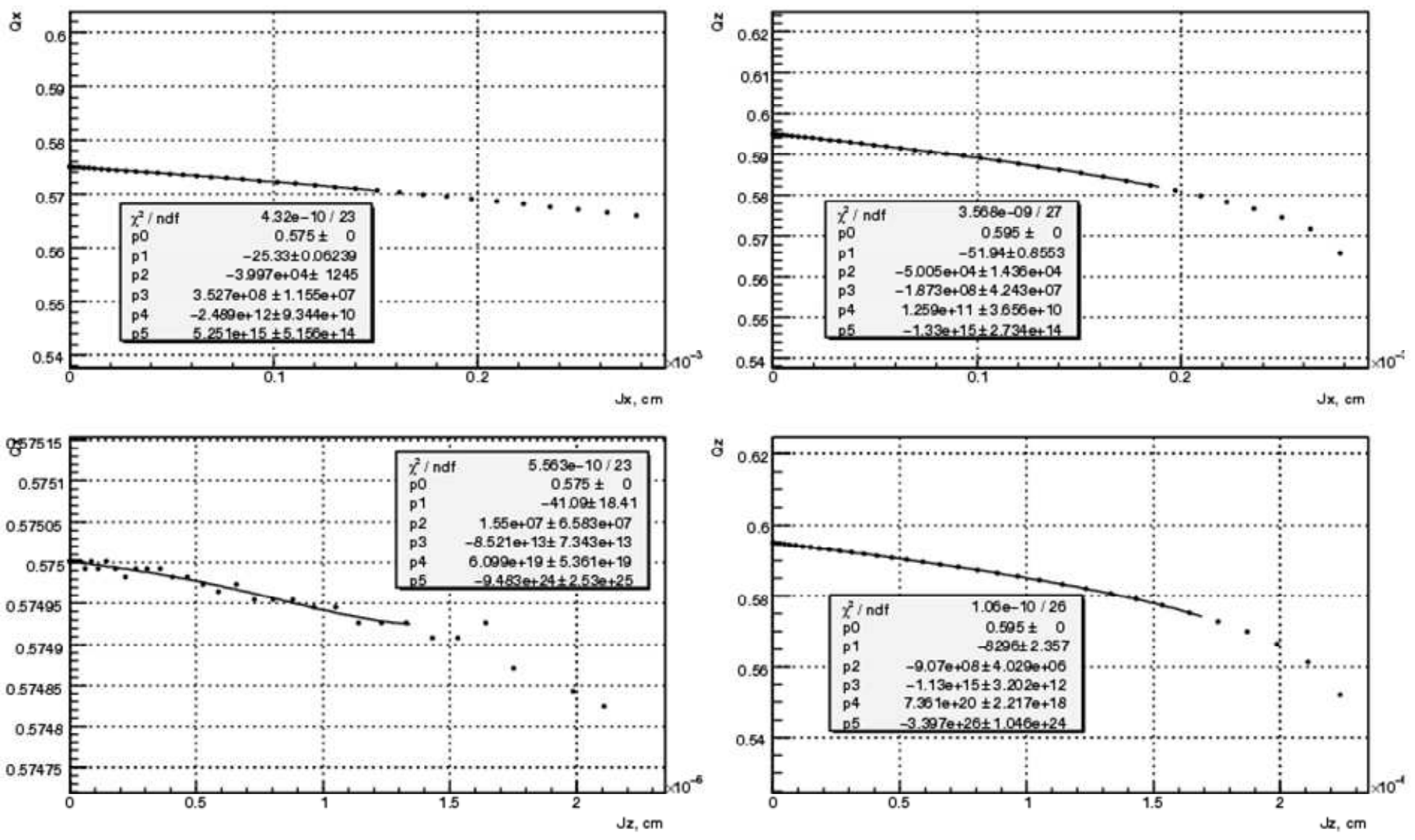

*IR sextupoles only*

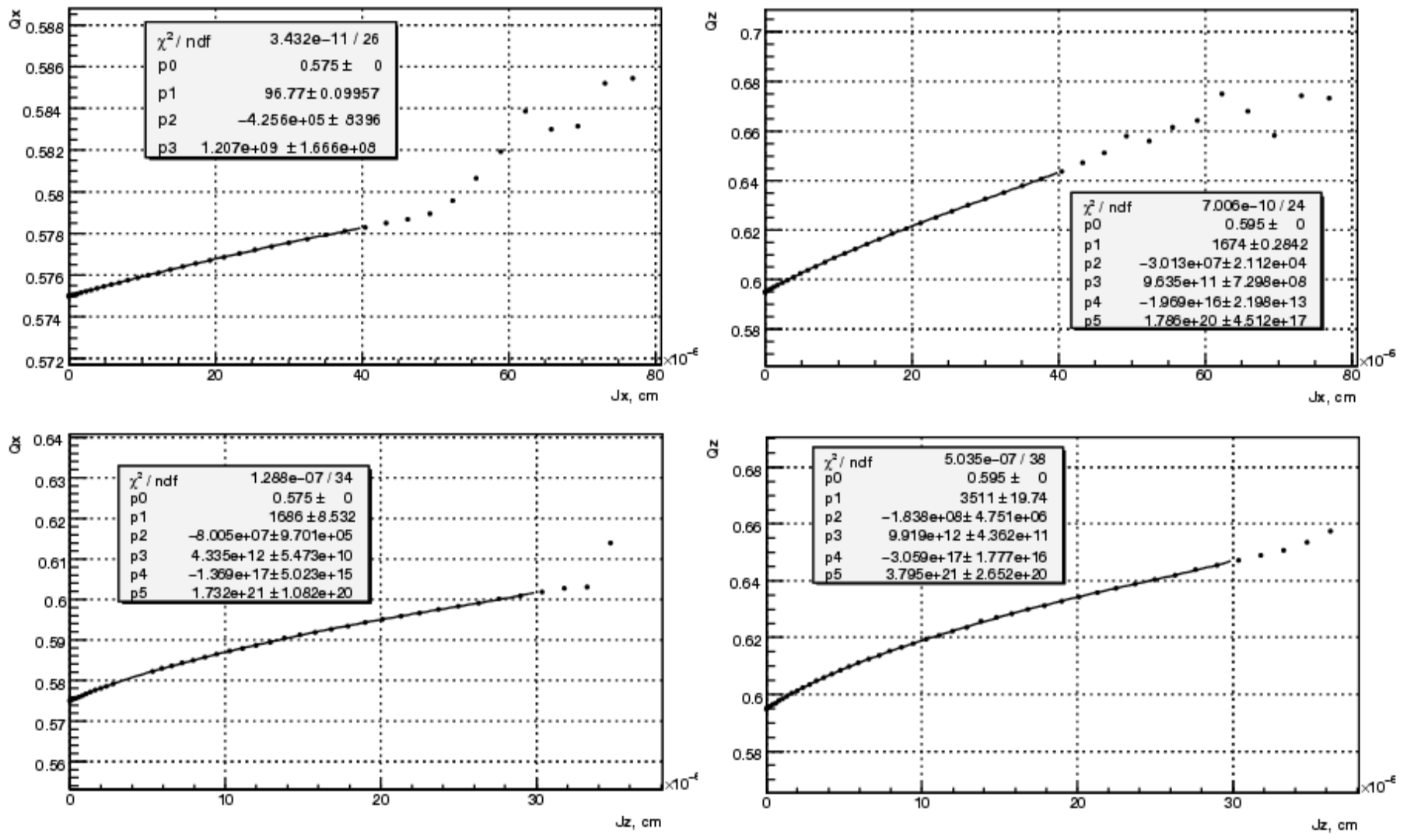

*IR sextupoles+Quad Fringe+Kinematics*

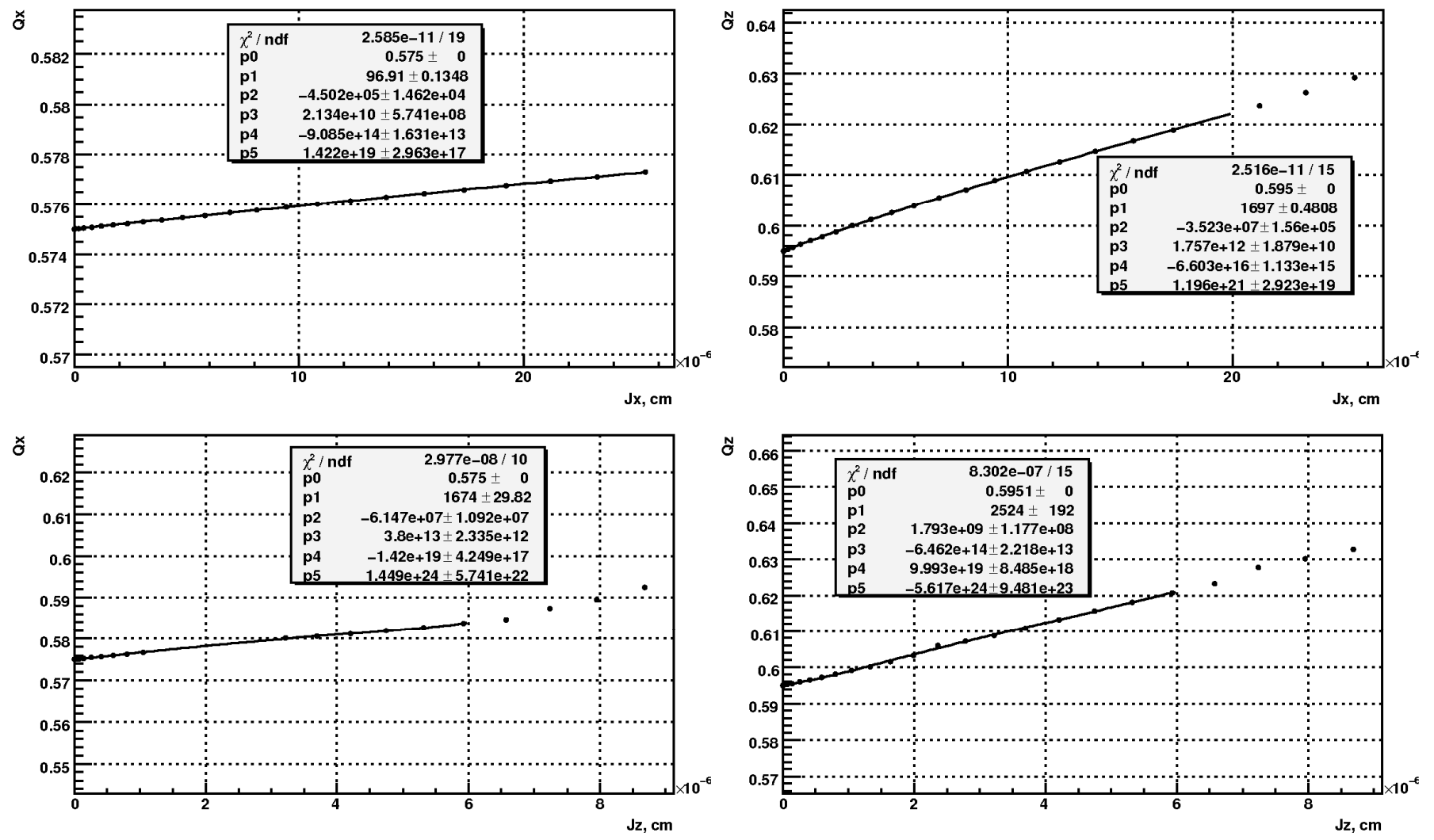

*IR sextupoles+Quad Fringe+Kinematics+Crab Sextupoles*

Figure 6.23: Tune dependence with particle amplitude for various non-linear contributions.

Table 6.3 Tune-action coefficients.

| Type | LER | | | HER | | | |
|---|---|---|---|---|---|---|---|
| | $C_{xx}$ | $C_{xz}$ | $C_{zz}$ | $C_{xx}$ | $C_{xz}$ | $C_{zz}$ | **Units** |
| IP Sextupoles | –25 | –51 | –8300 | –25 | –51 | –8300 | $cm^{-1}$ |
| Crab sextupoles | –0.6 | –5.5 | –114 | –0.6 | –5.5 | –114 | $cm^{-1}$ |
| Arcs sextupoles | 273 | –450 | –93 | 270 | –330 | –71 | $cm^{-1}$ |
| **Sub total: all sextupoles** | **247** | **–510** | **–8520** | **253** | **–402** | **–6510** | $\mathbf{cm^{-1}}$ |
| Octupoles | –120 | 112 | 384 | –124 | 142 | 365 | $cm^{-1}$ |
| Quadrupole fringe fields | 240 | 1440 | 5750 | 240 | 1510 | 5830 | $cm^{-1}$ |
| Kinematic term | 0.6 | 35 | 5090 | 0.6 | 35 | 4440 | $cm^{-1}$ |
| **Total** | **368** | **1205** | **3380** | **375** | **1320** | **3390** | $\mathbf{cm^{-1}}$ |

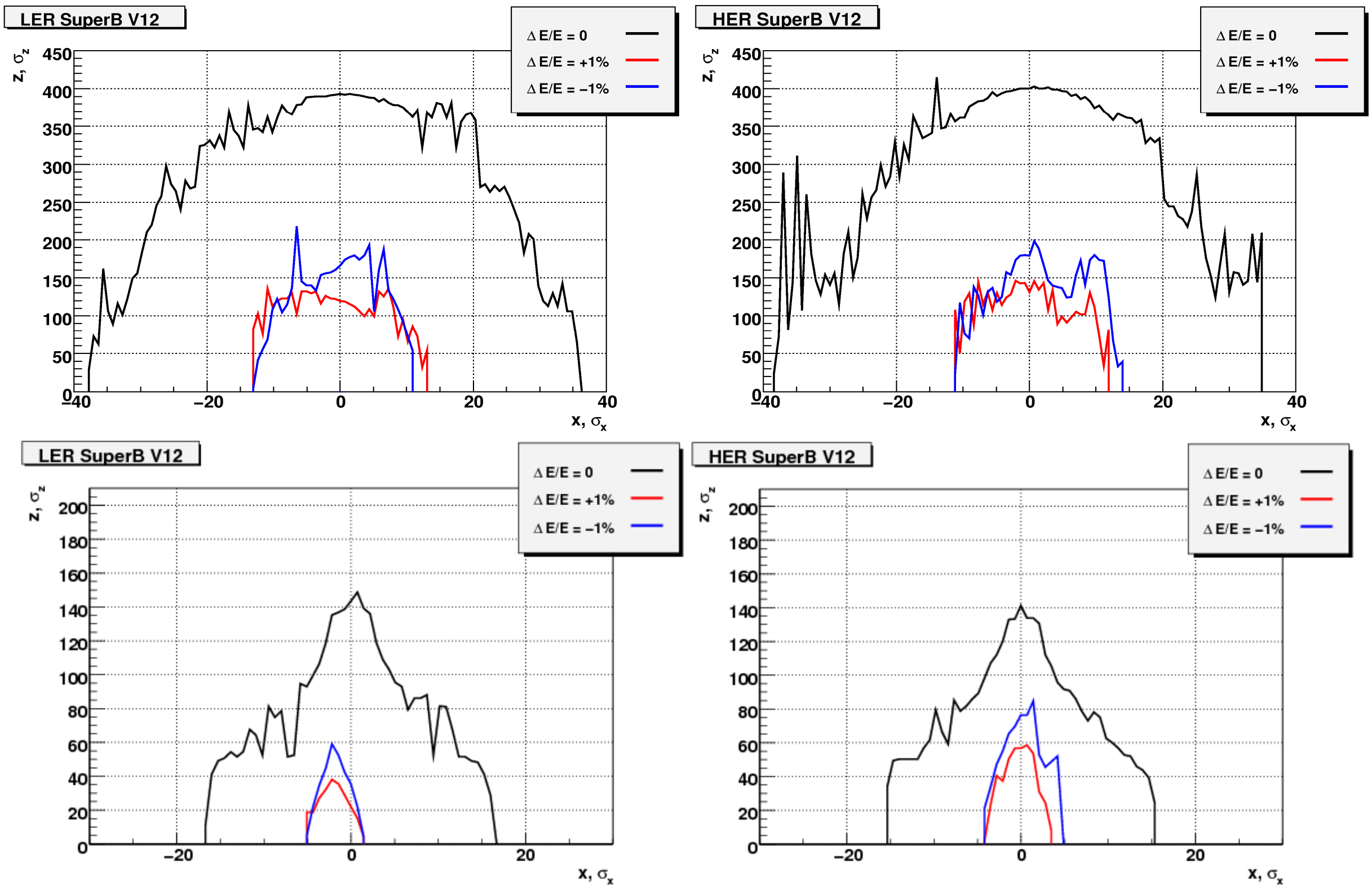

Figure 6.24: Off-energy dynamic aperture without (top) and with (bottom) crab sextupoles.

Table 6.4: RF parameters for off-energy tracking.

| Parameters | LER | HER | Units |
|---|---|---|---|
| RF stations | 8 | 28 | |
| RF Frequency | 479.1 | | MHz |
| RF voltage | 0.75 | 0.35 | MV |
| RF phase | 152.8 | 152.8 | degree |
| Synchrotron tune | 0.0129 | 0.0135 | |

Betatron tunes as a function of the beam energy deviation are depicted in Fig. 6.25. Even if there are stable tune points behind the positive momentum deviation of ~1% (below the half integer resonance in both direction), particles are unstable at the fractional tune of 0.5 and the value of $\Delta E / E_0 \approx 1\%$ limits the bandwidth. In absence of gradient errors, which may produce strong half-integer resonances, in our case the limitation comes from relatively weak (but still unstable) resonance generated by the third order perturbation terms.

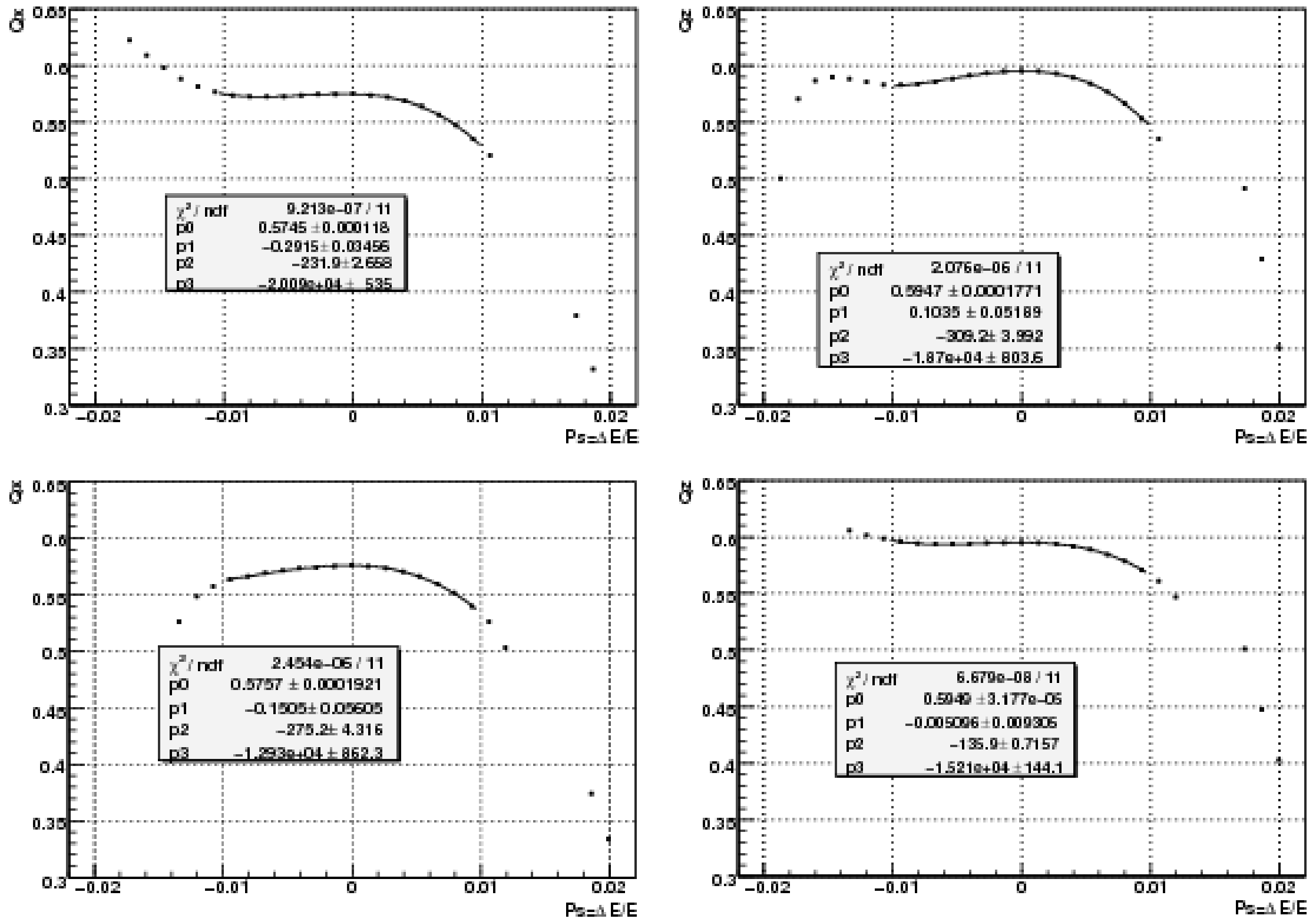

Figure 6.25: Tune-momentum dependence for LER (top) and HER (bottom). All sextupoles, fringes and kinematics are included.

The size of the horizontal dynamic aperture as a function of energy deviation is demonstrated in Figure 6.26, while Figure 6.27 shows dependence of this aperture on the distance from the half integer resonance. It is seen that the aperture tends to increase with betatron tune moving away from 0.5.

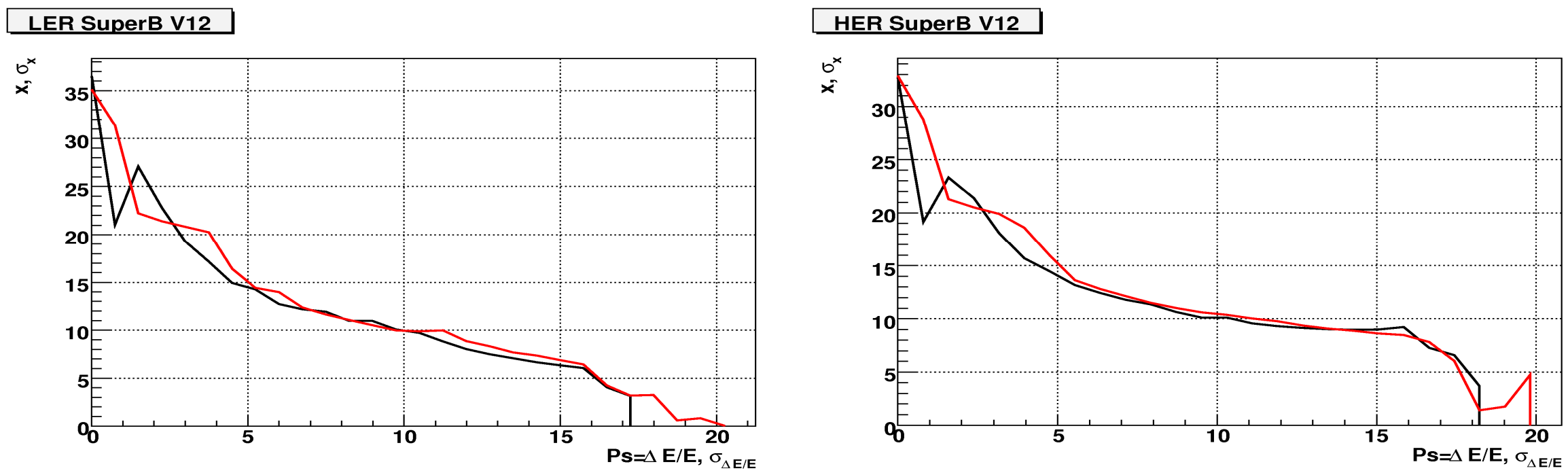

Figure 6.26: Horizontal DA vs. energy deviation: synchrotron oscillations are on (red) and off (black) for LER (top) and HER (bottom).

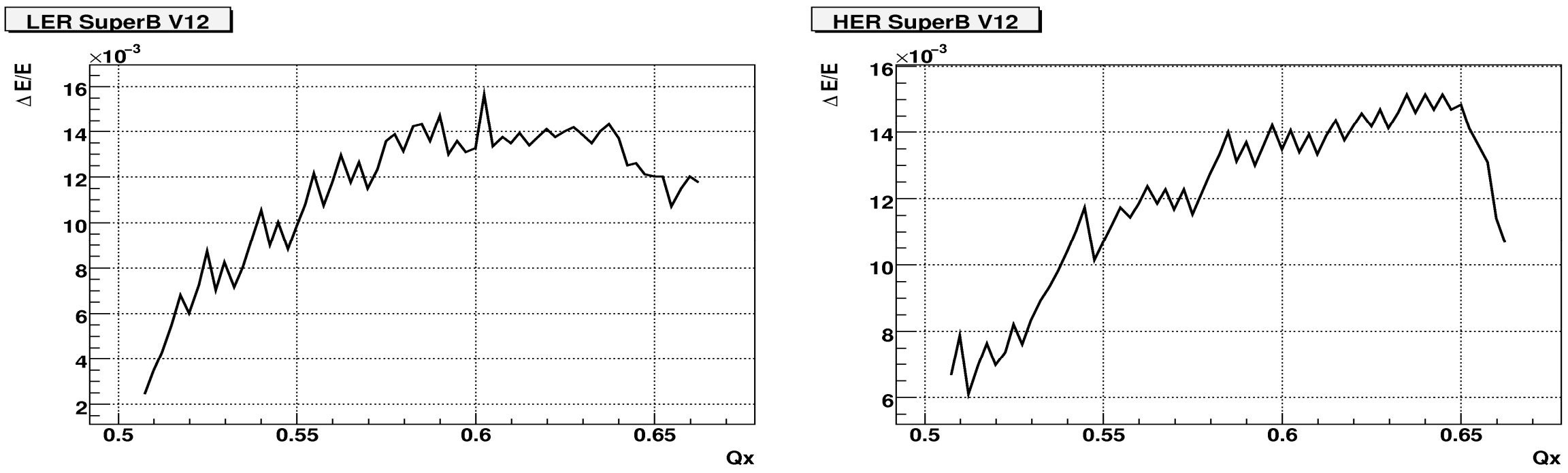

Figure 6.27: Energy acceptance vs. horizontal betatron tune for LER (top) and HER (bottom).

## Main sources of dynamic aperture limitation

An important issue of the dynamic aperture exploration is the analysis of the main perturbation sources in the rings. In our case we distinguish:

- Chromatic sextupoles in the Interaction Region,
- Crab sextupoles,
- Chromatic sextupoles in the arcs,
- Leading (third) order term in the quadrupole fringes,
- Leading kinematic terms due to the extremely low vertical beta at IP.

The following sextupole magnets can be found in the IR (and the complementary pairs symmetrically to the IP):

- SDY1 & SDY2 in the vertical chromatic section,
- SFX1 & SFX2 in the horizontal chromatic section,
- SDY0 & SFX0 for larger energy dependent dynamic aperture.

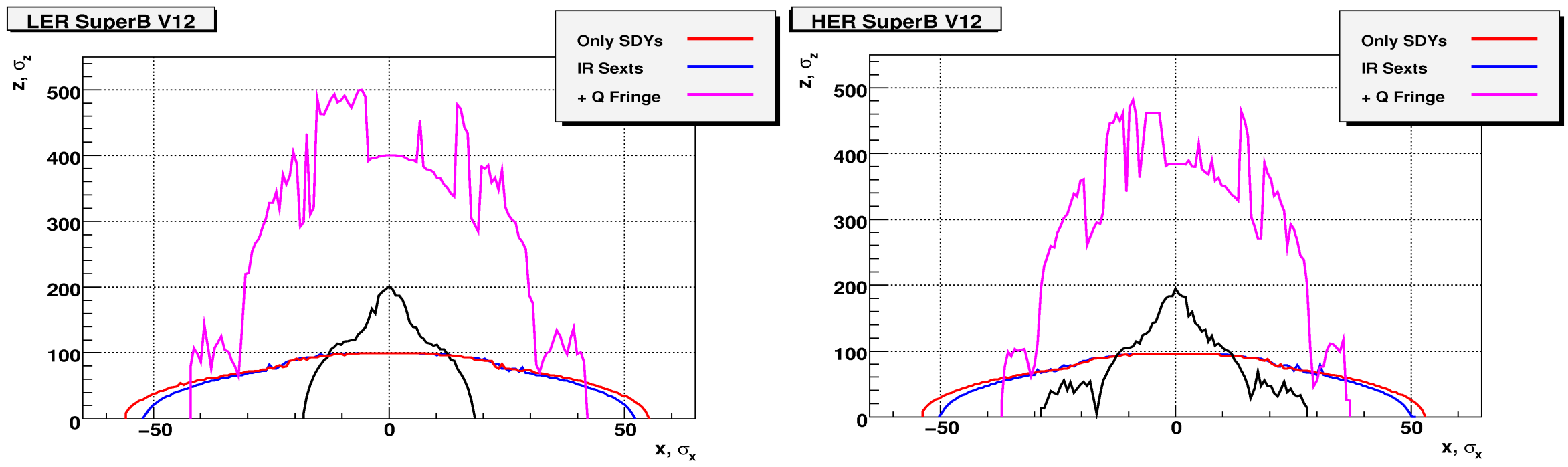

Fig. 6.28: On-energy DA due to the vertical IR sextupoles (red), all IR sextupoles (blue) and due to the combined effect of IR sextupoles + Quad Fringes + Kinematics (magenta). Here and below black curves refer to the DA with all nonlinearities included.

The strongest effect on dynamic aperture comes from the IR section of the vertical chromaticity correction (Fig. 6.28). The aperture size due to the horizontal section is much larger (~80 horizontal sigmas and ~12000 vertical sigmas).

Analysis shows that the DA limitation by the IR sextupoles is explained by a finite magnet length: when the length approaches zero (while the integrated strength is conserved) the dynamic aperture increases up to infinity. Including the quadrupole fringe fields and kinematic terms introduces positive tune shift with amplitude and recovers the vertical DA reduced by the IR sextupoles. The horizontal DA slightly shrinks but not critically. Such effect is possible because the leading term in all three kinds of perturbation (finite sextupole length, quadrupole fringe field and kinematics) comes from the third order aberration, which can interfere either constructively or destructively.

The origin of the DA limitation caused by the crab sextupoles is also explained by the finite magnet length. However, the crab sextupoles alone do not significantly reduce the dynamic aperture as it is illustrated in Fig.6.29. So the crab sextupole effect which is shown in Fig. 6.20 is caused by the interference between the crab sextupoles and the IR sextupoles. Also from Fig. 6.30 one may conclude that the arc sextupoles are not critical for the DA limitation.

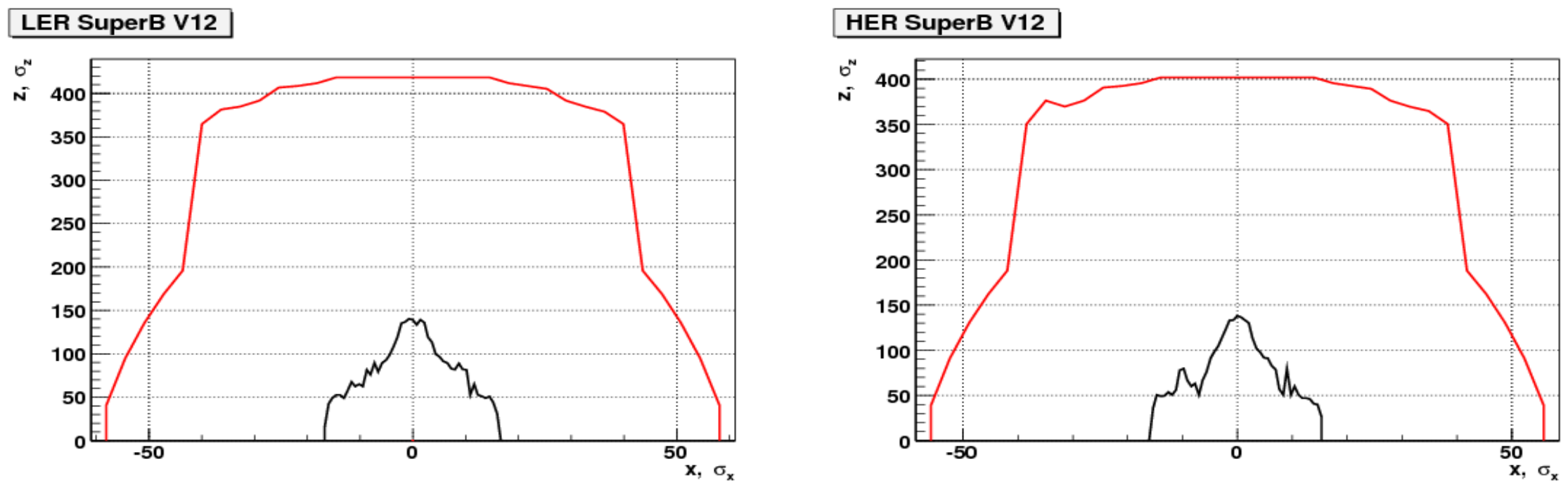

Fig. 6.29: On-energy dynamic aperture limited by the crab sextupoles only.

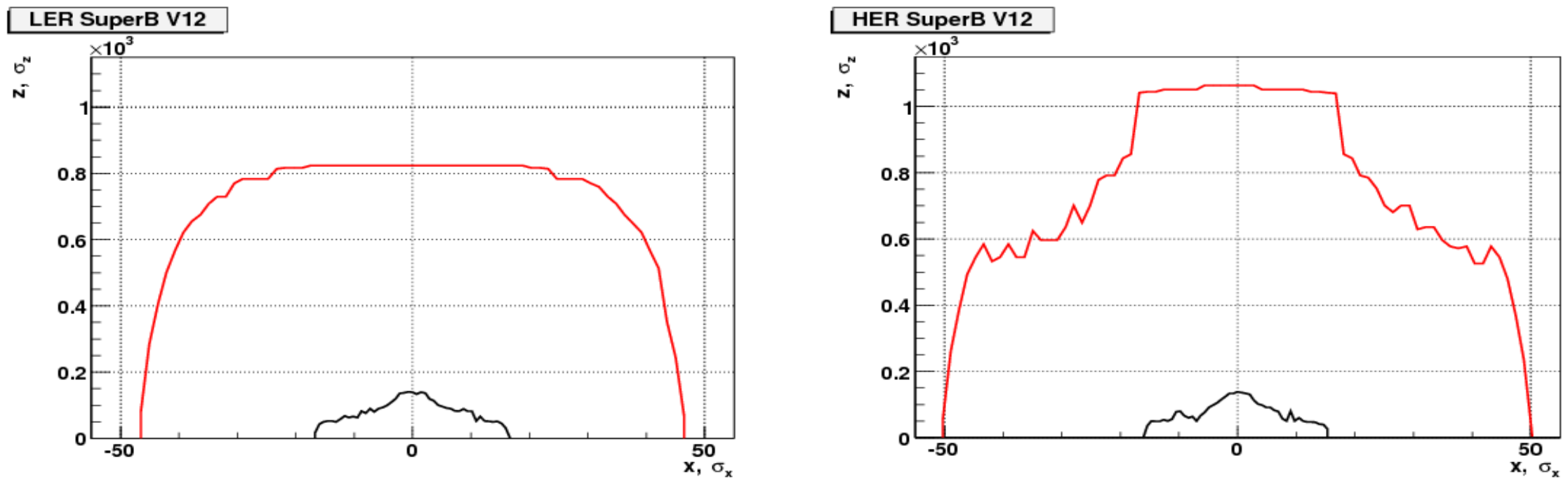

Fig. 6.30: On-energy dynamic aperture limited by the arc sextupoles only.

One more source of the nonlinear perturbation comes from the quadrupole fringe fields. Since this nonlinearity is rather weak, it can be only important when β-functions are high. And indeed, simulation shows that for SuperB the main contribution is caused by the FF quadrupoles, then by the quadrupoles in the chromatic correction sections, while all other quadrupole fringes do not influence the dynamic aperture at all (Fig. 6.31).

As for the kinematic terms, this effect alone does not limit the dynamic aperture but contributes to the tune-amplitude dependence, especially in the vertical direction.

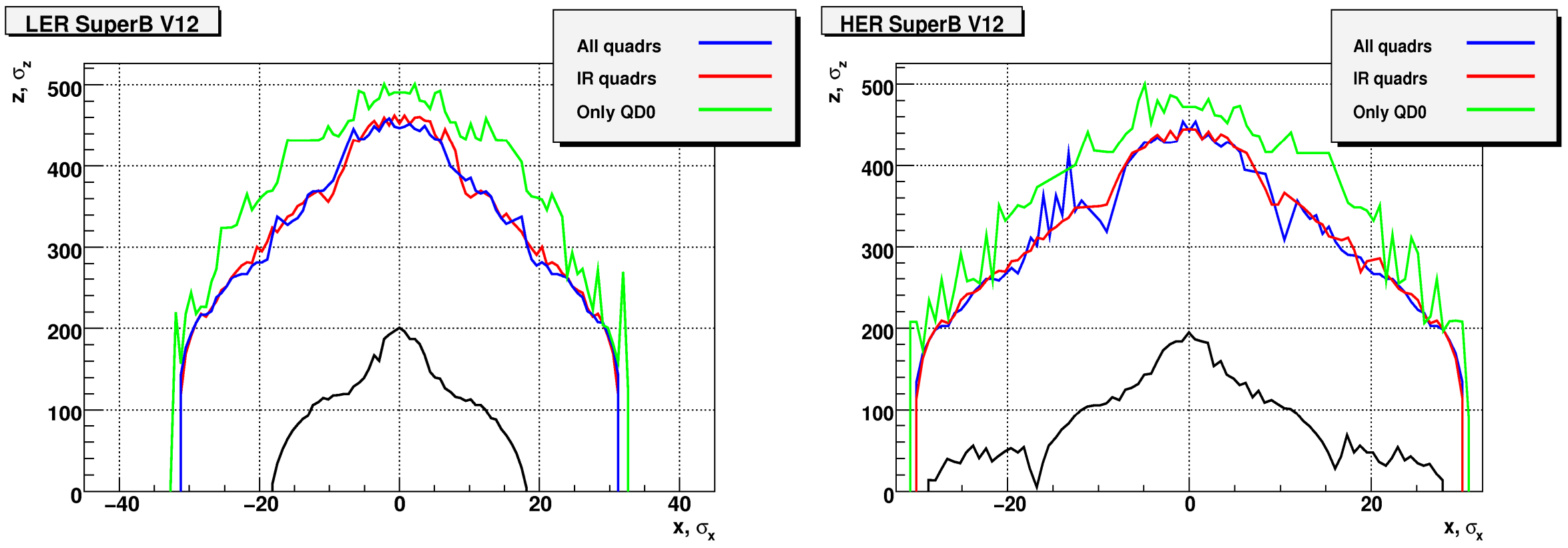

Fig. 6-31: Dynamic aperture limited by the quadrupole fringe fields. Green line corresponds to the first defocusing quadrupole ($\beta_z \approx 1200$ m), red line corresponds to all quadrupoles with $\beta > 200$ m and blue comes from all quadrupole fringe fields.

## Errors and imperfections

Since the lattices show similar nonlinear features, we study the errors influence for the LER only.

Taking into account high strength of the IR chromatic sextupoles, it is important to estimate sensitivity to perturbation of the $-I$ condition. Fig.6.32 shows the results for the vertical IR sextupole section (SDY1, SDY2) which is most critical for the DA limitation.

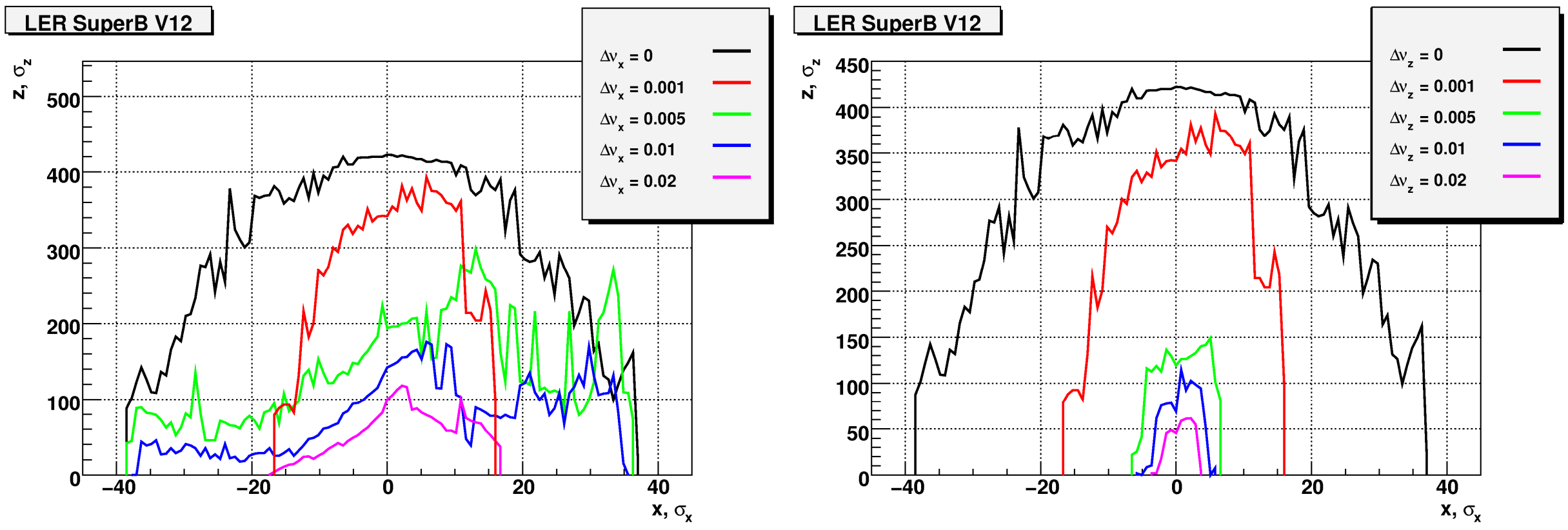

*Without the crab sextupoles*

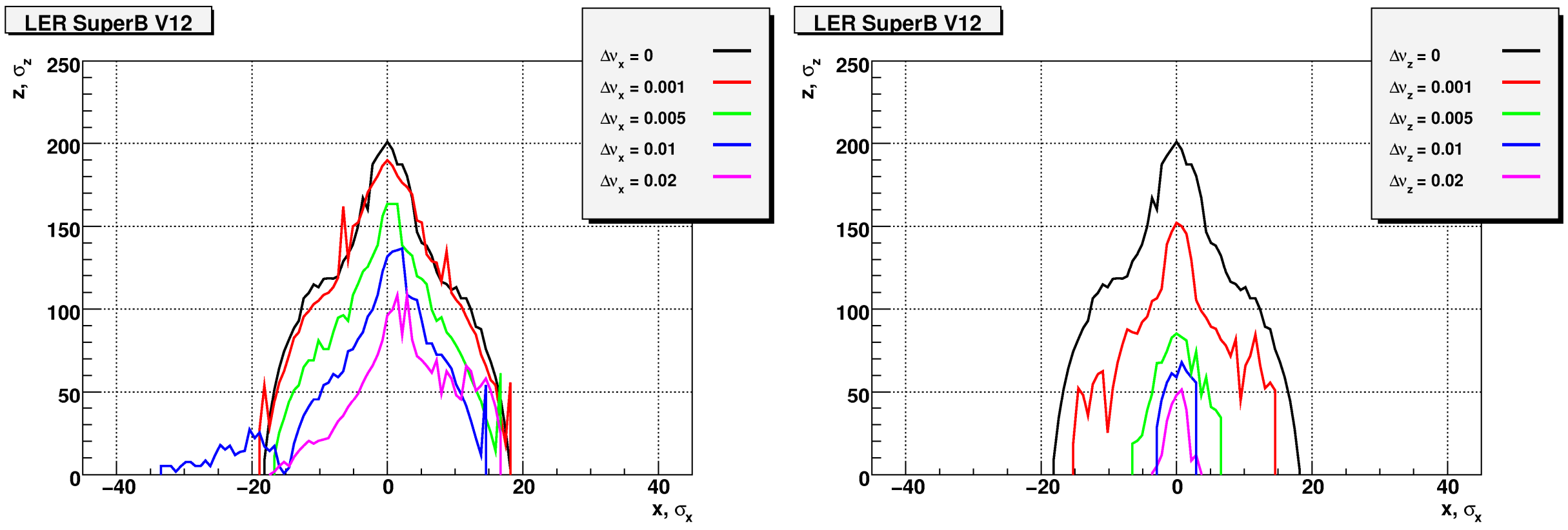

*With the crab sextupoles*

Fig. 6.32: Effect of $-I$ condition detuning for the IR vertical chromatic section.

In this simulation the phase advance between the sextupole centres is detuned while the betatron functions remain the same. Note that in Fig. 6.32 the betatron phase mismatch is given in terms of $\Delta\nu = \Delta\mu / 2\pi$ for one chromatic section; in the other section the phase advance is also changed identically.

Similar calculations, but for the crab sextupoles mismatch are illustrated in Fig. 6.33.

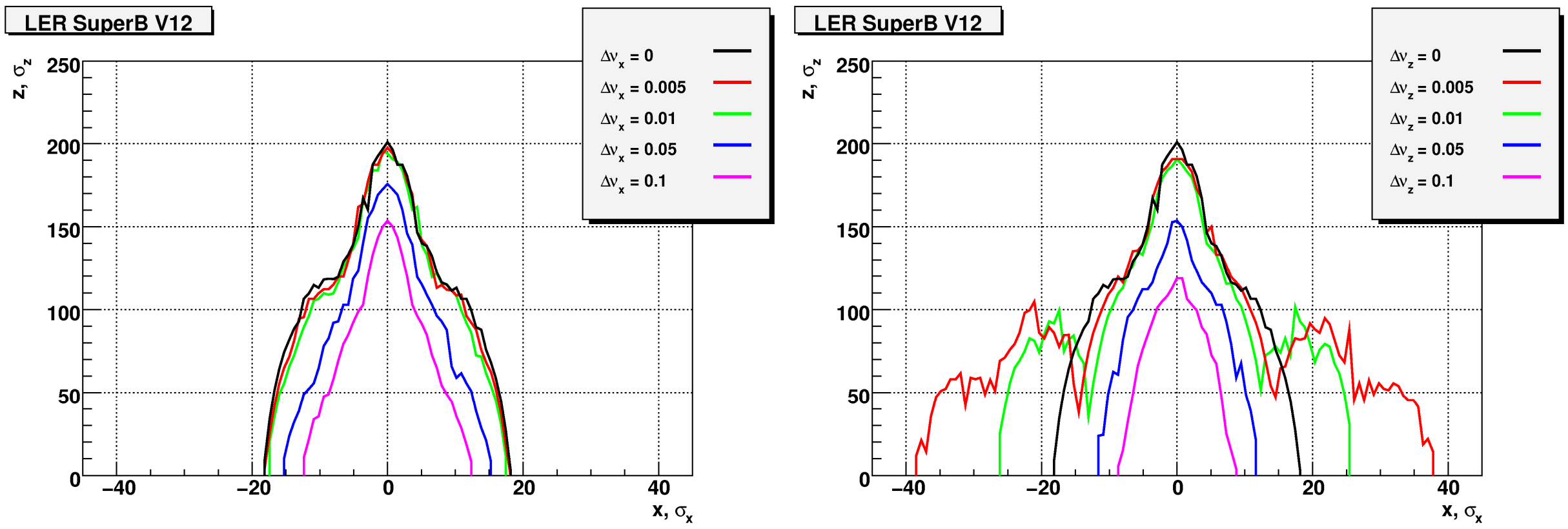

Fig. 6.33: Effect of –*I* condition detuning for the crab sextupoles.

It is worth noting that for some value of the phase advance mismatch the dynamic aperture increases. The question arises if this effect is reliable (and can not be destroyed by other errors) and if it is possible to use it for the DA correction.

The next point is the effect of linear optical errors which also distort the β-functions and create a mismatch of phase advance in the –*I* sextupole pairs. To test sensitivity of the dynamic aperture to such optical errors, we introduce random integrated gradient errors with the rms value of ±0.1%, first in the QD0 quadrupole alone, and then in all other quadrupoles. Unstable solutions and those with maximum beta distortions which exceeded 10% are sorted out. The results for 10 random seeds are shown in Fig. 6.34.

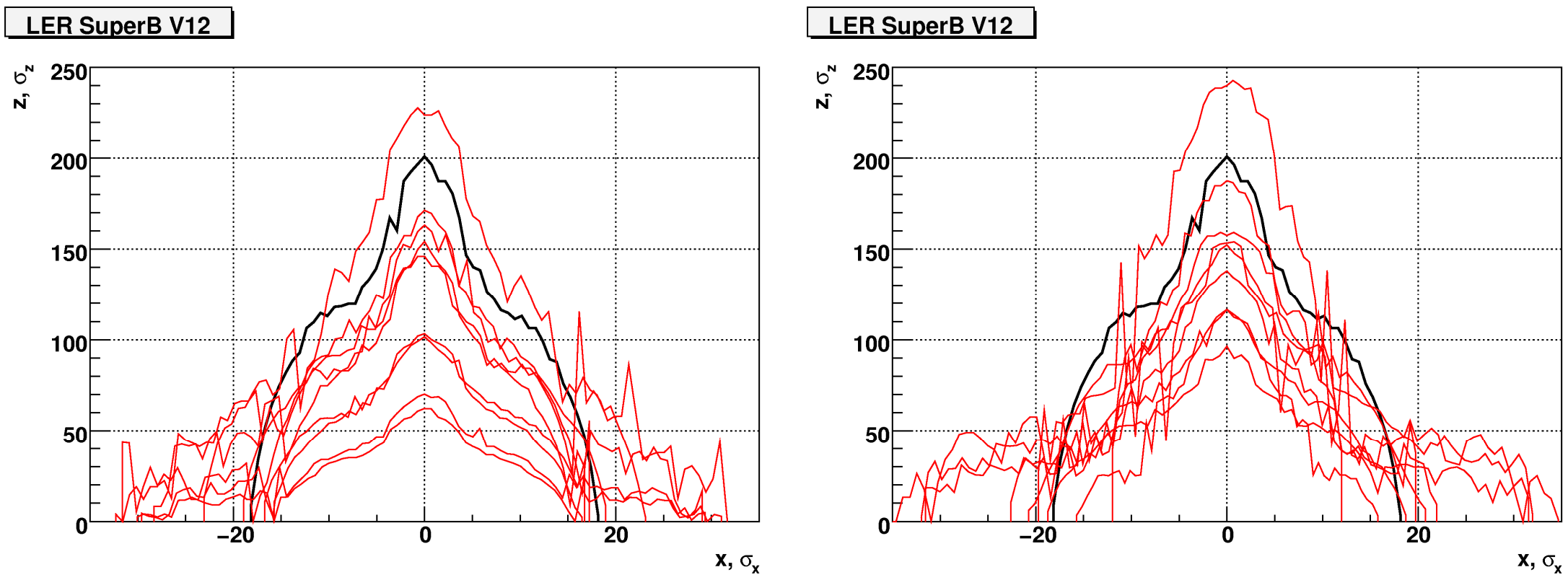

Fig.6.34: On-energy DA due to the $10^{-3}$ integrated gradient errors in QD0 (left plot) and in all quadrupoles in the ring (right plot).

The curves in Fig. 6.34 show that the errors in QD0 have dominant effect on dynamic aperture.

More sophisticated study of errors and imperfections, such as nonlinear components in magnets or influence of the polarization insertions, requires more detailed knowledge of the perturbation sources and strengths.

### Tune point selection

As the dynamic aperture is sensitive to the betatron tune point selection, it is of interest to simulate a scan of the DA size as a function of the betatron tunes. Preliminary results of these scans are given in Fig. 6.35, where the colour scale indicates the DA size.

The following can be concluded from this tune scan:

- DA vs. tune looks similar for both rings (LER and HER);
- a net of (rather strong) betatron resonances (mainly coupling ones) surround the chosen tune point;
- fine tune optimization should be done in the future. One of the obvious ways is the reduction of the vertical fractional tune to ~0.54 while keeping the horizontal tune at the same value;
- all identified resonances are of even order, which also confirms that they are produced by the third order (octupole-like) nonlinearities.

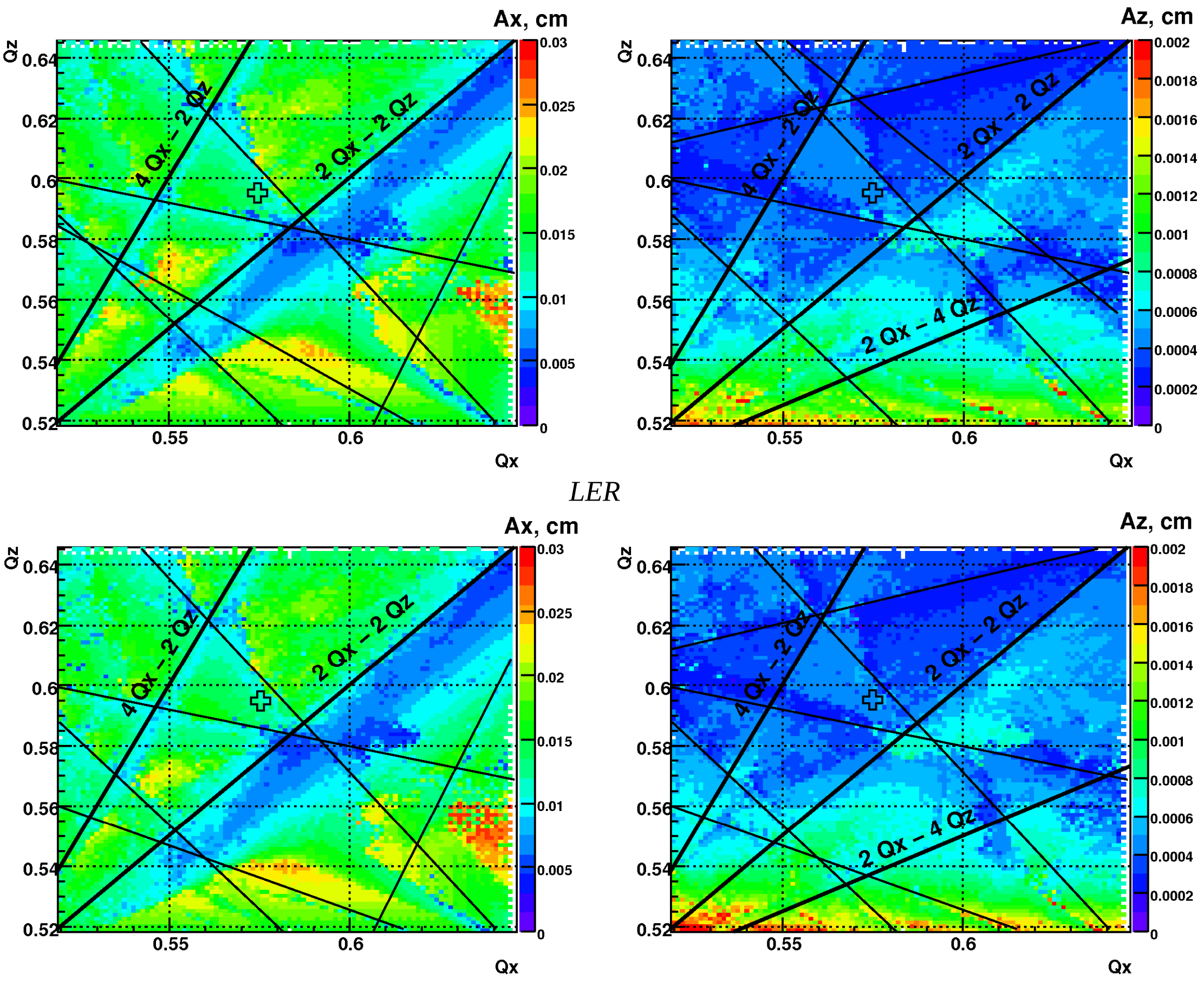

Fig. 6.35: DA scan (horizontal aperture is on the left and the vertical is on the right plot) as a function of the betatron tunes. Bold dot indicates the chosen tune point.

**Summary of DA calculations**

For a vertical space charge parameter ~0.1 the computed dynamic aperture seems at the moment large enough to get the design luminosity, even without active correction of the existing nonlinearities. For larger values the beam-beam (BB) footprint crosses the BB resonances and beam core blow up is observed.

The question is how the dynamic aperture will be influenced by the machine errors and by other effects, such as, for example, a strong solenoid fringe field, which is traversed by the particles at the angle, producing third order nonlinear terms, emphasized by the high value of the vertical betatron function. To include this effect in our simulation a realistic field map of the SuperB detector solenoid field is needed.

Moreover the sextupole –*I* sections mismatch and the focusing errors all other errors and imperfections should be taken into account. This work will be performed in the future. Special care is also needed for the nonlinear multipole field components in the FF quadrupoles due to the large β-functions there. In any scenario, moving the tune point away from strong betatron resonances is desirable; this should increase the dynamic aperture and make it less sensitive to the lattice errors. Interesting fact concerns the DA increase with the betatron phase mismatch in the –*I* sextupole pairs with finite length. This influence deserves special attention and an explanation in future. Clearly, the final conclusion for dynamic aperture budget may come only from the realistic simulation of the beam-beam effects in presence of the lattice nonlinearities, errors and imperfections, which is rather time consuming but it will be performed in the future.

# 7. Imperfections and errors

## 7.1 Tolerances, Vibrations and Stability

The movement of elements in the magnetic lattice of the SuperB accelerator will affect the equilibrium emittance of the beam. The horizontal emittance, and particularly the vertical emittance, are quite small and will require special care to achieve. We will first discuss errors in the rings outside the interaction region. The roll stability of quadrupole magnets, as well as the horizontal and vertical offset stability of the quadrupole magnets, are the most important sources of errors. There are several recent studies for the next generation low-emittance storage rings that have looked extensively at this stability issue. PETRA-III, NSLS-II, and the ILC Damping Rings all have lattice specifications that are similar to the arc and straight section magnets for SuperB. The design reports of these accelerators discuss these tolerances [1]. The total effect is estimated by including magnet errors around the complete ring with the appropriate betatron and phase weighting. The amplitude of fast magnet motion due to normal ground motion has only a small impact on the emittance. However, slow magnet motion can lead to an increased emittance, according to "*AT L*" models, which incorporate temporal and spatial correlations in reasonable agreement with observations. In the model, $< y^2 >= AT\,L$, where y is the transverse offset, $A$ is a constant about $4 \times 10^{-6}$ μm²/m/s, $T$ is the time and $L$ is the separation distance between points of interest, for example two adjacent quadrupoles. As a result, orbital steering corrections at the 5–10 μm level are required over timescale of a few minutes in order to keep the vertical emittance within specifications. BPM resolutions of order 1 μm are also needed.

The final quadrupole doublets adjacent to the IP have strong fields and the beams have large beta functions. Vibration tolerances for these magnets are especially tight. Typically, there is a one-to-one correspondence between the size and direction of vertical motion by final doublet quadrupole magnets and motion of the beam at the IP. The vertical beam size at the IR is 20–35 nm. Since we need to keep the beam in collision with tolerances at the 0.1 sigma level or less, quadrupole magnets must be be kept stable to 2–4 nm. The vibration of large objects such as quadrupoles depends on the design of the mechanical supports and the local ground excitation.

Typical motion is about 50 nm in the 50 Hz range. Since there are only a few of these magnets, active vibration controls in the mechanical supports can be employed to bring vibrations within specification. An active vibration suppression by a factor of 10–20 is within industry standards.

Bunches will collide at 476 MHz. To maintain luminosity it is important to keep the bunches transversely centered on one another. Feedback systems using the position monitors and the luminosity signal will be required. A discussion is presented in Section 5.

## 7.2 Coupling and Dispersion Tuning for Low Vertical Emittance Rings

A variety of collective effects can increase the vertical beam emittance at high currents; however, in the low-current limit, which we consider in this section, three effects dominate contributions to the vertical emittance. The non-zero vertical opening angle of the synchrotron radiation in dipole magnetic fields excites vertical betatron motion of particles as they "recoil" from photon emission. Vertical dispersion from steering errors generates vertical emittance, in the same way that horizontal dispersion from the bending magnets determines the horizontal emittance of the beam. Betatron coupling from skew quadrupole errors leads to a transfer of horizontal betatron motion (and hence horizontal emittance) into the vertical plane. The first of these effects, the non-zero vertical opening angle of the synchrotron radiation, places a fundamental lower limit on the vertical emittance that can be achieved in any storage ring; this can be calculated for a given lattice design. In most rings, including the SuperB rings, the lower limit is a fraction of a picometer, and is significantly smaller than the specified vertical emittance. The effects of vertical dispersion and betatron coupling, which arise from magnet alignment and field errors, invariably dominate the vertical emittance in an operating storage ring; reducing the vertical emittance in the SuperB rings to the value required to achieve the specified luminosity will require highly precise initial alignment of the machine, followed by careful tuning and error correction.

The lowest vertical emittance achieved in an operating storage ring is 2 pm in the Swiss Synchrotron Light Source at PSI (Zurich); the SuperB rings are specified to operate at 5-6 pm so the alignment and tuning issues require some attention.

Broadly speaking, we may characterize the behavior of the vertical emittance in a given lattice by calculating the vertical emittance generated by a variety of magnet alignment errors. The principal errors to consider, in this context, are vertical sextupole misalignments and rotations or tilts of quadrupoles around the beam axis, both of which generate unwanted skew quadrupole components. Also relevant is the closed orbit distortion generated by vertical misalignments of the quadrupoles, which results in vertical beam offsets in the sextupoles with the same consequences as vertical misalignments of the sextupoles themselves. Estimates of the sensitivity of a lattice to these errors can be made using analytical formulae [2] involving the magnet strengths and lattice functions; it is usually found that simulations support the results of these analytical calculations. However for SuperB a dedicated procedure for the so called "Low Emittance Tuning" (LET) has been expressly developed.

## 7.3 LET Procedure

### Tools

To implement the LET procedure we use MADX [3] and MATLAB[4]. MADX might be used alone, implementing misalignments, correction and iterations, but it does not allow complete freedom in plotting and correction may not be handled to include additional steering constraints. Moreover to change monitor or corrector pattern is slow and may lead to errors. Using Matlab a graphical interface was built that allows for:

- interactivity with MADX for input definition and elements installation
- analysis of any machine and/or error sequence
- definition of multiple errors in any element (including or excluding IR)
- showing and saving plots
- using user defined correction methods.

### Orbit and Dispersion Free Steering

Using only the information retrieved from monitors it is possible to correct the orbit generated by machine imperfections using Singular Value Decomposition to calculate a pseudo-inverse of the Response Matrix(ces). Following **[5]**, we use Dispersion Free Steering that allows constraining at the same time orbit and dispersion. In this work dispersion is computed at monitors via:

$$\eta_u = \frac{u_{+\frac{DE}{E}} - u_{-\frac{DE}{E}}}{2\frac{DE}{E}}.$$

The complete orbit-dispersion system is:

$$\begin{pmatrix} (1-\alpha)\vec{M} \\ \alpha\vec{\eta} \end{pmatrix} = \begin{pmatrix} (1-\alpha)ORM \\ \alpha DRM \end{pmatrix}\vec{K};$$

with *ORM* the Orbit Response Matrix, DRM the calculated Dispersion Response Matrix and $\alpha$ the relative weight between orbit and dispersion correction. Orbits are obtained by MADX with the input defined via the Matlab interface. Matlab then reads MADX output to build the matrices, and calculates the correction using the selected weights. All matrices are calculated without misalignments applied, so the correction needs to be reiterated including the effect of previously applied kicks. The kicks $K_{n+1}$ applied at $n+1$ iteration will be:

$$\vec{K}_{n+1} = svd(M)^{-1}\left(\vec{R} + M\vec{K}_n\right)$$

where $K_n$ are the previous kicks, $R$ is the readings vector and $M$ the Response Matrix used.

### Coupling and β-beating Free Steering

The same procedure may be further specialized. Without introducing additional correctors or skew quadrupoles it is possible to measure two new response matrices for coupling (CRM) and $\beta$-beating ($\beta$RM). The columns of the response matrices are calculated as follows:

$$\forall Y kick\, K_y^j\; CRM^j = \begin{pmatrix} \frac{\vec{x}_{+\Delta Y} - \vec{x}_{-\Delta Y}}{2\Delta Y} \\ \\ \frac{\vec{y}_{+\Delta H} - \vec{y}_{-\Delta H}}{2\Delta H} \end{pmatrix} \quad (1)$$

$$\forall X kick\, K_x^j\; \beta RM^j = \begin{pmatrix} \frac{\vec{x}_{+\Delta H} - \vec{x}_{-\Delta H}}{2\Delta H} \\ \\ \frac{\vec{y}_{+\Delta Y} - \vec{y}_{-\Delta Y}}{2\Delta Y} \end{pmatrix} \quad (2)$$

where $\Delta H$ and $\Delta V$ are two fixed kicks applied in the horizontal or in the vertical plane while $x$ and $y$ are column vectors of the orbit at the BPMs. For example the notation $x{-}\Delta H$ represents the x orbit in presence of a fixed kick in the Horizontal plane of value $-\Delta H$ and the response matrix for this vector is the top quadrant of $\beta$RM. The first matrix (CRM) is studied only varying Y correctors, while the second one ($\beta$RM) only varying X correctors. Calling the coupling orbit and $\beta$-beating orbit to be corrected $C$ and $\beta$ (calculated as the columns of the response matrix) the complete systems of equations for the two planes are now:

$$\begin{pmatrix} (1-\alpha-\omega)\vec{M}_x \\ \alpha\vec{\eta}_x \\ \omega\vec{\beta} \\ \omega\vec{\beta}_{\pi/2} \end{pmatrix} = \begin{pmatrix} (1-\alpha-\omega)ORM \\ \alpha DRM \\ \omega\beta RM \\ \omega\beta RM_{\pi/2} \end{pmatrix}\vec{K}_x \quad (3)$$

$$\begin{pmatrix} (1-\alpha-\omega)\vec{M}_y \\ \alpha\vec{\eta}_y \\ \omega\vec{C} \\ \omega\vec{C}_{\pi/2} \end{pmatrix} = \begin{pmatrix} (1-\alpha-\omega)ORM \\ \alpha DRM \\ \omega CRM \\ \omega CRM_{\pi/2} \end{pmatrix}\vec{K}_y \quad (4)$$

where $\pi/2$ indicates the use of a different corrector at a phase advance of approximately 90 degrees for both planes. Solving this system is now like selecting among all the possible orbits, the one that has the minimum rms dispersion and coupling, hence the minimum vertical emittance.

### Steering parameters

In Figure 7.1 a simulation for SuperB HER lattice (excluding the FF) shows how vertical emittance and rms kick strength vary using an increasing number of eigenvectors (ordered by decreasing eigenvalue). It is clear that 65 eigenvectors are a good guess to have optimal correction, maintaining at the same time small kick sizes. This value is also confirmed by the same plot for rms dispersion and rms orbit, not shown here. To determine the optimal values for $\alpha$ and $\omega$, a scan for different values of these parameters is performed. The selected values are $\alpha$ = 0.5 and $\omega$ = 0.01, being at the center of the optimal correction region.

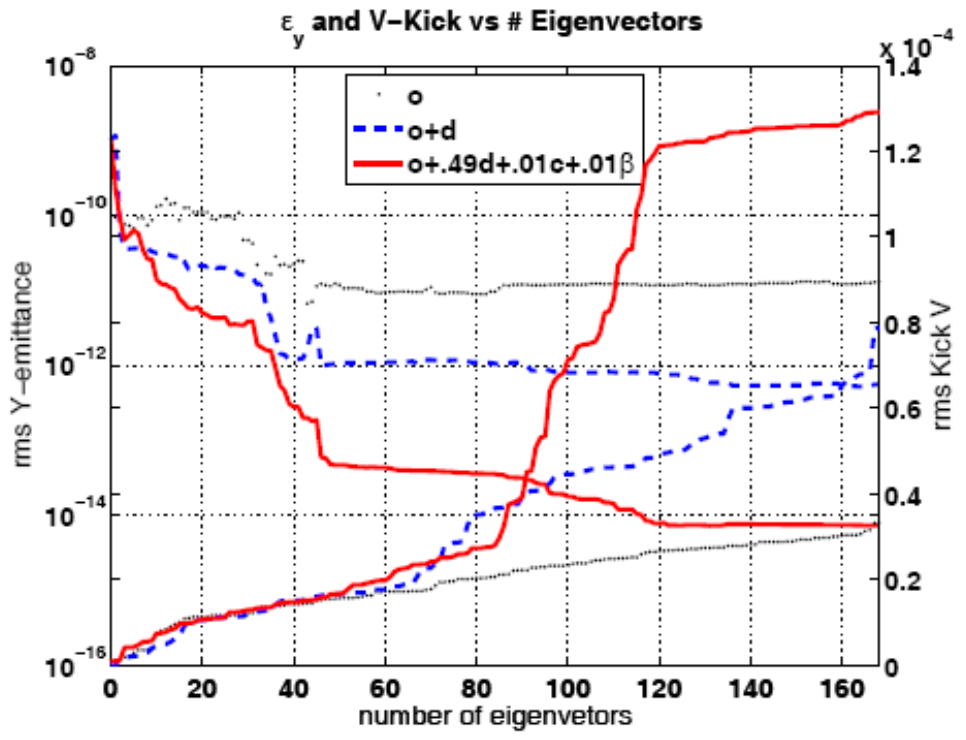

Figure 7.1: rms $\varepsilon_y$ (m) and rms Kick applied (rad) vs number of eigenvectors used (ordered by decreasing eigenvalue), after vertical correction for machines with 100 $\mu$m vertical misalignments in quadrupoles and sextupoles. Kick increases while emittance decreases.

## Simulations

All simulations presented are made for HER at 6.7 GeVwith 168 H and V correctors, and 168 H and V monitors, installed at every quadrupole, sextupole and octupole. Misalignments are applied with a gaussian distribution truncated at 2.5 $\sigma$. To determine the maximum tolerated misalignment, plots as that in Figure 7.2 are considered [6]. For 10 different values of error variance a summary of the distribution obtained is given. The central mark shows the average, while the error bars include the distribution from the 5th to the 95th percentile. The effect of BPM offsets of 300 μm, is also taken into account. A comparison of different correction scheme is also presented to give evidence of the improvement given by dispersion (D) and coupling and $\beta$-beating (C) free steering respect to pure orbit (O) correction.

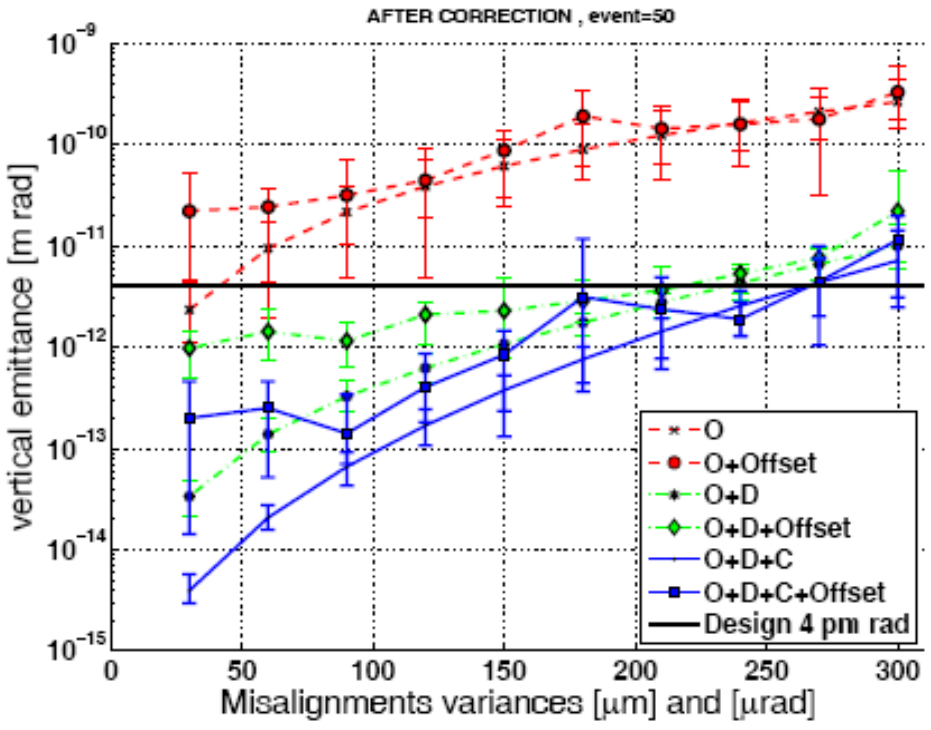

Figure 7.2: Vertical emittance (m) for machine misalignment from 30 to 300$\mu$m H and V for Sext and Quad and quadrupole Tilts of 30-300 μrad. Orbit (O), Dispersion (D) and Coupling and Beta-beating (C) Free Steering are compared.

## Tolerance Table

To summarize the result of LET a table (Table 7.1) of tolerated imperfections is built. To determine the tolerated value the following procedure is used:

1. misalignments of sextupoles and misalignments and tilts of quadrupoles are analyzed separately for increasing variance
2. an interval of variances that leads to emittances under 1pm is selected in both cases
3. these intervals of variances are applied together and the tolerated values are selected as those giving a 0.5 pm threshold
4. once the values of the previous step are fixed, the monitor offset variance is studied.

As a result of this analysis the combination of all the imperfections gives a vertical emittance of less than 1pm for the tolerated values. This low threshold is necessary to allow the subsequent introduction of errors in solenoids and FF magnets.

Correction is performed for every simulation in three steps: the first with sextupoles off and only orbit correction, the second and third using dispersion, coupling and $\beta$-beating free steering parameters mentioned above.

Figure 7.3 shows the effect of quadrupole displacements and tilts (red), sextupole displacements (blue) and monitor offsets (green). Using the new correction scheme errors like monitor offsets and quadrupole displacements influence less the final emittance and the tolerated values may be higher.

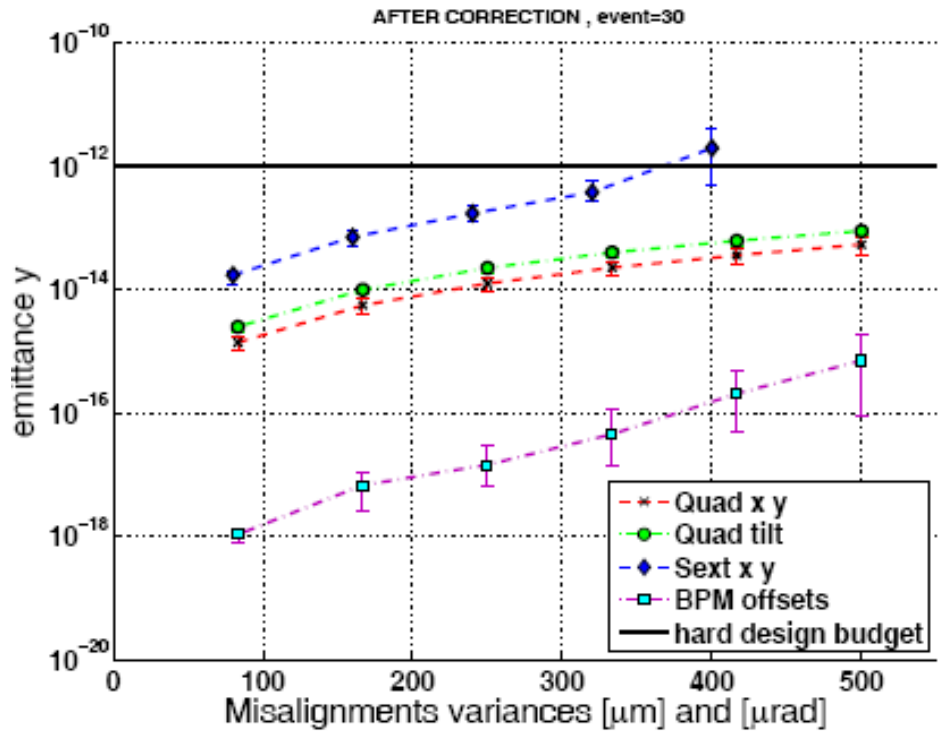

Figure 7.3: Misalignments tilt and BPM offset errors. Every point is the average of 5 simulations.

In Figure 7.4 is shown a histogram of the vertical emittance before and after correction for 50 different machine misalignments sets with the imperfections variances listed in Table 7.1.

Table 7.1: SuperB magnets tolerated imperfections

| Error | Tolerance |
|---|---|
| Quadrupole Y | 300 μm |
| Quadrupole X | 300 μm |
| Quadrupole Tilt | 300 μrad |
| Sextupole Y | 150 μm |
| Sextupole X | 150 μm |
| BPM offset | 400 μm |
| Vertical emittance | <1 pm rad |

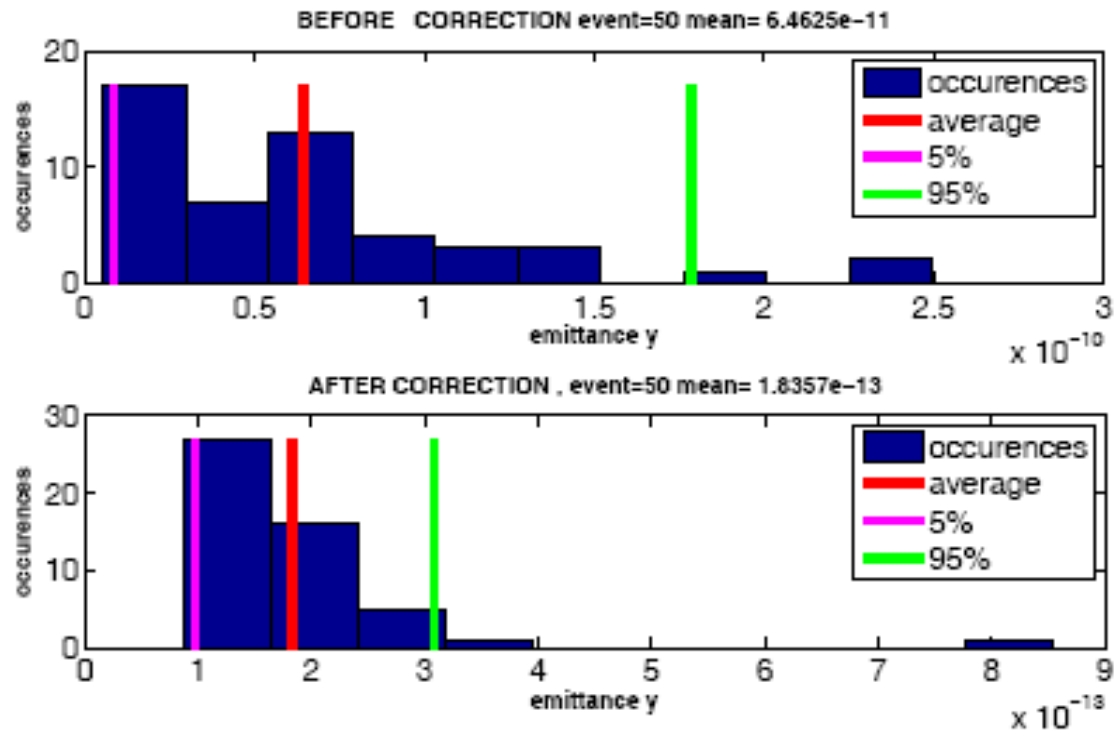

Figure 7.4: Vertical emittance for 50 simulation with misalignment and tilts from Table 7.1.

**Final Focus**

The same analysis can be applied to the ring with FF. A preliminary study was performed including 230 correctors and 250 monitors. The same weights and correction scheme are applied using 90 eigenvectors. In all the simulations the errors in the arcs are fixed to the values determined for the machine without final focus. However, for these values, the errors tolerated in the final focus are very little (< 30 μm). This work is preliminary and needs to be completed with a more realistic simulation of common errors for elements installed on the same support.

## 7.4 Final Focus Tuning

The FF "tuning knobs" are adjustments of magnet field and alignment to compensate the linear and non-linear beam aberrations and beam size growth at the IP caused by "slow" field or tilt errors in the FF quadrupoles. Sextupoles, octupoles and decapoles can be used in the tuning knobs. Alternatively, the normal and skew quadrupole correcting coils can be considered, which have the advantage of not creating second-order orbit distortions. This method has been studied for the FF systems of the NLC, ILC, ATF2 (see for example ref. [7]) since all these machines employ the same design principles. A short summary is provided here.

Very large peaks produce a characteristic 90° to-IP phase advance at most of the FF magnets. This 90° phase advance reduces the number of efficient tuning knobs, but also helps in correcting the FF errors, since the FF correctors are effectively at the same phase as the FF errors. However, this assumes that the out-of-90° phase aberrations propagating to the IP from the upstream optics can be corrected prior to the FF.

A number of linear and non-linear tuning knobs can be implemented. Examples of orthogonal linear knobs are:

- horizontal offset in a sextupole to correct the horizontal dispersion at IP and the longitudinal offset of $\beta_x^\star$, $\beta_y^\star$ waists (3 knobs);
- vertical offset in a sextupole to correct the vertical IP dispersion and the dominant (x, y) coupling term ($R_{32}$) at the IP (2 knobs).

Adjustment of field and tilt angles of the FF sextupoles can be used to correct second-order optical aberrations at the IP, which is needed as well. Additionally, adjustment of the octupole and decapole fields can be used for the third- and fourth-order corrections. These magnets create many high-order terms; "absolute" orthogonality between different terms is therefore typically not possible to achieve using a limited number of correctors. Hence the goal is to create approximately orthogonal knobs that excite one dominant term per knob, while keeping the other terms small. The sextupole knobs can be calculated with second-order matrix optimization using MAD code [3]. A simple octupole knob can correct the octupole field error, and two decapole knobs can correct the decapole field error and the field difference between the two decapoles. The fixed 90° phase to the IP limits the number of matrix terms (knobs) which can be created. To improve the orthogonality of knobs based on sextupole fields, extra sextupoles can be added to the lattice.

The effectiveness of these knobs depends on the set of the random machine errors, which cause the IP aberrations. Tracking of many sets of errors would show which aberrations are the largest at the IP, and therefore which correcting knobs are most important. An example of the iterative procedure for FF tuning can be found in Ref. [7]. An ideal initial beam distribution is first generated with a large number of particles, and tracking is done without magnet errors, thereby characterizing the ideal beam at the IP. Random field and alignment errors are then assigned to magnets and BPMs, and tracking with the errors before any correction and measurement of the beam at IP is performed. The initial orbit is corrected using the corrector quadrupole x, y offsets, and the known response matrix between the correctors and BPMs, and then tracking is performed again. The IP tuning correction is obtained by applying the tuning knobs one-by-one with the orbit correction after each knob, followed by tracking and measuring again. In the tuning loop, the linear knobs are applied first, then the second-order vertical and horizontal knobs. Finally, octupole and decapole knobs can be applied. This procedure can be iterated as needed, and various combinations of rms errors must be studied. An example of the efficiency of this method for the NLC Final Focus tuning simulation is shown in Fig. 7.5.

A good tuning efficiency requires that the residual orbit in the FF is well corrected. Therefore, a beam-based alignment (BBA) procedure is required to minimize the misalignments; and the orbit correction must be optimized. Tracking with various levels of misalignment will demonstrate the level of residual alignment error required for good tuning.

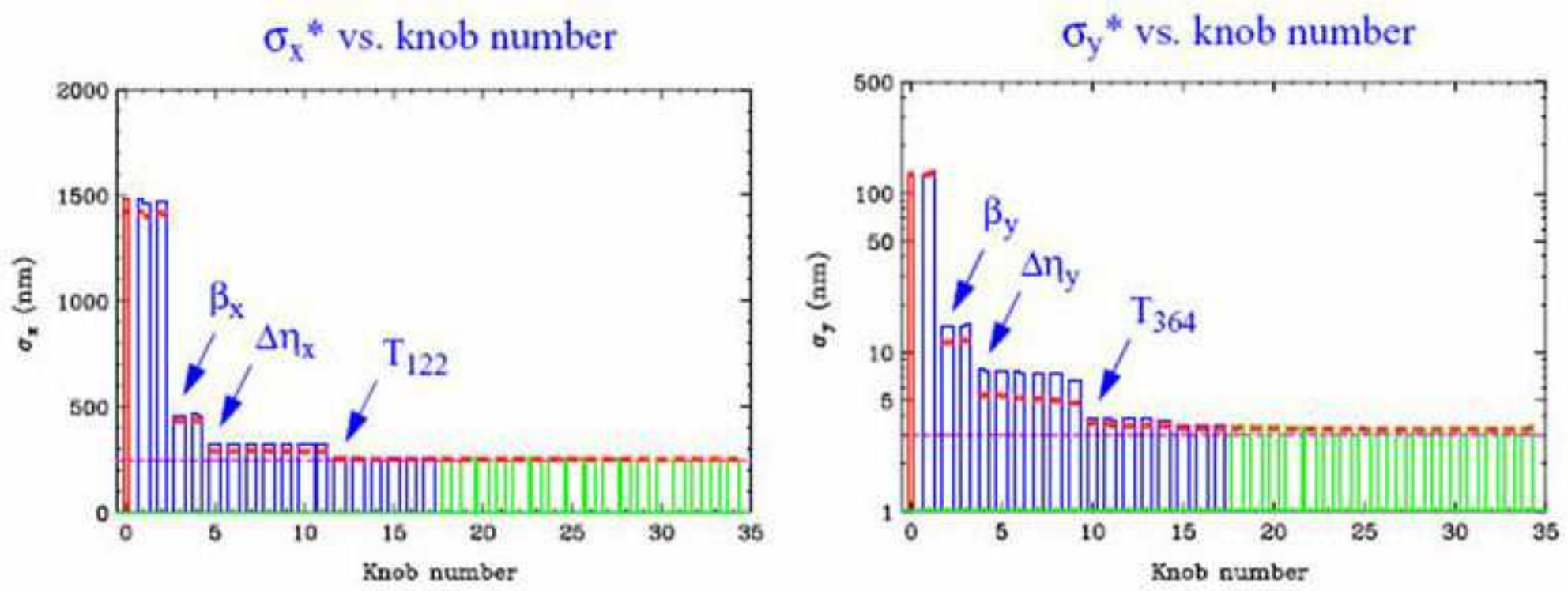

Figure 7.5: Example of IP beam size tuning in the NLC FF using 17 tuning knobs in presence of magnet field and tilt errors. The knobs are applied one-by-one in two loops (blue and green), where the effect of each knob is represented by a bin.

# 8. Intensity dependent effects

## 8.1 Beam-beam interaction

The Crab Waist collision scheme turned out to be very beneficial for beam-beam effects. Due to effective suppression of beam-beam induced resonances [1] it allows increasing the value of $\xi_y$ by a factor of about 3 as compared with the ordinary head-on collision. Accordingly, the same factor can be gained in the luminosity. In [2] the design value of $\xi_y$ was 0.17, appreciably below the limit and therefore widening the area of possible working points. The luminosity contour plot (old parameters) versus the betatron tunes is shown in Fig. 8.1. In these simulations the machine lattice was considered to be linear, without betatron coupling, and the vertical emittance was generated by a Gaussian noise which does not depend on the betatron tunes. In reality, of course, the vertical emittance is generated by coupling, so the working point must be above and not too close to the main coupling resonance – above the dashed line in Fig.1. The red colour corresponds to the luminosity $L \geq 0.95 \cdot 10^{36}$ $cm^{-2}c^{-1}$, and the distance between successive contour lines is $\Delta L = 0.05 \cdot 10^{36}$ $cm^{-2}c^{-1}$. It is worth mentioning that the tune shift limit increases when the betatron tunes are shifted closer to half-integer resonance. For instance at the working point (0.525, 0.545), marked by a white star in Fig. 8.1, $\xi_y$ can even reach the value of 0.25.

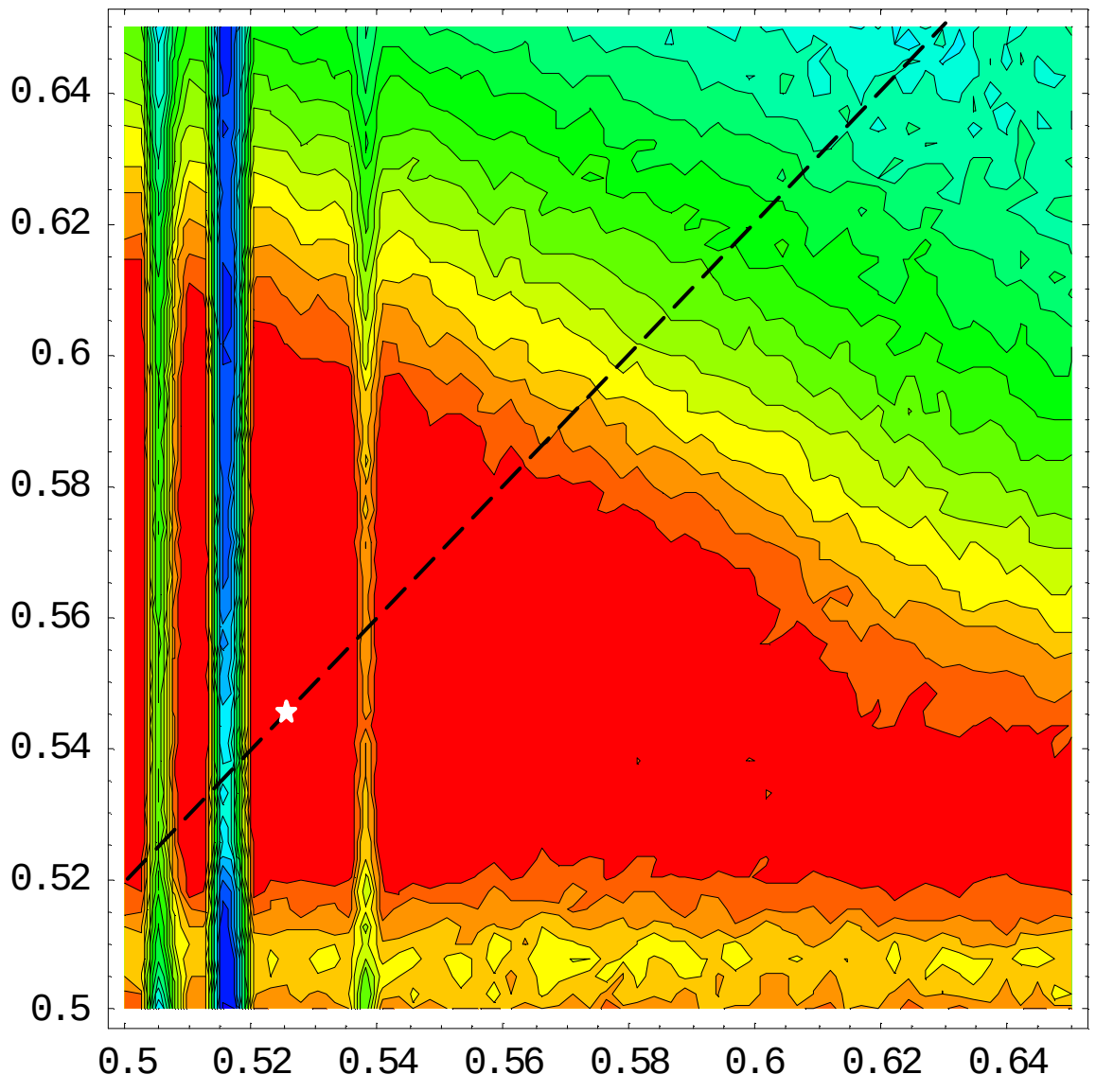

Figure 8.1: Luminosity contour plot vs. the betatron tunes. Parameters of December 2006, $\xi_y$=0.17, $\nu_s$=0.02. In the red area $L \geq 0.95 \cdot 10^{36}$ $cm^{-2}s^{-1}$. Working point should be chosen above dashed line.

Usually in colliders is desirable to keep $\xi_y$ close to the beam-beam limit, but in the current version of SuperB parameters its design value was lowered to about 0.1. As a result, the area of possible working points widens significantly, see Fig. 8.2. For the new set of parameters the horizontal sinchro-betatron resonances (satellites of half-integer), which appear as vertical lines of low luminosity, are also weaker. The main reasons are the decrease of the synchrotron tune and of the horizontal beam-beam tune shift $\xi_x$.

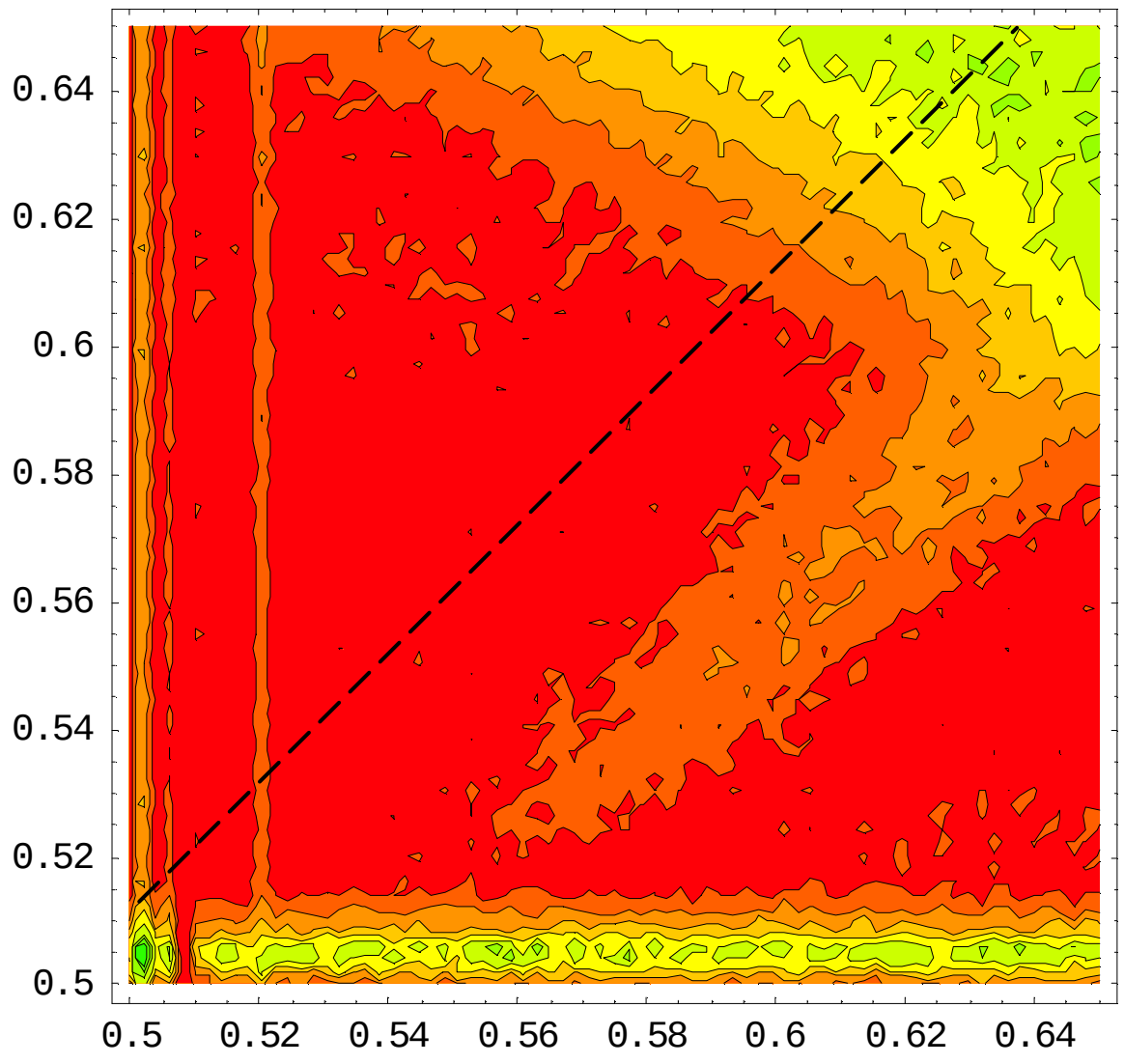

Figure 8.2. Luminosity contour plot vs. the betatron tunes. Parameters as of September 2009: ξy=0.097, νs=0.0118. In the red area L ≥ 0.95·1036 cm-2c-1. Working point should be chosen above dashed line.

Increasing the area of possible working points due to lowering the $\xi_y$ value can be very useful, since there are other factors limiting its choice, e.g. Dynamic Aperture optimization. One more advantage of working with small $\xi_y$ is that we avoid any beam-beam induced blow-up and long tails in the distribution density, thus improving the beam lifetime and detector background. On the other hand, when considering operation with higher tune shifts, we have to take into account the following aspects:

- First, it is not so easy to achieve higher values of $\xi_y$ even without beam-beam considerations. Possible ways are to decrease the vertical emittance, which is already extremely small, or decrease the bunch length, that is rather questionable, or increase the bunch current. Besides, in all these cases the IBS contribution to the vertical emittance grows and the Touschek lifetime diminishes so affecting the total lifetime.

- Second, increasing $\xi_y$ means increasing the luminosity for the same bunch current, so decreasing the luminosity lifetime in the same proportion. Being this rather short already, about 5 minutes only, a further decrease is not desirable.

On the other side, our goal is to maximize the luminosity, not the tune shift. And we do not need rising it up, if the designed value of $10^{36}$ $cm^{-2}s^{-1}$ can be achieved with a relatively small $\xi_y$. Besides, we always have a possibility to increase $\xi_y$ without incurring into beam-beam problems – if the other conditions allow.

Due to the asymmetry in emittances and beta-functions between HER and LER, the optimum waist rotations are different for the two beams: 0.8 of the nominal value for HER and 1.0 for LER. The beam-beam perturbations are more pronounced in LER, but $\xi_y$ is too small to make the difference significant. Simulation results of equilibrium beam density distributions (with linear lattice) are shown in Fig. 8.3. The working point was chosen not too close to half-integer, taking into account the possible DA issues: $\nu_x$=0.553, $\nu_y$=0.580. As shown in the pictures, both the core region and the beam tails remain actually unperturbed for both LER and HER.

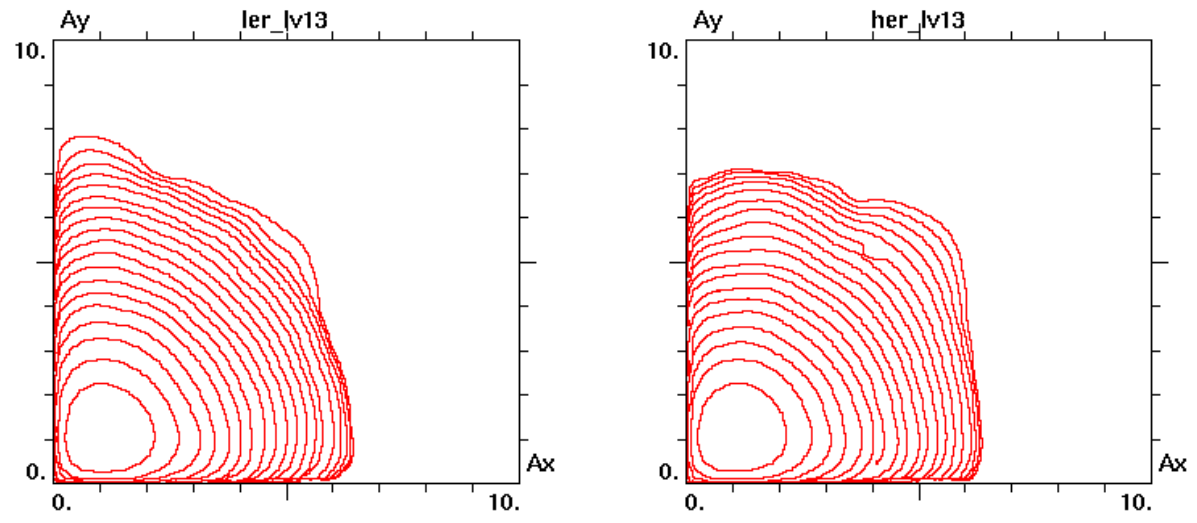

Figure 8.3. Equilibrium density contour plots in the plane of normalized betatron amplitudes for LER (left) and HER (right).

Of course, more realistic simulations with the nonlinear lattice are required. This will be done as soon as the lattice is finalized and DA optimized. But in general we do not expect any serious problems with beam-beam effects, since the designed value of $\xi_y$ is reasonably small.

## 8.2 Electron cloud effect and remediation

Under certain conditions, electrons can accumulate in the vacuum chamber of a positron storage ring. Primary electrons are generated by the interaction of beam synchrotron radiation with the chamber walls or by ionization of residual gas. These primary electrons produce secondary electrons after impact with the vacuum chamber walls. An electron cloud develops if beam and chamber properties are such to generate secondaries at a sufficiently high rate. Depending on the electron density level, the interaction between the cloud and beam may lead to detrimental effects such as single-bunch and coupled-bunch instabilities [1]. Electron cloud effects have been a limitation for the B-factories, requiring installation of solenoids to suppress the build-up of the cloud, and are expected to be a serious issue in the SuperB positron (HER) ring. For a complete evaluation, both the build-up of the cloud and its effects on the beam must be considered. In the following we present estimates, based on numerical simulations, of the cloud density at which single-bunch instability is expected to set in, and of the density levels of the electron cloud in the SuperB HER.

### *Numerical simulations*

In order to estimate with great accuracy the single-bunch instability threshold we performed simulation with the strong-strong code CMAD [2]. In this code both the bunch and the electron cloud are represented by macro-particles, and the interactions between them are determined by solving a two-dimensional Poisson equation using the particle-in-a-cell method.

Table 8.1: Input parameters for CMAD simulations.

| | |
|---|---|
| Beam energy E[GeV] | 6.7 |
| Circumference L[m] | 1370 |
| Bunch population Nb | 5.74x1010 |
| Bunch length $\sigma z$ [mm] | 5 |
| Horizontal emittance $\varepsilon x$ [nm] | 1.6 |
| Vertical emittance $\varepsilon y$ [pm] | 4 |
| Hor./vert. betatron tune $\nu_x/\nu_y$ | 40.57/17.59 |
| Synchrotron tune $\nu_s$ | 0.01 |
| Hor./vert. av. beta function | 20/20 |
| Momentum compaction $\alpha$ | 4.04 $10^{-4}$ |

Although the code can track the evolution of the instability trough a realistic lattice, here we assume that the interaction between beam and cloud is localized at 40 positions uniformly distributed around the ring, assuming a uniform value of the $\beta$ functions. Figure 8.4 shows emittance growth, for various cloud densities, due to the interaction of the electron cloud with a bunch in the SuperB HER as obtained by CMAD using the input parameters of Table 8.1. The threshold density is determined by the density at which the growth starts. From this numerical simulation, we determine that in SuperB HER the instability starts at $\rho_e = 4x10^{11}\ m^{-3}$.

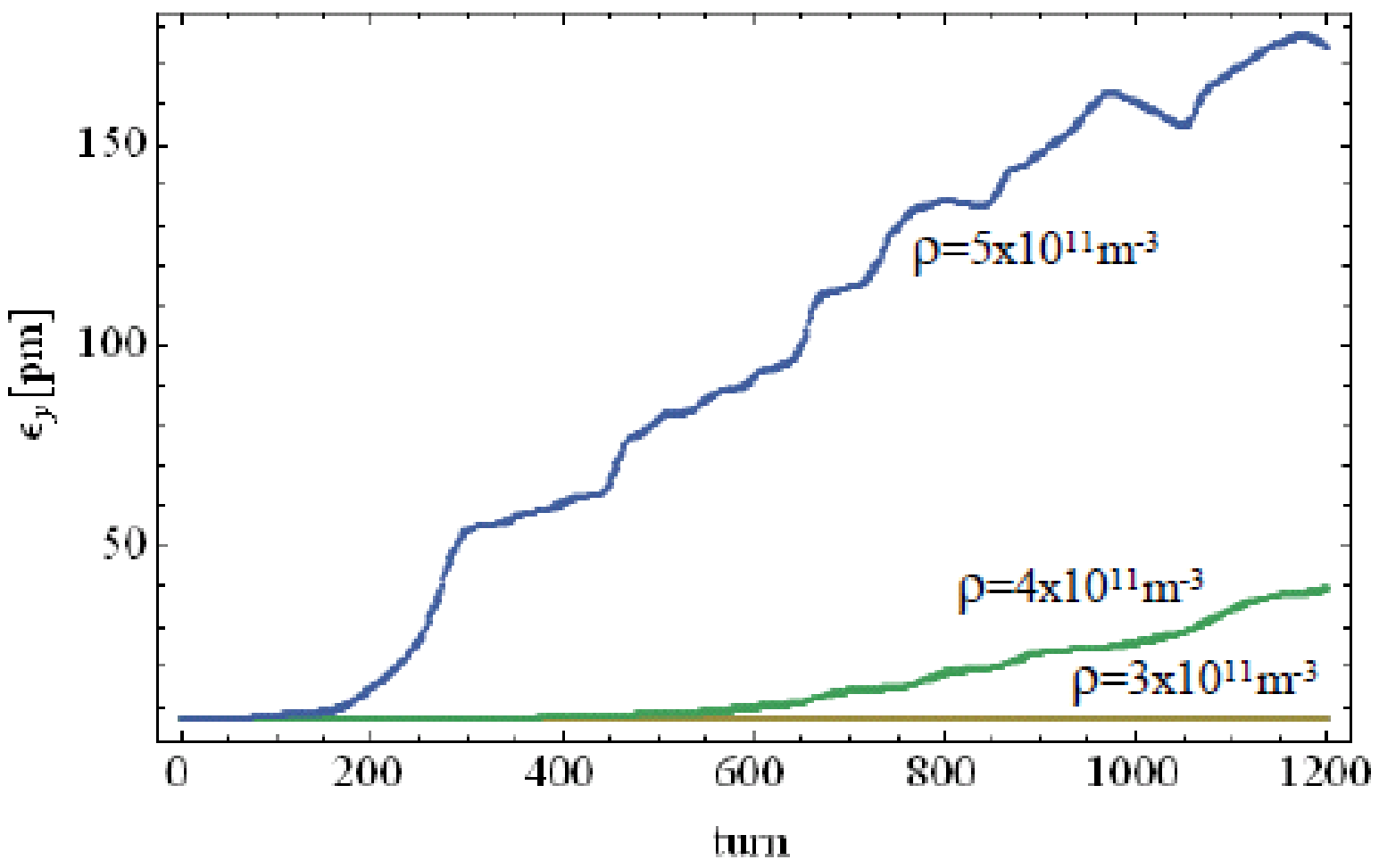

Figure 8.4. Emittance growth due to the fast head-tail instability caused by the electron cloud effect

### *Electron cloud density*

We have used the simulation code ECLOUD [3] to evaluate the contribution to the electron cloud build-up in the arc bends of SuperB. The KEKB and PEP-II B Factories have adopted external solenoid fields to mitigate the electron cloud effect in field-free regions, which constitute a large fraction of the rings. In magnetic field regions, external solenoid fields are not effective in suppressing the build-up of the electron cloud. Thus, we have focused our simulations on the build-up of an electron cloud in the arc bend regions. We have assumed a vacuum chamber with an antechamber design and, in order to take into account the reduction of electron yield by the ante-chamber, we used a reduced number of primary electrons:

$$e^- / e^+ / m = \frac{dn_\gamma}{ds} Y(1-\eta)$$

where $dn_\gamma / ds$ is the average number of emitted photons per meter per $e^+$, $Y$ is the quantum efficiency, and $\eta$ is the percentage of photons absorbed by the antechambers. In Table 8.2 are reported the saturation values of the electron cloud central densities (i.e., within a region of $10\sigma_x \times 10\sigma_y$ around the beam centre) as obtained from ECLOUD for different values of the peak secondary emission yield (SEY) and of the antechamber protection factor of $\eta$. Simulation were performed for a typical SuperB bending magnet, assuming a uniform vertical bending field $B_y$ = 0.5T and an elliptical chamber geometry with horizontal and a vertical aperture 95mm, and 55mm respectively.

Table 8.2: e-cloud densities from ECLOUD simulations.

| SEY | $\eta$ | $\rho_e$ $[10^{12}\ e^-/m^3]$ |
|---|---|---|
| 1.1 | 95% | 0.4 |
| 1.1 | 99% | 0.09 |
| 1.2 | 95% | 0.9 |
| 1.2 | 99% | 0.2 |
| 1.3 | 95% | 8.0 |
| 1.3 | 99% | 4.0 |

The density values given in Table 8.2 have to be scaled by the ”filling” factor of dipoles (i.e. the fractions they cover the ring), which amount to about 0.5. The results show that a that a peak secondary electron yield of 1.2 and 99% antechamber protection result in a cloud density close to the instability threshold.

### *Electron Cloud Remediation Techniques*

Possible remedies for the electron cloud formation considered recently include clearing electrodes and vacuum chamber grooves [4, 5]. Our simulations show that the insertion of clearing electrodes in the vacuum chamber is indeed a extremely powerful way to suppress electron cloud formation. We will describe the effect of clearing electrodes in the dipole magnetic field regions and the chamber layout.

### *Electron cloud build-up and clearing electrode effect*

The simulation code POSINST was used to evaluate the contribution to the electron cloud build-up in the arc bends of SuperB. The KEKB and PEP-II B-Factories have adopted external solenoid fields to mitigate the electron cloud effect in field-free regions, which constitute a large fraction of the rings [6, 7]. The SuperB rings typically do not have long field free regions. For the most part of the ring the beam pipe is surrounded by magnets, where large electron cloud densities may develop. In magnetic field regions, external solenoid fields are not effective in suppressing the build-up of the electron cloud. Thus, we have focused our simulations on the build-up of an electron cloud in the arc bend regions. To remove most of the synchrotron radiation emitted in the arc sections, we have assumed a vacuum chamber with an antechamber design. For these preliminary simulations, we have assumed the same bunch population of $2 \times 10^{10}$ particles per bunch but a reduced bunch spacing of 1.5 ns in comparison with the ILC DR (6.154 ns). Results for the electron cloud build-up are shown in Fig. 3. To mitigate the formation of an electron cloud, we have also simulated the effect of clearing electrodes installed in the bend vacuum chamber, and extending along the longitudinal direction of the magnet. The electrodes are biased with a positive potential. In a bend or wiggler magnet, the electrodes can be arranged along the top and bottom, since the electron cloud forms mostly along stripes directed along the vertical magnetic field lines [7]. The effect of the electrodes is to compensate, on average, for the electric field from the positron bunch, which tends to attract the electrons to the center of the chamber. The electrons at the wall are first accelerated to the center by the bunch, and then accelerated back to the surface by the electrodes, during the time interval between bunches. The effect of the two clearing electrodes is shown in Fig. 3-76. The average cloud chamber density and the central cloud density are plotted on the left and right side of the figure, respectively, for different electrode bias potentials. A bias voltage of 1 kV is sufficient to suppress electron cloud formation and drastically reduce the central cloud density near the beam. These simulations show the effect of the clearing electrode suppression in SuperB, although with beam parameters (bunch population and bunch spacing) that differ from the SuperB configuration.

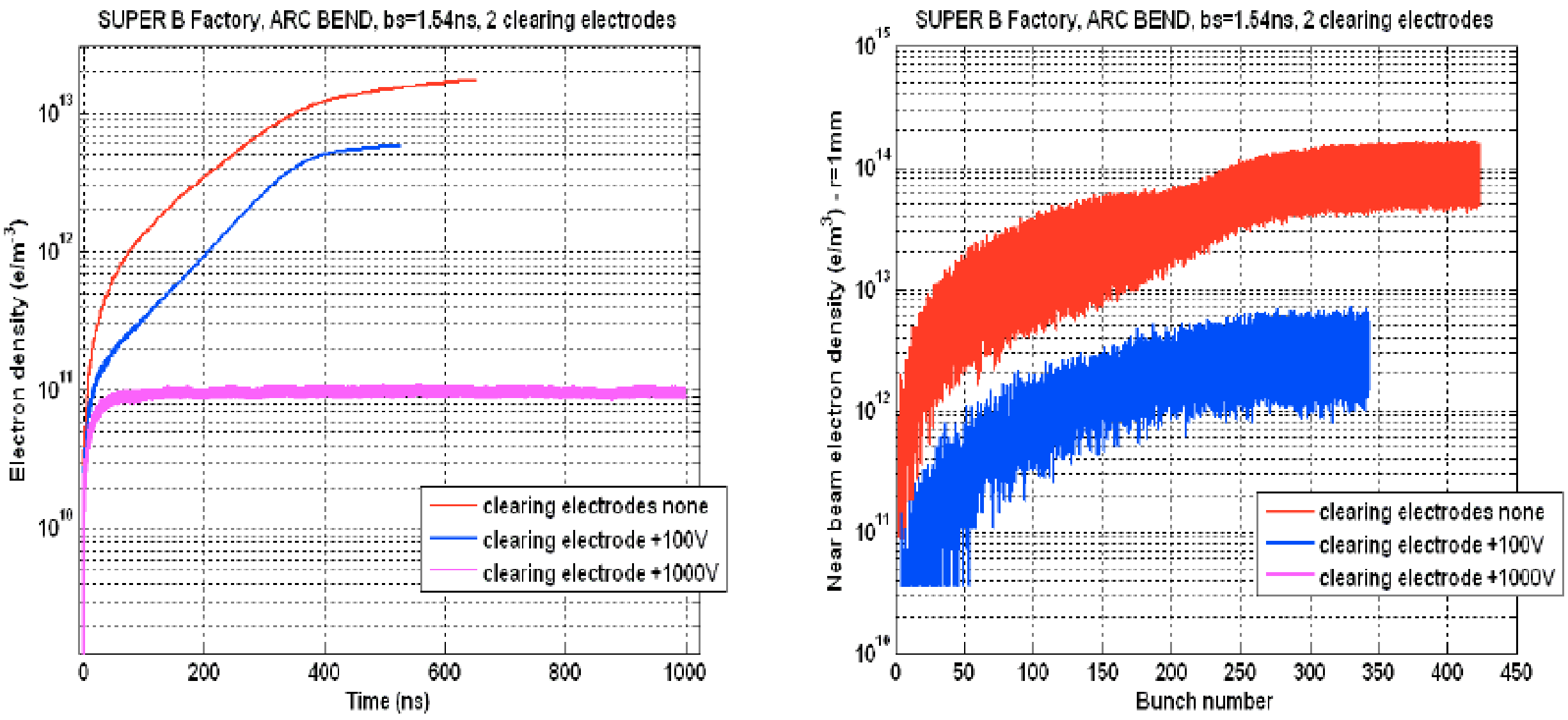

Figure 8.5: Simulation of electron cloud build-up in SuperB, using two clearing electrodes. Average (left) and central (right) electron density, with and without clearing electrodes are illustrated. Note: up to $1 \times 10^6$ macroparticles to represent the electrons were used.

## 8.3 Space charge effects in the LER

The large bunch population and small beam sizes result in appreciable space charge tune shifts in the SuperB rings, and in particular in the LER, as space charge effects scale inversely with the beam energy. For the LER at the design equilibrium and bunch population (N = 6.5 × $10^{10}$) linear theory (i = x, y):

$$\Delta\nu_i = -\frac{1}{4\pi}\frac{2r_e}{\beta^2\gamma^3}\int_0^C \frac{\lambda\beta_i}{\sigma_i(\sigma_x+\sigma_y)}ds$$

yields the following horizontal and vertical space charge tune shifts: $\Delta\nu_x=-0.002$, $\Delta\nu_y=-0.07$. This equation, in which β and γ are the relativistic factors, $\beta_x$, $\beta_y$ are the lattice functions, $\sigma_x$, $\sigma_y$ the horizontal and vertical rms beam sizes, $\lambda= N/(2\pi\ \sigma_z^2)^{1/2}$ the longitudinal peak density ($\sigma_z$=5 mm is the rms longitudinal bunch length), applies to particles undergoing infinitesimally small betatron and synchrotron oscillations about the centre of a gaussian bunch. Plots of the transverse beam sizes for the LER at equilibrium, as determined using the design emittances $\varepsilon_x$=2.46 nm-rad, and $\varepsilon_y$=6.15 pm-rad are shown in Figure 8.6.

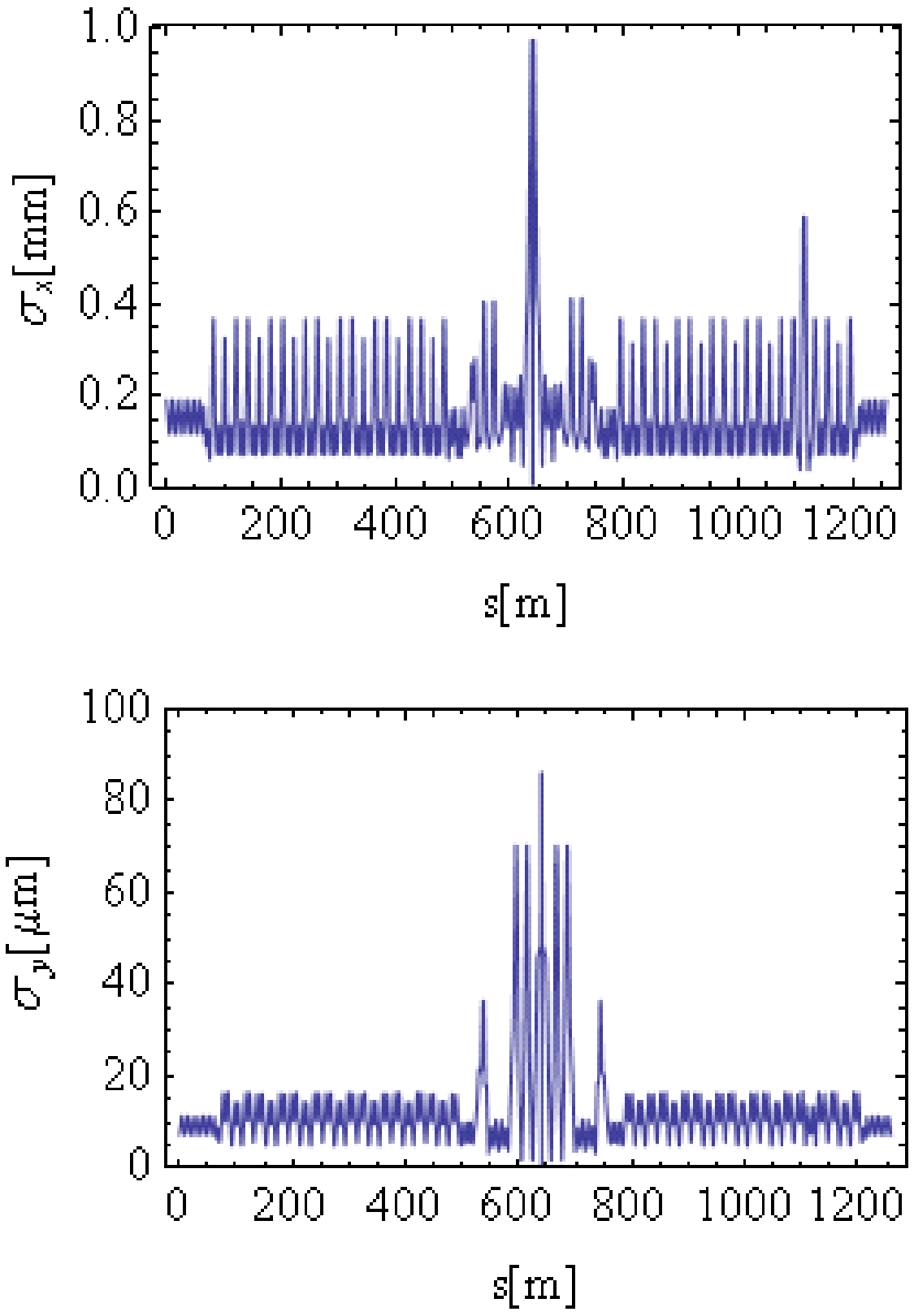

Figure 8.6: rms transverse beam sizes along the LER lattice at equilibrium.

While space charge should have little effect on injection efficiency, since its effects become noticeable only after several damping times, it could cause particle beam losses at later times, if the working point in tune-space is sufficiently close to an unstable lattice resonance. Proximity to stable resonances would be less damaging, but could also be detrimental, and could lead to unacceptable emittance degradation. Far from resonances, space charge may still compromise the target vertical equilibrium emittance, when its impact is considered in combination with radiation and linear coupling in a non-ideal lattice. The latter effect, however, should be small [1], and was neglected here.

A preliminary study was conducted for the CDR version of the SuperB LER lattice ($\varepsilon_x$=0.65 nm-rad, $\varepsilon_y$= 2.4 pm-rad, $\sigma_z$=2.46, N=6.12·$10^{10}$, $\Delta\nu_x=-0.004$, $\Delta\nu_y=-0.179$) using the weak-strong model for space charge implemented in an augmented version of the Marylie/Impact (MLI) code [2]. The code was validated during the ILC damping ring studies by calculations carried out independently using SAD [3, 4]. The space charge effects were assessed by producing tune space scans and looking for the rms emittance changes in the transverse plane. The results are reported as color-density plots showing the maximum value of the rms emittance experienced by the macroparticle beam within the indicated duration of tracking.

In Figure 8.7 the case with space charge is compared with the case without space charge. In the absence of space charge, the vertical emittance tune scan shows evidence of two third-order resonances at $2\nu_{0y}+\nu_{0x}$=n and $2\nu_{0y}-\nu_{0x}$=n, with the first being considerably stronger, and resulting in about 100% emittance growth over 300 machine turns. The other resonance resulted in a smaller ~10% growth over the same tracking time.

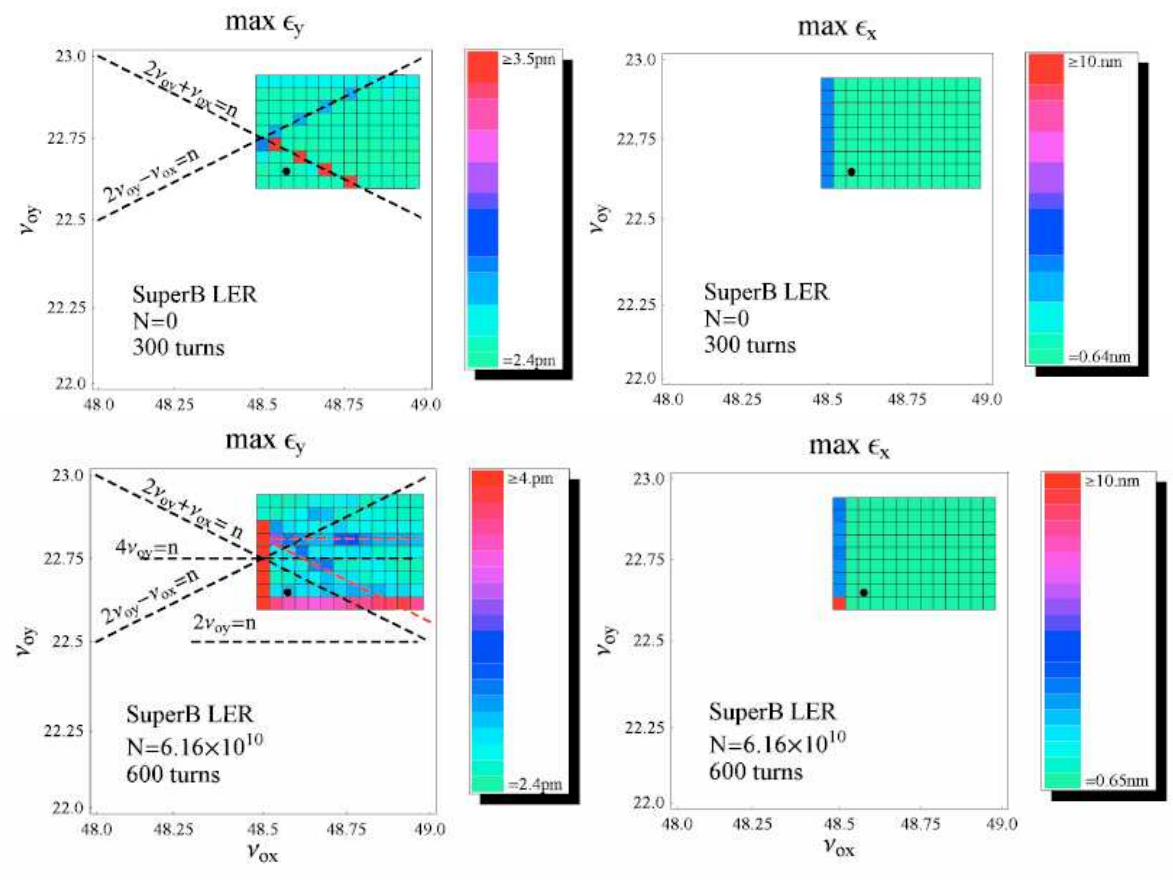

Figure 8.7: Tune scan of horizontal and vertical maximum rms emittance growth over 600 machine turns, with (top) and without (bottom) the effects of space charge. The color coding shows the vertical (left) and horizontal (right) emittance on a linear scale from minimum to maximum. The design working point is shown as a black dot.

Outside these narrow resonances, the vertical rms emittance appears to remain largely unchanged. Inclusion of space charge causes some additional degradation of the rms vertical emittance that is not apparent in short term-tracking. Not unexpectedly, the largest growth occurs along the half-integer $\nu_{0x}$=48.5 line. This resonance is already present in a bare lattice, but with visible consequences only on the horizontal motion. Its impact on the vertical motion is fostered by the x/y coupling introduced by space charge. The emittance growth detected along this line was very large, and for some choices of the vertical tune was found to lead to particle losses. Outside this resonance line and the upper part of the region affected by the $\nu_{0y}$=22.5 resonance we observe some smaller, but clearly noticeable, emittance growth up to about 30% over 600 turns (region with bluish shading).

In conclusion, this preliminary study indicates that space charge effects are noticeable in the low energy ring. One clear consequence is the enlargement of strong half integer structural lattice resonances present in the bare lattice, causing fast emittance growth and possibly, particle losses. This alone poses a significant limitation to the choice of the working point because of the sizeable space charge vertical tune shift. On a longer time scale, we encountered some areas of moderate, but clearly detectable, emittance growth. Encouragingly, however, calculations also show the existence of regions in the tune space that appear little affected by emittance growth. Further studies are needed to insure that motion stability persists on a longer timescale, up to a few damping times, and in the presence of lattice errors.

## 8.4 Fast ion instability

### *Model*

We consider $CO^+$ ions as the instability source, because the major components of residual gas in vacuum systems are CO and $H_2$, and the ionization cross-section of CO is 5 times higher than that of H2. The ionization cross-section is $1.9\times10^{-22}$ $m^{-2}$ for CO at the electron beam energy, E = 7 GeV. We assume that the partial pressure of CO gas is P = 3 × $10^{-8}$ Pa. The number of ions created by the electron beam with a population $N_e$ is expressed by:

$$n_i[\mathrm{m}^{-1}] = 0.046\, N_e P[\mathrm{Pa}]$$

In our case $n_i$ = 27$\mathrm{m}^{-1}$ for $N_e$ = 1.9 × $10^{10}$ and $P$ = 3 × $10^{-8}$ Pa. We investigate ion instabilities for various bunch filling pattern in SuperB. A simulation method based on the model shown in Fig. 8.8 is used.

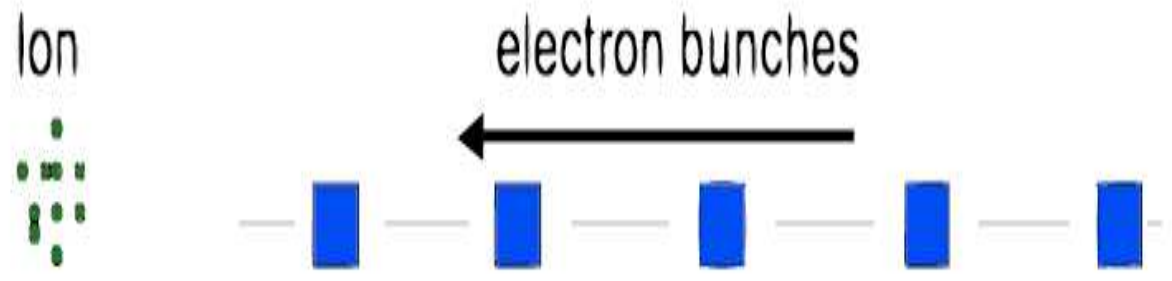

Figure 8.8: Model of beam-ion interaction.

Ions are represented by macro-particles, and each bunch is represented by a rigid transverse gaussian macro-particle. The beam size of the bunch is fixed, as determined by the emittance and β function, and only dipole motion is considered. Beam-ion interaction is expressed by the Bassetti-Erskine formula [1] for a beam with gaussian distribution in the transverse plane. The equations of motion for electrons and ions are expressed as:

$$\frac{d^2 x_{e,a}}{ds^2} + K(s)\, x_{e,a} = \frac{2r_e}{\gamma}\sum_{j=1}^{N_i} F(x_{e,a} - x_{i,j})$$

$$\frac{d^2 x_{i,j}}{dt^2} = \frac{2r_e c^2}{M_i / m_e}\sum_{a=1}^{N_i} F(x_{i,j} - x_{e,a})$$

where the suffixes $i$ and $e$ denote the ion and electron, respectively. $M_i$ and me are masses, and $N_i$ and $N_e$ are their number. γ and re are the Lorentz factor of the beam and the classical electron radius, respectively. *F(x)* is the Coulomb force expressed by the Bassetti-Erskine formula. These consist of $N_e$ + $N_i$ differential equations, where each electron couples to the motion of all ions, and each ion couples to the motion of all electrons.

It is easy to solve the equations simultaneously with a numerical method [2]. The structure of the bunch train and the $\beta$ function variation are also taken into account with this approach. The effect of a bunch-by-bunch feedback system is included in the simulation. The feedback system has a damping time of 50 turns and fluctuation of $0.02\ \sigma_y$. This gain is rather conservative with present technology.

### *Simulation of ion instability*

The simulation gives the position and momenta of every bunch, turn by turn. Figure 8.9 shows the vertical position of every bunch after 1000 turns. We use as filling parameters the bunch population (Ne = 5.5 × $10^{10}$), the bunch spacing ($L_{sp}$ = 4 ns), the number of bunches in a train ($N_b$ = 1000). Gaps between trains are simulated for three cases, $L_{gap}$ = 10, 20 and 90 × 4 ns. In Figure 8.9, the gap is removed: i.e., $y$ at *1–50*, *51–100* etc. are the vertical bunch positions of the first, second etc. trains, respectively. The amplitude of the head of the first train is exactly zero, because there is no ion effect, and the amplitudes of the first 50 bunches do not depend on the gap length.

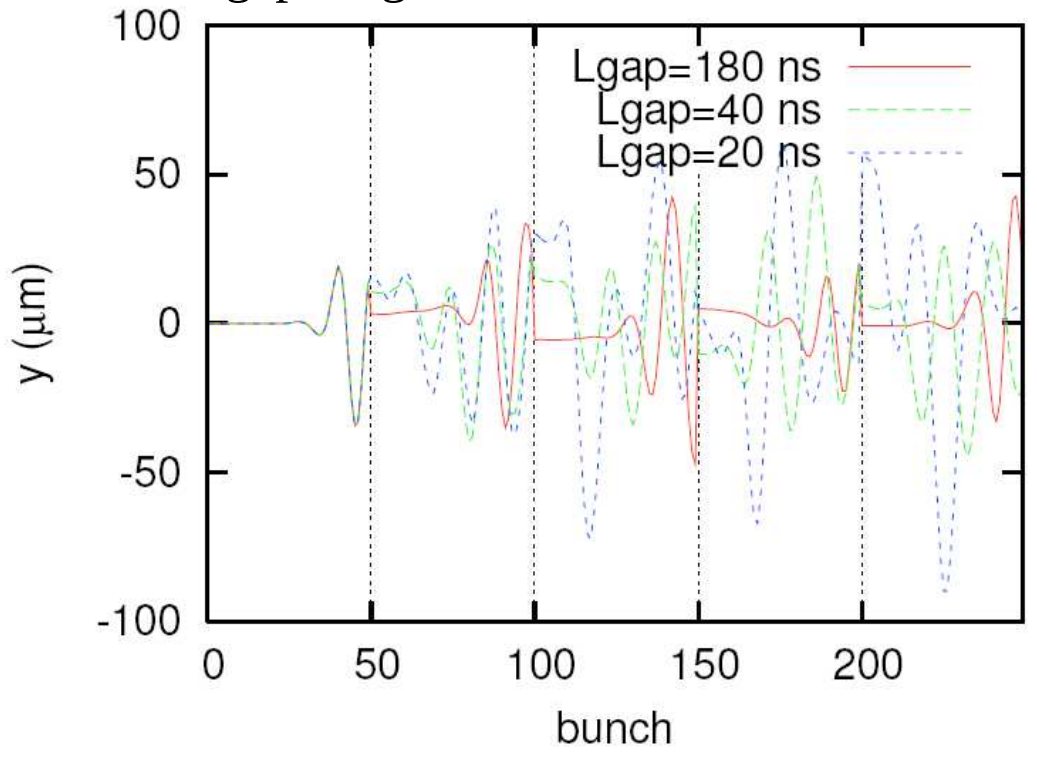

Figure 8.9: Vertical position of all bunches after 1000 turns for various train gap lengths.

Those of the second, third etc. trains are not zero, and depend on the gap length. Some ions remaining after the passage of previous trains affect the head part of the subsequent trains. The maximum amplitude is saturated for all trains at $L_{gap} \leq 40$ *ns*. This means that the gap length is efficient for clearing the ions. On the other hand, the maximum amplitudes increase along trains for $L_{gap} \leq 20$ *ns*; i.e., the gap length is not sufficient, and ions are built up.

The maximum amplitude for all bunches $\sqrt{J_y}$ is obtained turn-by-turn from the simulation. Figure 8.10 shows the evolution of $\sqrt{J_y}$ with turn number. The red and blue lines show the evolution with and without the bunch-by-bunch feedback system, respectively. From top to bottom, the amplitude growth is shown for the three gap lengths, $L_{gap}$ = *20, 40* and *180 ns*. Beam oscillations are suppressed by the feedback system for $L_{gap} \leq 40$ *ns*, while considerable residual oscillation remains for $L_{gap} \leq 20$ *ns*.

Figure 8.11 shows the variation in amplitude growth with the number of bunches in a train ($N_b$ = *100, 150, 200*), where $L_{gap}$ = *180 ns*. The instability for $N_b$ = *100* is suppressed by the feedback system, but it is not suppressed for longer trains, $N_b \geq 150$.

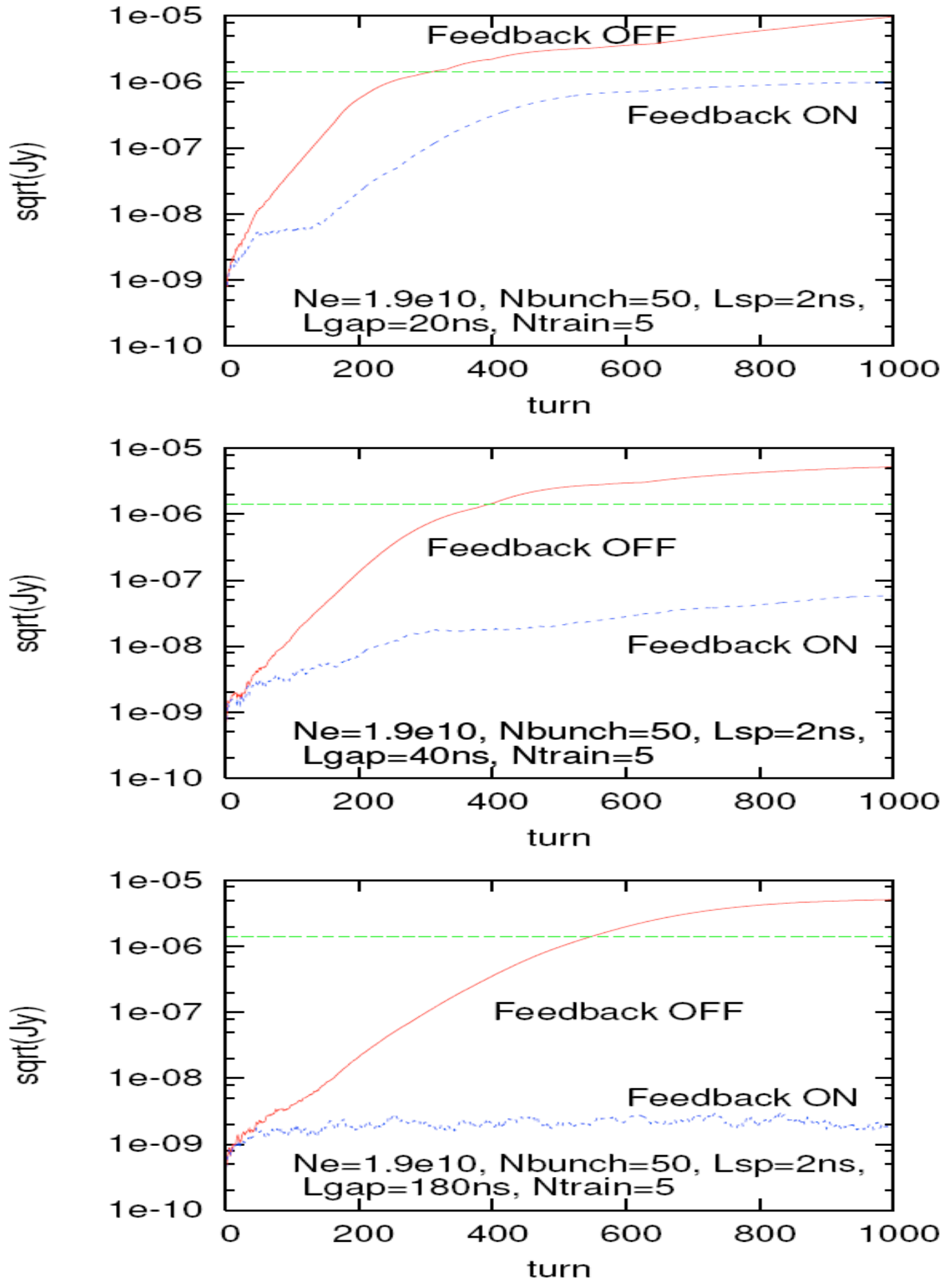

Figure 8.10. Evolution of maximum amplitude ($\sqrt{J_y}$) for train gap lengths (top to bottom) $L_{gap}$=20, 40 to 180 ns.

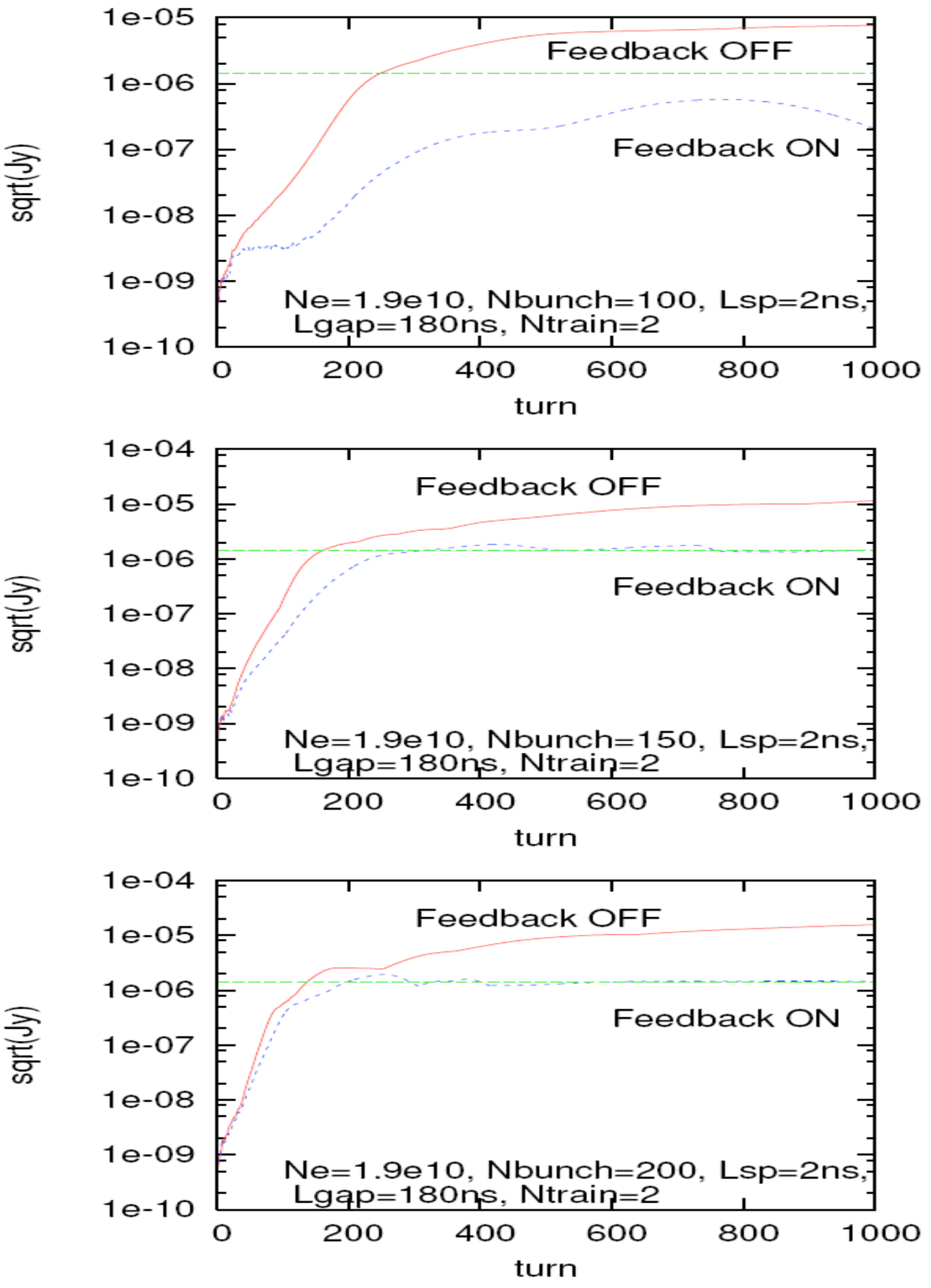

Figure 8.11. Evolution of maximum amplitude ($\sqrt{J_y}$) for various train lengths (top to bottom) $N_b$=100, 150, 200.

## 8.5 Intra Beam Scattering

Intrabeam scattering [1, 2] is associated with the Touschek effect; while single large-angle scattering events between particles in a bunch leads to loss of particles (Touschek lifetime), multiple small-angle scattering events lead to emittance growth, an effect that is well known in hadron colliders and referred to as intrabeam scattering (IBS). In most electron storage rings, the growth rates arising from IBS are usually very much longer than synchrotron radiation damping times, and the effect is not observable. However, IBS growth rates increase with increasing bunch charge density, and for machines that operate with high bunch charges and very low vertical emittance, the IBS growth rates can be large enough that significant emittance increase can be observed. Qualitative observations of IBS have been made in the LBNL Advanced Light Source [3], and measurements in the KEK Accelerator Test Facility (ATF) [4] have been shown to be in good agreement with IBS theory, IBS is expected to increase the horizontal emittance in the ILC damping rings by roughly 30% [5]; the SuperB rings will operate with comparable bunch sizes and beam energy, and with somewhat larger bunch charge, so we may expect similar emittance growth from IBS in SuperB to that in the ILC damping rings. There is a strong scaling with energy, with IBS growth rates decreasing rapidly with increasing energy. Therefore, we expect significantly larger IBS emittance growth in the SuperB low energy ring than in the high energy ring.

Several formalisms have been developed for calculating IBS growth rates in storage rings, notably those by Piwinski [1] and by Bjorken and Mtingwa [2]. IBS growth rates depend on the bunch sizes, which vary with the lattice functions around the ring; to calculate accurately the overall growth rates, one should therefore calculate the growth rates at each point in the lattice, and average over the circumference. Furthermore, since IBS results in an increase in emittance, which dilutes the bunch charge density and affects the IBS growth rates, it is necessary to iterate the calculation to find the equilibrium, including radiation damping, quantum excitation and IBS emittance growth. The full IBS formulae include complicated integrals that must be evaluated numerically, and can take significant computation time; however, methods have been developed [5, 6] to allow reasonably rapid computation of the equilibrium emittances, including averaging around the circumference and iteration.

For calculation of the IBS emittance growth in the SuperB rings, we use the formulae of Kubo et al. [6], which are based on an approximation to the Bjorken-Mtingwa formalism [2]. This approximation has been shown to be in good agreement with data on IBS emittance growth collected at the ATF [4, 7]. In our calculations, the average growth rates are found from the growth rates at each point in the lattice, by integrating over the circumference; we assume lattice natural emittances as equilibrium values at low bunch current and use iteration to find the equilibrium emittances in the presence of radiation and IBS.

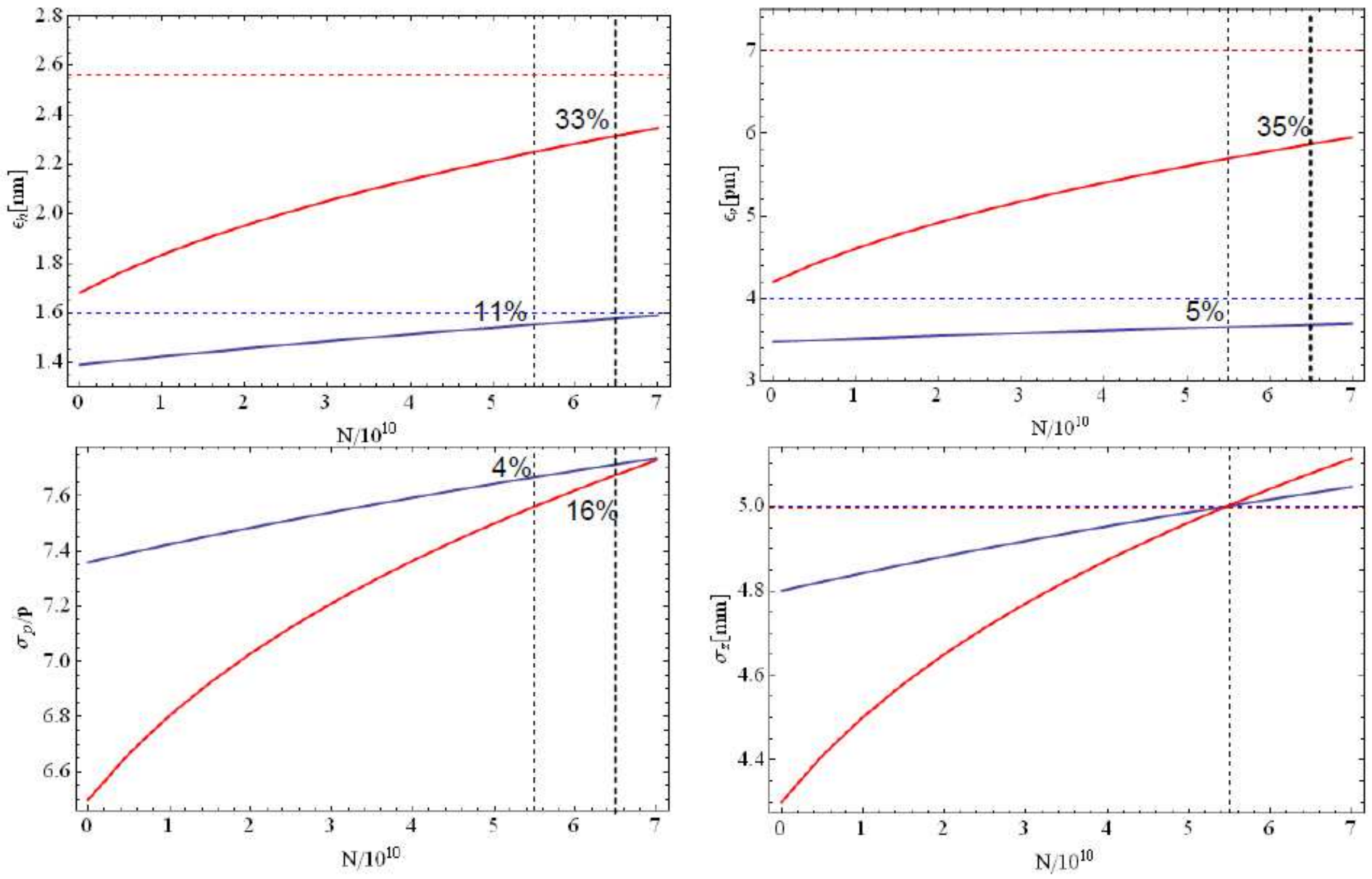

Figure 8.12: Transverse emittance growth, and growth in bunch length and energy spread in the SuperB LER, as functions of the bunch charge.

Figure 8.12 shows the equilibrium transverse emittances, bunch length and energy spread in the SuperB rings as functions of the bunch charge. In the LER at the nominal bunch charge of 6.5 $10^{10}$, the horizontal emittance is nearly 30% higher, there is also an increase in the vertical emittance 35%. The increase in transverse emittances is significant, but still below the design values indicated by the dashed lines in figure. The strong scaling of IBS growth rates with energy means that in the HER the emittance growth from IBS is much less than in the low energy ring; the effects of IBS are further mitigated by the lower bunch charge in the high energy ring. There is a 11% increase in horizontal emittance at the nominal bunch charge of 5.5 $10^{10}$ particles, and an increase in vertical emittance of about 5%.

## 8.6 Multibunch instability and feedbacks

### *Multibunch Instabilities*

Electromagnetic interaction between charged particle beam and its surroundings and between different bunches of the same beam can cause collective and coupled-bunch instabilities, which must be controlled to achieve the SuperB ambitious luminosity design goals. Control requires a combination of passive damping techniques and fast active feedbacks on an unprecedented technological scale. Solutions of multibunch instability control problems can be based on different approaches and steps:

a) theoretical analysis of the instability sources possibly using analog (i.e. real particle circular accelerators) and digital (i.e. software) models to foresee their effects versus beam specifications and bunch patterns;
b) implementation of efficient and powerful instability diagnostics for an accurate characterization of the actual problems;
c) identification of the instability sources;
d) cure or mitigation of the undesired effects;
e) passive damping techniques;
f) fast active bunch-by-bunch feedback systems in the two transverse planes and in the longitudinal one.

In the transverse plane, the resistive wall impedance is one of the most important sources of coherent multibunch growth rate. The resistive wall impedance can be estimated by the following formula:

$$Z_T(\omega) = (R * Z_0 / b^3) * \delta_s(\omega))$$

where R is the accelerator radius, $Z_0$ is the vacuum characteristic impedance, $\delta_s$ is the skin depth, which is proportional to the square root of the conductivity of the chamber wall, and b is the aperture radius. The most common materials for the vacuum chamber are copper, aluminum and stainless steel. The first two metals show much lower impedance than the third one, therefore it is better to avoid using stainless steel for the vacuum chambers to keep the transverse impedance $Z_T(\omega)$ as low as possible.

The beam also loses energy due to wake fields, which are excited in the beam pipe vacuum elements. Wake fields include short-range fields, like resistive wall and geometrical wake fields, and long-range fields like higher order modes (HOMs) excited in the RF cavities and kickers and possible low-Q geometrical cavities in the beam pipe, for example between in and out tapers.

A powerful and efficient way to analyze and measure (a posteriori) the beam modal growth rates in a circular accelerator is by switching off – switching on bunch-by-bunch feedback systems and recording data streams as shown in Figure 8.13. These results can be analyzed offline by using diagnostics programs [1], [2] developed during last 10-15 years mainly for feedback testing purpose.

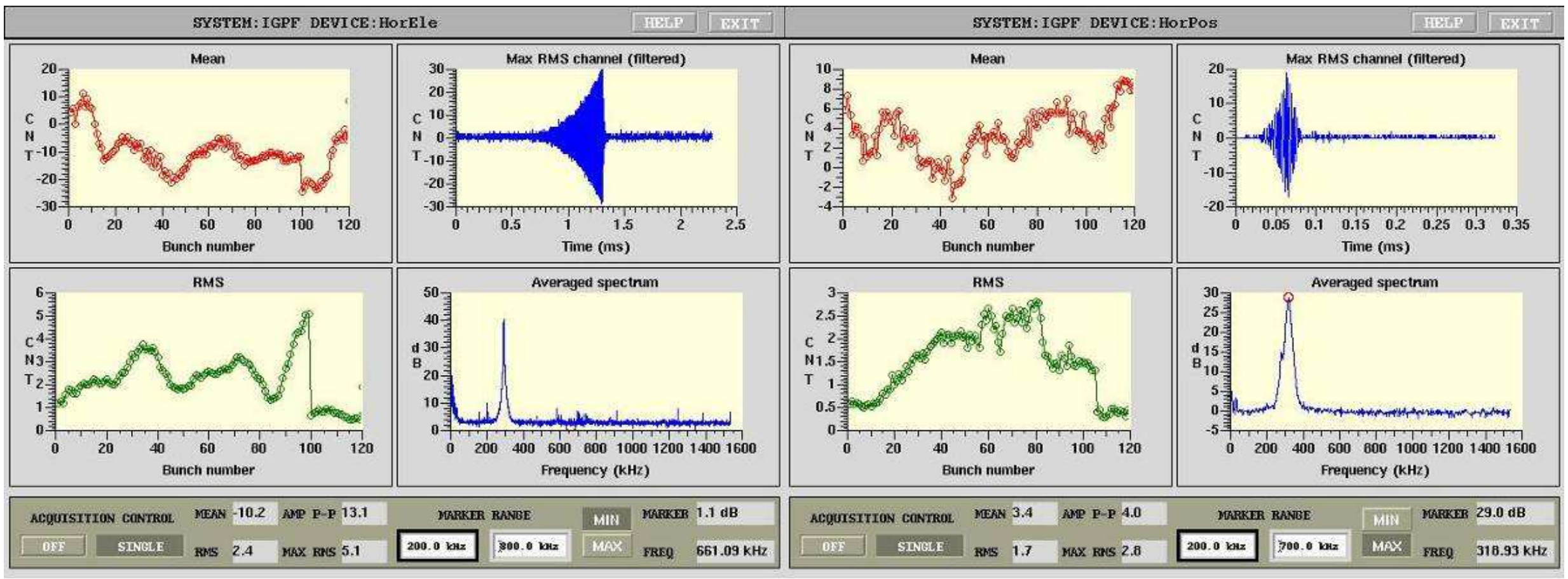

Figure 8.13: DAΦNE: e- (left) and e+ (right) real time feedback-off / feedback-on plots automatically generated by the "iGp" feedback systems installed in the horizontal e+ / e- planes.

Coupled bunch instabilities in the longitudinal and transverse planes can be also easily excited by mismatches in energy, by phase jitters or by trajectory errors of the injected charges. As consequence, the timing and the injection system specifications should limit the charge arrival time uncertainty to a peak-peak jitter <1 ps.

Parasitic electron clouds in the positron ring and positive charged ions in the electron ring are another source of coherent coupled bunch instabilities. For example in DAΦNE, as shown in Figure 8.14, the two main rings, in spite of their identical RF cavities and vacuum chambers, show very different behavior in terms of coherent coupled bunch instabilities [3]. In the positron ring, coherent instability growth rates have been measured with speed up to 10 μs (not in the picture), corresponding to ~ 30 revolution time. In the other twin ring, the electron beam has shown much slower coherent instability grow rate, at level of 140 μs.

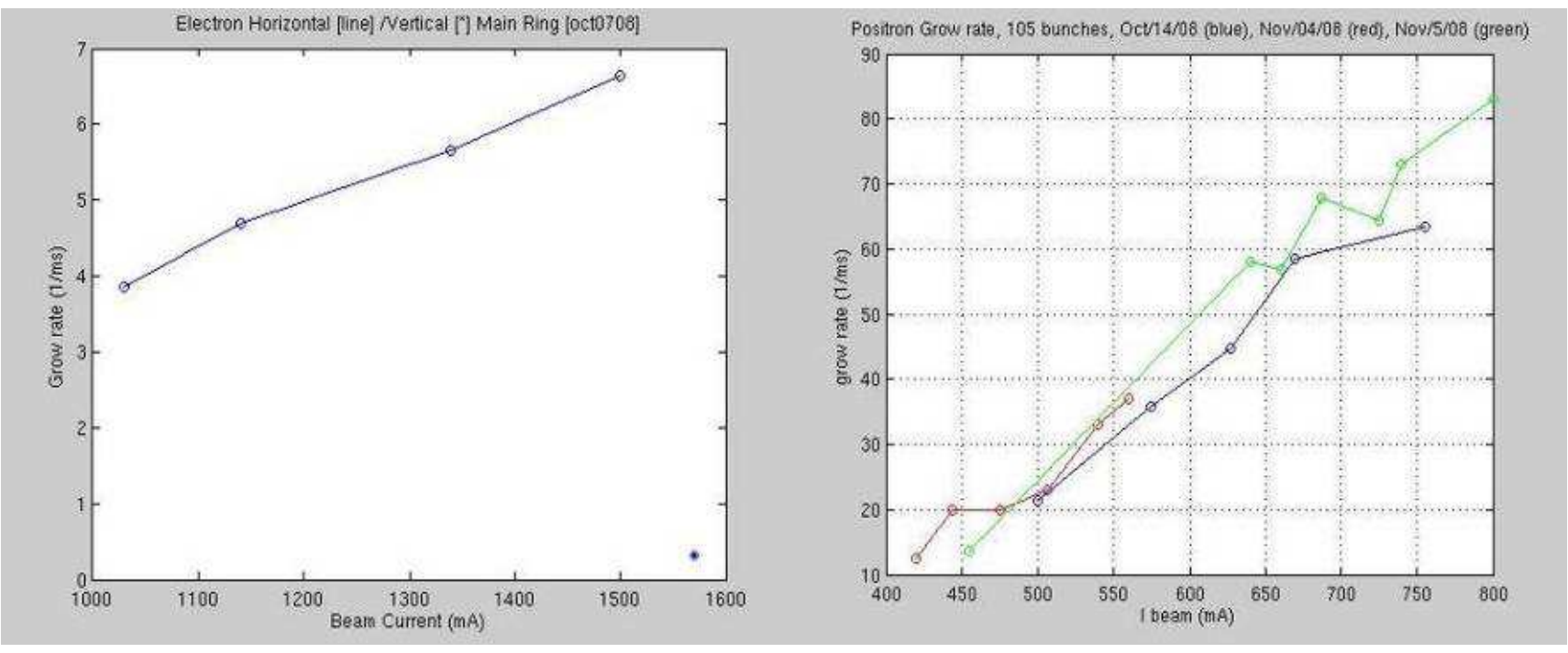

Figure 8.14: DAΦNE $e^-$ and $e^+$ horizontal inverse growth rate (1/ms) versus beam current. The $e^-$ inverse rate of 7 $ms^{-1}$ corresponds to 142 μs, while the e+ 83 $ms^{-1}$ corresponds to 12 μs. The e+ beam shows at least a factor 12 in the instability growth rate respect to the $e^-$ beam.

The past experience at PEP-II indicates that the actual inverse growth rate for transverse instabilities has been close to 1/ms, several times larger than originally estimated [4]. In term of revolution periods, this value gives a rough estimate of more than 100 turns for the PEP-II coherent growth rates. For the SuperB that have much shorter bunches and much lower transverse emittances, the hypothetical growth rates could reach speeds close to few revolution periods or even less. However from tests performed in the DAΦNE positron ring, we know that it is always possible manage more power in the feedbacks installing as many systems as necessary. Indeed it has been proved [5] that two separate feedback systems for the same oscillation plane can work in perfect collaboration doubling the feedback damping inverse time, as shown in Figure 8.15, where it is shown the DAΦNE single horizontal feedback (top plots, for I=560mA, mode -1 [=119], grow=34.5 $ms^{-1}$, damp=-104 $ms^{-1}$), and double horizontal feedback (bottom plots: I=712mA, mode -1 [=119] , grow=43.7 $ms^{-1}$, damp=-233 $ms^{-1}$). The instability damping time is 4.3 μs, i.e. ~13 revolution turns.

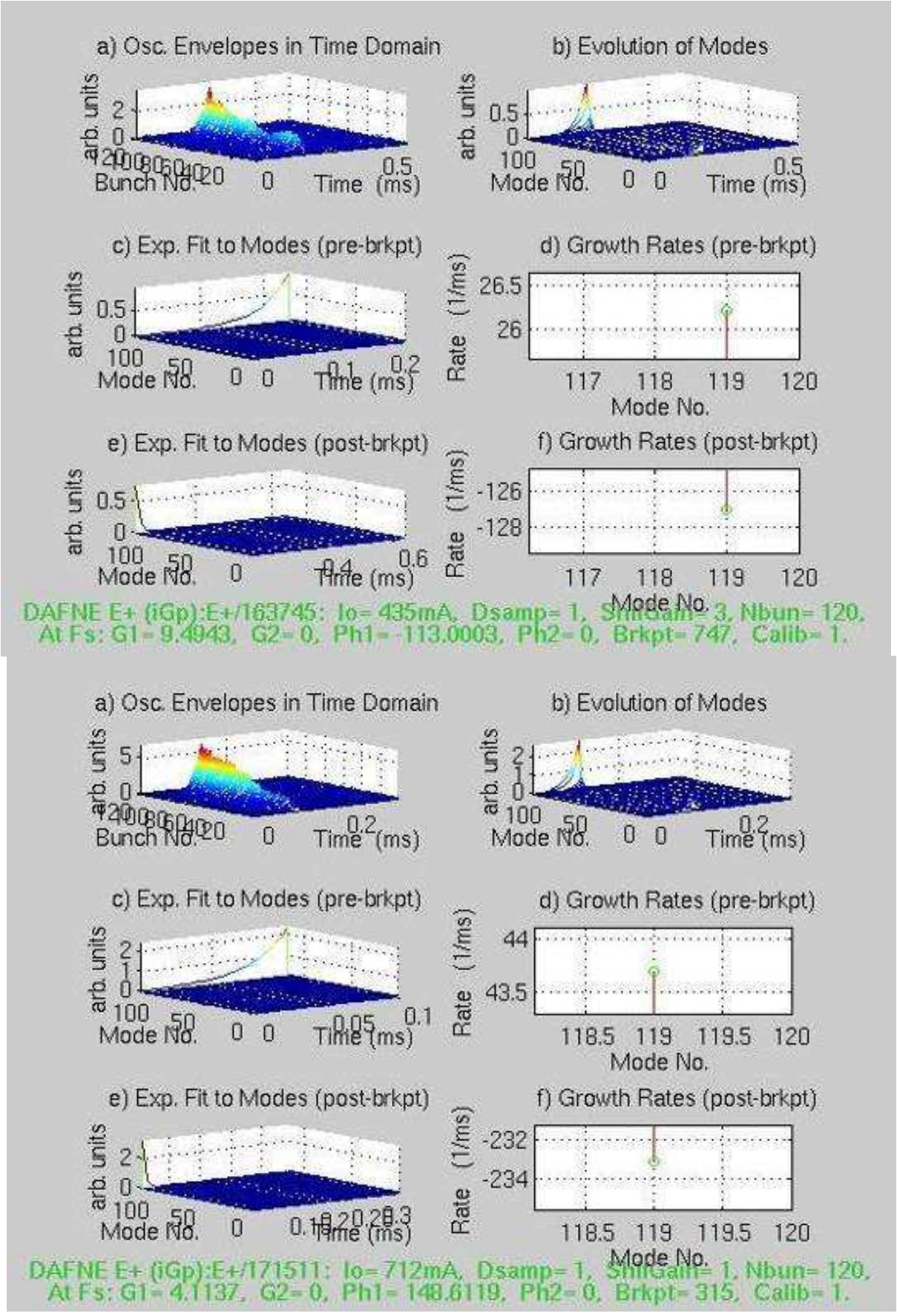

Figure 8.15: DAΦNE $e^+$ single horizontal feedback (top plots) and double horizontal feedback (bottom).

### *Transverse and Longitudinal Bunch-by-Bunch Feedback Systems*

The motivations for a new design of the bunch-by-bunch feedback are based on the following three points:

1. Dramatic acceleration of electronic component development, is making obsolete in short time all the signal processing modules of the existing PEP-II and DAΦNE transverse and longitudinal bunch-by-bunch feedback systems, and SuperB commissioning cannot start before 2014.
2. Moreover the present advancement of the electronic technology doesn't justify anymore two different designs for the transverse [6-10] and the longitudinal [11-14] feedbacks, as it was necessary in the past for PEP-II and other circular accelerators for many reasons. Two different designs bring of course also to more maintenance problems, both from hardware/software and from human resource points of view.
3. Low emittance beams ask for small impact feedback design. Horizontal and vertical emittances can be calculated using the following formula [15]:

$$\sigma_i^2 = \beta_i \varepsilon_i + (\eta_i \sigma_\varepsilon)^2$$

where $\sigma_i$ is the measured beam size in the horizontal or vertical plane (i= x, y), $\beta_i$ and $\eta_i$ are respectively the betatron and dispersion functions at the source point in the corresponding plane; and $\varepsilon_i$ and $\sigma_\varepsilon$ are the emittance and the relative energy spread of the e+ / e- beam [3]. From the above formula (8.5.2), it is evident that an increment of the beam size leads directly to an emittance growth and the feedback systems, that send the correction signals by powerful amplifiers, can increase the beam size, in particular the vertical one, pumping undesired noise even if minimal.

A new feedback design can't be just a software porting but it must be based on robustness, flexibility, scalability and innovation, and, as first consequence, the digital processing unit (DPU) should be the same for transverse and longitudinal feedback systems. New feedback systems needs internal and beam diagnostics tools and the legacy of the previous systems should be carefully implemented with the best compatibility. A preliminary scheme of the transverse and longitudinal bunch-by-bunch feed-backs is shown in Figure 8.16.

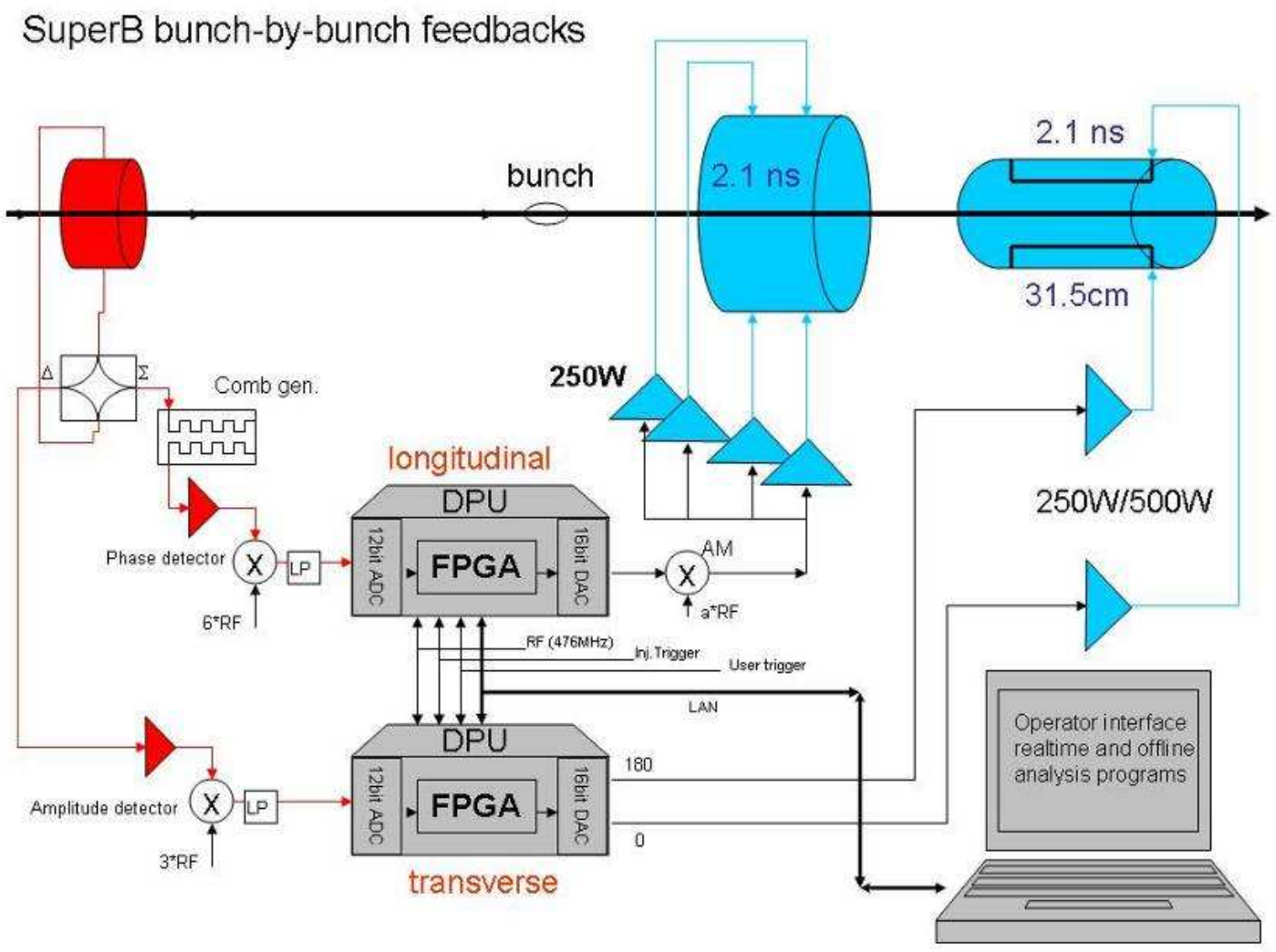

Fig. 8.16: SuperB bunch-by-bunch feedbacks are based on identical DPU (digital processing unit) for both longitudinal and transverse systems. The DPU core is implemented by a single powerful FPGA (field programmable gate array) containing >2000 DSP (digital signal processor).

The bunch-bunch feedback R&D list includes the following main upgrade points respect to the previous feedback versions:

1. have a very low noise analog front end @ n*RF, with n in a range between 3 and 6;
2. maintain low cross-talk between adjacent bunches under 40 dB (better 60 dB) in front end;
3. given that in the proposed DPU (digital processing unit) the betatron (or synchrotron) phase response is generated by a flexible F.I.R (finite impulse response) filter, only one pickup for each feedback system is at this point necessary. This simplifies design, installation and maintenance of the systems and helps to have less noise in the output correction signals;
4. digital processing unit with 12-bit ADC (analog to digital converter) and 16-bit DAC (digital to analog converter) for high dynamic range feedback loop >= 72dB and to have minimal quantization noise. Jitter on the sampling clock signal (476 MHz) must be less than 1ps (peak-peak);

5. “dual gain” approach to minimize residual beam motion and feedback noise on the beam: this feature can be implemented in DPU;
6. integrated beam-feedback model with easy code and parameter download to DPU;
7. test (at DAΦNE) 500W versus 250W power amplifiers with a bandwidth > 476MHz/2 to cope with fully populated buckets (2.1 ns spacing) even if, in the first commissioning times, this feature will not be strictly necessary. A 250MHz (>RF/2) bandwidth is enough large frequency band because every unstable mode to be damped has two sidebands and it is sufficient for damping to kick just one of them;
8. dual separated timing to pilot the transverse power stage and the stripline kickers by a more flexible timing scheme;
9. Cavity kickers (for longitudinal systems) and stripline kickers (for transverse systems) for 2.1 ns bunch spacing. Respect to the past PEP-II, working with double bunch spacing, shorter stripline kickers for the transverse systems are necessary, with a 31.5 cm length (half bucket) to allow electromagnetic filling of the kickers avoiding crosstalk between bunches.

## 8.7 Coherent synchrotron radiation

With a very short bunch length, coherent synchrotron radiation (CSR) emission can bring additional energy losses and can drive microwave instabilities. The physics of this effect can be seen from the pictures of electric force lines of a charged bunch moving in a magnetic field inside a vacuum chamber [1], see Figure 8.17.

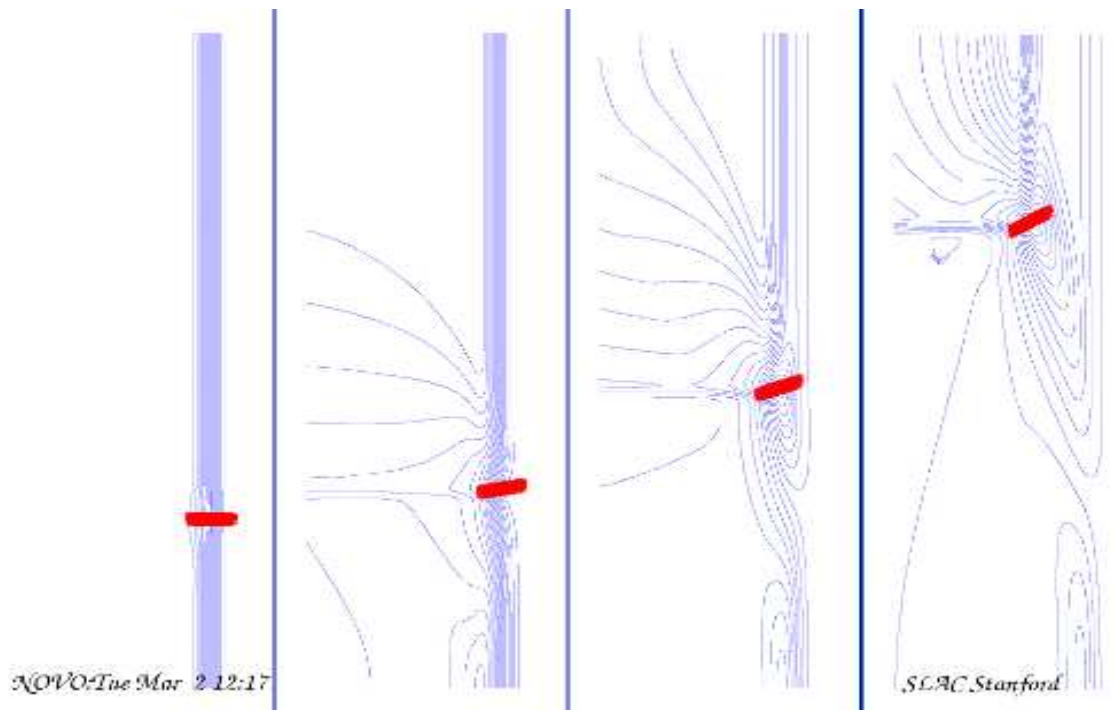

Figure 8.17: Snapshots of electric force lines of a charged bunch moving in a magnetic field inside a vacuum chamber.

In the open space the energy loss per turn due to CSR may be described by the formula [2]

$$U_{CSR} = \frac{Z_0 c Q}{2\rho}\left(\frac{\rho}{2\sigma}\right)^{4/3}$$

We use the following definitions: $Q$ is the bunch charge, ρ is the bending radius, $Z_0$ is the impedance of a free space, $c$ is the speed of light. This formula is valid if the bunch length satisfies the condition:

$$\sigma \le \frac{\rho}{\gamma^3}.$$

This condition is very well fulfilled in the Super-B case: HER energy is 6.7 GeV and LER energy is 4.18 GeV, corresponding to relativistic factors of 1.3 $10^4$ and 0.8 $10^4$. However it is surprising that energy loss per turn increases with the bending radius as $\rho^{1/3}$.

If the bunch length is comparable to the size of the beam pipe then metal walls will shield CSR emission. To take into account this effect we may use an approximate formula for the results, obtained in reference [3] for the parallel plate shielding ($h$ is half the distance between plates)

$$\tilde{U}_{CSR} \approx U_{CSR} \times \frac{\Sigma}{\sinh(\Sigma)} \qquad \Sigma = \frac{\sqrt{12}\sigma}{h}\sqrt{\frac{\rho}{h}}$$

In this case we have a more physical result for the energy loss: it decreases with the bending radius:

$$\tilde{U}_{CRS} \propto \left(\frac{h}{\sigma}\right)^3 \times \frac{1}{\sigma^{1/3}\rho^{2/3}}$$

The shielding function is shown in Fig. 18.8. Parallel plates shield the CSR emission by a factor of 10 when $\Sigma = 4.5$. Power loss of the beam with a current $I$ is.

$$P_{CSR} = U_{CSR} \times I$$

The bunch length is designed to be 5 mm in the Super-B HER and LER and the bunch charge $Q$=9.3 nC for the nominal current $I$=2.12 A and number of bunches of $N$=1018. The two rings have different type of bending magnets (see Figure 8.19), however the most of the dipole magnets have a bending radius of 85.2 m in HER and 28.4 m in LER. For comparison, PEP-II had dipole magnets with a bending radius of 164 m in HER and 13.8 m in LER.

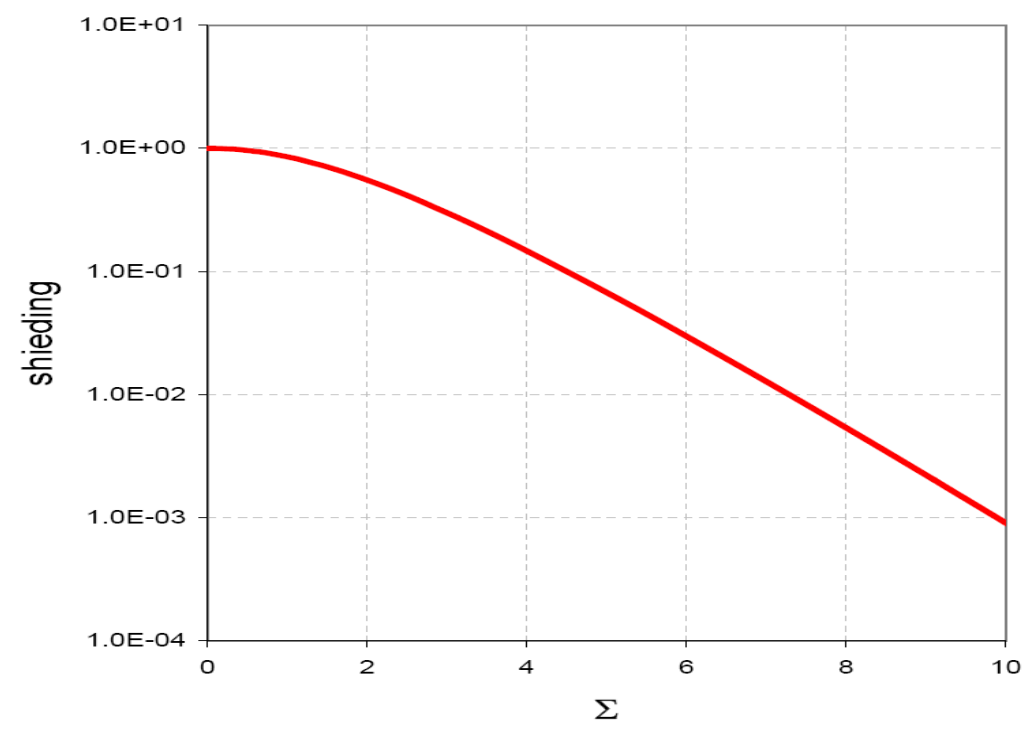

Figure 8.18: Parallel plate shielding

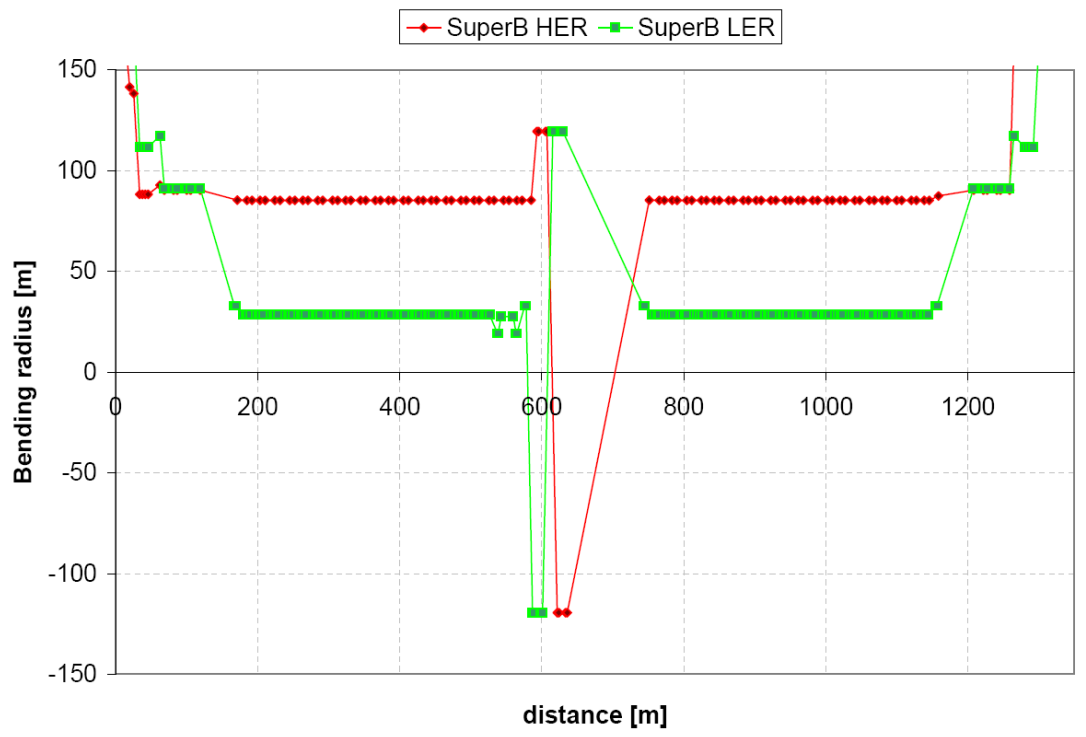

Figure 8.19: Bending fields in HER and LER.

The energy loss and power loss for nominal currents are shown in Fig. 8.20. We may state that the CSR effect will not play an important role in the Super-B project.

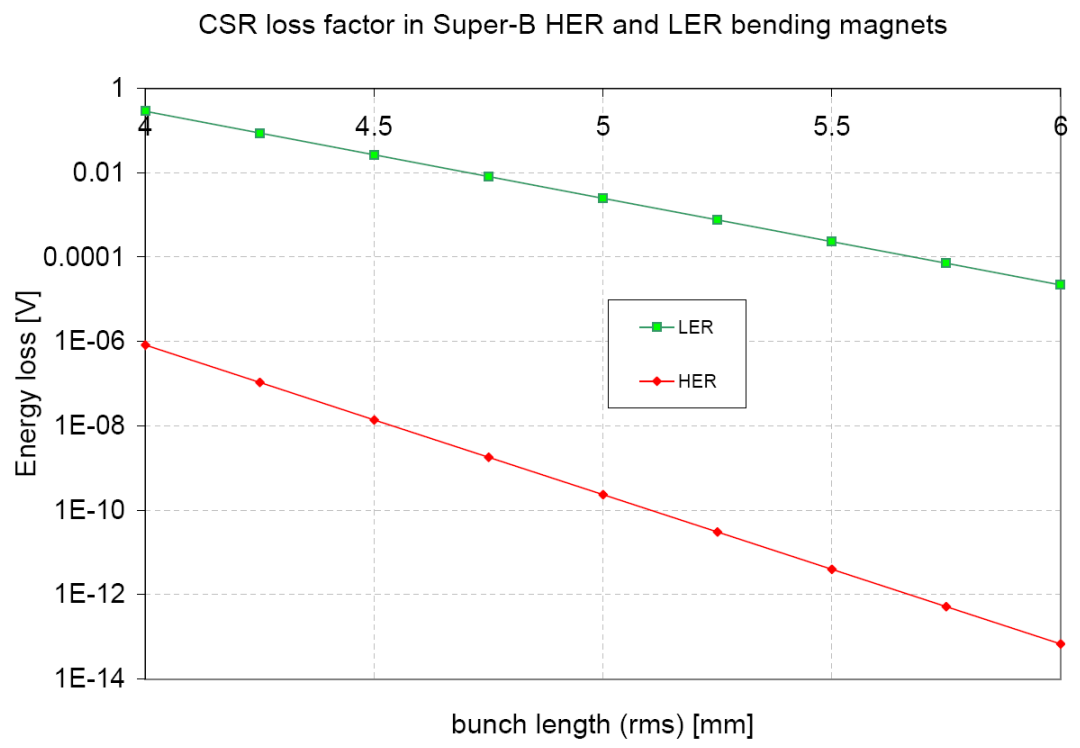

Figure 8.20: Energy loss due to CSR emission.

## 8.8 Single bunch impedance effects

The main effect, which determines the stability of the beam in a storage ring, is electromagnetic interaction of the beam particles with vacuum chamber elements. We describe this interaction by the wake potentials of all vacuum elements in the ring. Phase-space distribution function effectively describes the motion of particles in the bunch. This function is the solution of the Fokker-Planck equation. We use an effective numerical algorithm [1] to solve this non-linear equation together with wake field potentials.

### Wake Green function

We describe wake potential $W(s)$ as convolution of charge density distribution $\rho(s)$ and wake function $w(s)$

$$W(s) = \int_{-\infty}^{s} w(s-s')\rho(s')ds'$$

We use known analytical expressions for the wake (or Green) function. Green function is the wake potential of a point charge for a particular accelerator element. Direct numerical solution of the Maxwell equations gives wake potentials only for finite length bunches. We introduce a new wake function $\tilde{w} = \tilde{w}(s_q, s)$, which has an additional distance parameter $s_q$. We define this function as a "quasi-Green function". With this function we calculate directly the approximation $\tilde{W}(s)$ of the wake potential

$$\tilde{W}(s) = \int_{0}^{\infty} \tilde{w}(s_q, s')\rho(s+s_q-s')ds'$$

From the expression (2) one can see that the quasi Green function becomes real Green function, when the distance parameter takes zero value

$$\tilde{w}(s_q, s) \rightarrow w(s) \text{ when } s_q \rightarrow 0$$

### Fokker-Plank equation

To study the effect of the wake fields on the longitudinal beam dynamics in a storage ring we use the solutions of the Fokker-Planck equation for the phase-space distribution function $\psi = \psi(t,x,p)$ of momentum and coordinate:

$$\frac{\partial}{\partial t}\psi + \dot{x}\frac{\partial}{\partial x}\psi + \dot{p}\frac{\partial}{\partial p}\psi = \lambda\frac{\partial}{\partial p}\left\{p\psi + \frac{\partial}{\partial p}\psi\right\}$$

Time derivatives of the canonical coordinates are:

$$\dot{x} = p$$

$$\dot{p} = F = -\frac{\sin(k_{rf}x+\varphi_0)-\sin\varphi_0}{k_{rf}} + \frac{eNc}{V_{rf}\omega_{rf}\sigma_0}W(t,x)$$

$$k_{rf} = \frac{\omega_{rf}}{c\sigma_0} \qquad \lambda = \frac{2}{f_s\tau_{damp}}$$

Bunch density is:

$$\rho(t,x) = \int_{-\infty}^{\infty}\psi(t,x,p)dp$$

Wake potential is:

$$W(t,x) = \int_{-\infty}^{x+x_q}\rho(t,x')\tilde{w}(x_q, x+x_q-x')dx'$$

The coordinate and momentum are normalized by natural (zero-current) value of the bunch length $\sigma_0$ and momentum spread $p_0$. Time is measured in synchrotron periods. So for the description of the ring we need only:

- natural bunch length $\sigma_0$
- bunch charge $eN$
- RF voltage $V_{rf}$
- RF frequency $\omega_{rf}$
- synchrotron frequency $f_s$
- damping time $\tau_{damp}$

### Wake potentials

We consider two models of the beam vacuum chamber of the ring.

First model includes only the most important elements: RF cavities; longitudinal, transverse, injection and abort kickers; interaction region and collimators. We also include resistive wall wakes. Wake potentials were calculated for a bunch of 0.5 mm length. Several wake potentials are shown in Figures. 8.21 to 8.23.

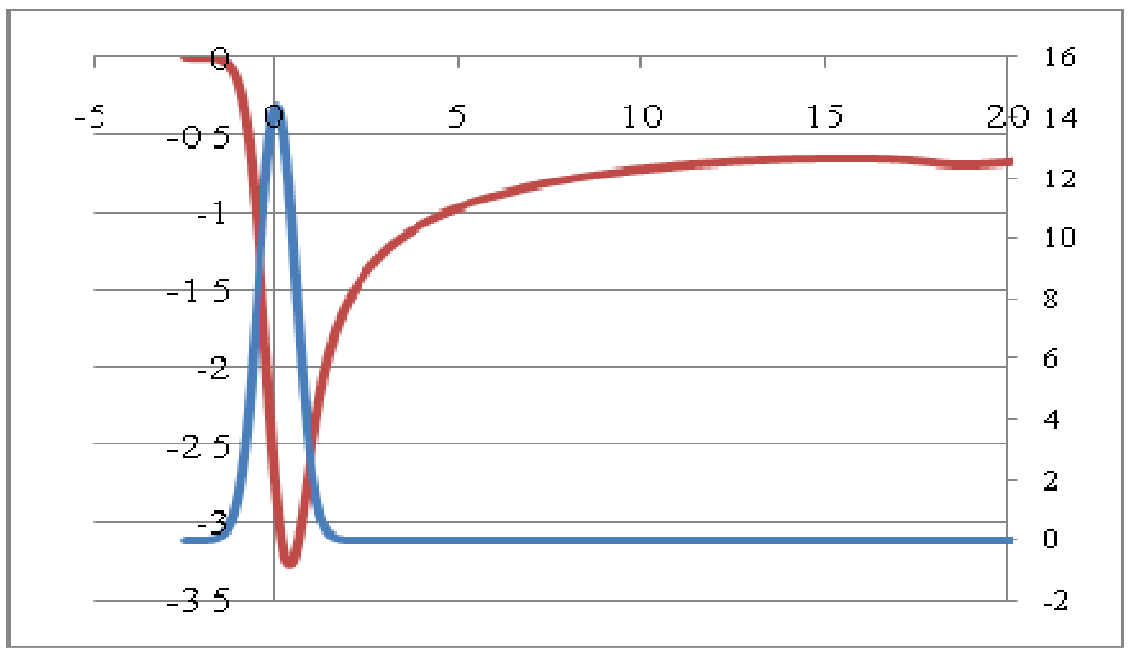

Figure 8.21: Cavity wake potential

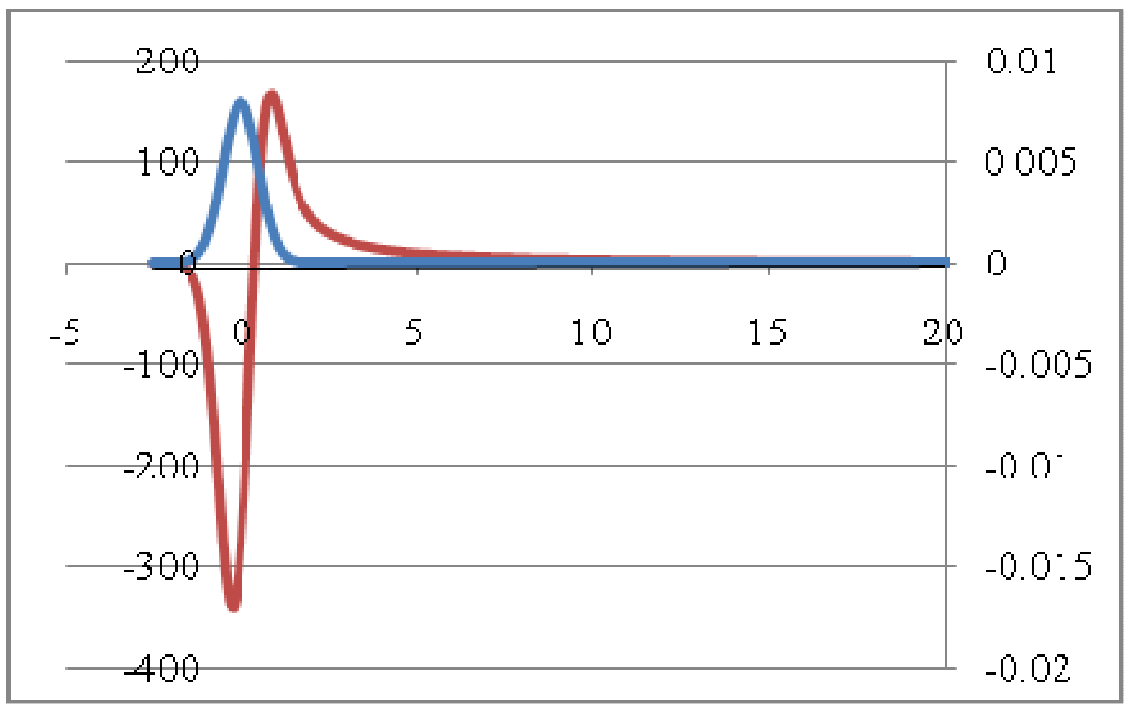

Figure 8.22: Resistive-wall wake potential

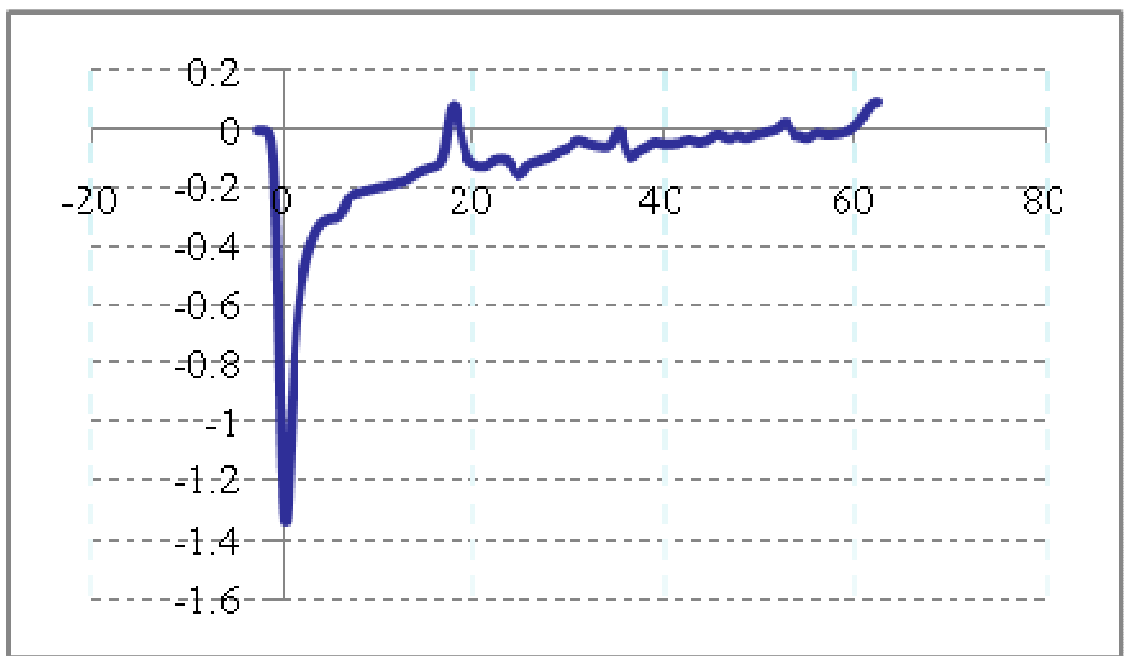

Figure 8.23: Transverse kicker wake potential.

Total wake potential for the Super-B low energy ring (LER) is shown in Figure 8.24. The wake potential for the nominal bunch length of 5 mm is shown in Figure 8.25. Loss factor is 11 V/pC.

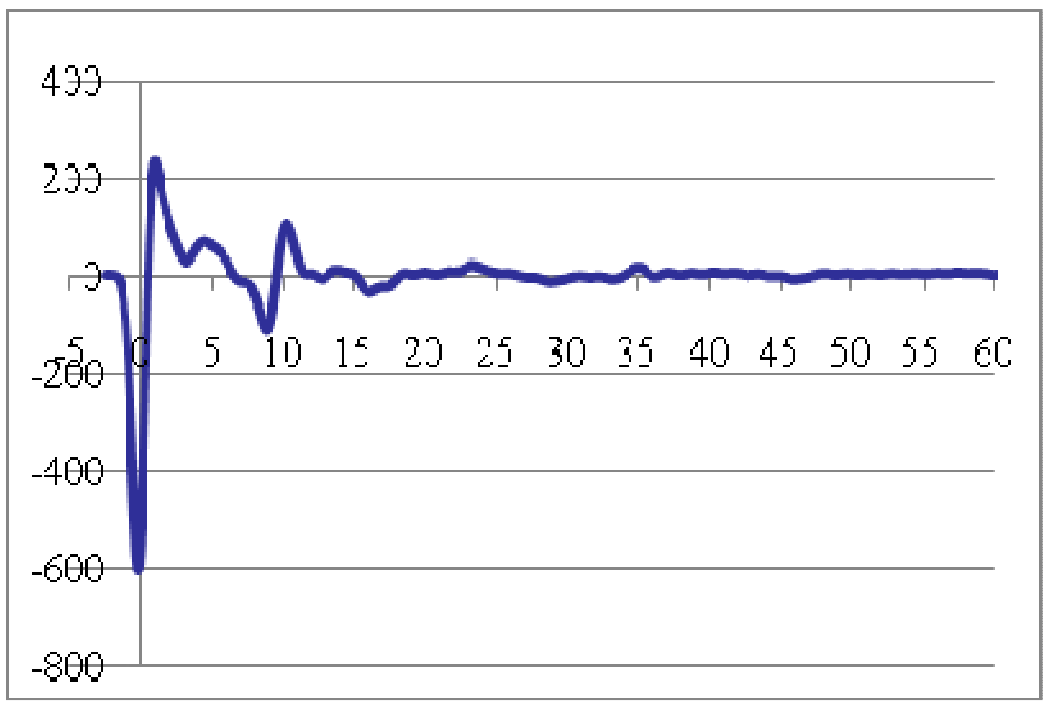

Figure 8.24: Total wake potential for LER

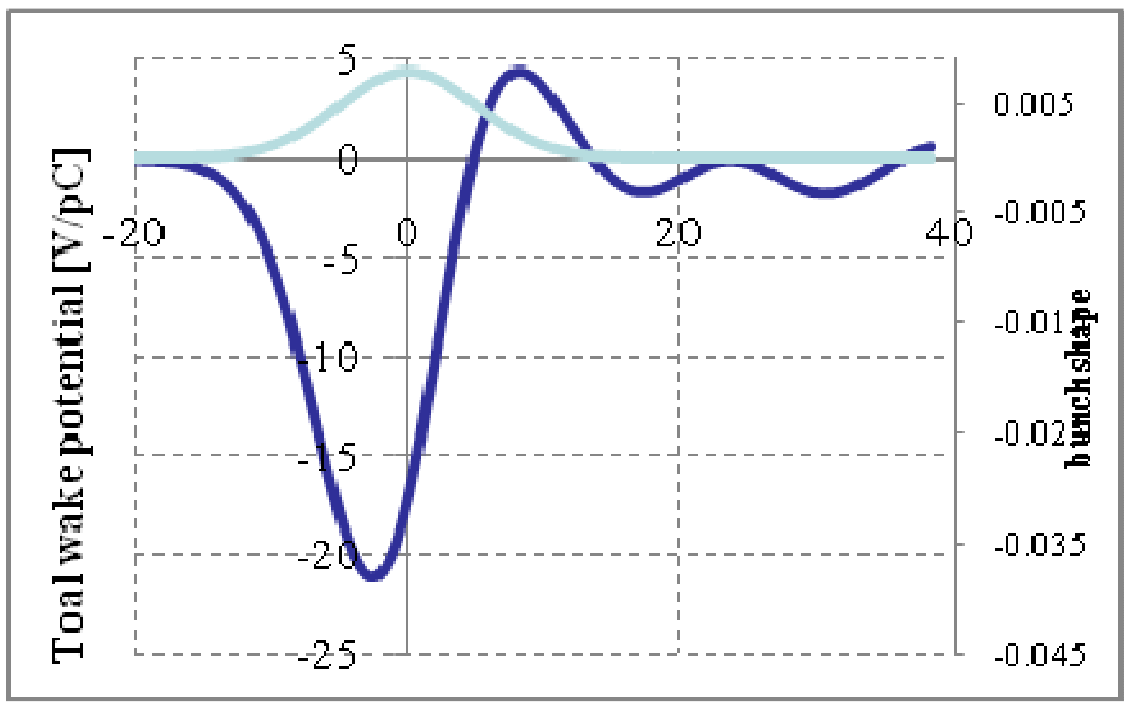

Figure 8.25: Wake potential for a 5 mm bunch .

Result of simulations of the bunch particle motion in the LER for the nominal parameters is shown in Figure 8.26.

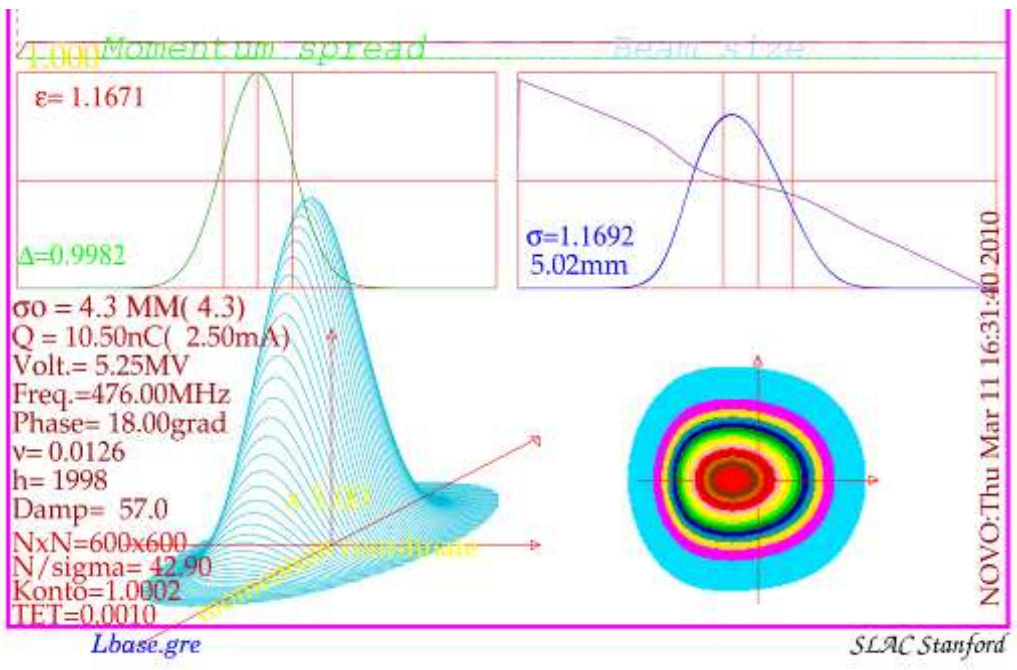

Figure 8.26: Distribution function and bunch shape for the nominal super-B parameters.

Bunch lengthening and energy spread is shown in Figure 8.27. Instability starts at 40 nC per bunch and has a turbulent character (see Figure 8.28).

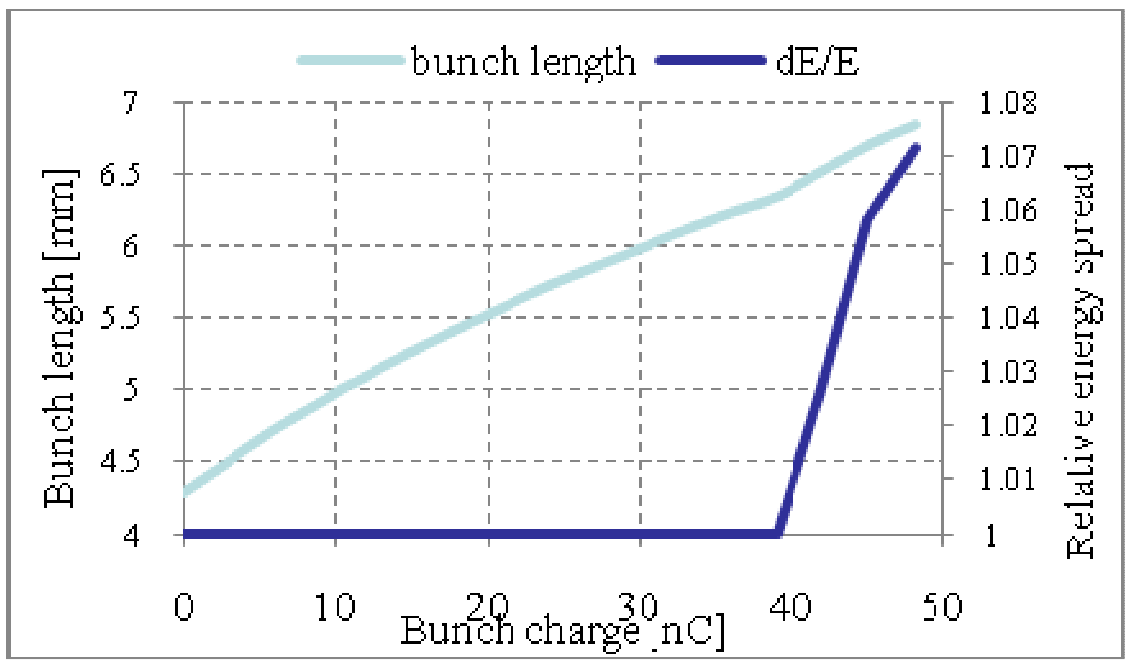

Figure 8.27: Bunch lengthening and energy spread.

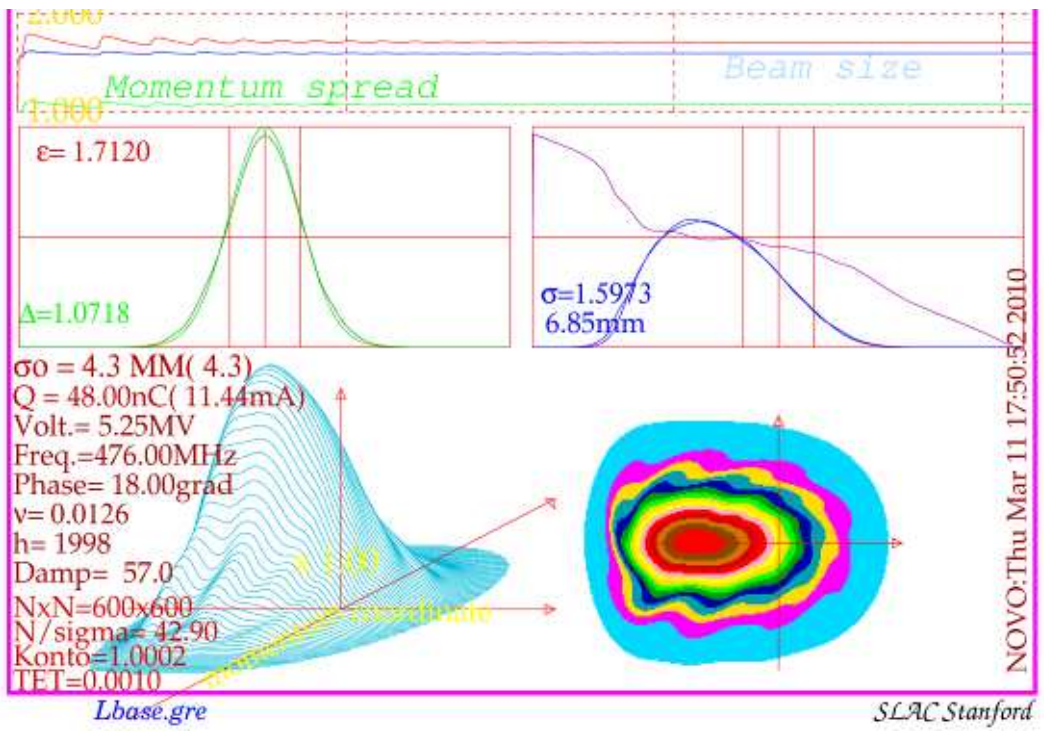

Figure 8.28: Turbulent instability at 11 mA per bunch.

Results for the high energy ring (bunch lengthening and energy spread) are shown in Figure 8.29. We did not find instability up to 50 nC per bunch.

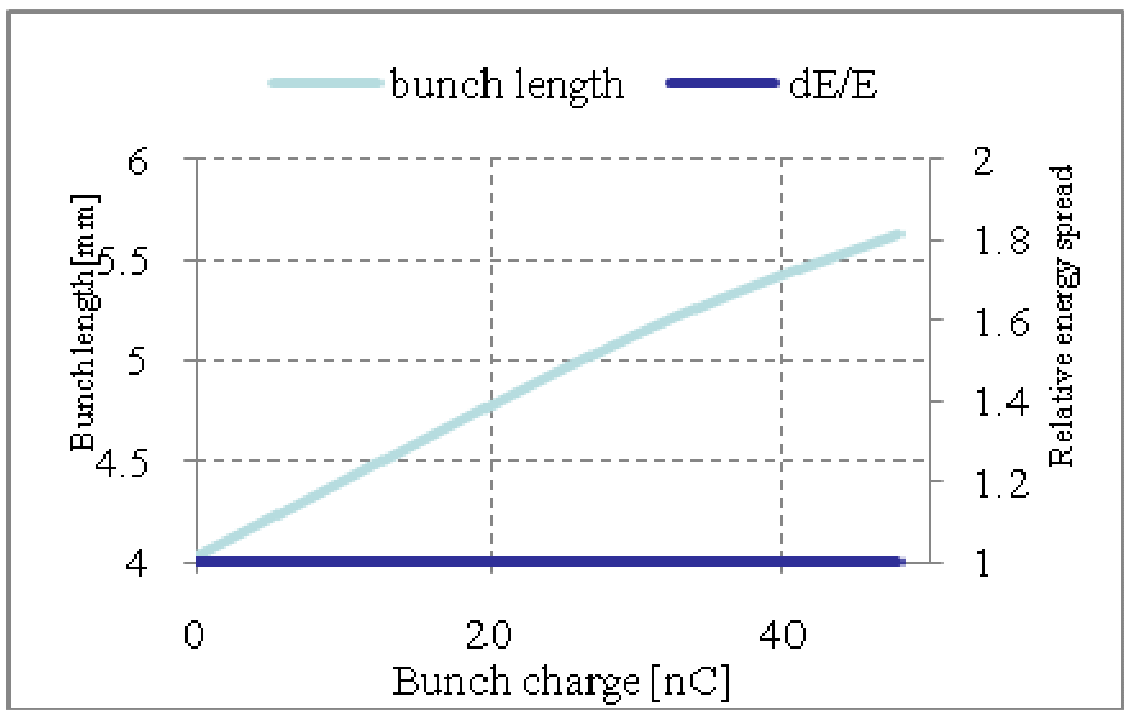

Figure 8.29: Bunch lengthening in the HER ring.

Other model is the vacuum chamber of the PEP-II rings. We found wake potentials for LER and HER rings from the results of the real measurement of the loss factor and the bunch length current dependence [2-4]. This wake potential describes real chamber, which includes a lot of different elements like vacuum ports, gap rings, and bellows and so on. We assume that for the Super-B the wake potential must be approximately two times smaller as the Super-b ring is smaller than PEP-II ring. Results of simulations for bunch lengthening and energy spread are shown in Fig. 8.301. Now instability starts earlier at 13 nQ, but still higher than the nominal bunch charge.

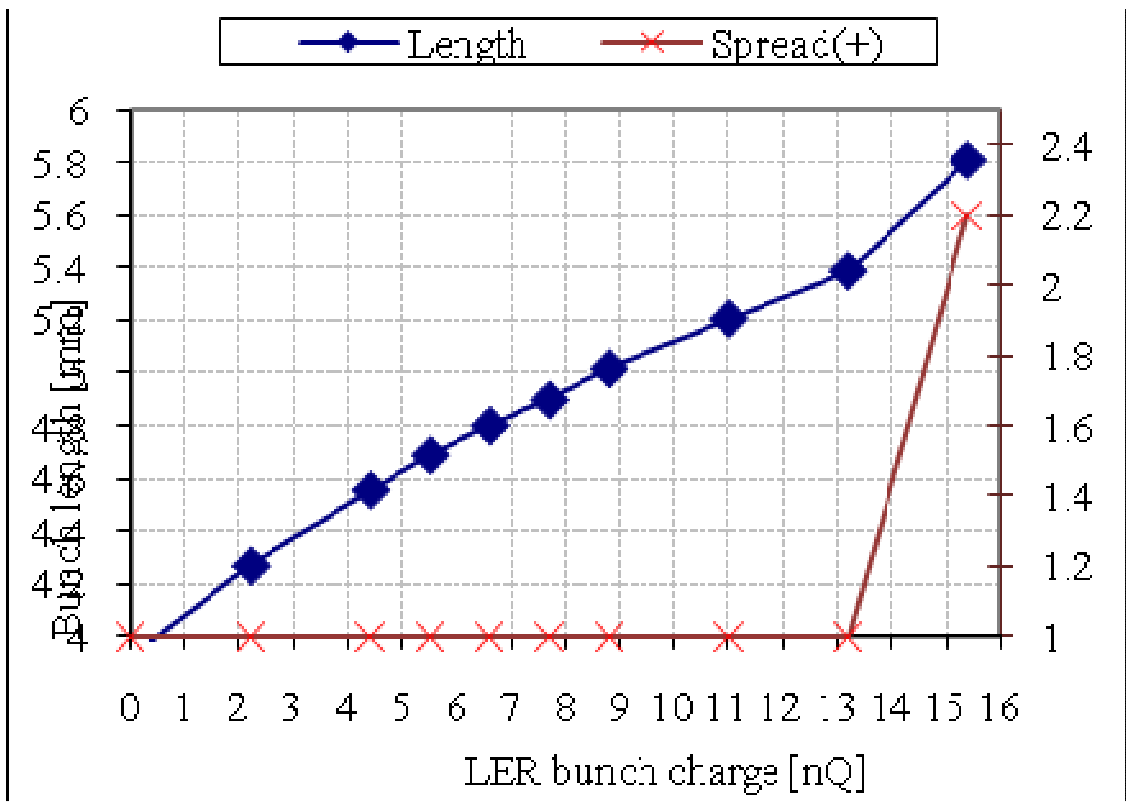

Figure 8.30: Bunch lengthening and energy spread based on the PEP-II model.

# 9. Lifetime overview

## 9.1 Touschek lifetime

The Touschek effect [1] in SuperB is expected to be relevant, particularly for the LER, because of the extremely small beam emittance. Dedicated studies have been performed for handling both Touschek lifetime and backgrounds, using a numerical code developed for DAΦNE firstly tested with the KLOE data [2] and, more recently, on the crab waist collision scheme [3].

The generation of the scattering events in the simulation code is done continuously all over the ring, averaging the Touschek probability density function on every three machine elements. Touschek particles are extracted randomly within one transversely Gaussian bunch with the proper energy spectra and beam sizes, and then tracked over many turns or until they are lost, checking at every turn whether they exceed the RF or the physical acceptance. If the scattered particle is lost during tracking, then its transverse positions and divergences are recorded all the way from the longitudinal position where the scattering takes place to where it gets lost. The Touschek lifetime is evaluated from $\frac{1}{\tau} = \frac{1}{N}\frac{dN}{dt}$ where N is the initial number of particles and dN/dt the losses of the Touschek scattered particles during tracking. We used the Touschek probability density function given by Le Duff [4]

$$\frac{1}{\tau} = \frac{\sqrt{\pi}\, r_0^2 c\, N}{(4\pi)^{3/2}\, \varepsilon^2 \sigma'_x \gamma^3\, \sigma_x \sigma_y \sigma_s} C(u_{min}),$$

where $\varepsilon$ is the momentum acceptance $\sigma_i$ the beam size in the three planes, $\gamma$ is the Lorentz factor,

$$u_{min} = \left(\frac{\varepsilon}{\gamma \sigma'_x}\right)^2$$

and

$$\sigma'_x = \sqrt{\frac{\varepsilon_x}{\beta_x} + \sigma_p^2\left(D'_x + \frac{D_x \alpha_x}{\beta_x}\right)^2}$$

is the beam angular divergence.

For the function $C(u_{min})$ we use the Bruck's approximation [5], valid for $u_{min}<0.01$:

$$C(u_{min}) = \ln\left(\frac{1}{1.78 u_{min}}\right) - \frac{3}{2}.$$

The total machine acceptance $\varepsilon$ is the minimum between the RF acceptance and the lattice acceptance, due to physical or dynamic aperture, $\varepsilon_{machine} = \min(\varepsilon_{RF}, \varepsilon_{lattice})$.

The RF acceptance is evaluated from:

$$\varepsilon_{RF} = \sqrt{\frac{V_0}{\pi \alpha_c h E_0}} F\left(\frac{V_{RF}}{U_0}\right)$$

with $F(q) = 2\left(\sqrt{q^2 - 1} - \arccos(1/q)\right)$.

The machine parameters relevant for the evaluation of the Touschek effect are listed in Table 9.1 and are referred to the V12 lattice. The SuperB RF acceptance is about 4% for the LER and 3% for the HER. However, the minimum momentum acceptance is given by the dynamic aperture that is intrinsically taken into account by including sextupoles and octupoles in the tracking, resulting about 1%. Figure 9.1 shows the Touschek probability loss as a function of the particle energy deviation ΔE/E for the first five machine turns.

Table 9.1.Nominal SuperB beam parameters for the V12 lattice

| | HER | LER |
|---|---|---|
| Beam Energy (GeV) | 6.7 | 4.18 |
| Bunch length (mm) | 5 | 5 |
| Nominal horizontal emittance (nm) | 1.97 | 1.80 |
| Horiz. emittance (nm) including IBS | 2.00 | 2.46 |
| Coupling (%) | 0.25 | 0.25 |
| Particles/bunch | $5.08 \times 10^{10}$ | $6.56 \times 10^{10}$ |

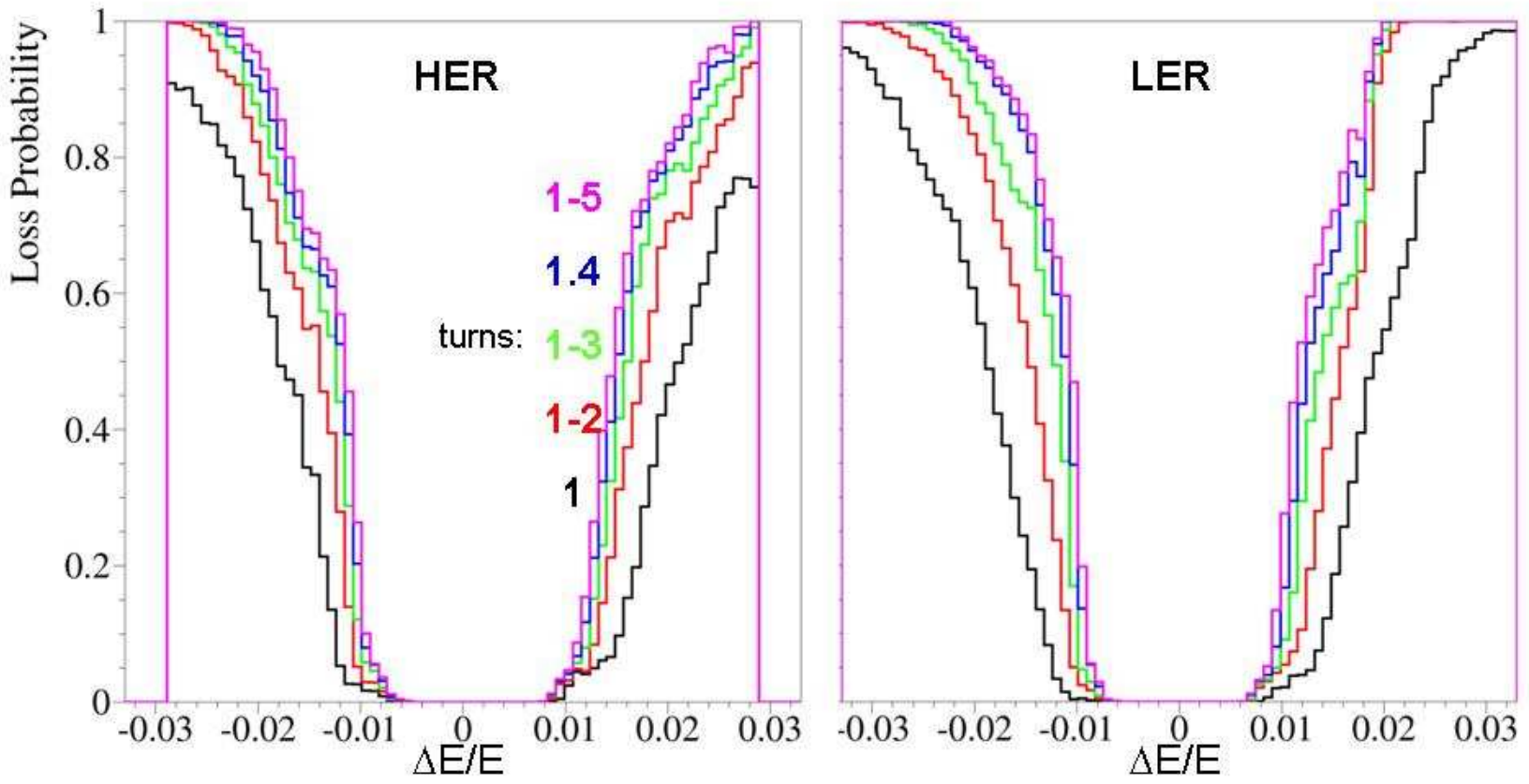

Figure 9.1: Energy acceptance of Touschek scattered particles for the HER and LER for the first five machine turns as resulting from tracking simulations.

The LER Touschek lifetime is 7.8 minutes, with $\varepsilon_x$=2.4 nm due to the IBS effect, diminishing to 5.9 minutes with the nominal horizontal emittance of $\varepsilon_x$=1.8 nm. The HER Touschek beam lifetime is 40 minutes without collimators and it is reduced to 33.2 minutes after the collimators insertion. Table 9.2 summarizes the Touschek beam lifetimes for both rings at their design energy showing the dependence on the emittance enlargement due to intra beam scattering (IBS) for the LER. Moreover, the proposed collimators set that minimizes IR particle losses reduces lifetime by a factor 20% for both HER and LER.

Table 9.2.Summary of SuperB Touschek lifetime

| | $\tau_{Tou}$ HER [min] | $\tau_{Tou}$ LER [min] |
|---|---|---|
| No collimators, $\varepsilon_x$ including IBS | 40.0 | 7.8 |
| No collimators, nominal $\varepsilon_x$ (no IBS) | 39.8 | 5.9 |
| Optimal set of Collimators, $\varepsilon_x$ including IBS | 33.2 | 6.6 |

The larger lifetime value obtained for the HER is due to higher energy, lower beam current and higher horizontal emittance. In collision, however, since the luminosity beam lifetime will be lower for the HER than the LER, due to the smaller number of particles present in an HER bunch, the actual beam lifetimes are expected to be similar; a few minutes for each ring.

The IR particle losses due to Touschek scattering can be analysed in detail, determining upstream and downstream rates, transverse phase space and energy deviation of these off-energy particle losses as a function of different beam parameters, of different optics and for different sets of movable collimators. In fact, the simulation code gives the longitudinal positions where Touschek particles are generated, showing also the longitudinal positions corresponding to large radial oscillations of the scattered particles. Assuming that each collimator has an external and an internal jaw that can be separately inserted in the vacuum pipe, the optimal radial jaw opening ca be found with numerical studies. A circular vacuum chamber of 4 cm has been assumed all over the rings but for the IR.

Collimation studies have been performed with the goal of reducing as much as possible IR particle losses while keeping the subsequent lifetime reduction within 20%. The most effective location for collimators would be at longitudinal positions corresponding to large radial oscillation of scattered particles that is at high $\beta_x$ and $D_x$ locations in the final focus upstream the IR.

The proposed solution is to have three primary collimators in the final focus that would intercept most of the particles that would otherwise be lost in the IR. Their longitudinal positions are at s=-49 m for COL2, s=-67.7 m for COL3 and s=-85.5 m for COL4 far from the IP, as shown in Figure 9.2. A secondary collimator at s=-21 m (COL1) would stop the remaining Touschek scattered particles generated so close to the IR that secondary collimators cannot be effective (see lower left plot of Figure 9.4 for the HER and upper right plot of Figure 9.5 for the LER).

Longitudinal position of collimators is the same for the HER and the LER, even if positions of the two radial jaws have been optimized separately for each ring, and also for each collimator.

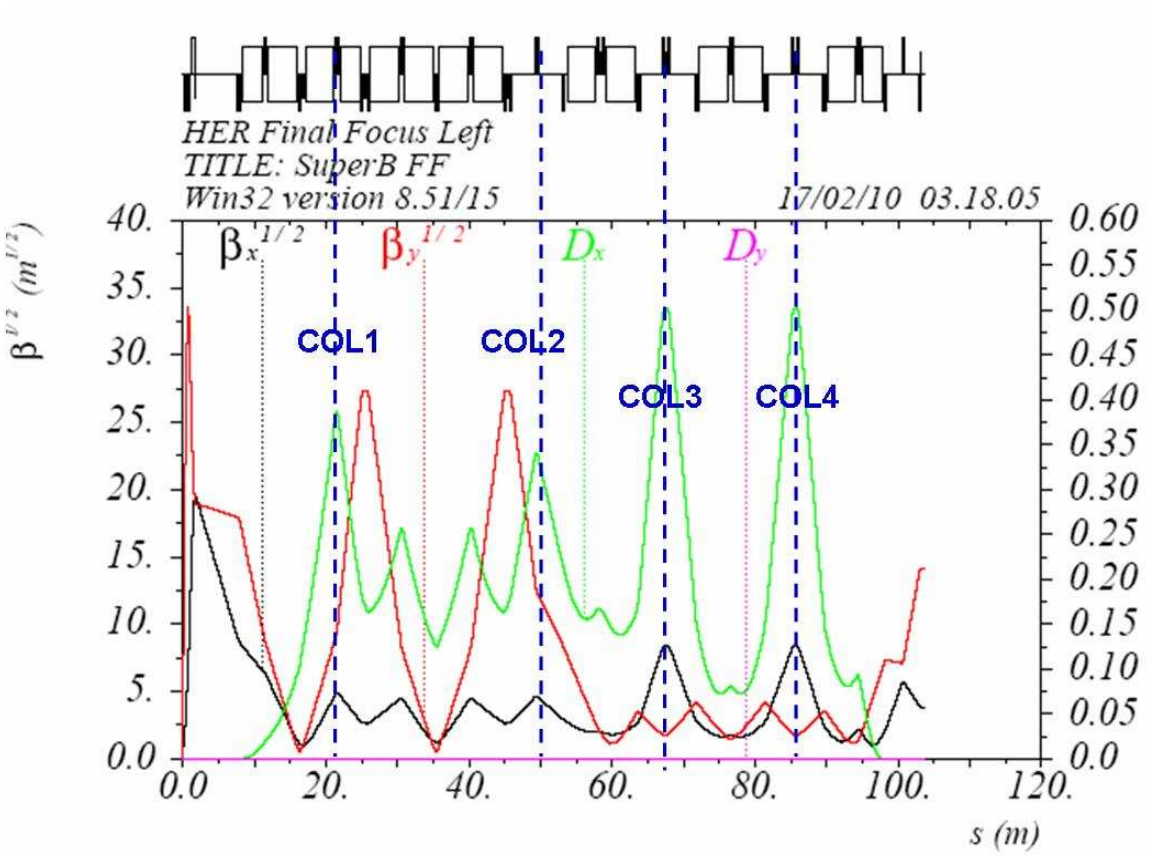

Figure 9.2: Longitudinal position of collimators in the final focus of the V12 lattice, all at high $\beta_x$ and $D_x$ locations. COL1 is the secondary collimator, being the closest to the IP (S=0 m) while COL2, COL3 and COL4 are the primary ones.

A rough estimate of the horizontal opening position of the two collimators jaws can be obtained from the condition that they should intercept particles that would be lost at the QF1 physical aperture (PhA), obtaining the following condition:

$$Aperture(COL) = 0.9 \cdot PhA(QF1) \cdot \sqrt{\frac{\beta_x(COL)}{\beta_x(QF1)}}$$

where the 0.9 factor has been cautiously added.

For example, for the two HER collimators COL3 and COL4 we get a radial opening of 1.5 cm for the two jaws just by substituting the QF1 physical aperture of 4 cm, $\beta_x(QF1) \approx 390$ m and $\beta_x \approx 70$ m. However, starting from the jaws opening positions found with this criterion, the two radial jaws of each collimator have been optimized by simulations finding the best trade off between lifetime and IR losses reduction. The final optimized values found by simulation for COL3 and COL4 are close to ±1.5 cm, being -1.8 cm and +1.4 cm for COL3 and -1.4 cm and +1.8 cm for COL4. The complete collimators sets for HER and LER are reported in Table 9.3.

Figure 9.3 summarizes some results on HER Touschek background simulations. Left plots show the trajectories of scattered particles that are eventually lost at the IR for the HER without (upper) and with (lower) collimators; right upper plot show the radial position of the IR losses with collimation while right lower their energy deviation for the corresponding loss position.

Table 9.3: Final set of collimators external and internal radial jaws

| Collimator name | HER (cm) | LER (cm) |
|---|---|---|
| COL1 | -1.0/+1.2 | -1.1/+1.4 |
| COL2 | -1.0/+4.0 | -1.4/+4.0 |
| COL3 | -1.8/+1.4 | -4.0/+1.6 |
| COL4 | -1.4/+1.8 | -1.4/4.0 |

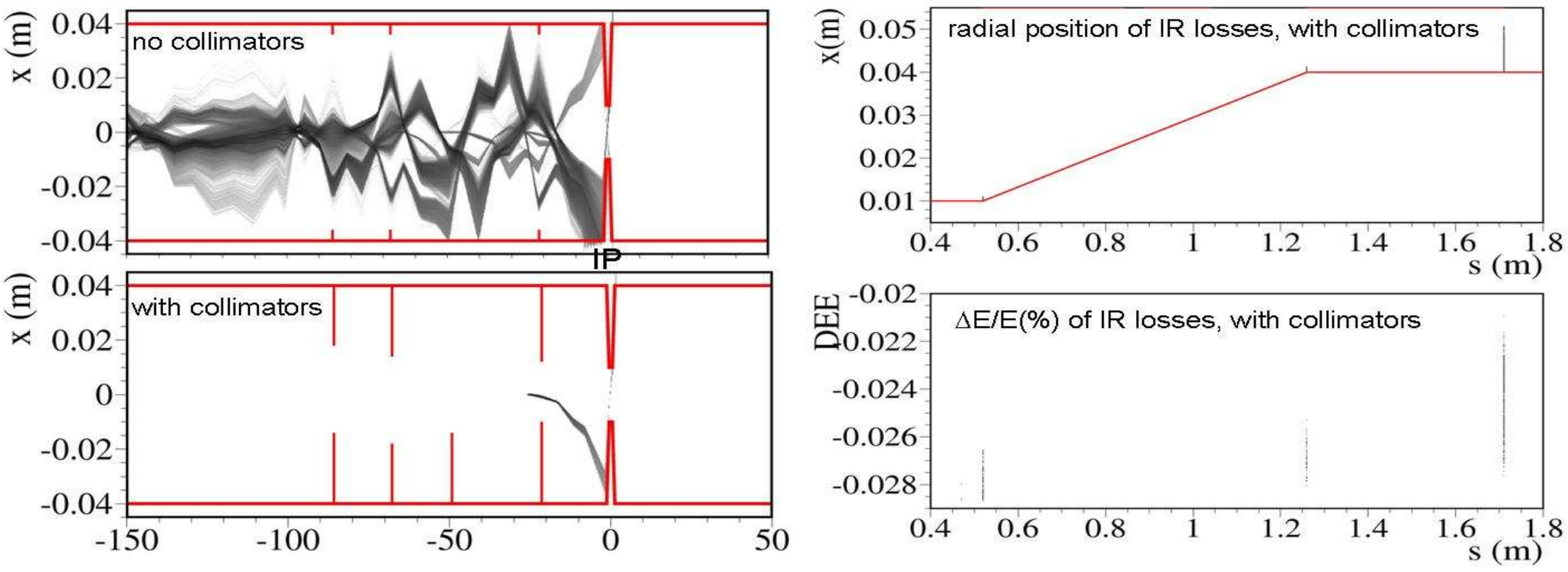

Figure 9.3: Trajectories of HER Touschek particles eventually lost at IR (|s|<2 m) in the first five turns without (upper) and with (lower) collimators. Right plots show the radial position and the energy deviation of the particles not intercepted by the collimators and lost at the IR.

Upper plots of Figure 9.4 show the HER Touschek trajectories zoomed at the IR without (left) and with (right) collimators insertion. Corresponding distributions of IR particle losses are reported in the lower plots, showing that collimators are very effective, reducing rates by a large factor. Moreover, it can be noticed that when collimators are in only losses downstream the IP are foreseen, with a calculated rate of 0.45 KHz for one bunch at nominal current $I_b$ = 1.9 mA.

Figure 9.5 summarizes LER Touschek simulation studies. Left plots show the trajectories and distribution of particle losses in the final collimators set configuration zoomed at the IR. Expected loss rates are higher than for HER, as expected, resulting 14.5 KHz for a single bunch a nominal current $I_b$ = 2.5 mA. Some particle losses downstream the IP, corresponding to the QF1 position, are foreseen even with collimation system on. Generation points and trajectories of these particles are shown in upper right plot of Figure 9.5. Lower right plots indicates their energy deviation.

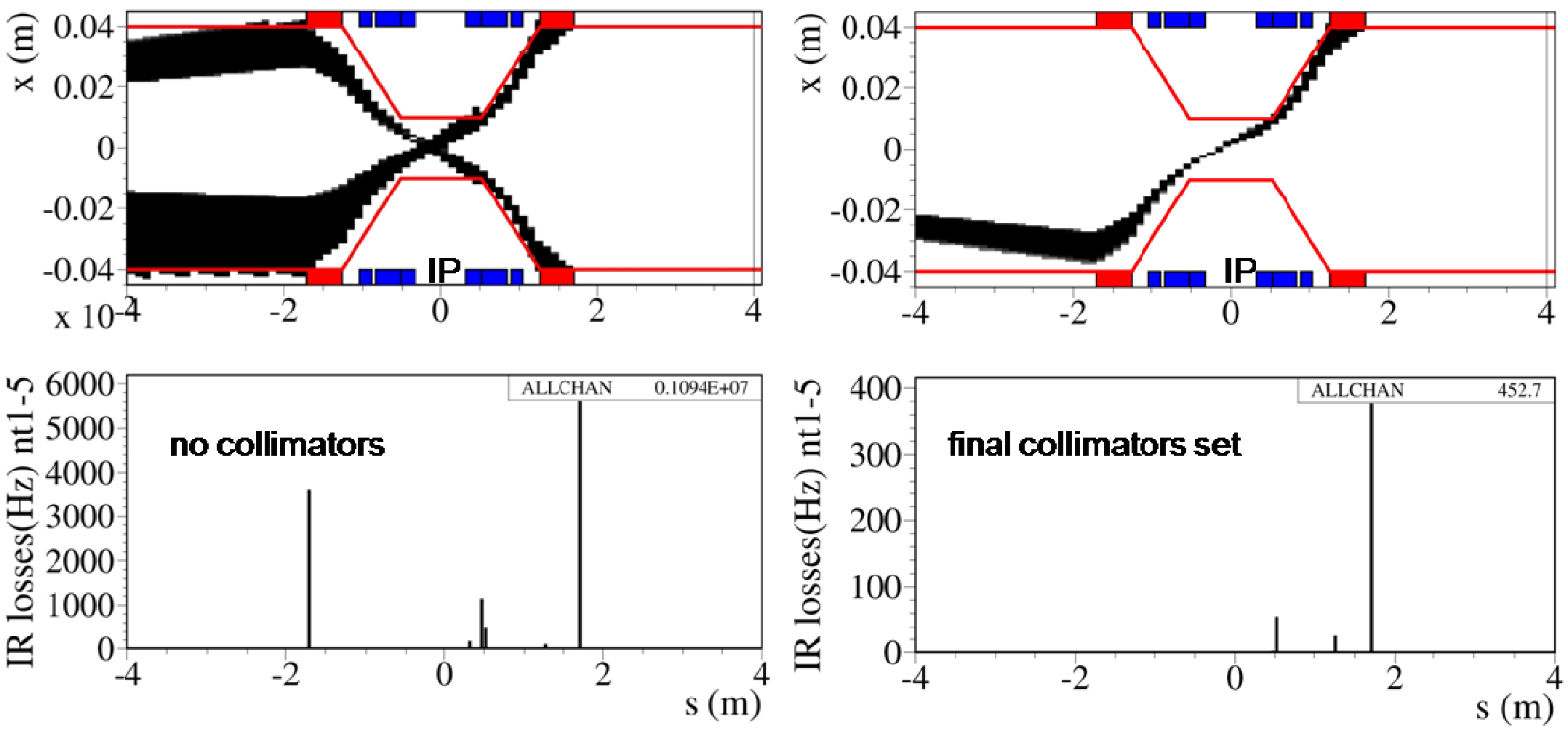

Figure 9.4: Trajectories and distribution of HER Touschek particles lost at IR (|s|<2 m) in the first five turns without (left) and with (right) collimators inserted. Calculated rates of IR particles losses are also indicated correspondingly in the upper right plot, referred to a single bunch at nominal current. With the final collimator set only downstream losses are foreseen, with strongly reduced rates.

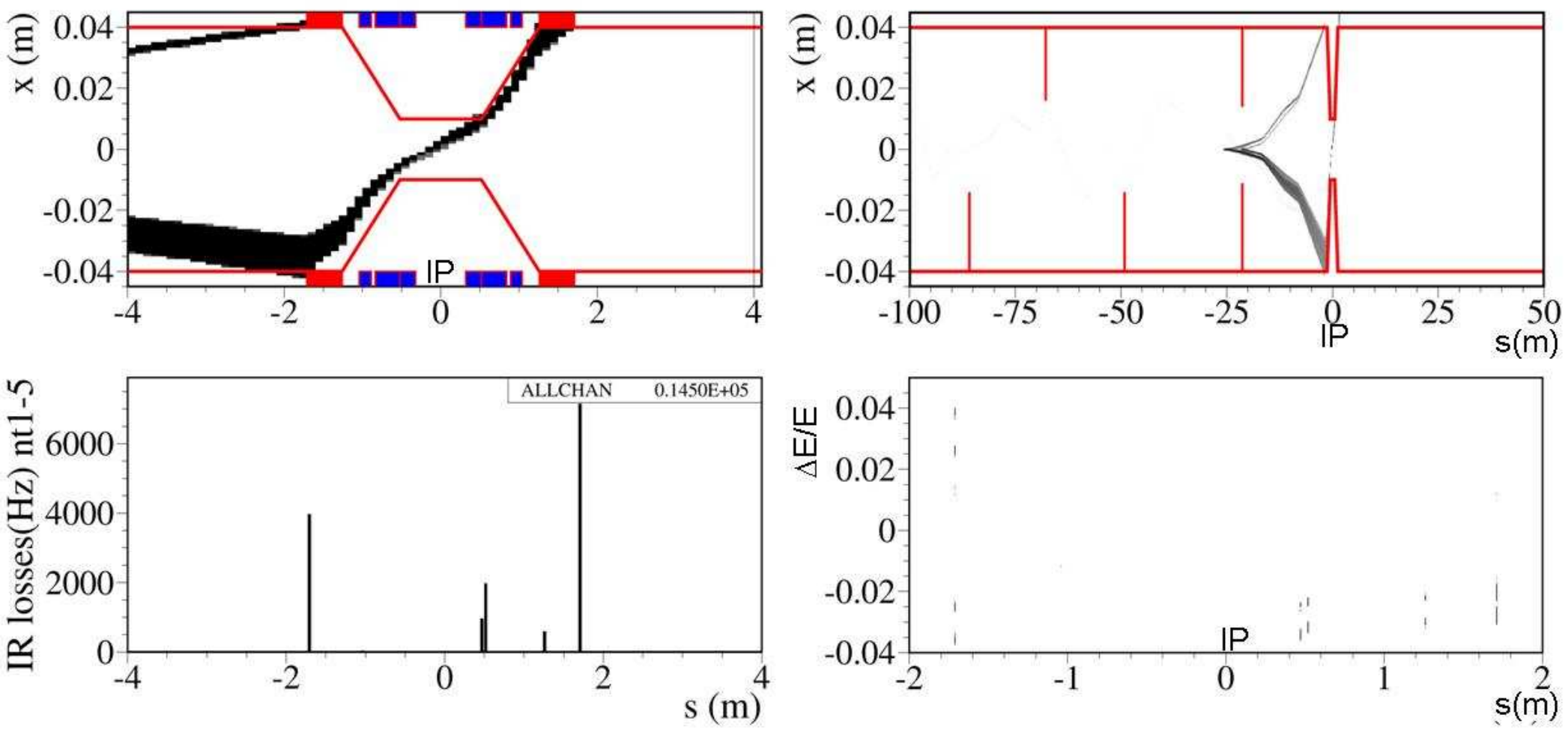

Figure 9.5: Summary of LER Touschek simulations with the optimized collimators set. Left plots: IR particles trajectories and distribution of IR losses. Right plots: trajectories of particles not intercepted by the collimators and lost at the IR with their energy deviation (upper and lower, respectively).

In conclusions, Touschek effect is relevant at SuperB, especially for the LER, but it is not the limiting effect for lifetime. However, special care is needed to properly control Touschek particle losses and reduce possible showers in the detectors. A proper set of collimators that fulfils this requirement has been found.

## 9.2 Radiative and elastic Bhabha lifetime

Electro-magnetic scattering occurring at the interaction point are among the principal causes of beam particle losses. The huge cross section of these processes together with the unprecedented luminosity of Super*B* lead to an extremely short beam lifetime.

The luminosity loss rate for the ring "*i*" depends on the luminosity *L* and on the "particle loss" cross section $\sigma_i$ according to:

$$\frac{dN_i}{dt} = -\sigma_i L$$

where $N_i$ are the total number of particles in the ring *i*.

Assuming *L* constant, the following approximation holds:

$$N_i(t_0 + \Delta t) \approx N_i(t_0)\, e^{-\frac{\sigma_i L}{N_i(t_0)}\Delta t}$$

The beam lifetime $\tau_i$ of the ring "*i*" is defined as:

$$\tau_i \equiv \frac{N_i(t_0)}{\sigma_i L}$$

The principal processes contributing to $\sigma_i$ are the radiative and the elastic Bhabha scatterings, i.e. $e^+e^- \to e^+e^-\gamma$ and $e^+e^- \to e^+e^-$.

The photon emitted by the beam particle can carry away enough energy to bring the radiating lepton outside the energy acceptance region of the storage ring. The cross section of this process is given with good approximation by [1]:

$$\sigma \cong \frac{16\alpha r_e^2}{3}\left[\left(\frac{1}{2} - \ln\frac{s}{m_e^2}\right)\left(\frac{5}{8} + \log\Delta\varepsilon\right) + \frac{1}{2}(\ln\Delta\varepsilon)^2 - \frac{3}{8} - \frac{\pi^2}{6}\right]$$

where $\Delta\varepsilon$ is the fractional energy aperture of the ring, $\alpha$ is the fine structure constant, *s* is the total energy squared in the center of mass reference frame, $r_e$ is the classical radius of the electron and $m_e$ is its mass.

For a 1% energy acceptance the previous formula gives us a cross section at the Υ(4S) peak of 265 mb.

Actual measurements of this cross section [2] found a smaller value with respect to prediction. This reduction can be ascribed to the effect of finite bunch density. To correctly model this effect the BBBREM Monte Carlo generator [3] was used. The predicted cross section as a function of the energy acceptance is shown in Fig. 9.6 together with the best fitting function:

$$\sigma = -43.9mb * \log(2\Delta\varepsilon)$$

The cross section predicted by BBBREM for a ring energy acceptance of 1% is 170 mb, corresponding to radiative Bhabha beam lifetimes reported in Table 9.4 for the various Super*B* configurations

Table 9.4: Radiative Bhabha beam lifetimes for several SuperB options.

| | **Base Line** | | **Low Emittance** | | **High Current** | |
|---|---|---|---|---|---|---|
| | **HER** | **LER** | **HER** | **LER** | **HER** | **LER** |
| τ (min) | 4.87 | 6.29 | 3.76 | 4.85 | 7.96 | 10.3 |

A different treatment of the finite size and finite density of the colliding bunches reported in [4] predicts a radiative Bhabha cross section of 166mb.

Despite the fairly good agreement of this result with the BBBrem one it is worth while to note that this is a mere coincidence holding as long as the infrared cut-off parameter used in [4] (i.e., the vertical beam size at the waist position) is comparable with the BBBrem cut-of parameter [3] (i.e. the typical distance among neighborhood particles at waist).

The other loss mechanism connected with scattering occurring at the IP is the elastic Bhabha. An electron and a positron can knock each other hard enough to be deflected outside the transverse ring acceptance.

Assuming a mechanical aperture of 30 $\sigma_x$ on the radial plane and a dynamical aperture of 50 $\sigma_y$ on the vertical plane the contribution to the cross section $\sigma_i$ can be evaluated at tree level by:

$$\sigma_{el} \approx \frac{8\pi(\hbar c\alpha)^2}{s}\frac{E_j}{E_i}\left(\frac{1}{\vec{\vartheta}_x^2} + \frac{1}{\vec{\vartheta}_y^2}\right)$$

where $E_i$ is the energy of the beam for which the lifetime is under evaluation, $E_j$ is the energy of the opposite beam and $\vec{\vartheta}_x$ ($\vec{\vartheta}_y$) are the radial (vertical) angular deflection beyond which scattered particles would be lost. The expected cross section loss is 10 *mb* for the HER and 21 mb for the LER. The Bhabha (radiative plus elastic) scattering beam lifetime is reported in Table 9.5.

Table 9.5: Bhabha (radiative and elastic) beam lifetimes for several SuperB options.

| | **Base Line** | | **Low Emittance** | | **High Current** | |
|---|---|---|---|---|---|---|
| | **HER** | **LER** | **HER** | **LER** | **HER** | **LER** |
| τ (min) | 4.6 | 5.6 | 3.6 | 4.3 | 7.5 | 9.2 |

.

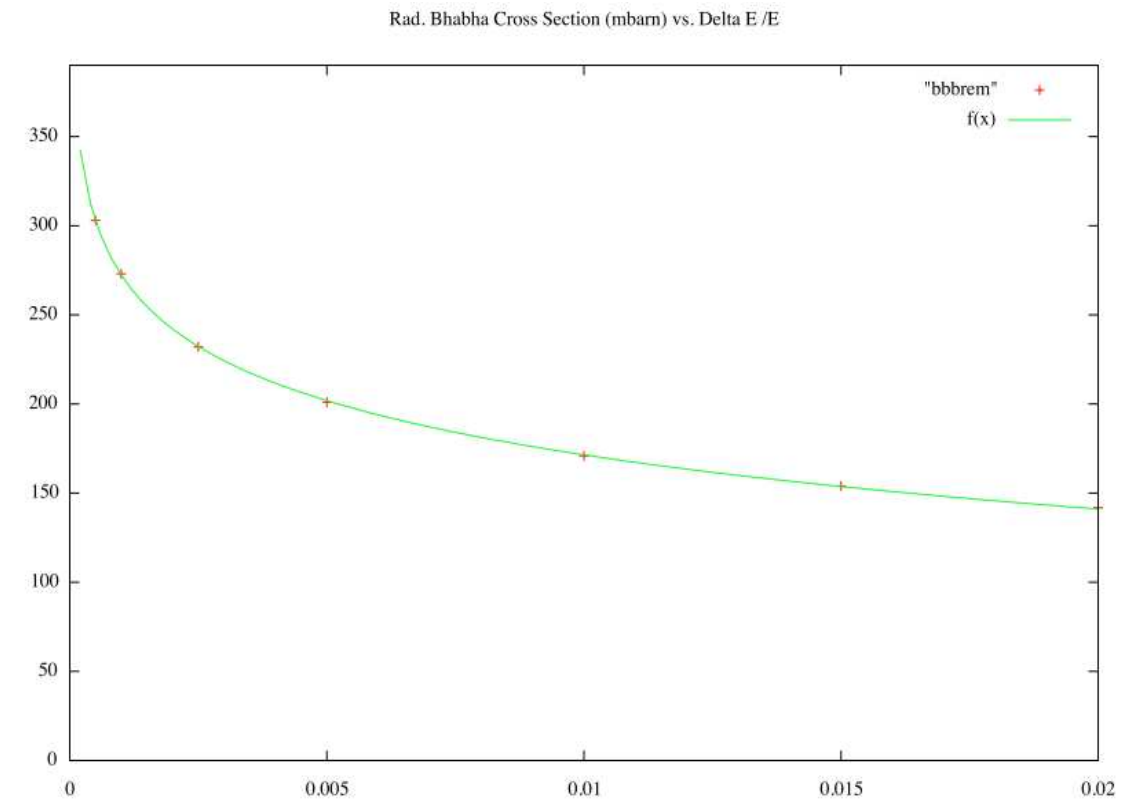

Figure 9.6: Radiative Bhabha cross section in mbarn as a function of the minimum fractional radiated energy. BBBrem predictions are represented by the crosses. The best fit is represented by a continuous line.

# 10. High Order Modes heating

## 10.1 HOM in RF cavities

The main contribution to the narrow-band impedance comes from the RF cavities. This means that HOMs trapped in the cavity must be very well damped like it was done in the PEP-II cavities. Measured and calculated frequencies and Q-values of longitudinal higher order modes in the PEP-II cavity are shown in the Table 10.1 (from [1]). Table 10.2 presents the impedance of the transverse higher order modes. PEP-II spectrum, calculated from the wake potential of a 4 mm bunch is shown in Fig. 10.1 (from [2]).

TABLE 10.1: Impedance and *Q*'s of monopole modes estimated from calculations and measurements. Shunt impedance definition $R = V^2/2P$.

| $f_{meas}$ [MHz] | $R/Q_{meas}$ [Ohm] | $Q_{meas}$ | $R_{meas}$ [Ohm] | $f_{calc}$ [MHz] | $R_{calc}$ [Ohm] | $Q_{calc}$ |
|---|---|---|---|---|---|---|
| 476 | $117.3^{+0.00}_{-18.5}$ | 32469 | $3.809 \times 10^6$ | 476 | | |
| 758 | 44.6 ±13.4 | $18^{+0.0}_{-4.0}$ | $809^{+241}_{-362}$ | 758 | 879 | 15 |
| 1009 | $0.43^{+0.00}_{-0.05}$ | $128^{+0.0}_{-3.0}$ | $55^{+0.0}_{-7.0}$ | 1010 | 35 | 100 |
| 1283 | $6.70^{+6.4}_{-0.00}$ | $259^{+47}_{-92}$ | $1736^{+2272}_{-617}$ | 1291 | 1013 | 88 |
| 1295 | 10.3 ±2.1 | $222^{+0.0}_{-88}$ | $2287^{+455}_{-1184}$ | 1307 | 1831 | 203 |
| 1595 | $2.43^{+0.00}_{-2.14}$ | $300^{+0.0}_{-170}$ | $729^{+0.0}_{-691}$ | 1596 | 214 | 52 |
| 1710 | 0.44 ±0.11 | $320^{+125}_{-0.0}$ | $141^{+104}_{-35}$ | 1721 | 476 | 54 |
| 1820 | 0.13 ±0.013 | $543^{+0.0}_{-120}$ | $70^{+7.0}_{-21}$ | | | |
| 1898 | 0.17±0.043 | $2588^{+0.0}_{-1693}$ | $442^{+111}_{-328}$ | 1906 | 715 | 685 |
| 2121 | 1.82 ±0.18 | $338^{+69}_{-100}$ | $616^{+199}_{-226}$ | 2113 | 1346 | 163 |
| 2160 | 0.053 ±0.011 | $119^{+10}_{-35}$ | $6^{+2.0}_{-3.0}$ | 2153 | 293 | 300 |
| 2265 | 0.064 ±0.016 | $1975^{+0.0}_{-1314}$ | $126^{+32}_{-95}$ | 2263 | 450 | 306 |
| 2344 | | $693^{+0.0}_{-511}$ | | | | |

TABLE 10.2: Transverse impedance and *Q*'s of dipole modes estimated from calculations and measurements

| $f_{r\,meas}$ (MHz) | $R/Q_{meas}$ [Ω] | $Q_{meas}$ | $R_{\perp meas}$[kΩ/m] | $f_{rcalc}$ [MHz] | $R_{\perp calc}$ [Ω/m] | $Q_{calc}$ |
|---|---|---|---|---|---|---|
| 792 | 9.69±0.997 | 115 | 42.0±4.2 | 800 | 38.7 | 96 |
| 1063 | 50.4 ±10.1 | 27 | 38.0±7.6 | 1071 | 40.1 | 34 |
| 1133 | 1.29±0.65 | 54 | 1.82±0.91 | | | |
| 1202 | 0.56±0.56 | 871 | 12.2 | 1218 | 17.6 | 642 |
| 1327 | 0.56±0.56 | 611 | 76.7 | 1335 | 99.5 | 510 |
| 1420 | 5.58 ±0.28 | 1138 | 126.9 | 1417 | 143.7 | 554 |
| 1542 | 0.50±0.50 | 92 | 0.89 | 1553 | 2.0 | 130 |
| 1595 | 0.51±0.21 | 145 | 1.39 | 1611 | 11.6 | 180 |
| 1676 | 4.63 ±0.46 | 783 | 64.5 | 1672 | 33.9 | 265 |
| 1749 | 0.10±0.01 | 1317 | 2.31 | 1774 | 9.15 | 1234 |

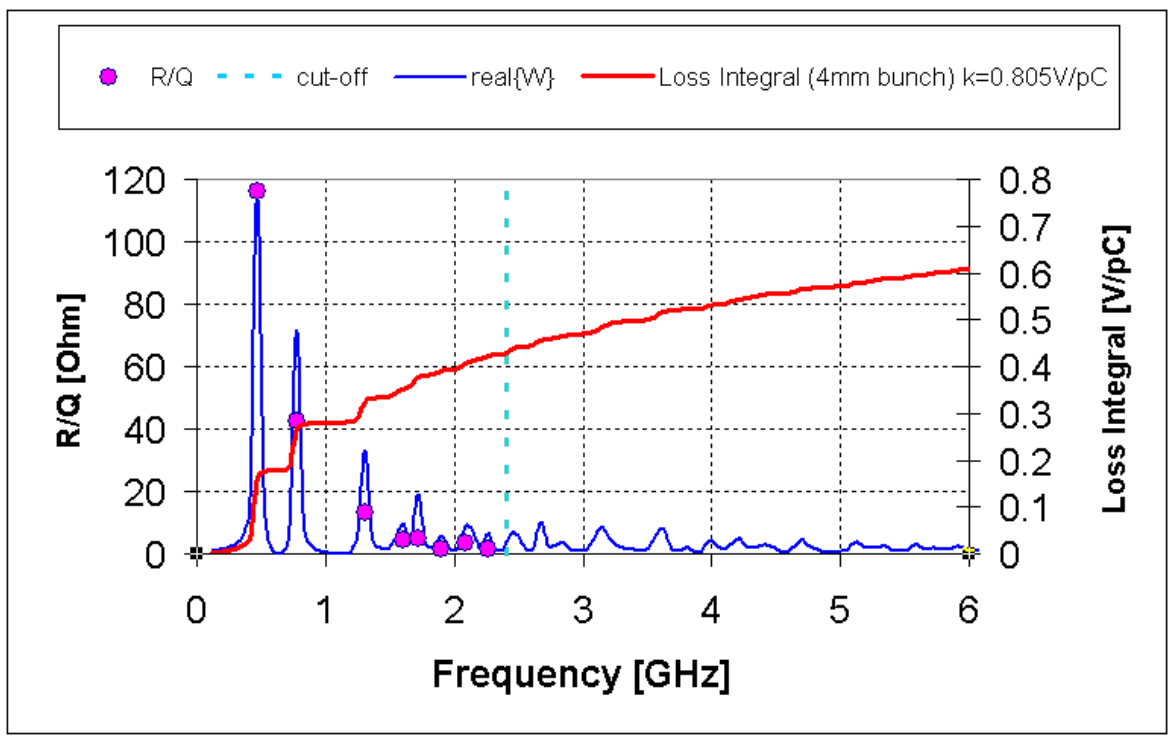

Fig. 10.1: PEP-II cavity spectrum, R/Q and loss integral.

### HOM power for bellow cut-off frequencies

Power loss into n[th] mode in a cavity according to [3, 4] is:

$$P_n = I^2 \times \tau \times k_n$$

$$P_n = I^2 \times \tau \times \frac{\omega_n}{2}\left(\frac{R}{Q}\right)_n \frac{1-\exp\left(-2\frac{\tau}{\tau_{l,n}}\right)}{1-2\exp\left(-\frac{\tau}{\tau_{l,n}}\right)\cos(\omega_n\tau)+\exp\left(-2\frac{\tau}{\tau_{l,n}}\right)}$$

$$\tau_{l,n} = 2\frac{Q_{l,n}}{\omega_n}$$

Definitions: $k_n$ is los factor, $I$ is beam current, $\tau$ is bunch spacing, $Q_{l,n}$ and is a loaded Q.

Table 10.3 shows, for a beam current of 1 A, the HOM power into each bellow cut-off modes, and the total loss of a PEP-II cavity.

Table 10.3: HOM power in a PEP-II cavity for modes bellow cut-off for the current of 1 A.

| Mode frequency GHz | R/Q Ohm | Qload | Loss factor V/pC | Filling time mks | cos() | exp() | Bunch spacing nsec | Power loss for I= 1 A kW |
|---|---|---|---|---|---|---|---|---|
| | | | | | | | | |
| 0.475997 | 117.3 | 8000 | 0.1754 | 2.675 | 1.000 | 0.9969 | 4.202 | 0.0000 |
| 0.758 | 44.6 | 18 | 0.1062 | 0.004 | 0.398 | 0.1082 | 4.202 | 0.4701 |
| 1.009 | 0.43 | 128 | 0.0014 | 0.020 | 0.066 | 0.6595 | 4.202 | 0.0013 |
| 1.283 | 6.7 | 259 | 0.0270 | 0.032 | -0.774 | 0.7699 | 4.202 | 0.0083 |
| 1.295 | 10.3 | 222 | 0.0419 | 0.027 | -0.933 | 0.7349 | 4.202 | 0.0140 |
| 1.595 | 2.43 | 300 | 0.0122 | 0.030 | -0.299 | 0.7552 | 4.202 | 0.0055 |
| 1.71 | 0.44 | 320 | 0.0024 | 0.030 | 0.398 | 0.7542 | 4.202 | 0.0023 |
| 1.82 | 0.13 | 543 | 0.0007 | 0.047 | -0.602 | 0.8378 | 4.202 | 0.0002 |
| 1.898 | 0.17 | 2588 | 0.0010 | 0.217 | 0.988 | 0.962 | 4.202 | 0.0065 |
| 2.121 | 1.82 | 338 | 0.0121 | 0.025 | 0.850 | 0.718 | 4.202 | 0.0519 |
| 2.16 | 0.053 | 119 | 0.0004 | 0.009 | 0.889 | 0.3835 | 4.202 | 0.0033 |
| 2.265 | 0.064 | 1975 | 0.0005 | 0.139 | -0.994 | 0.9412 | 4.202 | 0.0000 |
| | 184.4370 | | 0.3811 | Total | HOM | power | | 0.5635 |

### HOM power for above cut-off frequencies

Calculated loss factor [5, 2] for different bunch length is shown in Fig. 10.2.

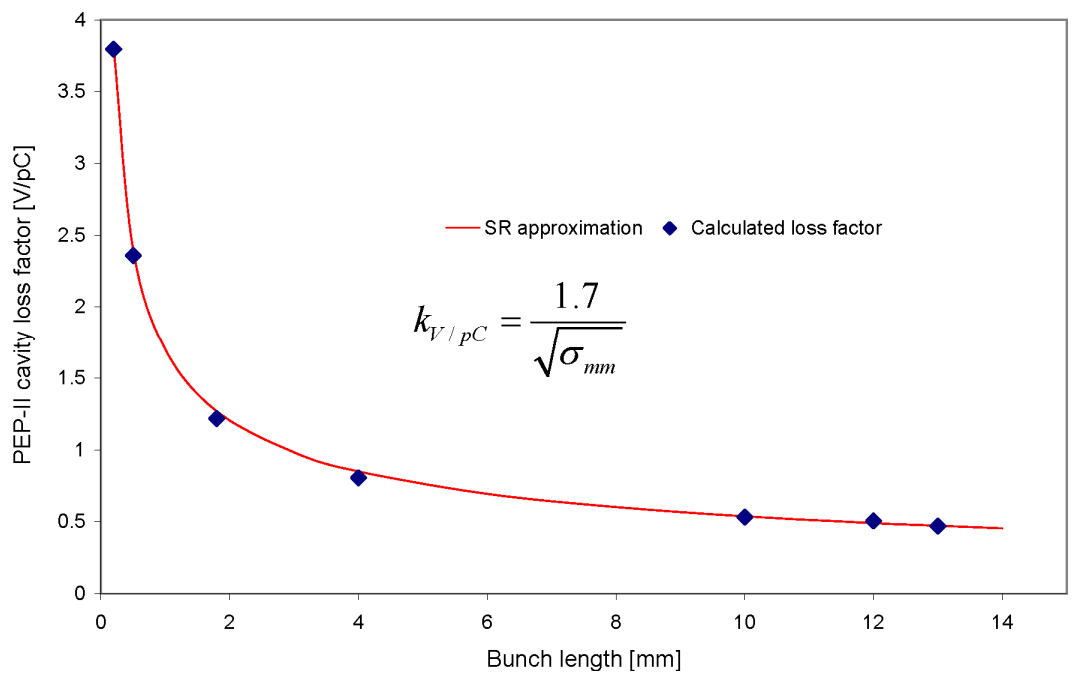

Fig. 10.2: Wake field Loss factor.

HOM power above cut-off frequency for a PEP-II cavity [2]

$$P_{kW} = \left(\frac{1.7}{\sqrt{\sigma_{mm}}} - .3811\right) \times \tau_{spacing} \times I^2$$

HOM losses for bellow and above cut-off frequencies are shown in Table 10.4. Total HOM all cavity losses for PEP-II and Super-B parameters are also shown there. At the “High current” regime the bunch spacing is two times smaller i.e. 2.1 ns.

Table 10.4: Total all cavities HOM power.

| | | Beam current A | Bunch length mm | Number of cavities | Power bellow cut-off kW | Power above cut-off kW | Total HOM cavity power kW |
|---|---|---|---|---|---|---|---|
| LER | Base line | 2.447 | 5 | 8 | 26.99 | 76.28 | 103.28 |
| | High current | 4 | 4.4 | 12 | 54.10 | 173.11 | 227.21 |
| HER | Base line | 1.892 | 5 | 12 | 24.21 | 68.41 | 92.61 |
| | High current | 3.094 | 4.4 | 20 | 53.94 | 172.62 | 226.56 |

## 10.2 HOM in vacuum system

### Vacuum chamber geometry and resistive wall wake fields

We will assume that Super-B vacuum chamber is approximately same as PEP-II vacuum chamber for LER and HER, as we use PEP-II magnets; however the circumference is smaller and it is 1258.4 m

Table 10.5: PEP-II LER and HER vacuum chamber

| | Material | % | pipe Radius [m] | resistivity [Ohm m] |
|---|---|---|---|---|
| LER | Cu | 10 | 0.025 | 1.69E-08 |
| | Al | 50 | 0.035 | 2.86E-08 |
| | SS | 40 | 0.045 | 7.14E-07 |
| HER | Cu | 60 | 0.025 | 1.69E-08 |
| | SS | 40 | 0.045 | 7.14E-07 |

Resistive wall loss factor [6]:

$$s_0 = \left(2a^2 \frac{\rho}{Z_0}\right)^{1/3} \text{ when } \frac{s_0}{\sigma_z} << 1$$

$$k_{RW} \approx 0.2 * \frac{Z_0 c}{4\pi a^2} * \left(\frac{1}{\sigma_z}\right)^{3/2} * \sqrt{2\frac{\rho}{Z_0}} * F(a,b)$$

For these beam pipe geometries loss factor is almost the same (Fig. 10.3), because Al part of LER has larger size. Calculated resistive wall losses for LER and HER rings are shown in Table 10.5 for PEP-II and Table 10.6 and 10.7 for Super-B.

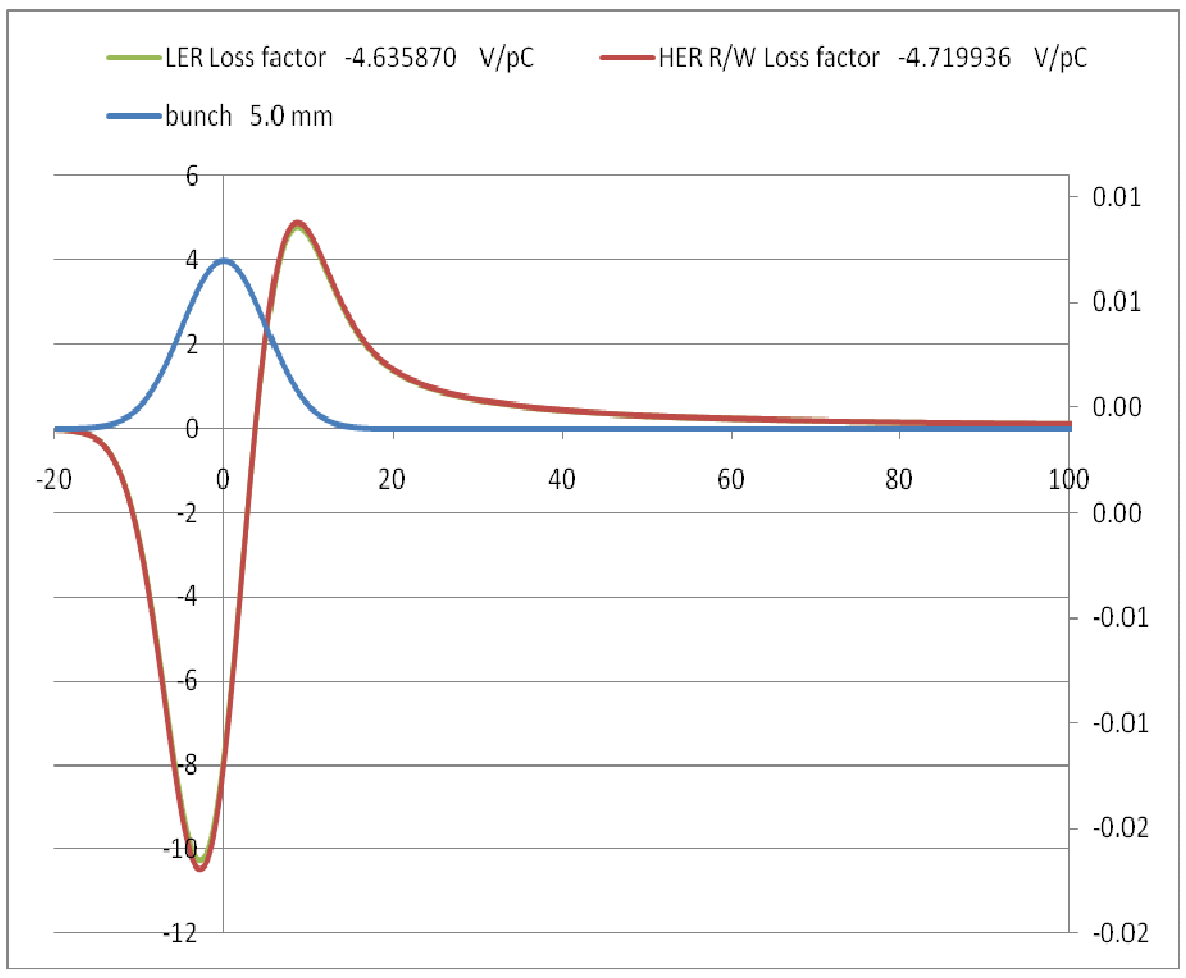

Figure 10.3: Resistive wake potential for LER and HER

Table 10.6: LER resistive wall losses.

| | Base line | High current |
|---|---|---|
| Bunch length [m] | 0.005 | 0.0044 |
| Bunch spacing [nsec] | 4.2 | 2.1 |
| Beam current [A] | 2.447 | 4 |
| Power (10/50/40) [kW] | 122.49 | 198.24 |

Table 10.7: HER resistive wall losses.

| | Base line | High current |
|---|---|---|
| Bunch length [m] | 0.005 | 0.0044 |
| Bunch spacing [nsec] | 4.2 | 2.1 |
| Beam current [A] | 1.892 | 3.094 |
| Power (60/0/40) [kW] | 74.55 | 120.76 |

### Other beam chamber elements

#### *Longitudinal kickers*

Longitudinal kicker spectrum and loss factor as a function of bunch length from the azimuthally symmetric model are shown at Fig. 10.4. Fig. 10.5 shows measured single bunch spectrum of the kicker at PEP-II and wake potential for a 5 mm bunch. Wake field power in longitudinal kickers of LER and HER for Super-B parameters is given in Table 10.8.

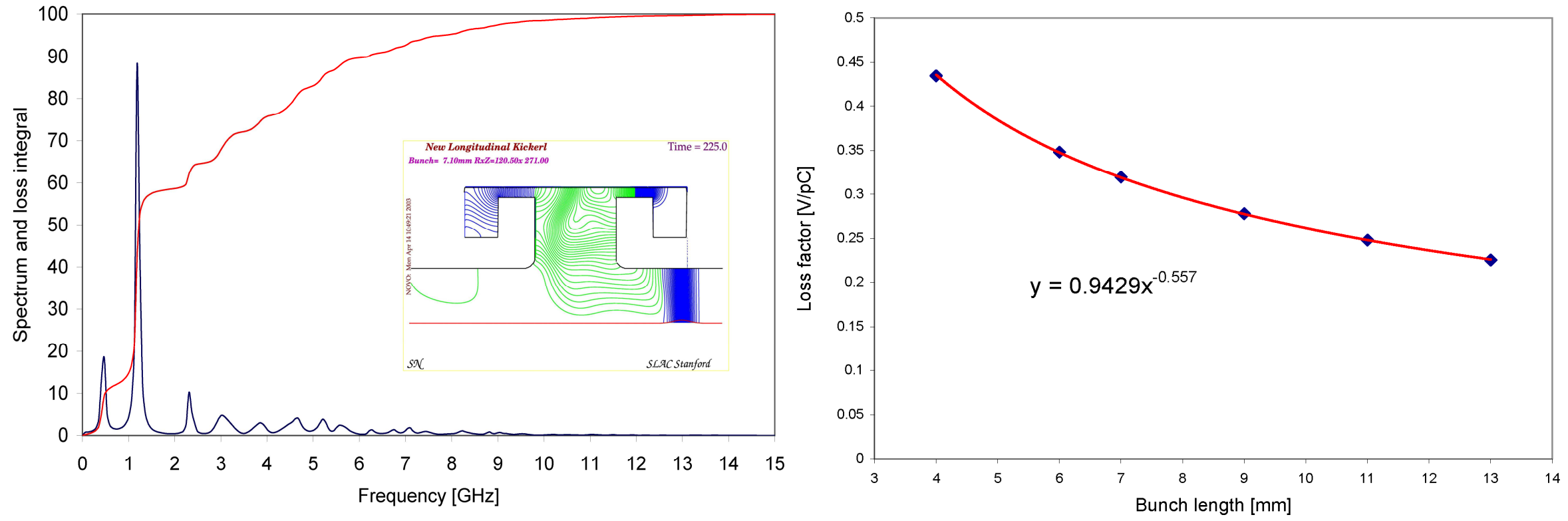

Figure 10.4: Longitudinal kicker spectrum (left) and loss factor as a function of bunch length.

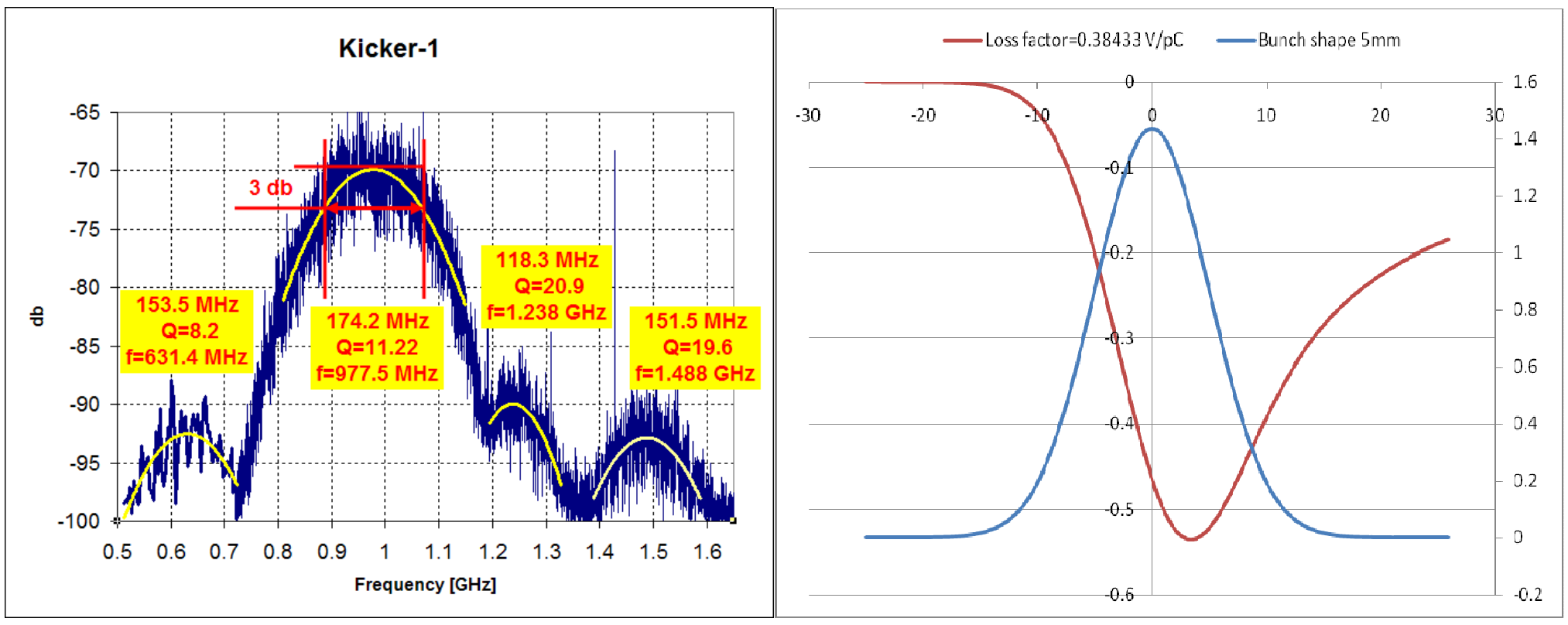

Figure 10.5: Measured single bunch spectrum and wake potential of a longitudinal kicker.

Table 10.8: Wake field power in longitudinal kickers

| | | Beam current (A) | Bunch length (mm) | Number of long. Kickers | Wake field power (kW) |
|---|---|---|---|---|---|
| LER | Basic Line | 2.447 | 5 | 2 | 10.40 |
| | High currents | 4 | 4.4 | 2 | 31.34 |
| HER | Basic Line | 1.892 | 5 | 2 | 6.22 |
| | High currents | 3.094 | 4.4 | 2 | 18.75 |

### *Transverse kickers*

Transverse kicker loss factor as a function of bunch length from the azimuthally symmetric model are shown at Fig. 10.6. Wake field power in the transverse LER and HER kickers for Super-B parameters is given in Table 10.9.

Table 10.9: Wake field power in the transverse kickers

| | | Beam current A | Bunch length mm | number of long. Kickers | Wake field power kW |
|---|---|---|---|---|---|
| LER | Basic Line | 2.447 | 5 | 2 | 12.57 |
| | High Currents | 4 | 4.4 | 2 | 18.27 |
| HER | Basic Line | 1.892 | 5 | 2 | 6.22 |
| | High Currents | 3.094 | 4.4 | 2 | 9.38 |

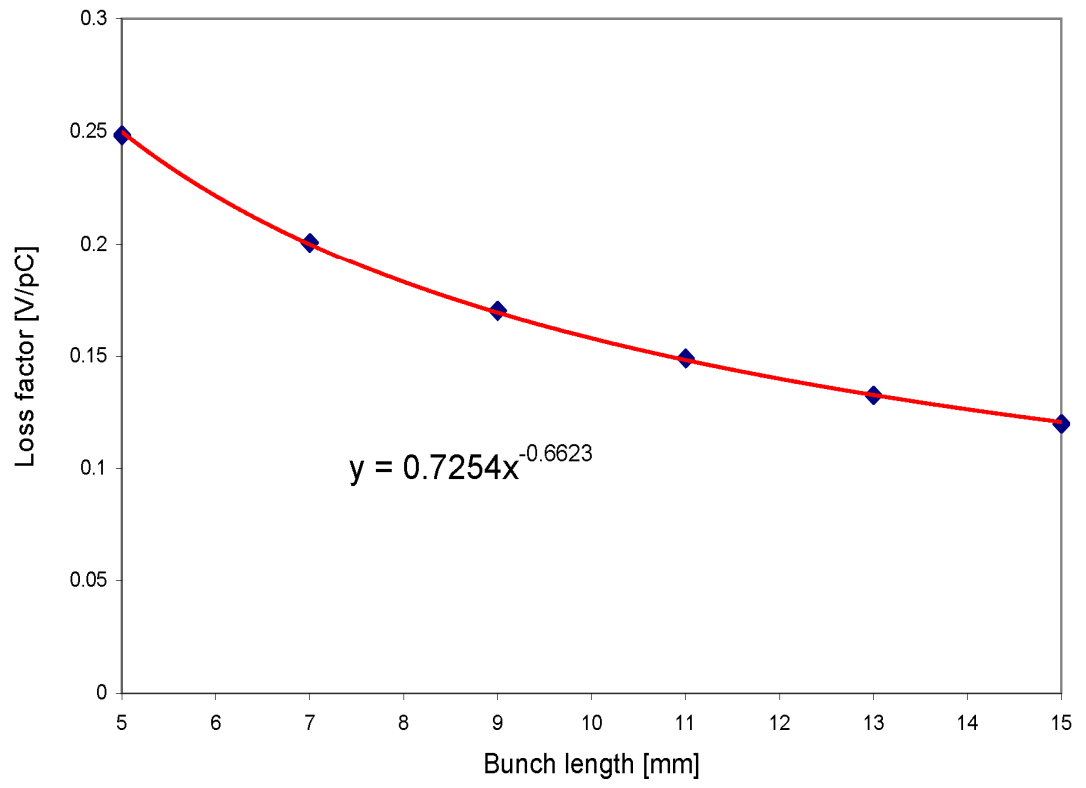

Figure 10.6: Transverse kicker loss factor as a function of bunch length.

### *Abort kickers*

Fig. 10.7 shows beam current dependence of the power dissipated in the injection and abort kickers in LER PEP-II. At beam current of 3 A the power in these 4 LER kickers reaches 2 kW at bunch spacing of 2.1 ns. We may assume that bunch dependence goes as resistive wall wake field dependence $\sigma^{-3/2}$. Estimated wake field power for Super-B parameters is listed in Table 10.10.

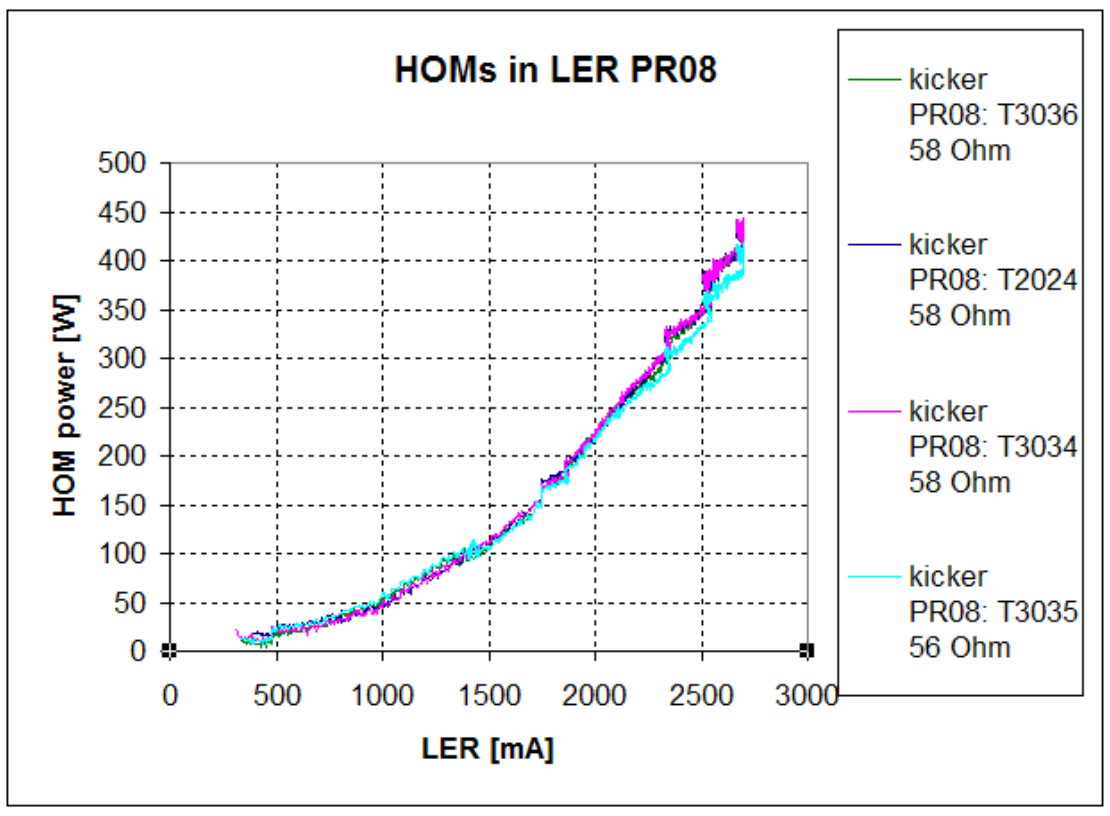

Figure 10.7: Dissipated power in injection and abort kickers in LER PEP-II.

Table 10.10: Wake field power in injection and abort kickers

| | | Beam current A | Bunch length mm | number of long. Kickers | Wake field power kW |
|---|---|---|---|---|---|
| LER | Basic Line | 2.447 | 5 | 4 | 9.89 |
| | High Currents | 4 | 4.4 | 4 | 16.01 |
| HER | Basic Line | 1.892 | 5 | 4 | 5.92 |
| | High Currents | 3.094 | 4.4 | 4 | 9.58 |

### *PEP-II collimator*

We may use in Super-B the beam collimators of PEP-II type, presented in Fig. 10.8.

The loss factor bunch length dependence is described by a formula:

$$k_{[V/pC]} = \frac{5.5}{\sigma^2_{z[mm]}}$$

Wake field collimator losses for Super-B parameters are shown in Table 10.11.

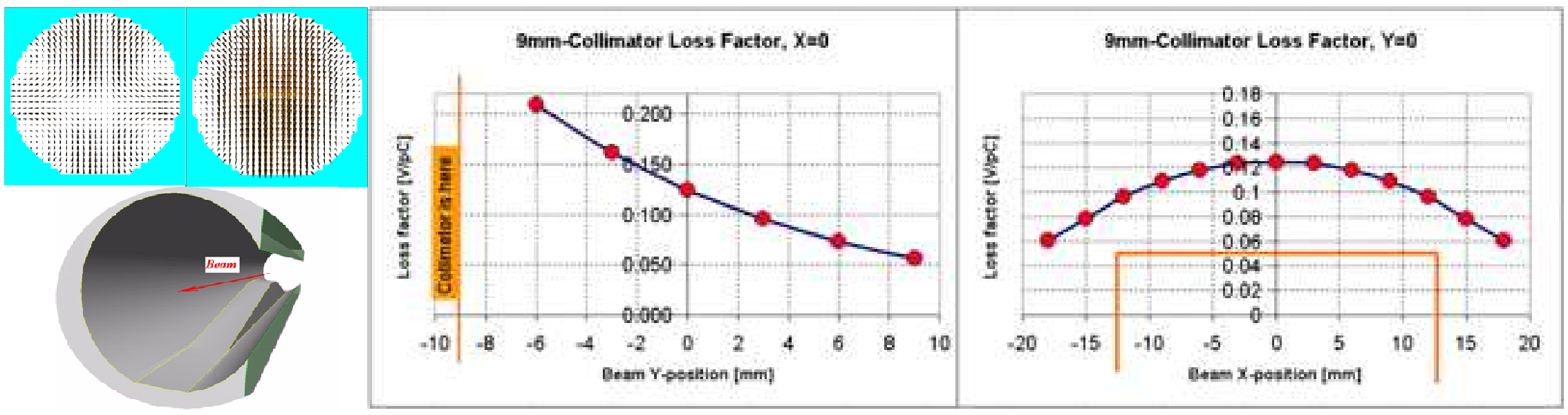

Figure 10.8: PEP-II collimator and calculated loss factor.

Table 10.11: Wake field power from collimators

| | LER | | HER | |
|---|---|---|---|---|
| | Base Line | High Currents | Base Line | High Currents |
| Beam current (A) | 2.447 | 4 | 1.892 | 3.094 |
| Bunch length (mm) | 5 | 4.4 | 5 | 4.4 |
| Number of collimators | 7 | 7 | 6 | 6 |
| Wake field power (kW) | 19.36 | 66.82 | 19.85 | 34.27 |

## 10.3 HOM in IR

### Geometrical wake fields in the Super-B Interaction Region

The geometry of storage ring collider interaction regions presents an impedance to beam fields resulting in the generation of additional electromagnetic fields (higher order modes or wake fields) which affect the beam energy and trajectory. These affects are computed for the Super B interaction region by evaluating longitudinal loss factors and averaged transverse kicks for short range wake fields. Results indicate at least a factor of 2 lower wake field power generation, in comparison with the interaction region geometry of the PEP-II B-factory collider. Wake field reduction is a consideration in the Super B design. Transverse kicks are consistent with an attractive potential from the crotch nearest the beam trajectory. The longitudinal loss factor scales as the -2.5 power of the bunch length. Figure 10.9 is a cutaway orthographic view of the model of the Super B IP region used for this study. The geometry spans 0.11 by 0.04 by 1.5 meters. One of the beam chamber axis is oriented along the z-axis, the direction a simulated bunch propagates. This model is meshed with up to 23 million points. The chamber material is considered infinitely conductive. Field solvers Gdfidl [7] and MAFIA [8] are used to evaluate wake field loss factors and averaged transverse wake field kicks.

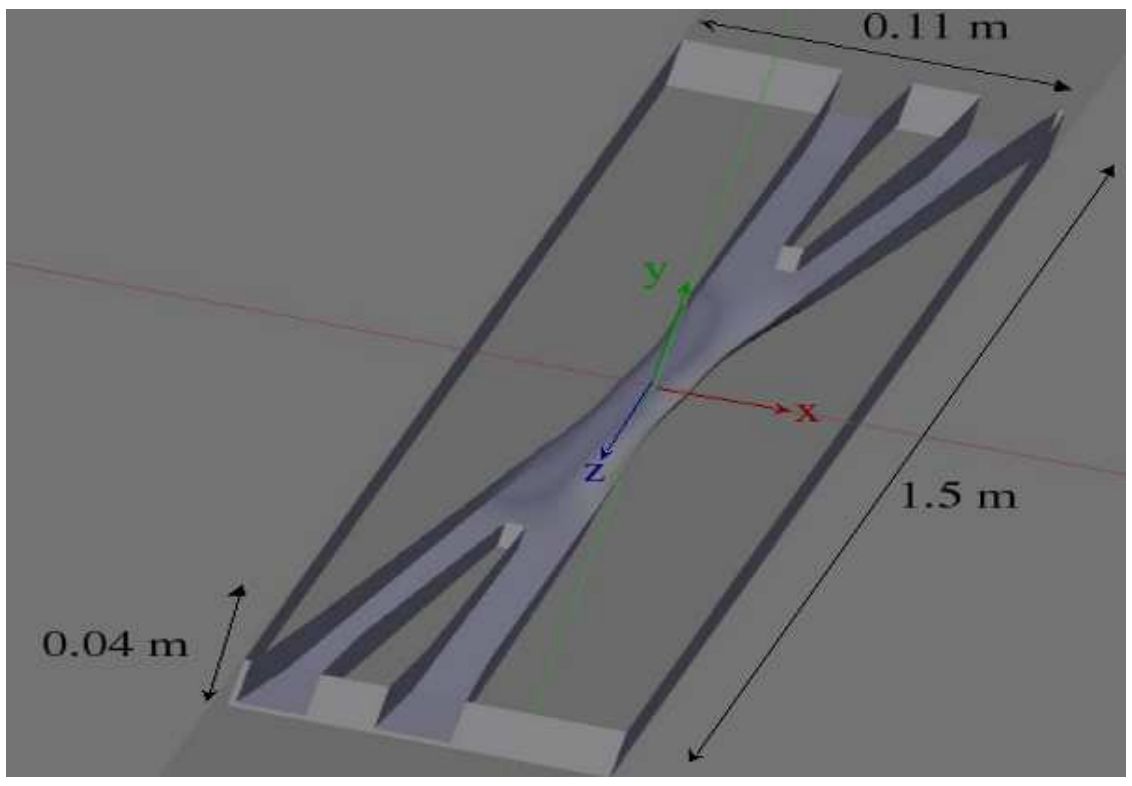

Figure 10.9: Cutaway model of the Super B interaction region. Dimensions are in meters. Beam direction is along the z-axis.

HOM generation is affected by beam trajectory. This is investigated by computing the loss factors for horizontally displaced beam trajectories. Figure 10.10 displays the calculated longitudinal loss factors, wake potentials, and the bunch profile for horizontal displacements of -1.6, 0 and + 1.6 mm from the nominal beam trajectory for a 6 mm long bunch. Longitudinal loss factors are respectively -0.379, -0.114 and -0.037 V/pC, decreasing with positive offset. With reference to the geometry of Figure 10.2, a positive horizontal offset brings the beam closer to the upstream crotch. A negative offset means a trajectory closer to the downstream crotch. Loss factors increase with proximity to the downstream crotch.

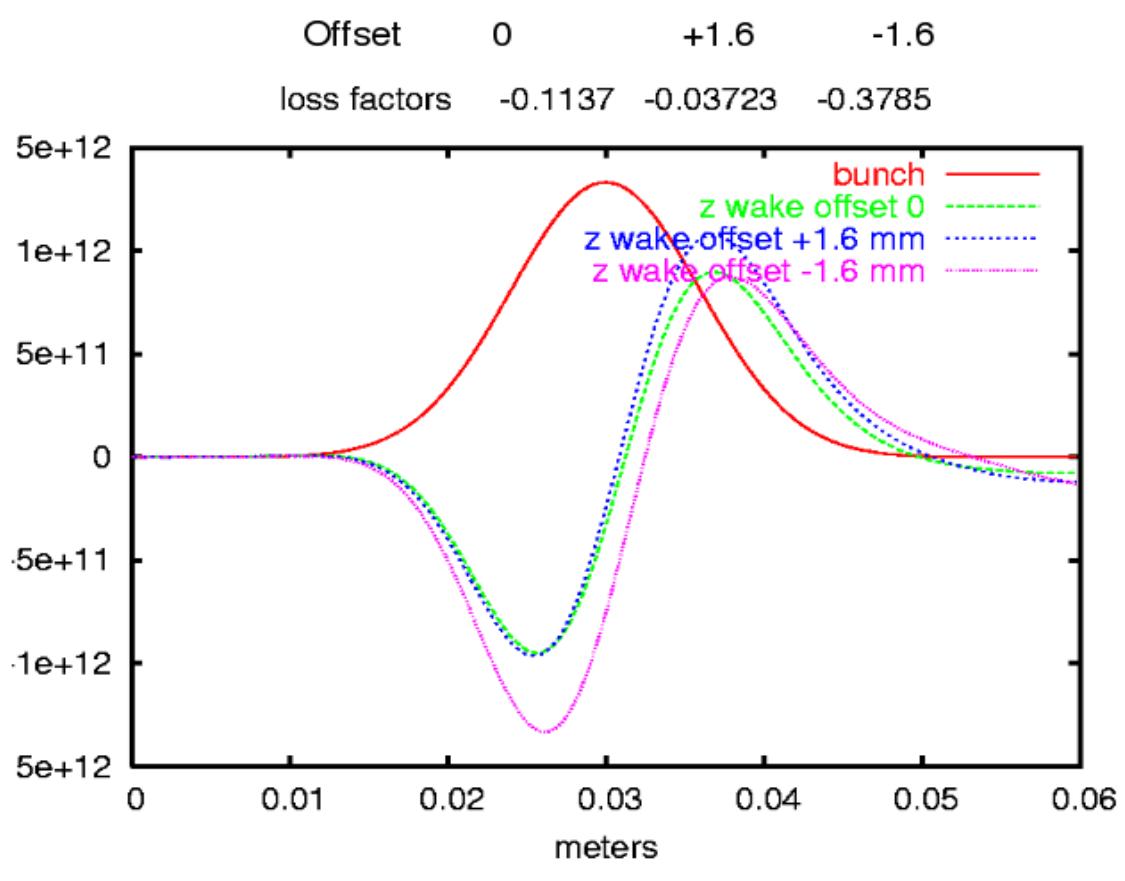

Figure 10.10: Longitudinal wake fields and loss factors for a 6 mm long bunch at various horizontal beam trajectory offsets from the ideal beam path: -1.6 mm, 0, and +1.6 mm.

For a given total charge shorter bunch length increases the peak current and excites higher frequencies in the wake field spectrum. In general this leads to higher loss factors. Loss factor vs bunch length is shown in Figure 10.1 for the case of the Super B interaction region. This increase in loss factor is non-linear and varies as the -2.55 power of the bunch length.

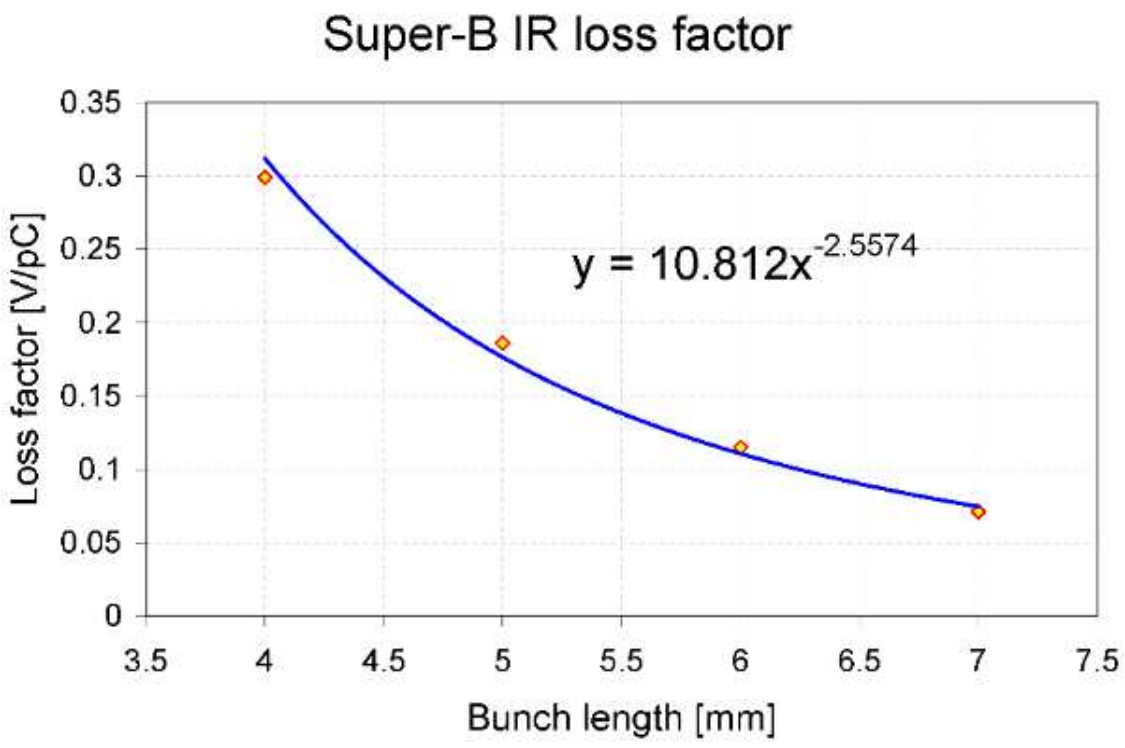

Figure 10.11: Loss factor dependence on bunch length is nonlinear.

For one 5 mm long bunch the Super B interaction region loss factor is $\kappa$=0.186 V/pC. Both beams contribute to wake field generation under nominal colliding conditions. For a 5 mm bunch length the wake field power $P_{beam}$ generated with both LER and HER current $I_{\pm}$=2.12 Amperes, and bunch spacing of □=4.2 ns is

$$P_{beam} = \kappa \cdot \tau \cdot (I_+^2 + I_-^2) \sim 7.0 \text{ kW} .$$

HOM generated power at IP as a function of a bunch length is shown at Fig. 10.12

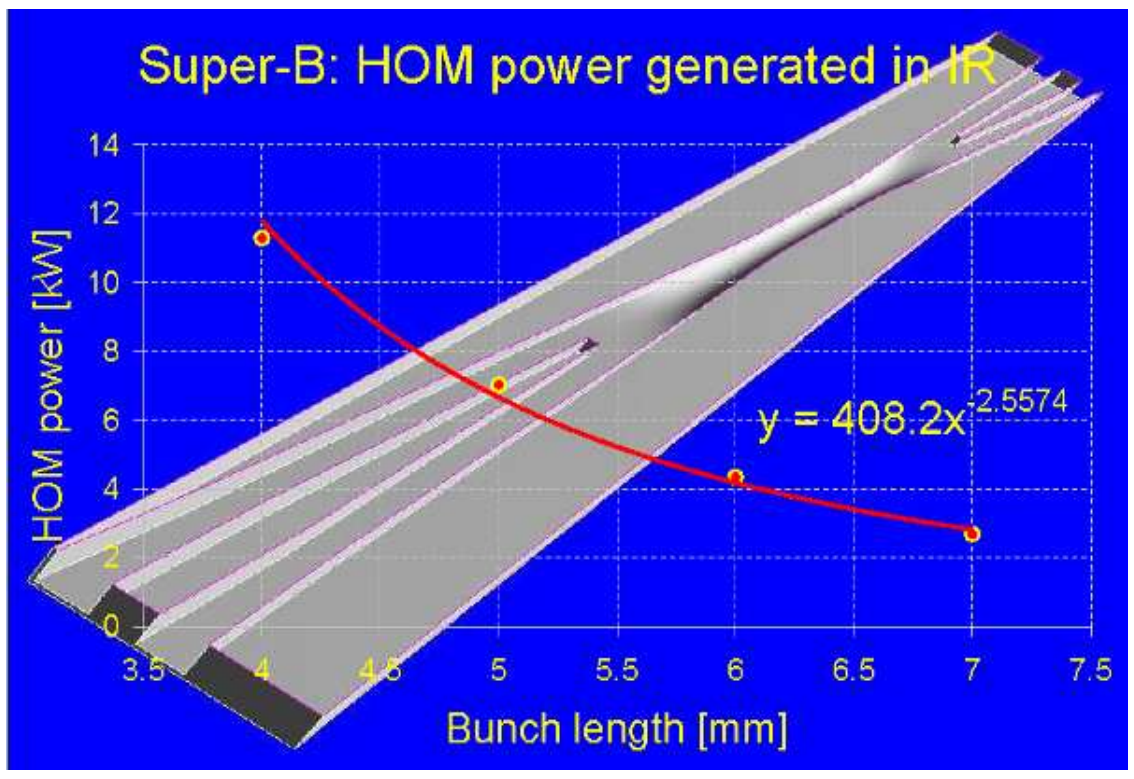

Figure 10.12: HOM generated power at IP as a functi of a bunch length.

To make a comparison we show the PEP-II interaction region model at Fig. 10.13. This model includes only a small part of the IR, however it may generate much more power. Tapers, masks and offsets are the dominant contribution to wake field generation in this case. For a bunch length of 13 mm in this PEP-II interaction region the loss factor was computed to be $\kappa$=0.06 V/pC. Scaled to a 5 mm bunch length with both LER and HER current $I_{\pm}$=2.12 Amperes, and bunch spacing of $\tau$=4.2 ns, this small part of the PEP-II interaction region generates $P_{beam}$ = 14 kW, which is a factor of two larger than wake field power generated in the Super-B interaction region for the same operating parameters.

Based on this comparison with PEP-II, the Super B interaction region design presents a smaller impedance to the beam. This helps preserve emittance and allows the Super-B factory IR to sustain the short bunches at the currents required to produce high luminosity.

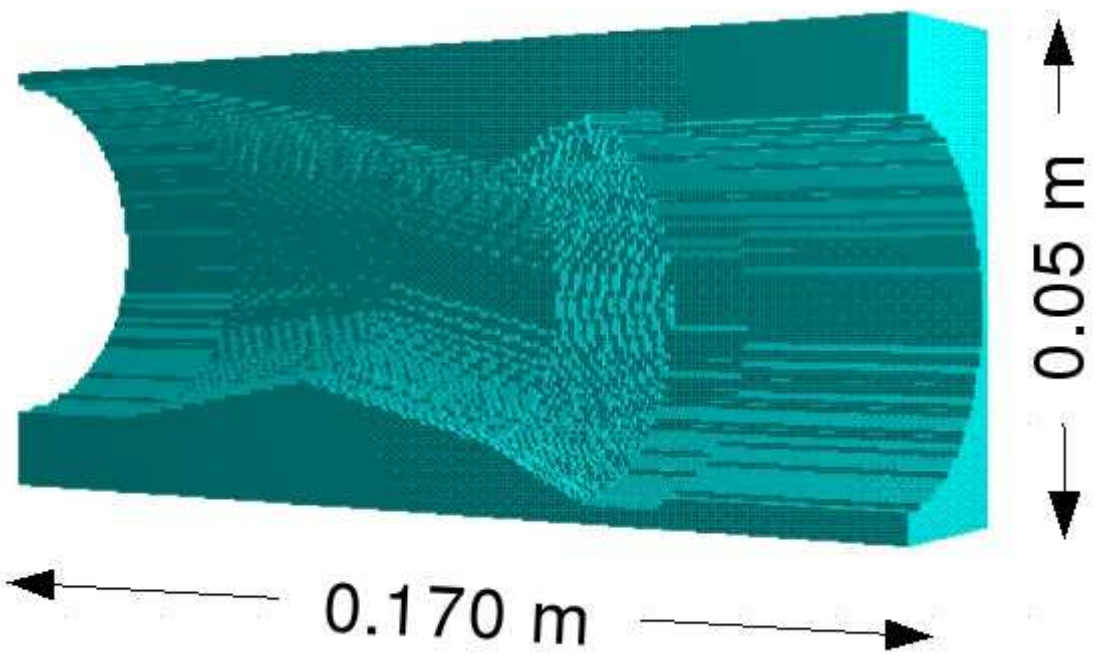

Figure 10.13: Section of PEP-II interaction region used for loss factor comparison.

# 11. Rings Magnet System

The SuperB rings will be built with room-temperature magnets. The lattice has been designed to take maximum advantage of the potential availability of the PEP-II ring magnets without compromising perfor-mances. This is possible, since the energies and the circumference of the PEP-II rings are quite comparable to those of SuperB.

In order for PEP-II magnets to be suitable for SuperB, the magnet apertures must be sufficient, but not too much larger than needed to avoid excessive power consumption. At SuperB, with its small beam sizes, the apertures will be dominated by impedance and vacuum conductance considerations, rather than the size of the beams, and the apertures required will be similar to those of PEP-II. Thus, there is a good match between the size of the PEP-II magnet apertures and the anticipated SuperB requirements.
Table 11.1 lists the magnet inventory of the PEP-II HER. In many cases, these magnets are capable of higher field strengths than used operationally at PEP-II, since they were originally designed for the 18 GeV PEP-I rings. This has been taken into account in Table 11.1.

The PEP-II LER magnet inventory is listed in Table 11.2. These magnets were built specifically for the PEP-II LER. In most cases, the maximum field was specified such that the PEP-II LER can reach 3.5 GeV in energy, while for SuperB LER the request is for 4.18 GeV. In the Tables sagittas and gradients have been computed at the nominal SuperB energies.

Table 11.1: PEP-II High Energy Ring magnets.

| **Dipoles (Location)** | **Length (m)** | **Aperture (mm)** | **Field (T)** | **Int. Stregth (T m)** | **Sagitta @ 6.7 GeV (mm)** | **Current (A)** | **Quantity (#)** |
|---|---|---|---|---|---|---|---|
| Arc | 5.4 | 60. | 0.27 | 1.4 | 44. | 950 | 194 |
| IR Soft | 2.0 | 150x100 | 0.092 | 0.184 | 2. | 170 | 6 |
| **Quadrupoles (Location)** | **Length (m)** | **Aperture (mm)** | **Gradient (T/m)** | **Int. Stregth (T)** | **K2 @ 6.7 GeV ($m^{-2}$)** | **Current (A)** | **Quantity (#)** |
| Arc | 0.56 | R 50 | 16.96 | 9.5 | 0.76 | 350 | 202 |
| Injection section | 0.45 | R 50 | 11.11 | 5 | 0.5 | 200 | 4 |
| Straight | 0.73 | R 50 | 17.53 | 12.8 | 0.79 | 350 | 81 |
| IR | 1.5 | - | 6.67 | 10 | 0.3 | 650 | 2 |
| IR | 1.5 | - | 10 | 15 | 0.45 | 1150 | 2 |
| Global skew | 0.3 | R 90 | 2.33 | 0.7 | 0.1 | 250 | 4 |
| IR Skew | 0.2 | R 50 | 0.32 | 0.064 | 0.014 | 50 | 4 |
| IR Skew | 0.3 | R 50 | 1.33 | 0.4 | 0.06 | 12 | 4 |
| **Sextupoles (Location)** | **Length (m)** | **Aperture (mm)** | **Gradient ($T/m^2$)** | **Int. Stregth (T/m)** | **K3 @6.7 GeV ($m^{-3}$)** | **Current (A)** | **Quantity (#)** |
| Arc SF, SD1 | 0.3 | R 60 | 210 | 63 | 0.94 | 400 | 104 |
| **Correctors (Location)** | **Length (m)** | **Aperture (mm)** | **Field (T)** | **Int. Stregth (T m)** | | **Current (A)** | **Quantity (#)** |
| Arc X | 0.3 | 90x50 | 0.018 | 0.0054 | - | 12 | 96 |
| Arc Y | 0.3 | 90x50 | 0.01 | 0.003 | - | 12 | 96 |
| Straight | 0.3 | R 50 | 0.012 | 0.0036 | - | 12 | 91 |

Table 11.2: PEP-II Low Energy Ring magnets.

| Dipoles (Location) | Length (m) | Aperture (mm) | Field (T) | Int. Strength (T m) | Sagitta @ 4.18 GeV(mm) | Current (A) | Quantity (#) |
|---|---|---|---|---|---|---|---|
| Arc | 0.45 | 63.5 | 0.93 | 0.42 | 1.7 | 750 | 192 |
| Straight BB+/- | 0.45 | - | - | - | - | - | 10 |
| Straight BM,BV | 0.5 | - | 0.56 | 0.28 | 1.26 | 850 | 10 |
| Straight, BC | 1.5 | - | 0.37 | 0.562 | 7.5 | 175 | 10 |
| **Quadrupoles (Location)** | **Length (m)** | **Aperture (mm)** | **Gradient (T/m)** | **Int. Strength (T)** | **K2 @4.18 GeV ($m^{-2}$)** | **Current (A)** | **Quantity (#)** |
| Arc, Q58Al4 | 0.43 | R 5 | 9.5 | 4.1 | 0.682 | 160 | 196 |
| Straight Q58Al4 | 0.43 | R 5 | 9.5 | 4.1 | 0.682 | 160 | 127 |
| IR 2 Q58Cu4 | 0.43 | R 5 | 11.9 | 5.1 | 0.854 | 200 | 30 |
| Insertion QF2 | 0.5 | - | 13.6 | 6.8 | 0.976 | 1200 | 2 |
| Insertion QD1 | 1.2 | - | - | - | - | (pm) | 2 |
| Insertion SK1 | 0.2 | - | - | - | - | (pm) | 2 |
| Skew | 0.2 | - | 2.6 | 0.52 | 0.187 | 12 | 15 |
| **Sextupoles (Location)** | **Length (m)** | **Aperture (mm)** | **Gradient ($T/m^2$)** | **Int. Strength (T/m)** | **K3 @4.18 GeV ($m^{-3}$)** | **Current (A)** | **Quantity (#)** |
| Arc SF, SD1 | 0.25 | R 60 | 192 | 48.1 | 13.78 | 310 | 76 |
| Arc SD2 | 0.35 | R 60 | 245 | 85.6 | 17.58 | 500 | 8 |
| IR2 | 0.25 | R 60 | - | - | - | - | 7 |
| **Correctors (Location)** | **Length (m)** | **Aperture (mm)** | **Field (T)** | **Int. Strength (T m)** | | **Current (A)** | **Quantity (#)** |
| Arc X | 0.233 | 130x90 | 0.0365 | 0.0085 | - | 12 | 96 |
| Arc Y | 0.312 | 250x90 | 0.0212 | 0.0066 | - | 12 | 92 |
| Arc X wide | - | - | - | 0.012 | - | 12 | 4 |
| Straight | 0.3 | - | 0.0252 | 0.00755 | - | 12 | 104 |

## 11.1 Dipoles

### HER dipoles

Positrons will be stored in the HER. The PEP-II HER dipoles have C-shaped yokes and 2.2 cm sagitta based on their design 165 m bending radius. For SuperB, the bending radius for the main arc dipoles is 80 and 91 m, for the two different cells, and the sagitta will range from 4. to 4.5 cm. This value may be tolerable given the more than 5 cm total width for the good-field region and the fact that the magnets can always be centered on the average beam orbit. Figure 11.1 shows a sketch of a PEP-II HER dipole. It has to be checked if there is significant space available in the horizontal plane to accommodate an antechamber for the vacuum system.

While the PEP-II HER dipoles are the original PEP dipoles, the magnets were completely overhauled and refurbished during construction of PEP-II, serialized, and mechanically and magnetically measured. The measurement data – 1 Bdl, field harmonics at 0.9 Tm and gap height vs. s – are available in the archives of the Magnetic Measurement Group at SLAC [1].

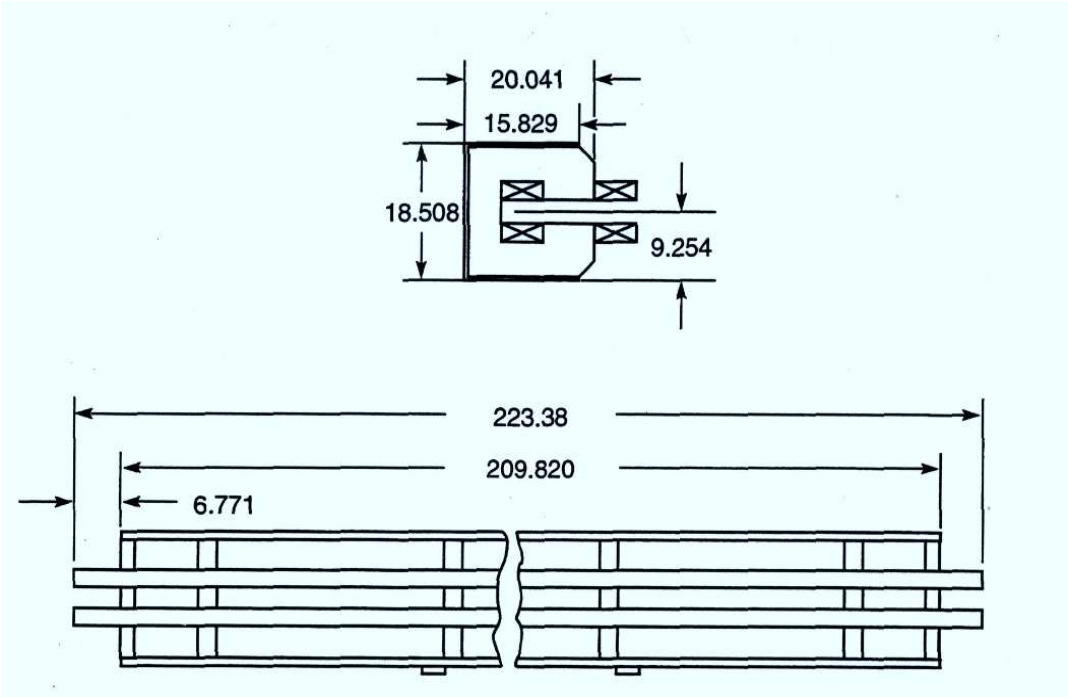

Figure 11.1: Cross section of a PEP-II HER main dipole. All dimensions are in inches.

They have been in constant use since PEP-II commissioning began. Despite the high beam current, the radiation environment in the PEP-II arcs is actually quite benign, and no evidence for significant radiation damage to the magnet coils has been seen. We therefore, at present, see no need to re-measure or refurbish the dipole magnets, although each magnet coil will be carefully inspected for signs of aging.

### LER dipoles

The 0.45 m-long PEP-II LER dipoles are box-type magnets. Because of their short length, there is no issue with the different sagitta at any reasonable bending angle. In SuperB the actual dipole length will be 0.9m, however it is probably possible to put side-to-side two PEP-II LER dipoles. The missing dipoles will be newly built, likely using the laminations cut for the existing PEP-II LER arc dipoles. The PEP-II LER dipole magnets were built new at the time of PEP-II construction. They were measured at the factory at that time however an individualized set of measurements does not exist for each magnet. There is also a certain variation of field shape with excitation in these magnets. We therefore anticipate re-measuring each of the dipoles at the operating field for SuperB, before installation in the SuperB LER. As in case of the HER dipoles, however, there is no need to refurbish the PEP-II LER dipole magnets; a careful inspection should suffice. Figure 11.2 shows the cross section of the LER dipoles.

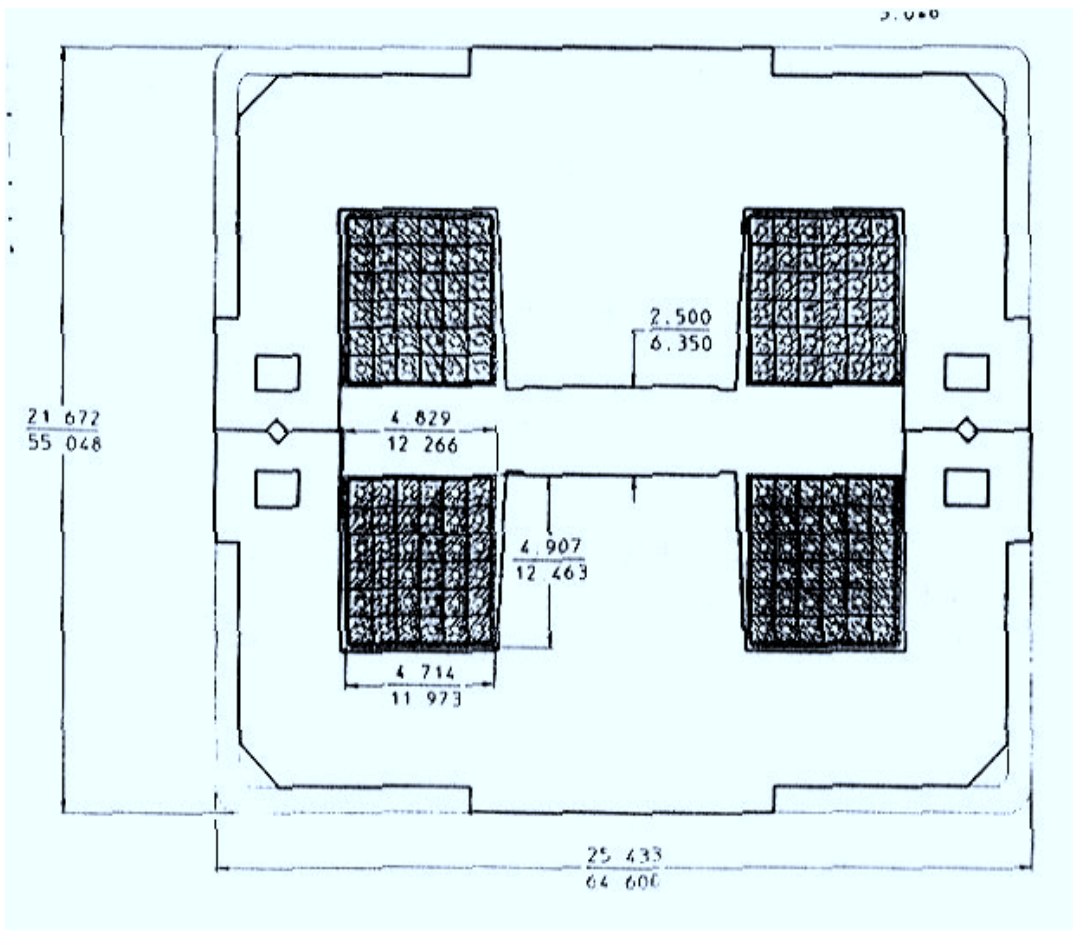

Figure 11.2: Cross section of a PEP-II LER main arc dipole. Dimensions are given in inches (upper numbers) and cm (lower numbers).

### Final Focus dipoles

For the Final Focus brand new dipoles, 4 m and 2.8 m long, will be needed.

## 11.2 Quadrupoles

It is anticipated that most PEP-II quadrupoles will find use in SuperB. Most of the 0.56 m quadrupoles of the PEP-II HER will be used in the SuperB HER. For the 0.43 m-long SuperB quadrupoles the needs are covered by the existing PEP-II LER quadrupoles, with spares available. The latter come in three different coil configurations with somewhat different maximum excitation, so care will be taken in matching the coil type to the requirements. In addition, about forty new 0.215 m long quadru-poles are needed; however with some adjustments in the cells layout design we may well use the exceeding 0.43 m long LER quadrupoles. For the 0.73 m long ones the need is well satisfied by the existing PEP-II HER ones. A complete audit trail exists for the measurements of the PEP- II HER quadrupoles, while for the PEP-II LER quadrupoles only a sparse data set is available. As a result, we will need to re-measure the PEP-II LER quadrupoles as well. Careful inspection of all coils will detect any sign of aging, and there is a significant number of spare coils available in case it is decided to replace some of the coils. There may, however, be cases of quadrupoles in SuperB being excited at higher current than in PEP-II. In these cases we will change the cooling circuits to connect all coils in parallel, thus minimizing the total temperature increase during operation.

Figure 11.3 shows a cross sectional and side view of a PEP-II HER quadrupole, while Fig. 11.4 shows a PEP-II LER quadrupole.

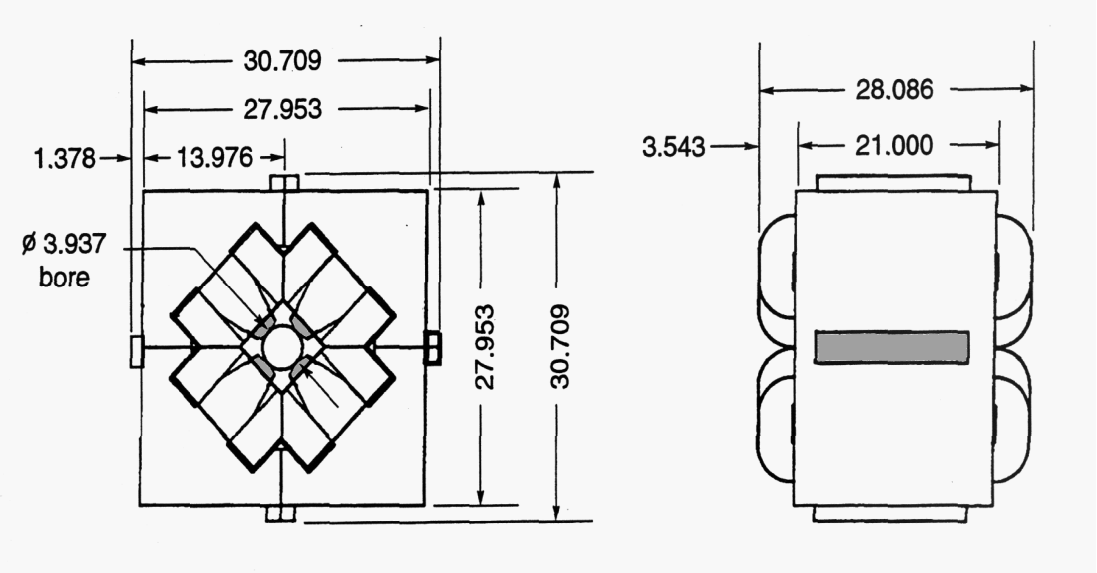

Figure 11.3: Cross section and side view of a PEP-II HER main arc quadrupole. All dimensions are in inches).

The IR quadrupoles have been described in the Interaction Region Chapter 5.

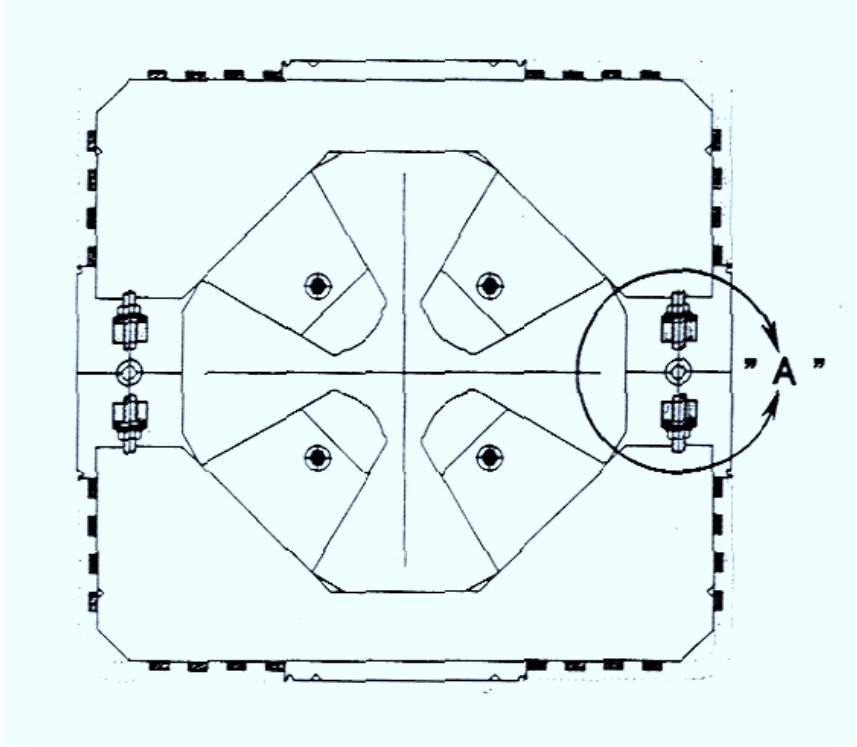

Figure 11.4: Cross section of a PEP-II LER main arc quadrupole.

## 11.3 Sextupoles

Altogether, the two SuperB rings will use about 200 sextupoles, with magnetic length ranging from 0.25 to 0.4 m. Most of them are available from PEP-II, The longer ones in the FF (0.4 m) are quite strong and will need a new design. Eight additional magnets 0.35m-long will be needed, and may be built using the original PEP-II lamination die, same for the 12 0.25m long ones which are missing and maybe replaced in the design by can use the 0.3 m-long ones that we have in excess.

## 11.4 Correctors

The basic orbit-corrector dipoles exist in three types for each ring: horizontal arc-type, vertical arc-type and straight-section type, which are mounted either as horizontal or vertical correctors around the chamber of circular cross section. Where prudent these magnets will be reused. However, the SuperB vacuum system will differ from that of PEP-II, which may prevent reuse of some of these magnets. Since the cost for orbit correctors is fairly modest, it appears prudent to avoid compromising the vacuum chamber geometry in order to reuse existing orbit corrector magnets. The same principle applies to other correction magnets, such as skew quadrupoles. At the moment the number of correctors used in the Low Emittance Tuning procedure is 168/ring, but this number can be reduced further.

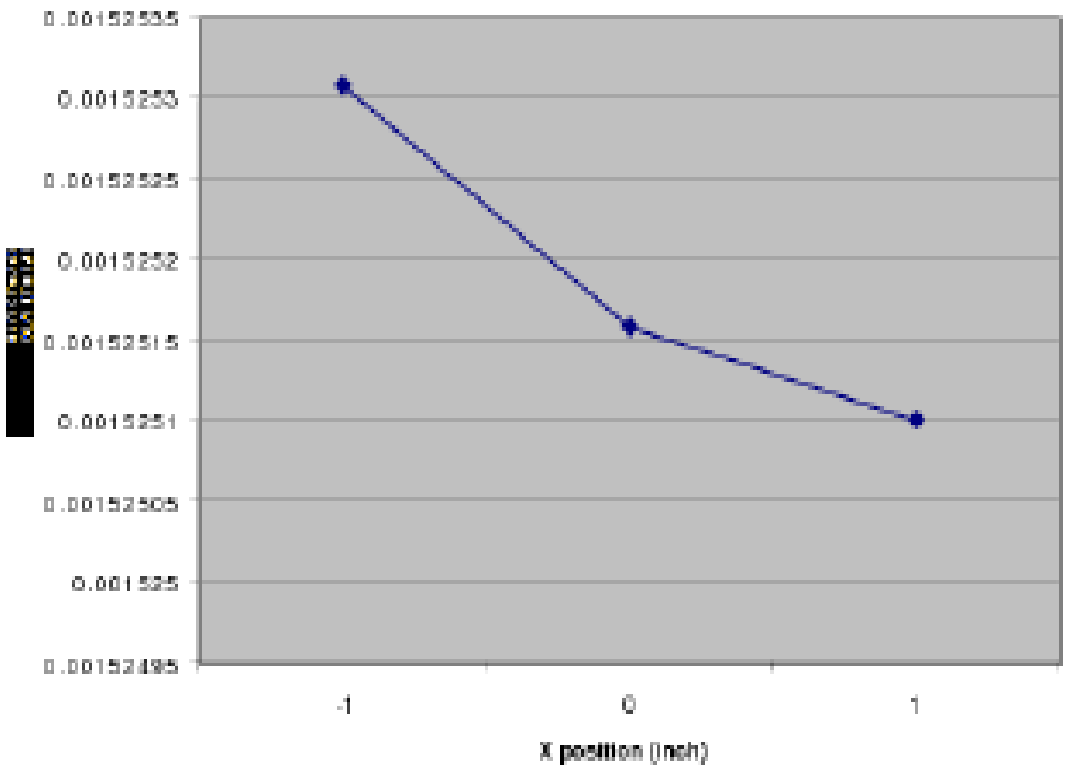

Figure 11.4: Field uniformity of a sample of PEP-II HER 5.4-m dipole magnets.

## 11.5 Field quality

Field uniformity requirements for SuperB magnets will be determined following more detailed tracking studies. However, since the beam sizes are small and orbit excursions will have to be tightly controlled in order to preserve the small emittances, the beams do not sample field regions far from the nominal center line. We therefore expect the field uniformity tolerances of the PEP-II magnets to be sufficient for SuperB applications.

The field uniformity of the PEP-II HER dipole is shown in Fig. 11.4; the field harmonics of the PEP-II HER 0.56 m quadrupoles magnets are shown in Fig. 11.3. Since we will have individual measurement data for each magnet, sorting algorithms will be employed as necessary to mitigate the effect of field differences between the magnets in a family, as was done for PEP-II.

The magnet errors were based on those observed for the PEP-II ring magnets (see Table 11.3 for the HER and Table 11.4 for the LER). They are parameterized in terms of a multipole expansion:

$$\frac{(B_y + iB_x)}{B_0(r)} = \sum_{n=1} (b_n + ia_n)\left(\frac{x}{r} + i\frac{y}{r}\right)^{n-1}$$

where r is the reference radius and $B_0$ is the main field of the magnets.

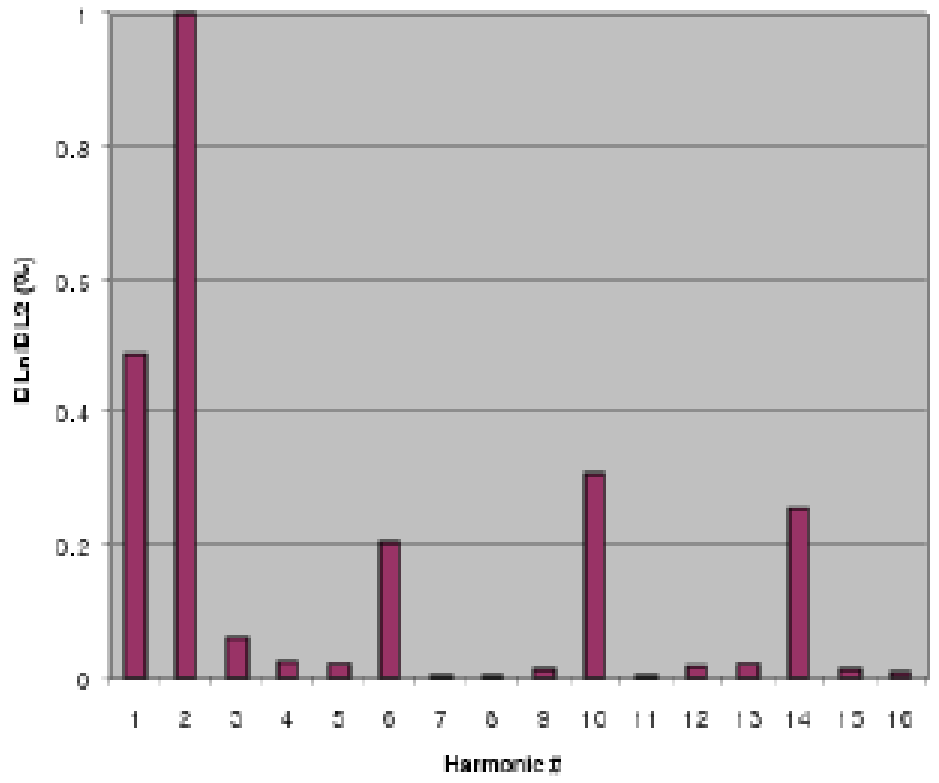

Figure 11.5: Harmonic spectrum of a sample of PEP-II HER 0.56m quadrupoles. Harmonic # 2 is the gradient.

Table 11.3: PEP-II HER magnet errors

| Multipole index(n) | n | Systematic: $b_n$ | Random: $b_n$ |
|---|---|---|---|
| Dipole (r=0.03m) | 3 | $1.00 \times 10^{-5}$ | $3.20 \times 10^{-5}$ |
| | 5 | - | $3.20 \times 10^{-5}$ |
| | 7 | - | $6.40 \times 10^{-5}$ |
| | 9 | - | $8.20 \times 10^{-5}$ |
| Quadrupole (r=0.05m) | 3 | $1.02 \times 10^{-4}$ | $4.63 \times 10^{-5}$ |
| | 4 | $1.91 \times 10^{-4}$ | $8.09 \times 10^{-5}$ |
| | 5 | $1.89 \times 10^{-5}$ | $8.86 \times 10^{-6}$ |
| | 6 | $5.69 \times 10^{-4}$ | $2.80 \times 10^{-5}$ |
| | 7 | $6.60 \times 10^{-6}$ | $3.45 \times 10^{-6}$ |
| | 8 | $9.60 \times 10^{-6}$ | $5.72 \times 10^{-6}$ |
| Sextupole (r=0.05652m) | 5 | - | $2.20 \times 10^{-3}$ |
| | 7 | - | $1.05 \times 10^{-3}$ |
| | 9 | $-1.45 \times 10^{-2}$ | - |
| | 15 | $-1.30 \times 10^{-2}$ | - |

Table 11.4: PEP-II LER magnet errors

| Multipole index (n) | n | Systematic: $b_n$ | Random: $b_n$ |
|---|---|---|---|
| Dipole (r=0.03m) | 3 | $-0.50 \times 10^{-4}$ | $1.00 \times 10^{-4}$ |
| | 5 | $3.00 \times 10^{-4}$ | $1.00 \times 10^{-4}$ |
| | 7 | - | $1.00 \times 10^{-5}$ |
| | 9 | - | $1.00 \times 10^{-5}$ |
| Quadrupole (r=0.05m) | 3 | $1.02 \times 10^{-4}$ | $4.63 \times 10^{-5}$ |
| | 4 | $1.91 \times 10^{-4}$ | $8.09 \times 10^{-5}$ |
| | 5 | $1.89 \times 10^{-5}$ | $8.86 \times 10^{-6}$ |
| | 6 | $5.69 \times 10^{-4}$ | $2.80 \times 10^{-5}$ |
| | 7 | $6.60 \times 10^{-6}$ | $3.45 \times 10^{-6}$ |
| | 8 | $9.60 \times 10^{-6}$ | $5.72 \times 10^{-6}$ |
| | 9 | $7.14 \times 10^{-6}$ | $3.85 \times 10^{-6}$ |
| | 10 | $3.37 \times 10^{-4}$ | $5.62 \times 10^{-6}$ |
| | 11 | $6.08 \times 10^{-6}$ | $3.32 \times 10^{-6}$ |
| | 12 | $5.34 \times 10^{-5}$ | $6.20 \times 10^{-6}$ |
| | 13 | $1.10 \times 10^{-5}$ | $6.53 \times 10^{-6}$ |
| | 14 | $6.65 \times 10^{-5}$ | $8.20 \times 10^{-6}$ |
| Sextupole (r=0.05652m) | 5 | - | $2.20 \times 10^{-3}$ |
| | 7 | - | $1.05 \times 10^{-3}$ |
| | 9 | $-1.45 \times 10^{-2}$ | - |
| | 15 | $-1.30 \times 10^{-2}$ | - |

## .11.6 Power Conversion

The DC power supplies for the magnets represent a significant share of the overall cost of the magnet system. The approach taken in PEP-II is to power long strings of identical magnets with 500V, 400A chopper units fed from a bulk power supply (one for each ring), which in turn is fed by a 480V ac line. Shorter strings (or \families") are fed from smaller individual supplies operating on 208 or 480V ac feeds. All of these supplies are of relatively modern switching type and therefore, in principle, capable of being operated at the 50 Hz ac frequency used in Europe (as opposed to the 60 Hz used in North America) including the large bulk power supplies for the choppers [97]. While the details must worked out, reuse of components of the PEP-II magnet power system appears to be feasible.

## 11.7 Summary of Regular Lattice Magnets

SuperB magnet requirements are well within the performance envelope of the PEP-II magnets, and almost all PEP-II magnets, with the possible exception of specialty magnets such as the insertion quadrupoles, will be reused.

Additional magnets will be built to existing designs wherever feasible. Only a limited number of SuperB magnet designs have no PEP-II counterpart and will be of a new design. The summary of the overall magnet number needed for SuperB LER and HER is in Table 11.5. To have a summary of the fields and gradients needed for the different magnets, in Table 11.6 and 11.7 are the summary of needed gradients and fields for SuperB HER and LER, while in Figures 11.5 to 11.10 are summarized the LER and HER types of magnets for different families (IR quadrupoles, QD0 and Qf1, are not here)..

There will be also need of some small octupole and decapole magnets in the IR.

Table 11.5: Summary of SuperB magnets needed (* 59 if new are built 0.9 m long)

| | SuperB HER+LER | Existent @ PEP-II | Needed | Design |
|---|---|---|---|---|
| Dipoles (L =5.4 m) | 84 | 194 | - | |
| Dipoles (L =4. m) | 44 | - | 44 | Soft bends |
| Dipoles (L =2.8 m) | 16 | - | 16 | Soft bends |
| Dipoles (L =0.45 m) | 320 | 202 | 118 or 59* | PEP-II (lamin.) |
| Quads (L =0.56 m) | 198 | 202 | - | |
| Quads (L =0.73 m) | 54 | 81 | - | |
| Quads (L =0.215 m) | 36 | - | 36 | PEP-II |
| Quads (L =0.43 m) | 264 | 353 | - | |
| Sexts (L =0.25 m) | 86 | 76 | 10 | PEP-II |
| Sexts (L =0.30 m) | 84 | 104 | - | |
| Sexts (L =0.35 m) | 16 | 8 | 8 | New |
| Sexts (L =0.4 m) | 8 | - | 8 | New |
| Octupoles | 8 | - | 8 | New |

Table 11.6: SuperB High Energy Ring magnets.

| Dipoles (Location) | Length (m) | Field (T) | Int. Strength (T m) | Radius (m) | Sagitta (mm) | Quantity (#) |
|---|---|---|---|---|---|---|
| Arc | 5.4 | 0.28 | 1.5 | 80.44 | 45.3 | 52 |
| Arc | 5.4 | 0.245 | 1.3 | 91.33 | 40. | 28 |
| Disp. Supp. | 3.13 | 0.28 | 1.5 | 80.44 | 15.2 | 4 |
| IR | 2.8 | 0.42 | 0.184 | 52.8 | 18.6 | 2 |
| IR Soft | 2.8 | 0.08-0.18 | 0.22-0.5 | 265-165 | 3.7-7.8 | 4 |
| IR | 4.0 | 0.22-0.3 | 0.9-1.2 | 101-75.5 | 20.26.5 | 12 |
| IR Soft | 4.0 | 0.11-0.17 | 0.44-0.7 | 201-133 | 10.-15 | 10 |
| **Total** | | | | | | |
| **Quadrupoles (Location)** | **Length (m)** | **Gradient (T/m)** | **Int. Strength (T)** | | **$K2_{max}$ ($m^{-2}$)** | **Quantity (#)** |
| Arc | 0.56 | 12 | 6.7 | | 0.5 | 107 |
| Arc | 0.43 | 10 | 4.4 | | 0.46 | 90 |
| Straight | 0.73 | 18 | 13 | | 0.8 | 54 |
| IR | 0.43 | 22.4 | 9.6 | | 1.0 | 18 |
| IR | 0.215 | 22.4 | 4.8 | | 1.0 | 18 |
| **Sextupoles (Location)** | **Length (m)** | **Gradient ($T/m^2$)** | **Int. Strength (T/m)** | | **$K3_{max}$ ($m^{-3}$)** | **Quantity (#)** |
| Arc SF/SD | 0.3 | 356 | 107 | | 16.0 | 84 |
| IR crab | 0.35 | 740. | 260. | | 33. | 2 |
| IR SD | 0.35 | 360. | 126. | | 16 | 6 |
| IR SF | 0.25 | 180. | 45 | | 8. | 2 |
| IR SD | 0.40 | 560. | 225. | | 25. | 4 |

Table 11.7: SuperB Low Energy Ring magnets.

| Dipoles (Location) | Length (m) | Field (T) | Int. Strength (T m) | Radius (m) | Sagitta (mm) | Quantity (#) |
|---|---|---|---|---|---|---|
| Arc | 0.9 | 0.46 | 0.42 | | 3.3 | 56 |
| Arc | 0.9 | 0.52 | 0.47 | | 3.8 | 104 |
| Straight | 0.522 | 0.52 | 0.27 | | 1.27 | 4 |
| IR | 2.8 | 0.05-0.11 | 0.14-0.31 | | 3.7-7.8 | 4 |
| IR | 2.8 | 0.26 | 0.73 | | 18.6 | 2 |
| IR | 4.0 | 0.15-0.19 | 0.6-0.76 | | 22-26.5 | 10 |
| IR | 4.0 | 0.05-0.14 | 0.2-0.56 | | 3.7-19.8 | 16 |
| **Quadrupoles (Location)** | **Length (m)** | **Gradient (T/m)** | **Int. Strength (T)** | | **$K2_{max}$ ($m^{-2}$)** | **Quantity (#)** |
| Arc | 0.56 | 12. | 6.7 | | 0.9 | 91 |
| Straight | 0.43 | 9. | 3.9 | | 0.8 | 156 |
| IR | 0.43 | 14. | 6. | | 1. | 18 |
| IR | 0.215 | 14. | 3. | | 1. | 18 |
| **Sextupoles (Location)** | **Length (m)** | **Gradient ($T/m^2$)** | **Int. Strength (T/m)** | | **$K3_{max}$ ($m^{-3}$)** | **Quantity (#)** |
| Arc SF/SD | 0.25 | 165. | 42. | | 12. | 84 |
| IR SF | 0.35 | 230. | 81. | | 16.5 | 6 |
| IR crab | 0.35 | 460. | 160. | | 33. | 2 |
| IR SD | 0.4 | 265. | 106. | | 19. | 4 |

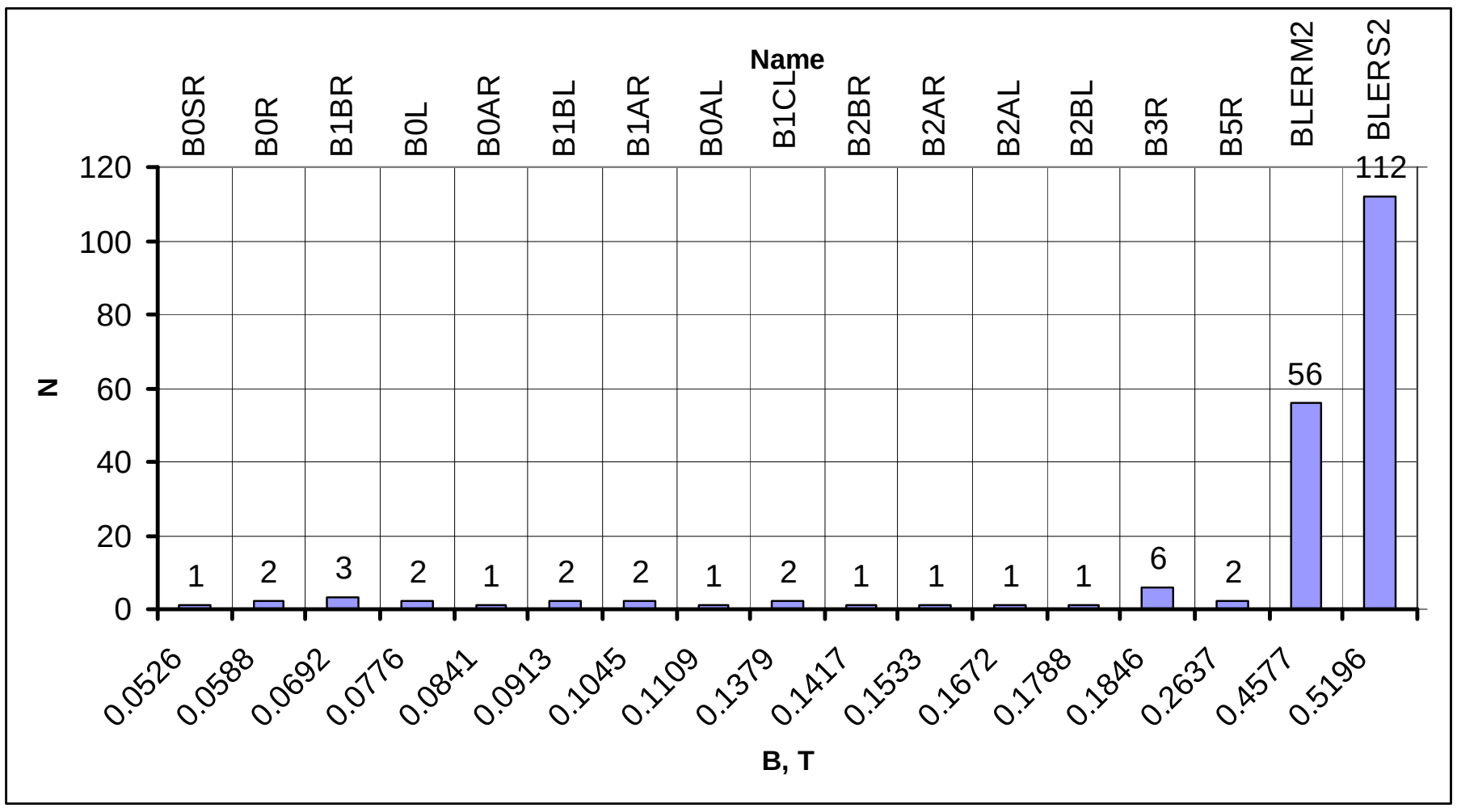

Figure11.5: Number of bending magnet vs. magnet families for LER.

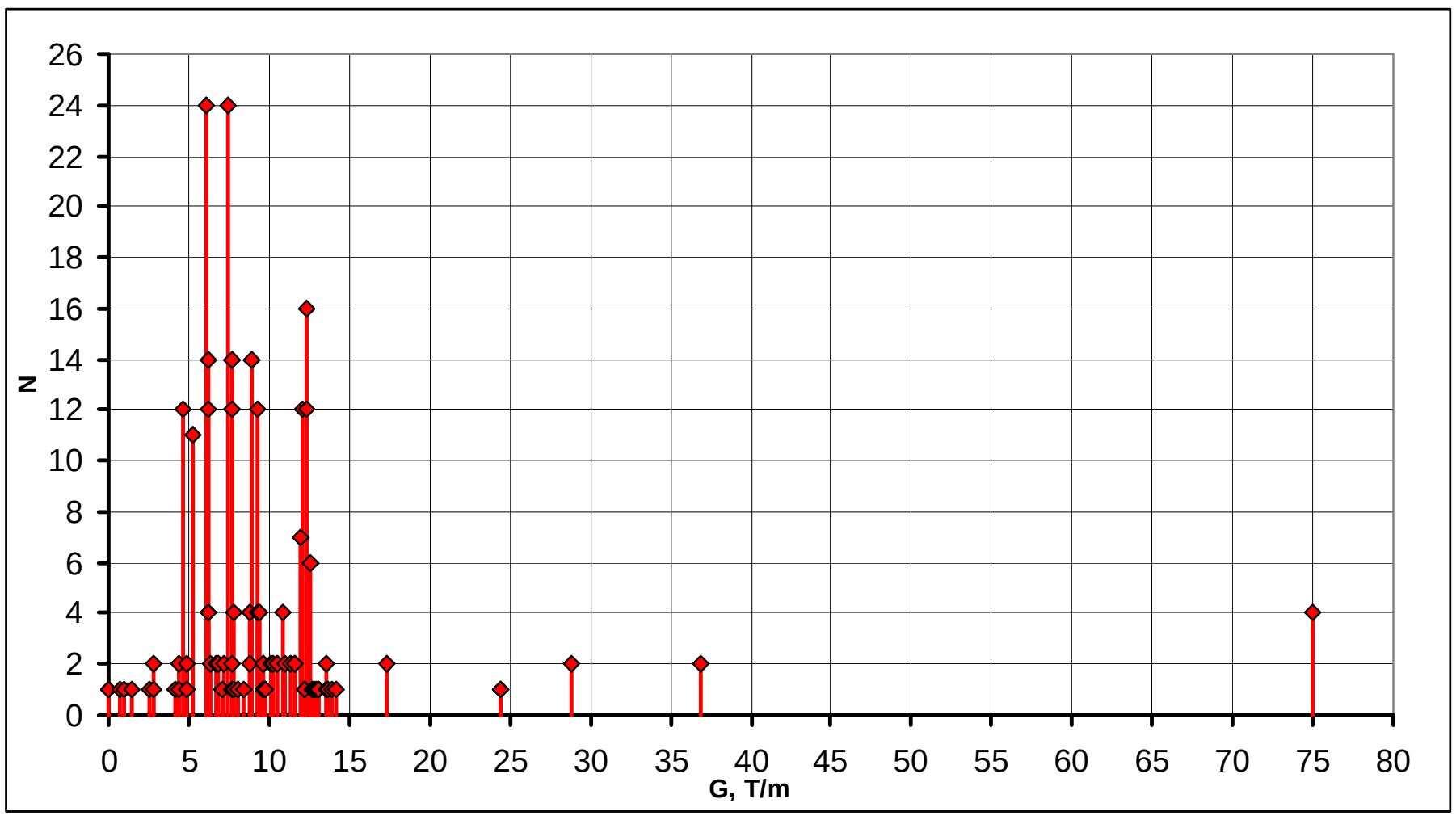

Figure11.6: Number of quadrupole lenses vs. quadrupole families for LER.

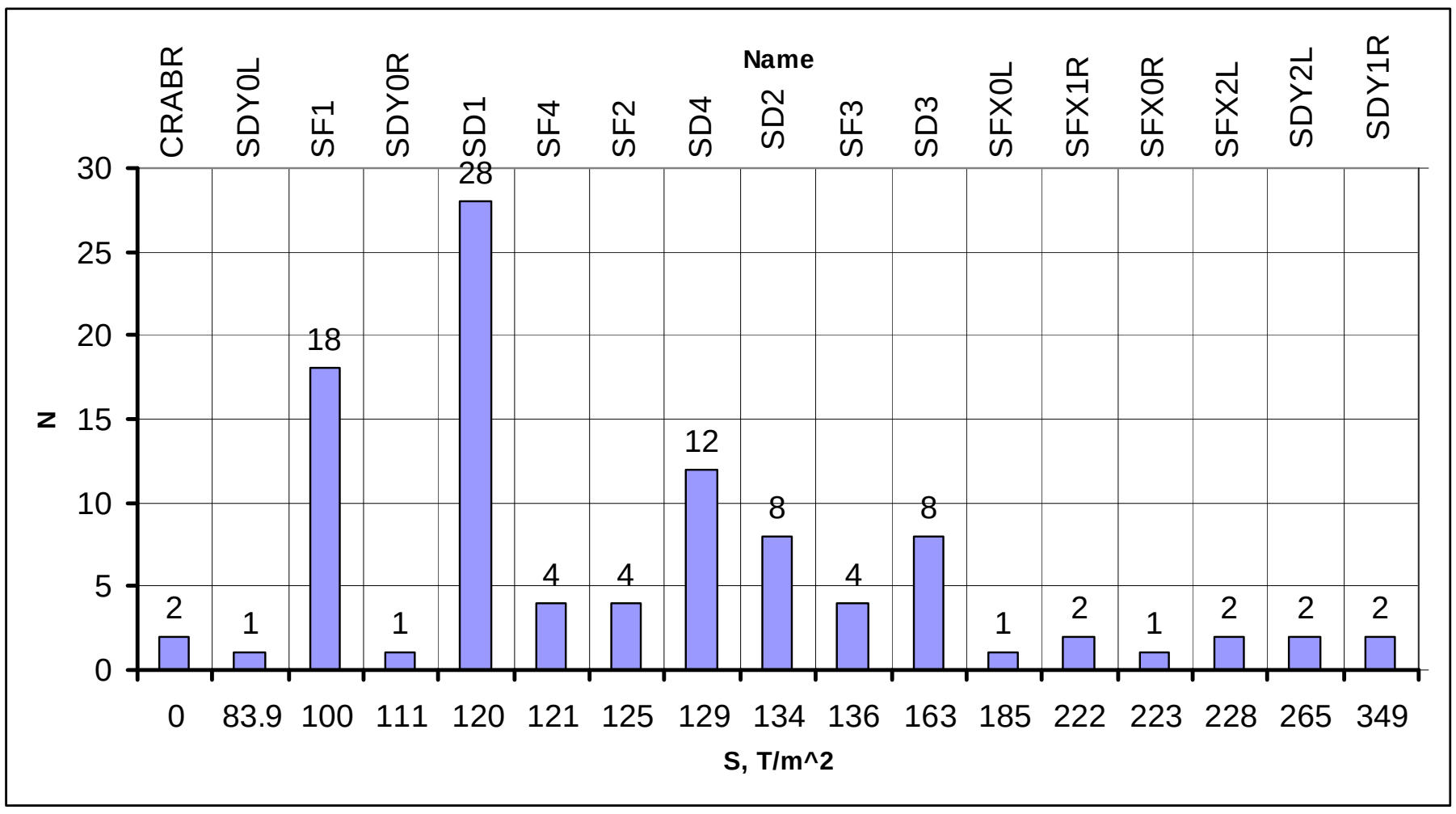

Figure 11.7: Number of sextupole lenses vs. sextupole families for LER.

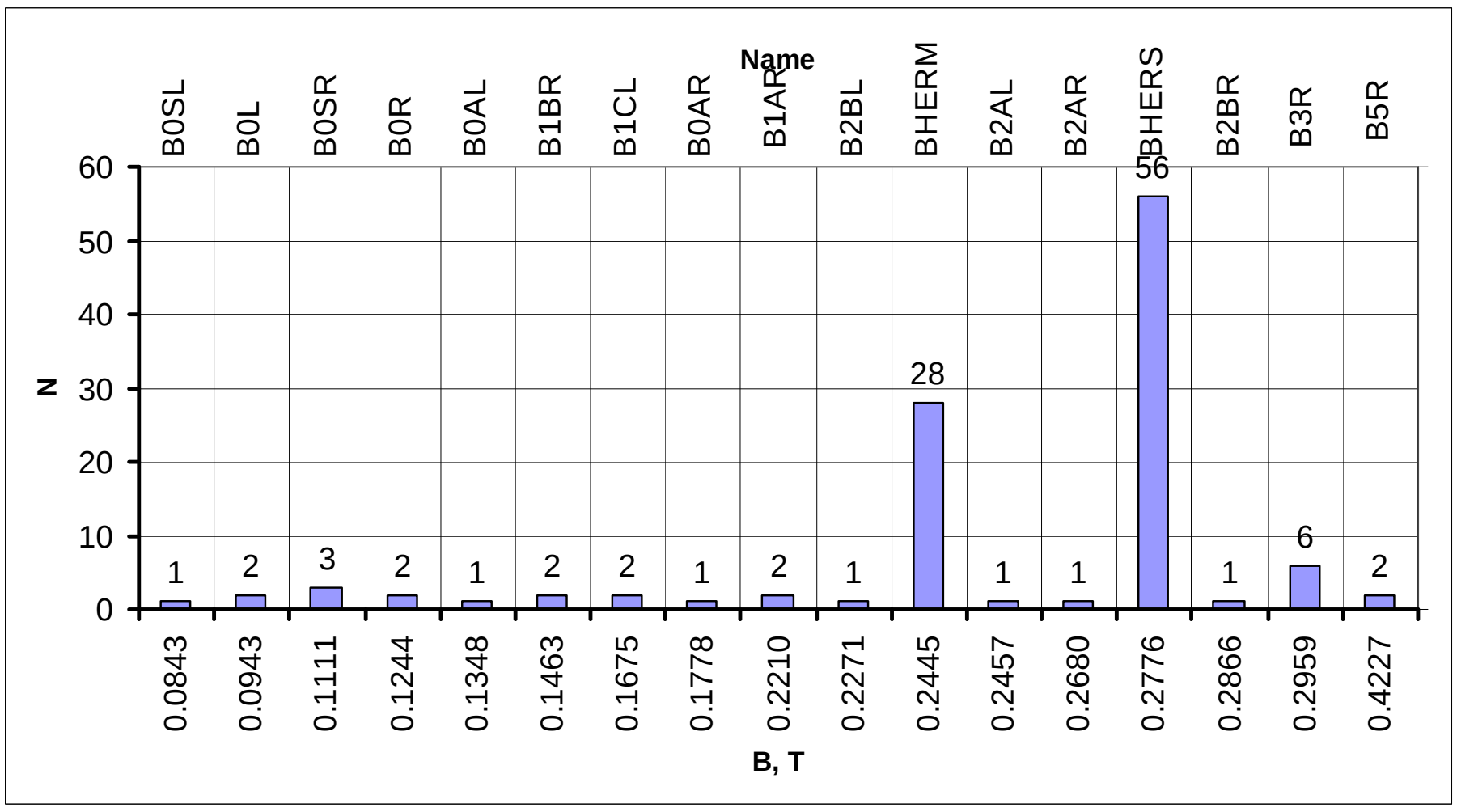

Figure11.8: Number of bending magnet vs. magnet families for LER.

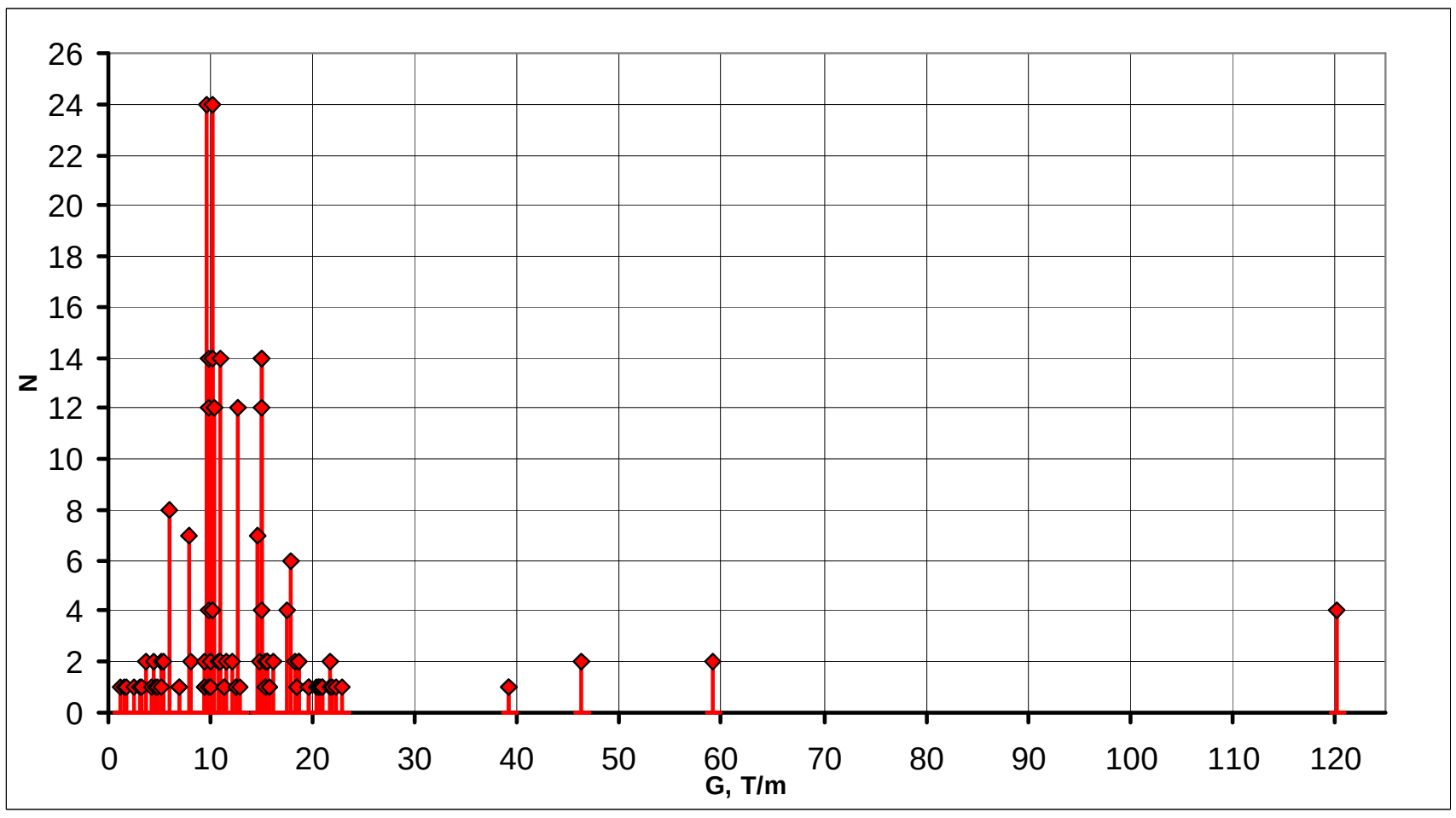

Figure11.9: Number of quadrupole lenses vs. quadrupole families for LER.

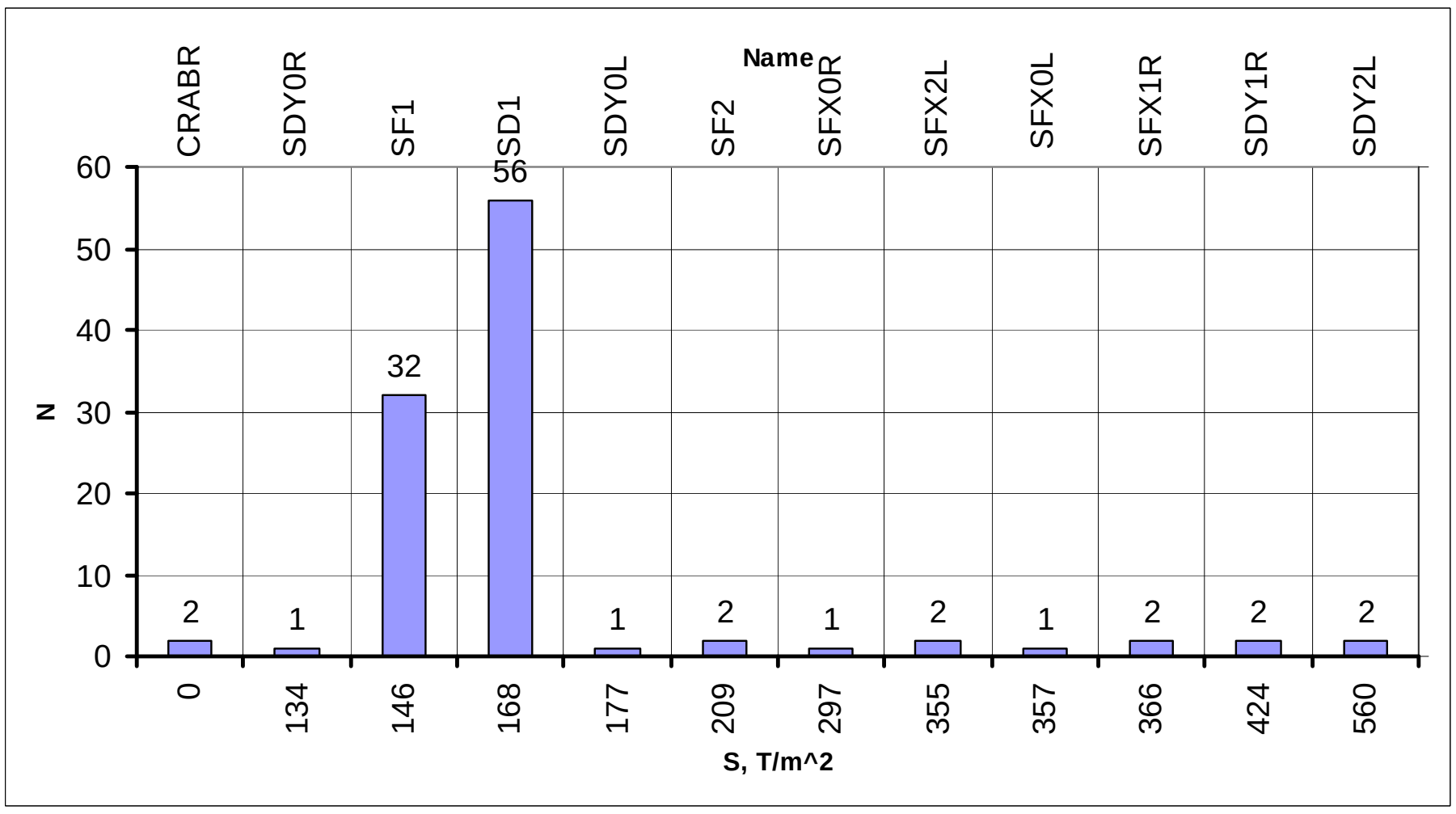

Figure 11.10: Number of sextupole lenses vs. sextupole families for HER.

# 12. RF systems

## 12.1 Super-B parameters and RF specification

The main task of the Super-B RF system is to provide power to the beam necessary to compensate the beam energy loss and to control the longitudinal beam stability in the ring. The main parameters of the machine, which will be used in this chapter, are shown in Table 12.1.

Table 12.1. Main parameters of the machine

| Parameter | Symbol | value | value | Units |
|---|---|---|---|---|
| | | HER | LER | |
| Beam Energy | $E$ | 6.7 | 4.18 | GeV |
| Beam Current | $I$ | 2.12 | 2.12 | A |
| RF frequency | $f_{RF}$ | 476 | 476 | MHz |
| Revolution frequency | $f_{ref}$ | 227 | 227 | kHz |
| Bunch spacing | $\tau_b$ | 4.2 | 4.2 | ns |
| Harmonic number | $h$ | 2100 | 2100 | |
| Number of bunches | $N_b$ | 1018 | 1018 | |
| S.R. Energy loss per turn | $U_{SR}$ | 2.03 | 0.83 | MeV |
| Momentum compaction | α | 4.04 x10$^{-4}$ | 2.24 x10$^{-4}$ | |
| Relative Energy spread | $\delta_E$ | 6.15x10$^{-4}$ | 6.57x10$^{-4}$ | |
| Longitudinal damping time | $\tau_b$ | 14.5 | 22 | ms |
| Total RF Voltage | $V_{RF}$ | 5.7 | 4.1 | MV |
| Longitudinal damping time | $\tau_s$ | 10.8 | 10.8 | ms |

The majority of the beam energy loss comes from the synchrotron radiation in bending magnets. This is mainly incoherent radiation power, which is proportional to the beam current and the fourth power of the beam energy. There is also a small amount of synchrotron radiation loss due to coherent synchrotron radiation (CSR).

The beam also loses energy due to wake fields, which are excited in the beam pipe vacuum elements. Wake fields include short-range fields, like resistive wall and geometrical wake fields, and long-range fields like higher order modes (HOMs) excited in the RF cavities and kickers and possible low-Q geometrical cavities in the beam pipe, for example between in and out tapers.

The power of the wake fields, like power of CSR, is proportional to square of the beam current. Total beam losses are:

$$P_{beam} = \underbrace{U_{S.R.} \times I}_{\text{incoherent radiation}} + \underbrace{Z_{HOMs} \times I^2}_{\text{coherent radiation}}$$

The averaged HOM impedance is proportional to the bunch spacing and the loss factor of the ring:

$$Z_{HOMs} = \tau_b \times K$$

The ring loss factor must not include the loss factor of the cavity main mode. The loss factor strongly depends upon the bunch length. The natural (zero current) bunch length may be calculated using the formula:

$$\sigma_0 = \frac{c}{f_{RF}} \times \delta_E \times \sqrt{\frac{\alpha h}{2\pi} \times \frac{E}{\cos\left(90^\circ - \phi_S\right)}}$$

where the synchronous phase should satisfy the equation:

$$\sin\left(90^\circ - \phi_S\right) = \frac{U_{SR}}{V_{RF}}$$

The synchrotron frequency and synchrotron tune are calculated using these formulas:

$$f_S = f_{RF}\sqrt{\frac{\alpha}{2\pi h} \times \frac{V_{RF}\cos\phi_S}{E}} \qquad \nu_S = h\frac{f_S}{f_{RF}}$$

Values for these parameters and synchrotron loss power, calculated from the ring parameters (Table 12.1) are shown in Table 12.2.

Table 12.2. Other parameters of the machine

| Parameter | Symbol | value | value | Units |
|---|---|---|---|---|
| | | HER | LER | |
| Synchronous phase | $\phi_s$ | 69.1 | 78.3 | degrees |
| Synchrotron frequency | $f_s$ | 2.355 | 2.652 | kHz |
| Synchrotron tune | $\nu_s$ | 0.01033 | 0.001163 | |
| Bunch length | $\sigma_0$ | 5.0 | 5.0 | mm |
| S.R. Power | $P_{S.R.}$ | 4.3 | 1.76 | MW |

There must also be additional power to compensate the main mode Joule losses in the room-temperature cavities. This power is proportional to the square of the total RF voltage and inversely proportional to the shunt impedance of the cavity and the number of cavities:

$$P_{cav} = \frac{V_{RF}^2}{2N_c Z_{sh}}$$

With unmatched conditions, when beam is not perfectly coupled to the cavity, some power will be reflected back from the cavity. We must also include this in the total power consideration. The reflection coefficient can be described by a formula:

$$\Gamma = 1 - \frac{\alpha_{cav}}{1 + \frac{P_{S.R.} + P_{HOM}}{(\beta+1)P_c}}$$

where $\alpha_{cav}$ , $\beta$ are geometrical parameters of a cavity; β represents a coupling coefficient or coupling factor.

The reflected power is proportional to the incident power and reflection coefficient squared:

$$P_{ref} = P_{in}|\Gamma|^2$$

So the total incident power will be the sum of beam power loss, cavity losses and reflected power

$$P_{in} = P_{beam} + P_{cav} + P_{ref}$$

## 12.2 Beam and RF power

The choice of the RF voltage and number of cavities is based on the bunch length, the maximum operational voltage in a cavity and the maximum transmitted and reflected power through a cavity RF window necessary to separate the cavity vacuum from the waveguide. The existing coupling factor may limit the total beam current because of large reflected power with unmatched conditions.

For the Super-B RF system we propose to re-use the main elements of the PEP-II RF system as klystrons, modulators, circulators and cavities with coupling boxes [1-8]. SLAC PEP-II RF operational experience shows that the power limit for each cavity window is 500 kW. Stable operational voltage in one cavity should be limited to 750-800 kV to avoid cavity arcs [9-12]. One klystron may supply power for two cavities. Parameters of a PEP-II cavity are shown in Table 12.3. Detailed information about calculated and measured parameters of the longitudinal and transverse modes of the PEP-II cavity is given in reference [1].

Table 12.3. PEP-II RF cavity parameters.

| Parameter | value | units |
|---|---|---|
| RF frequency | 476 | MHz |
| Shunt impedance | 3.8 | MOhm |
| Unloaded Q | 32000 | |
| R/Q | 118 | Ohm |
| Coupling factor | 3.6 | |
| Maximum incident power | 500 | kW |
| Maximum cavity voltage | 750-900 | kV |

For a given coupling factor we may optimize the transmitted power to the beam. The ratio of the incident power to the beam loss, as a function of a ratio of the beam losses to PEP-II cavity losses, is shown in Fig. 12.1.

With the PEP-II coupling factor, the minimum reflected power is achieved when beam losses are 2.2 times larger than the cavity losses. However, the minimum incident power is achieved with a higher ratio of beam to cavity power (4 to 6). Based on this optimization for the Super-B parameters and taking into account power and voltage limits we can calculate the necessary number of cavities and klystron (stations), and the supply power. For HOM power calculation we use the PEP-II LER impedance.

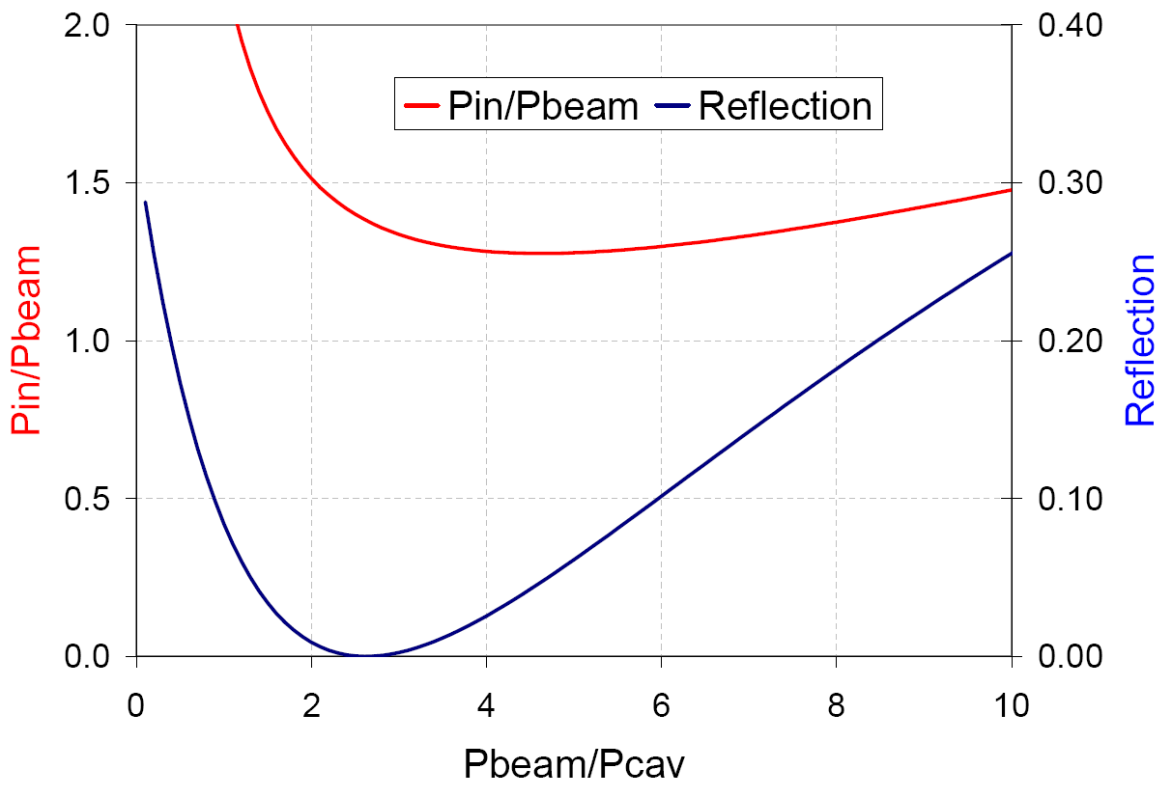

Figure 12.1: Efficiency of the transmitted power to the beam and reflected coefficient squared as a function of ratio of beam losses to cavity losses. Coupling factor β=3.6.

We can change the coupling factor in order to decrease the reflected power and power consumption by modifying only the coupler box of a cavity assembly (Fig 12.2.)

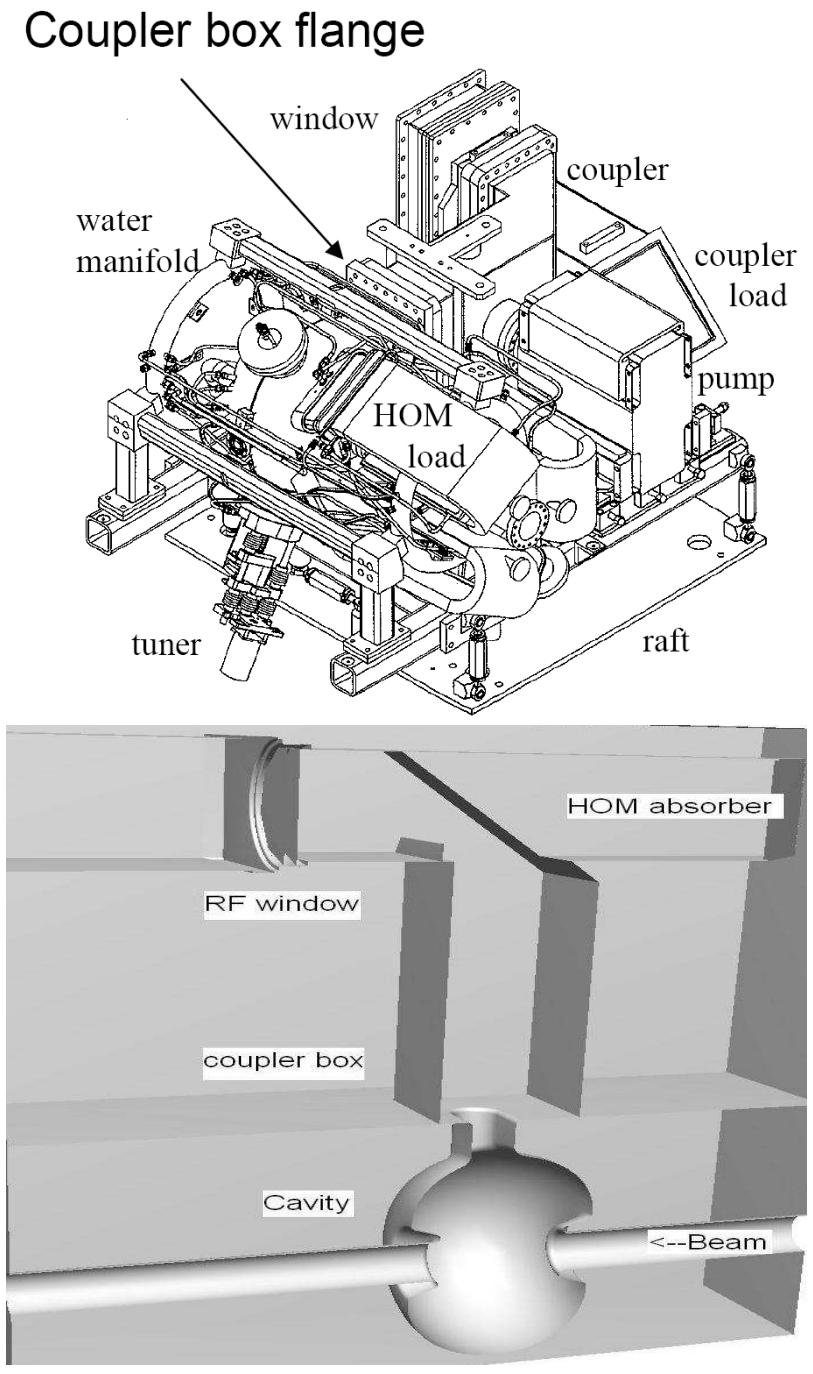

Figure 12.2: PEP-II cavity assembly (top), a cavity, coupler box and RF window (bottom).

We will change the small dimension of the waveguide leading to the coupler slot for 1/4 of a wavelength, forming a quarter-wave transformer. The waveguide impedance varies directly with this dimension. A β up to 6 is achievable without changes to the cavity itself. Since the quality of the match varies only slowly with β, we may optimize for a common coupling factor for all cavities.

RF parameters for the Super-B case are shown in Table 12.4. We assume that klystrons have 50% efficiency.

Table 12.4. RF parameters (coupling factor 6.0)

| | | | | | | | | | | | | |
|---|---|---|---|---|---|---|---|---|---|---|---|---|
| HER | HER | HER | HER | HER | HER | HER | HER | HER | HER | HER | HER | HER+ |
| Total | Zero I | | Max | Number | | | Total | Total | Total | Power fo | reflected | LER |
| RF | Bunch | Bunch | voltage | of | S.R. | HOM | cavity | reflectec | forward | to one | from | Total |
| voltage | length | spacing | er cavit | cavities | power | power | loss | power | power | cavity | one | forward |
| MV | mm | nsec | MV | klystro | MW | MW | MW | MW | MW | MW | MW | MW |
| | | | | | | | | | | | | |
| 5.78 | 5.00 | 4.20 | 0.51 | 12.00 | 4.30 | 0.45 | 0.37 | 1.30 | 6.41 | 0.53 | 0.11 | 9.03 |
| | | | | 6.00 | | | | | | | | |
| | | | | | | | | | | | | HER+ |
| LER | LER | LER | LER | LER | LER | LER | LER | LER | LER | LER | LER | LER |
| Total | Zero I | | Max | Number | | | Total | Total | Total | Power fo | reflected | Plug |
| RF | Bunch | Bunch | voltage | of | S.R. | HOM | cavity | reflectec | forward | to one | from | Power |
| voltage | length | spacing | er cavit | cavities | power | power | loss | power | power | cavity | one | eff.~50% |
| MV | mm | nsec | MV | klystro | MW | MW | MW | MW | MW | MW | MW | MW |
| | | | | | | | | | | | | |
| 4.10 | 5.01 | 4.20 | 0.58 | 8.00 | 1.76 | 0.40 | 0.28 | 0.17 | 2.61 | 0.33 | 0.02 | 18.05 |
| | | | | 4.00 | | | | | | | | |

We consider using 6 stations for HER and 4 stations for LER. We may install one spare station in each ring, assuming that the impedance of the four detuned cavities will not bring instability problems. The klystrons required, plus several spares, exist at SLAC, although more klystrons may eventually be built to replenish the supply as tubes age.

## 12.3 Gap transient and frequency detuning

The existence of an ion-clearing gap in the electron bunch train causes a change in cavity voltage and phase along the bunch train. The cavity voltage change further causes a change in synchronous phase of the electron bunches. The result is a turn-by-turn ``phase transient'' or ``gap transient''. The phases of the electron bunches will be modulated at the revolution harmonics causing the bunch phases to vary in a quasi-sawtooth fashion along the bunch train.

The $\beta_y$* at the SuperB IP is significantly shorter than the bunch length, and the beams cross with a non-zero angle. As a result, the HER and LER bunches must overlap exactly at the IP or the luminosity will suffer. For a 5 mm 1σ bunch length, a 1.0 mm relative shift in z-position (corresponding to about 0.5 degrees of RF phase) between the HER and LER bunches will reduce luminosity by about 1%. The HER and LER phase transients must match to about 0.5 degrees RMS to avoid more than a 1% reduction in luminosity. A large mismatch in phase transients will also cause stability problems due to tune shifts along the bunch train.

The magnitude and shape of the phase transients are functions of cavity beam loading and synchronous phase, which are functions of the number of cavities and the beam currents. With fixed (equal) beam currents in the HER and LER, the number of LER cavities can be adjusted to approximately match the phase transients, yielding the phase transients shown in Fig. 12.3. The RMS phase error is only 0.2 degrees, resulting in a negligible luminosity reduction.

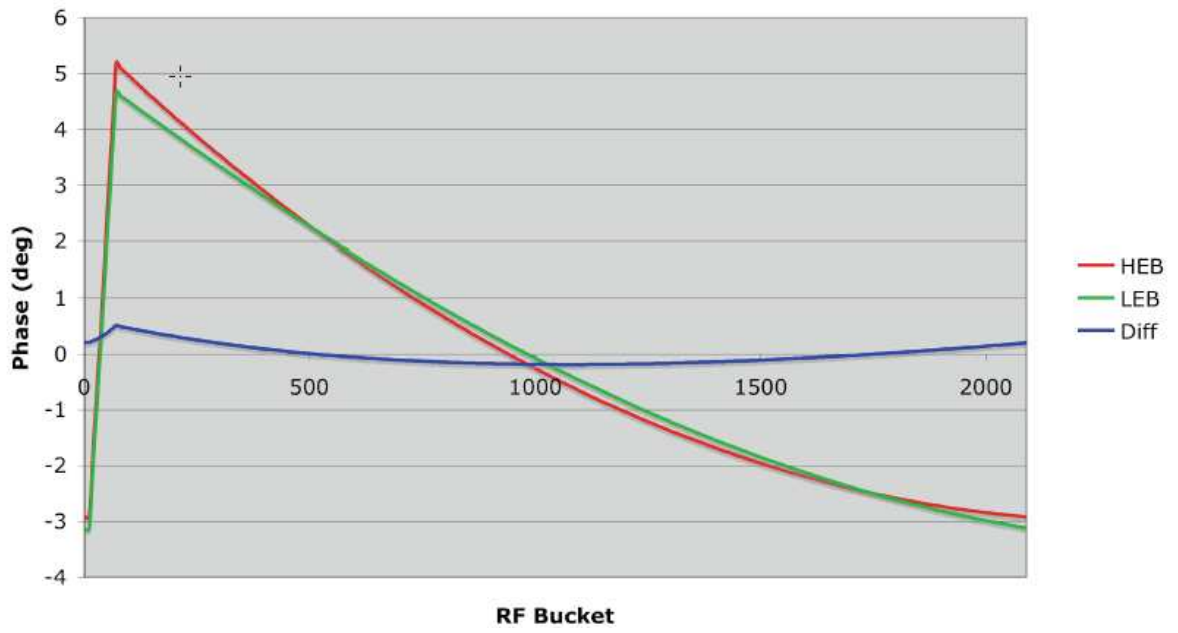

Figure 12.3: Phase transient in HER and LER and difference. The RMS phase error is 0.2 degree.

To avoid resonant instability at the main frequency (i.e. to compensate for beam loading), the RF cavities must be detuned from resonance according to the following formula

$$\delta f = -f_{RF} \times \frac{Z_{sh}}{Q} \times \frac{I}{V_{RF}} N_C$$

For the Super-B parameters the detuning in HER is 252 kHz and 233 kHz in LER. These numbers are near the revolution frequency (227 kHz). The feedback system must be designed to damp this -1 mode.

We can check beam stability for higher order modes using the same approach as in reference [14]. HOM cavity impedance must be less than the stability threshold defined by the beam and ring parameters including radiation damping time

$$Z_{th}(\omega) = \frac{4\pi E \nu_s}{\alpha \tau_s N_c I \omega}$$

Fig. 12.4 shows the impedance of a PEP-II cavity [1] and the threshold for super-B LER and HER beams. The left peak is the fundamental mode at 476 MHz. For comparison we present also the thresholds achieved at PEP-II, which are several times lower. If we use PEP-II type feedback systems we may increase the beam currents at Super-B several times.

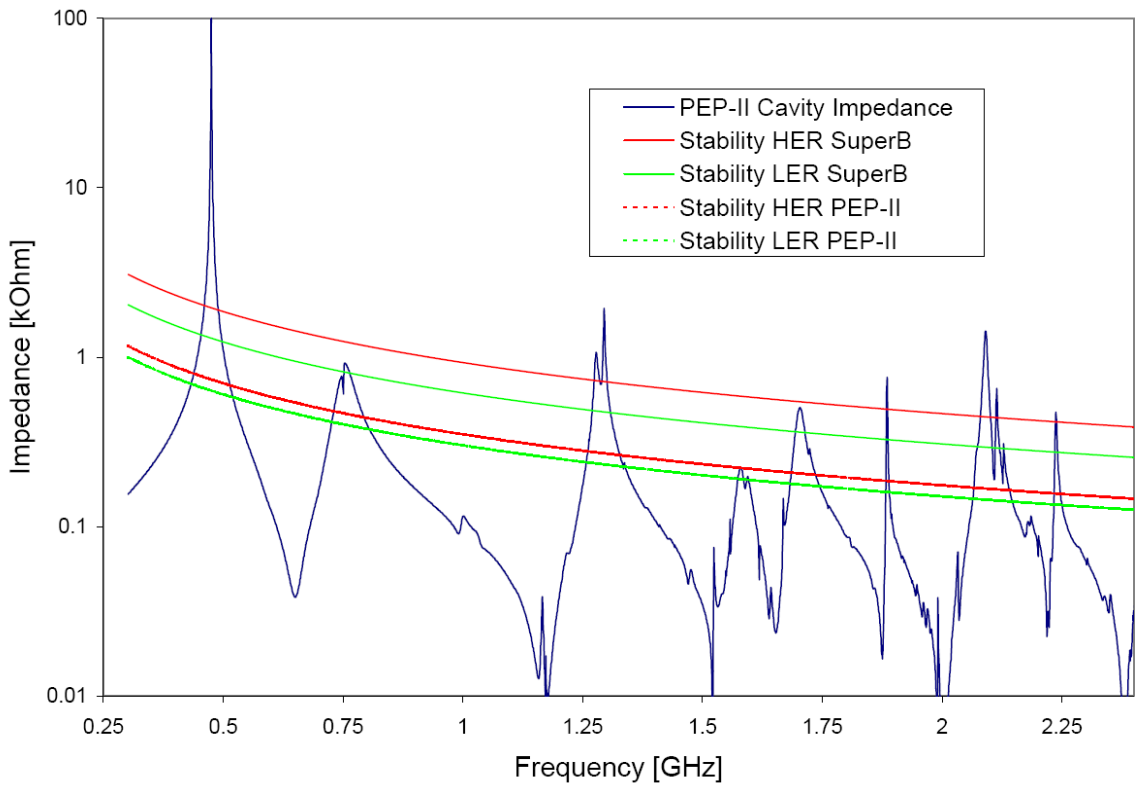

Figure 12.4: Impedance of a PEP-II cavity (blue line) and thresholds for Super-B HER (red upper line) and LER (green upper line) and for the PEP-II HER and LER rings (down lines).

Increasing the currents we need to increase the number of cavities, increase the coupling factor and total voltage. Plots for number of cavities and voltage as functions of the beam current are shown in Fig. 12.5. We assume that HER and LER have the same currents.

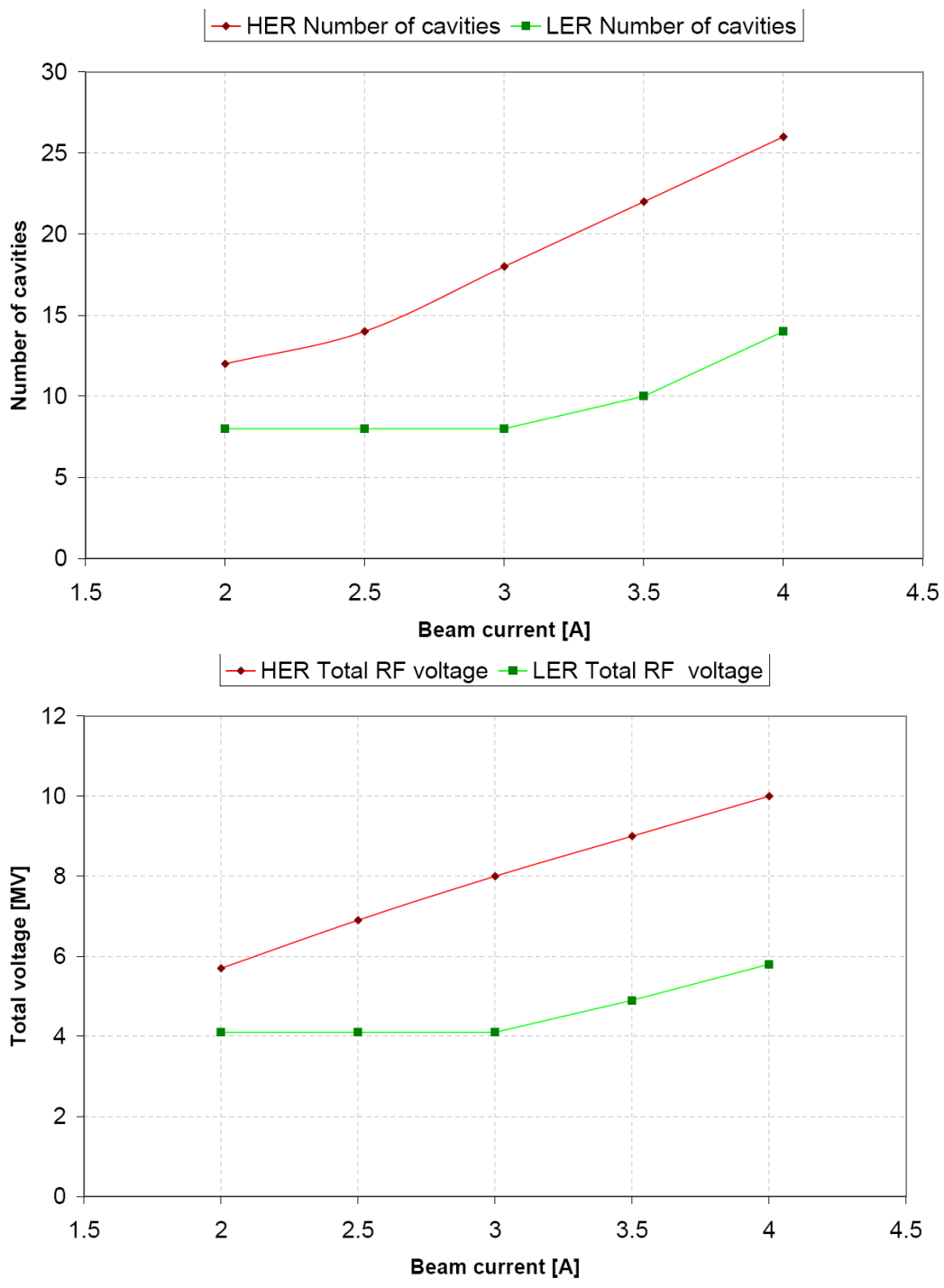

Figure 12.5: Number of cavities and voltage as functions of the beam current.

The needed wall plug power is shown in Fig. 12.6.

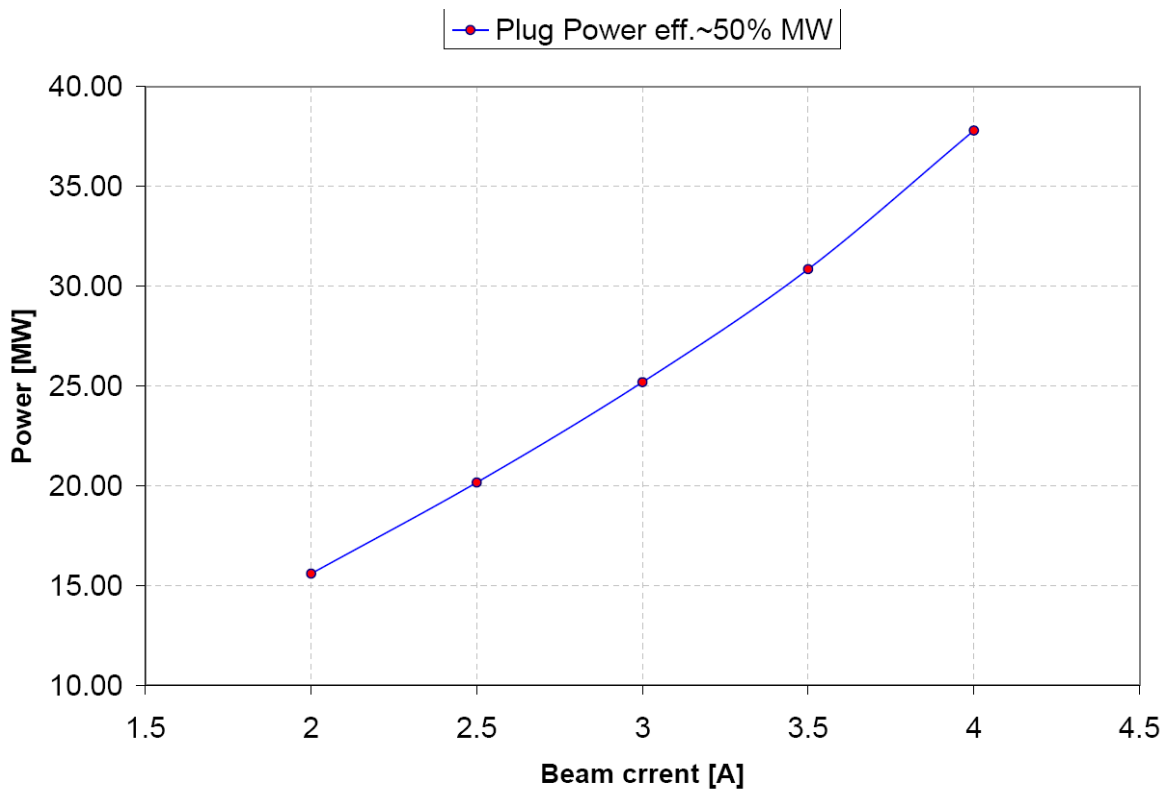

Figure 12.6: Wall plug power as a function of the beam current.

## 12.4 RF environment

The RF stations are located in the support building (Fig. 12.7). Each station consists of a 2 MW (90 kV, 23 A) high voltage power supply (HVPS), a 1.2 MW klystron amplifier with a high-power circulator for protection of the klystron from reflected power, a power splitter (Magic-tee with a 1.2 MW RF load), followed by waveguide distribution system from surface level down to the tunnel ending in two cavities. The RF distribution is via WR2100 waveguide, chosen primarily for low group delay. Each cavity has three HOM loads. For safety these loads were specified for up to 10 kW dissipation each [15-16].

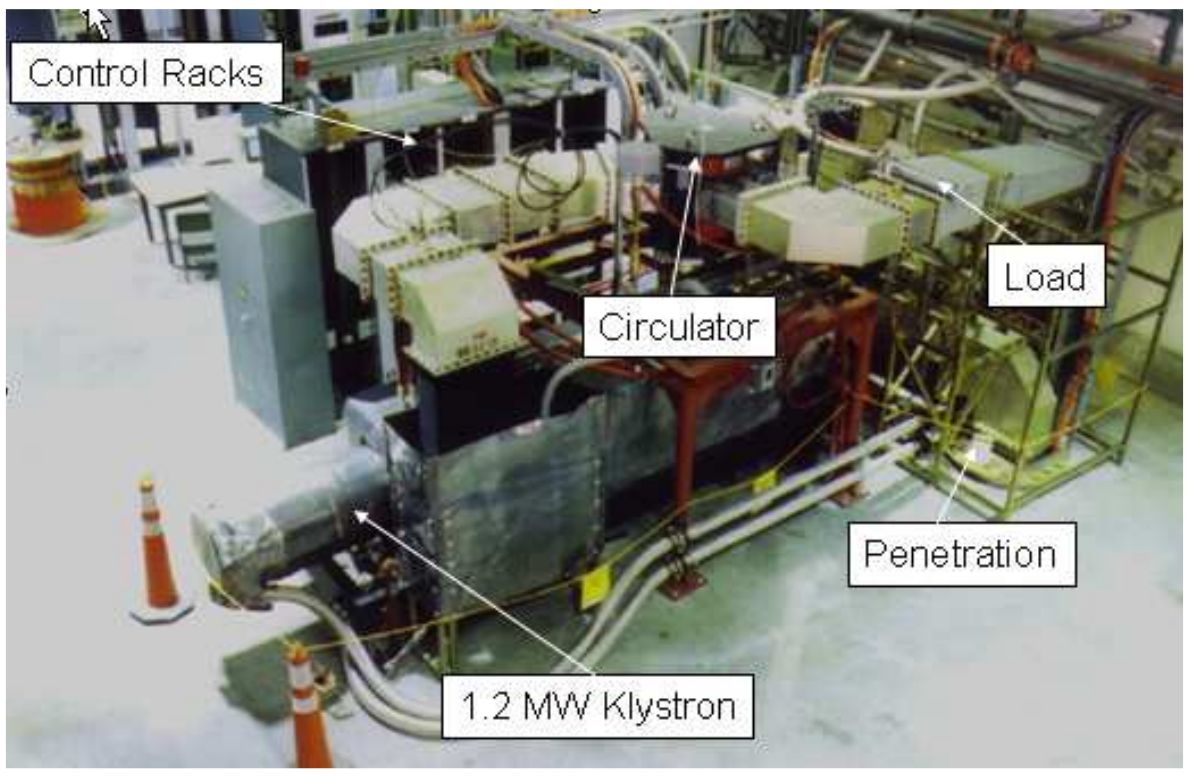

Figure 12.7: PEP-II HER RF station 12-3

## 12.5 Low Level RF System

A low-level RF system provides control and feedback for stable multi-bunch high current operation. There are several feedback loops [17-20] (see Fig. 12.8).

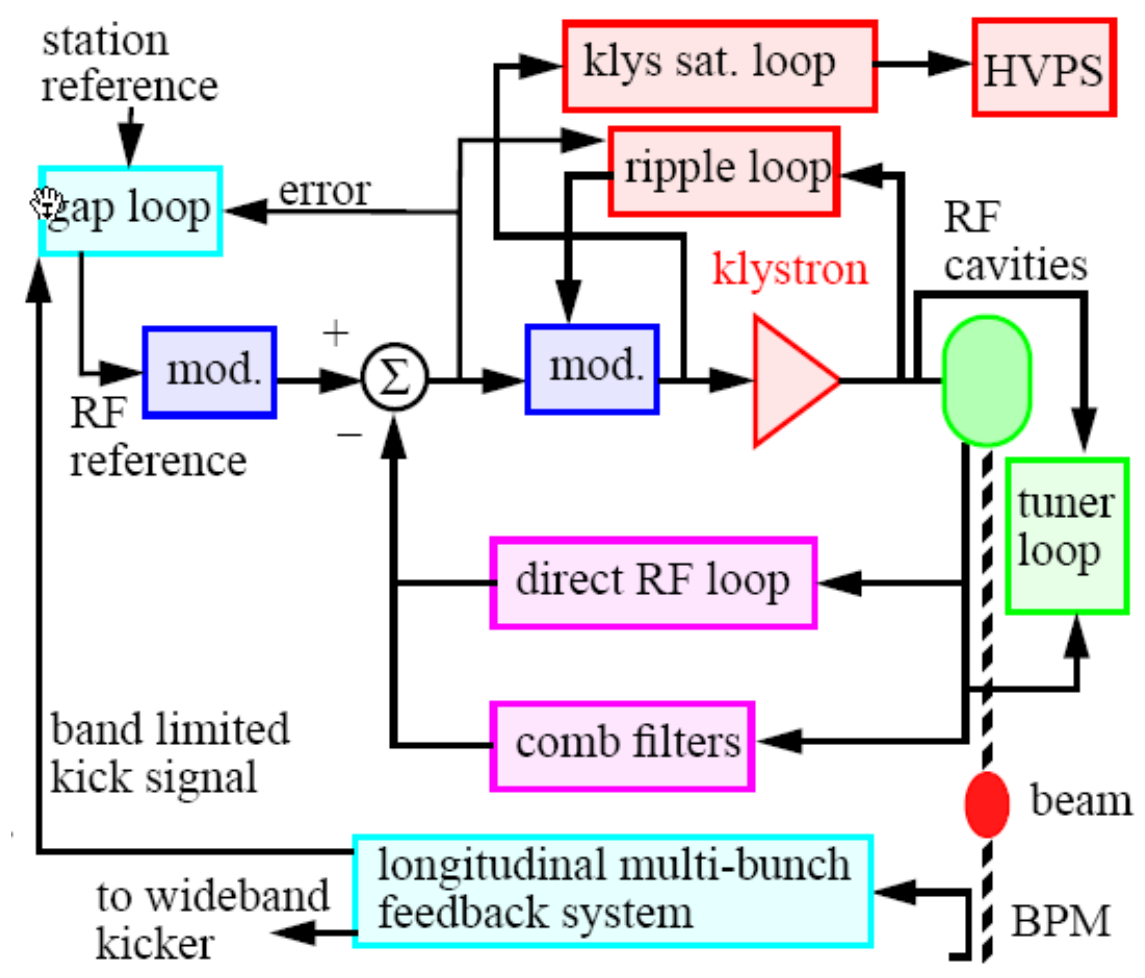

Figure 12.8: Block diagram of RF feedback circuits

The direct loop is required for lowering the cavity impedance to reduce multi-bunch oscillations of the beam. Functionally the direct loop keeps the gap voltage constant as set by a DAC reference over an 800 kHz bandwidth. The loop compares the combined baseband field signals of a station's cavities to the reference generated by the gap module. The resultant error signal is up-converted to RF and drives the klystron. The direct loop contains a PID controller with an integral compensation for smoothing out the ripple caused by the klystron high voltage power supply and lead compensation that increases the bandwidth and gain of the loop. The direct feedback loop options control the optional functions of the direct loop: frequency offset tracking, integral compensation and lead compensation. The frequency offset tracking loop takes out the phase shift caused by detuning of the cavities during heavy beam loading. It is used as a diagnostic for adjusting the waveguide network. The comb loop provides additional impedance reduction for the cavities at specific synchrotron frequency sidebands around the revolution harmonics of the beam. It operates over a bandwidth of 2 MHz and includes a 1 turn delay. The tuner loop tunes and maintains each cavity at resonance. It corrects for thermal frequency variations and compensates cavity beam loading by keeping the phase relationship between forward power and cavity field, as seen by the cavity probe, constant. The relevant phases are measured by digital IQ detectors and the loop is completed in software controlling the tuner position via a stepping motor. The HVPS loop adjusts the voltage to the klystron to provide sufficient output power to operate the station under whatever gap voltage or beam loading is requested. Functionally the loop keeps the klystron operating at about 10% below saturated output power. The loop measures the drive power at the input to the klystron and compares it to the ON CW drive power set-point. Based on the error the set-point for the high voltage power supply is adjusted up for excessive drive and down for insufficient drive. This is a slow loop with about a 1 Hz bandwidth. The DAC loop is a slow (0.1 Hz bandwidth) loop in software which functionally keeps the measured gap voltage of the station equal to it's requested "Station Gap Voltage" by adjusting the DAC in the gap voltage feed-forward module. The ripple loop is intended to remove amplitude and phase ripple in the klystron output power but at the time it is only utilized to keep the low bandwidth phase across the klystron and drive amplifier constant as the klystron voltage is varied. The gap feed forward loop is required to tell the direct loop to ignore the effects of the ion-clearing gap in the beam bunch train. Functionally the loop learns about the variation in the klystron drive caused by the beam gap and adds an equal variation in the reference signal so that the error signal driving the klystron stays unchanged. This loop adapts fully in about 1000 beam revolutions. The longitudinal feedback woofer is the third cavity impedance reduction loop along with the direct loop and the comb loop. It derives its information from the lowest beam oscillation modes detected by the longitudinal bunch-by-bunch feedback system and uses one RF station in each ring as a powerful longitudinal kicker.

## 12.6 Synchronization and timing

The goal of the synchronization and timing system is to assure that all the RF systems and the other timed devices will be able to work with signal and frequencies locked in phase within the ranges defined by the specifications. A master sinusoidal oscillator at the RF frequency (476 MHz) including a phase continuity feature must be considered, and it must be able to provide a $10^{-11}$ short term stability. Small change of frequency in a range <100 KHz (by steps of 1 or 5 kHz) must be accepted without loose of signal phase. The distribution of the RF main signal must be assured with a peak-peak jitter < 0.5 ps. Very low jitter phase shifters must be implemented to synchronize, separately for each ring, beam collisions and bunch injections. The synchronization and timing system must also provide sinusoidal frequencies for the LINAC cavities, typically 6 and/or 12 times the main RF sinusoidal signal. Generation of other (m/n)*RF frequencies, with m and n integer, could be considered if necessary. The utmost peak-peak jitter for these devices can be within 2 ps. The injection triggers have to be locked to the main RF frequency and to the 50 Hz of the main power supplies. Diagnostics and injection triggers must include at least the "Fiducial" (a reference revolution frequency given by main RF frequency divided by the harmonic number) and bunch number triggers, all locked in phase with the RF main frequency within a 2 ps peak-peak jitter.

# 13. Vacuum systems

## 13.1 Introduction

The main constraints on the ultimate performances of the vacuum system are given by the cells layout of the LER and HER and from the geometry of the vacuum chamber. Due to the charge exchange in SuperB rings with respect to PEP-II, we will assume to use for SuperB HER the PEP-II LER chambers and vice versa. The geometrical configuration imposes the pumping ports and so the static vacuum level. At the same time the magnetic strength and the curvature radius of the dipoles will impact on the dynamical vacuum due to the synchrotron radiation emission.

Pressure simulations have been performed on the LER and HER cells. The simulation program used is Vacuum Stability Code VASCO since it can take into account the ions cross out gassing of one gas for different ions species. This implies that the equation for one gas density is linked to the other gas densities.

The vacuum system is split into different segments. Each segment is characterised by different set of parameters (uniform out gassing rate, conductance, gas flow and holding pump at each extremity, uniform distributed pumping, uniform ion stimulated desorption, uniform photon stimulated desorption, uniform electron stimulated desorption, etc…).

As far as the dynamical vacuum simulations are concerned, at present, we will take into account only the synchrotron radiation induced degassing. So the other effects like ions or electron desorption is not considered.

## 13.2 Static vacuum in HER

### Basic vacuum parameters under static condition

The schematic magnetic layout of one HER Arc cell is shown in Figure 13.1. In a first approximation, we apply the geometry of HER (arc) quadrupole chamber to all the chambers. The linear conductance is considered approximately at 43 l m/s for the CO at 20°C and the perimeter at 54 mm (see Figure 13.2, PEP-II LER dipole chamber used for SuperB HER).

The simulation has been performed with 4 gases ($H_2$, $CH_4$, CO, $CO_2$ ) because those are main gases in a baked installation.

We consider for each segment that the vacuum chamber is in copper (OFE) or copper plated. The out gassing rate of baked copper in situ has been taken at $1.33x10^{-12}$ mbar l $s^{-1}$ $cm^{-2}$ for $H_2$ (20°C), at $6.65x10^{-15}$ mbar l $s^{-1}cm^{-2}$ for $CH_4$ (20°C), at $1.33.10^{-14}$ mbar.l.$s^{-1}$.$cm^{-2}$ for CO (20°C) and $6.65.10^{-15}$ mbar.l.$s^{-1}$.$cm^{-2}$ for $CO_2$ (20°C) [1]. The holding pumps are situated on both dipoles sides (Figure 13.3), in first approximation we have taken as effective pumping speed 60 l/s (20°C) for un-saturated "starcell" ion pump and for all gases. At each extremity of HER Cell, we took a half holding pumping.

In these conditions, a first simulation has been performed (see Figure 13.4 left). Most of the present gas is the hydrogen with an average pressure of $3x10^{-10}$ mbar. The pressure of other gases is about two orders of magnitudes smaller.

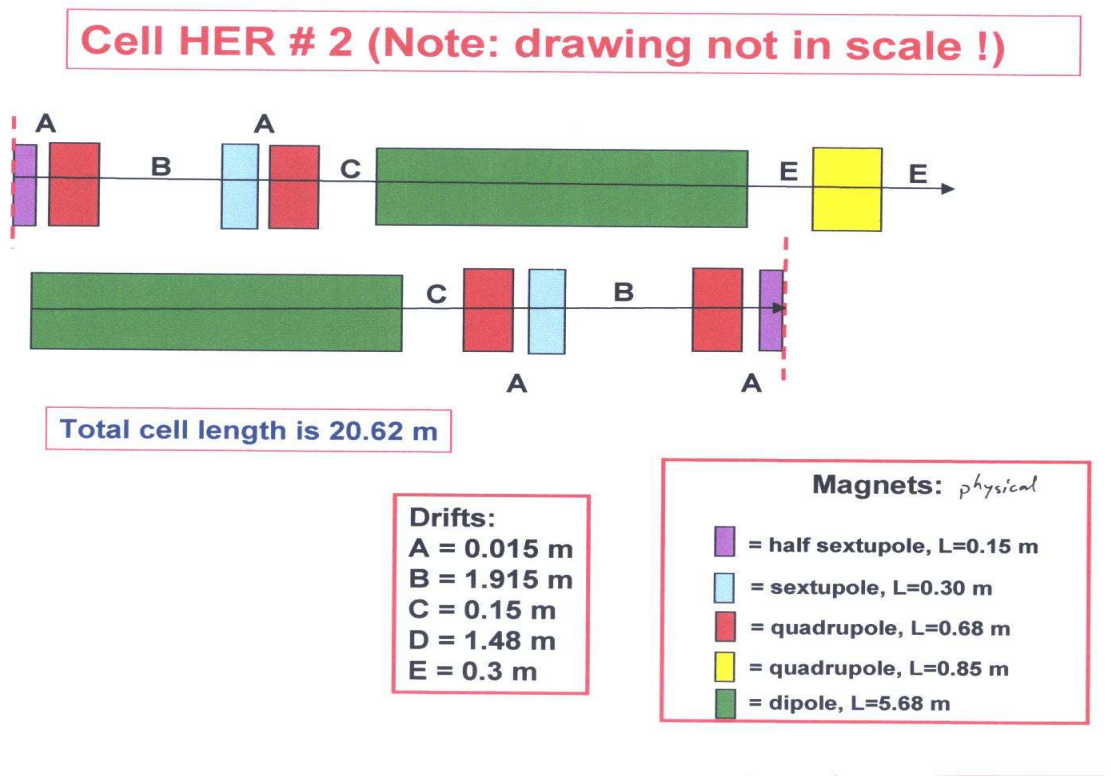

Figure 13.1: Magnetic layout of HER cell #2

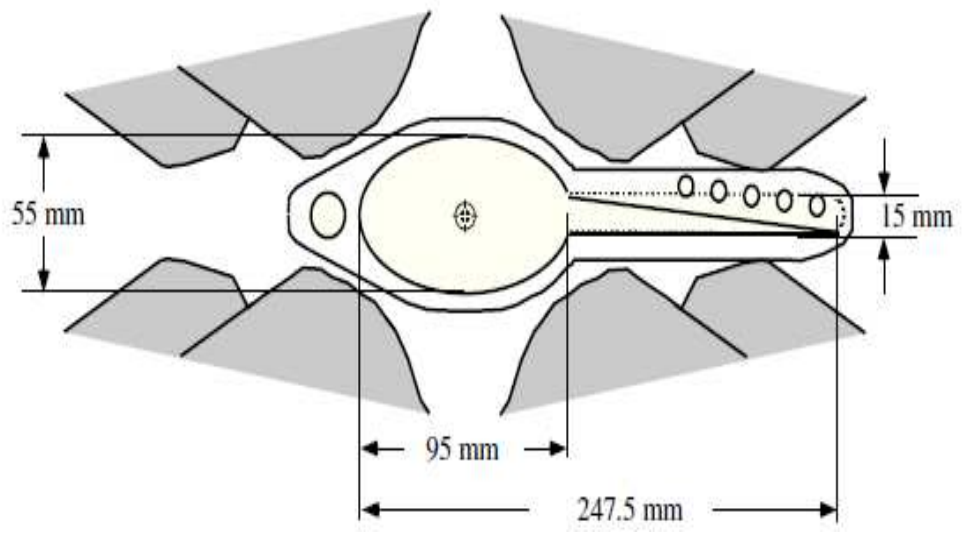

Figure 13.2: PEP-II LER Arc quadrupole vacuum chamber. Shaded part is the envelope of magnet poles and coils.

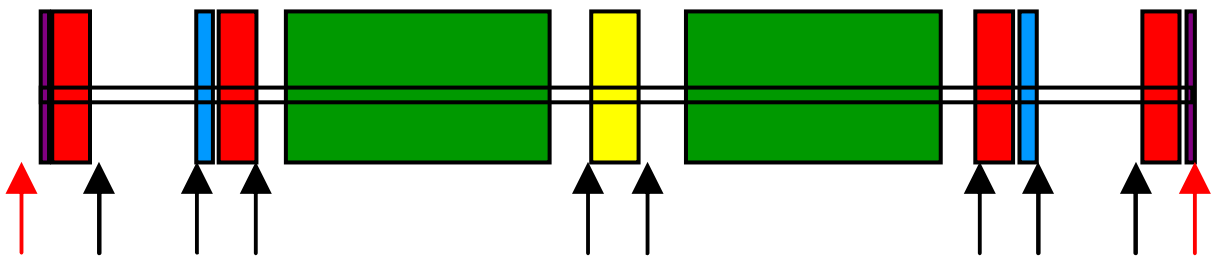

Figure 13.3: Positions of holding pumps in HER cell (black arrows) and half holding pump (red arrows).

### Simulation with distributed pumping HER

We have simulated the pressure distribution with NEG strip on dipoles and drift sections antechamber, not on the quadrupoles and the sextupoles chambers. This pumping adds to the holding pumping. According to CDR data [2], the linear pumping speed has been taken at 200 l $s^{-1}$ $m^{-1}$. We considered that this pumping speed was given for the CO and $CO_2$ and respectively for the $H_2$ and the $CH_4$ we set as speed 40 l/s/m and 0 (NEG doesn't pump $CH_4$). In this case, most of gas is the hydrogen with an average pressure of $1.2x10^{-10}$ mbar. With distributed pumping, the average pressure decreased by a factor of 2.5 (Figure 13.4 right). The $CH_4$ becomes the second majority gas with an average pressure of $2x10^{-12}$ mbar.

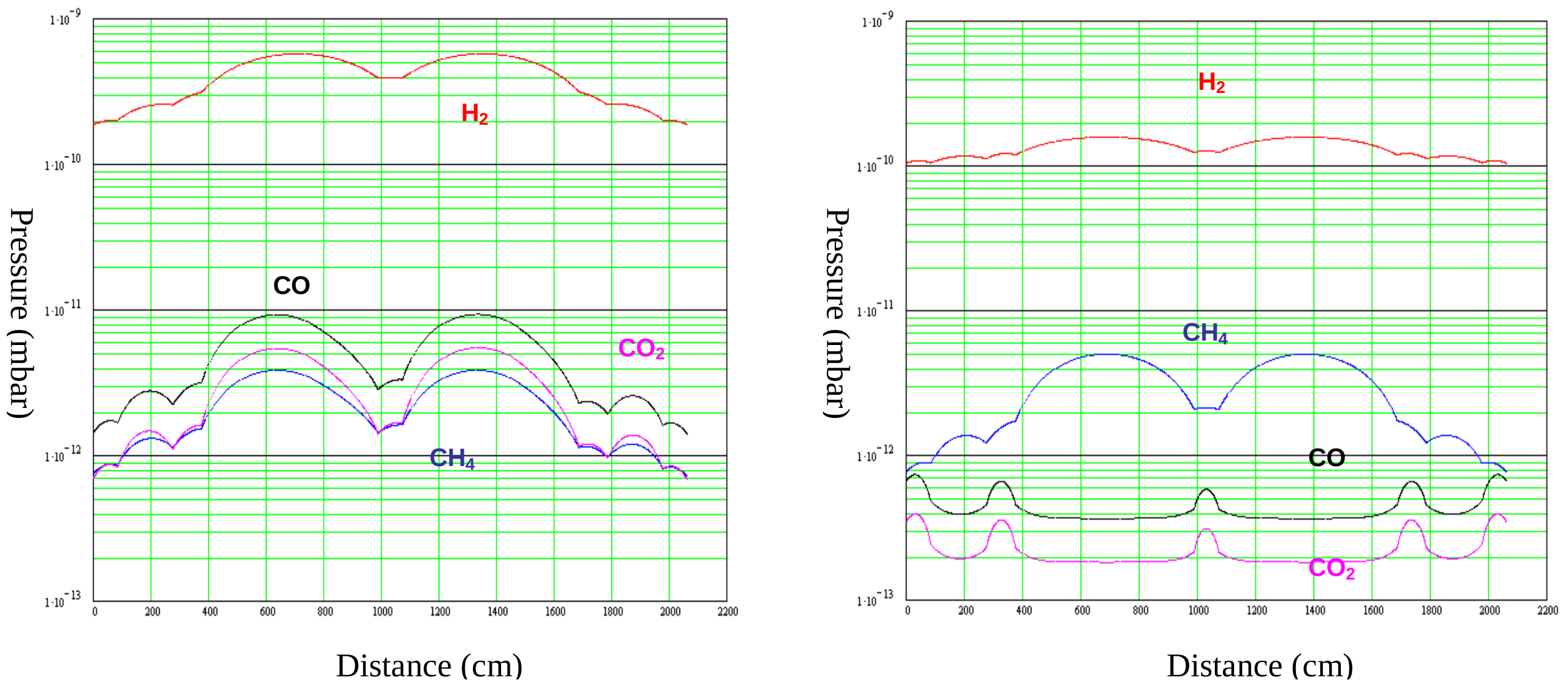

Figure 13.4: Pressure distribution in HER without (left) and with (right) distributed pumping.

## 13.3 Static vacuum in LER

### Basic vacuum parameters under static condition

In a first approximation, the geometry of LER (arc) quadrupole chamber to all the chambers is applied. The linear conductance is considered approximately at 35 l m/s for the CO and at 20°C and the perimeter at 28 mm. (Figures 13.5 and 13.6).

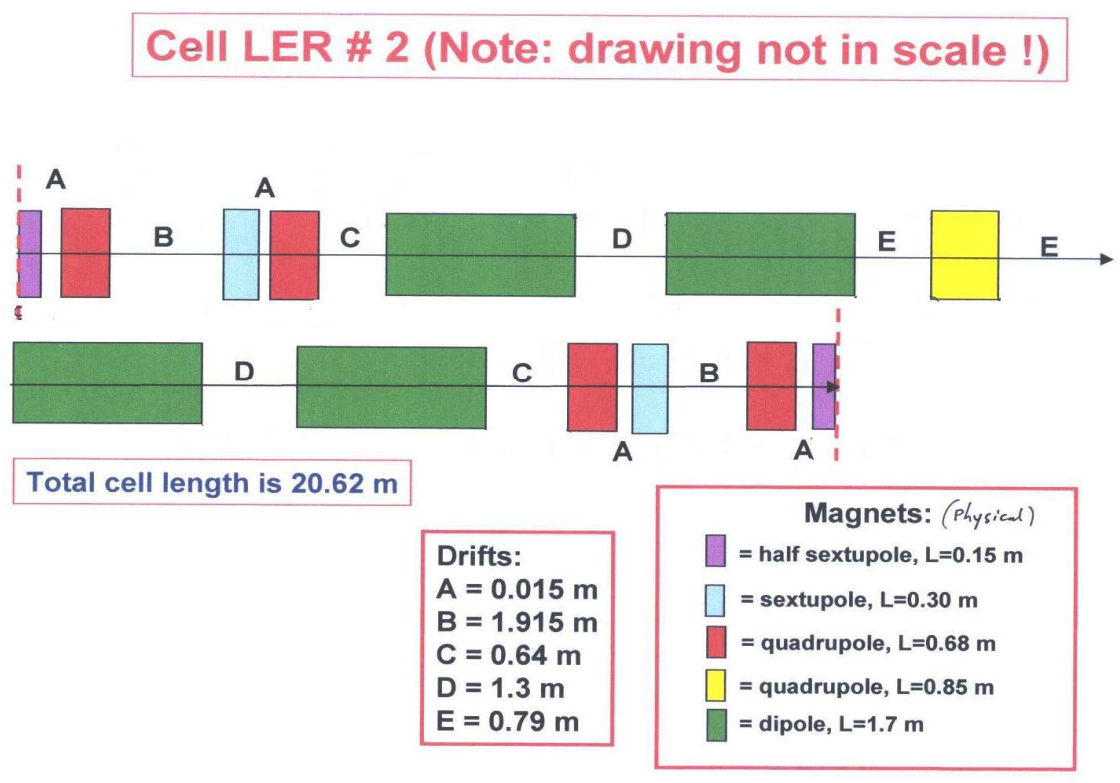

Figure 13.5: Magnetic layout of LER cell #2

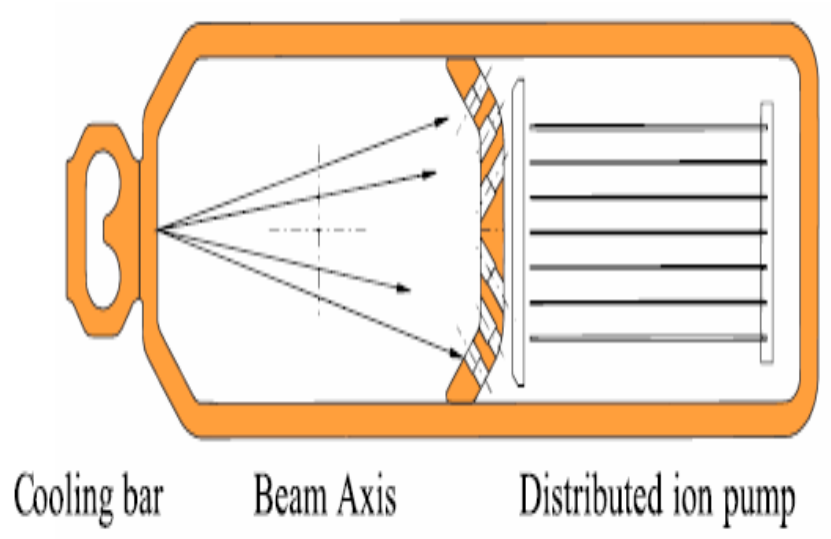

Figure 13.6: Cross section of PEP-II HER dipole vacuum chamber (used for SuperB LER).

The simulation has been performed with the same four gases of the LER ($H_2$, $CH_4$, CO, $CO_2$ ).Also in this case the vacuum chamber is in copper (OFE) or copper plated, with the same out gassing rates. The holding pumps are situated on both sides of the dipoles (Figure 13.7), in first approximation we have taken as effective pumping speed 60 l/s (20°C) for unsaturated starcell ion pump and for all gases. At each extremity of Cell HER, we took a half holding pumping.

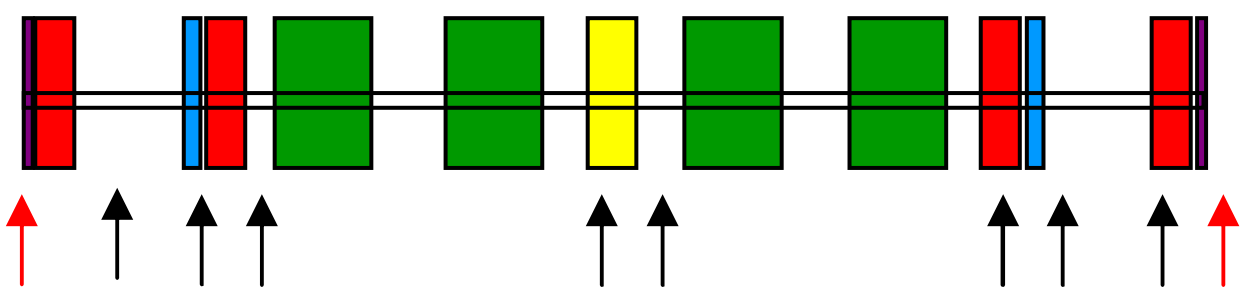

Figure 13.7: Positions of holding pumps in LER cell (black arrows) and half holding pump (red arrows).

In these conditions, a first simulation has been performed (Figure 13.8 left). The majority gas is the hydrogen with an average pressure of 2x$10^{-10}$ mbar. The pressure of other gases is about two orders of magnitudes smaller.

### Simulation with distributed pumping LER

The pressure distribution with NEG strip on drift sections antechamber (not on quadrupoles and sextupoles chambers) and ion pump on dipole antechamber has been evaluated. This pumping adds to the holding pumping. Also in this case the linear pumping speed for the NEG strips has been assumed to 150 l $s^{-1}$ $m^{-1}$ for the CO and $CO_2$ and respectively 30 l/s /m and 0 for the $H_2$. The distributed pumping inside dipole antechamber is an ionic pumping of 120 l/s/m for all the gases

In this case, the majority gas is the hydrogen with an average pressure of 5x$10^{-11}$ mbar (Figure 13.8 right).

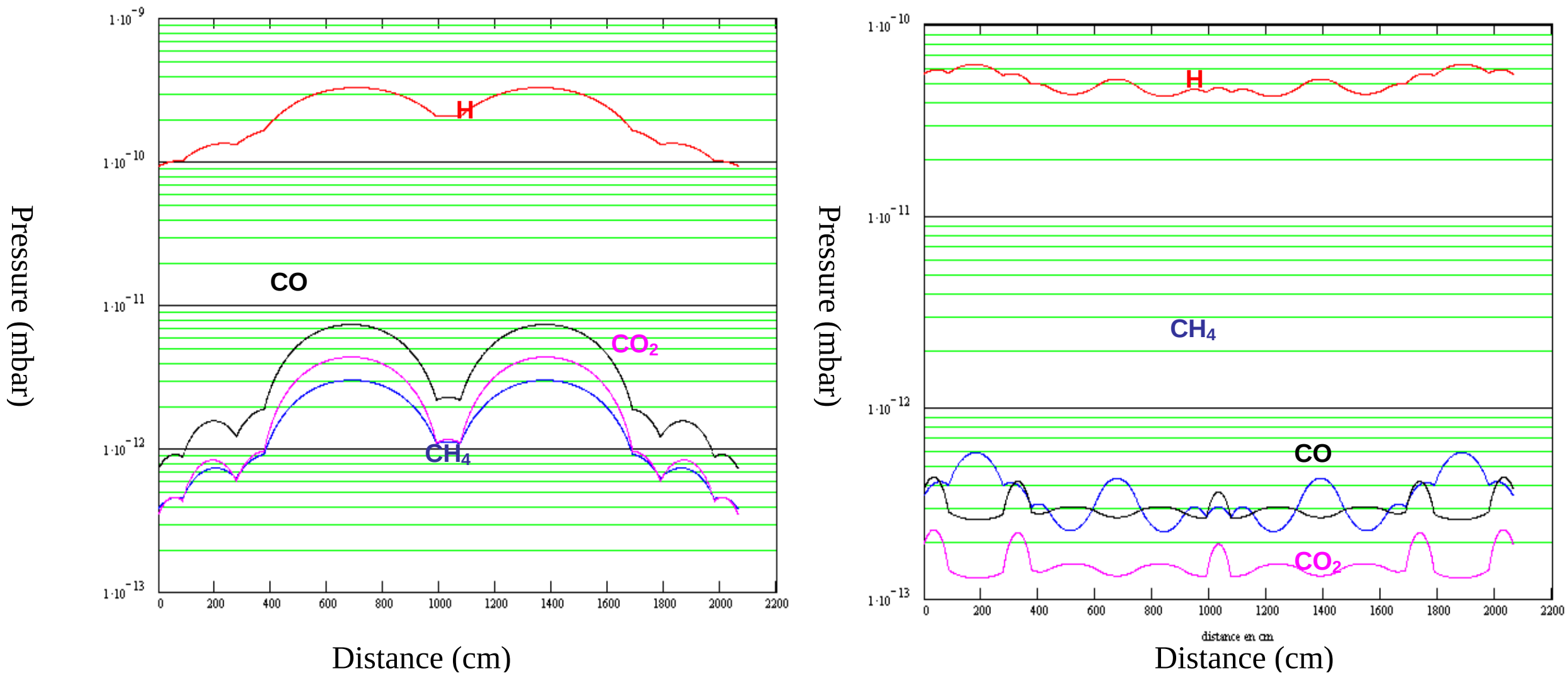

Figure 13.8: Pressure distribution in HER without (left) and with (right) distributed pumping.

## 13.4 Dynamic vacuum in LER and HER

The previous estimations and boundary conditions have been taken into account to estimate the equilibrium vacuum in dynamical regime, with the presence of the beam. In this case the important rate of synchrotron radiation emission affect the out gassing rate and so the ultimate vacuum level.

It was previously estimated that, under static vacuum, the majority gas is the hydrogen with an average pressure of $5 \times 10^{-11}$ mbar. The pressure of other gases is about two orders of magnitudes smaller. The lower contribution to synchrotron radiation is given in the HER with respect to the LER.

### Dynamic vacuum in HER

Simulations input for the HER: the synchrotron radiation flux from the HER dipoles (ρ=148.97 m), at 6.7 GeV and with a beam average current of 1892A, is $\Gamma = 1.1 \times 10^{19}$ ph/s/m with $E_c \sim 4.5$ keV and P = 2.4 kW/m.

The photo-desorption rate after conditioning has been obtained from the PEP-II results: $10^{-7}$ molecule/photon for $H_2$, CO and $CO_2$, and $10^{-8}$ for $CH_4$.

In Figure 13.9 the emitted photon flux in the cell arc (left) and the resulting vacuum pressure (right) are illustrated. In these conditions a Hydrogen equilibrium pressure of $5 \times 10^{-10}$ mbar is estimated and $1.7/1.2 \times 10^{-10}$ mbar for the CO or $CO_2$ and $CH_4$. The total pressure is close to $10^{-9}$ mbar.

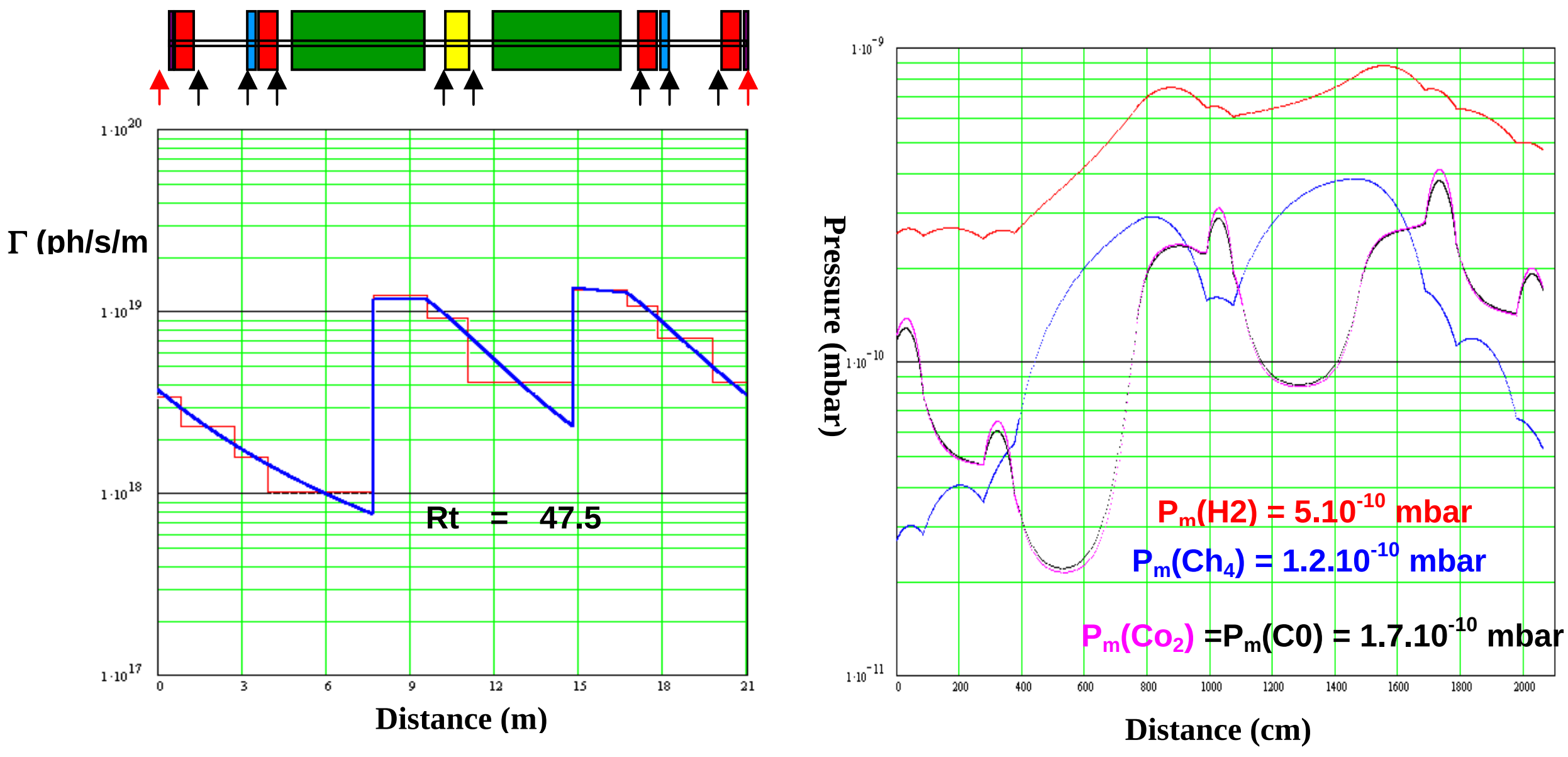

Figure 13.9: Emitted photon flux in the HER cell arc (left) and the resulting vacuum pressure (right).

**Dynamic vacuum in LER**

In the LER the situation is more difficult due to the short dipoles and their curvature radius. The synchrotron radiation flux from the LER dipoles (ρ=29.8 m), at 4.18 GeV and with a beam average current of 2447A, is $\Gamma = 4.42\text{x}10^{19}$ ph/s/m with $E_c \sim 5.45$ keV and P = 11.8 kW/m.

The photo-desorption rate after conditioning has been obtained from the PEP-II results: $10^{-7}$ molecule/photon for $H_2$, CO and $CO_2$ and $10^{-8}$ for $CH_4$. In Figure 13.10 the emitted photon flux in the cell arc (left) and the resulting vacuum pressure (right) are illustrated.

Under this condition an average pressure of $8\text{x}10^{-10}$ mbar for the Hydrogen and respectively 5 and $1\text{x}10^{-10}$ mbar for the CO or $CO_2$ and the $CH_4$ is evaluated. So the total pressure is approximately of $1.9\text{x}10^{-9}$ mbar. This pressure is near a factor three higher than the expected one (see [2]).

To obtain a better performance it will be possible to coat NEG on the drift chambers at the place of the NEG strip. This will significantly reduce CO and $CO_2$ pressures in the drifts chambers but it will not have a strong impact on the dipoles chambers were the Hydrogen partial pressure will decrease from 8 to $5\text{x}10^{-10}$ mbar. The other possibility is to increase the pumping speed in the dipoles chambers or to have longer dipoles with a longer curvature radius. These estimations are based on the CDR hypothesis as far as the pumping speed is concerned. Different detailed simulations, taking into account the effective pumping speed for each gas species, should provide a better estimation.

This is the results for HER and LER after a full scrubbing of the vacuum chambers walls. At the beginning of the injection in the rings it will be impossible to inject the full current due to the very strong out gassing given by the synchrotron radiation flux. A strategy for the gradual process of the vacuum improvement should then be established

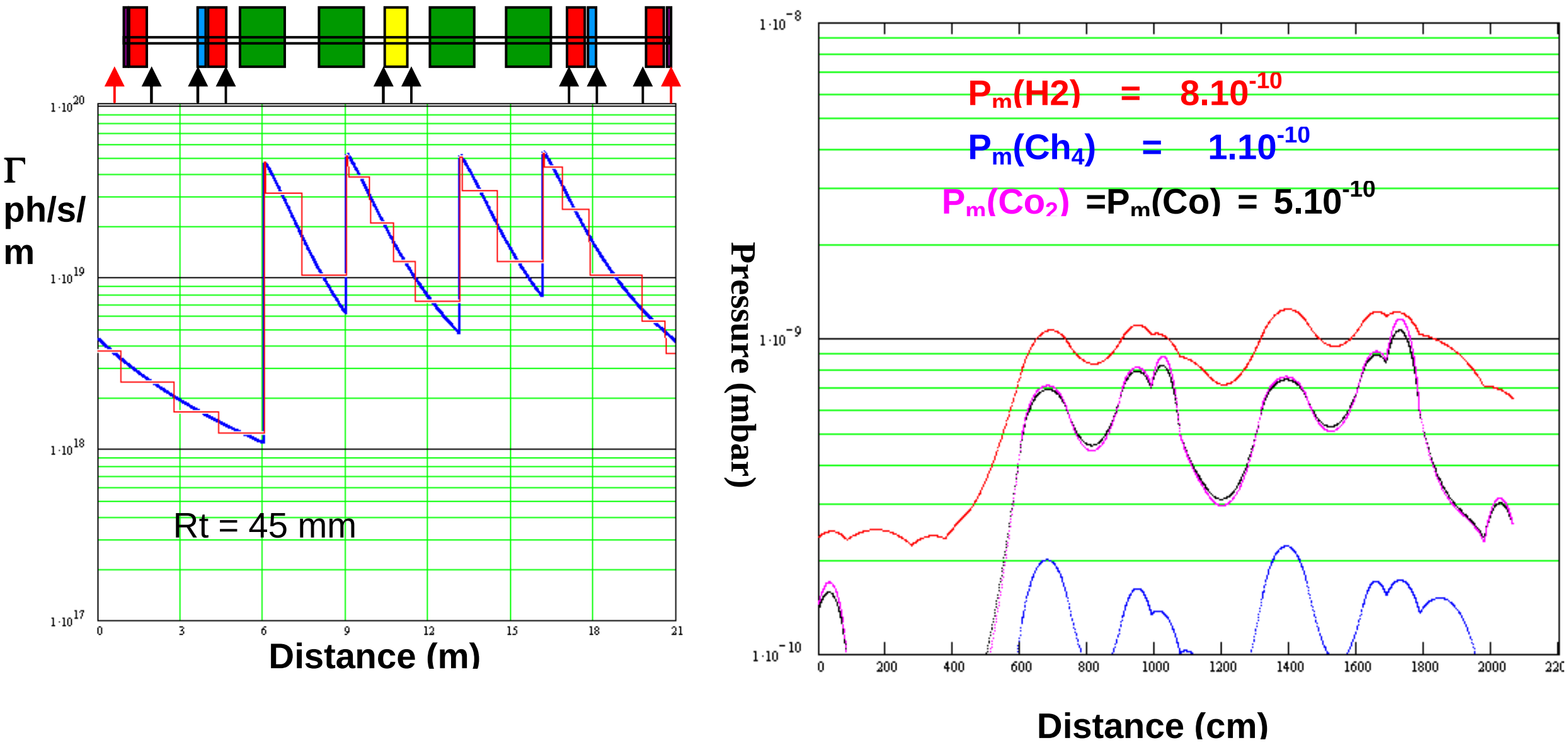

Figure 13.10: Emitted photon flux in the LER cell arc (left) and the resulting vacuum pressure (right).

# 14. Instrumentation and controls

## 14.1 Beam-Position Monitors

### Requirements

The beam-position monitors (BPMs) for SuperB will benefit from the experiences of other rings. Heating by the unprecedented beam currents in PEP-II and KEKB drove improvements in electrode ("button") design. The growing number of synchrotron light sources, with their demanding requirements for orbit stability, has led to impressive commercial processors that support fast orbit feedback.

The BPMs must serve a range of conditions, from tracking the orbit of a small injected charge on its first turn, with an accuracy of 100 µm, to measuring a stable orbit to 200 nm in a full ring with over 2.4 A of circulating beam. The measured orbit must be insensitive to the fill pattern. Measurements such as the phase advance require turn-by-turn beam positions for 1000 or more consecutive turns all around the ring. The position history of the last 1000 or more turns must be available after a beam abort for post-mortem investigation. Data must be available on a speed compatible with global orbit feedback.

### Buttons

PEP-II used 15-mm-diameter buttons mounted flush with the chamber walls to measure beam position. Identical buttons were used as pick-ups for the transverse and longitudinal feedbacks, and the tune and bunch-current monitors. The buttons were mounted at approximately 45 degrees to the horizontal and vertical axes (with variations for the different cross-sections of the vacuum chambers) to avoid direct hits from synchrotron radiation. These buttons are stainless steel, mounted on molybdenum pins that pass through a ceramic feedthrough to an SMA connector outside the chamber. For copper (LER near the IP, HER arcs) and stainless-steel (standard straights for both rings) vacuum chambers, the button assemblies were electron-beam welded into place. However, they were not suitable for welding to aluminum chambers (LER arcs and wiggler straights), and so the buttons there were mounted on flanges.

In June 2005, the RF voltage in the LER of PEP-II was increased from 4.05 to 5.4 MV to shorten the bunch length. The additional high-order mode heating, combined with typical currents of 2.4 A, caused some buttons on the upper half of a few chambers to fall off within a week. The end of the molybdenum pin was captured inside a socket on the back surface of the stainless button with a press fit requiring some spring force. This force appeared to have weakened after years of thermal cycling, and gave way completely with the increased high-order-mode power from the shorter bunches.

The flanged buttons in PEP were replaced with the 7-mm-diameter buttons shown in Fig. 14.1(a). These buttons and pins are made together from a single piece of molybdenum. Another suitable design (Fig. 14.1(b)) was developed for SuperKEKB and uses 6-mm buttons [1]. The choice for SuperB will be decided after comparative modeling and testing. Given the difficulty in replacing welded buttons, all SuperB buttons should be mounted on flanges.

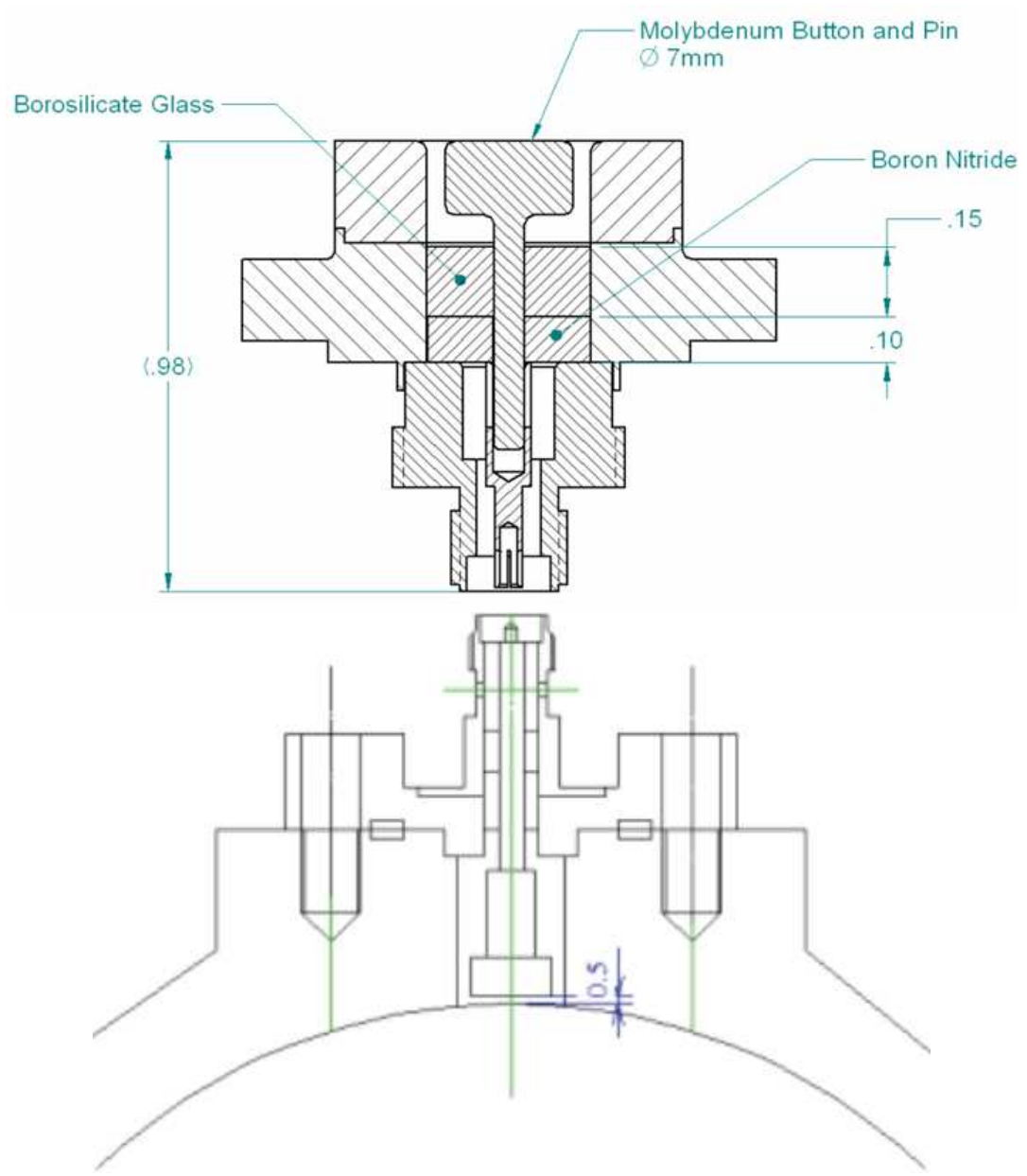

Figure 14.1: Top (a): New PEP-II BPM 7-mm button assembly, mounted in a vacuum flange. Note the integral molybdenum button and pin. Bottom (b): 6-mm test button developed for SuperKEKB, flanged and attached to chamber.

It is also essential that the mechanical position of each BPM is fixed relative to the adjacent quadrupole. With only 0.25% coupling, the vertical beam size is only a few microns. There can be little tolerance for thermal and diurnal motion of the magnets or BPMs, and each 4-button set must have negligible roll.

#### Processors

The growing number of light sources around the world in recent years has stimulated the development of commercial BPM processors that should also satisfy the requirements for SuperB. Electronics change rapidly, and so it is too early to select a processor for this project, but the performance available commercially is illustrated by the Libera Brilliance processor [2] from Instrumentation Technologies in Slovenia.

Each Brilliance is a 1-unit-high rack-mounted chassis that receives the four coaxial cables from the buttons, a ring-turn clock (238 kHz for SuperB), an acquisition trigger, and a beam-abort trigger. The analog input for each button includes a 19-MHz bandpass filter and automatic gain control for a wide dynamic range. The four buttons are reassigned to different input channels using a crossbar switch at 13 kHz, so that the effect of differences among the channels is removed from the averaged position. The signals are digitized at a frequency near 120 MHz (adjusted for each ring's RF) and downconverted digitally. For SuperB, the appropriate frequency would be $f_{RF}/4$ = 119 MHz.

The beam position is computed by firmware in an FPGA. An internal processor can run EPICS to serve the measurements to the control system over ethernet (see Section 10). The data may be read at various rates: sample by sample, turn by turn, 10 kHz for fast orbit feedback, or 10 Hz for position monitoring. Depending on the requested rate, digital filters further narrow the bandwidth to reduce noise and to remove dependence on the fill pattern. In turn-by turn mode, the processor records data from up to hundreds of thousands of consecutive turns following the acquisition trigger, which can be synchronized either with stored beam or with an injection fiducial. Similarly, the abort trigger freezes a 16,000-entry circular buffer of turn-by-turn beam positions measured prior to the abort.

Global-orbit feedback can be run at rates up to 1 kHz using the 10-kHz data, which is streamed through fast gigabit ethernet from each Brilliance unit to a processor doing feedback computations for a cluster. The cluster topology must be carefully planned by taking into account the ring's symmetries and the locations of fast correctors.

## 14.2 Beam-Size Monitors

In storage rings, synchrotron radiation from bend magnets provides the standard measurement of beam size. However, the coupling of 0.25% in SuperB will lead to a very small vertical size, below 10 µm in the arcs, even near defocusing quadrupoles, and simple imaging does not have sufficient resolution. We first choose lattice locations for the monitors, and then evaluate several techniques.

#### Monitor Locations

The source point for the synchrotron light should be at a point where the vertical beam size is as large as possible, and so should be in a dipole close to a (horizontally) defocusing quadrupole (QD). It should be outside the coupled region around the IP, where coupling due to the detector solenoid is corrected. Each ring should have two monitors, where the horizontal dispersion makes a small and large contribution to the horizontal size, in order to measure the emittances and energy spread. Then the sizes found at the detectors can be transferred to the IP using a fully coupled model and taking the beam-beam interaction at the IP into account [3].

Table 14.1 and Fig. 14.2 shows two appropriate locations in an arc, with low and with high dispersion. The third possibility is a location near the IP, close enough to benefit from the large beta functions approaching the final focus, but far enough from the IP to be outside the coupled zone. There we need to measure a vertical beam size of 40 µm, while a monitor in an arc must resolve 8 µm. However, we need the second location only for the smaller contribution of horizontal dispersion. It is sufficient to resolve the vertical size only where it is larger, near the IP. Next we examine several techniques to determine their suitability.

Table 14.1: Lattice parameters and beam sizes at possible locations for synchrotron-light monitors.

| | **HER** | | **LER** | |
|---|---|---|---|---|
| | ***x*** | ***y*** | ***x*** | ***y*** |
| Energy $E$ [GeV] | 6.7 | | 4.18 | |
| Energy spread $\delta$ [$10^{-4}$] | 6.31 | | 6.68 | |
| Emittance $\varepsilon$ [pm] | 2000 | 5 | 2460 | 6.15 |
| **1. Arc at Low Dispersion** | | | | |
| $s$ [m from IP, clockwise] | 568.9 | | 562.2 | |
| Critical energy $E_c$ [keV] | 8.29 | | 6.04 | |
| Beta function $\beta$ [m] | 1.59 | 15.6 | 1.61 | 17.5 |
| Dispersion $D$ [m] | 0.028 | 0 | 0.030 | 0 |
| Size without $D$ [µm] | 56 | 8.8 | 63 | 10.4 |
| Beam size with $D$ [µm] | 59 | 8.8 | 66 | 10.4 |
| **2. Arc at High Dispersion** | | | | |
| $s$ [m from IP, clockwise] | 508.0 | | 510.7 | |
| Critical energy $E_c$ [keV] | 8.29 | | 6.04 | |
| Beta function $\beta$ [m] | 1.65 | 14.7 | 3.21 | 29.6 |
| Dispersion $D$ [m] | 0.065 | 0 | 0.090 | 0 |
| Size without $D$ [µm] | 57 | 8.6 | 89 | 13.5 |
| Beam size with $D$ [µm] | 70 | 8.6 | 107 | 13.5 |
| **3. Near IP at High Dispersion** | | | | |
| $s$ [m from IP, clockwise] | 42.7 | | 1215.7 | |
| Critical energy $E_c$ [keV] | 6.60 | | 1.60 | |
| Beta function $\beta$ [m] | 12.1 | 323 | 12.1 | 323 |
| Dispersion $D$ [m] | 0.251 | 0 | 0.251 | 0 |
| Size without $D$ [µm] | 156 | 40.2 | 173 | 44.6 |
| Beam size with $D$ [µm] | 222 | 40.2 | 241 | 44.6 |

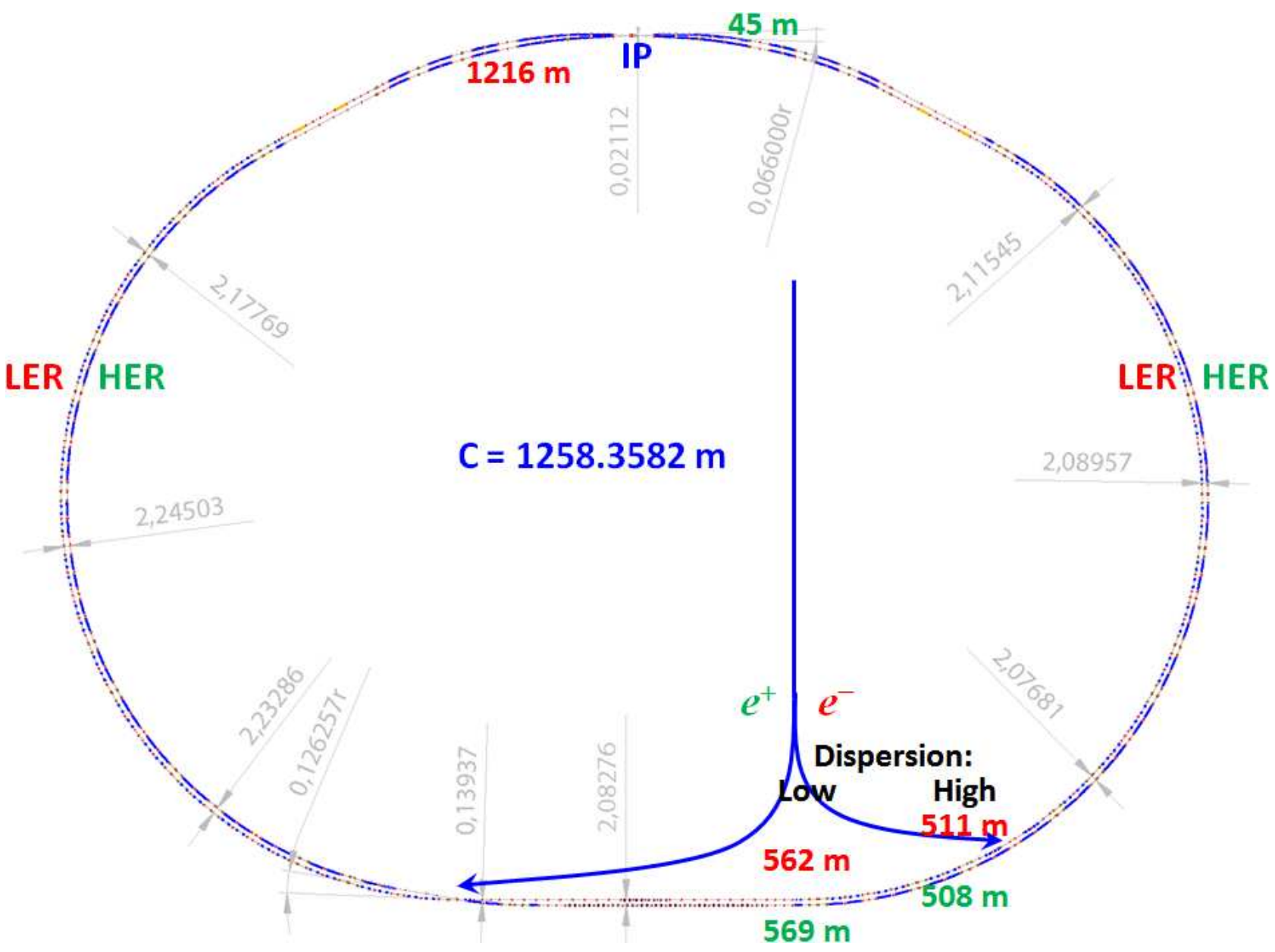

Figure 14.2: The locations listed in Table 14.1 for synchrotron-light monitor

**Interferometry with Visible Light**

A two-slit interferometer using visible synchrotron light can be a useful alternative to imaging. This technique, originally devised by Michelson to measure the diameter of a star, was much later adapted to beam-size measurements by Mitsuhashi [5]. As in the classic experiment by Young, a monochromatic point source is imaged by a lens onto a screen. When a narrow horizontal slit is placed in front of the lens, the screen shows a $(\sin y/y)^2$ intensity pattern. The second slit imposes a $\cos^2$ modulation with minima where the optical path from each slit differs by half a wavelength (see Fig. 14.3(a) for a calculated example for the HER of SuperB). When the source has a finite size, the minima from various points on the source fall at different places, and so the contrast between maximum and minimum—the visibility of the fringes—drops as the beam gets larger (Figs. 14.3(b) and 14.4). A wide bandwidth also washes out visibility (Figs. 14.3(c) and (d)), since the distance between adjacent minima scales with wavelength, and so a narrow optical bandpass filter must be used. Expressions including source size and bandwidth are given in Ref. [6].

Fig. 14.4 shows the visibility of the fringes as a function of beam size, measured under the conditions of Fig. 14.3(c). We see that the 40-nm beam size expected with 0.25% coupling for the source point near the IP should be measurable, but not the size at the source points in the arcs. However, two measurements are needed only in the horizontal; to separate the effects of emittance and energy spread, but in the vertical one interferometer may be sufficient.

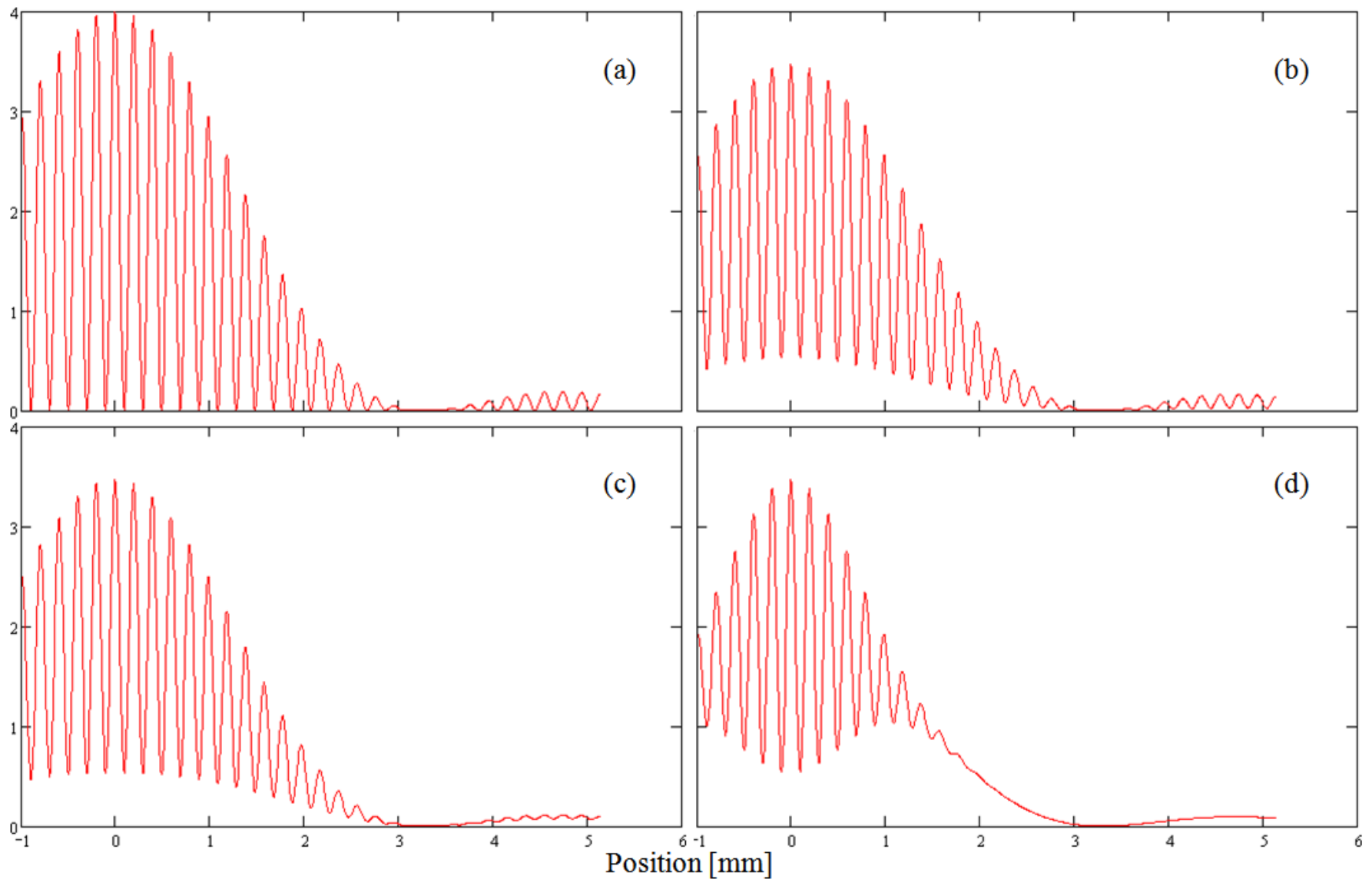

Figure 14.3: Calculated interference pattern for the HER of SuperB following Ref. [6]. This is a projection of the pattern on the camera onto one axis for 1.2-mm-wide slits, 20 mm apart and 10 m from the source, using a 5-m lens and 400-nm light. (a) A monochromatic point source. (b) Widen the previous beam to a 25-µm-RMS Gaussian. (c) Widen the bandwidth to 10 nm (full width at half maximum). (d) Widen the bandwidth to 40 nm.

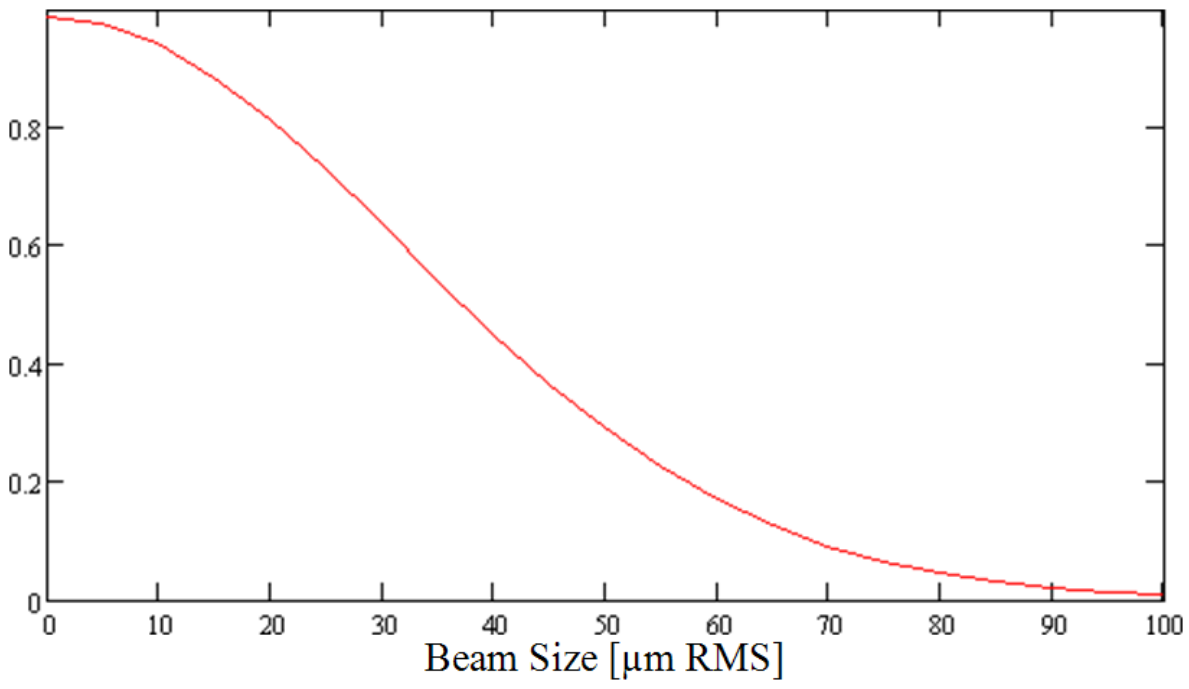

Figure 14.4: Visibility of fringes of Fig. 14.3(c) for different beam sizes.

Does blurring due to depth of field, discussed above for imaging, also affect this interferometric measurement? Considering the vertical direction (as in Fig. 14.3), a point $y$ on the image plane receives photons emitted at different positions $s$ along the orbit and at different heights $Y(s)$ on the plane at $s$. Since all these photons are uncorrelated, the fringes arise from the interference of the two slits for any emitting point $Y(s)$. If we extend the arguments of [6] to a source point $s$ that is not at the nominal source plane $s = 0$, then the optical-path difference between a ray from $Y$ to $y$ via the center of the upper slit compared to a ray through the center of the lower slit is:

$$\Delta s = -\frac{b}{z_{0s}}\left[Y(s)+\frac{y}{m}\right]\left(1+\frac{s}{z_{0s}}\right)$$

Here $b$ is the distance between the two slits, $z_{0s}$ is the distance from the nominal source plane to the slit plane (10 m here), and $m$ is the optical magnification from source to image plane. We see from Fig. 3.10.5 that the phase difference $|k\Delta s|$ for $s = 0$ runs up to perhaps 10 fringes. The factor $(1+s/z_{0s})$ can reduce fringe visibility, but since $|s| < 5$ cm, this correction is no more than 0.005, or up to 0.05 fringes. Consequently, depth of field does not significantly affect this measurement.

### Null in the Vertical Polarization

The vertically polarized component of synchrotron emission, viewed as a function of vertical angle, has odd symmetry about the horizontal plane, with a null on that plane [4]. With vertical polarization, an electron orbiting on axis must image to a null on axis, as the equal and opposite components from emission above and below the midplane cancel. Diffraction produces some light above and below center, and so a projection of the image has two peaks separated by a null. In a manner that resembles the interferometer, the contrast from peak to valley is reduced as the source size grows, because the nulls from different heights on the source are imaged to different heights on the camera. This technique, first developed at MAX-Lab in Sweden [7], has been used with visible light at the Swiss Light Source (SLS) to resolve beams as small as 1 μm from a vertical emittance as low as 2.8 pm [8,9], as shown in Fig. 14.7.

As with interferometry, the vertical size cannot be seen directly from the image, but is determined by running a complete model of the emission and the optical system, including diffraction, using the code SRW [10, 11]. The smallest beams measured at SLS appear to be near the limit of the technique, but this resolution is suitable for all the emission points proposed in Table 14.1 for SuperB.

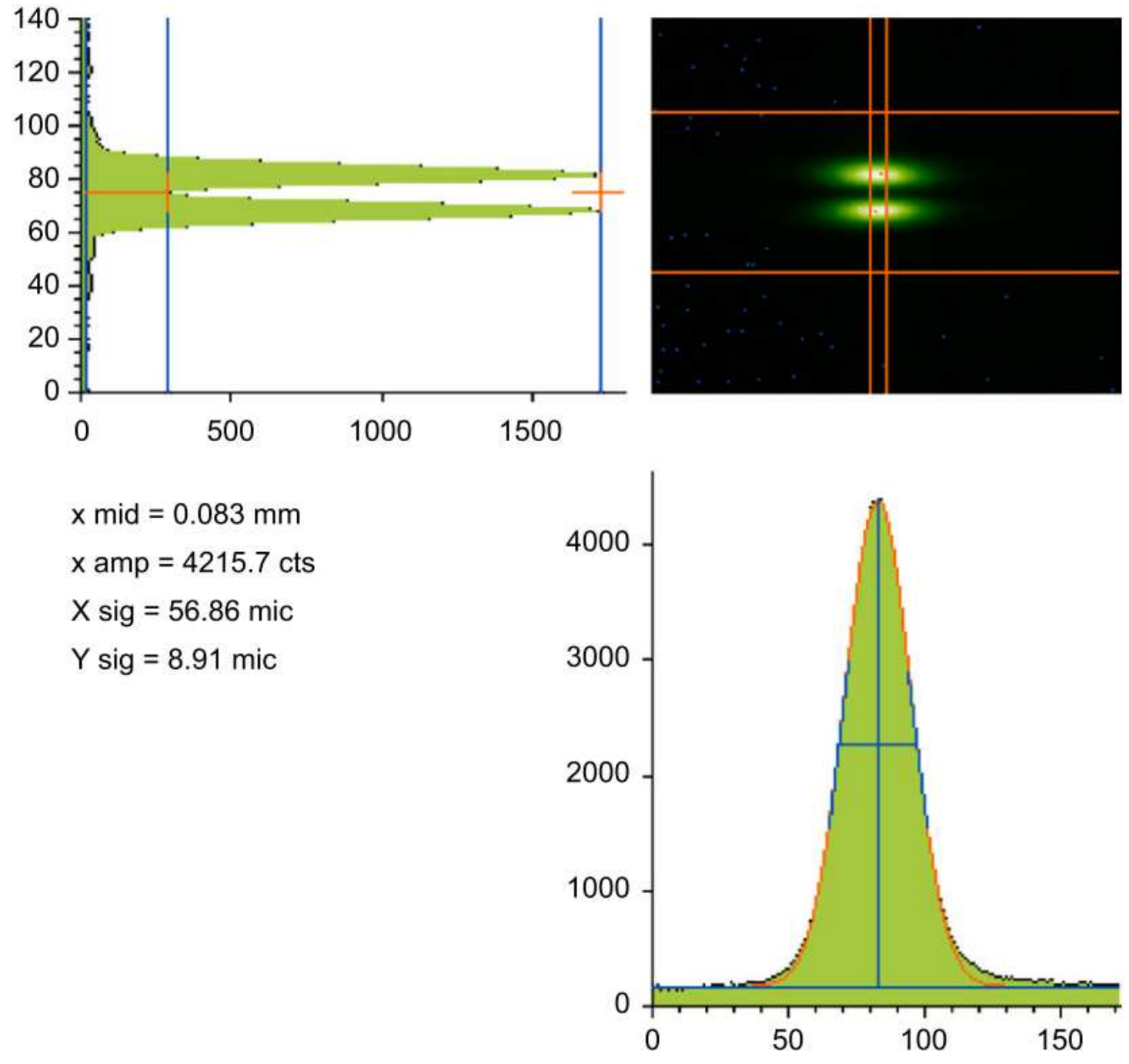

Figure 14.7: Image of the SLS beam using vertically polarized, visible synchrotron light [9]. Distances on the vertical and horizontal projections are in μm.

### X-Ray Pinhole Camera

The resolution of a pinhole is limited by geometric optics for large holes, and by diffraction for small holes. If $d_1$ is the distance from the source to a pinhole of radius $r$ and $d_2$ is the distance from the pinhole to the image, then geometric optics gives a resolution on the *image* plane of:

$$\sigma_g = \frac{r}{\sqrt{3}}\frac{d_1+d_2}{d_1} \qquad (0.0.1)$$

Diffraction limits the resolution on the image plane to:

$$\sigma_d = \frac{5}{8\pi}\frac{\lambda d_2}{r} \qquad (0.0.2)$$

The overall resolution is the quadrature sum of these two effects. The optimal pinhole radius is:

$$r_{\mathrm{opt}}^2 = \frac{5\sqrt{3}}{8\pi}\frac{\lambda d_1}{1+d_1/d_2} \qquad (0.0.3)$$

and the corresponding resolution—now adjusted by the magnification $d_2/d_1$ to give the resolution on the *source* plane—is:

$$\sigma_{\text{opt}}^2 = \frac{5}{4\pi\sqrt{3}}\left(1+\frac{d_1}{d_2}\right)\lambda d_1 \quad (0.0.4)$$

The resolution improves as the wavelength gets shorter, but the power emitted drops quickly below the critical wavelength, limiting $\lambda$ to about $\lambda_c/5$. The distance $d_1$ to the pinhole should be as short as possible, but is limited by the length of the magnet and the need for sufficient drift space to separate the photons from the particles. There must be additional room between the magnet and pinhole for x-ray filters to remove longer wavelengths, since they have poorer resolution. Filters also help to remove heat that would distort the pinhole, since half the synchrotron power is at $\lambda > \lambda_c$. Collimation before the filters and pinhole is useful too, by blocking heat at larger radii. The pinhole camera for the LER of PEP-II in Ref. [12] shows a typical layout. Finally, $\sigma_{\text{opt}}$ improves with a large magnification $d_2/d_1 > 2$, but $d_2$ can be limited by available space in the tunnel.

As an example, a measurement at 0.1 nm (12.4 keV), with a pinhole 7 m from the source and a magnification of 2, has an optimal pinhole radius of 13 µm and a resolution of 16 µm. The optimal resolution on the source plane is somewhat better with a square pinhole:

$$\sigma_{\text{opt}}^2 = \frac{1}{2\pi}\left(1+\frac{d_1}{d_2}\right)\lambda d_1 \quad (0.0.5)$$

By lowering the wavelength to 0.05 nm, putting a square pinhole 5 m from the source, and using a magnification of 3, the resolution becomes 7.3 µm.

However, Ref. [13] notes that the expression (0.0.2) for the diffraction resolution applies only in the Fraunhofer diffraction zone, while x-ray pinhole cameras with a layout typical of light sources are in the Fresnel regime, or on the boundary between the two. Their full diffraction calculation gives a resolution that is roughly half that calculated with the expression above.

The scintillator can also coarsen image resolution. A thick scintillator spreads the source of visible light longitudinally, adding depth of field considerations, and transversely, with isotropic emission followed by refraction at the scintillator's surface. But a thin scintillator converts only a fraction of the energy of hard photons. In addition, the image must span many camera pixels to avoid resolution loss.

With some care, the resolution of a pinhole camera can be adequate to image the beams at all source points in Table 14.1.

### Fresnel Zone Plate

A zone plate is essentially a lens that focuses using diffraction rather than refraction or reflection [14-16]. An x-ray-opaque metal, typically gold, is deposited in a pattern of $N$ (typically hundreds) of narrow (~1 µm) circular rings (Fig. 14.8) onto a thin membrane of x-ray-transparent material, such as $Si_3N_4$. The thickness and separation of the rings vary systematically so that, when illuminated by a collimated and monochromatic x-ray beam, each ring forms a first-order diffraction maximum that adds in phase at a focal point downstream. Zone plates are produced commercially by firms such as Xradia [17] for use at synchrotron light sources.

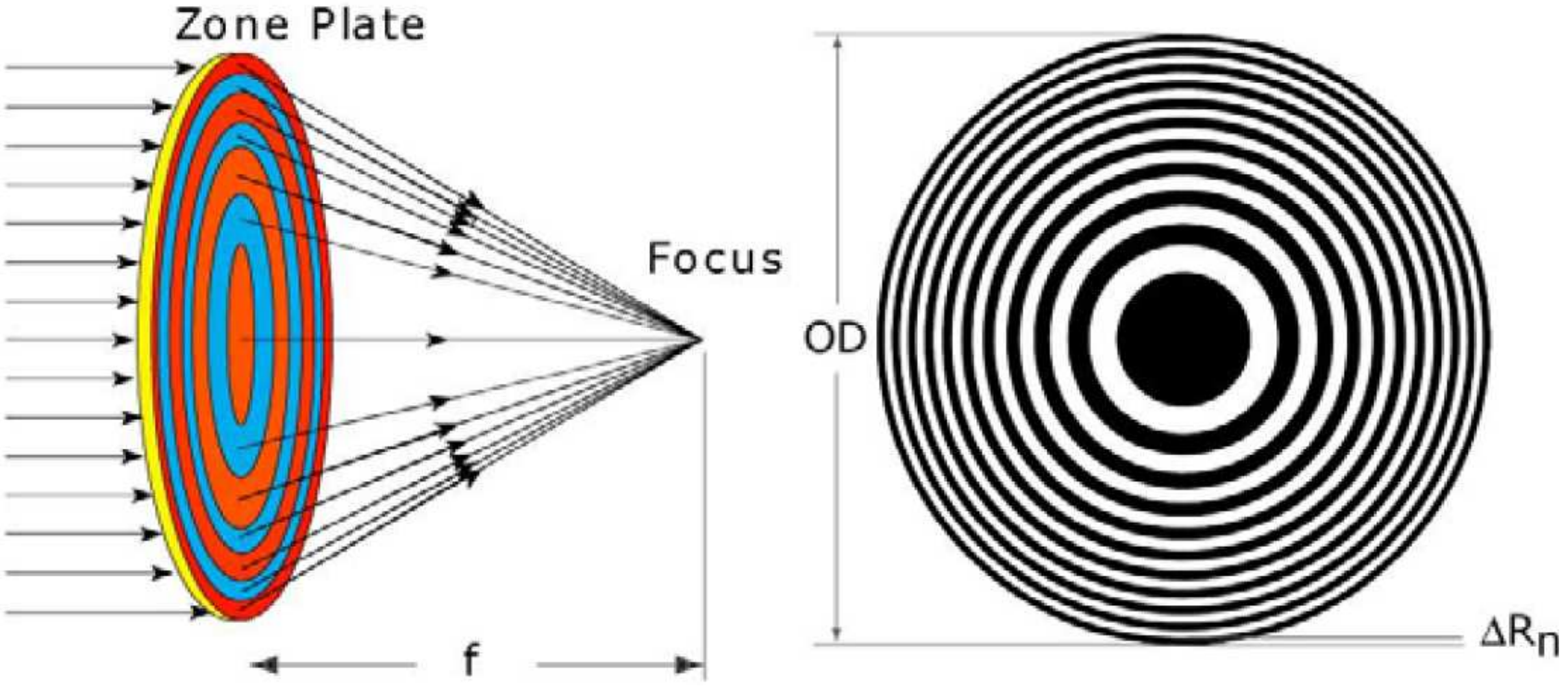

Figure 14.8: A monochromatic x-ray beam focused by a zone plate [16].

The focal length of a zone plate of radius $r$ is [4]:

$$f = \frac{r^2}{N\lambda}$$

Since $f$ depends on the wavelength $\lambda$, the lens is strongly chromatic, and so the bandwidth of x rays from a dipole's synchrotron emission must be first narrowed. Bandwidth selection also helps with high beam currents by reducing the power striking the zone plate to a safe level for this delicate structure.

As an example, Fig. 14.9 shows the imaging system at the Accelerator Test Facility at KEK in Japan. This uses two zone plates to resolve vertical sizes below 6 µm, with an overall system resolution of 0.7 µm [18].

Common x-ray monochromators, such as that used at ATF, are based on Bragg diffraction from a single crystal, with a typical bandpass $\Delta E/E$ of $10^{-5}$. This is costly in terms of flux and is far narrower than needed for imaging. Instead, a bandwidth of about 1% can be obtained with a grazing-incidence multilayer mirror, a substrate coated with alternating thin layers of light and heavy materials. Fig. 14.10 shows the calculated reflectivity of a mirror with alternating layers of $B_4C$ and Mo [19]. The center of the band may be tuned by small variations in the angle of incidence. To preserve the direction of the incident beam while tuning, the mirrors are commonly used in pairs, with the outgoing beam parallel to the incoming beam, but displaced slightly. Like zone plates, such mirrors are available commercially for use at light sources, from firms such as Rigaku/Osmic [20].

The high heat load now strikes the first multilayer mirror rather than the zone plate. Although the mirror is far more robust than a zone plate, it is important to reduce the surface heating to maintain the flatness and thickness of the layers. As with the pinhole camera, collimation and filtering are necessary before the mirror. Grazing incidence at 1° spreads the remaining heat. Water-cooling channels in the substrate (silicon or silicon carbide) are carefully designed to conduct this heat away with minimal distortion.

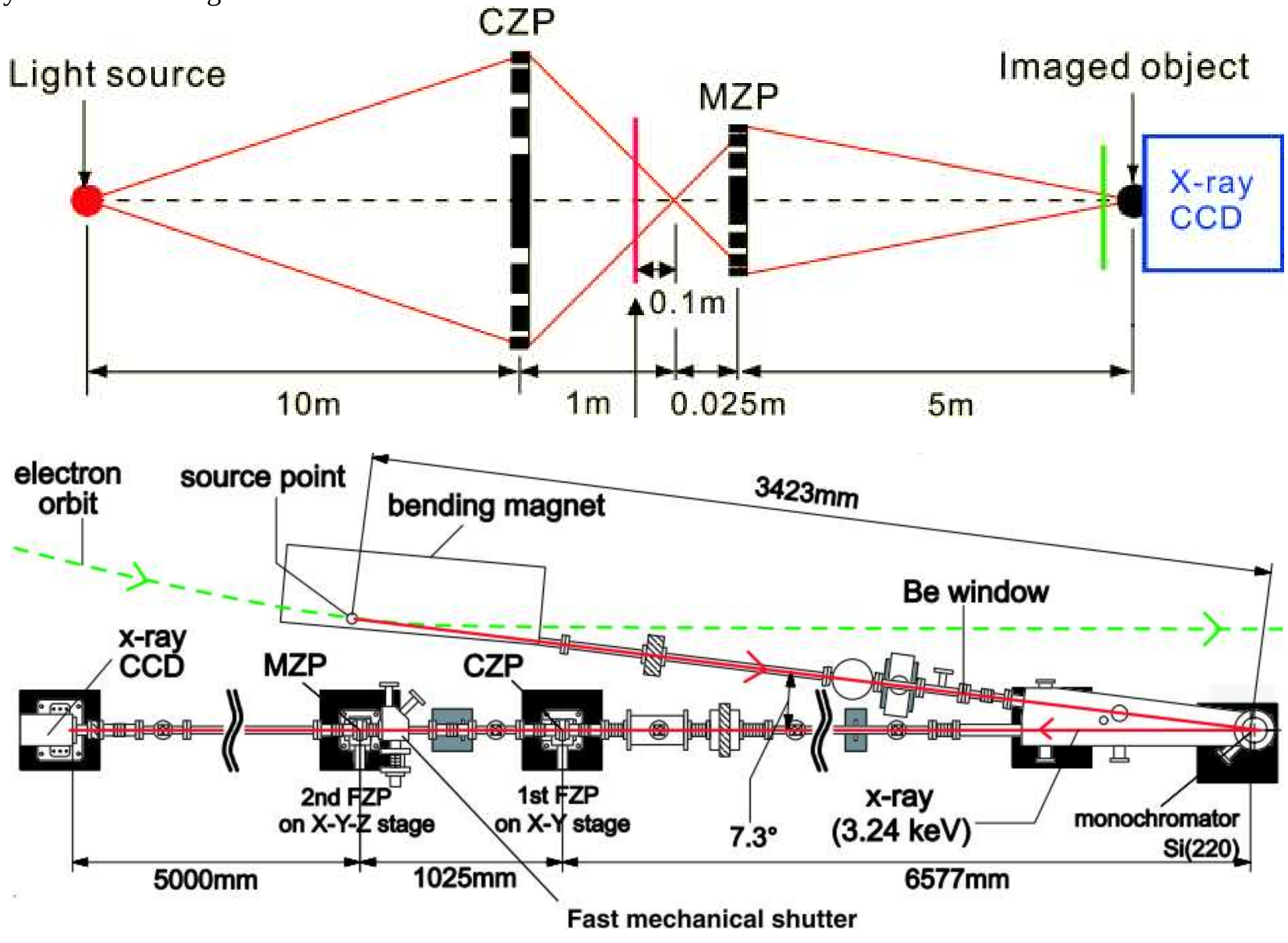

Figure 14.9: Top (a) Imaging with two zone plates at the ATF [18]. Bottom (b) Layout of the ATF beamline.

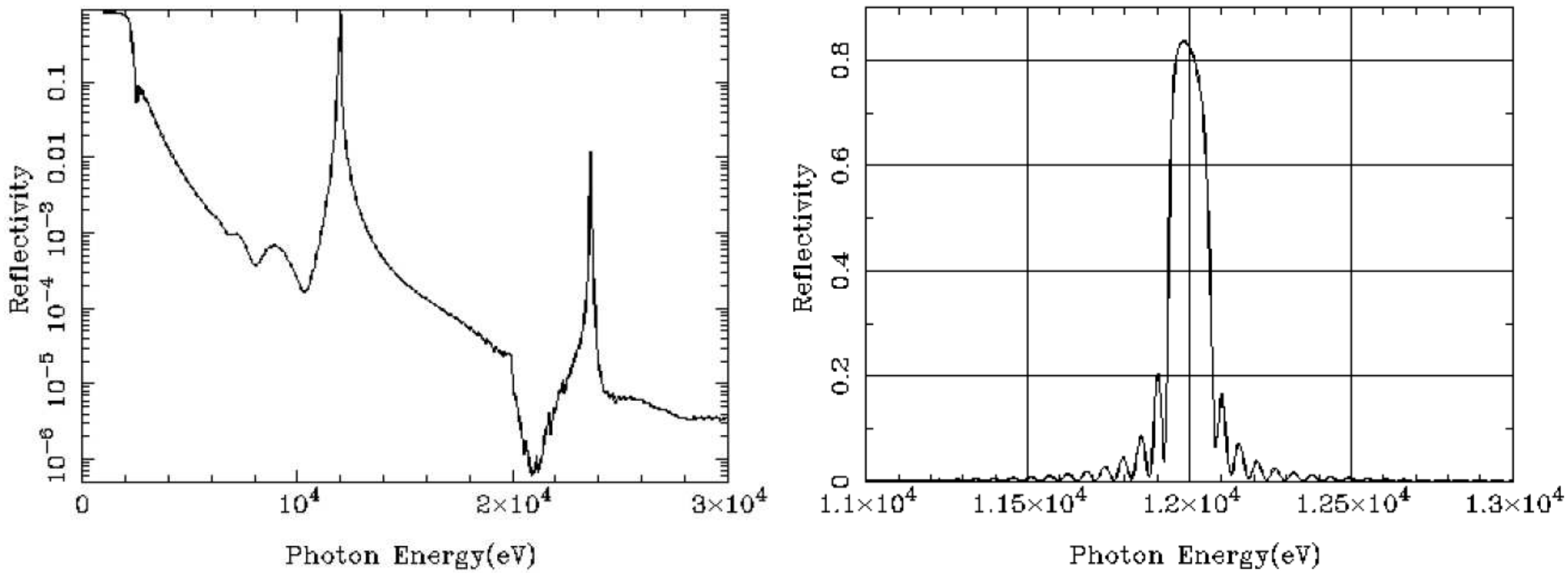

Figure 14.10: Calculated reflectivity [19] versus energy for a single multilayer mirror. P-polarized x-rays incident at 1.007° to grazing on a mirror with 200 layers each of 2.1 nm of $B_4C$ and 0.9 nm of Mo, and with an interdiffusion thickness of 0.5 nm, deposited on a silicon substrate.

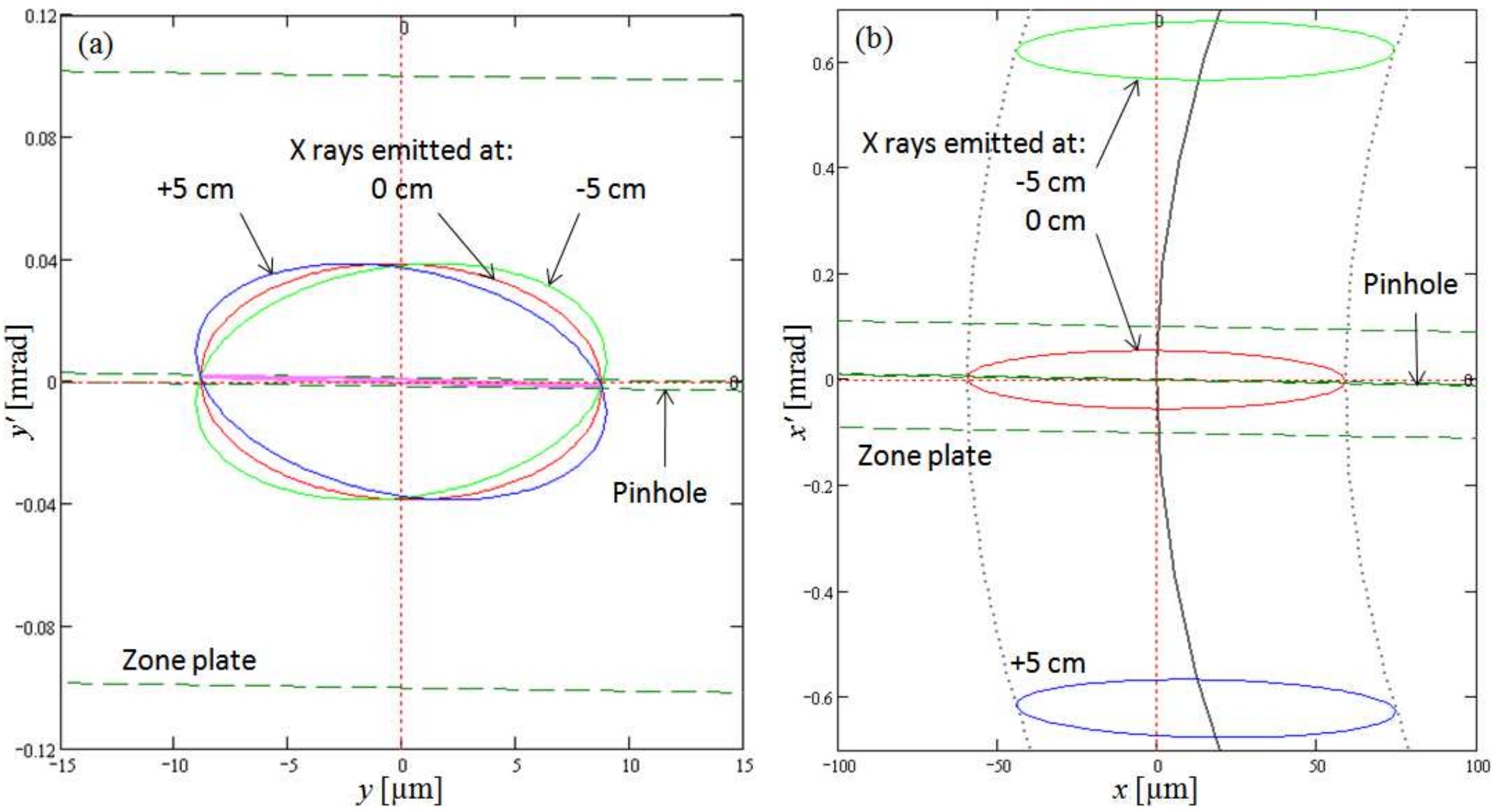

Figure 14.11: (a) Vertical and (b) horizontal phase space of 0.1-nm x rays emitted in the HER arc at the low-dispersion point. Lines show the edges of a zone plate with a 1-mm radius and of a circular pinhole with the optimum radius of Eq. (0.0.4).

The blurring effect of depth of field in the dipole is significant for visible imaging (Sec. 3.10.2.2), but not for x rays, with their narrow opening angle. Fig. 3.10.11 plots the phase space of x rays emitted in the HER arc.

Unlike Fig. 3.10.3(a) and 3.10.4(a), x rays emitted ±5 cm off the nominal source plane are easily rejected, and no significant blurring is expected

### Laser Techniques

Two other techniques [21] use laser light to measure small beams: the laser wire and the laser interferometer. The incoming laser photons Compton scatter from the electrons (or positrons), producing gammas that are measured with a scintillator and photomultiplier.

The laser wire uses a lens with a small F-number to make a tight focus in the middle of the beampipe. As the electrons are gradually scanned horizontally or vertically across the focus, the PMT signal maps the bunch profile along the scan direction. The resolution is limited by the size of the focal spot, which for a small F-number is essentially the wavelength, and so argues for near-ultraviolet light (~250 nm). However, the smallest focus is not always best: for a projection, the laser light should remain essentially parallel as it passes through the electrons, and so its Rayleigh length should be at least twice the RMS bunch size in the laser's direction. For the SuperB arcs, these criteria give a resolution of 2 µm.

If the laser light is split into two beams that cross as they pass transversely through the beampipe, they form a standing wave with a peak spacing that depends on the crossing angle. As the electron bunch is scanned across the standing wave, the modulation of gamma signal depends whether the electron bunch is narrow compared to peak spacing. This technique has measured 60-nm beams in the FFTB at SLAC.

The resolution of these techniques comes at a cost: both require a substantial laser, with harmonic generation to produce ultraviolet, and both involve scanning the electron beam through the light. For the beam size in SuperB, neither is truly needed. Several concepts discussed earlier—interferometry, the null in the vertically polarized image, the x-ray pinhole camera and x-ray imaging with zone plates—have demonstrated the necessary resolution.

## 14.3 Beam-Abort System

At design currents, the stored energy in the HER and LER beams of SuperB is 53 and 42 kJ respectively. A sudden beam loss depositing this energy into a small region could melt the beam pipe, and so we require a system that quickly detects faults and then extracts the beam into a dump. Moreover, a gap in the bunch pattern is mandatory for avoiding ion-trapping mechanism in the electron ring.

Several types of faults should trigger beam aborts, such as a trip of a main magnet string, a fault in an RF station, a rapid loss of beam current, or excessive background radiation in the detector. Table 14.3 provides a detailed list.

Table 14.3: Triggers for beam aborts.

| | |
|---|---|
| Manual abort from control room | Fault in beam-abort trigger system |
| Beam-stopper insertion (Personnel Protection) | HV on abort kicker < 80% |
| Vacuum-valve insertion | Rapid drop in beam current |
| Fault in dipole or major quadrupole string | Sudden large orbit excursion |
| Fault in an RF station | High radiation level at Super *BABAR* |
| Fault in longitudinal feedback | Temperature over limit on a thermocouple |
| Fault in transverse feedback | Trip of a klixon (thermal switch) |

The fastest of these mechanisms is a loss of RF, causing beam to spiral inward and scrape within some tens of turns. A suitable response speed can be attained only with a hard-wired system that bypasses the latency inherent to the network of control-system computers. Other fault processes are substantially slower. Magnet trips are slowed by inductance, but the response time (milliseconds) is still fast enough to hard-wire the trigger. Thermocouple trips are still slower due to heat capacity, and so can be detected by the control system, which then triggers the abort.

In a large machine, abort triggering is necessarily distributed, with processing electronics at several stations around the ring. At PEP-II, these were connected together in a bidirectional loop for each ring. Each station passes on a request for an abort to the next station. For fail-safe operation, this abort-request line normally propagates a fast clock (the “heartbeat”) that is halted to initiate an abort. The loop starts and ends at the controller for the abort kicker, which monitors the heartbeat.

The triggering hardware must latch the source of the abort and pass this information along to the control system, so that the source of an abort triggered by a momentary excursion can be determined. Also, an abort often causes the firing of other abort triggers. For example, RF stations will indicate high reflected power after the beam is dumped. The automatic recording of precise time stamps for each trigger is essential to determine the sequence of events.

The dump itself need not be under vacuum. In PEP-II, the beam exited the vacuum through a thin aluminum window on a chamber downstream of the kicker. Then it was stopped by blocks of graphite, aluminum, and finally copper in a meter-long dump. To ensure that the beam had not burned a hole through the dump, there was a small pocket of gas, at a pressure somewhat above ambient, trapped between the second and third layers. If the pressure in this burn-through monitor dropped, then an interlock would halt all further injection. (The PEP-II dumps never showed such damage.)

The abort kicker must dump the beam within one turn. Since a bunch passing through the kicker magnet while its field is rising would not get a sufficient kick to exit into the dump, but instead would instead start a large orbit oscillation, the kicker must have a fast rise that is synchronized with a short gap in the fill pattern.

Then the field must decay slowly over the course of one ring turn, so that all bunches strike the dump, but each deposits its energy at a different point, in order to avoid damage to the dump window and to the dump itself.

## 14.4 Control System

The control system outlined here takes advantage of the considerable body of experience from other accelerator laboratories, while leaving the flexibility to draw upon new technology. In particular, the global EPICS collaboration provides a standard architecture, with a distributed database and a large collection of software tools that are continually developed, shared, supported and upgraded by the many participating labs. The collaboration is large, mature, and invaluable, since it is no longer necessary to write custom code for tasks that are common to many machines.

The architecture of the control system has three tiers of distributed computing. At the front end, EPICS IOCs (input-output controllers) communicate with instruments, process the measurements, and serve this data by way of gateway computers and middleware to user applications at the top layer.

#### Front-End Designs

Older instrumentation commonly involves modules in VME and VXI crates, or in CAMAC crates for even older installations. Stand-alone instruments like oscilloscopes communicate through short-range GPIB connections to a local computer or to a GPIB-to-ethernet interface, allowing control by a distant machine.

This arrangement is substantially changed in new installations. CAMAC, VXI and GPIB are gone, and the need for VME is greatly reduced. Some devices interface to an IOC through PLCs (programmable logic controllers). Newer instruments communicate directly over ethernet and often include embedded processors, arranged with one for each device or for a collection of like devices. Gigabit Ethernet and network-industry-standard fast busses such as ATCA provide another possibility for special applications.

The EPICS collaboration has developed drivers for a wide range of hardware and instruments, such as motors, video cameras, and oscilloscopes. A scope now is essentially a computer hidden behind a front panel with the usual oscilloscope knobs and display. EPICS communicates with the scope and gathers data through its ethernet port. It is interesting to note that these instruments often allow remote control via a web browser, using a web page served by the scope itself. While this method is of limited use for our application, since it is not integrated with the control system, the concept illustrates the evolution of instrument architecture.

Some devices, such as the BPM processors discussed in Section 3.10.1.3, can run EPICS on their embedded processors, turning the device itself into part of the control system. These also have the capability to save data from many ring turns and to work jointly with other processors and higher-level applications to implement fast orbit feedback.

Other diagnostics need special hardware for bunch-by-bunch data capture. For example, transverse and longitudinal feedback, and bunch-current monitoring, all begin with a task-specific analog front end that combines signals from beam pick-ups, mixes the result with an appropriate harmonic of the ring’s RF, and outputs a signal suitable for digitizing at the RF rate or faster. All bunch-by-bunch tasks can use identical digital hardware, starting with a fast digitizer, followed by an FPGA (field-programmable gate array), and finally a fast DAC (digital-to-analog converter) to drive the feedback correction signal. A computer, either nearby or on an additional board in the same box, loads the FPGA with firmware written for the specific job, reads the data accumulated by the FPGA, and serves as an IOC to communicate with the rest of the control system. The FPGA data includes both the essential results (such as the charge in each bunch) and a considerable body of supplemental beam-diagnostic information (such as the spectrum of modes being corrected by feedback). All of this can be monitored by the user over EPICS.

As always, video is needed in many places, such as at screens on the injection line, or for measuring beam size with synchrotron light. In older systems, analog cameras send signals over coaxial cable either to modulators for a closed-circuit cable television system that brings multiple channels to users in the control room, or to digitizers on frame-grabber boards in computers outside the tunnel. Digital cameras have also been available, but with interfaces that do not allow transmission over the long distances typical of large rings or linacs.

Recently, a new camera standard has been introduced that replaces the coaxial analog video output with a gigabit ethernet port. The output is all digital and can be transmitted over 100 m with no loss of resolution. Once on the network, the image can be displayed or analyzed by any computer. Many such images might overwhelm the capacity of the network, delaying communications with other instrumentation. One way to preserve network bandwidth is to set the cameras for a lower update rate for slowly changing images. A more thorough approach gives the cameras a separate gigabit network.

This progress can be seen in the current support for the LCLS at SLAC. In addition, there are strong standardization efforts underway for controls-oriented ATCA hardware and software.

### High-Level Applications

Many applications that will be fundamental to running SuperB are similar to those at other accelerators and so are available from the EPICS collaboration, with only modest modification. This category might include BPM orbit displays, steering, orbit feedback, video and oscilloscope displays, an archiver (recording all signals on change or periodically, typically at 1 Hz), and an error log (a recording of each change to a setting of an accelerator component, such as a magnet, or state, such as an excessive temperature).

A high-level mathematical language such as Matlab is useful for writing applications, but tools must be added to provide access to EPICS data, ideally in a manner structured by physical devices to organize the many EPICS channel names. SNS as well as the LCLS, for example, are using XAL, a Java class hierarchy providing a programming framework based on the physical layout of the accelerator. The user interfaces for broadly used applications should be designed with input from operators and physicists. For less elaborate tasks, the tools should allow the accelerator physicists themselves to write the necessary code.

### Management

Several items must be organized at an early stage. For example, a relational database of control items must be set up at the outset, along with a well-planned naming convention that includes both an overall scheme and many examples. Another early need is an environment for developing, versioning, and testing code. This provides a basis for code management and bug tracking, as well as for code testing and release.

Also, the timing system should be carefully planned and started early. Timing includes both a means of generating triggers and a means of distributing pulse information to devices or processes which need that information. This combination allows triggering and data acquisition linked to events like the travel of an electron or positron bunch along the linac, to the injection of a bunch into a ring, or to one or more turns of a stored bunch in either ring.

### 3.10.6.4 Safety and Security

The computers on the control-system network must be highly secure, but still must allow remote users to connect and control the machine. These requirements need secure firewalls and gateways restricting outside access, and also good security even within the firewalls.

Safety systems, in the sense of a subsystem of the general control system, both for machine protection and for personnel protection need special attention. These two, and especially the latter, must be kept distinct from the rest of the control system and designed to meet rigorous standards.

# 15. Injection system

## 15.1 Overview

The injection system for SuperB is capable of injecting electrons and positrons into their respective rings at full energies. The HER requires positrons at 6.7 GeV and the LER 4.18 GeV polarized electrons. At full luminosity and beam currents, up to 4 A, the HER and LER have expected beam lifetimes in the range 3÷8 minutes. Thus, the injection process must be continuous, called top-up injection, to keep nearly constant beam current and luminosity. Multiple bunches (~five) will be injected on each linac pulse into one or the other of the two rings. Positron bunches are generated by striking a high charge electron bunch onto a positron converter target and collecting the emergent positrons.

The transverse and longitudinal emittances of the electron bunches and, especially, of the generated positron bunches are larger than the LER and HER acceptances and must be pre-damped. A specially designed damping ring at 1 GeV is used to reduce the injected beam emittances. This damping ring is shared for the beams to reduce costs. A sketch of the injection system is shown in Figure 15.1. The transport lines into and out of the damping ring are shown in Figure 15.2.

Electron to positron conversion is done at about 0.6 GeV using a newly designed capture section to produce a yield of more than 10%.

Electrons from the gun source are longitudinally polarized. The particle spins are rotated to the vertical plane in a special transport section downstream of the gun. The spins now remain vertical for the rest of the injection system and injected in this vertical state into the LER.

The specific injection parameter values are described here. The linac operates at 50 Hz. A short train of 5 bunches (1 to 20 possible) at a time are produced for each beam type, stored for 20 msec in the damping ring, and then extracted and accelerated to full injection energy. The nominal stored beam current in the rings is ~2 A, but the injector is designed to provide a maximum positron current of 3 A. At 3 A the total number of stored positrons are about $10^{14}$. Taking into account the beam lifetime (~7 minutes), ~2 x $10^{11}$ particles are lost per second per ring. With 5 injected bunches per pulse and an injection rate of 25 Hz per ring, each injected bunch must provide a charge of about 300 pC (2 x $10^{9}$ particles/bunch). This charge is about 3 % of the SuperB stored positron bunch charge. The injection of electrons in the respective ring is less critical thanks to the larger margin in e- generation and transport. The vertical polarization averaged over the electron bunch is expected to be about 88%. The details of the injection system are described in the following sections.

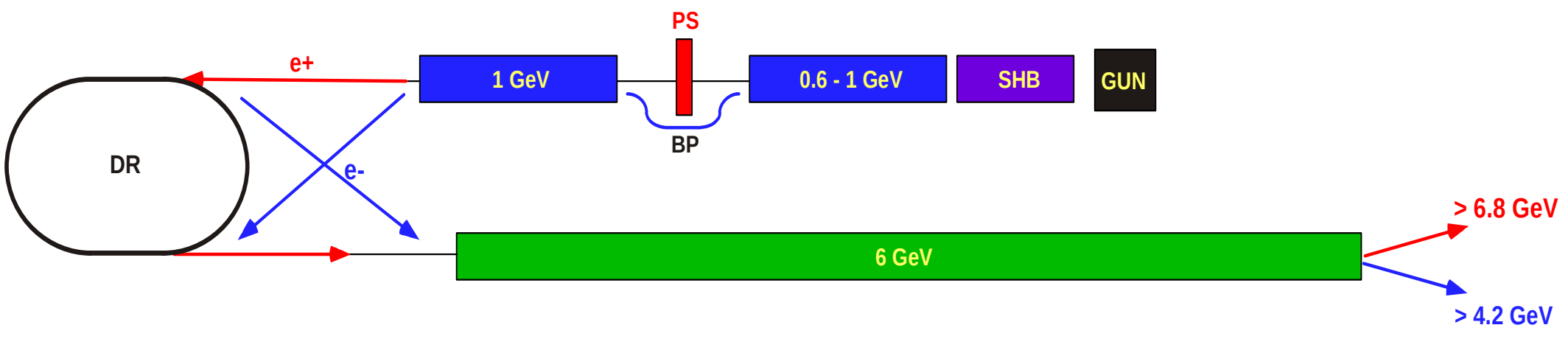

Figure 15.1: Overall layout of the SuperB Injection System.

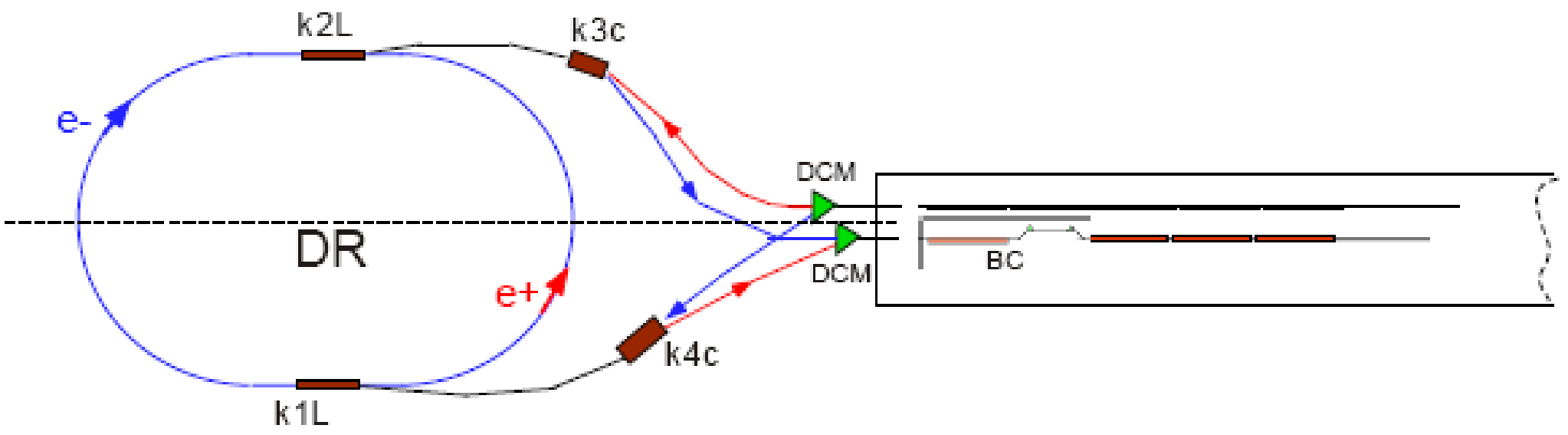

Figure 15.2: Injection and extraction from the shared Damping Ring.

## 15.2 Electron source

The electron source for the injector linac of SuperB will be a nearly identical copy of the source used by the SLC collider at SLAC [1]. The SuperB source needs to produce single or a few electron bunches that are longitudinally polarized. The polarization is expected to be above 80% which was routine during the SLC operation. An overview of this source is shown in Figure 15.3 below.

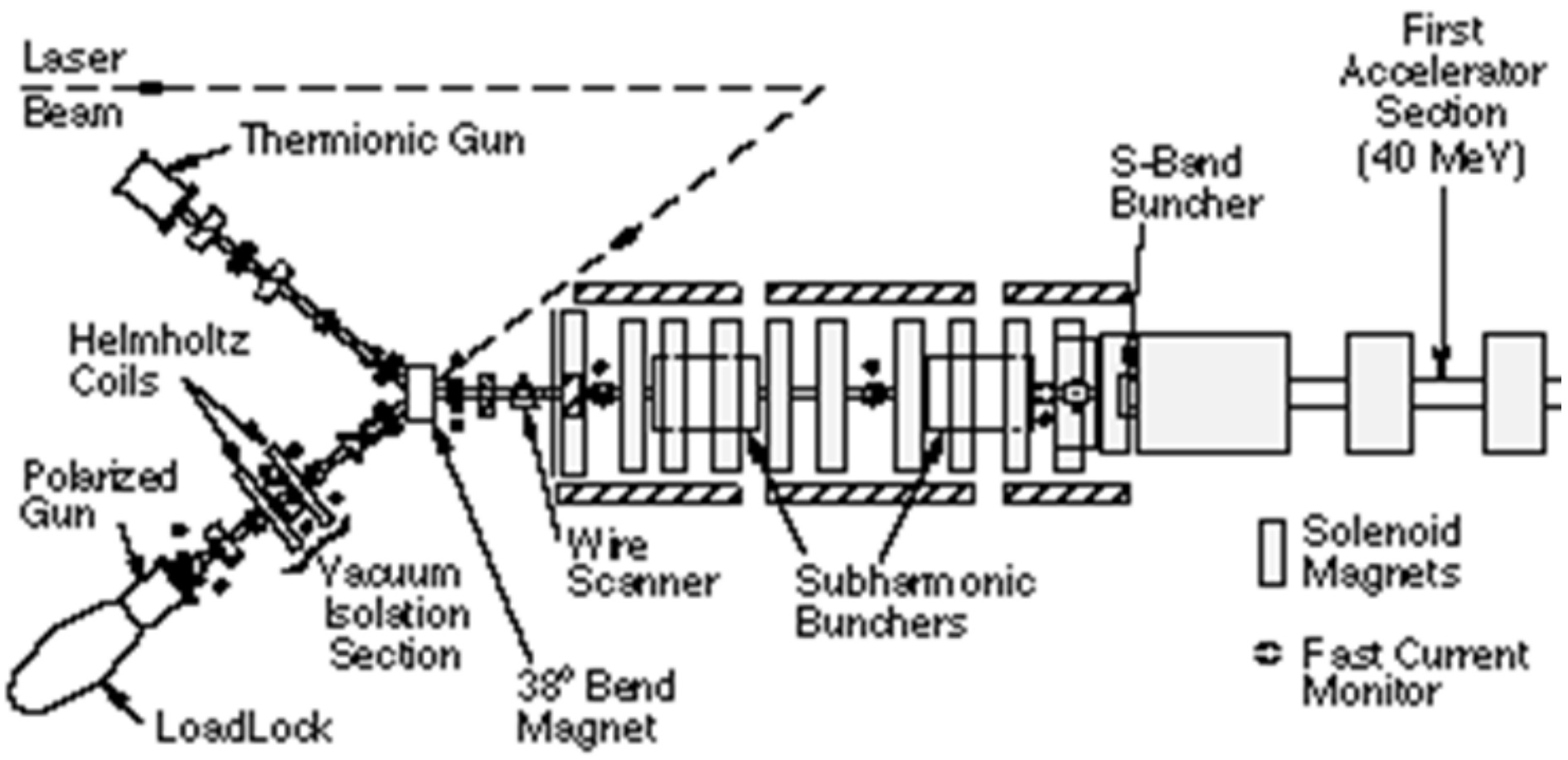

Figure 15.3: Overview of the polarized electron source for SuperB including initial bunch compression.

The polarized electrons are emitted from a GaAs/GaAsP strained layer super-lattice cathode after being struck by a pulsed polarized laser beam. Single or multiple bunches can be emitted depending on how the laser pulse is sub-divided. The properties and capabilities of the gun are listed in Table 15.1. The gun at SLAC is presently available for use in SuperB if desired, although some modest refurbishment is needed to bring the controls to modern standards. Spare parts are also available. The construction of the photocathode is illustrated in Figure 15.4.

Table 15.1: Properties of the polarized electron source

| Parameter | Units | SLC |
|---|---|---|
| Electron charge per bunch | nC | 16 |
| Bunches per pulse | | 2 |
| Pulse rep rate | Hz | 120 |
| Cathode area | $cm^2$ | 3 |
| Cathode bias | kV | -120 |
| Gun to SHB1 drift | cm | 150 |
| Gun $\varepsilon_n$ rms (fm EGUN) | $10^{-6}$ m | 15 |
| RF frequency | MHz | 475 |

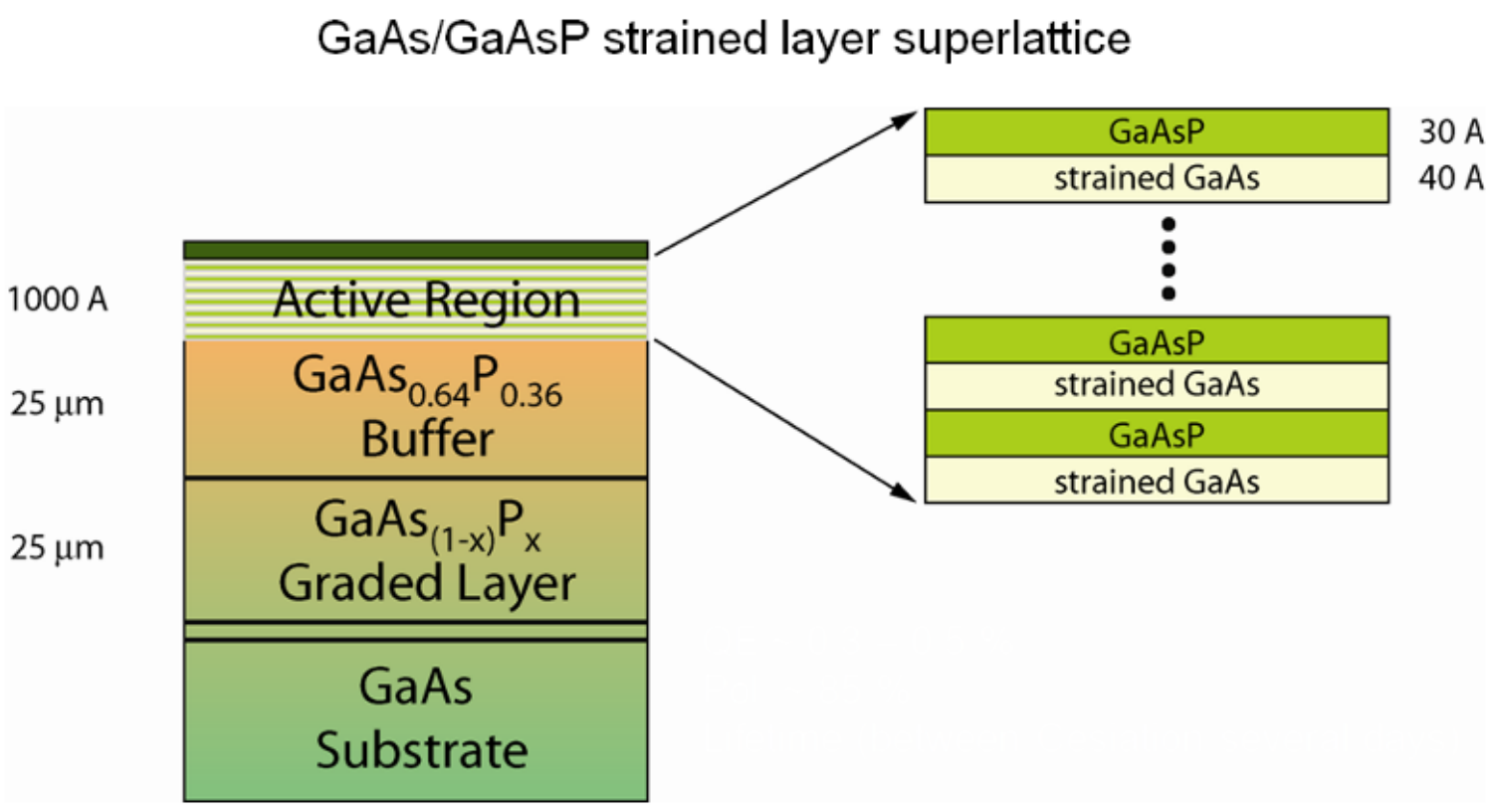

Figure 15.4: Construction details of the strained layer super-lattice of the polarized cathode.

## 15.3 Linac

The SuperB injector will make use of the Polarized Electron Source (PES) developed at SLAC for SLC [1]. It consists of a polarized laser system which drives a GaAs photocathode. Two different trains are produced at 25 Hz repetition rate:

1) high intensity electron bunches (up to 10 nC) for positron production;

2) low intensity electron bunches for injection in the HER.

Bunches are compressed by 2 sub-harmonics cavities and a β-graded S-band short section. The beam is then accelerated along a series of S-band travelling wave structures up to the positron converter area. Here, the high charge electron bunches hit the positron target while the low charge ones, after 20 msec, pass through a 5 mm hole made off axis in the target. Both particle types, at a rate of 50 Hz, are alternately accelerated up to 1 GeV and injected/extracted into/from the Damping Ring (DR) by means of the system of kickers k2L-k3C and k1L-k4C, as shown schematically in Fig. 15.2. The injected bunches are damped for 20 ms, extracted and accelerated to the final energy for injection into the main rings. In the electron mode, 12 out of 29 RF stations of the high energy linac are switched off so it is possible to accelerate the beam to 4.2 GeV without an intermediate extraction and a dedicated transfer line. Similar procedure is adopted in the low energy section, when the injector operates in electron mode. A wall plug power reduction by about 15 % can be achieved in this way.

To have better positron capture efficiency, an L-band 1 GeV linac is used after the positron target.

### *Low energy conversion linac*

The solution proposed in [2] allows producing the positrons at energy between 0.6 and 1 GeV. In this case, as shown in fig.2, after the bunching system, the beam is accelerated up to 0.6÷1 GeV by means of a S-band linac. The high charge electron bunches hit the tungsten target and produce positron bunches that are accelerated to 1 GeV and injected into the DR. In order to increase the capture efficiency of the positrons going out from the target, it is proposed to use an L-band 1428 MHz linac up to 1 GeV before the injection in the DR.

The low charge electron bunches, generated in the next phase, after 20 msec, by-pass the positron target and travel along the linac to the damping ring.

Electron and positron bunches are generated and injected in each main ring every 40 msec (25 Hz) in an alternate sequence while the klystron stations operate at 50 Hz. The injector timing is schematically shown in Figure 15.5.

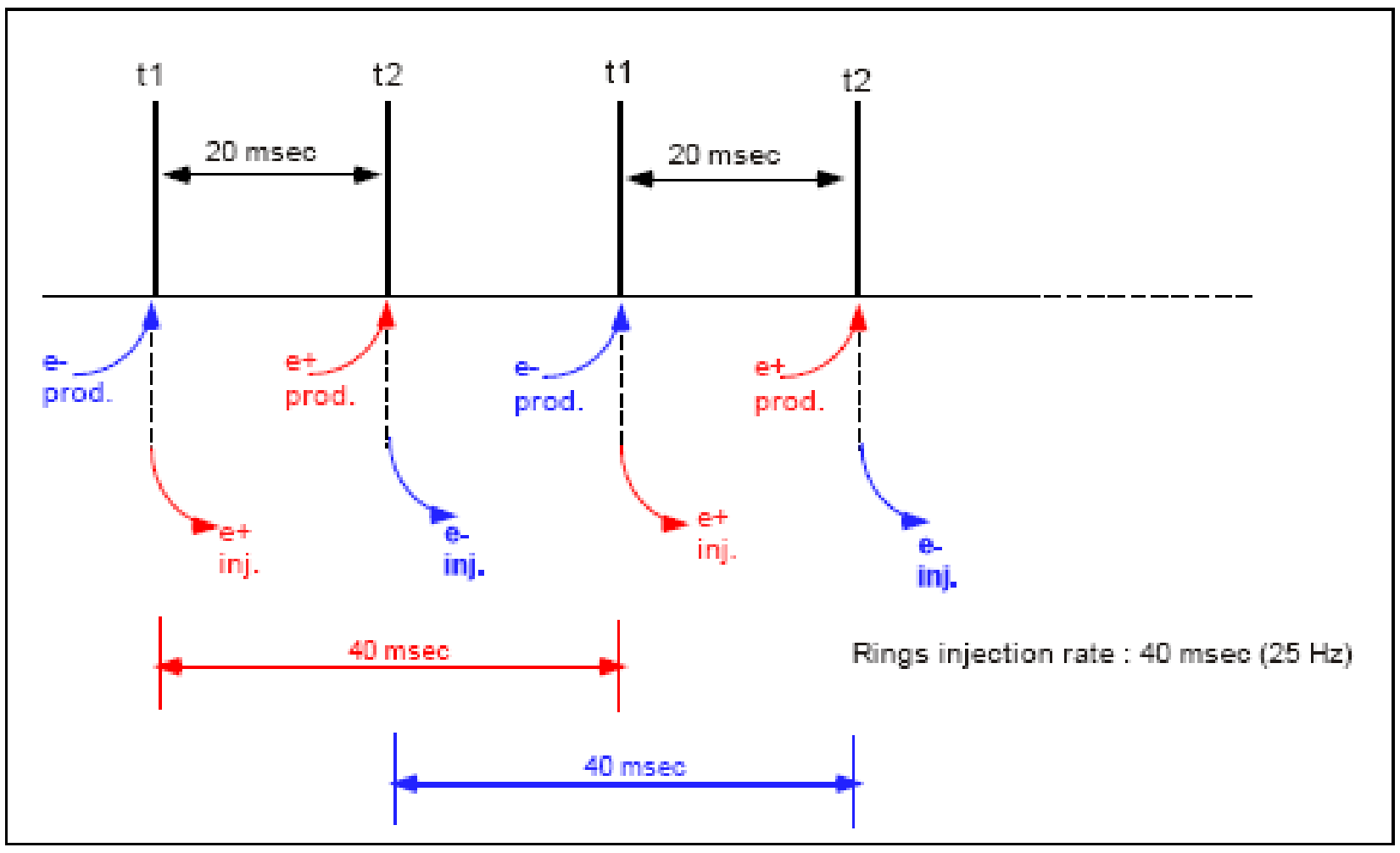

Figure 15.5: Injector timing.

### *High energy conversion linac option*

An alternative option to produce the positron bunches, in case the low energy conversion solution shouldn't guarantee enough positron particles, is shown in Figure 15.6. The positron conversion energy is 6 GeV. Three operation phases, schematically shown in Figure 15.7, are foreseen in this case.

First, high charge electron bunches are accelerated up to 6 GeV and hit the target bulk. The generated positron bunches, collected and accelerated to 1 GeV, are transferred back with an additional transfer line and injected into the DR.

The second phase is the extraction of the damped e+ bunches from the DR and the acceleration, up to the nominal energy, through the target hole. Meanwhile, low charge electron bunches are generated from the gun, accelerated up to 1 GeV and injected into the DR.
Finally, the electron bunches are extracted from the DR and accelerated to the nominal energy passing through the target hole. The three phases are alternated at 50 Hz but each ring is filled every 60 msec, because one phase is used to produce the positron beam.

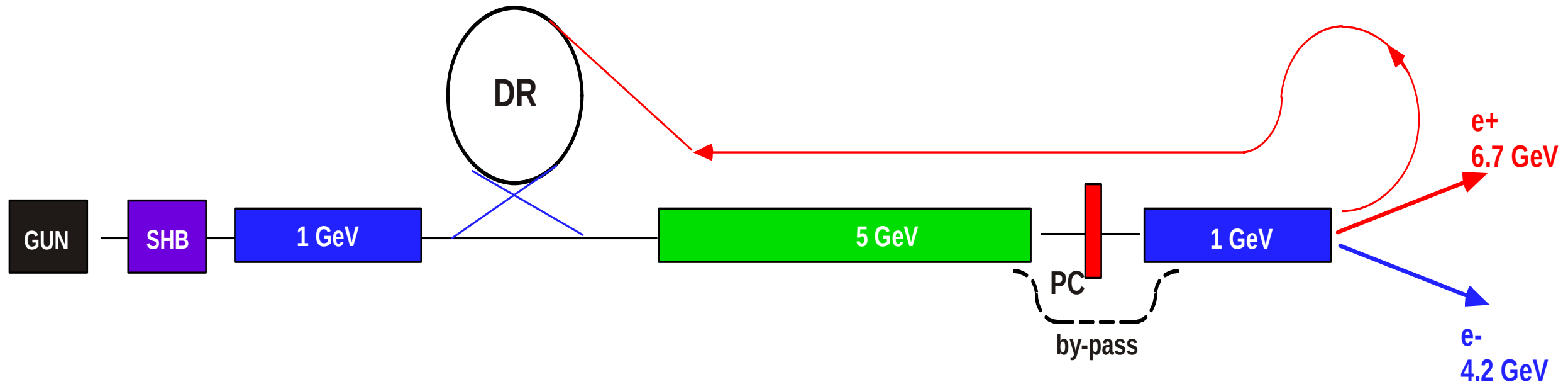

Figure 15.6: Overall layout of the SuperB Injection System for the high energy conversion option.

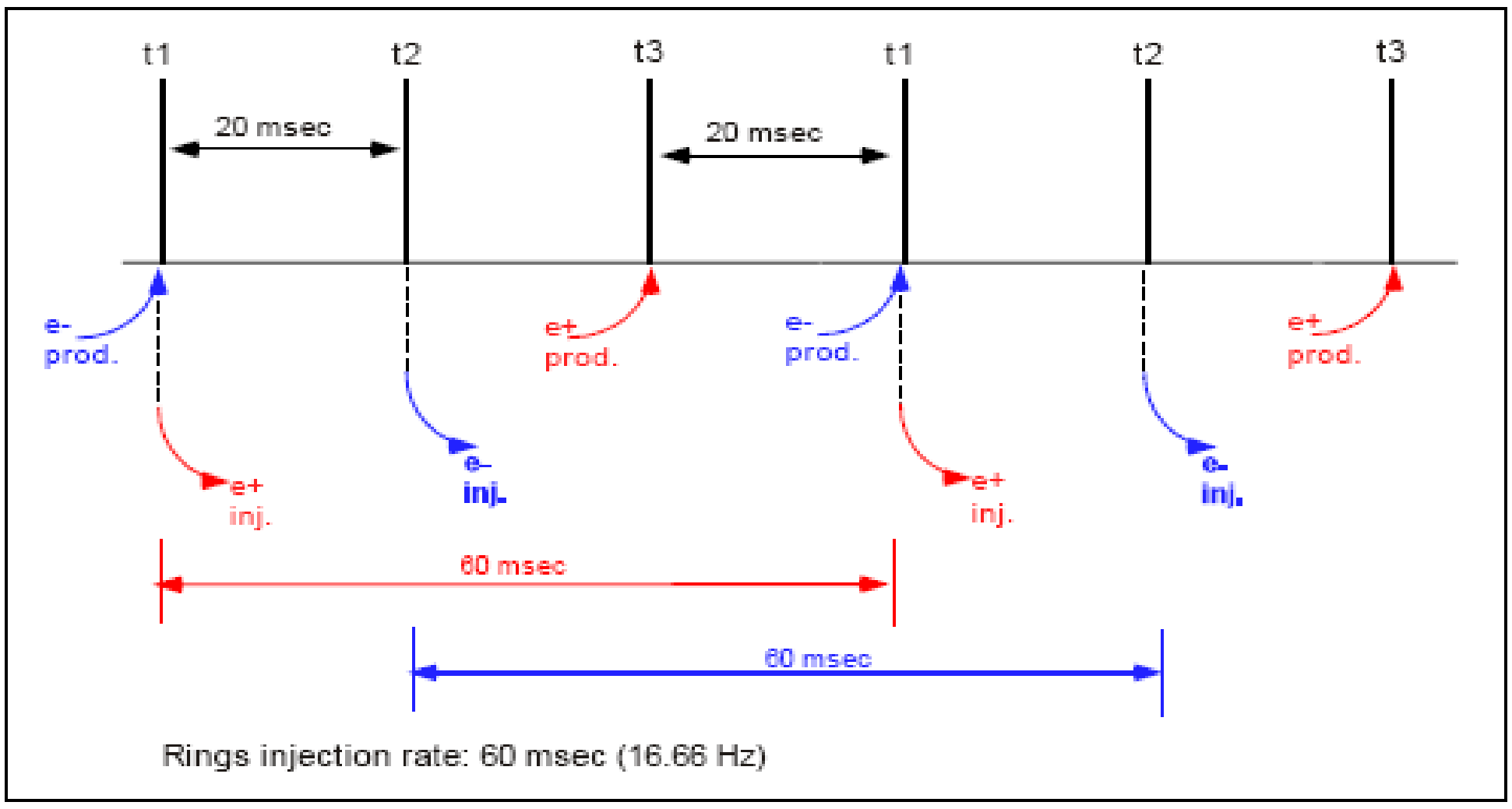

Figure 15.7: Timing of the high energy conversion injector option.

### *Buncher RF system*

The polarized gun can generate electron bunches with charge > 10 nC and length of 10 ns FWHM. In order to accelerate the beam with a S-band linac, the bunches must be compressed down to 10 psec FWHM. To meet this requirement a bunching system must follow the gun before the S-band linac.

Figure 15.8 shows the layout of the proposed Sub-Harmonic pre-buncher (SHB) for the SuperB injector. It consists of 2 standing wave cavities operating at 238 MHz and 476 MHz, the 12$^{th}$ and the 6$^{th}$ sub-harmonics of 2856 MHz respectively, followed by a S-band buncher.

The SH cavities are room temperature, re-entrant type, copper resonators. This type of bunching system has been already adopted or proposed in other laboratories and projects [2, 3].

Solid state or IOT pulsed amplifiers, needed to supply the SH cavities, are commonly available by the broadcasting market since the frequencies lay in the TV UHF-band. The output power of the UHF amplifiers is between 10 and 20 kW.

The SH pre-bunching cavities compress initially the bunches to ≈ 20 psec FWHM. A β-matched, S-band TW buncher, captures the beam coming out from the SH cavities and compresses the bunches to 10 psec FWHM. The β-graded buncher and the first few cells of the following TW pre-accelerator are immersed in a solenoidal field to focus the beam. A 2856 MHz, 3 meters, 25 MV/m, β = 1, TW section increases the beam energy to ≈ 80 MeV. The S-band buncher and the following TW section are supplied with a 20 MW peak klystron.

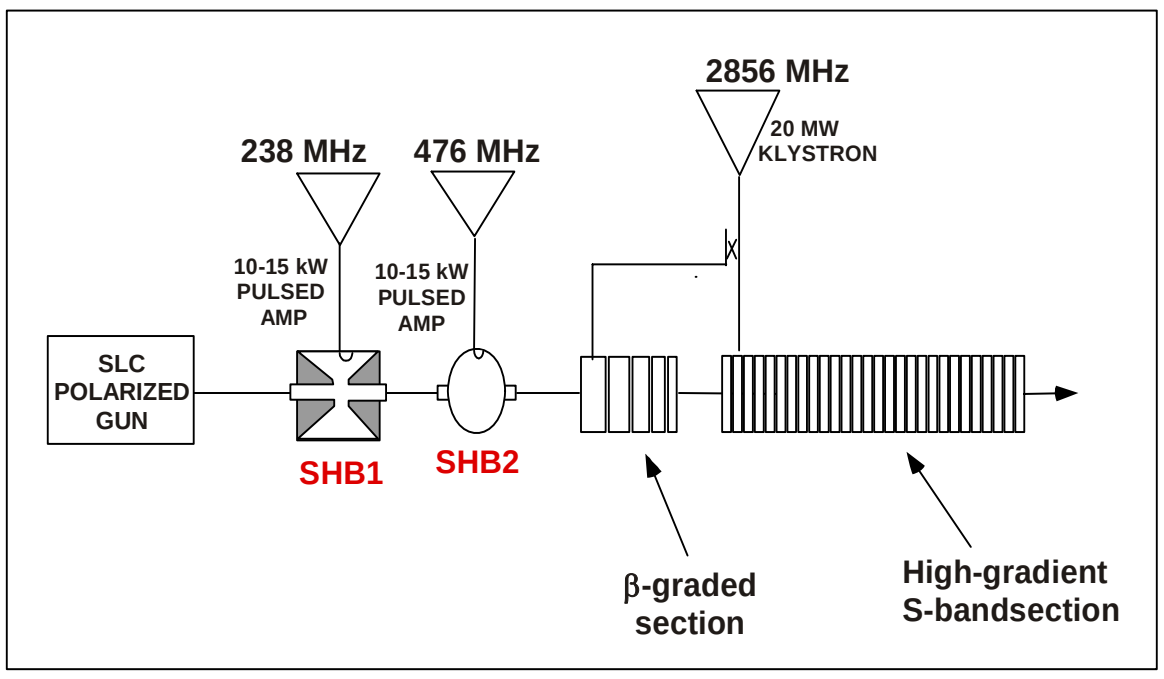

Figure 15.8: RF Layout of the bunching system.

### *S-band accelerating structures*

The accelerating structures of the SuperB linac are travelling wave (TW), constant gradient (CG), 2π/3, 3 mt. long, 2.856 GHz units. They are made of a series of 86 RF copper cells, joint with a brazing process performed in high temperature, under vacuum furnaces.

The cells are coupled by means of on-axis circular irises with decreasing diameter, from input-to-output, to achieve a constant-gradient configuration. The RF power is transferred to the accelerating section through a rectangular slot coupled to the first cell. The power not dissipated in the structure (about 1/3rd) is coupled-out from the last RF cell and dissipated on an external load.

The industrial companies, which can develop the accelerating structures, are only a few in the world. The fabrication is a complex task that requires specialized know-how, availability of very advanced equipment and facilities and top-level organization.

The maximum achievable accelerating gradient is the most important parameter of such devices. The SuperB operates at an average gradient of 23.5 MV/m that is a medium level accelerating field. Nevertheless, it requires the use of selected materials, precise machining, high-quality brazing process, surface treatments and cleaning, ultra-pure water rinsing, careful vacuum and RF low power tests. Table 15.2 gives the main parameters of the sections. Figure 15.9 shows an S-band accelerating structure before being installed on the beam-line.

In order to maintain the structure tuned to the 2π/3 mode, that guarantees the necessary cumulative energy gain for the beam particles, the accelerating sections are kept at very constant temperature (ΔT = ± 0.1°C) by means of regulated cooling water systems.

Table 15.2: S-band sections parameters

| | |
|---|---|
| Frequency | 2.856 GHz |
| Type | TW,CG |
| Structure | Disk-loaded |
| Mode of operation | 2π/3 |
| Phase velocity | c |
| Period | 3.499 cm |
| Number of cells | 86 (including couplers) |
| Attenuation constant | 0.57 nepers |
| Normalized group velocity | 0.0202 to 0.0065 ($V_g$/C) |
| Shunt impedance | 53 to 60 MΩ/m |
| No load energy (50 MW input) | 70 MeV (theoretical) |
| Bandwidth (VSWR ≤ 1.2) | ≥ 4 MHz |
| Phase shift per cell | 120 |
| Filling time | 0.85 μsec |
| Q of structure | 13400 (approx) |
| In/Out VSWR | ≤ 1.1 |

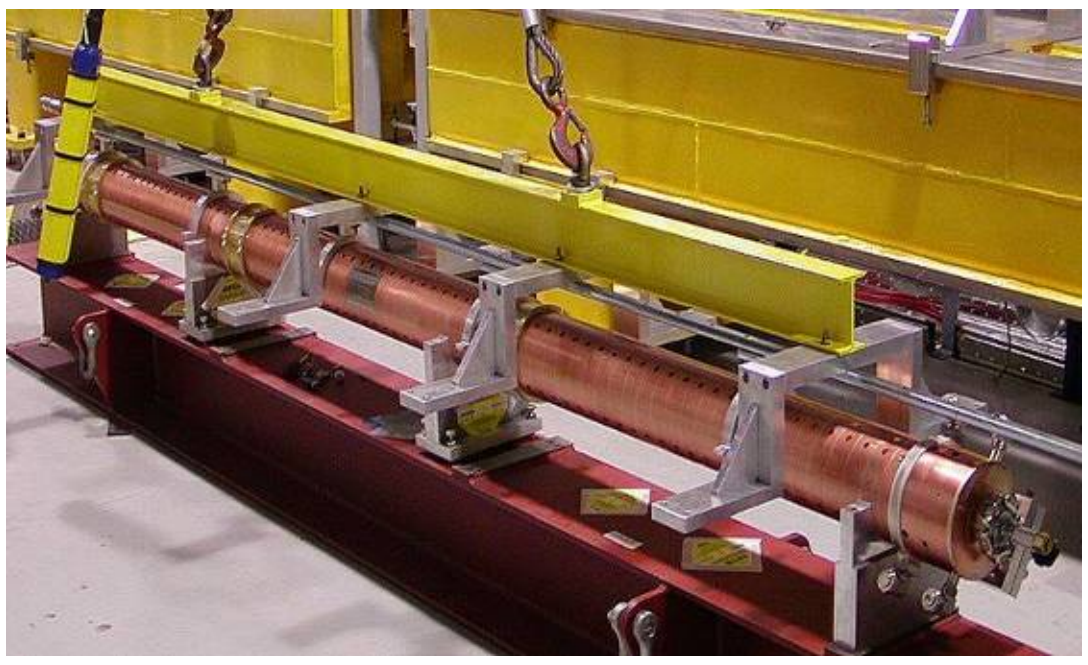

Figure 15.9: S-band accelerating structure.

### RF Power Sources

The RF power sources for the SuperB injector consist of 60 MW peak S-band klystrons. Klystrons that meet the requirements of the SuperB linac are available on the market. A set of klystron parameters is given in Table 15.3. RF power sources of similar specifications are used in other accelerator laboratories. Each klystron, equipped with beam focusing coils, will be supplied by a High Voltage (HV) Modulator and installed in the upper tunnel of the linac.

Table 15.3: SuperB klystron main specifications

| | |
|---|---|
| Frequency | 2.856 GHz |
| RF pulse duration | 4 μsec |
| Repetition rate | 50 pps |
| Cathode voltage | 350 ÷ 370 kV |
| Beam current | 400 ÷ 420 A |
| HV Pulse width FWHM | 6 μsec |
| RF Peak Power | 60 MW |

Basically, a pulsed Modulator consists of a HV charging unit, a Line-Type Pulse Forming Network (PFN) and a 1/n HV pulse transformer, immersed in a tank filled with insulating oil. The system, schematically shown in Figure 15.10, generates almost rectangular HV pulses, applied to the klystron cathode, after the PFN discharge that occurs when the HV switch, that can be a thyratron or a solid state device, is operated by a trigger signal. The nominal HV pulse duration is 6 μsec FWHM with rise and fall time, determined by the PFN parameters, respectively of 0.5 and 1 μsec.

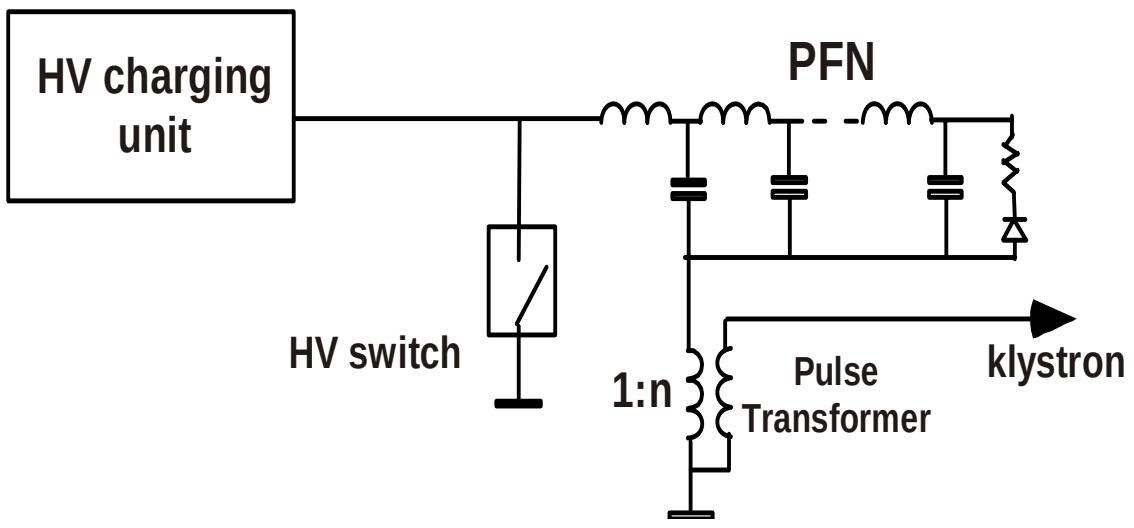

Figure 15.10: Schematic layout of a Pulsed Power Modulator.

Some manufacturers replace the line-type PFN with a series of solid-state switching boards. The goal is to make the modulator size compact and lower the primary voltages but the long term reliability of this solid state solution is still being proved. The HV Modulators for scientific applications are very special systems produced by a few specialized industries. The main features are the very high voltage of the modulator pulse needed to drive the klystron and some tight requirements regarding amplitude and pulse-to-pulse phase stability. Table 15.4 gives the parameters for the SuperB injector modulator.

Table 15.4: HV Pulsed Modulator parameters

| | |
|---|---|
| Pulse primary transformer voltage | 25 kV |
| Pulse secondary transformer voltage/klystron gun | 370 kV |
| Pulse secondary transformer current/klystron gun | 420 A |
| Pulse transformer ratio | 1/15 |
| High voltage pulse duration (FWHM) | 6 μsec |
| High voltage rise/fall time (0 to 90%) | 0.5 /1 μsec |
| Pulse flatness during flat-top | ± 0.1 % |
| Pulse to pulse Voltage fluctuation | ± 0.3 % |
| Pulse repetition rate | 50 Hz |

### RF Power Distribution

The layout of the S-band power station for the SuperB is shown in Figure 15.11. The 60 MW klystron sources feed, through the KEK-type energy compressor, 3 TW accelerating structures. To divide equally the RF power among 3 units, a 4.8 dB directional coupler is used to draw 1/3rd of the full power for feeding the first section. A 3 dB directional coupler split into two halves the remaining power to the following sections.

About 50 MW peak are available to each accelerating structure after the pulse compressor. Such a power level produces an average accelerating gradient of 23.5 MV/m that is an energy gain of about 210 MeV per RF station. A network of rectangular WR284 copper waveguides distributes the RF power to the Pulse Compressors and to the accelerating structures. The waveguides are pumped down to $10^{-8}$ mbar with a distributed pumping system and are connected to the accelerating structures with ceramic RF windows to protect the beam line vacuum.

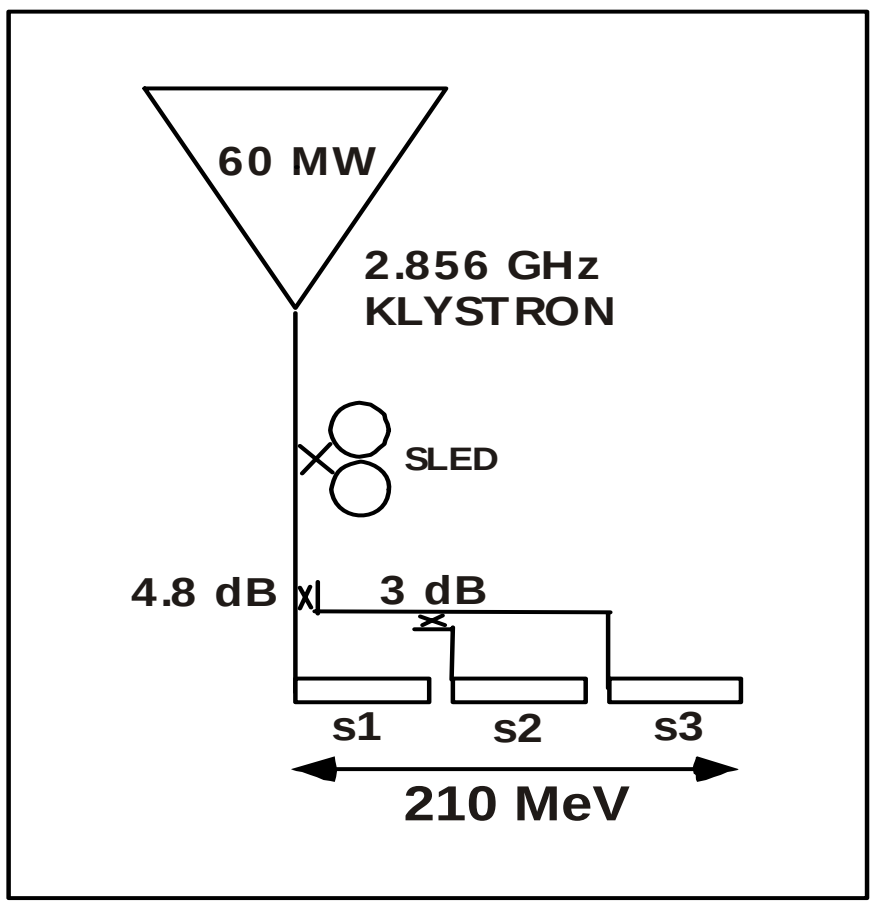

Figure 15.11: Layout of the SuperB injector RF Power Station.

### The L-band Linac

The need to improve the positron production efficiency requires to use an L-band linac, just after the positron converter. Since the beam consists of a train of 5 bunches at 4.2 nsec apart, the L-band frequency must be a sub-harmonics of the 2856 MHz, that is 1428 MHz. L-band room temperature linacs are not common in the accelerator laboratories. However, the design and the manufacturing of 1428 MHz accelerating structures shouldn't be a hard task. By scaling the 1300 MHz Linac of the ISIR, Osaka University [4], a possible set of L-band structure parameters is shown in Table 15.5.

Table 15.5: L-band sections parameters

| | |
|---|---|
| Frequency | 1.428 GHz |
| Type | TW, Constant Gradient |
| Structure | Disk-loaded |
| Mode of operation | 2π/3 |
| Length (L) | 3 m |
| Period | 7.00 cm |
| Attenuation constant (τ) | 0.3 nepers |
| Shunt impedance (Z) | 45 to 50 MΩ/m |
| Filling time | 1.7 μsec |
| Q of structure | 18000 (approx) |
| In/Out VSWR | ≤ 1.2 |

With the above parameters, the energy gain per section is:

$$V_{MeV} = [PZL \bullet (1-e^{-2\tau})]^{1/2} \approx 7.8 \bullet (P_{MW})^{1/2}$$

Therefore, with a 30 MW - 4 μsec klystron, two L-band, 1428 MHz structures can be driven in pairs, obtaining:

$$E_{Kly} = 2 \cdot 7.8 \cdot (15_{MW})^{1/2} \approx 60 \text{ MeV}$$

per power station, i.e. 10 MV/m.

The energy gain per power station could be 100 MeV (16.6 MV/m) with the use of a pulse compressor that should be developed ad hoc. A 1 GeV L-band linac for the SuperB injector will consist therefore of 17 power stations and 34 L-band accelerating sections (or 11 stations and 22 sections with the pulse compressor). The high energy section of the linac, after the DR, is made of S-band structures (see Fig. 15.12).

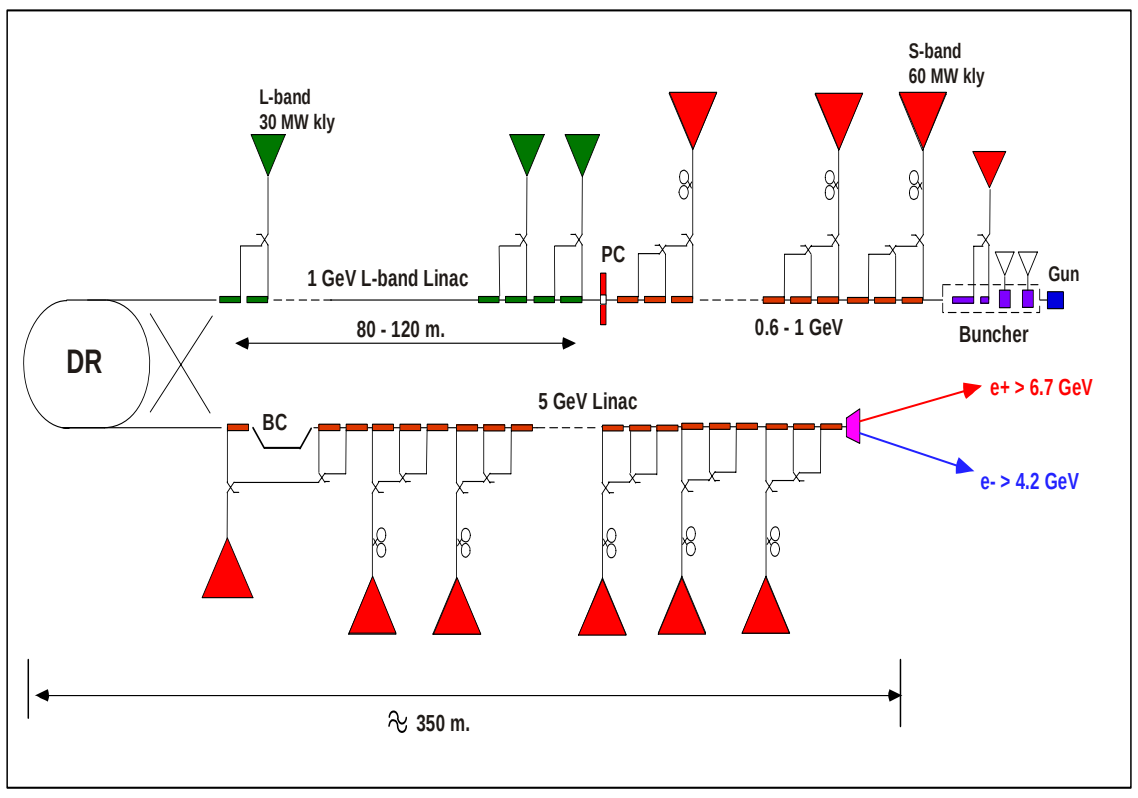

Figure 15.12: RF Layout of the injector with the L-band linac.

### Low Level RF

The Low level RF system (LLRF) provides:

a) RF power to the RF stations, with proper phase and amplitude.

b) interface to set and monitor the operating parameters

c) interlocks to protect people, machine, klystrons, windows, etc.

The RF distribution system is based on the use of a 476 MHz Main Drive Line made of a rigid, low losses, low thermal variations coaxial cable which transports the reference signal along the Linac. The 476 MHz signal is synchronized with the general master oscillator of the SuperB main rings. The 1428 and the 2856 MHz frequencies are derived from the master signal by means of x3 and x6 multiplications.

In addition, Phase Reference Lines for the L-band and S-band systems, cover the linac length with low-thermal cables.

The LLRF must guarantee the beam energy stability within a few tens of %. Therefore, in addition to working with the klystrons in saturation, phase loops around the klystrons will be implemented.

Moreover, one S-band cavity at the linac-end, driven by a dedicated low power klystron, can be used for beam energy control. In this station, the klystron would not work in saturation but the output power would be modulated by a signal proportional to the beam energy deviation.

Extensive use of the digital technology is foreseen for signal monitoring and feedback loops implementation. A simplified scheme of the LLRF is shown in Figure 15.13.

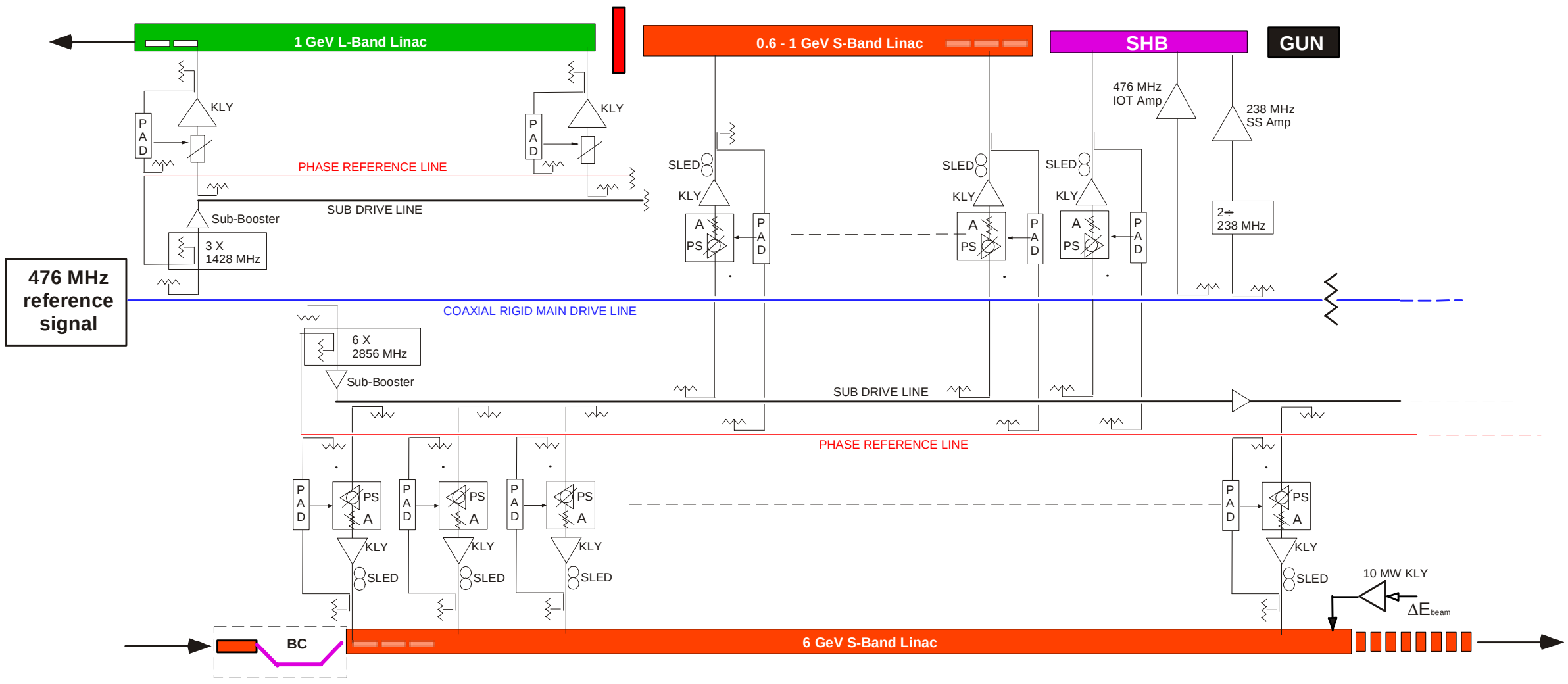

Figure 15.13: Synthetic scheme of the Linac LLRF system.

## 15.4 Positron production

In SuperB, the positron source performances are constrained by the required injection rate in the LER and HER ring and by the injection acceptance of the Damping Ring. At present, the injection scheme foresees a 50 Hz operation of the injection complex, with half of the repetition frequency to be used for the positron production, capture and injection.

Every 40 ms a train of five positron bunches will be sent in the DR to reduce their emittance. This implies that a charge per bunch of ~ 300 pC is required to fulfill the SuperB constraints imposed by the beam lifetime. The phase space volume of the injected bunches must fit the DR acceptance of ±1% energy spread (± 10 MeV) and an emittance of 3 $10^{-6}$[m rad].

The reasonable required charge per bunch, the impressive performances of the SLAC electron gun (up to 16 nC per bunch) and the necessity to reduce the injector total length and costs suggested to propose a positron source based on a low energy drive beam. In the studied configuration the positrons are generated by a train of 5 $e^-$ bunches – 10nC each, impinging on an amorphous tungsten target at 600 MeV. In the target the electrons create gammas by bremsstrahlung that in the nuclear field of the target will generate $e^+e^-$ pairs. At the target exit the produced positrons show an important angular and energy spread. To reduce the angular dispersion an Adiabatic Matching System (AMD) is foreseen, where a decreasing magnetic field (from 6 to 0.5 T in 0.5 m) transforms beam divergences in positions. Immediately after the AMD the positrons are bunched and accelerated by a pre-injector system of accelerating cavities, surrounded by coils which guarantee the transversal confinement with a constant 0.5 T magnetic field. The accelerating capture section takes the beam up to the energy of ~ 300 MeV, downstream which, the transverse confinement is assured by a FODO lattice.

In Figure 15.14 the schematics of the positron source system is illustrated.

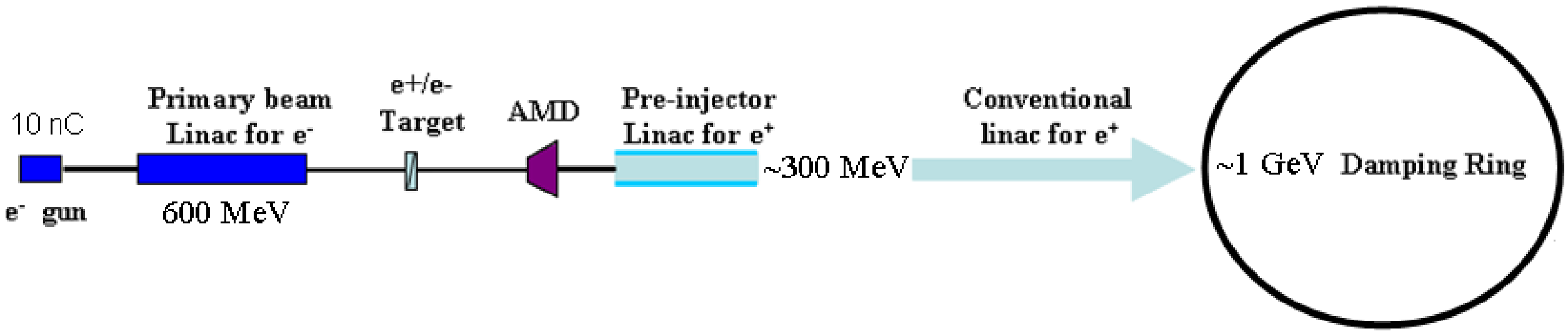

Figure 15.14: Scheme for the SuperB low energy drive beam positron source. A drive beam of 600 MeV impinges on a tungsten target. The AMD collects the positrons and injects them in the pre injector, which accelerates the bunches up to ~ 300 MeV.

Several simulations, with different methods, were carried out to estimate the efficiency of this positron source. The pairs production is estimated at 600 MeV by GEANT4 and is optimized at 1.7 e+/e- for a target thickness of 1.04 cm. The produced positrons are then given as inputs to ASTRA [1] simulation for the positrons capture and acceleration.

Optimization studies were carried out taking into account different AMD configuration and acceleration section structures. The first approach was to use the standard SLAC cavities for capture and acceleration. This strongly reduces the geometrical acceptance of the system due to the small radius of the cavity iris. Therefore, to increase the geometrical acceptance, the capture in a L band system was proposed. To maintain the harmonic of the SLAC cavities a frequency of 1.428 GHz, requiring a new design, was adopted.

Different scenarios were studied to optimize the positron yield at the injection depending on the frequency and on the phase of the capture cavities. Both S and L band were tested in full acceleration scenario, as well as in the scheme proposed at SLAC [2], where the first cavity is decelerating to improve the bunching. In the end a new idea was proposed, where the L band capture is performed, but the first decelerating cavity is operated in the TM020 mode. This allows for preserving the large geometrical acceptance of the L band cavity but at the same time for increasing the bunching effect of the decelerating phase.

Results are summarized in Table 15.6 below. For comparison purposes the results are given, for each scenario, with a similar cavity peak gradient of 25 MV/m. More studies [3] have been performed including lower peak gradient for the 1.428 GHz case and have shown negligible differences to the results of this table.

Table 15.6: Different scenarios for the positrons capture with similar maximum cavity peak gradient for acceleration.

| Scenario | 1 | 2 | 3 | 4 |
|---|---|---|---|---|
| RF (MHz) | 2856<br>full acceleration | 2856<br>Deceleration and acceleration | 1428<br>Deceleration and acceleration | 3000 deceleration<br>1428 acceleration |
| Mean Energy (MeV) | 302 | 287 | 295 | 333 |
| $E_{rms}$ (MeV) | 21.4 | 32.3 | 16.83 | 5.2 |
| $Z_{rms}$ (mm) | 2.7 | 6.4 | 8.89 | 3.5 |
| $X_{rms}$ (mm) | 3.8 | 4.4 | 8.0 | 8.1 |
| $X'_{rms}$ (mrad) | 1.02 | 1.11 | 1.69 | 1.4 |
| $\varepsilon_x$ =X'X (mm.mrad) | 3.8 | 4.6 | 13.0 | 11.4 |
| Total Yield (%) | 2.8 | 7.53 | 32.3 | 31.9 |
| Yield ± 10 MeV (%) | 1.3 | 3.9 | 19.6 | 29.3 |

It is worthwhile to note that, especially for the last case, the positron yield accepted in a low energy spread is important. In fact, taking into account a 10 nC drive beam, this will result in ~ 3 nC per bunch in the accepted energy spread (also if some % can be lost in the transport to 1 GeV). Regardless, this is a factor of 20 more than the required bunch intensity in SuperB. The longitudinal phase space of the fourth scenario is showed in Figure 15.15.

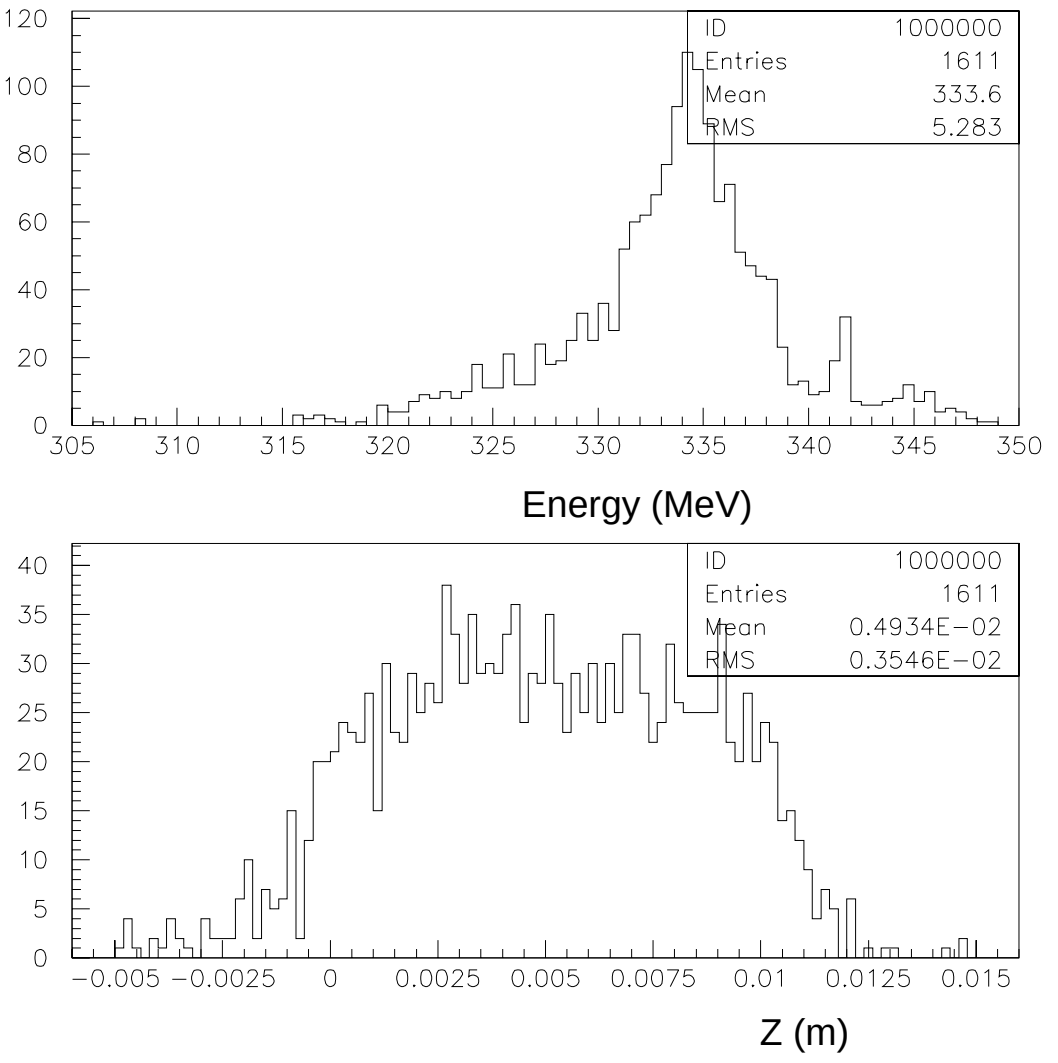

Fig. 15.15: Longitudinal phase space of the captured positrons, at the exit of the pre-injector in the case of deceleration in the TM 020 mode and acceleration at 1.428GHz. One can appreciate the very good bunching.

This gives very good confidence as far as the transport losses and transverse emittance selection are concerned, before injection. It is important to remember there are still some 'safety knobs' as far as the total positron rate is concerned, like the number of bunches per train and the energy of the drive beam.

These studies demonstrate that the low energy solution for the SuperB positron source is feasible.

## 15.5 Damping ring

The tight tolerances on the injected beam due to the small acceptance of the collider main rings led to the design of a small emittance, small energy spread and short bunch length Damping Ring which can store alternatively both electrons and positrons at 1 GeV.

The lattice is based on the same kind of cell adopted for the SuperB main rings. The phase advance per cell of 0.75*2π in the horizontal plane and 0.25*2π in the vertical one helps in cancelling the non linear contributions of the chromaticity correcting sextupoles. This kind of cell, due to the small dispersion in the dipoles, yields a low value of the momentum compaction, and therefore a short bunch length, which, however, must be further reduced by a bunch compression system at extraction. Betatron damping times slightly larger than 7 msec ensure adequate reduction of the injected beam emittance and energy spread at the maximum repetition rate of 50 Hz foreseen by the injection scheme. Figure 15.16 shows a schematic layout of the ring, while Table 15.7 summarizes the main parameters of the damping ring.

Electrons are injected at one of the two septa indicated with S in the figure and extracted from the other one. Positrons follow the opposite path with the same fields in the ring. A train of 5 consecutive bunches from the Linac are stored in the damping ring at each injection pulse by means of the fast kicker, indicated with K in the figure, downstream the septum, damped and finally extracted by the second kicker upstream the extraction septum. Fig.15.17 shows the optical functions of half ring, the second half being identical. Each half ring consists of three cells surrounded by two half dispersion suppressors providing enough space for injection septa and kickers with the correct horizontal phase advance between them.

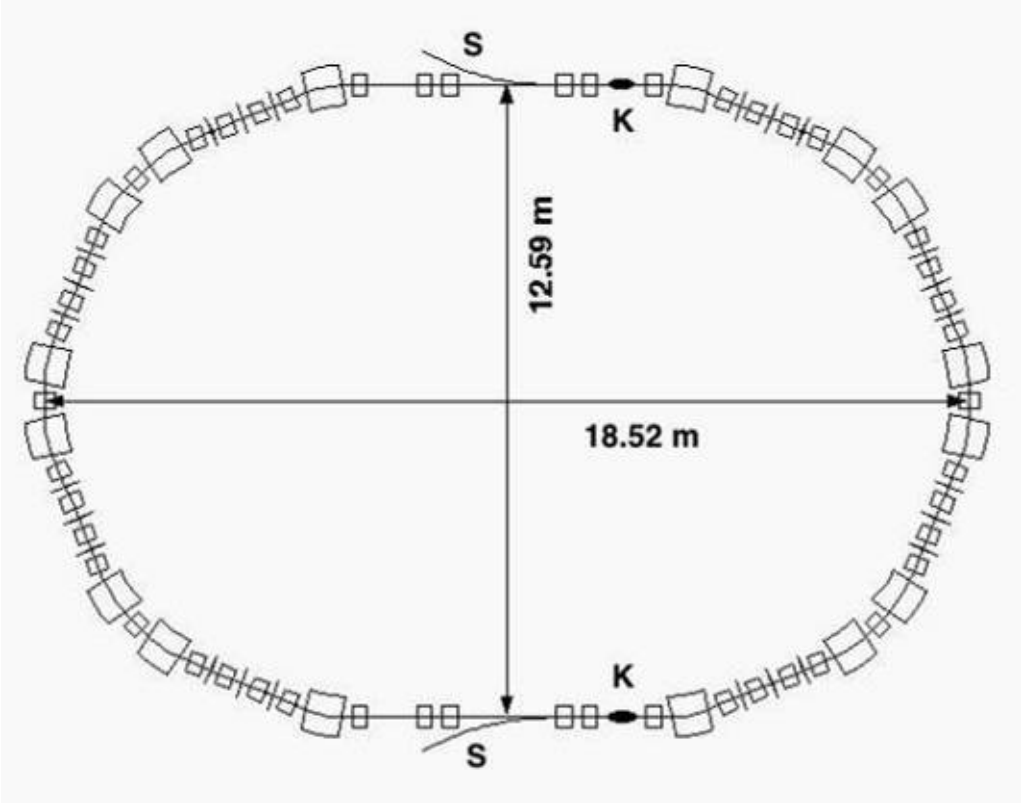

Figure 15.16: Damping ring layout.

Table 15.7: Damping ring main parameters

| Parameter | Units | |
|---|---|---|
| Energy | GeV | 1.0 |
| Circumference | m | 51.1 |
| Horizontal betatron tune | | 7.403 |
| Vertical betatron tune | | 2.717 |
| Horizontal chromaticity | | -11.5 |
| Vertical chromaticity | | -6.5 |
| Horizontal phase advance/cell | degrees | 270 |
| Vertical phase advance/cell | degrees | 90 |
| Maximum horizontal beta | m | 7.9 |
| Maximum vertical beta | m | 7.3 |
| Maximum dispersion | m | 0.77 |
| Equil. horizontal emittance | nm | 23 |
| Momentum compaction | | 0.0057 |
| Hor. betatron damping time | msec | 7.26 |
| Vert. betatron damping time | msec | 7.36 |
| Synchrotron damping time | msec | 3.70 |
| Equilibrium energy spread | | $6.2x10^{-4}$ |
| RF frequency | MHz | 475 |
| Harmonic number | | 81 |
| RF peak voltage | MV | 0.5 |
| Bunch length | cm | 0.48 |

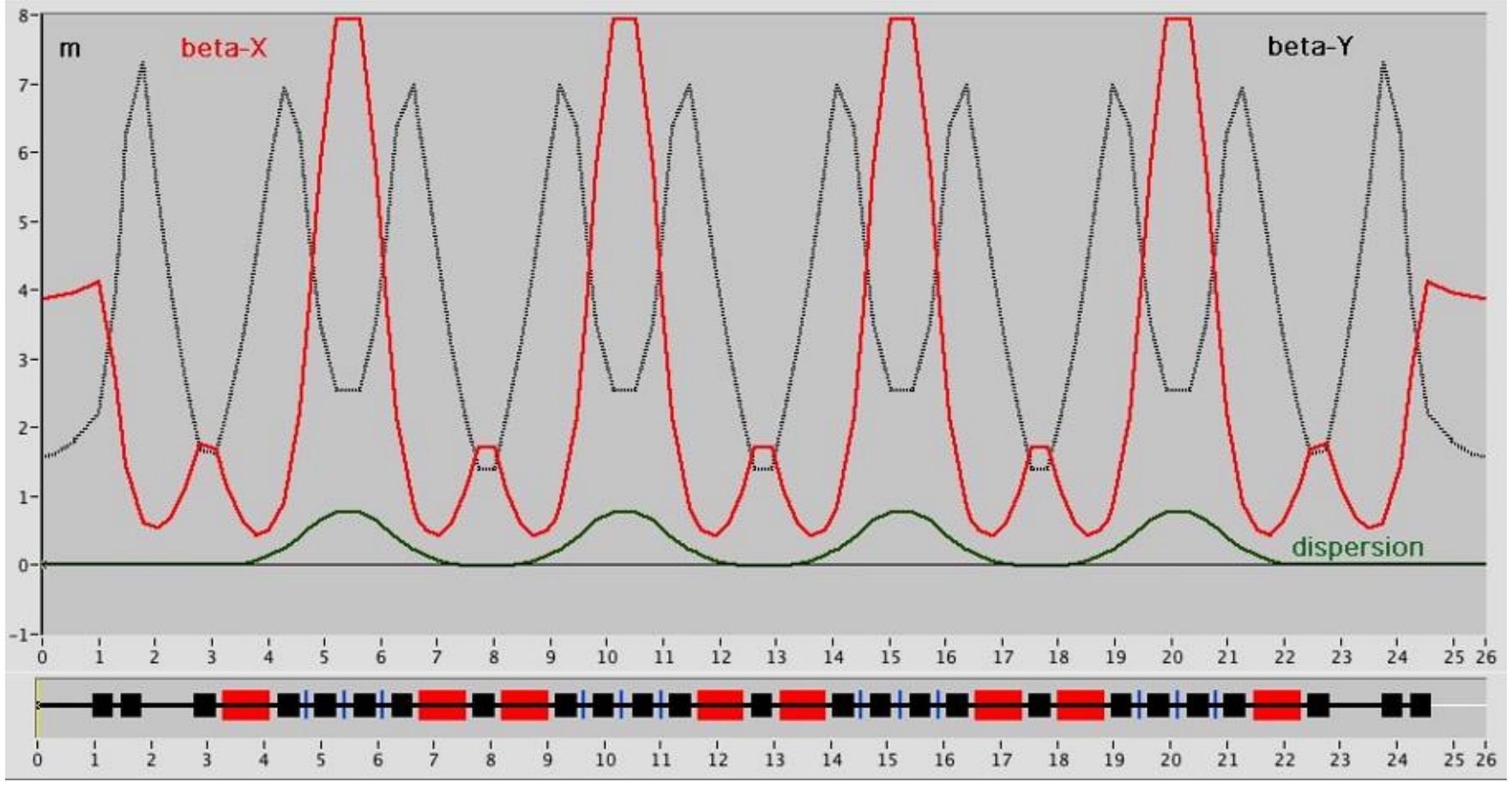

Figure 15.17: Optical functions of half damping ring. Dipoles are red, quadrupoles black, sextupoles blue.

The quadrupoles in the cells are powered in three independent families, while five other power supplies are needed for the dispersion suppressors. Table 15.8 summarizes the requirements for the magnet system.

Table 15.8: Magnet system parameters

| | Units | |
|---|---|---|
| Number of dipoles (rectangular) | | 16 |
| Dipole length | m | 0.75 |
| Dipole field | T | 1.745 |
| Number of quadrupoles | | 50 |
| Quadrupole length | m | 0.3 |
| Maximum gradient | T/m | 23.3 |
| Number of independent power supplies | | 8 |
| Number of sextupoles | | 24 |
| Maximum integrated gradient $\int(\partial^2B/\partial x^2)ds$ | T/m | 19.0 |
| Number of independent power supplies | | 2 |

Bunches will be injected and extracted at a repetition rate of 50 Hz, synchronized with the Linac and the RF system of the Main Rings. The emittance of the extracted beam is:

$$\varepsilon = \varepsilon_i e^{-\frac{2t}{\tau}} + \varepsilon_o \left(1 - e^{-\frac{2t}{\tau}}\right)$$

where $\varepsilon_i$ is the emittance of the injected beam (3.0 mm.mrad for $e^+$, much smaller for the $e^-$), $\varepsilon_0$ the equilibrium emittance of the damping ring, t the storage time (20 msec) and $\tau$ the betatron damping time (7.26 msec). The result is 35 nm (12. nm from the first term and 23. nm from the second) for $e^+$, which after adiabatic damping due to the acceleration in the Linac to 6.7 GeV, becomes 5.2 nm. For $e^-$ the emittance injected in the damping ring is much smaller and at the exit it will be equal to the DR equilibrium emittance; after acceleration in the Linac to 4.2 GeV, it becomes 5.5 nm.

With a RF voltage of 0.5 MV, the low current bunch length has an r.m.s. value of 4.8 mm. The corresponding energy spread at the end of the Linac would be ≈4% at one standard deviation of the bunch distribution. A bunch compressor at extraction is therefore required, with a compression factor of 4 to reach a final energy spread of 0.3%, about half the off-energy dynamic aperture of LER.

The phase advance of the basic lattice cell ensures complete cancellation of the aberrations induced by the sextupoles if they are separated by two cells. Due to the small number of cells (3 in each half ring) only two sextupoles can be used to correct chromaticity in each plane in a non-interleaved scheme, leading to extremely large required gradients. An interleaved scheme with 8 horizontally focusing sextupoles placed at the boundary of each cell and 16 vertically focusing ones inside each quadrupoles doublet in the cell has been adopted, obtaining a smooth distribution of moderate gradient sextupoles.

Figure 15.18 shows the energy and oscillation amplitude dependence of the tunes with and without sextupoles, while Figure 15.19.represents the dynamic aperture on energy and at ±1% energy deviation.

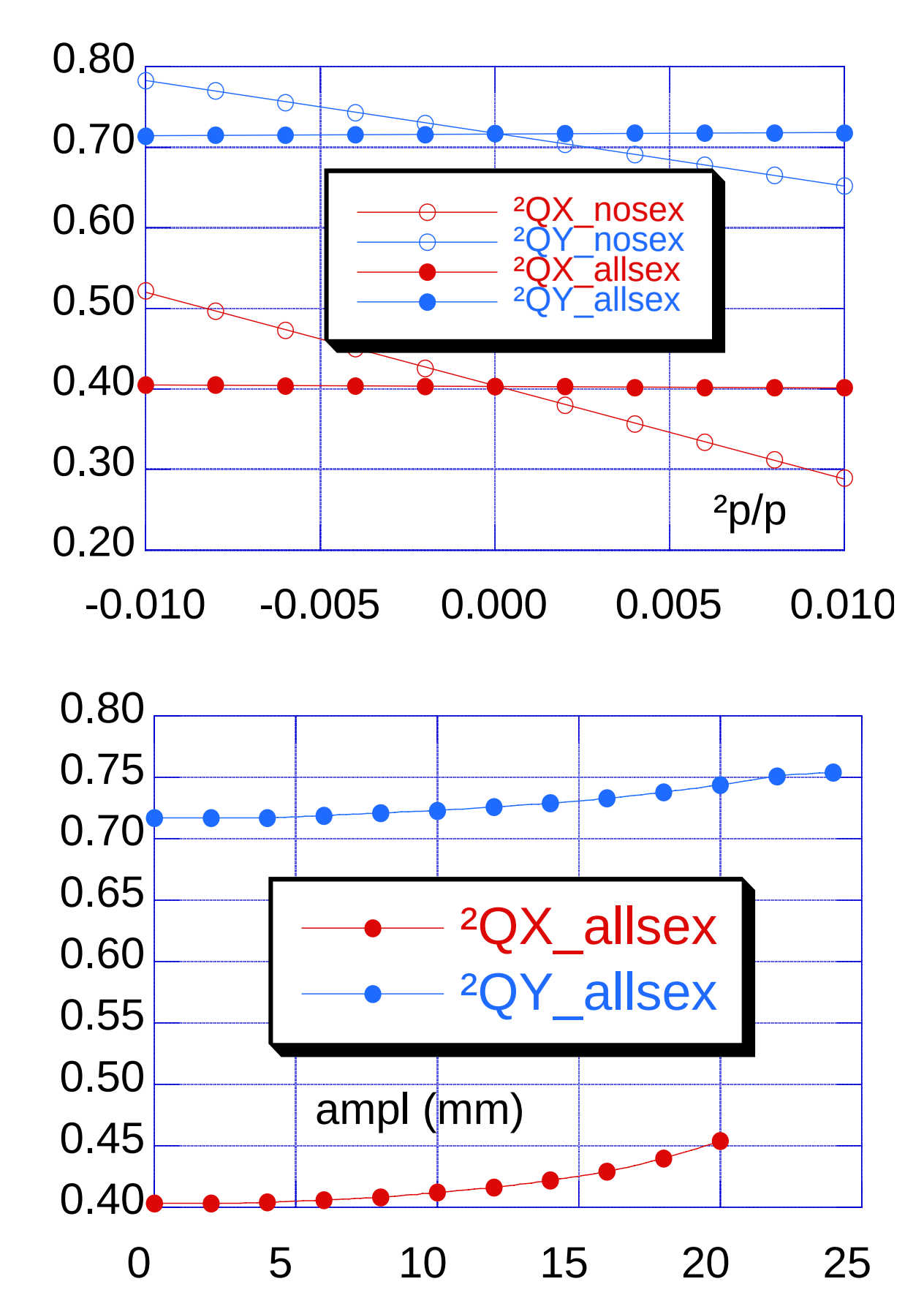

Figure 15.18: Tune dependence on energy deviation (top) and oscillation amplitude (bottom).Red: horizontal, blue: vertical. Empty dots: without sextupoles, full dots: with sextupoles.

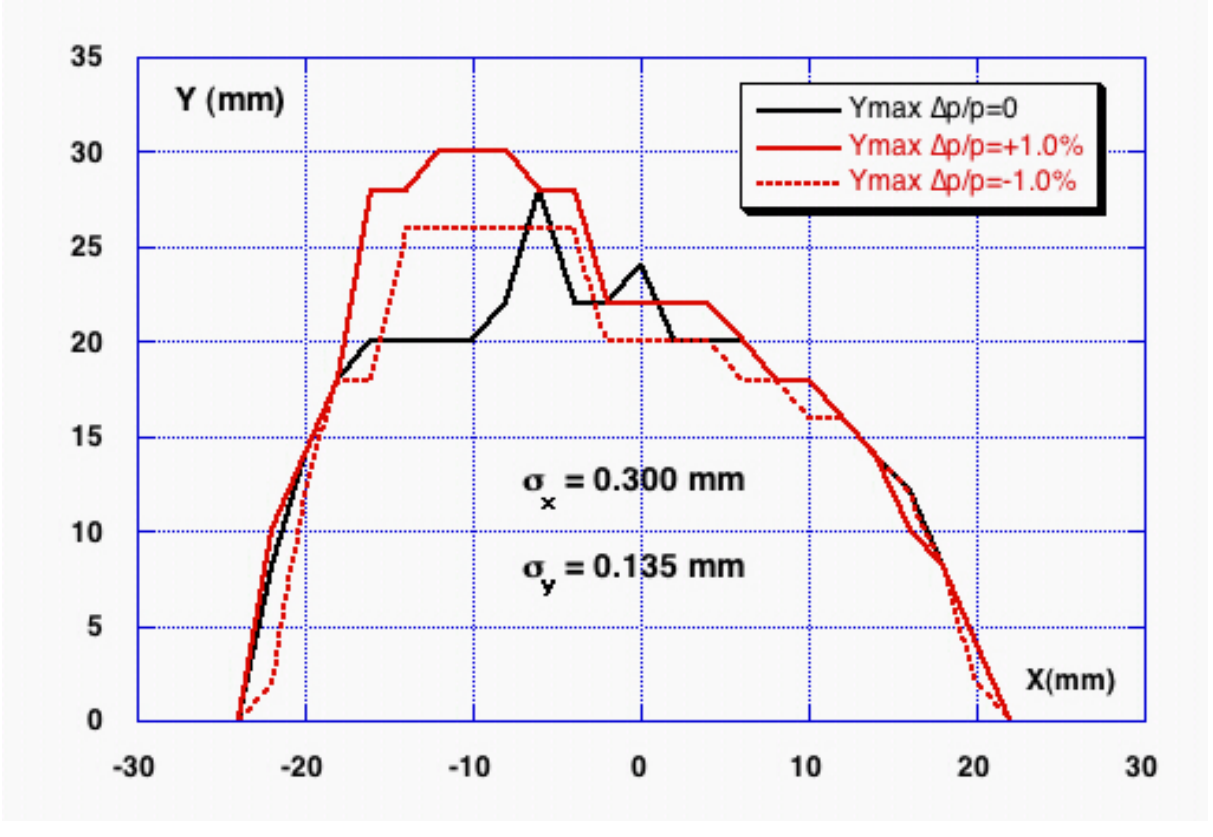

Figure 15.19: Damping ring dynamic aperture (black on energy, red ±1% energy deviation).

### 15.6 Bunch compressor

After the damping ring, the electron and positron beams are transported to a linac where they are accelerated up to their nominal energy and injected in the main rings. Due to the acceleration, the longitudinal distribution of the beam will be changed and the energy spread may increase. Therefore, the beam is compressed before the linac by means of a RF cavity running on the zero phase and a magnetic chicane in order to minimize the rms momentum spread. The acceleration in the linac is assumed on crest and the phase of the particle does not change during the transport in the linac (ultra-relativistic beam). The different parameters used for the optimization of the chicane are summarized in Table 15.9. The beam is assumed to be uncorrelated in the longitudinal phase space when it crosses the RF cavity before the chicane.

Table 15.9: Parameters for the chicane

| | Units | Notation | $e^-$ | $e^+$ |
|---|---|---|---|---|
| After the damping ring | | | | |
| Energy of the reference particle | GeV | $E_{d,0}$ | 1 | 1 |
| rms bunch length | mm | $l_d$ | 4.8 | 4.8 |
| rms phase extension | rad | $\theta_d$ | 0.287 | 0.287 |
| rms momentum spread | $10^{-4}$ | $\delta_d$ | 6.2 | 6.2 |
| After the linac | | | | |
| Energy of the reference particle | GeV | $E_{l,0}$ | 4.18 | 6.7 |
| RF frequency | GHz | $f_l$ | 2.856 | 2.856 |
| Main ring | | | | |
| Total energy | GeV | $E_S$ | 4.18 | 6.7 |
| Slipping factor | $10^{-4}$ | $\eta$ | 4.86 | 4.33 |
| RF voltage | MV | $V_{RF}$ | 4.0 | 5.7 |
| Synchronous phase | rad | $\theta_S$ | 0.156 | 0.372 |
| RF frequency | MHz | $f_{RF}$ | 476 | 476 |
| Harmonic number | - | $h$ | 2015 | 2015 |
| Synchrotron period | ms | | 0.35 | 0.40 |
| Longitudinal damping time | ms | | 13.4 | 28 |

The synchrotron period is small compared to the longitudinal damping time. Therefore, during the first synchrotron periods, the synchrotron damping can be neglected. After the damping ring, the longitudinal distribution is Gaussian. At first order, the matrix of the chicane in the longitudinal phase space is a 2x2 matrix with 1 on the diagonal and a non-zero term $R_{56}$ in the top right. After acceleration in the linac, the rms energy spread of the injected beam $\delta_l$ is linked to the RF cavity voltage $V_c$, to the matrix term $R_{56}$ of the chicane and to the total voltage $V_l$ in the linac by:

$$\delta_l^2 = \Delta^2 + \frac{E_{d,0}^2}{E_S^2}\delta_d^2 + \frac{V_c^2}{2E_S^2}\left[1-\exp\left(-2\left(\frac{f_c}{f_l}\theta_d\right)^2\right)\right]$$
$$-2\frac{V_l\Delta}{E_S}\exp\left(-\frac{1}{2}\left(\frac{R_{56}\omega_l}{\beta_c c}\delta_d\right)^2\right)F\left(1,\frac{R_{56}\omega_l V_c}{\beta_c c E_S},\frac{\omega_c}{\omega_l}\right)+$$
$$\frac{V_l^2}{2.E_S}\cdot\left[1+\exp\left(-2\left(\frac{R_{56}\omega_l}{\beta_c c}\delta_d\right)^2\right)\right]F\left(2,\frac{R_{56}\omega_l V_c}{\beta_c c E_S},\frac{\omega_c}{\omega_l}\right)$$

with $\Delta = \dfrac{E_S - E_{d,0}}{E_S}$

$$F(x,y,z) = \frac{1}{\sqrt{2\pi}\theta_d}\int_{-\infty}^{\infty}\cos(x\theta + y\sin z\theta)\exp\left(-\frac{\theta^2}{\theta_d^2}\right)d\theta$$

By using the parameters given in Table 15.9, we obtain Figure 15.20 which links the rms energy spread of the beam to the compressor RF cavity voltage. For the LER (HER), the minimum is reached for $V_c$=17 MV (21 MV) and then $\sigma_{E,m}$=0.14% (0.11%). A good compromise for the cavity voltage is then a value of ≈19 MV. The corresponding value of $R_{56}$ is 0.92 m and the total voltage in the linac $V_l$≈3.18 GV (5.70 GV).

An idea to correct the non linearities was to add another X-band cavity at the frequency 11.424 GHz. The results showed that the energy spread does not significantly change. The best compromise was for the voltages $V_c$=19 MV for the S-band cavity and $V_X$=-0.6 MV for the X-band cavity. The value of $R_{56}$ is 0.97 m.

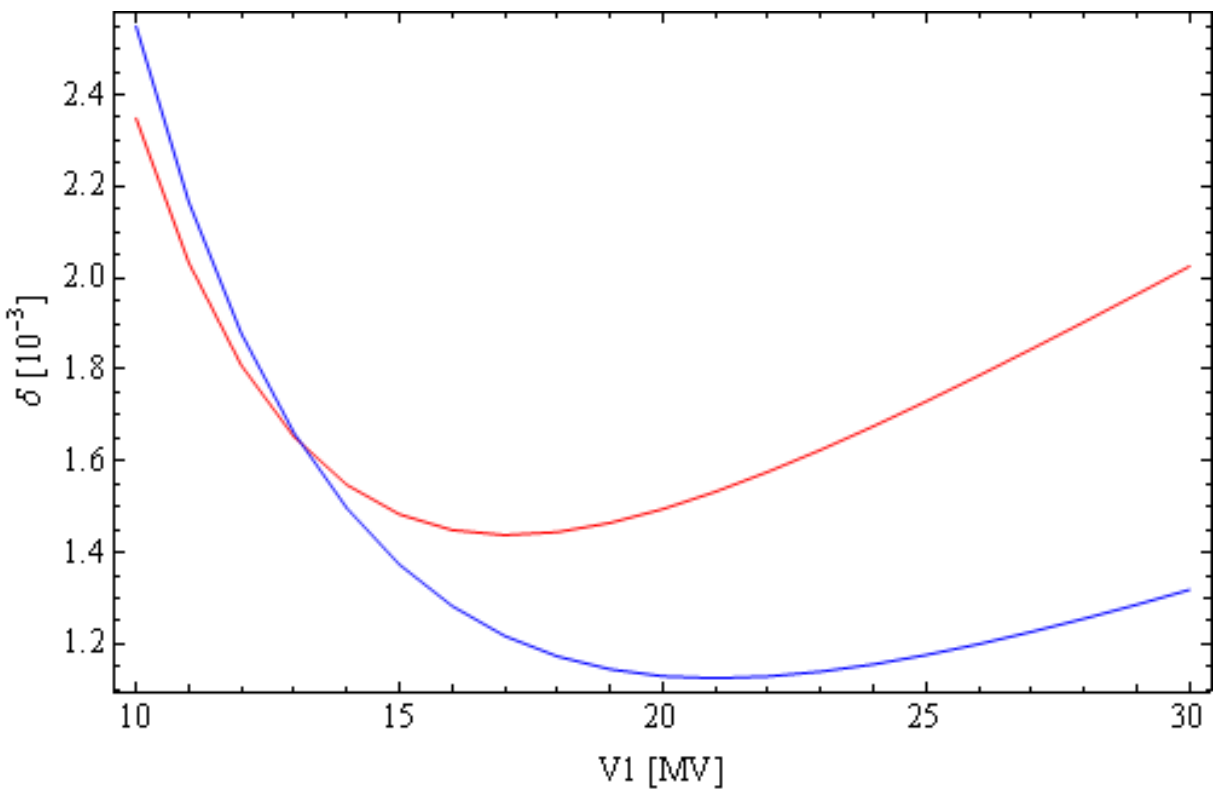

Figure 15.20: rms energy maximum vs the RF cavity voltage in the LER (red) and in the HER (blue).

The bunch compressor chicane consists in a RF cavity of frequency $f_c$=2.856 GHz, of voltage $V_c$, at the phase 180° (particles in the head of the bunch lose energy) followed by a chicane made of several dipoles which give a path length which depends on the energy. The needed voltage of the cavity is $V_c$=19 MV whereas the value of $R_{56}$ is then $R_{56}$=0.97 m. We have chosen to use a C-chicane because it is naturally achromatic [1]. The edges of the rectangular dipoles are then perpendicular to the linac axis. Let α be the angle of one of the dipoles in the chicane. The layout of the chicane is given in Figure 15.21.

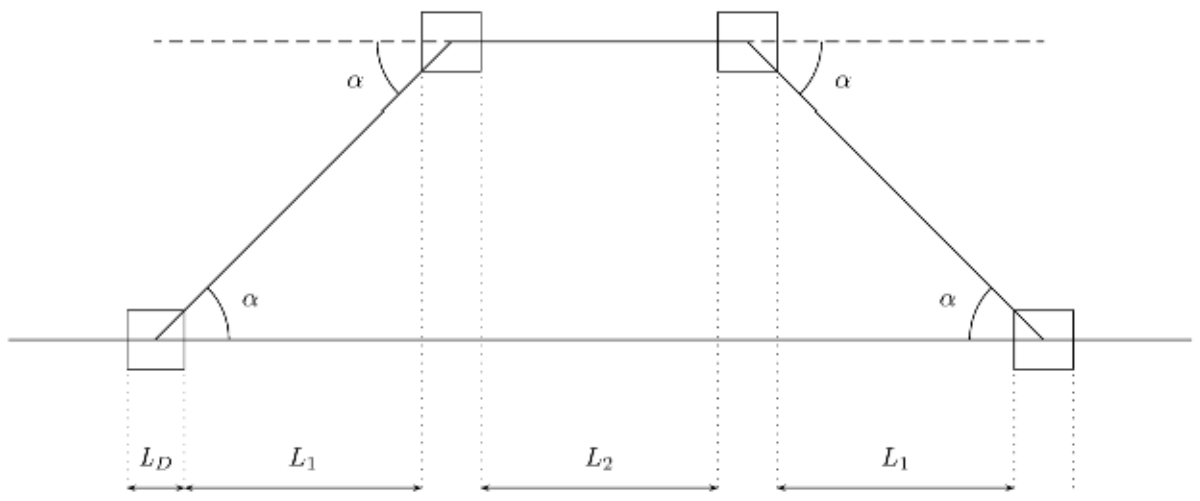

Figure 15.21: Layout of the bunch compressor chicane

To realize the C-chicane, we have four groups of dipoles. It was chosen to use the same dipoles of the PEPII LER. The maximum field of these dipoles is 0.93 T and their gap is 63.5 mm. The magnetic length is $L_{dip}$=0.45 m. For a 1-GeV electron beam, the maximum angle is then $\alpha_{max}$=7.19°, but, to keep a margin of 10% on the field, we chose to use an angle of $\alpha_{dip}$=90°/14=6.23°. The magnetic field in the dipoles is then 0.83 T. For each group of dipoles, the best compromise was to use three PEP-II dipoles. The distance between the first two groups is then $L_1$=2.8 m and the one between the two middle groups is $L_2$=1 m. The distance between two dipoles of the same group was set to 0.5 m. In fact, we need two chicanes which have the first and last dipole in common. The first dipole is used as a separator between the incoming electron and positron beams, the last as a recombiner. Therefore, we cannot use the PEPII magnets for this dipole and its design will be specific. Nevertheless, for geometrical reasons, the total angle of this dipole must be equal to $3\alpha_{dip}$. To simplify, we use the same magnetic field in this dipole as in the other dipoles of the chicane. Its length is then 1.35 m. The gap between the centres of the electron and positron beams must then be sufficient at the second dipole. The horizontal aperture of one of these dipoles is around 646 mm. The distance between the centres of these two beams must be greater than this value at the second dipole.

The distance between the two centres is about 528 mm after the separator. At the end of the $L_1$-long drift, the separation is then 2.7 m, which is sufficient. The obtained lattice is given in Figure 15.22 where the initial betatron functions were arbitrarily chosen to have a waist at the middle. The betatron functions in the chicane are sufficiently small to make the insertion of quadrupoles unnecessary. The total length of the C-chicane is 14.0 m.

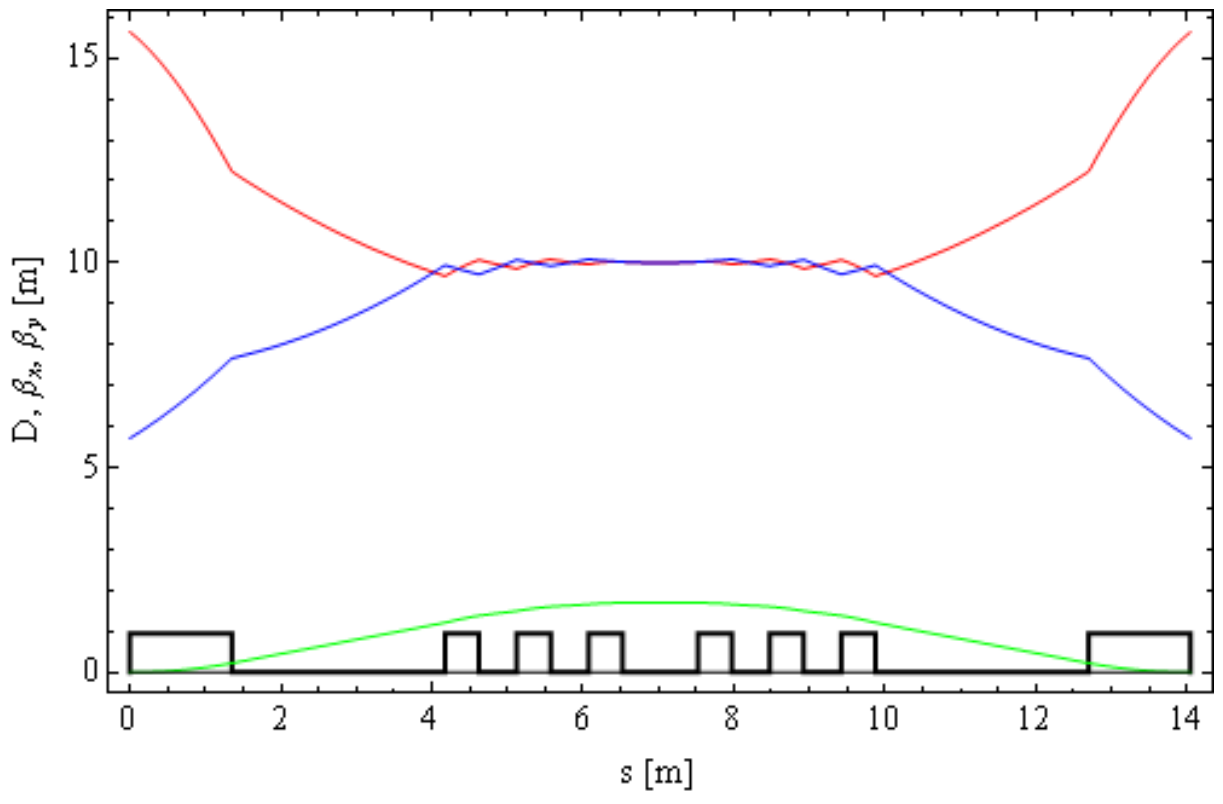

Figure 15.22: Optical functions in the chicane. In red $\beta_x$, in blue $\beta_y$, and in green $D_x$. The square boxes correspond to the rectangular dipoles.

## 15.7 Transport lines

In the present scheme, two linacs are used. The first one accelerates up to 1 GeV electrons and positrons, which are then injected into the damping ring. After damping, they are extracted and transported to the second linac to be accelerated up to the main rings.

### Transport line from/to the damping ring

Let be respectively (E1), (P1) the transport lines which go from the low energy linac to the damping ring and (E2), (P2) those going from the damping ring to the second linac for the electron and positron beams. The layout of these transport lines is given in Figure 15.23. The transport lines (E1) and (E2) are symmetric as well as (P1) and (P2). In order to keep the two parts of the linac in the same building, the distance between both linacs must be ≈ 5 m (4.76 m in the present design).

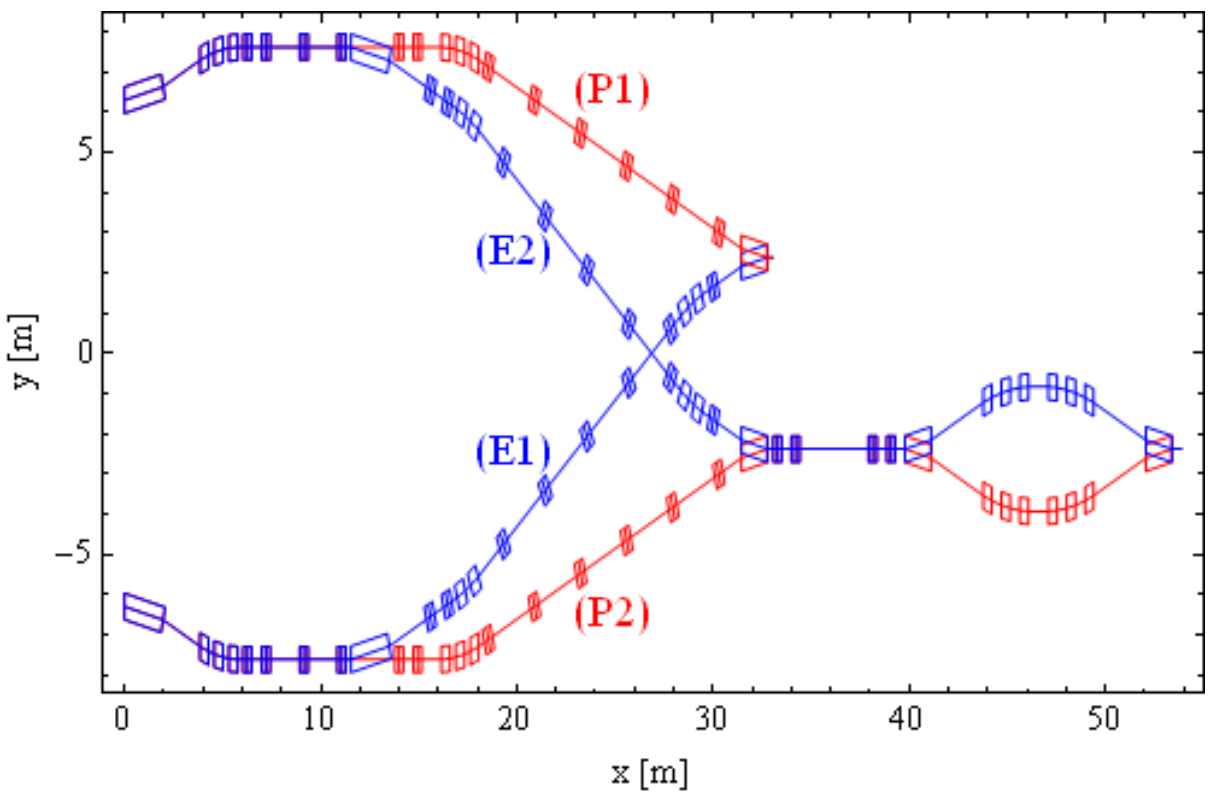

Figure 15.23: Synoptic of the transport lines going from/to the damping ring. The transport lines for the electrons are in blue and those for the positrons in red.

The dipoles and quadrupoles used for the transport lines are assumed to be the same as those of PEP II. The beam is extracted from the damping ring by means of a septum magnet. Another set of dipoles is then put as near as possible to cancel the deviation angle. A kicker and a septum magnet are inserted behind them to separate the positron beam from the electron one. The separator magnet at the beginning of (E1), (P1) and at the end of (E2) and (P2) are assumed to be the same as those in the bunch compressor chicane.

The transport lines must be achromatic. The lines (P1) and (E2) have some quadrupoles in common, as (P2) and (E1). Since the particles have opposite charges and opposite velocities, the same force is applied in both transport lines. However, in the linacs, the strengths of the quadrupoles are opposite because the charges are opposite and the velocities are the same. The betatron functions at the end/beginning of the linac in the horizontal (vertical) plane for positrons are then equal to those in the vertical (horizontal) plane for electrons. A 3.5 m-long drift is put at the end of (E2) and (P2) to enable the insertion of the compressor RF cavity.

The optical functions for (E1), (E2), (P1) and (P2) are given in Figures 15.24, 15.25, 15.26 and 15.27. The maximum for the betatron functions is less than 60 m.

The minimum for the betatron functions is ≈0.3 m. The maximum gradient for the quadrupoles is equal to 7.12 T/m, which is inside the range of PEP II quadrupoles. The extreme values for the optical functions and the value of $R_{56}$ are summarized in Table 15.10.

Table 15.10: Summary of the optical functions in the transport lines (E1), (E2), (P1) and (P2).

| | (E1) | (E2) | (P1) | (P2) |
|---|---|---|---|---|
| Max. $\beta_x$ (m) | 26.623 | 32.594 | 28.349 | 32.228 |
| Min. $\beta_x$ (m) | 0.4556 | 0.5329 | 0.3150 | 0.5775 |
| Max. $\beta_y$ (m) | 48.438 | 57.655 | 57.655 | 42.627 |
| Min. $\beta_y$ (m) | 1.5603 | 0.4175 | 0.3504 | 0.4957 |
| Max. $D_x$ (m) | 2.6576 | 4.9404 | 1.9158 | 2.3301 |
| Min. $D_x$ (m) | -3.371 | -3.469 | -3.855 | -2.810 |
| Length (m) | 35.749 | 57.018 | 34.108 | 55.376 |
| $R_{56}$ (m) | -0.803 | -0.501 | -0.306 | 0.7607 |

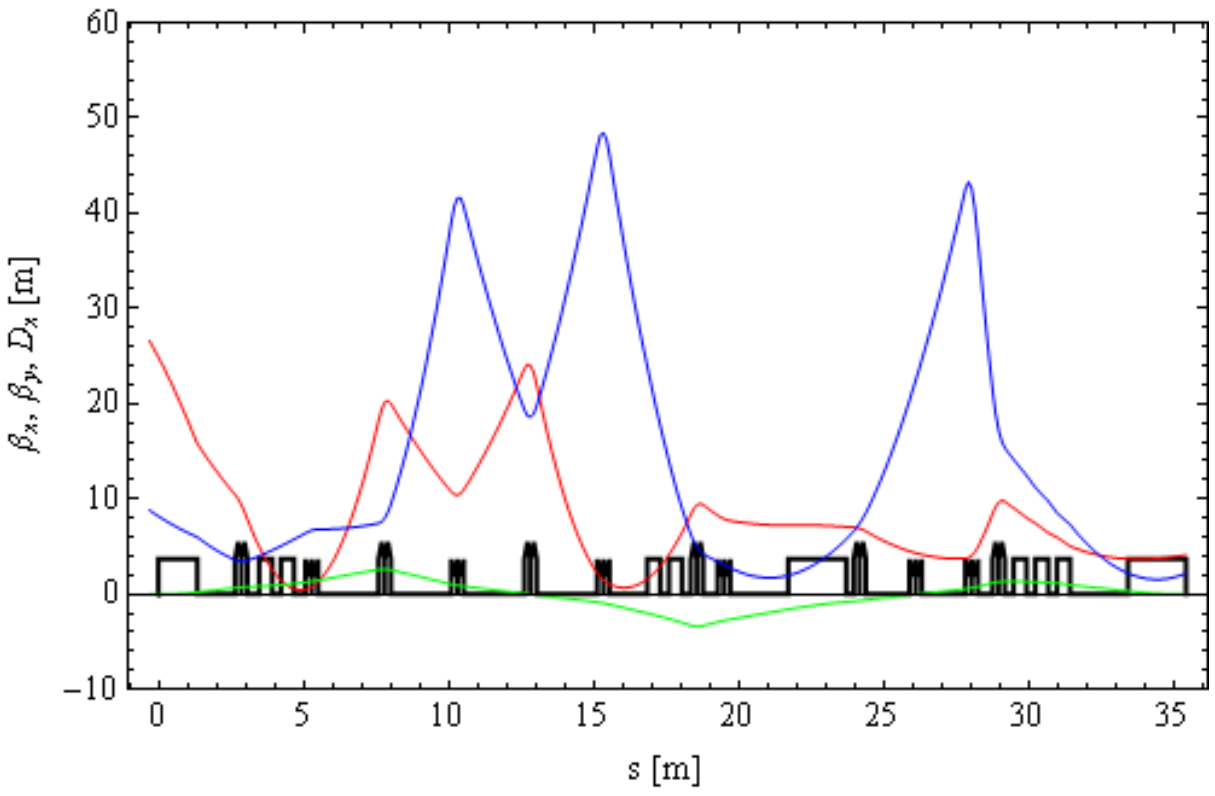

Figure 15.24: Optical functions in the transport lines (E1). The horizontal betatron function is in red, the vertical one in blue and the horizontal dispersion in green.

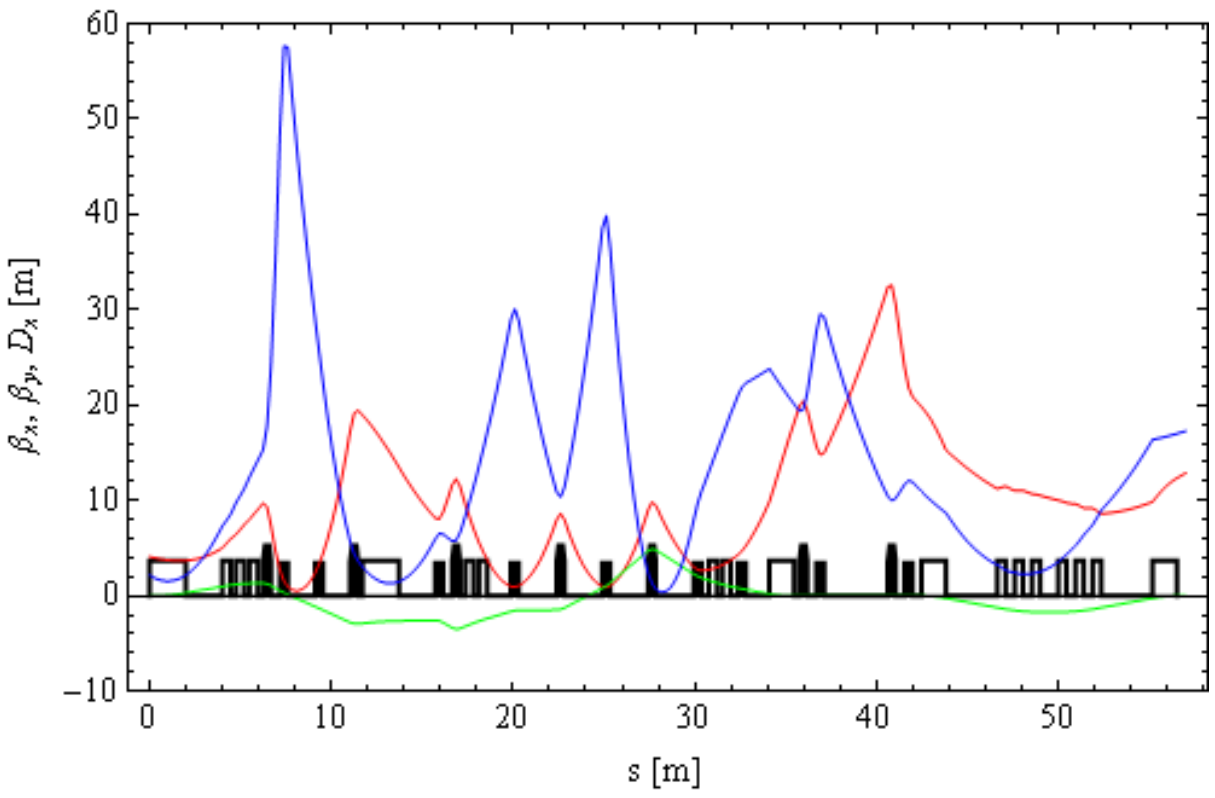

Figure 15.25: Optical functions in the transport lines (E2).

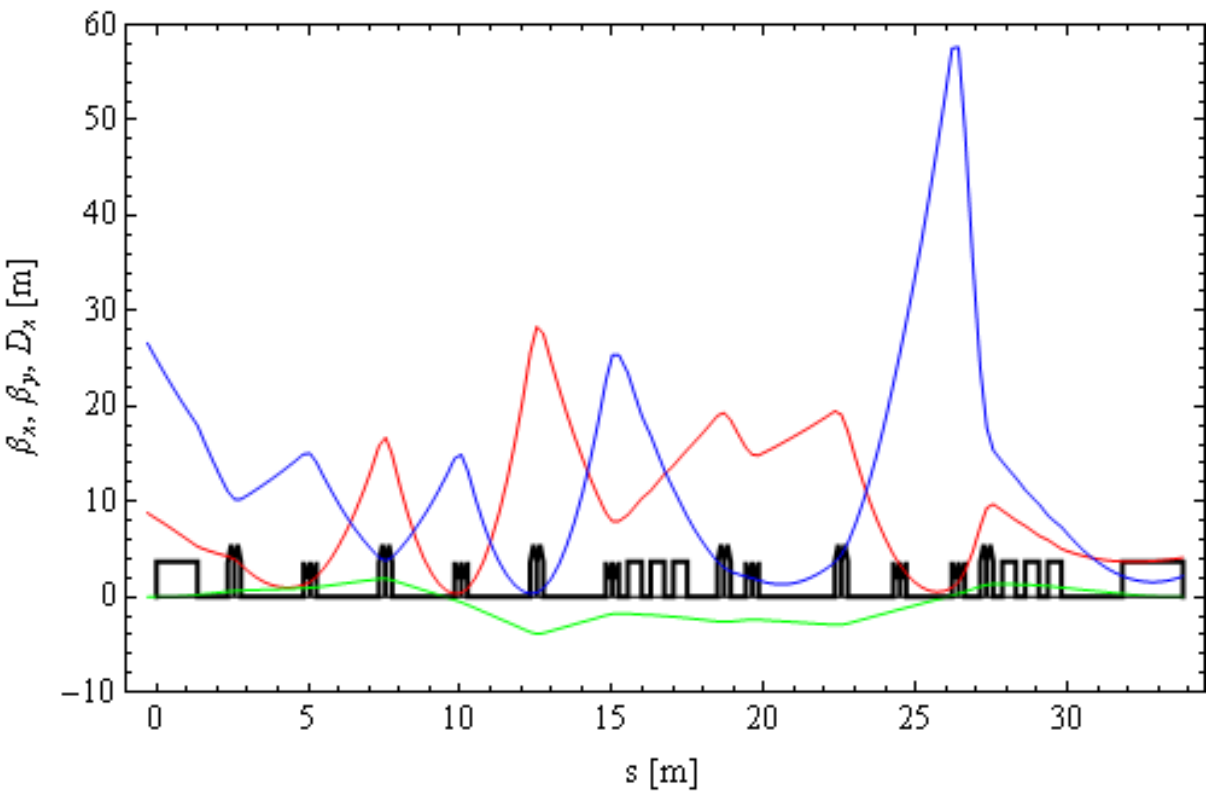

Figure 15.26: Optical functions in the transport lines (P1).

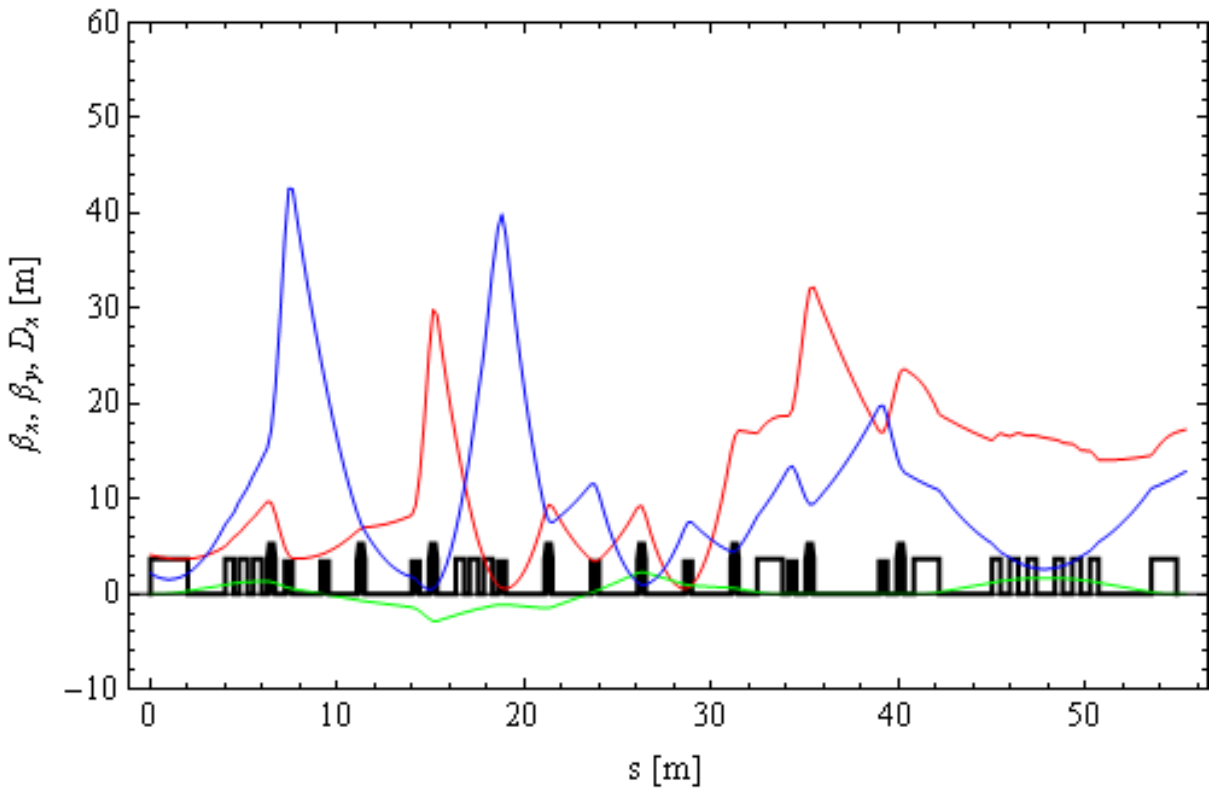

Figure 15.27: Optical functions in the transport lines (P2).

### Transport line to the main rings

The electron and positron beams are accelerated in the second linac up to their nominal energy and injected into the main rings. Since the damping ring and the main rings are not at the same ground level, a ≈10-m vertical bump must be foreseen in the transport lines. The top view layout of the transport lines between the linac and the injection points in the main rings is given in Figure 15.28 (positrons on the left, electrons on the right).

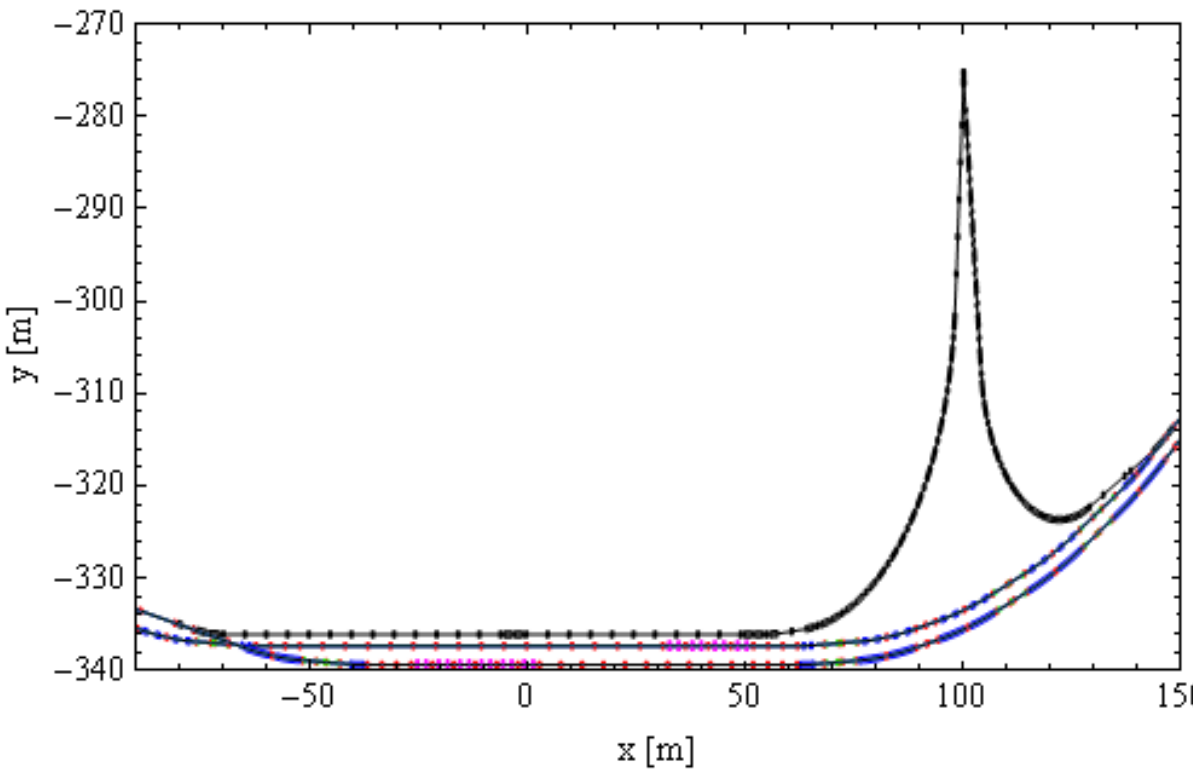

Figure 15.28: Schematic top view of the transport lines going from the linac to the main rings.

The design of the transport line to the LER was made to keep the orientation of the polarization vector despite the vertical dipoles. The requirement of injecting downstream the RF cavities implies a long straight section in the transport line to the HER, which must lie in the same tunnel as the LER.

A side view layout of the transport line to the LER is given in Figure 15.29. The vertical bump can be made by a ≈30-m long dog-leg. In the transport line to the HER, the vertical bump can be made in the long straight section. The location of the vertical dipoles is given in Figure 15.30 and Figure 15.31. Although performing the vertical bump in the transport line seems to be feasible, an alternative could be to do it at lower energy in the linac. The advantages of this solution is to use weaker dipoles or/and to shorten the vertical bump. Moreover, there is less constraint on the transport lines, which enables a better optimization of the betatron functions.

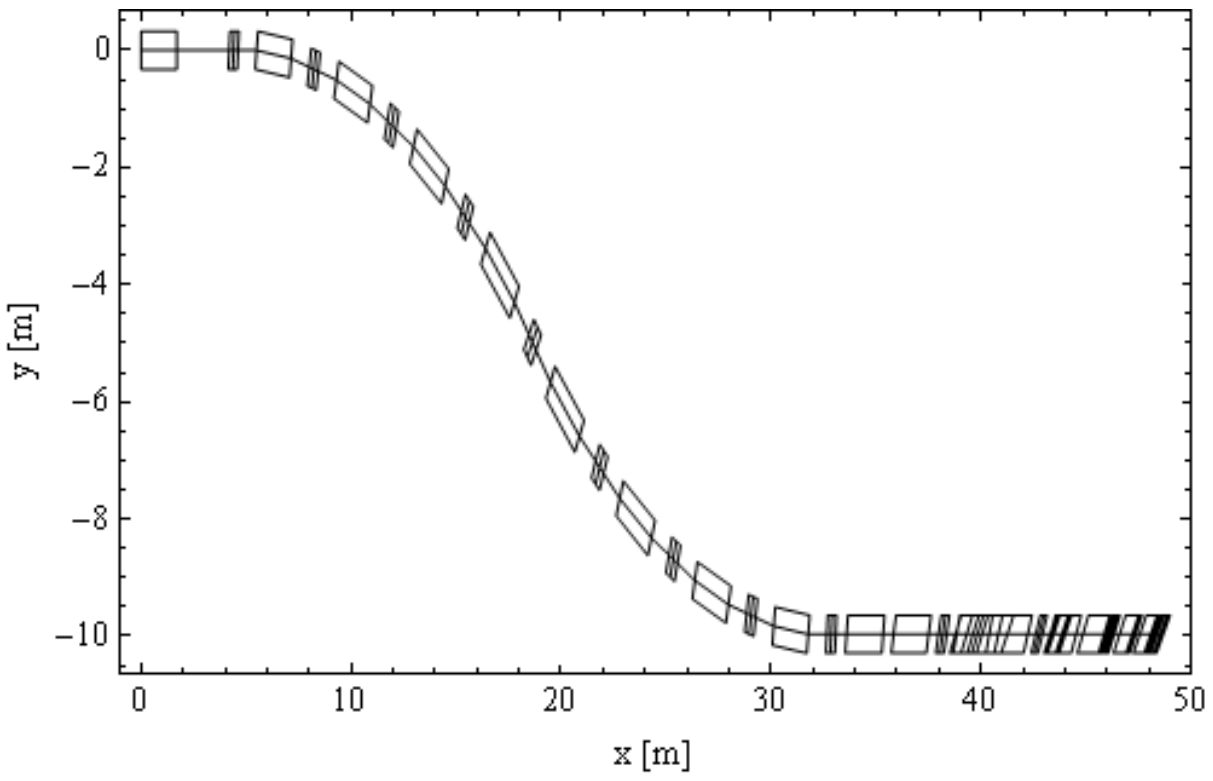

Figure 15.29: Schematic side view of the transport lines going to the LER.

The injection is performed by using a set of two septum magnets. The first one is 4 mm thick and has a magnetic field of ≈0.2 T. The deflection given by this septum is not sufficient and the incoming beam must go through the quadrupole near the septum. The effect of this quadrupole was taken into account for the design. The second septum is more than 1 cm thick and has a field of 0.8 T.

The beam dynamics in the linac is not completely defined at this stage of the project. The design of the transport lines was then made by considering the initial betatron functions as a free parameter.

The optical functions of the transport lines are given in Figure 15.30 for the LER and in Figure 15.31 for the HER. The maximum and minimum values of the optical functions and the value of $R_{56}$ are summarized in Table 15.11. The maximum needed field for the dipoles and the maximum gradient of the quadrupoles are respectively 1.4 T and 16.7 T/m for the transport line to the LER and 1.2 T and 13.7 T/m for the transport line to the HER.

Table 15.11: Summary of the optical functions to the main rings.

| | LER | HER |
|---|---|---|
| Max. $\beta_x$ (m) | 113.459 | 120.019 |
| Min. $\beta_x$ (m) | 0.436 | 0.973 |
| Max. $\beta_y$ (m) | 76.827 | 73.695 |
| Min. $\beta_y$ (m) | 0.475 | 4.466 |
| Max. $D_x$ (m) | 2.321 | 1.285 |
| Min. $D_x$ (m) | 0.000 | 0.000 |
| Max. $D_y$ (m) | 2.731 | 2.079 |
| Min. $D_y$ (m) | 2.731 | -2.079 |
| Length (m) | 89.472 | 224.004 |
| $R_{56}$ (m) | -3.484 | -1.204 |

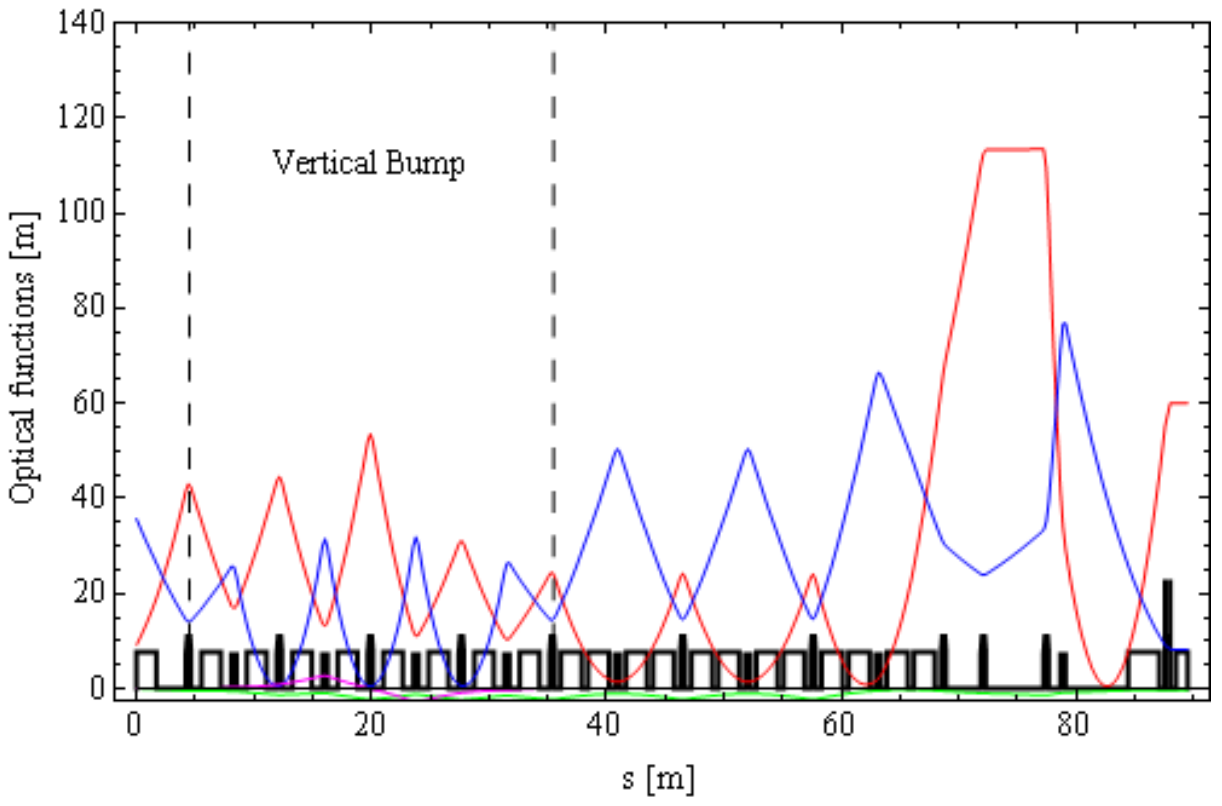

Figure 15.30: Optical functions in the transport line from the linac to the LER.

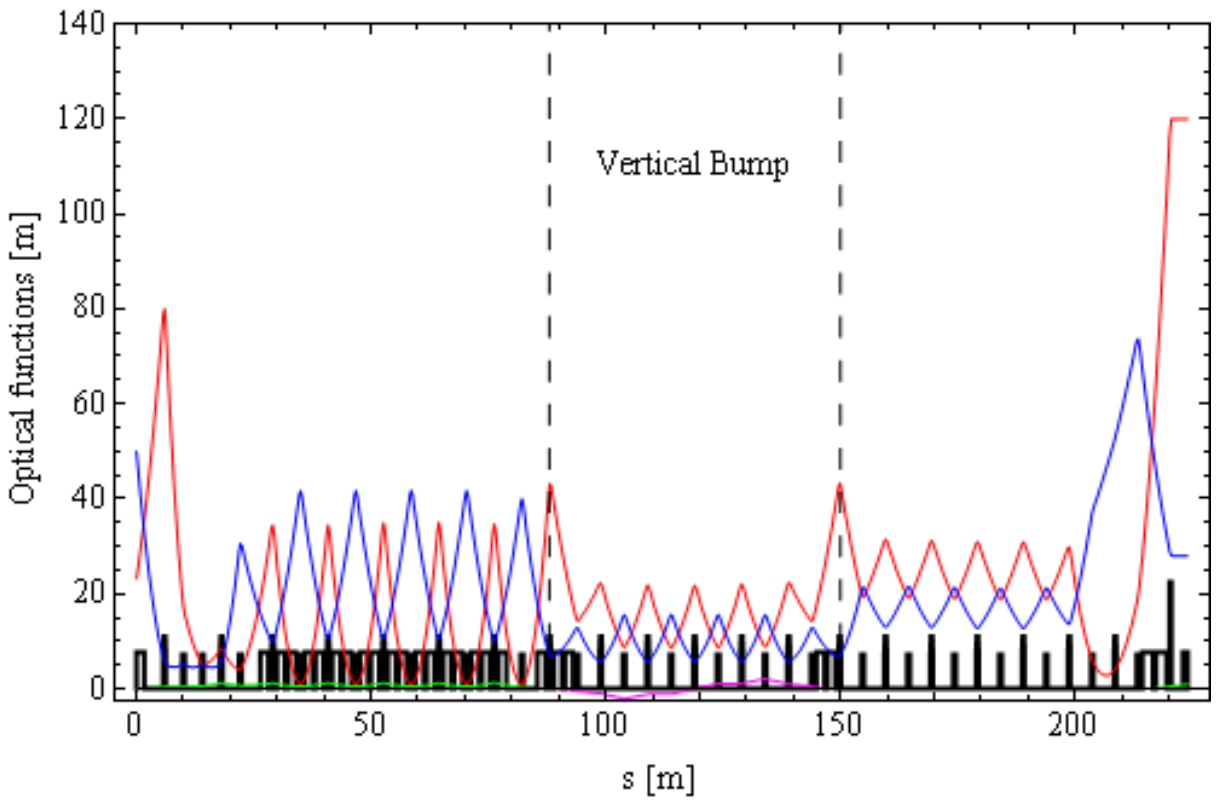

Figure15.31: Optical functions in the transport line from the linac to the HER.

## 15.8 Injection into the rings

Injection into the main rings is done at full energy in a continuous mode, called top-up injection, to keep nearly constant beam current and luminosity. Both single and multiple bunches (~five) will be injected on each linac pulse into one or the other of the two rings. The transverse and longitudinal emittances of the electron and positron bunches are damped at 1 GeV in a damping ring.

The injection system is made of the first septum magnet (4 mm thick), used to deflect the injected beam on an orbit parallel to the ring orbit, and two pulsed stripline kickers to make a fast orbit bump used for the injection. To make a closed orbit bump with its maximum at the septum the phase advance between the kickers is a multiple of π and the phase advance between kicker and septum is π/2+2kπ/2.

Since we inject with colliding beams we want to keep the betatron oscillation of the injected beam as low as possible to avoid any perturbation to luminosity and detector backgrounds. Therefore we propose a configuration with non-zero dispersion at injection and an energy offset of the injected beam. This allows smaller oscillations of the injected beams at the interaction point, where dispersion is zero.

A schematic view of the injected and stored beam layout in the horizontal phase space, with the indication of the main injection parameters is shown in Figure 15.32.

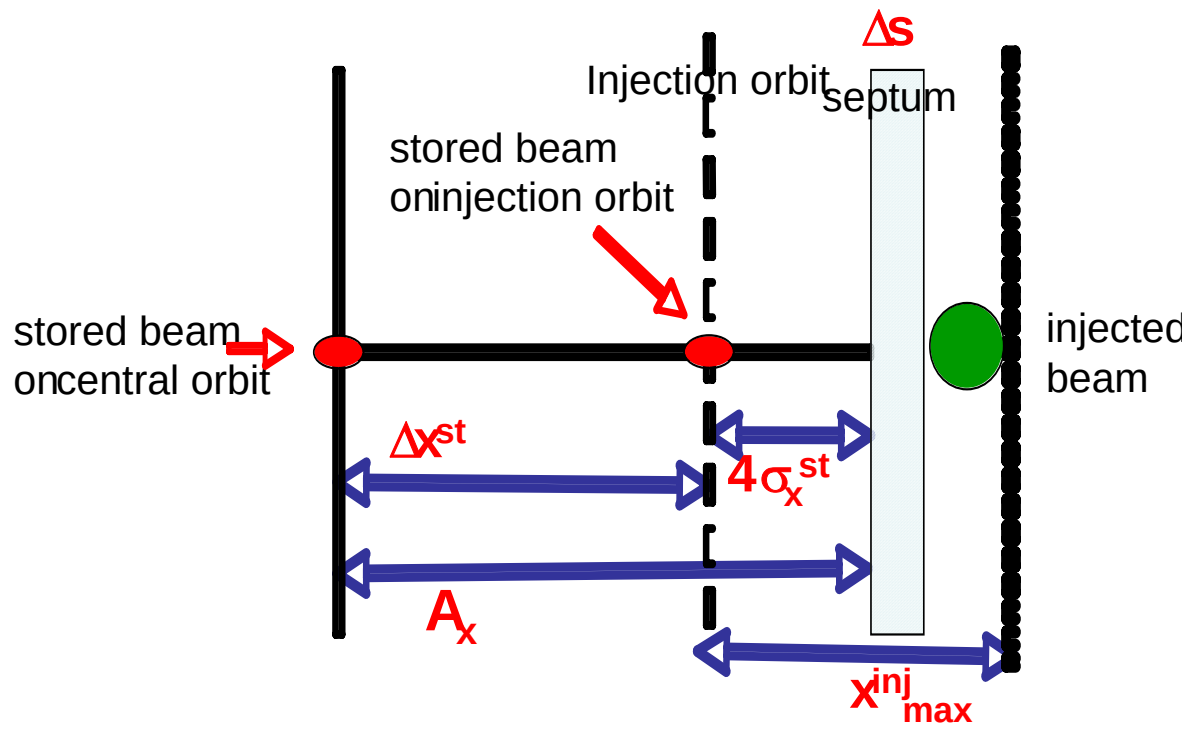

Fig.15.32 - Schematic view of the injected and stored beam layout in phase space

Let's indicate with

$$\sigma_x^{st} = \sqrt{\varepsilon_x^{st}\beta_x^{st} + \left(\sigma_p^{st}D_x^{st}\right)^2}$$

and

$$\sigma_x^{inj} = \sqrt{\varepsilon_x^{inj}\beta_x^{inj} + \left(\sigma_p^{inj}D_x^{inj}\right)^2}$$

the beam sizes at the ring entrance for the stored and injected beam respectively.

First a configuration with zero dispersion $D_x^{st} = D_x^{inj} = 0$ is considered. The value of the beam stay-clear $A_x$ is determined by beam lifetime considerations.

At the entrance of the injected beam the closed orbit is displaced by the fast closed bump $\Delta x^{st}$. To avoid losses of the stored beam on the septum it has to be:

$$\Delta x^{st} \le A_x - 4\sigma_x^{st}$$

and to avoid losses on the injected beam:

$$\Delta x^{st} \ge \Delta s + 2k\sigma_x^{inj}$$

To calculate the maximum oscillation amplitude of the injected beam in the ring in units of the stored beam size let's take the largest possible value of the orbit displacement:

$$\Delta x^{st} = A_x - 4\sigma_x^{st}$$

Finally a check that it is inside the beam stay clear aperture is needed:

$$\frac{x_{max}^{inj}}{\sigma_x^{st}} = \frac{\Delta s}{\sigma_x^{st}} + 2k\frac{\sigma_x^{inj}}{\sigma_x^{st}} + 4 \le \frac{A_x}{\sigma_x^{st}}$$

The parameters for injection in both LER and HER rings are listed in Table 15.12, columns 1 and 3. The value of the betatron function in the ring is large to reduce the contribution of the septum thickness. The value of the betatron function of the injected beam has been calculated to match the ring acceptance.

Table 15.12: Parameters for injection in LER and HER

| | LER | | HER | |
|---|---|---|---|---|
| | $D_x^{inj} = 0$ | $D_x^{inj} \neq 0$ | $D_x^{inj} = 0$ | $D_x^{inj} \neq 0$ |
| $D_x^{st} = D_x^{inj}$ (m) | 0 | 0.33 | 0 | 0.63 |
| $\delta$ | 0 | 3.00E-03 | 0 | 3.00E-03 |
| $\beta_x^{st}$ (m) | 140 | 140 | 270 | 270 |
| $\beta_x^{inj}$ (m) | 60 | 60 | 120 | 120 |
| $\sigma_x^{st}$ (m) | 5.80E-04 | 6.08E-04 | 7.35E-04 | 8.35E-04 |
| $\sigma_x^{inj}$ (m) | 5.75E-04 | 7.17E-04 | 7.92E-04 | 9.39E-04 |
| $\sigma_{x\beta}^{inj}$ (m) | 5.75E-04 | 5.75E-04 | 7.92E-04 | 7.92E-04 |
| $A_x/\sigma_x^{st}$ | 30 | 30 | 30 | 30 |
| $\Delta s$ (m) | 0.004 | 0.004 | 0.004 | 0.004 |
| $\theta_k$ (mrad) | 0.17 | 0.18 | 0.16 | 0.18 |
| $x^{inj}_{max}$ (m) | 9.8E-03 | 8.9E-03 | 1.2E-02 | 1.0E-02 |
| $x^{inj}_{max}/\sigma_x^{st}$ | 16.8 | 14.6 | 15.0 | 12.2 |

If a configuration with non zero dispersion $D_x^{st} = D_x^{inj} \neq 0$ and an energy offset δ of the injected beam are taken into account, all the same relations hold but now the injected beam oscillates with respect to a closed orbit displaced by $\Delta x^{st}$ - $D_x\delta$ with respect to the central orbit. Its maximum oscillation amplitude in units of the stored beam size is now:

$$\frac{x_{max}^{inj}}{\sigma_x^{st}} = \frac{\Delta s}{\sigma_x^{st}} + 2k\frac{\sigma_{x\beta}^{inj}}{\sigma_x^{st}} + 4 - \frac{D_x\delta}{\sigma_x^{st}}$$

where $\sigma_{x\beta}^{inj} = \sqrt{\varepsilon_x^{inj}\beta_x^{inj}}$. The injection parameters for this configuration, are listed in Table 1, columns 2 and 4 for LER and HER ring respectively.

The angle of the injection kicker is given by:

$$\theta_k = \Delta x^{st} \Big/ \sqrt{\beta_x^k \beta_x^{st}}$$

also listed in Table 15.12 for all the configurations.

The configuration with non-zero dispersion allows smaller betatron oscillations of the injected beams (15 $\sigma_x$ for LER and 12 $\sigma_x$ for HER) and should give enough safety margin when beam-beam and nonlinear effects are taken into account. A possibility that could be explored to reduce this oscillation is a thinner septum, which poses more challenges on the septum magnet design.

The kickers strength is nearly the same as the DAΦNE kickers and therefore it is possible to use the same type of fast pulsers, allowing single bunch injection with a small, even negligible, perturbation of the neighbouring bunches. In this case it is possible to share the oscillation amplitude between the stored and the injected beam by increasing the strength of the kicker on the injected beam with respect to the value needed for the closed orbit bump. This would allow values of the oscillation amplitude as low as 6÷8$\sigma_x$. for LER (6$\sigma_x$ for HER). To inject trains of up to 5 bunches it is necessary to have also kickers pulses with 20 ns flat top. This can be achieved by using two stripline kickers for each position, a short one (~30 cm) with a fast pulser and a longer one (~1m) for the longer pulse.

A simulation tracking the distribution of injected particles through the ring, taking into account the effect of the beam-beam kick and the machine errors and nonlinearities, will be performed to set the tolerances on the injection parameters.

# 16. Spin polarization

Before describing concepts for attaining electron spin polarization for SuperB we present a brief overview of the theory and phenomenology. We can then draw on this later as required. This overview is necessarily brief but more details can be found in [1, 2].

## 16.1 Self polarization

The spin polarization of an ensemble of spin -$\frac{1}{2}$ fermions with the same energies travelling in the same direction is defined as

$$\vec{P} = \left\langle \frac{2}{\hbar}\vec{\sigma} \right\rangle \qquad (1)$$

where $\vec{\sigma}$ is the spin operator in the rest frame and <> denotes the expectation value for the mixed spin state. We denote the single particle rest-frame expectation value of $\frac{2}{\hbar}\vec{\sigma}$ by $\vec{S}$ and we call this the "spin". The polarization is then the average of $\vec{S}$ over an ensemble of particles such as that of a bunch of particles.

In a storage ring the spins are not stationary but precess in the external fields In particular, the motion of $\vec{S}$ for a charged particle travelling in electric and magnetic fields is governed by the Thomas–BMT equation $\frac{d\vec{S}}{ds} = \vec{\Omega} \times \vec{S}$ where $s$ is the distance around the ring [2, 3]. The vector $\vec{\Omega}$ depends on the electric $(\vec{E})$ and magnetic $(\vec{B})$ fields, the energy and the velocity which evolve according to the Lorentz equation. Thus $\vec{\Omega}$ depends on $s$ and on the position of the particle in the 6–d phase space of the motion. For a purely transverse $\vec{B}$ field and vanishing $\vec{E}$ field, the Thomas-BMT equation describes spin precession about $\vec{B}$ with the angle $\gamma G\Theta = \gamma\frac{(g-2)}{2}\Theta$ (relative to the orbit). Here $\Theta$ is the deflection angle; for electrons, $\gamma G \approx 0.0011596$.

In any storage ring with or without misalignments, there exists a 1-turn periodic "stable spin direction" attached to the closed orbit at each location in the ring, denoted by the unit vector $\hat{n}_0$. For particles on the closed orbit and in the absence of synchrotron radiation, a polarization vector $\vec{P} \,||\, \hat{n}_0$ will remain stationary turn after turn. Any component of $\vec{P}$ normal to $\hat{n}_0$ will precess about $\hat{n}_0$ and the number of such spin revolutions per turn is called the "spin tune". For particles away from the closed orbit executing synchro-betatron oscillations, the concept of $\hat{n}_0$ must be generalized to a vector $\hat{n}$ to be discussed later. In a perfectly aligned flat ring with no solenoids $\hat{n}_0$ is vertical everywhere and the spin tune is $\gamma G$.

Relativistic electrons circulating in the guide field of a storage ring emit synchrotron radiation and a tiny fraction of the photons can cause spin flip from ''up'' along an initial direction to ''down'', and vice versa. However, the up–to–down and down–to–up rates differ, with the result that the beam can become spin polarized. In a perfectly aligned flat ring with no solenoids the polarization is anti–parallel to the vertical guide field, reaching a maximum polarization, $P_{st}$, of $\frac{8}{5\sqrt{3}} = 92.4\%$. This, the Sokolov–Ternov (S–T) polarizing process [4], is very slow on the time scale of other dynamical phenomena occurring in storage rings, and the inverse time constant for the exponential build up in a uniform dipole field is [4]:

$$\tau_{st}^{-1} = \frac{5\sqrt{3}}{8}\frac{r_e\gamma^5\hbar}{m_e|\rho^3|} \qquad (2)$$

where $r_e$ is the classical electron radius, $\gamma$, the Lorentz factor, $\rho$, the radius of curvature in the field and the other symbols have their usual meanings.

In a simplified picture the majority of the photons in the synchrotron radiation do not cause spin flip but tend instead to randomize the orbital motion in the magnetic fields due to the presence of dispersion. Then the spin-orbit coupling embodied in the Thomas–BMT equation together with the non-uniformity of the quadrupole fields can cause spin diffusion, i.e. depolarization. Compared to the S–T polarizing effect the depolarization tends to rise very strongly with beam energy. The equilibrium polarization is then less than 92.4% and will depend on the relative strengths of the polarization and depolarization processes. Even without depolarization, certain dipole layouts can reduce the equilibrium polarization to below 92.4 %.

Analytical estimates of the attainable equilibrium polarization are best based on the Derbenev-Kondratenko (D–K) formalism [5, 6]. This implicitly asserts that the value of the equilibrium polarization in an electron storage ring is the same at all points in phase space and is given by

$$P_{dk} = \mp\frac{8}{5\sqrt{3}}\frac{\oint ds\left\langle \frac{1}{|\rho(s)|^3}\hat{b}\cdot\left(\hat{n}-\frac{\partial\hat{n}}{\partial\delta}\right)\right\rangle_s}{\oint ds\left\langle \frac{1}{|\rho(s)|^3}\left(1-\frac{2}{9}(\hat{n}\cdot\hat{s})^2+\frac{11}{18}\left(\frac{\partial\hat{n}}{\partial\delta}\right)^2\right)\right\rangle_s} \qquad (3)$$

where $\langle\ \rangle_s$ denotes an average over phase space at azimuth $s$, $\hat{s}$ is the direction of motion and $\hat{b}$ is the magnetic field direction. $\hat{n}$ is a unit 3–vector field over the phase space satisfying the Thomas–BMT equation along particle trajectories $u(s)$ (which are assumed to be integrable) and it is 1–turn periodic: $\hat{n}(u, s+C) = \hat{n}(u,s)$ where $C$ is the circumference of the ring. $\delta$ is the fractional energy offset due to synchrotron oscillations.

The vector field $\hat{n}(u,s)$ is a key object for systematizing spin dynamics in storage rings. It provides a reference direction for spin at each point in phase space and it is now called the *"invariant spin field"* [2, 7, 8]. At zero orbital amplitude, i.e. on the closed orbit, $\hat{n}(0,s)$ is just $\hat{n}_0(s)$. For electron rings, and away from so-called spin–orbit resonances [1], $\hat{n}$ is normally at most a few milliradians away from $\hat{n}_0$.

A central ingredient of the D–K formalism is the implicit assumption that the equilibrium electron polarization at each point in phase space is parallel to $\hat{n}$ at that point. In the approximation that the particles have the same energies and are travelling in the same direction, the polarization of a bunch measured in a polarimeter at s is then the ensemble average

$$\vec{P}_{ens,dk}(s) = P_{dk}\langle\hat{n}\rangle_s . \quad (4)$$

In conventional situations in electron rings, $\langle\hat{n}\rangle_s$ is very nearly aligned along $n_0(s)$. The *value* of the ensemble average, $P_{ens,dk}(s)$, is essentially independent of s.

In the presence of radiative depolarization the rate in Eq. 16.2 must be replaced by:

$$\tau_{dk}^{-1} = \frac{5\sqrt{3}}{8}\frac{r_e\gamma^5\hbar}{m_e}\frac{1}{C}\oint ds\left\langle\frac{1-\frac{2}{9}(\hat{n}\cdot\hat{s})^2+\frac{11}{18}\left(\frac{\partial\hat{n}}{\partial\delta}\right)^2}{|\rho(s)|^3}\right\rangle_s \quad (5)$$

The quantity $d^2 = \left(\frac{\partial\hat{n}}{\partial\delta}\right)^2$ is a key parameter in evaluating the expected polarization. Large values of $d^2$ cause low equilibrium polarization $P_{ens,dk}(s)$ and small time constants $\tau_{dk}$, thus reducing the polarization attainable. $d^2$ can become very large at the spin-orbit resonances [1]. The polarization build-up time $\tau_{st}$ of a real ring is obtained by setting $d^2$ to zero in the above equation. It depends on the layout of the ring and it is usually in the range of a few minutes to a few hours.

## 16.2 Polarization in SuperB

Quantitative evaluation of $\tau_{st}$ for SuperB gives about 5…7 hours for either ring (the bending radii of the dipoles in the LER are much smaller than those for the HER, compensating for the lower energy). Such large times are not useful in practice. Therefore SuperB will achieve polarized beams by injecting polarized electrons into the LER. We chose the LER rather than the HER because the spin rotators (see below) employ solenoids which scale in strength with energy.

In SuperB at high luminosity the beam lifetime will be only 3…5 minutes and continuous-injection (“trickle-charge”) operation is a key component of the proposal. By injecting at a high rate with a polarized beam one can override the depolarization in the ring as long as the depolarization time constant is not too small. The equilibrium polarization under continuous injection is given by

$$P = P_i\frac{\tau_{dk}}{\tau_{dk}+\tau_b} + P_{dk}\frac{\tau_b}{\tau_{dk}+\tau_b},$$

where $\tau_b$ is the beam lifetime of the ring and $P_i$, the polarization of the injected beam. As long as $\tau_b < \tau_{dk}$, the first term dominates (for high $P_i$).

## 16.3 Spin Rotators

In the ring arcs, the vector $\hat{n}_0$ which gives the direction of the polarization must be close to vertical to minimize depolarization. In order to obtain longitudinal polarization at the IP, a net rotation of $\hat{n}_0$ by 90° about the radial axis is required. To do this directly with vertical dipoles [9], significant vertical bending would be required causing vertical dispersion and emittance growth, which is not acceptable in SuperB. A series of interleaved horizontal and vertical dipoles can achieve the required spin rotation [10, 11], but still at too strong an effect on the vertical emittance. A rotation of 90° in a solenoid followed by a spin rotation of 90° in the horizontal plane also provides the required net rotation about the radial axis without any vertical bending and is therefore adopted for SuperB. In addition, the solenoid rotator is more compact. The solenoid field integral required is 21.88 Tm for 90° spin rotation, well within the technical capabilities of superconducting solenoids of the required aperture.

After the IP $\hat{n}_0$, and with it the polarization, has to be restored to vertical by a second spin rotator. Two geometries are possible: an antisymmetric geometry where the dipoles and solenoids after the IP have polarities opposite to those before the IP and a symmetric geometry, where the polarities are all the same. The two solutions have significantly different properties [12]:

With the antisymmetric geometry and perfect alignment, $\hat{n}_0$ is vertical in the arcs at all energies. However it is exactly longitudinal at the IP for just the design energy. Moreover the whole interaction region, seen from the arc, is spin transparent [1] for synchrotron motion. This is because the rotator fields and the horizontal dispersion are antisymmetric so that the effects on spin due to small energy offsets $\delta$ cancel. Then if spin transparency for horizontal betatron motion can be arranged, $d^2$ at the dipoles in the arcs can be small unless $\hat{n}_0$ is strongly tilted in the arcs because of misalignments. On the other hand, the dipole bending, being a net 0°, does not contribute to the overall bending required. However, the spin-rotator insertion causes significant increase in the length of the ring.

With the symmetric geometry $\hat{n}_0$ is vertical in the arcs and longitudinal at the IP at just the design energy and the whole interaction region will normally not be spin transparent since deviations in the spin rotation due to any energy offset δ will add up. Then it is likely that $d^2$ is large at the dipoles in the arcs. However, by bending in the same direction, the dipoles now become part of the overall bending. Thus the additional length required for the spin rotators is limited to that of the solenoids and associated optics.

For SuperB at high luminosity, the LER beam lifetime is about 3…5 min. Under these conditions it turns out that a symmetric spin-rotator scheme is feasible and can achieve 70% polarization or better (see below).

Coupling induced by two solenoids needs to be compensated somehow. The simplest and at the same time very convenient way to do this was suggested by V.Litvinenko and A.Zholents [13]. If matrices of the FODO lattice, which is inserted between solenoids, satisfy the requirement:

$$T_y = -T_x$$

then the horizontal and the vertical betatron oscillations became fully decoupled. Additional requirement comes from the spin transparency condition [7] (see Fig. 16.1):

$$T_x = -T_y = \begin{pmatrix} 1 & 0 \\ 0 & 1 \end{pmatrix}$$

For a spin rotation by the total angle $\varphi \le \pi$, this expression transforms into:

$$T_x = -T_y = \begin{pmatrix} -\cos(\varphi) & -2r\cdot\sin(\varphi) \\ (2r)^{-1}\cdot\sin(\varphi) & -\cos(\varphi) \end{pmatrix} \quad , \quad r=pc/eB$$

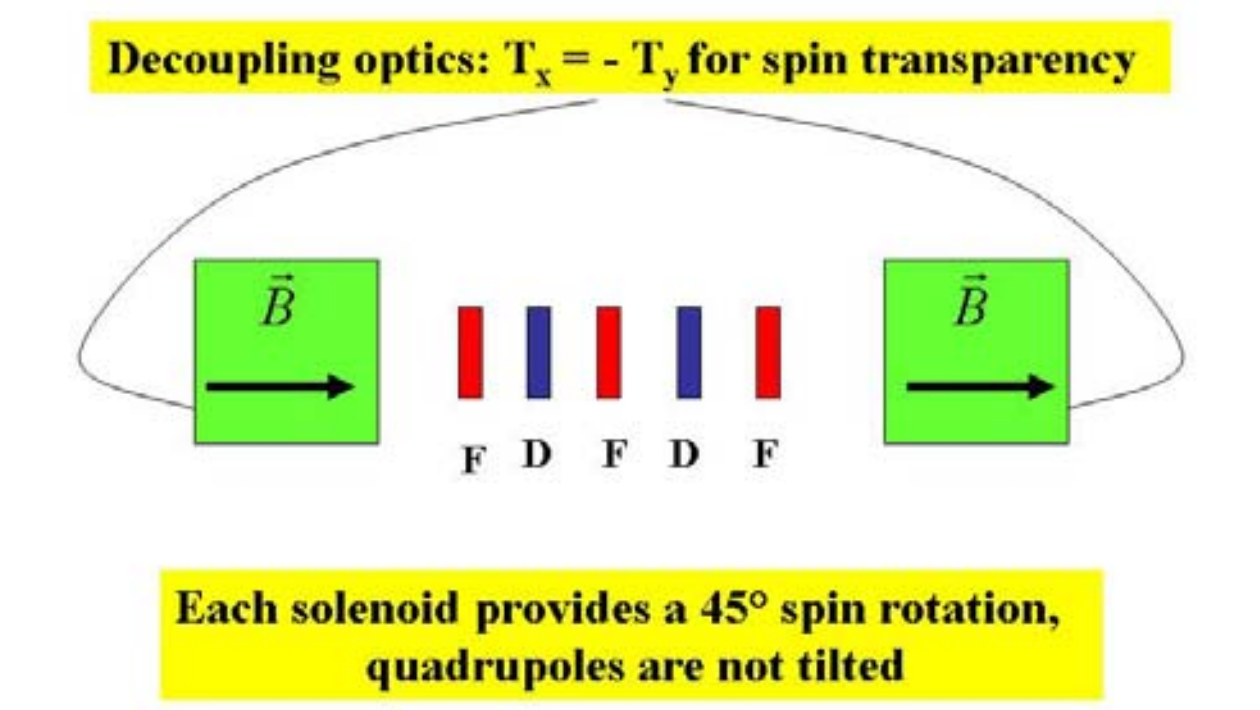

Figure 16.1: FODO lattice decouples horizontal and vertical motion in the spin rotator.

However, it might still be necessary to find settings for the quadrupoles in the interaction region and between the half solenoids to obtain sufficient spin transparency of the whole interaction region (IR).

## 16.4 LER Spin rotator layout

Figure 16.2 shows the IR of the LER with spin rotators. The rotator parameters are given in Table 16.1.

Table 16.1: Parameters of spin rotator

| Parameter | Value | Unit | Comment |
|---|---|---|---|
| Design energy | 4.18 | GeV | |
| Spin rotation of solenoids | 90 | ° | one side |
| Solenoid field integral | 4*10.94 | Tm | 4 individual solenoids |
| Solenoid field | 2.39 | T | |
| Total length of solenoid section | 23.07 | m | includes decoupling optics |
| Spin rotation of dipoles | 270 | ° | one side |
| Bending of dipoles | 28.4 | ° | one side |

Note that the dipole section rotates the spin by 270° instead of 90°; this was done in order to integrate the rotators with the local chromaticity correction needed in the IR. With 90° dipole angle, either the total bending would have been too small and the dispersion insufficient for effective chromaticity correction, or one would have to place the solenoids in the middle of the chromaticity-correcting section (between the vertical and the horizontal subsections). An attempt to do this was made but it was found that the constraints did not allow optimization of the optical properties of the whole section, leading to unacceptable compromises in transverse beam dynamics. The price paid is a further increase in $d^2$. On the positive side, the chosen layout provides a space with nearly longitudinal polarization away from the detector, for inclusion of a precision polarimeter (see section 16.6 below).

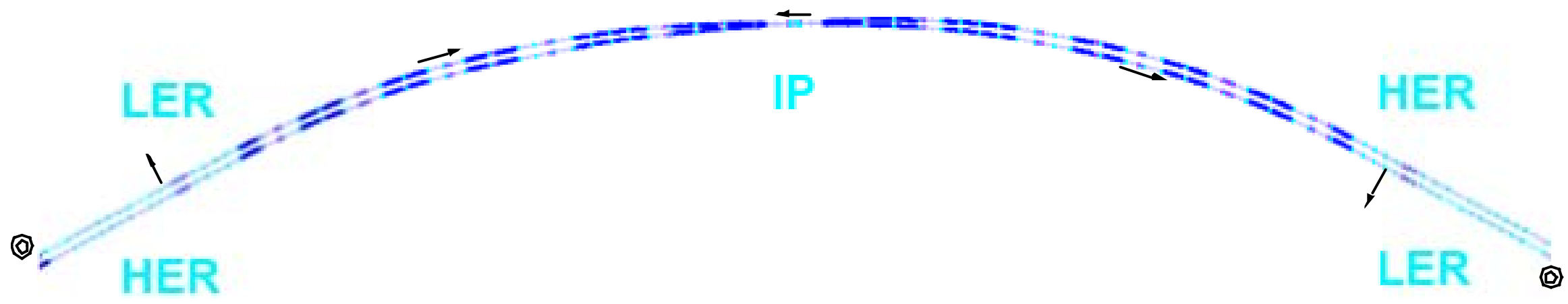

Figure 16.2: Layout of LER IR with spin rotators. The arrows and circles indicate the direction of P.

## 16.5 Spin dynamics

In order to evaluate the rate of depolarization we used the code SLICKTRACK [14]. This code is an extension of the code SLICK which can perform analytic first-order thick-lens evaluations using the SLIM [15] formalism. The extension comprises a Monte-Carlo spin-orbit tracking algorithm for simulating full 3-d spin-orbit motion in the presence of synchrotron radiation.

The following results are based on the MAD lattice model of the LER to which accelerating cavities have been added. A limited set of misalignments (in the arcs only) was implemented. Orbit correction was done in SLICKTRACK using a reduced set of correctors. SLICKTRACK calculates beam emittances and the values obtained are close to the design values in the horizontal plane, while larger than the design in the vertical plane (due to limited attempts at orbit correction). The energy spread and synchrotron tune are close to the design values. The differences to the MAD parameters are explained by the different treatment of RBENDs in MAD *vs* SLICKTRACK (MAD takes the hardware length as orbit length whereas SLICKTRACK corrects for the sagitta) as well as the different misalignment and orbit-correction setups.

Figure 16.3 shows a plot of the (de-) polarization time $\tau_{dk}$ *vs* ring energy (which is $0.441 \cdot \gamma G$ GeV).

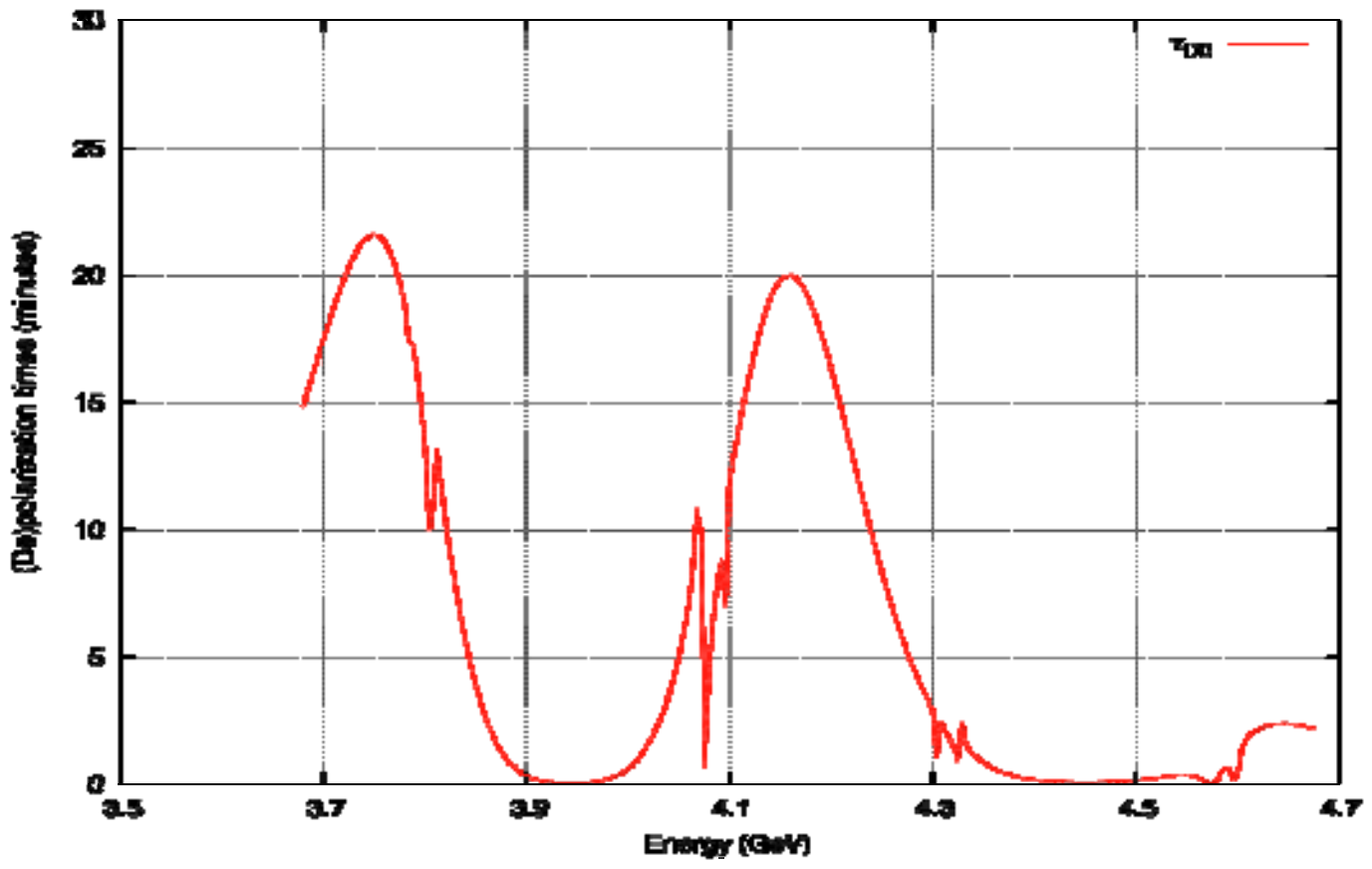

Figure 16.3: Depolarization time *vs* ring energy in the LER.

The behavior seen mainly reflects the variation in $d^2$ in the arcs as the angle of tilt of $\hat{n}_0$ from the vertical in the arcs varies with the ring energy as a result of using the symmetric rotator. Moreover since the interaction regions are not spin transparent, $d^2$ is not small even when $\hat{n}_0$ is vertical in the arcs. Consequently $\tau_{dk}$ is always much smaller than the Sokolov-Ternov time.

Figure 16.4 shows the attainable equilibrium polarization due to the pure Sokolov-Ternov effect. While the Sokolov-Ternov polarization would eventually reach values above 90% (except at energies close to integer values of the spin tune $\gamma G$), the expected equilibrium polarization in the LER is very small due to the large values of $d^2$.

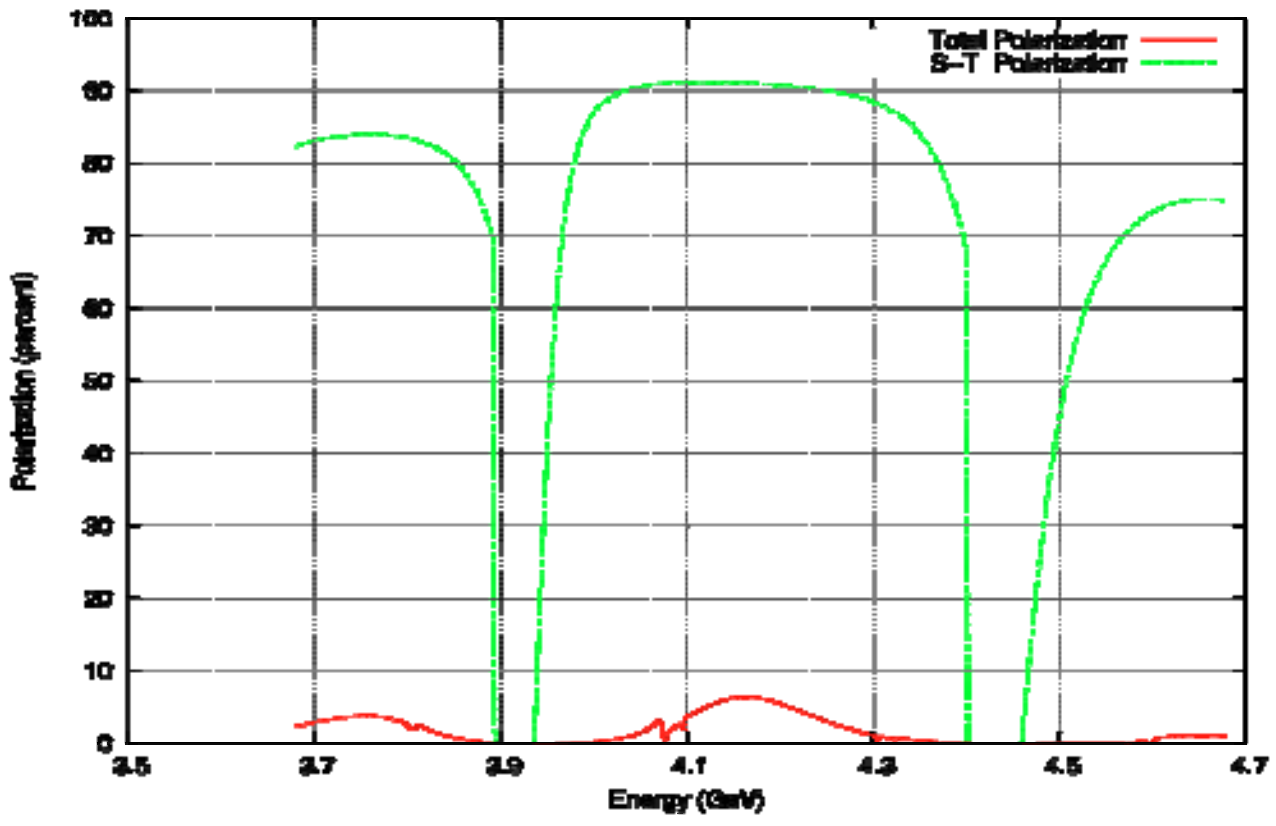

Figure 16.4: Equilibrium polarization in the LER. The green curve is the pure Sokolov-Ternov polarization, the red curve includes spin diffusion.

In all these plots, narrow dips can be seen near 4.1 GeV. These reflect the presence of spin-betatron resonances [1] where harmonics of the spin tune and the betatron tunes coincide. Their widths depend on the quantum excitation in the ring and their location on the betatron tunes. The wide dips are due to spin-synchrotron resonances. These are always strong when $\hat{n}_0$ is tilted in the arcs for then, electron spins with non-zero δ couple to the vertical quadrupole fields because of the horizontal dispersion.

For the above parameters we can evaluate the expected degree of polarization under continuous injection. Figure 16.5 shows the result for 90% polarization at injection and a beam lifetime of 3.5 min (i.e. at full luminosity). There is a significant band in energy where the polarization is expected to exceed 70%. We also evaluated the deviation of $\hat{n}_0$ from the longitudinal direction at the IP *vs* beam energy, shown in Figure 16.6. Given that the longitudinal component of the polarization scales with the cosine of this angle, there is a wide plateau where effectively $P_{long}=P$. This dependence may be important in assessing any systematic effect for the precision polarimetry.

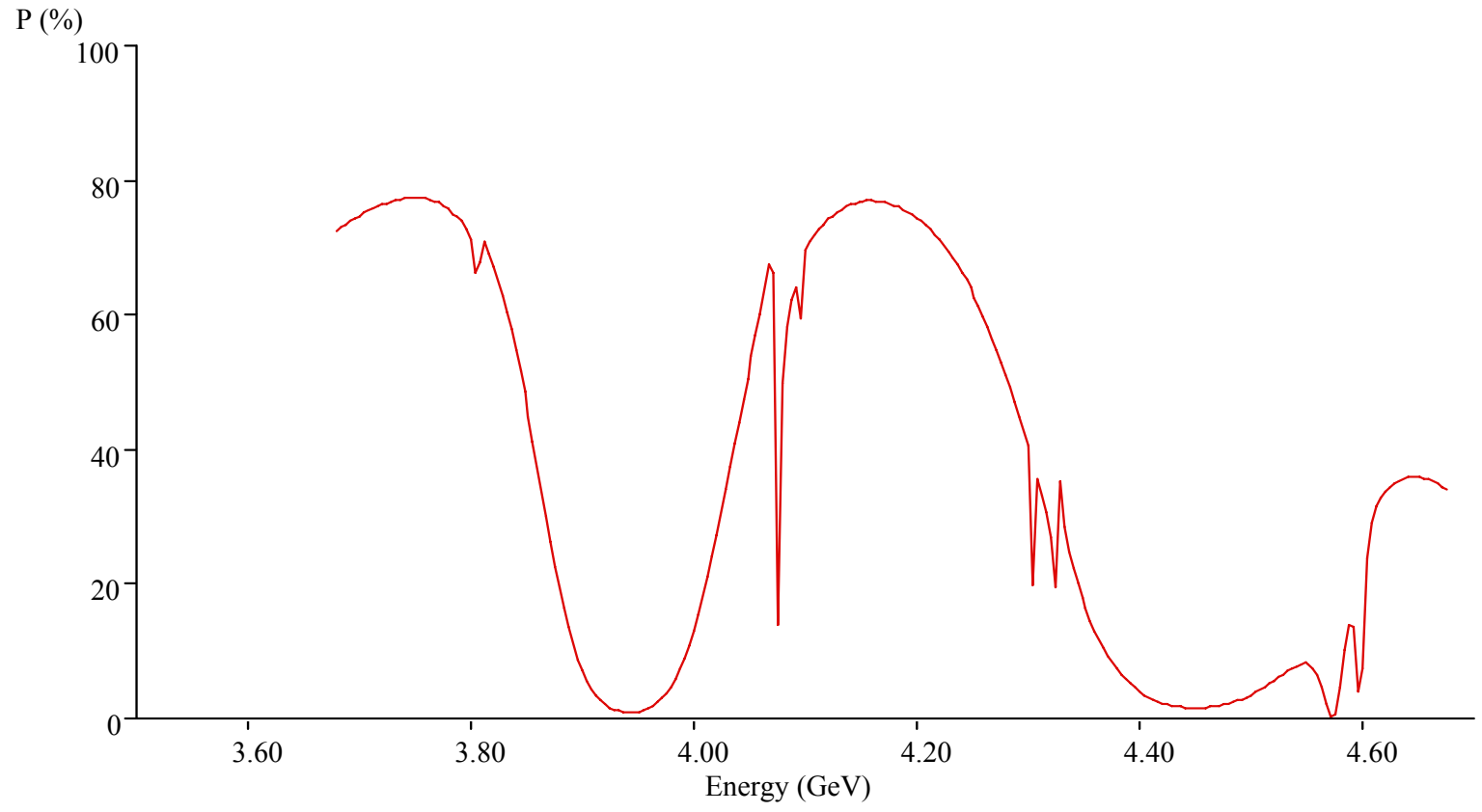

Figure 16.5: Equilibrium polarization in the LER under continuous injection at full luminosity.

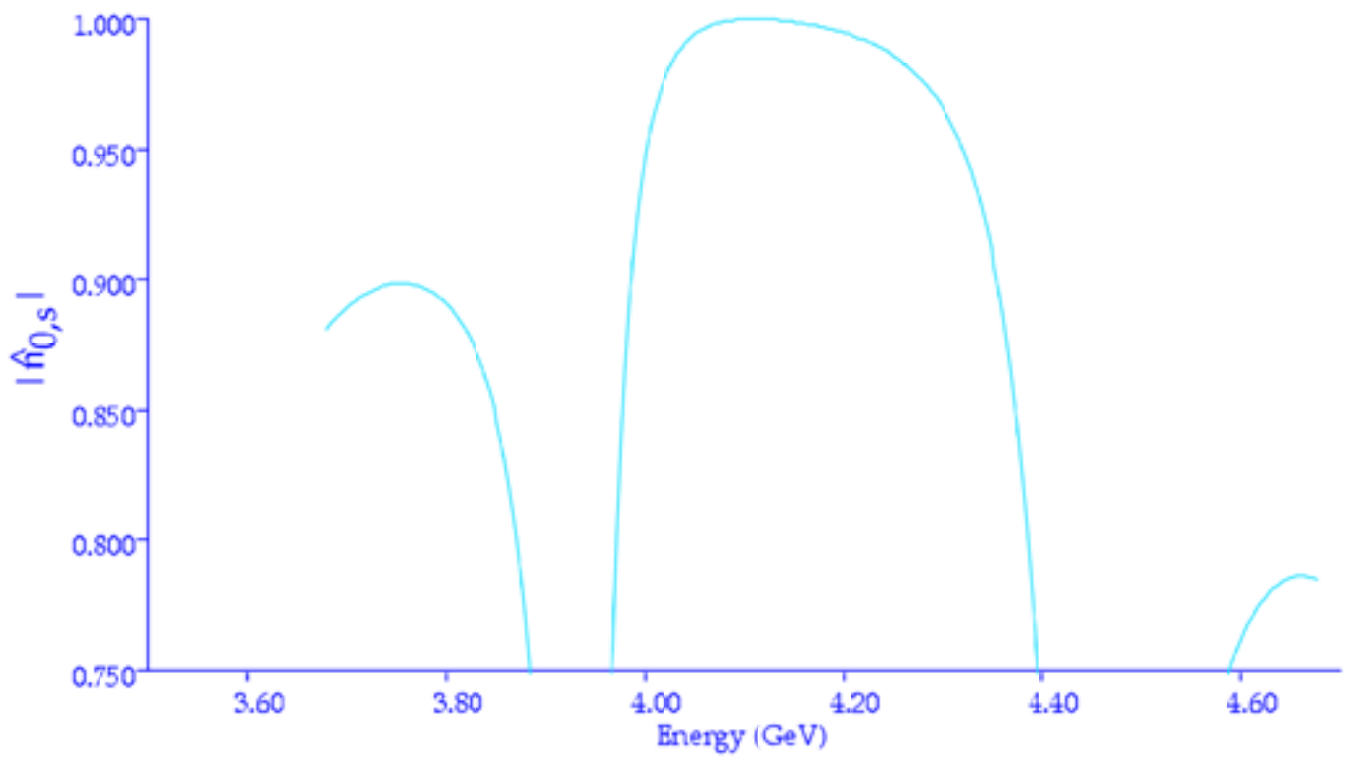

Figure 16.6: Tilt of $\hat{n}_0$ axis against the beam direction at the IP. The longitudinal component is plotted.

The estimates presented here are based on 1st-order analytic evaluation. Initial tracking runs indicate that any higher-order effects resulting from the full 3-d motion of spins should be quite small and should not significantly restrict the operating range in energy. Higher-order effects would manifest themselves in dips in the depolarization time at energies corresponding to the higher-order depolarizing resonances. The studies have been carried out without the detector solenoid field. This reflects our plan to compensate the detector field to a high degree. Moreover, so far there has been no attempt to satisfy some specific conditions on the optics among the rotator solenoids aimed at achieving spin transparency [1] in the transverse and longitudinal planes.

The results obtained so far give confidence that a polarization in excess of 70% at high luminosity can be achieved in the SuperB LER with the injection of polarized beams. Further studies will focus on the assessment of the need to improve the spin matching [1] of the interaction region including the rotator solenoid sections, the effect of imperfections including residual detector solenoid fields, and spin-tracking to include the higher order effects. The preservation of the polarization during the injection process should also be studied.

## 16.6 Polarimetry

### Overview

The physics program of the SuperB [1] demands precise polarimetry with <1.0% accuracy. The polarization measurement will be performed using a Compton polarimeter. An accuracy of (ΔPe_ / Pe_) = 1.0% should be achievable. Compton polarimetry is chosen for several reasons:

- The physics of the scattering process is well understood in QED, with radiative corrections less than 0.1%;[2]
- Detector backgrounds are easy to measure and correct by using laser off pulses;
- Polarimetry data can be taken simultaneously with physics data;
- The Compton scattering rate is high and small statistical errors can be achieved in a short amount of time (sub-1% precision in 30 seconds is feasible);
- The laser helicity can be selected every ~100 msec;
- The laser polarization is readily determined with 0.1% accuracy.

### Compton scattering basics

One defines $E_0$ and $\omega_0$ to be the incident energies of the electron and photon, and E and ω to be the scattered energies of the electron and photon. The dimensionless *x*, *y* and *r* scattering parameters are defined by:

$$x = \frac{4E_0\omega_0}{m^2}\cos^2(\theta_0/2) \approx \frac{4E_0\omega_0}{m^2} \qquad (1)$$

$$y = 1 - \frac{E}{E_0} = \frac{\omega}{E_0} \qquad (2)$$

$$r = \frac{y}{x(1-y)} \qquad (3)$$

where *m* is the mass of the electron and $\theta_0$ is the crossing angle between the electron beam and the laser beam. For polarimeters with small crossing angles at the Compton interaction point, $\cos^2(\theta_0/2) \approx 1$.

The spin-dependent differential Compton cross section is given by:

$$\left(\frac{d\sigma}{dy}\right)_{Compton} = \left(\frac{d\sigma}{dy}\right)_{unpol}\left[1 + P\cdot\lambda\cdot A_z(x,y)\right] \qquad (4)$$

$$\left(\frac{d\sigma}{dy}\right)_{unpol} = \frac{0.49barn}{x}\left[\frac{1}{1-y} + 1 - y - 4r(1-r)\right] \qquad (5)$$

$$A_z(x,y) = rx(1-2r)(2-y) \qquad (6)$$

where P is the longitudinal polarization of the electron and λ is the circular polarization of the laser photon. The Compton asymmetry analyzing power, $A_z$ (x; y), is maximal at the kinematic endpoint, corresponding to 180o backscattering in the center-of-mass frame, with

$$E_{\min} = E_0\frac{1}{1+x} \qquad (7)$$

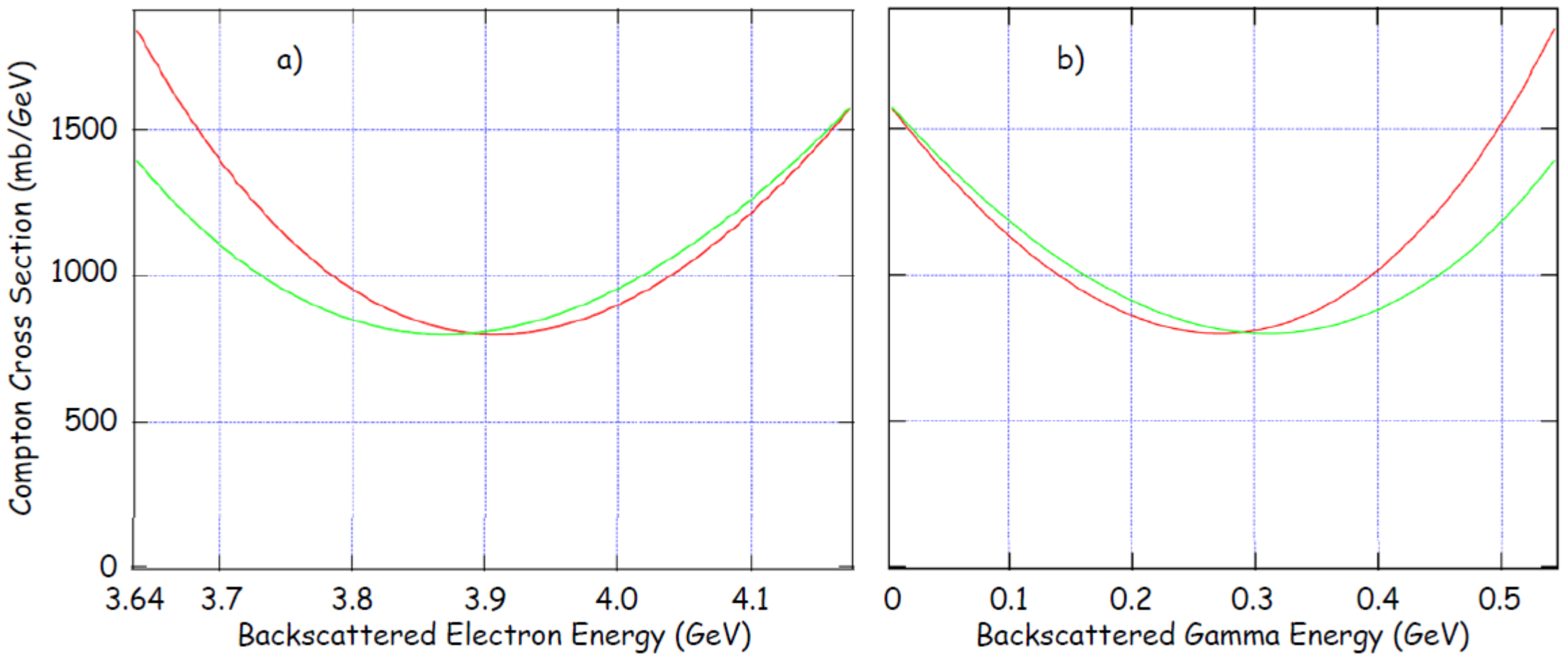

Figure 16.7: Compton cross section for scattering of 532 nm photons with a 4.18 GeV electron beam a) electron energy (b) Gamma energy. The $J_z = 3/2$ ($J_z = 1/2$) cross section for electron and photon spins aligned (anti-aligned) is shown in red/darker line (green/lighter line).

For a 4.18 GeV electron beam colliding with a 532-nm laser, the Compton-scattered electrons have their kinematic endpoint at $E_{min}$ = 3.64 GeV. Figure 16.7 shows the resulting $J_z = 3/2$ and $J_z = 1/2$ Compton cross sections. Table 16.2 gives the maximum Compton gamma energy and asymmetries for two different laser energies on 4.18 GeV electrons. The analyzing power at the Compton edge is 0.137 for the present default laser giving green light at the Compton IP. A larger analyzing power occurs for UV light and a laser system giving light in the UV is being evaluated.

Table 16.2: Compton polarimeter asymmetries (A) and cross section for two laser systems with 2.33 eV (green light) and 3.45eV (UV light) on 4.18 GeV electrons.

| $E_{beam}$ (GeV) | $E_{photon}$ (eV) | $W_{max}$ (GeV) | $A_{\gamma max}$ | $A_{\gamma\, flux\, wt}$ | $A_{\gamma\, E\, wt}$ | $\sigma_{unpol}$ (mbarn) |
|---|---|---|---|---|---|---|
| 4.18 | 2.33 green | 0.537 | 0.137 | 0.030 | 0.064 | 1089 |
| 4.18 | 3.45 UV | 0.756 | 0.197 | 0.040 | 0.088 | 731 |

### Spin Alignment

The electron beam spin direction is normal to the ring at injection and stays in the vertical direction for most of the orbit in the SuperB ring as shown in Figure 16.8. The physics program requires longitudinal polarization of the electron beam at the electron-positron Interaction Region. The spin is rotated from the vertical to longitudinal in a system of solenoids and dipole magnets on each side of the interaction region. There are 1½ π spin rotations in the horizontal plane between the solenoid and the IR. In groups of 5 the helicity of the 1011 electron bunches in the ring will be randomly selected to be left or right-handed at the polarized electron gun and be topped off in the ring with the correct polarization every few seconds.

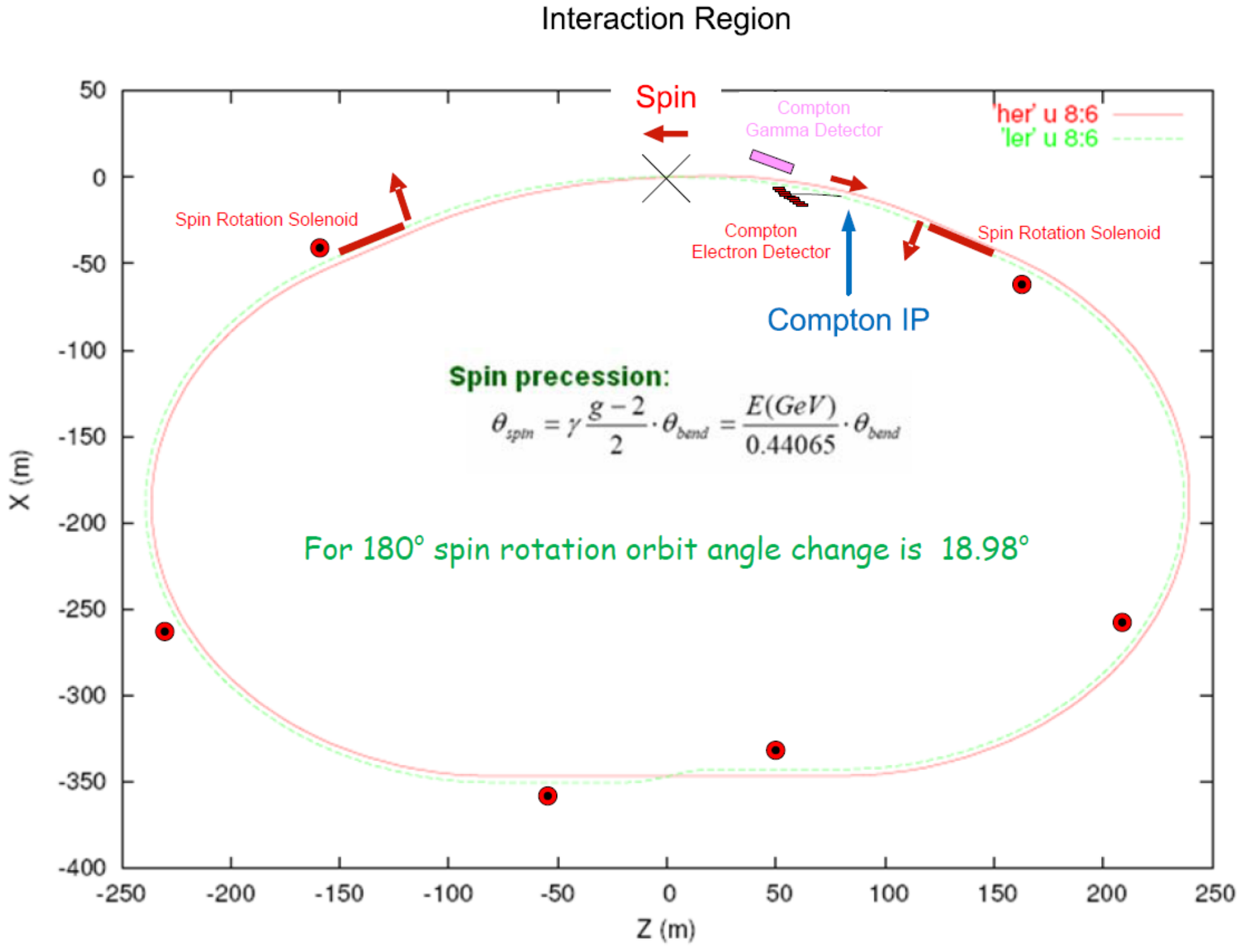

Figure 16.8: SuperB ring showing the location of spin rotation solenoids and the Compton polarimeter at z = 87 m upstream of the Interaction Region.

### Compton Polarimeter

The preferred location of the Compton Polarimeter is immediately downstream of the IR where the direction of the electron beam is the same as at the IR. However, the space at this location is minimal to locate the Compton IP and severe backgrounds from the $e^+e^-$ collisions would be intolerable in the Compton gamma and electron detectors. As a result the Compton Polarimeter will be located upstream of the IR where the spin rotation is close to 180 degrees from the spin orientation at the IR. An ideal location is where the spin orientation is longitudinal and exactly $\pi$ rotation from that at the IP. However, that point occurs inside a dipole magnet of the SuperB lattice.

The orbit angle change for $\pi$ spin rotation is -0.3312 radians at 4.18 GeV. The selected location of the Compton IP in a magnetic field free region has an orbit angle change of -0.3580 radians between the Compton IP and the Interaction Region resulting in the spin direction ~14 degrees from longitudinal. At the Compton IP the longitudinal spin projection is 0.968. The longitudinal polarization at the IR will be larger by 1/0.968 than that measured in the Compton polarimeter. A systematic error will be introduced in the extrapolation due to uncertainty in the beam direction at the Compton IP with respect to that at the IR. An uncertainty of 1 mrad in the orbit will give an uncertainty in the polarization at the IR of 0.25%. A beam energy uncertainty of 20MeV from 4.18 GeV will give a 0.2% error in the polarization at the IR from the measurement at the Compton IP. The layout of the Compton polarimeter is shown in Figure 16.9.

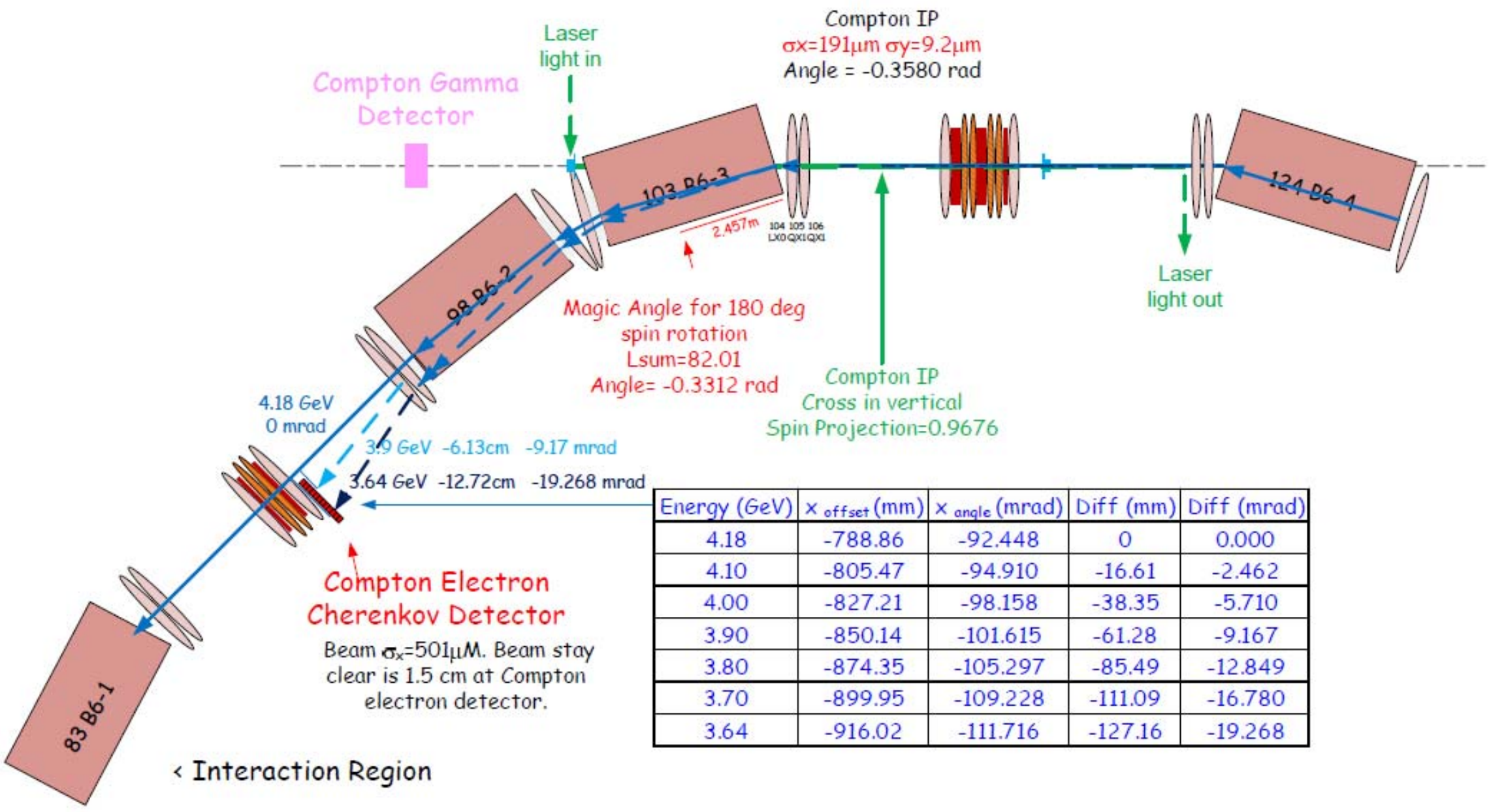

| Energy (GeV) | $x_{offset}$ (mm) | $x_{angle}$ (mrad) | Diff (mm) | Diff (mrad) |
|---|---|---|---|---|
| 4.18 | -788.86 | -92.448 | 0 | 0.000 |
| 4.10 | -805.47 | -94.910 | -16.61 | -2.462 |
| 4.00 | -827.21 | -98.158 | -38.35 | -5.710 |
| 3.90 | -850.14 | -101.615 | -61.28 | -9.167 |
| 3.80 | -874.35 | -105.297 | -85.49 | -12.849 |
| 3.70 | -899.95 | -109.228 | -111.09 | -16.780 |
| 3.64 | -916.02 | -111.716 | -127.16 | -19.268 |

Figure 16.9: Layout of the Compton polarimeter.

The Compton polarimeter is considered using a 1 watt mode locked Nd:YLF circularly polarized laser at 119 megahertz which provides short pulses of 10ps length of 2.3 x $10^{10}$ photons with 2.33 eV. The laser beam enters ~4 cm above the beam line upstream of bend magnet 103 B6-3 and exits 14 m later near bend magnet 124 B6-4 with a crossing angle in the vertical of ~5.7 mrad. The Compton IP is in a no field region as shown in Figure 3. The small crossing angle allows the laser light to see all electron bunches even as they arrive early or late due to the sawtooth timing effect. A crossing angle in the vertical is required to avoid synchrotron radiation damage on the input optics window. Special design features for the beam pipe are needed to avoid RF heating on the input/output optical windows. The Compton electron and gamma detector must have time resolution < 4.2 nsec.

Compton electrons generated at the Compton IP will propagate essentially along the electron beam direction. Two dipole magnets and three quadrupole magnets fan out the Compton electron energy spectrum at the location of the Compton detector shown in Figure 16.9. The segmented electron detector samples the Compton electron flux at energies between 4.06 GeV and the Compton kinematic edge at 3.64 GeV. The Compton electron detector must discern the Compton edge electrons and must be located outside a 1.5cm beam stay clear. The first cell will start at ~2.5 cm from the beam. The Compton electron distance from the beam for different Compton electron energies is given in the insert of Figure 16.9 with the Compton edge electrons occurring at 12.7 cm from the beam. The detector shown in Figure 16.10 is a hodoscope of 30 quartz bars on a movable stage so the Compton edge can be determined with high precision. Each quartz bar is 5mm x 5 cm by 2.5 cm long. Each channel covers ~21 MeV of backscattered electrons. The 1mm aluminum beam pipe is flared and angled at ~200mrad. The silica bars are staggered to allow photomultiplier tubes to match the pitch of the counters and will give roughly 12 photoelectrons per track. Fused silica is a good match to the radiation dose in the quartz bars which will absorb ~50 megarads per year from the signal itself.

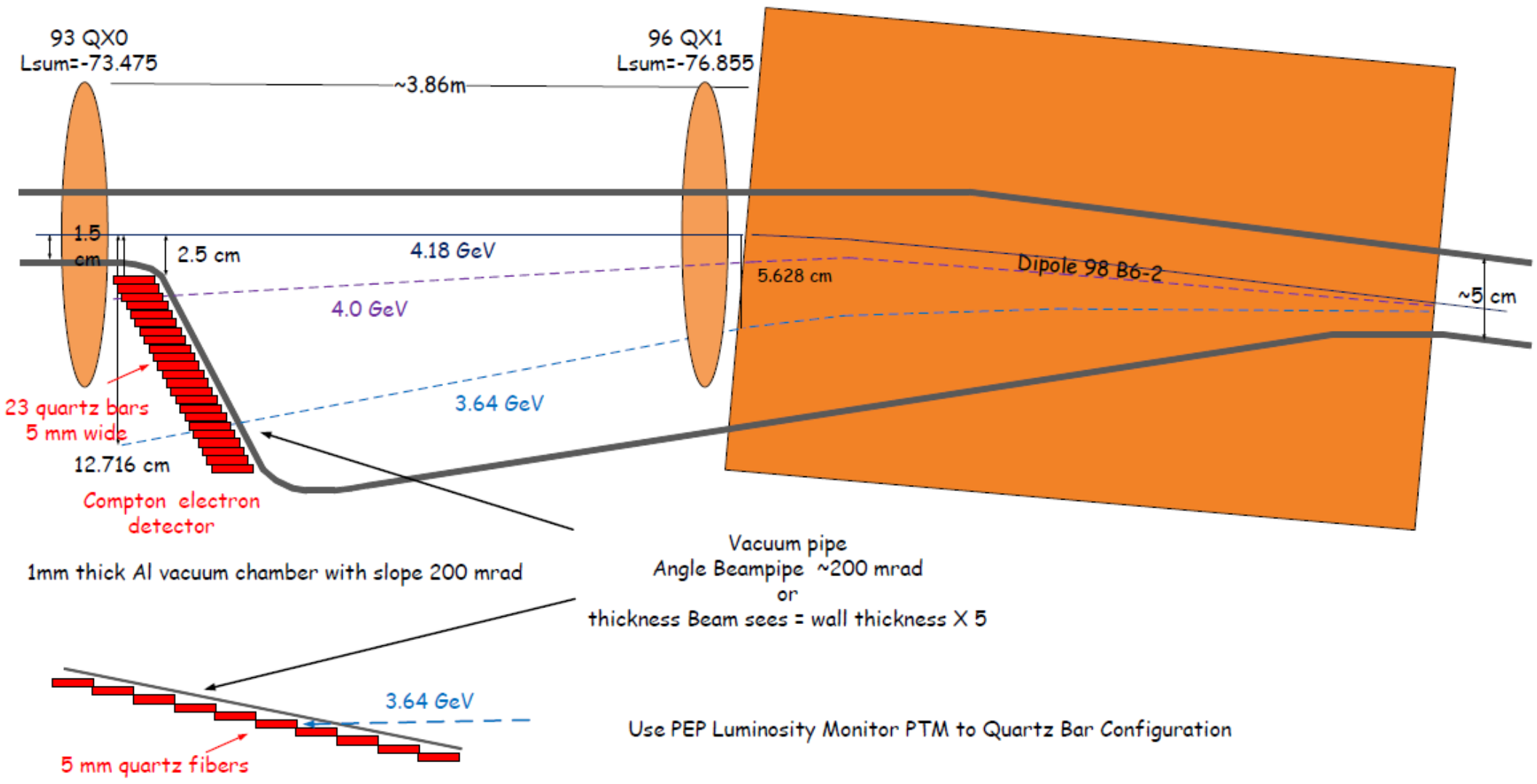

Figure 16.10: Detail for the Compton electron detector located in the drift region after quadrupole QX1.

The forward Compton gammas are detected in a 5 x 5 x 2.5 cm deep quartz plate. The Compton gammas exit through ~1.5 RL water cooled Al window to remove the heat from absorbed synchrotron radiation. The shower is rejuvenated using a local plate of tungsten of ~2 RL with the fused silica plate behind it. Cherenkov light is taken out through a slanted roof into a light pipe and matched to a fast PMT. The calorimeter will be shielded from backgrounds.

The Compton laser is pulsed with a pattern that matches the pulse and bunch structure of the electron bunches in the SuperB ring. Table 2 gives polarimeter parameters.

Table 16.3: Compton polarimeter parameters at 4.18 GeV.

| **Beam Parameter** | **Electron Beam** | **Laser Beam** |
|---|---|---|
| $\sigma_x$ | 500 μm | 100 μm |
| $\sigma_y$ | 5 μm | 100 μm |
| $\sigma_z$ | 5 mm | 1.3 mm |
| # particles/bunch | $5.7 \times 10^{10}$ | $2.3 \times 10^{10}$ |

The unpolarized Compton cross section for head-on collisions of 4.18 GeV electrons with 2.33 eV photons is 1.09 barns giving a rate of

$$R^0_{Compton} = \sigma_{Compton} \cdot \frac{N_{electrons} \cdot N_{photons}}{\pi \cdot \sigma_x \cdot \sigma_y} = 0.9 / bunch$$

The small vertical crossing angle, coupled with the electron bunch length, will increase the effective vertical spotsize of the colliding beams. This is parameterized by $f_{geom}$ which for small crossing angles is given by:

$$R^{eff} = R^0_{Compton} \cdot f_{geom}$$

where

$$f_{geom} = \frac{\sigma_y}{\sqrt{(\sigma_y)^2 + (\theta_y^{Compton} \cdot \sigma_z)^2}} =$$

$$= \frac{100\mu m}{\sqrt{(100\mu m)^2 + (5.7mrad \cdot 5mm)^2}} = 0.96$$

giving an effective rate for Compton scatters of 0.66 scatters per collision from a 1W laser beam at 119 MHz. The 0.87 scatters per collision are high enough to be non-linear in a counting mode giving a larger systematic error. For this reason it may be desirable to run at lower laser power.

Each of the 1011 electron bunches goes around the 1323 m ring 226,597 revolutions per sec giving a rate in the Compton gamma detector of ~196,980 gammas/sec for each of the bunches sampled (i.e. a rate of 78 MHz). Each cell of the Compton electron detector will see ~8360 Compton electrons/second per bunch sampled for a 3 MHz rate. The mode locked 119 MHz Compton laser pulses will collide with every other electron bunch in the ring. The timing of the Compton laser pulses can be varied so as to sample the other 505 electron bunches. The electron beam polarization at the Compton IP is determined from the rate in the Compton detectors by:

$$\sigma(E/E_0) =$$
$$= \sigma_0(E/E_0) \cdot \left[1 + P_e^{ComptonIP} \cdot P_\gamma \cdot A(E/E_0)\right]$$

with the measured asymmetry in the i[th] channel being

$$A_i^m = \frac{N_i^{\to\to} - N_i^{\to\leftarrow}}{N_i^{\to\to} + N_i^{\to\leftarrow} - 2\cdot N_i^{laseroff}} = $$
$$= a_i \cdot P_e^{ComptonIP} \cdot P_\gamma .$$

The analyzing power is calculated from the Compton cross section and the channel response function, $R_i$.

$$a_i = \frac{\int \frac{d\sigma_0}{dx}\cdot A(x)\cdot R_i(x)\cdot dx}{\int \frac{d\sigma_0}{dx}\cdot R_i(x)\cdot dx}.$$

The polarization at the IR is:

$$P_e^{IR} = P_e^{ComptonIP} / \cos\left(\theta_{spin}^{ComptonIP}\right) =$$
$$= P_e^{ComptonIP} / 0.9676 .$$

**Luminosity-weighted beam Polarization at the IR**

The luminosity-weighted beam polarization may differ from the measured polarization due to disruption and radiation in the beam-beam collision process. There are also effects from polarization spread and spin transport. The spin motion of a deflected electron or positron beam in a transverse magnetic field follows from the familiar T-BMT expression

$$\theta^{spin} = \gamma \frac{g-2}{2}\theta^{orbit} = \frac{E_0}{0.44065GeV}\theta^{orbit} \qquad (10)$$

where $\theta^{orbit}$ and $\theta^{spin}$ are the orbit deflection and spin precession angles, $E_0$ is the beam energy, $\gamma = E_0/m$, and (g-2)/2 is the famous g-factor anomaly of the magnetic moment of the electron. The difference between the luminosity-weighted beam polarization and the polarimeter measurement is written as $dP = P_z^{lum-wt} - P_z^{CIP}$. To minimize $dP$, it is required that the beam direction at the Compton IP be known with the collision axis at the $e^+e^-$ IR to within 1mrad. Orbit misalignments between the polarimeter IP and the collision IP are expected to be below 1mrad, which would give $dP$ < 0.25%. The effect of Sokolov-Ternov spin flips is expected to be small. Effects from the angular divergence of the beam at the Compton IP and the IR are expected to be negligible. Effects from chromatic aberrations are expected to be negligible. Table 3 gives estimates for the systematic errors on the polarization at the IR that can be expected from the polarization measurement. The measurement of polarization at the 1% systematic error level is feasible based on SLD experience [3] and at Jefferson Laboratory [4].

Table 16.4: Systematic errors expected for the polarization measurement.

| **Item** | **δP/P** |
|---|---|
| Laser Polarization | <0.1% |
| Background uncertainty | <0.25% |
| Linearity of phototube response | <0.25% |
| Uncertainty in dP (Difference between the luminosity weighted polarization and the Compton IP polarization. Includes uncertainties due to beam energy and direction uncertainties.) | <0.4% |
| Uncertainty in asymmetry analyzing power | ~0.5% |
| **Total Systematic Error** | **<1.0%** |

**Summary**

A scheme for measuring the electron beam polarization at SuperB near the IR has been described. The Compton polarimeter has been designed to fit into the existing lattice of the SuperB ring and results in the Compton IP measuring the polarization where the beam is almost longitudinal with opposite helicity to that at the IR. The polarization at the IR can be determined from the measurement at the Compton IP provided the beam direction at the electron-positron Interaction Region and the Compton IP are well known and the beam energy is measured to better than 20 MeV.

# 17. Site geology

The SuperB complex is proposed as an extension of the Frascati National Laboratories (LNF). The Laboratories, founded in the early 1955, are located on the slope of the Volcano Laziale. Site geology started more than 730000 years ago, when from the Mediterranean sea raised up a volcano, whose activity ended about 30000 years ago. Some gases, mainly carbon-dioxide, are still exhaled from the subsoil today. See Figure 17.1 for a pictorial view of the “Volcano Laziale”, its cross-section and a seismic map of Italy.

The underground composition reflects the evolution of the volcano activity, as it is possible to see in Figure 17.2, where a cross section of the ground in the north-south direction, in the central part of the Frascati Laboratory is reported up to a depth of 50 m. Below the vegetable soil, a few meters deep, there is a first layer of solid lava (layer 7 in Figure 17.2) having a thickness variable from 5 to 7.8 m. Then, different layers of pyroclastic grounds, with different types of inclusions, are present. They have a thickness of about 20 m, with one or more layers of lava rocks, variable from 1 to 5 m, have been again found (layer 3 in Figure 17.2).

These pyroclastic grounds are a mixture of sand and clay able to damp vibrations coming from natural sources like seismic activity or human noise coming, e.g., from surface roads and railways, as confirmed by vibration measurements.

From a seismic point of view the area is quite stable even if it is located near the boundary of a very active region (the central part of Italy). The maximum ground acceleration here is between 0.17g to 0.15g where g is the gravitational acceleration (see Figure 17.1).

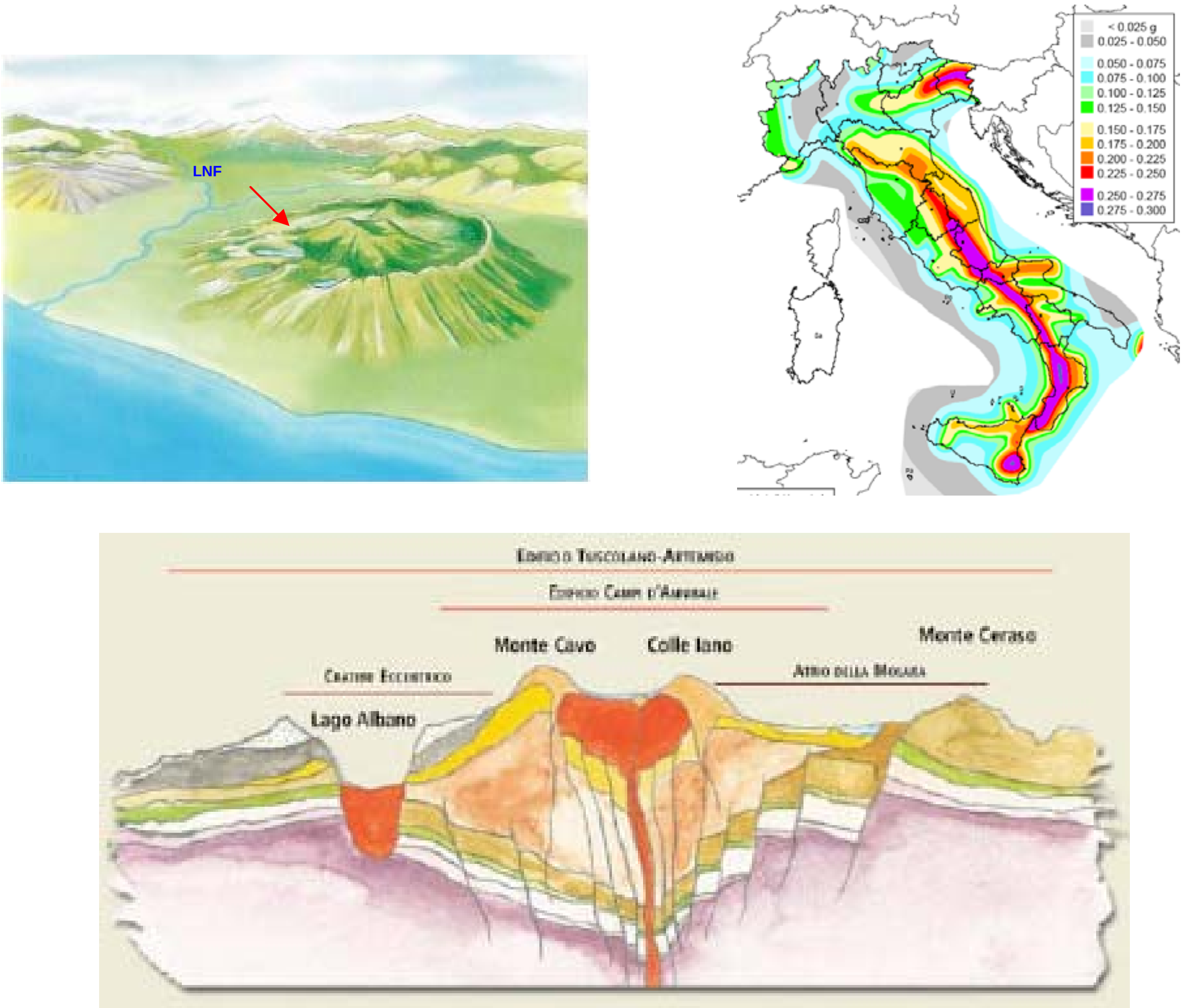

Figure 17.1: Volcano Laziale (top left), Seismic map of Italy (top right), Volcano Cross section (bottom).

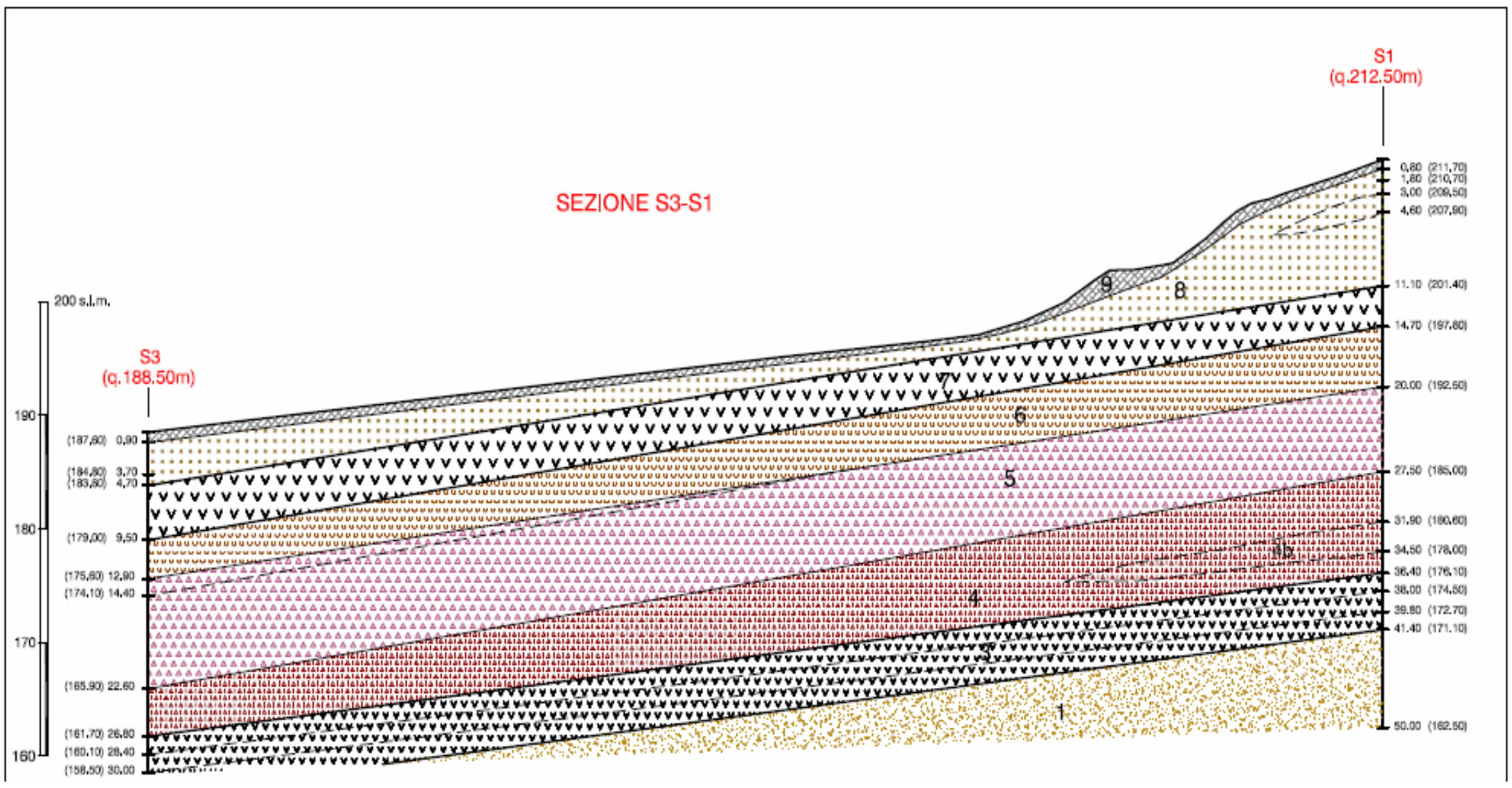

Figure 17.2: Laziale geological cross-section in the south part of the Frascati laboratory.

# 18. Potential Synchrotron Radiation Light Source Beam Lines

The possibility of state of the art synchrotron radiation beam lines being added to the SuperB collider was envisioned in the design process. Both the low energy and high energy rings have transverse and longitudinal emittances that are comparable to world class light sources, such as the recently commissioned PETRA3 and the NSLS-2 under construction. Furthermore, SuperB will have beam currents that are several factors higher than these other light source rings. Thus, the x-ray flux levels and spectral brilliance should compete favorably with these other sources.

The expected photon brilliance for SuperB is presently under study with many possibilities given the breadth of wiggler and undulator types available now.

In the following preliminary results for the spectral central intensity, spectral flux and brightness, as calculated analytically for bend magnets and undulators, will be shown. The radiation characteristics of bend magnets and undulators have been described in numerous publications (e.g. X-Ray Data Booklet [1]) where the fundamental properties are described which have been used to generate the following plots. In order to estimate the properties of possible SuperB synchrotron beam lines, several light sources have been considered. The parameters used in this comparative study for the light coming from bend magnets are shown in Table 18.1, and for undulators in Table 18.2. The sources for these parameters shown in the tables are either from the individual light source web pages or their published Design Reports.

Table 18.1: Parameters for different light sources used to compare figures of merit of synchrotron radiation generated from the bend magnets.

| Parameters | SuperB HER | SuperB LER | NSLS II | APS | ESRF | ELETTRA | ALS |
|---|---|---|---|---|---|---|---|
| E [GeV] | 6.7 | 4.18 | 3.0 | 7.0 | 6.03 | 2.0 | 1.9 |
| I [mA] | 1892 | 2447 | 500 | 100 | 200 | 320 | 500 |
| $\rho$ [m] | 69.64 | 26.8 | 24.975 | 38.961 | 23.623 | 5.55 | 4.81 |
| $\varepsilon_x$ [m rad] | 2.0e-9 | 2.46e-9 | 0.55e-9 | 2.514e-9 | 4.0e-9 | 7.0e-9 | 6.3e-9 |
| $\varepsilon_y$ [m rad] | 5.0e-12 | 6.15e-12 | 8.0e-12 | 22.6e-12 | 25.0e-12 | 70.0e-12 | 50.0e-12 |
| $\gamma_y$ [$m^{-1}$] | 0.334 | 0.537 | 0.050 | 0.101 | 0.10 | 0.50 | 0.740 |
| $\sigma_x$ [mm] | 82.1e-3 | 92.1e-3 | 125.0e-3 | 81.7e-3 | 77.0e-3 | 139.0e-3 | 101.8e-3 |
| $\sigma_y$ [mm] | 8.66e-3 | 9.11e-3 | 13.4e-3 | 27.0e-3 | 29.5e-3 | 28.0e-3 | 8.2e-3 |

The flux from bend magnets for different storage rings is shown in Figure 18.1. The SuperB LER and HER rings have the highest flux in this comparison due to the high beam current.

Fig. 18.2 shows a comparison of bend magnets spectral brightness from different storage rings. The calculation was performed analytically using the parameters from Table 18.1. From this comparison both Super-B storage rings show very good performances. The difference between SuperB and the other machines considered is here reduced, since the source size parameters in the light source storage rings have been of course optimized.

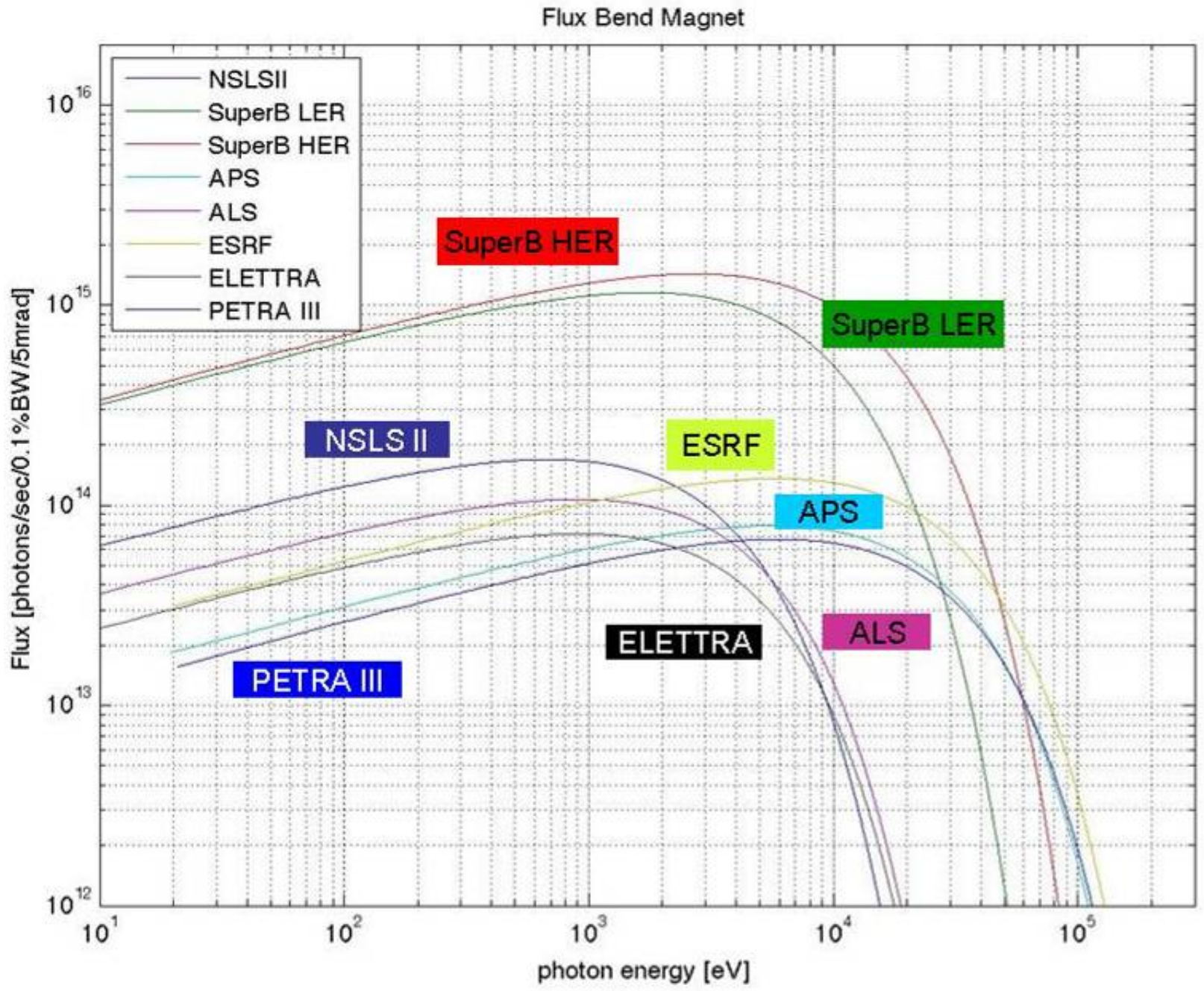

Figure 18.1: Comparison of bend magnets flux from different storage rings. The calculation was performed analytically using the parameters from Table 18.1.

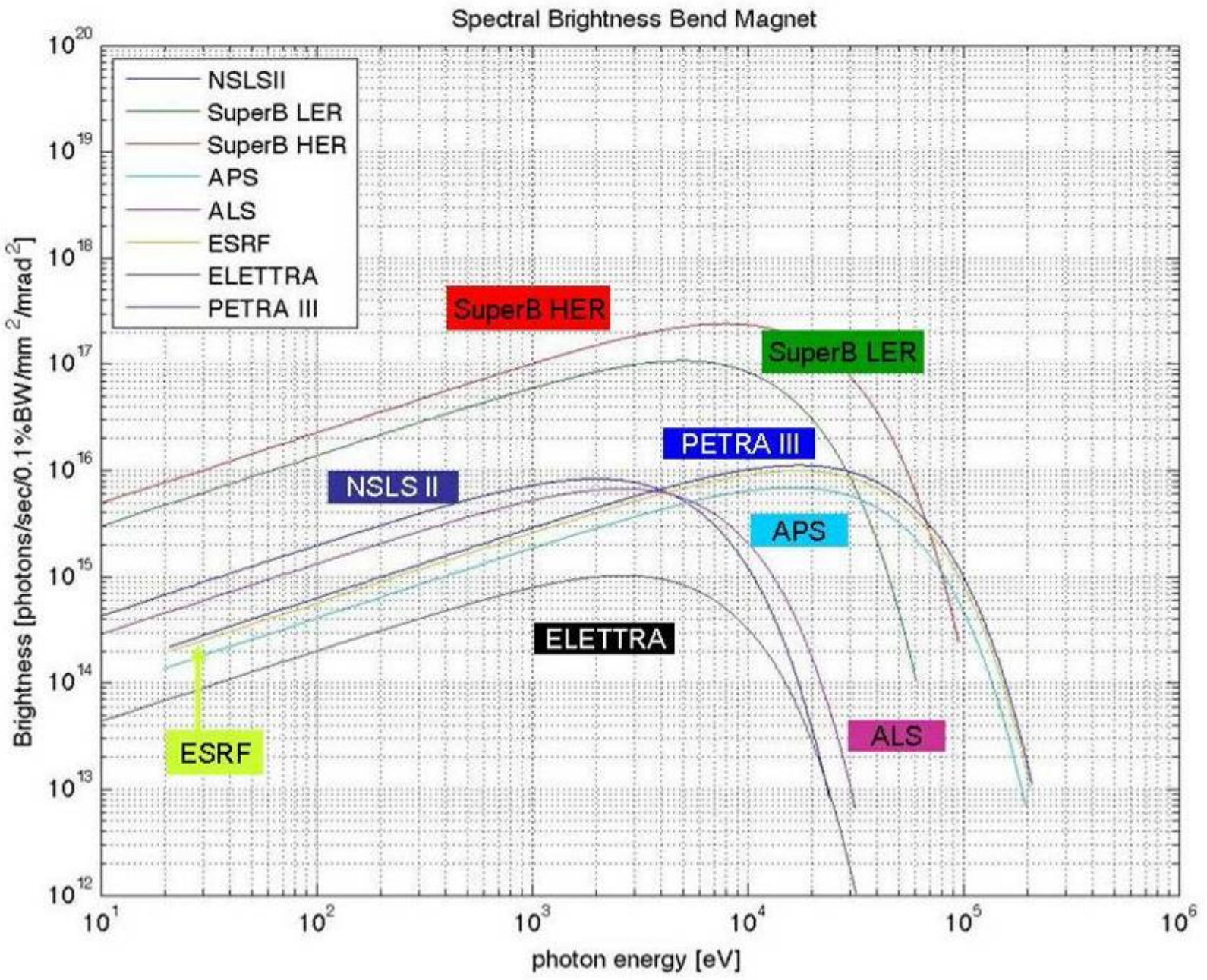

Figure 18.2: Comparison of bend magnets spectral brightness from different storage rings. The calculation was performed analytically using the parameters from Table 18.1.

New designs for light sources are optimized for synchrotron radiation from insertion devices. The performance of these devices is used to benchmark individual facilities. To allow for a comparison with the LER and HER a standard undulator was used to in combination with the ring and optics parameters as shown in Table 18.2. The resulting flux and brightness comparison is shown in Figures 18.3 and 18.4. The flux of both SuperB LER and HER exceed those of all other sources due to the large beam current. Only the spectral brightness of PEPX is larger, due to its small horizontal source size.

Table 18.2: Parameters for different light sources used to compare figures of merit of synchrotron radiation generated from undulators.

| Parameters | SuperB HER | SuperB LER | NSLS II | APS | PEPX | Soleil | Spring8 | PetraIII |
|---|---|---|---|---|---|---|---|---|
| Undulator | U20 | U20 | IVU20 | U33 | U23 | U20 | U24 | U29 |
| E [GeV] | 6.7 | 4.18 | 3.0 | 7.0 | 4.5 | 2.75 | 8.0 | 6.0 |
| I [mA] | 1892 | 2447 | 500 | 100 | 1500 | 500 | 100 | 100 |
| $\sigma_x$ [mm] | 60.0e-3 | 67.0e-3 | 33.3e-3 | 278e-3 | 22.2e-3 | 388e-3 | 286e-3 | 140e-3 |
| $\sigma_y$ [mm] | 3.0e-3 | 3.0e-3 | 2.9e-3 | 8.9e-3 | 7.0e-3 | 8.08e-3 | 6.0e-3 | 5.6e-3 |
| $\sigma_x$' [mrad] | 33.3e-3 | 37.0e-3 | 16.5e-3 | 11.3e-3 | 7.4e-3 | 14.5e-3 | 11.0e-3 | 7.9e-3 |
| $\sigma_y$' [mrad] | 2.7e-3 | 2.4e-3 | 2.7e-3 | 3.3e-3 | 1.2e-3 | 4.6e-3 | 1.0e-3 | 4.1e-6 |
| N [1] | 148 | 148 | 148 | 72 | 150 | 90 | 186 | 172 |
| $\lambda_u$ [mm] | 20 | 20 | 20 | 33 | 23 | 20 | 24 | 29 |
| $K_{max}$ [1] | 1.83 | 1.83 | 1.83 | 2.75 | 2.26 | 1.9 | 2.21 | 2.2 |
| $K_{min}$ [1] | 0.1 | 0.1 | 0.1 | 0.1 | 0.1 | 0.1 | 0.1 | 0.1 |

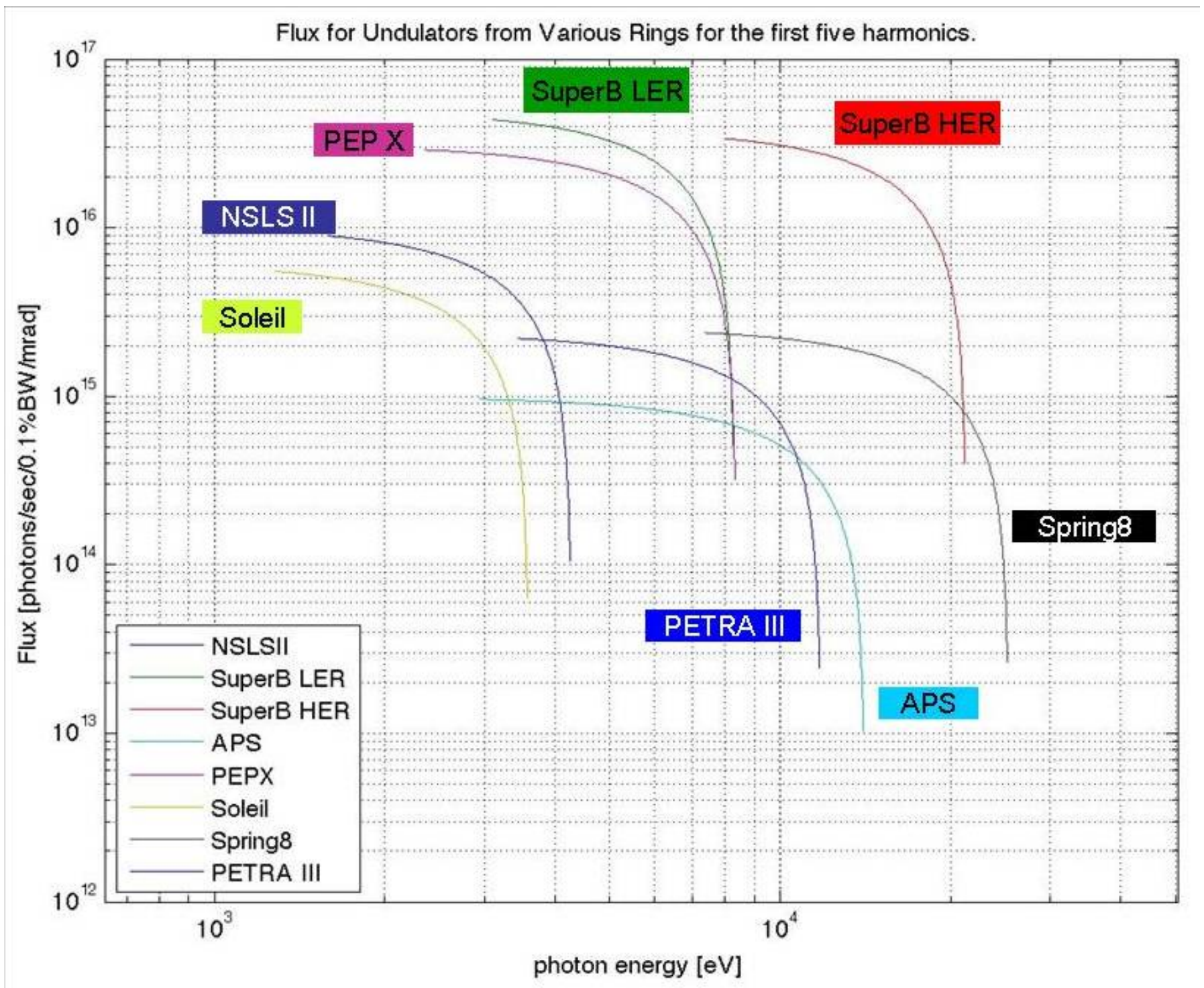

Figure 18.3: Comparison of undulator flux from different storage rings. The calculation was performed analytically using the parameters from Table 18.2.

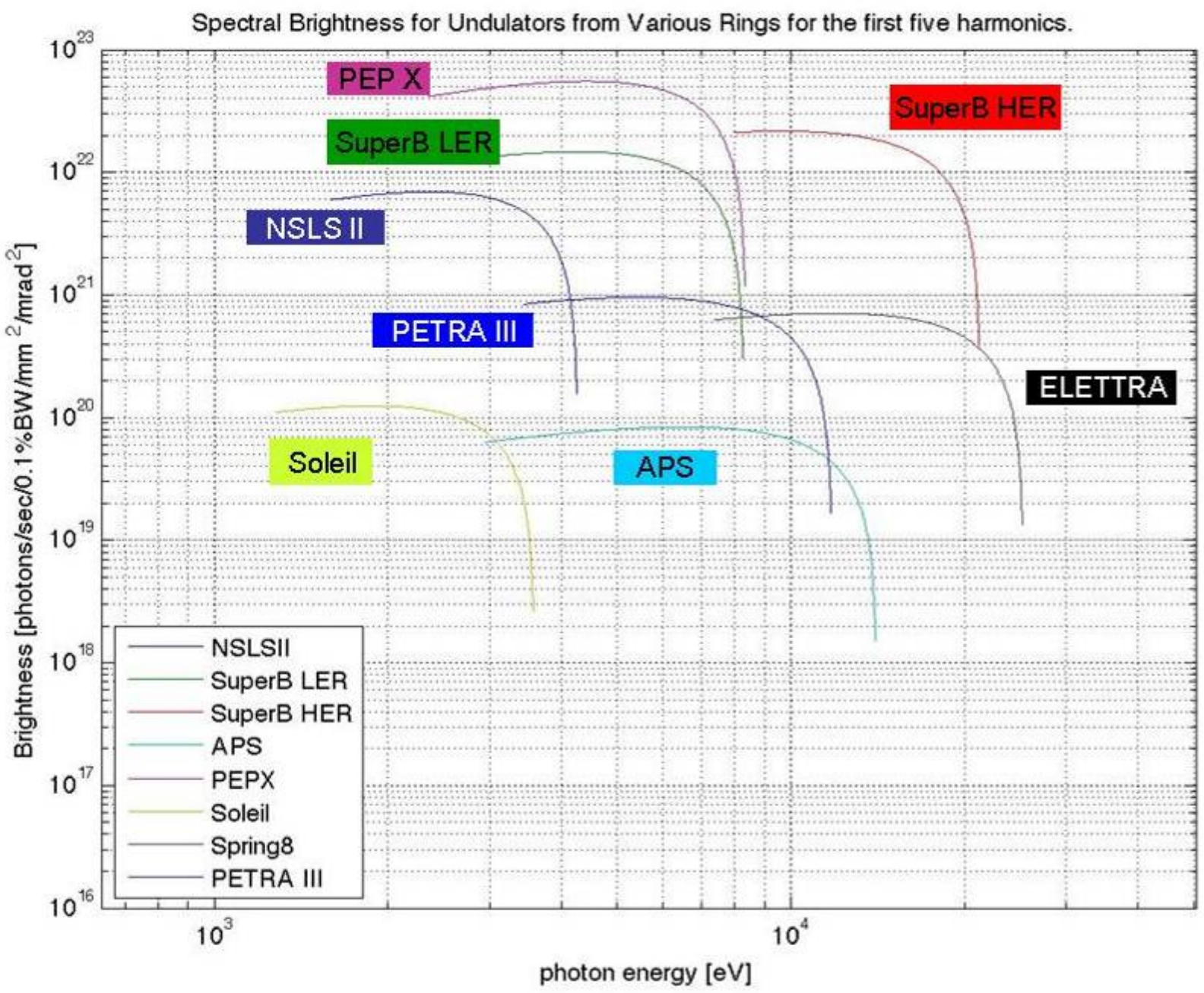

Figure 18.4: Comparison of undulator spectral brightness from different storage rings. The calculation was performed analytically using the parameters from Table 18.2.

It would be natural to have beam lines on both the high energy ring for hard x-ray users and on the low energy ring for soft x-ray users. Both sets of beam lines could operate during colliding beam operation. A possible layout of x-ray beam lines is shown in Figure 18.5 where, in this example, SuperB is located on the Frascati INFN site. Since the HER is on the outside over half the circumference and the LER the other half, having respective beam lines on these halves makes sense.

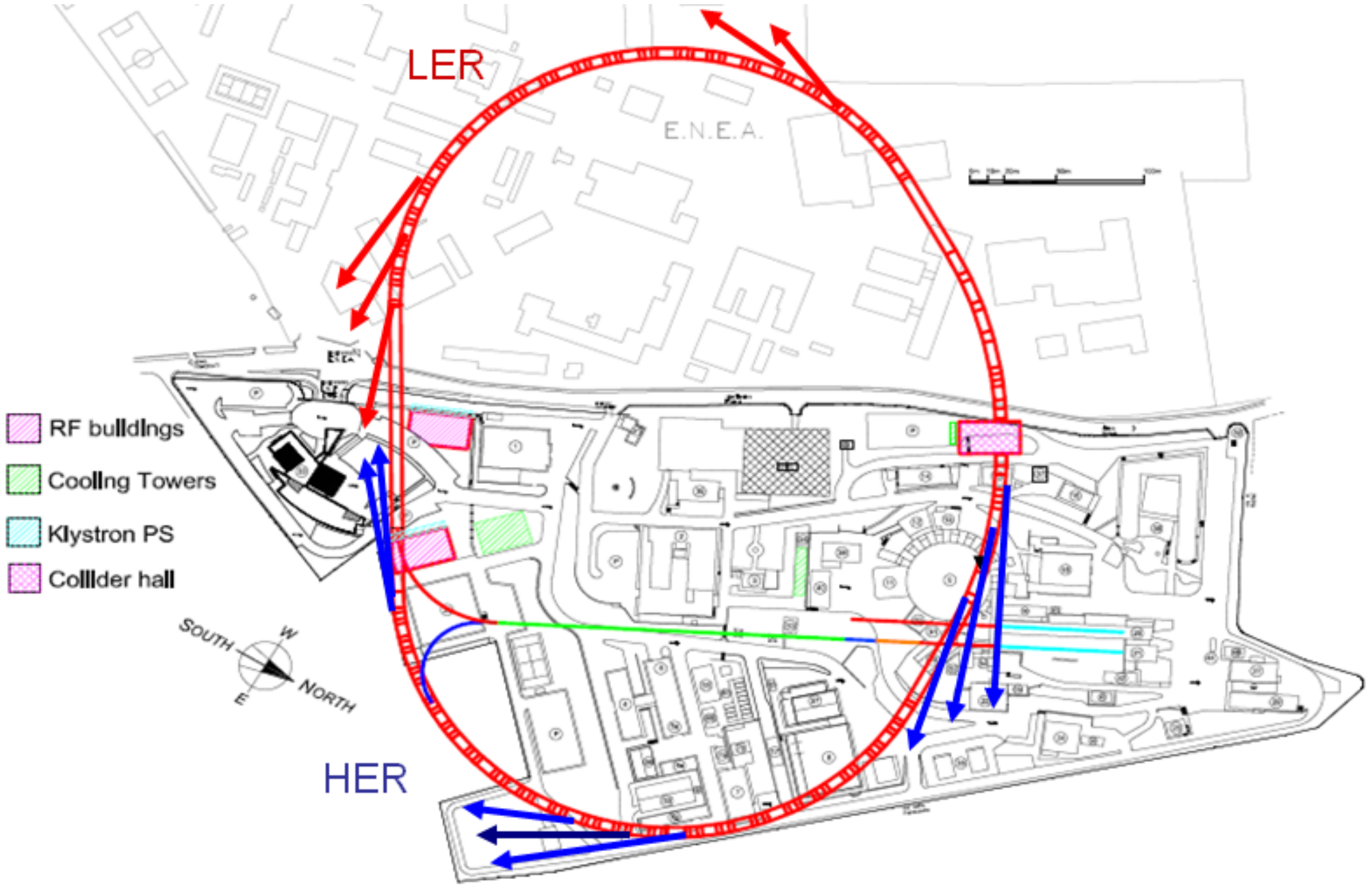

Figure 18.5: Possible locations of synchrotron x-ray lines at SuperB.

# 19. Ground motion measurements at LNF

Similar to the studies of the Virgo team [1], more detailed ground motion measurements have been performed at the LNF site for the SuperB project in collaboration with the French group from LAPP.

## 19.1 Measurement locations

A potential location of about 1.3 Km SuperB on the LNF site is shown in Fig. 19.1.

Measurements have been performed at different locations of the LNF site (see Fig. 19.1). Each location presents various properties and has been chosen in order to compare the influence of various vibrations sources (such as traffic, air cooling, railway track...) and the influence of the quality of the concrete.

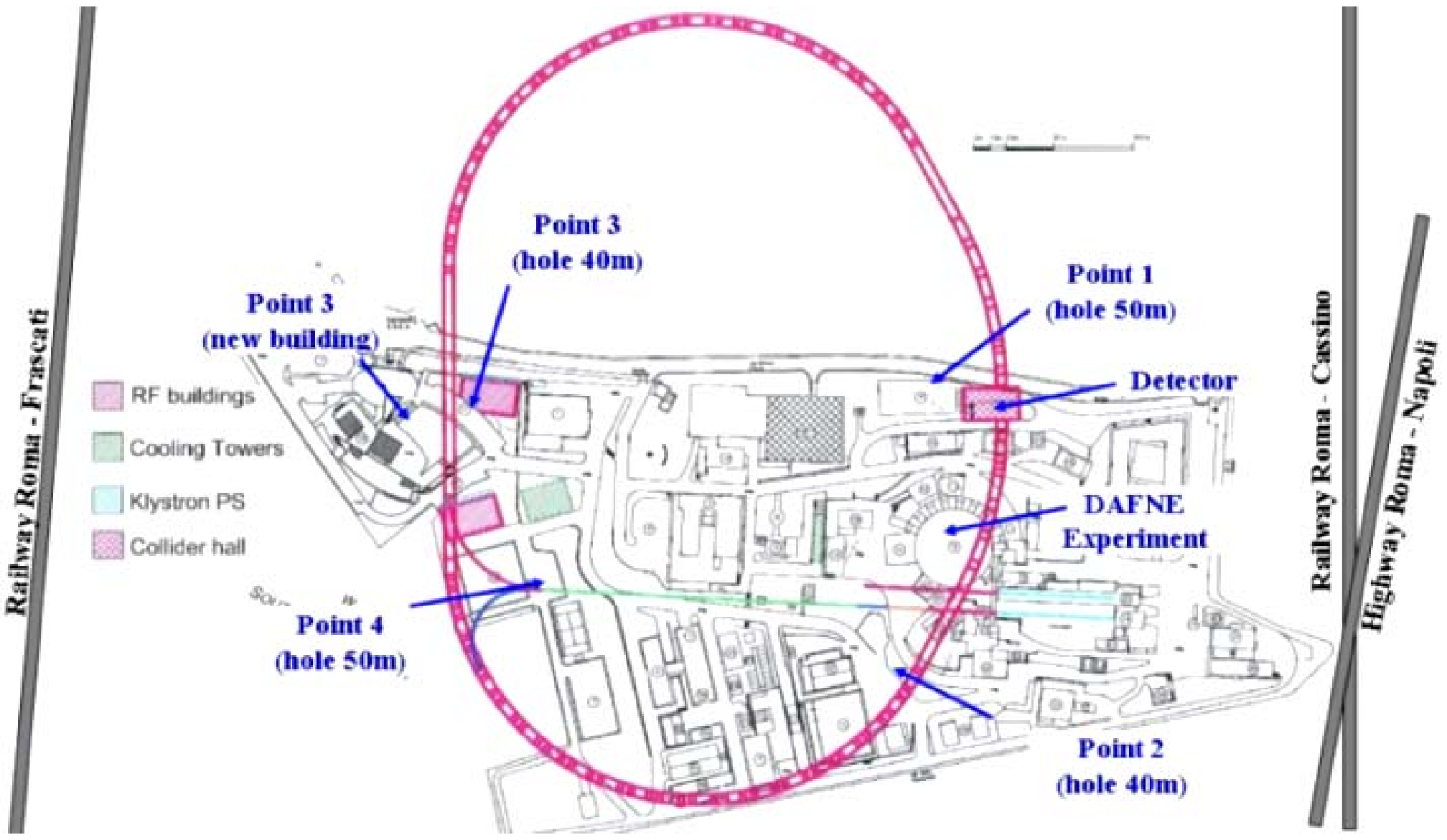

Figure 19.1: Layout of the LNF site showing the points where vibration measurements were performed.

The first point where measurements were performed corresponds to the location of the future collider hall. It is situated in the proximity of the main road (E. Fermi) where there is heavy traffic during the day and a power electrical substation is situated. Measurements have been done both at the surface and in the bottom of a hole at a depth of 50m (see Fig. 19.2).

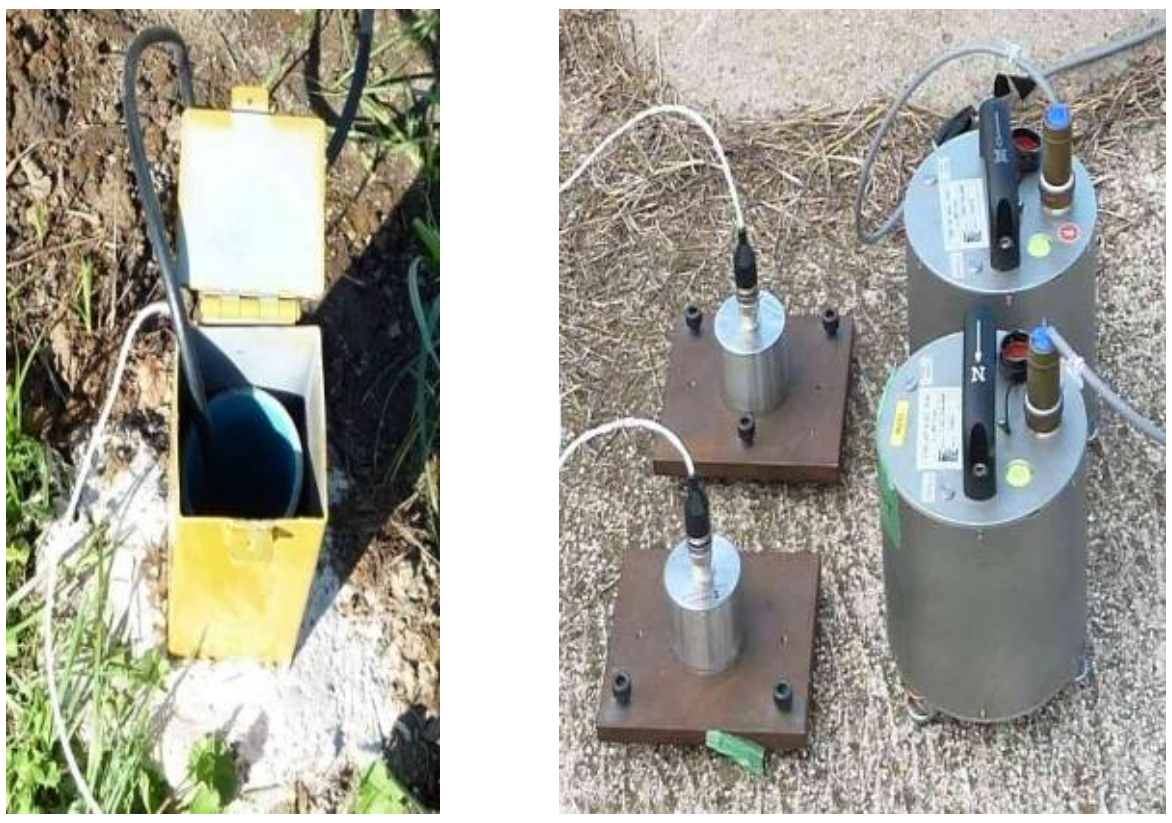

Figure 19.2: Layout of set of measurements at point 1.

The second point of measurements was located beside the DAΦNE damping ring and not far from the main pumping station of the DAΦNE cooling plant. The original plan was to do a set of measurements at the surface and also in a dedicated hole (40m depth), but due to a large quantity of water accumulated in the hole, only surface measurements have been performed. The last point of measurements was the point 3, where coherence measurements were performed on two different types of floor close to each other (see Fig. 19.3): on surface in the parking (soft floor) and on the concrete basement of the new guest house building (rigid floor). The last set of measurements was taken on the basement of this new guesthouse building in an acquisition session of 18 hours. Note that this point is also situated in the proximity of a main road as the point 1. No measurements have been performed in the holes located at the points 3 and 4 because one was blocked by a stone (point 3) and the other was filled with water (point 4).

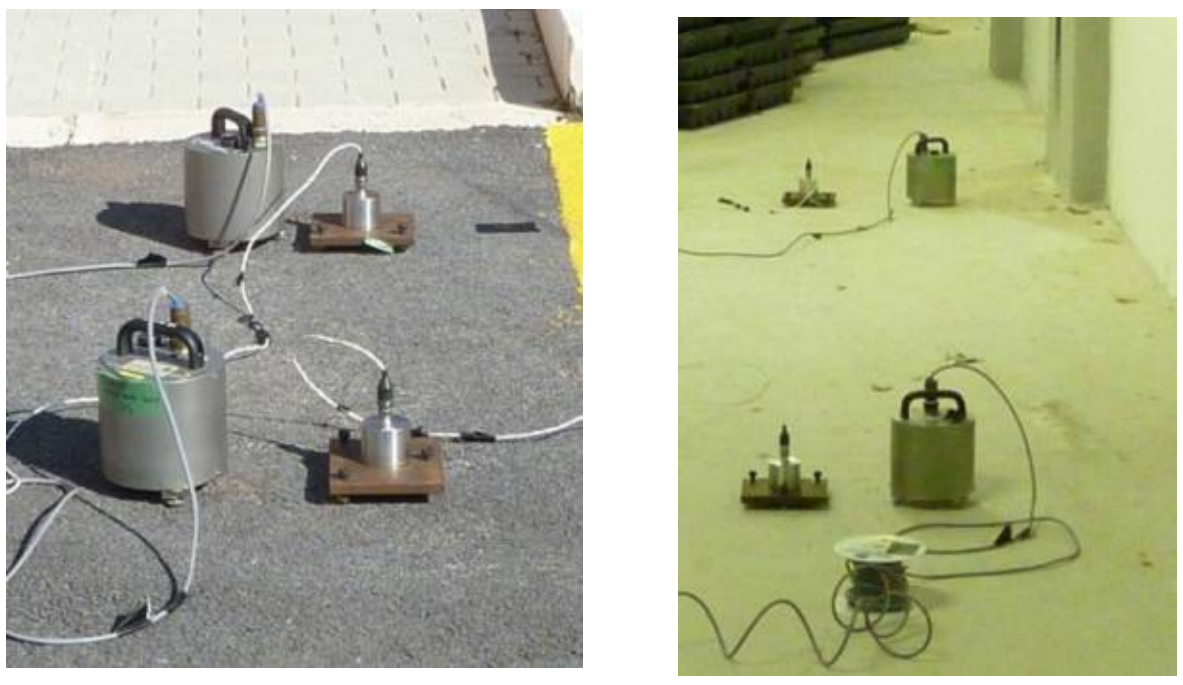

Figure 19.3: Layout of the set of measurements done at the point 3 on two different types of floor (at left: soft floor and at right: rigid floor).

## 19.2 Ground motion amplitude with time

Measurements were done during an 18 hours period in the vertical direction near a main road on the surface. They show that earth motion (from 0.2Hz to 1Hz) is around 70 nm, and that cultural noise (from 1Hz to 100Hz) varies from about 12nm to 35 nm between 17h40 and 8h00 and from 38nm to 65 nm between 09h40 and 11h40 on average, but increases significantly between 8h00 and 09h40 due to rush hours traffic ([3 to 30] Hz) up to 240nm.

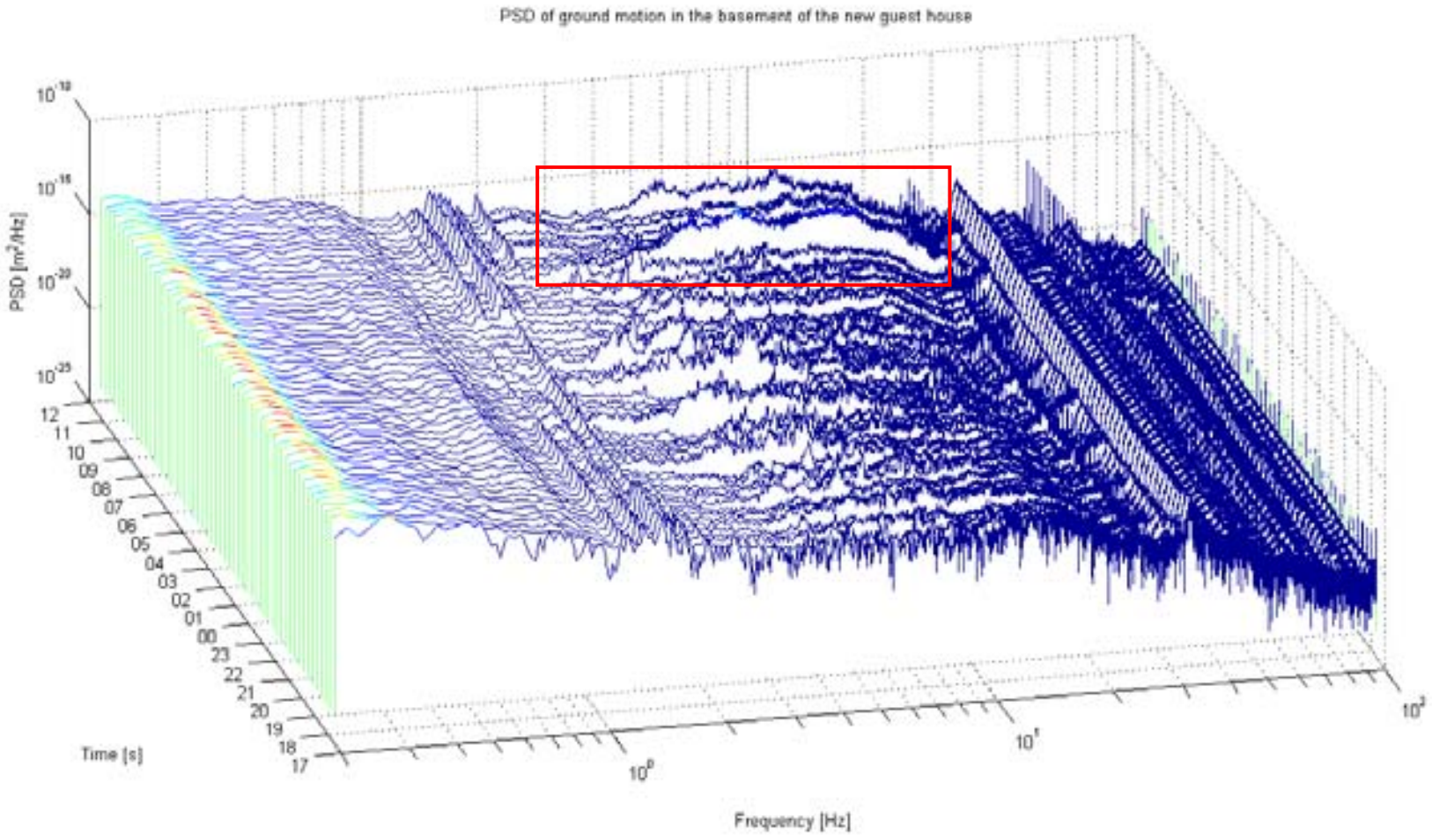

Figure 19.4: PSD of ground motion at point 3 (basement of the new guest house) versus time and frequency.

Many of the power spectral density (PSD) measurements have similar spectra with time, except between 8h00 and 9h40 where their amplitude increases much in the frequency range [3 to 30] Hz (see red rectangle in 19.4). This increase is certainly due to traffic since the time corresponds to rush hour and other studies have shown that vibrations due to traffic are exactly in this frequency range [2]. In order to have a better view of the amplitude increase, the PSDs are shown in Figure 19.5 (data reliable above 0.2Hz) in 2 dimensions (amplitude, frequency) only in the time area where the amplitude increases

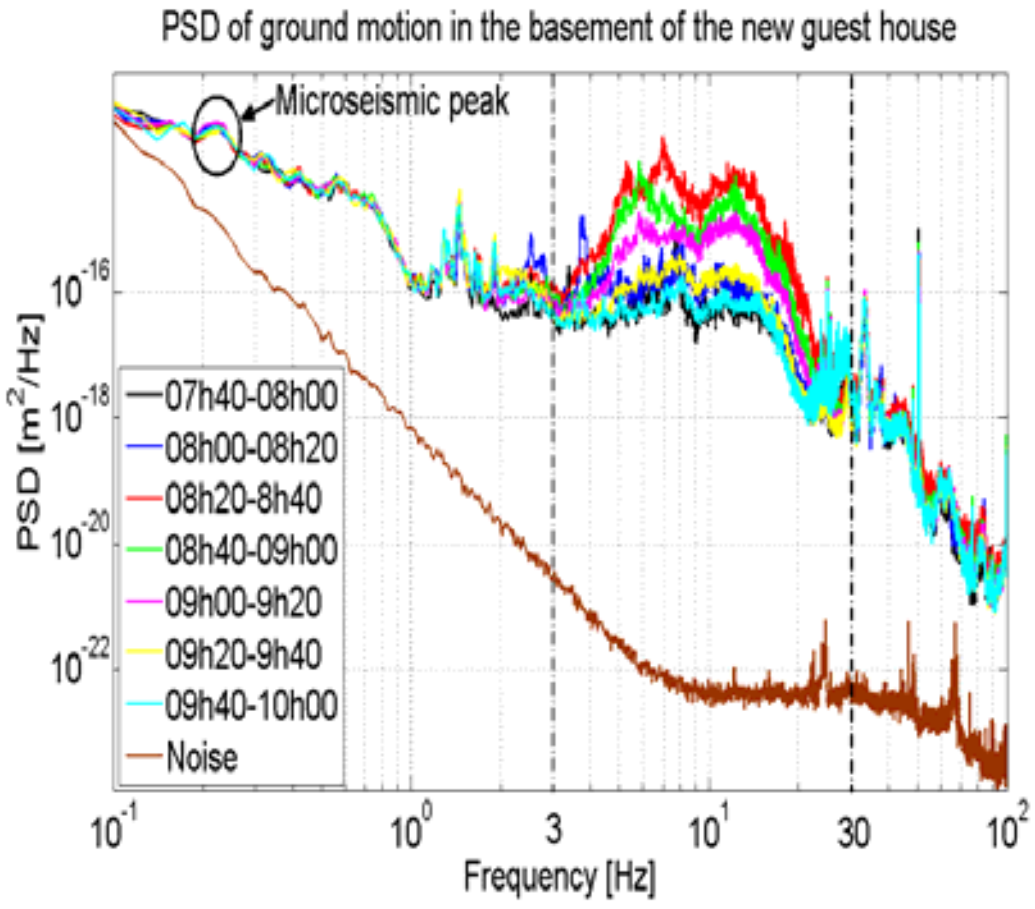

Figure 19.5: PSD of ground motion at point 3 (basement of the new guest house) versus frequency.

In the frequency range [3; 30] Hz, it is clearly seen that the amplitude slowly increases from 08h00 to 08h20, then highly increases from 08h20 to 8h40, and finally slowly decreases from 8h40 to 10h00 down to the same amplitude before 08h00. Note that below 1Hz, ground motion is due to earth motion and that above 1Hz, ground motion is due to cultural noise, that is to say human activities [3]. Especially in the frequency range [0.1; 1] Hz, ground motion is mostly due to the micro-seismic peak (motion of waves in the ocean), whose frequency can be seen around 0.2Hz in Figure 19.6. In order to have values of the amplitude of ground motion with time, the PSDs shown above have been integrated in different bandwidths (integrated RMS calculations): from 0.2Hz to 100Hz, from 1Hz to 100Hz, from 10Hz to 100Hz and from 50Hz to 100Hz. Results are shown in Fig. 19.6 below.

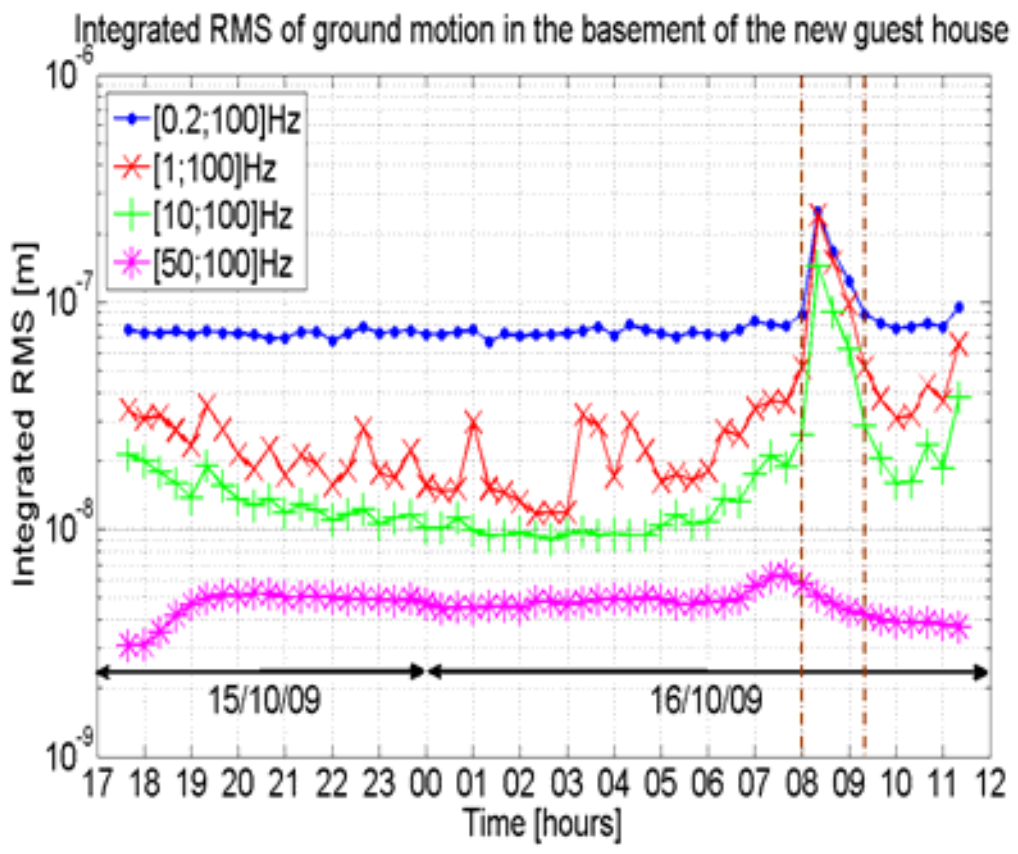

Figure 19.6: RMS of ground motion versus time at point 3 (basement of the new guest house) integrated in different bandwidths.

In the bandwidths [1; 100] Hz and [10; 100] Hz, amplitudes of ground motion vary from 12nm to 35nm and from 9nm to 21nm respectively between 17h40 and 8h00 (smallest values the night due to reduced human activities), and increase from 38nm to 65nm and from 20nm to 38nm respectively between 09h40 and 11h40 (increase due to the beginning of the day). However, the amplitudes highly increase up to 240nm above 1Hz and up to 144nm above 10Hz between 8h00 and 9h40. This time period corresponds to the peaks observed on the PSDs between 3Hz and 30Hz. These results consequently show that traffic can highly increase ground motion (more than a factor 10).

## 19.3 Comparison between surface and underground

At point 1, ground motion has been measured simultaneously on the surface and in a hole of 50m depth. From these measurements, ground motion PSDs on the surface and inside the hole as well as the vibration transfer function between the surface and the bottom of the hole have been calculated and are plotted in Fig. 19.7, top and bottom respectively. Results are shown above 1.3Hz, frequency from where data are reliable (high signal to noise ratio).

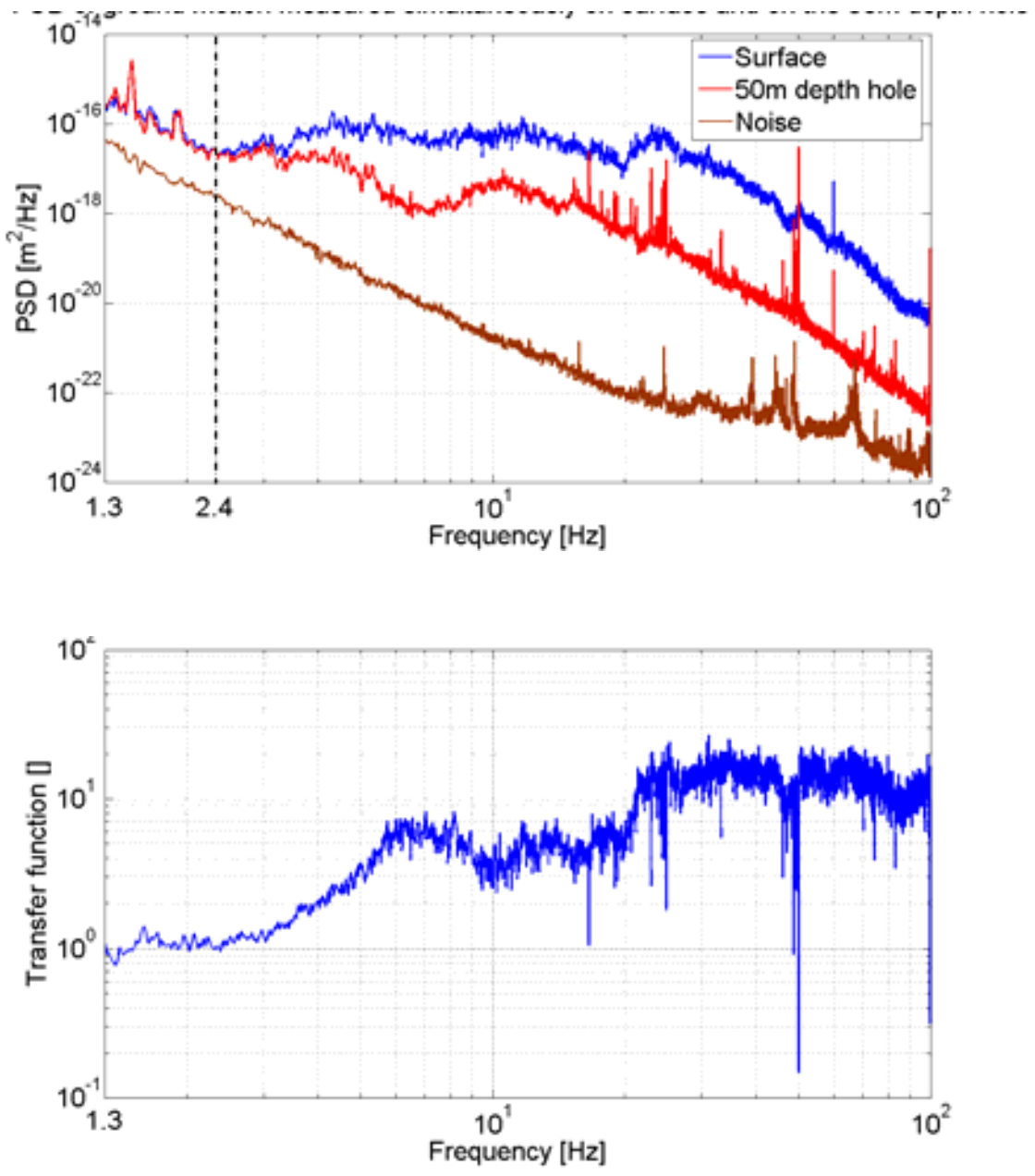

Figure 19.7: PSD of ground motion measured simultaneously on the surface and inside the 50 m depth hole (top) and vibration transfer function between the surface and the hole.

It can be clearly observed that vibrations are damped in the hole above 2.4Hz (beginning of human activities). Above 20Hz, the factor of damping goes up to 20. In order to get values in nanometer, the integrated RMSs of ground motion on the surface and inside the hole has been calculated and is plotted in Figure 19.8.

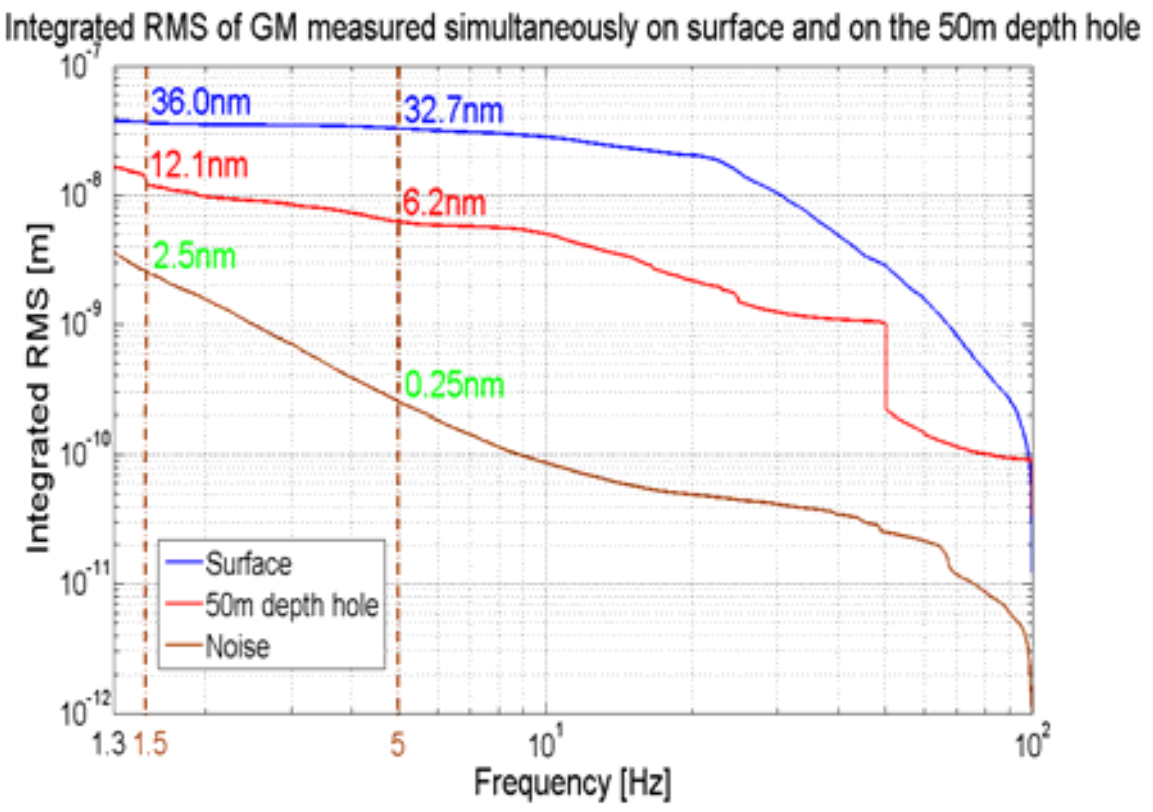

Figure 19.8: Integrated RMS of ground motion measured simultaneously on the surface and inside the 50m depth hole.

Above 1.5Hz, ground motion is of 36.0nm on surface and of 12.1nm inside the hole, that is to say a factor 3.0

of damping. Above 5Hz, it is of 32.7nm on surface against 6.2nm inside the hole, which gives a factor 5.3 of damping. This factor would be probably well higher during rush hours since cultural noise is much more important. All these results clearly show that cultural noise is well attenuated in depth.

## 19.4 Ground motion coherence

Measurements of ground motion coherence were performed in the vertical direction only at point 3 on the parking (soft floor) and on the basement of the new guesthouse building (rigid floor) in order to confirm the importance of a rigid floor for stability [4]. These measurements have been done up to 10m since coherence is lost down to low frequencies and above this distance. In Fig. 19.9, results are shown at top for the parking and at bottom for the basement of the new guesthouse. Results are shown above 3Hz since coherence was lost below this frequency (problem with one of the Guralp geophone). However, coherence is still at 1 for the highest distance (10m) from 3Hz to 6Hz (left) and from 3Hz to 4Hz (right) and is thus in reality at 1 below 3Hz. Fig. 19.10 shows ground motion coherence measurements done on the ATF2 beam line where a special floor was built for stability (same data analysis performed than for the LNF site). Results are shown above 0.3Hz, frequency from where data are reliable (high signal to noise ratio).

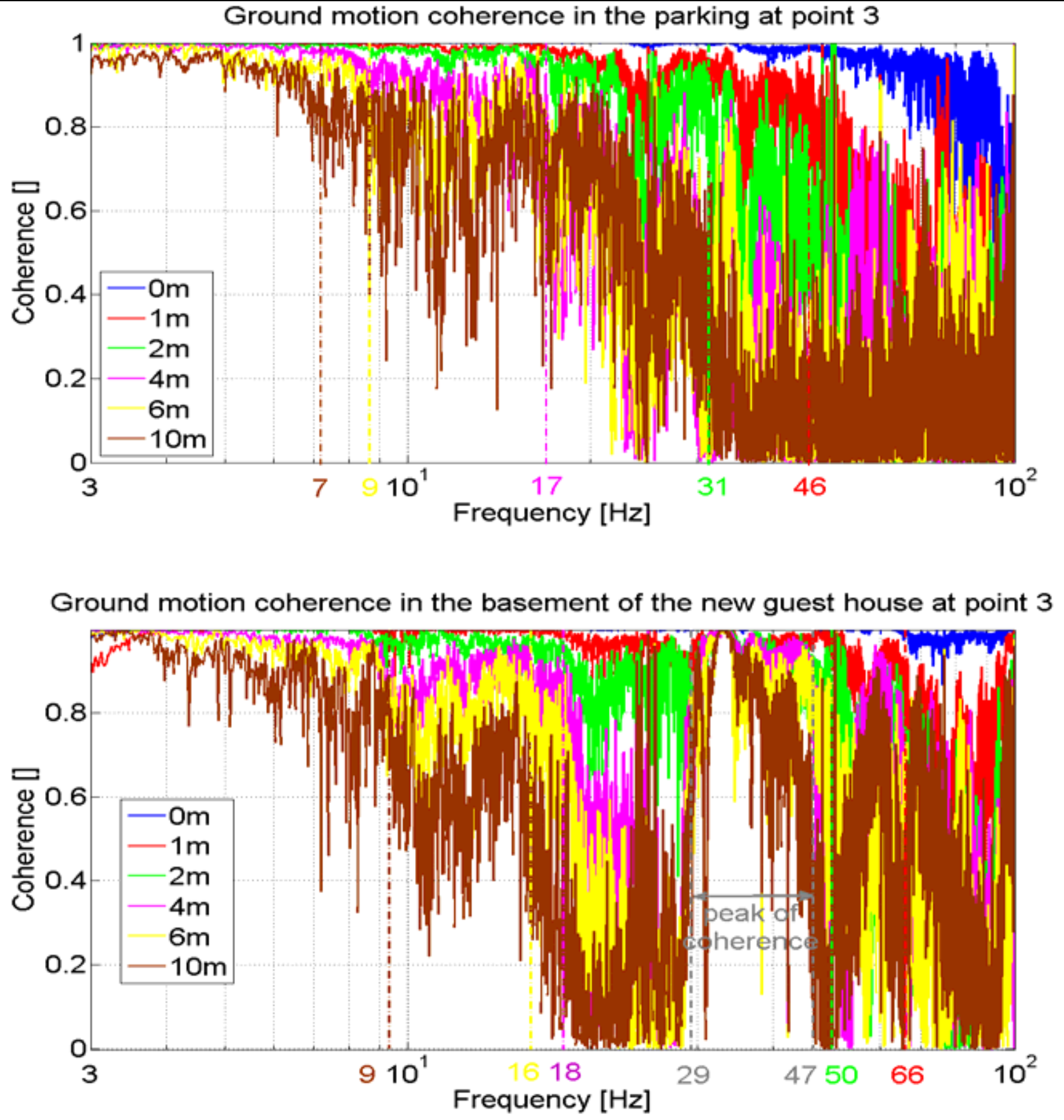

Figure 19.9: Ground motion coherence for different distances at point 3 on the parking whose floor is soft (top) and in the basement of the new guesthouse whose floor is rigid (bottom).

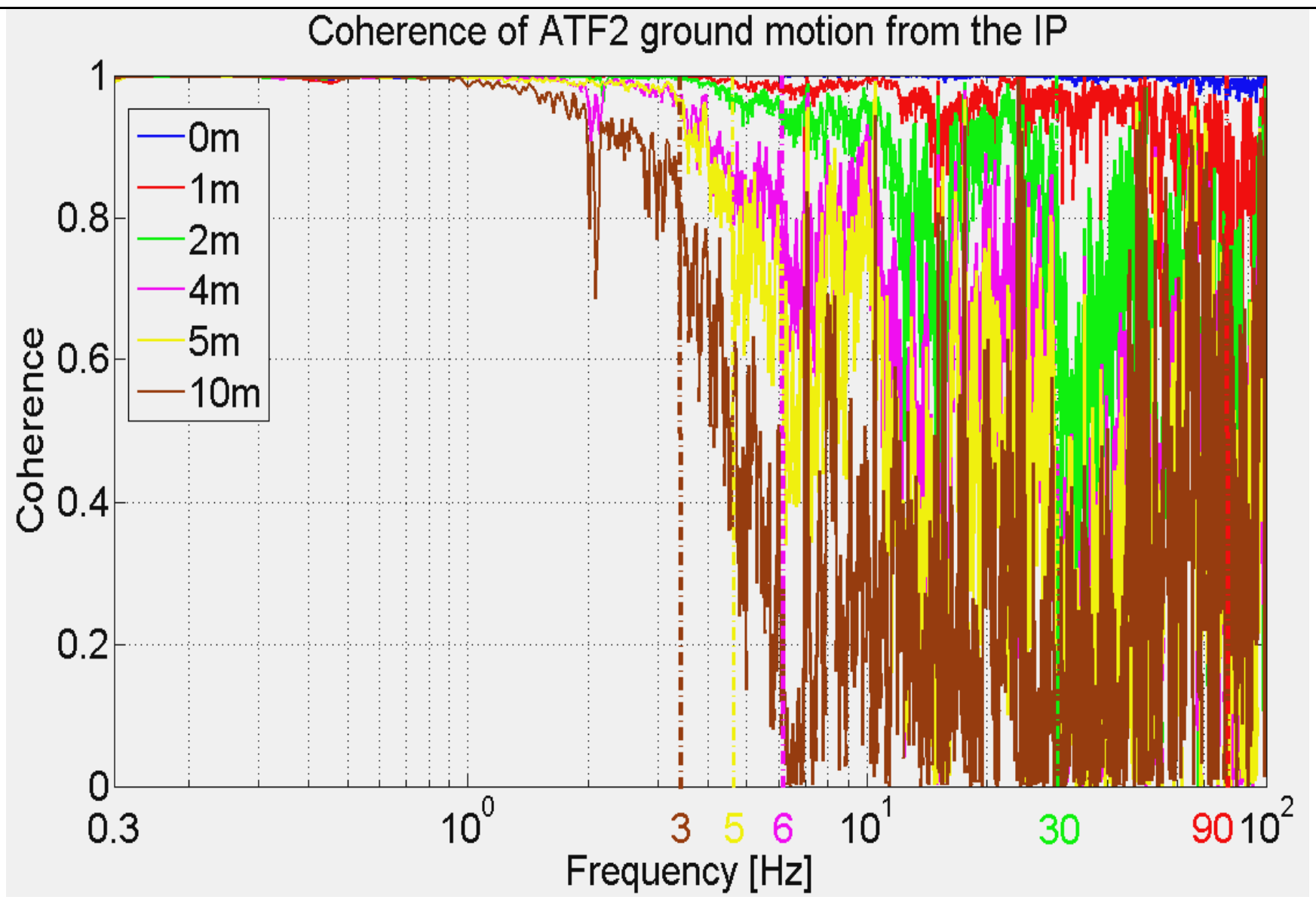

Figure 19.10: Ground motion coherence from the IP for different distances in the ATF2 beam line.

In the three plots above, the frequency where the coherence highly falls under a value of 0.8 is indicated for each distance in order to make a comparison between these 3 different floors. Note that for the basement of the new guesthouse, a peak of coherence appears between 29Hz and 47Hz even if the coherence has already fallen below this frequency range due to the distance. This peak of coherence may be due to the pylons (see Fig. 19.11) which transmit vibrations as seen in the measurements of ground motion coherence in the LHC tunnel [5]. This peak of coherence was consequently not taken into account to determine the frequency where the coherence highly falls.

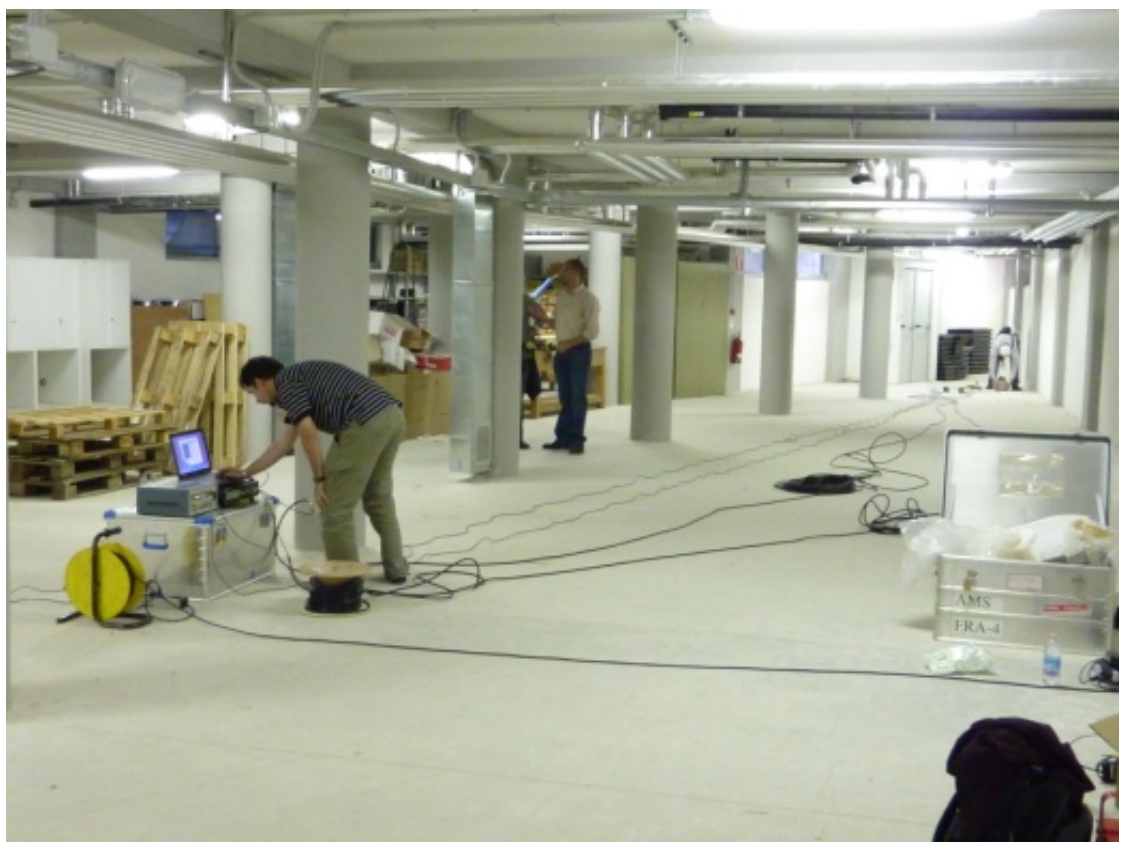

Figure 19.11: Basement of the new guest house with pylons (point 3).

# 20. Tunnels

The 1.35 km-circumference tunnel for the SuperB Factory has a horse shoe cross section type of about 5m wide and 3.4 m high to accommodate the two accelerator rings, trays for the power and control cables, cooling water pipes, access path for equipments, and room for safety egress, see Fig. 20.1. Other cross section shapes, e.g. circular, have been considered for the storage ring tunnels but at the moment the horse shoe type seems to meet the major requirements coming from the accelerator group, safety, etc. The two accelerator rings will be placed side-by-side in one tunnel to keep both rings in the same plane. This reduces or eliminates vertical bends that tend to increase the vertical emittance. The magnet power supplies, cooling water conditioners, RF power supplies, diagnostic and controls will be housed in the straight section buildings. The ring tunnel will be fully underground. Several digging techniques have been investigated for the construction of the SuperB storage ring tunnel, one possible method is the traditional one: drill and blast or mechanical machine when stiff rock layers are encountered and the punctual milling devices for soft layers made of sand and clay. The boring rate in the optimistic case is about 10m per day. A Tunnel Boring Machine (TBM) cannot be used because the length of the tunnel is less than 3Km making it not economically feasible and also because, as seen in the geology section, the underground type soil changes suddenly and the head of the TBM cannot be chosen properly and unambiguously.

For the LINAC tunnel, the diaphragm wall and cut and cover methods can be considered since the area to dig is not built yet, in fact it is located under an existing road. Two overlapped tunnels are foreseen for the LINAC component allocation, see Figure 20.2. The upper one is used to house modulators and klystrons while the lower one for accelerating structures. For the first 70m the DAΦNE LINAC civil infrastructure will be reused. In the next 70m the existing building will be modified and will play the role of connection between the old and new underground part. In the area of the present DAΦNE damping ring, a new building is foreseen about 20x20m$^2$ to allocate the new damping ring. In order to provide radiation containment, the soil surrounding the tunnels must have a thickness of about 3 to 6 meters, depending on detailed radioprotection calculations. A detailed study is underway. A floor drainage system with sump pumps will be provided in the tunnels to contain, collect and treat any free-running water.

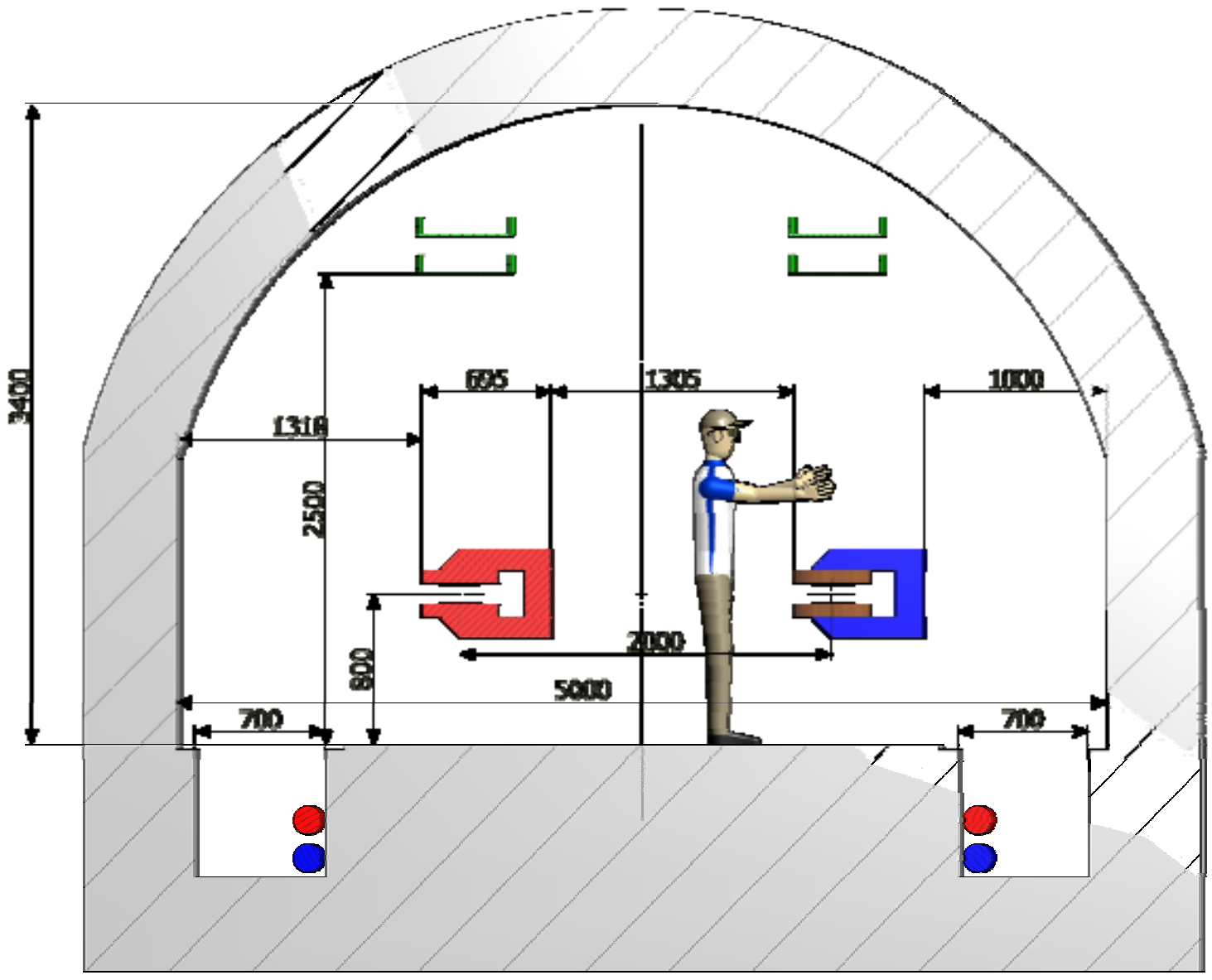

Figure 20.1: Tunnel occupancy.

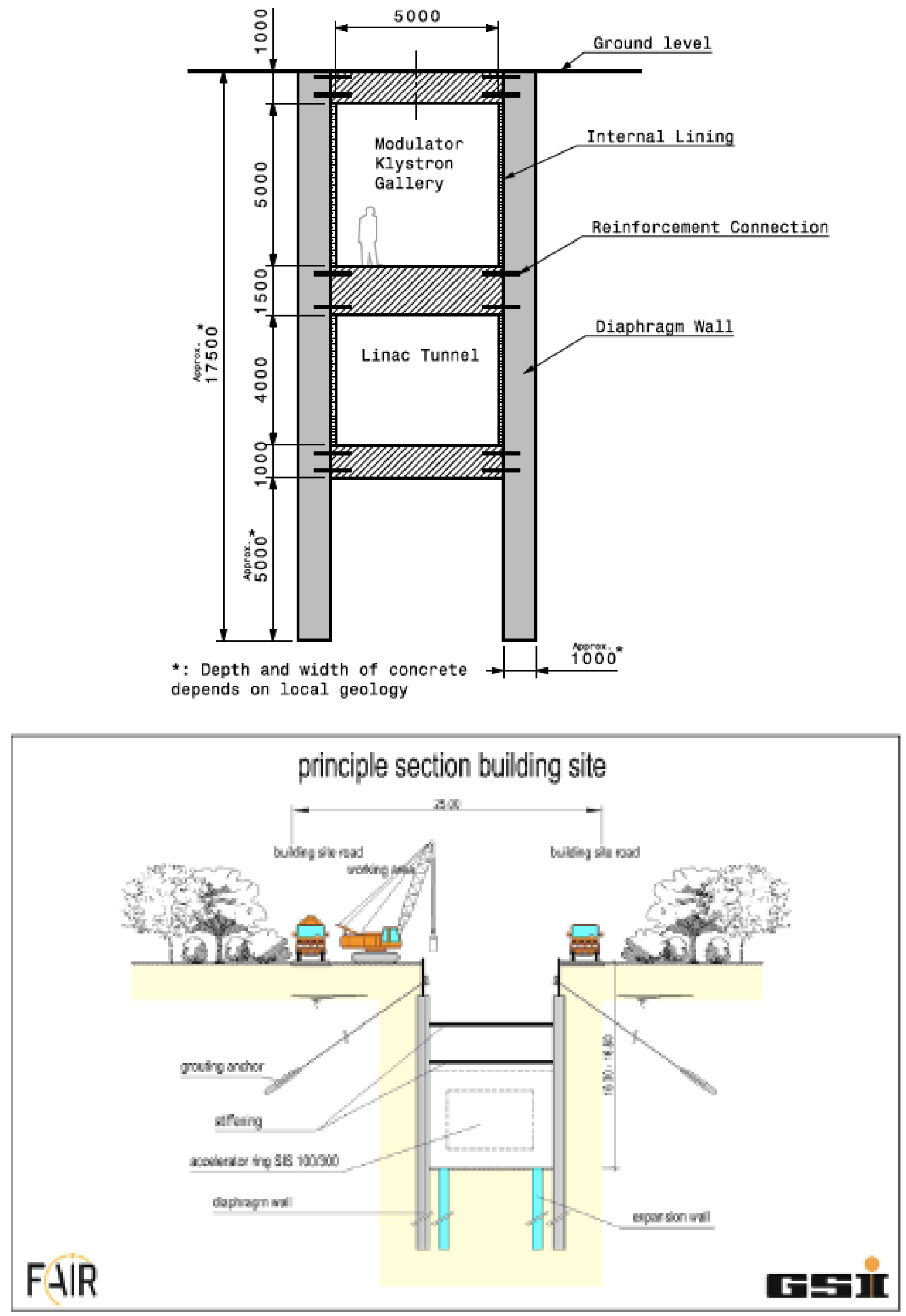

Figure 20.2: Diaphragm wall technique.

# 21. Alignment

The HER and LER of SuperB need to be aligned well in order to allow the very low horizontal and vertical beam emittances to be produced. The quadrupoles of both rings must be aligned to about 50 microns in the vertical direction and 100 microns in the radial direction, whereas the dipoles need to be aligned to about 200 microns in both directions. These specifications are attainable, for relative alignment of components over sliding windows of approximately 100 m, with standard alignment methods and instruments. The very tight tolerances for the interaction region magnets also need careful attention.

The alignment of the accelerator components is carried out using laser alignment trackers, exploiting their accuracy and good versatility. Optical levels and theodolites will be used in order to have redundancy of respectively vertical and horizontal measurements.

For the accelerator, a set of survey markers are mounted on the floor and walls of the tunnel about every 5 m. The laser trackers are then used to measure where these markers are situated relative to each around the whole ring in three dimensions. This forms a global survey grid. Optical levels are used to perform a cross check of the vertical positions of survey markers, which is necessary due to the lower accuracy of laser measurements in the vertical direction. A high density of network nodes together with the fact that the tunnel closes on itself are powerful constraints to reduce errors. The necessity and the opportunity of connecting this local grid to other external survey monuments on site is still under study. The survey grid using laser trackers is used to align the mounting bases for the magnets during installation. Finally the magnets are installed and aligned with the appropriate accuracy. A complete mapping of component positions is then carried out and, if the needed tolerances are not met, some subsequent local smoothing operations are iterated until the requirements are reached.

The alignment of components in the interaction region requires special attention as typically sightings can not be done directly through the detector. In this case a more extensive array of grid markers are mounted on the floor and walls of the detector hall to provide a local grid with enhanced sensitivity. This enhanced grid is then used to align the accelerator components on either side of the detector as well as the detector itself. The final alignment of the final doublet will be anyway performed with Beam Based Alignment techniques.

# 22. AC power

The accelerator requires power for electromagnets, RF systems, diagnostics and controls, and air handling systems. The largest power contribution is from the RF system used to replace the energy lost by the beams due to synchrotron radiation in the bending magnets and wigglers. The power requirements for SuperB are shown in Table 3.1 in Chapter 3. The table includes RF power including inefficiencies of the klystrons and power supplies, magnet power for the two rings, power for water distribution and cooling, control power, injector power and the total estimated requirement. The power required depends on the beam energies, since the beam current, synchrotron radiation and injector energy change with the energy of each ring. The needed power for three possible combinations of HER and LER beam energies are shown. Within this range of configurations, the minimum site power is about 34 MW and the maximum is 43 MW. The minimum wall power requirement is achieved with the design asymmetry of 4 on 7 GeV.

## 22.1 Electrical substation

The Electrical substation at LNF is connected to the national grid via two 150KV electric lines. Presently the substation allocate two 10 MVA transformers plus all the equipments needed. A preliminary study showed that an upgrade can be done and two transformers of 63 MVA each (the red box on the left of Figure 22.1), 150 kV/20 kV, can easily located without any particular difficulty, reusing at maximum all the aerial components. If more power will be needed, special solutions with vented transformers or forced oil cooling have to be found. Enough room is also in the nearest building (the right one), where the Medium Voltage breakers for the MV distribution are located, so that another 24 MV breakers can find place, see Figure 22.1. From there the MV voltage, 20 kV, distribution cables will bring the electric power directly to the 20 kV utilities or to the secondary MV/LV (0,4 kV) distribution sub-stations for the capillary power distribution. Where to put these sub-station, how many will be needed and how to realize the LV electric power distribution will be subject of a study as soon as the footprint of the Super-B and the positioning of the various systems and components will be finalized and frozen.

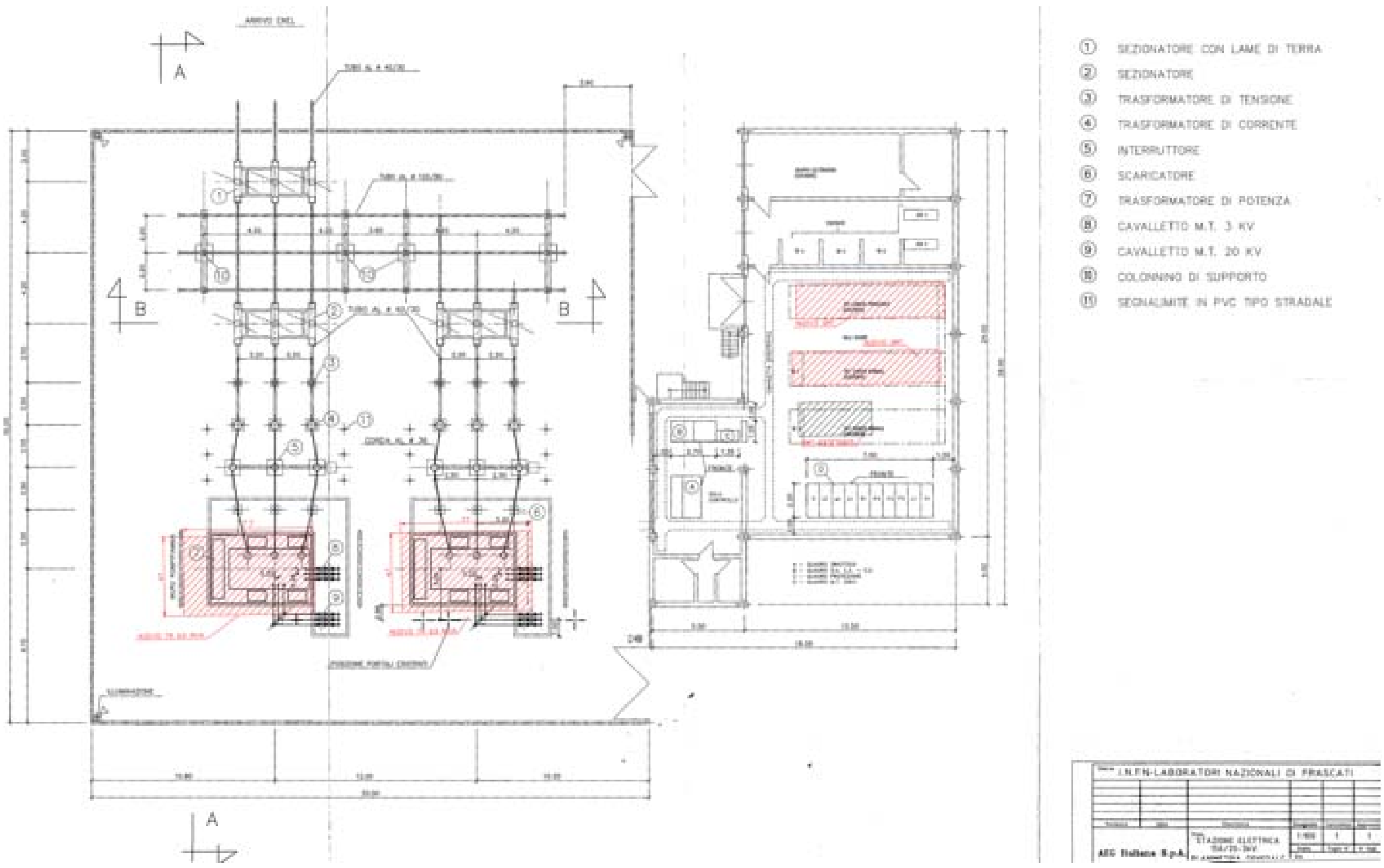

Figure 22.1: Electrical substation.

## 23. Cooling system

The electromagnets and RF systems require cooling water to operate at constant temperature. The cooling water must be pumped around the ring with two supply lines and two return lines. Each subsystem will tap into these lines. The cooling water will be chilled with cooling towers, dry coolers, pumps, and heat exchangers outside the tunnel. The preliminary study turns out that the covered areas needed are about four of about 700m$^2$ each while the no covered areas required are about four 1000m$^2$ each.

The covered rooms will be allocated in the two RF buildings in the south part of LNF, in the collider hall building and in the hangar where presently the DAΦNE main pumping units are allocated. The no covered rooms can be allocated on the roof of the RF and collider hall buildings and other areas near the existing buildings. Approximately 80% of the entire power must be removed using cooling towers or dry coolers. The remainder will be dissipated external to the tunnel, mostly by the high-power high-voltage power supplies for the RF system.

## 24. Air conditioning

The water-cooling system is needed to provide a steady temperature environment. All effort will be made to remove excess heat from equipments using the cooling water system, but the remaining heat will be removed using an air conditioning system, which will need a capacity of about 2.5 MW to remove the power unavoidably transferred to the building air via current-carrying cables etc.

## 25. Construction schedule

The schedule for SuperB requires a period of about four years for construction and installation and an additional six months for commissioning of the two rings before physics delivery to the detector can start. The injector will be commissioned during the last six months of ring installation. During the first year, the tunnel will be drilled, lined and finished. Then, in the following three years, the tasks of installing ring water systems, controls, supports, magnets, vacuum systems, RF systems, and the interaction region will be carried out. A rough schedule is included below in Table 25.1.

Table 25.1: Construction schedule

| Year | Quarter 1 / Quarter 2 | Quarter 3 / Quarter 4 |
|---|---|---|
| 1 | ▪ Tunnel design completed<br>▪ Injector components designed<br>▪ Ring component studied<br>▪ Tunnel contracts awarded<br>▪ Injector components ordered<br>▪ Ring components designed | ▪ Ring tunnel digging started<br>▪ Injector tunnel digging started<br>▪ Injector components started manufacturing<br>▪ Ring components designed<br>▪ Tunnel digging continued<br>▪ Injector components are in manufacturing<br>▪ Ring components orders started |
| 2 | ▪ Ring tunnel digging continues<br>▪ Injector tunnel finished<br>▪ Injector components start to arrive<br>▪ Ring components orders finished | ▪ Ring tunnel is completed<br>▪ Injector installation starts<br>▪ Ring components start to arrive for installation<br>▪ PEP-II components shipped from SLAC |
| 3 | ▪ Injector installation continues<br>▪ Ring component installation starts | ▪ Injector installation is completed<br>▪ Ring installation continues |
| 4 | ▪ Injector checkout starts<br>▪ Ring installation continues | ▪ Injector beam commissioning starts<br>▪ Ring installation is completed<br>▪ Ring checkout starts |
| 5 | ▪ Ring beam commissioning starts | ▪ SuperB beam delivery to detector starts |

## 26. SuperB Accelerator and Facility Budgets

The SuperB accelerator and infrastructure facility costs are presented in this chapter. The cost estimates have been determined using various techniques including local work estimates in the Rome and Frascati regions, estimates for PEP-II dismantlement at SLAC (Menlo Park, California), costs for LCLS construction at SLAC, costs for NSLS-II construction at Brookhaven (Upton, New York), and local shop refurbishment rates.

The reuse and refurbishment of existing equipment and components from PEP-II have been shown to significantly reduce the cost of the SuperB accelerator. The use of PEP-II components has been built into the design from the start and has not affected the desired technical outcome. Although the cost estimates presented here are based on existing knowledge of recent work, the estimates still need additional study to be sufficiently sound to be used for detailed final project planning. Thus, further scrutiny is needed. There are still technical decisions that need to be made that affect the overall cost. The site selection is one of these choices. The possible inclusion of synchrotron radiation beam lines is another.

Given the complexity of this project and the multinational nature of the work distribution, there are several challenging general issues to determine the costs including fluctuating currency rates, escalation of raw material costs, global and local economic factors, and the workings of the European ERIC governance model for the project.

The costing model used in this chapter is similar to that used for the CDR produced in 2007. The components are estimated in four different general categorizes: 1) EDIA (Engineering, Design, Inspection, and Acceptance), 2) Labor, 3) M&S, and 4) Replacement Value. All values are listed in man-months or in FY2010 kilo-Euros. The EDIA and Labor are monetized by using a generic rate of 12 keuros per man-month including overhead and benefits. A final and accurate monetary conversion can only be attempted after institutional responsibilities have been identified and the project time schedule and has been specified. Thus, the final total cost can be determined once the responsibilities are identified.

The cost for each component includes the design, procurement, construction, transportation, and laboratory testing costs. The installation and checkout costs of all these components are included in the Labor and M&S columns. The reuse column would be used for "in-kind" value determination only. The replacement values of the reused components represent how much money it would take to build them from scratch and were obtained by escalating the original cost of production for PEP-II and checked with recent costs for LCLS and NSLC-II. The replacement value costs do not include the costs for removal of the reused PEP-II components from the SLAC tunnel, refurbishment, transportation, testing, costs, and final installation. These costs are included in the EDIA, Labor and M&S columns for the respective components. Contingency for these all these work areas has been estimated at about 50% and have been added only to the total project cost at the roll up line at the top of the Tables.

The SuperB accelerator budget costs are shown in Table 26.1 listed to Work Breakdown Structure WBS level 2. These costs will depend somewhat on the specific site chosen and economic factors at the time of project approval.

The SuperB facility costs for the site and utilities are shown in Table 26.2 listed to WBS level 3. These costs will depend somewhat on the specific site chosen for the collider and local economic factors at the time of project approval.

Table 26.1: Accelerator budget estimate

| WBS | Item | Number of units | EDIA (mm) | Labor (mm) | M&S (k€) | Total (k€) | Repl. Value (k€) (not in total) |
|---|---|---|---|---|---|---|---|
| | | | | | | | |
| **2.00** | **Overall SuperB Accelerator total** | | **3159** | **2852** | **285350** | **357476** | **85760** |
| | | | | | | | |
| **2.01** | **Contingency and VAT (50%)** | | **1053** | **951** | **95117** | **119159** | **0** |
| | | | | | | | |
| **2.02** | **Overall Super B Project Sub-total** | | **2106** | **1901** | **190233** | **238317** | **85760** |
| | | | | | | | |
| 2.03 | Project management and admin | 15 man-yr | 180 | 0 | 400 | 2560 | 0 |
| 2.03 | Accelerator physics | 10 man yr | 120 | 0 | 200 | 1640 | 0 |
| | | | | | | | |
| **2.10** | **HER Ring Total** | | **275** | **300** | **30976** | **37876** | **15690** |
| 2.11 | Dipole magnets | 112 | 15 | 19 | 2265 | 2673 | 5100 |
| 2.12 | Quadrupole magnets | 289 | 35 | 40 | 3760 | 4660 | 6300 |
| 2.13 | Sextupole magnets | 98 | 24 | 20 | 722 | 1250 | 2200 |
| 2.14 | Dipole steering correctors | 290 | 8 | 12 | 90 | 330 | 310 |
| 2.15 | Special magnets | 8 | 15 | 13 | 350 | 686 | 200 |
| 2.16 | Vacuum chambers | 1250m | 50 | 85 | 13163 | 14783 | 180 |
| 2.17 | Power supplies and cables | 400 | 48 | 45 | 7967 | 9083 | 250 |
| 2.18 | Supports | 995 | 55 | 36 | 2485 | 3577 | 600 |
| 2.19 | Abort system and trigger | 1 | 25 | 30 | 174 | 834 | 550 |
| | | | | | | | |
| **2.20** | **LER Ring Total** | | **311** | **352** | **35209** | **43165** | **17070** |
| 2.21 | Dipole magnets | 356 | 30 | 34 | 5260 | 6028 | 8060 |
| 2.22 | Quadrupole magnets | 303 | 38 | 46 | 3540 | 4548 | 5200 |
| 2.23 | Sextupole magnets | 98 | 24 | 20 | 575 | 1103 | 2020 |
| 2.24 | Dipole steering correctors | 310 | 8 | 12 | 90 | 330 | 310 |
| 2.25 | Special magnets and spin rotators | 12 | 27 | 45 | 670 | 1534 | 250 |
| 2.26 | Vacuum chambers | 1250m | 50 | 85 | 13163 | 14783 | 180 |
| 2.27 | Power supplies & cables | 500 | 44 | 40 | 7857 | 8865 | 200 |
| 2.28 | Supports | 1085 | 65 | 40 | 3880 | 5140 | 300 |
| 2.29 | Abort system and trigger | 1 | 25 | 30 | 174 | 834 | 550 |
| | | | | | | | |
| **2.30** | **Interaction Region Total** | | **139** | **147** | **10020** | **13452** | **0** |
| 2.31 | QPM | 4 | 12 | 13 | 420 | 720 | 0 |
| 2.32 | QD0 | 4 | 15 | 17 | 1150 | 1534 | 0 |
| 2.33 | QF1 | 4 | 15 | 17 | 1240 | 1624 | 0 |
| 2.34 | Solenoids | 4 | 10 | 12 | 1100 | 1364 | 0 |
| 2.35 | Vacuum chambers | 5 | 24 | 24 | 1245 | 1821 | 0 |
| 2.36 | Power supplies and cables | 12 | 16 | 12 | 1085 | 1421 | 0 |
| 2.37 | Mech supports & vibration control | 26 | 14 | 14 | 1510 | 1846 | 0 |
| 2.38 | Cryostat and He plant and controls | 2 | 13 | 17 | 1720 | 2080 | 0 |
| 2.39 | Lumi&polar monitor & IP feedback | 1 | 20 | 21 | 550 | 1042 | 0 |
| | | | | | | | |
| **2.40** | **RF System Total** | | **119** | **116** | **4378** | **7198** | **41150** |
| 2.41 | Cavities | 36 | 14 | 14 | 540 | 876 | 18200 |
| 2.42 | Klystrons | 15 | 15 | 17 | 420 | 804 | 12000 |
| 2.43 | Circulators | 15 | 8 | 9 | 135 | 339 | 3000 |
| 2.44 | Wavequides and Ts | 300 m | 15 | 13 | 270 | 606 | 900 |
| 2.45 | RF loads | 30 | 6 | 6 | 80 | 224 | 200 |
| 2.46 | Supports | 15 | 12 | 14 | 613 | 925 | 100 |
| 2.47 | Low level RF controls | 15 | 24 | 22 | 910 | 1462 | 300 |
| 2.48 | High voltage power supplies | 15 | 11 | 13 | 780 | 1068 | 6000 |
| 2.49 | High voltage switch gear | 15 | 14 | 8 | 630 | 894 | 450 |
| | | | | | | | |
| **2.50** | **Ring Controls and Diagnostics Total** | | **252** | **237** | **12465** | **18333** | **6170** |
| 2.51 | Control computers & distribution | 4 | 120 | 80 | 1600 | 4000 | 250 |
| 2.52 | Power supply controllers | 900 | 18 | 12 | 1350 | 1710 | 0 |
| 2.53 | Beam position monitor system | 640 | 16 | 20 | 7200 | 7632 | 0 |
| 2.54 | Current monitor & Ibun controller | 4 | 10 | 8 | 35 | 251 | 270 |
| 2.55 | Transverse feedback | 4 | 24 | 30 | 520 | 1168 | 2400 |
| 2.56 | Longitudinal feedback | 2 | 24 | 32 | 470 | 1142 | 1900 |
| 2.57 | Thermo monitor system | 1700 | 14 | 17 | 450 | 822 | 350 |
| 2.58 | Tune & synch rad monitor system | 6 | 20 | 27 | 760 | 1324 | 780 |
| 2.59 | Beam loss monitor system | 200 | 6 | 11 | 80 | 284 | 220 |
| | | | | | | | |
| **2.60** | **e-/e+ Sources, Damping Ring Total** | | **216** | **234** | **21300** | **26700** | **2680** |
| 2.61 | Laser for source | 1 | 12 | 14 | 350 | 662 | 100 |
| 2.62 | e- polarized source | 1 | 14 | 16 | 190 | 550 | 350 |
| 2.63 | Buncher | 1 | 8 | 8 | 380 | 572 | 650 |
| 2.64 | e+ target & capture section | 1 | 14 | 9 | 780 | 1056 | 880 |
| 2.65 | Damping ring magnets & supports | 60 | 48 | 40 | 8700 | 9756 | 0 |
| 2.66 | Damping ring vacuum chambers | 1 | 28 | 40 | 3500 | 4316 | 0 |
| 2.67 | Damping ring RF | 1 | 16 | 20 | 400 | 832 | 400 |
| 2.68 | Transport lines, kickers, septa | 1 | 36 | 37 | 3300 | 4176 | 300 |
| 2.69 | Controls, pwr supplies, diag, cable | 1 | 40 | 50 | 3700 | 4780 | 0 |
| | | | | | | | |
| **2.70** | **Linac Total** | | **164** | **186** | **48235** | **52435** | **300** |
| 2.71 | Accelerating structures | 100 | 36 | 48 | 20000 | 21008 | 0 |
| 2.72 | Klystrons | 33 | 10 | 18 | 6600 | 6936 | 0 |
| 2.73 | Waveguides, splitters, loads | 800 m | 15 | 12 | 2000 | 2324 | 0 |
| 2.74 | Vacuum system | 400 m | 18 | 20 | 2300 | 2756 | 0 |
| 2.75 | Mechanical supports | 380 | 20 | 10 | 2600 | 2960 | 0 |
| 2.76 | Quadrupole magnets | 32 | 12 | 14 | 640 | 952 | 200 |
| 2.77 | Steering dipoles | 32 | 5 | 6 | 65 | 197 | 100 |
| 2.78 | Klystron modulators | 33 | 18 | 18 | 8250 | 8682 | 0 |
| 2.79 | Controls, pwr supplies, diag, cable | 33 | 30 | 40 | 5780 | 6620 | 0 |
| | | | | | | | |
| **2.80** | **Injection Transport Total** | | **123** | **124** | **9350** | **12314** | **2700** |
| 2.81 | Dipole magnets | 30 | 16 | 16 | 1200 | 1584 | 450 |
| 2.82 | Quadrupole magnets | 60 | 14 | 18 | 1800 | 2184 | 350 |
| 2.83 | Vacuum system | 250 m | 18 | 20 | 2500 | 2956 | 0 |
| 2.84 | Mechanical supports | 100 | 16 | 9 | 1300 | 1600 | 0 |
| 2.85 | Collimators | 4 | 6 | 6 | 90 | 234 | 0 |
| 2.86 | Injection kickers and septa | 8 | 16 | 12 | 420 | 756 | 1800 |
| 2.87 | Injection diagnostics | 10 | 12 | 14 | 700 | 1012 | 100 |
| 2.88 | Ring collimators for inj losses | 4 | 7 | 5 | 240 | 384 | 0 |
| 2.89 | Controls, pwr supplies, cables | 2 | 18 | 24 | 1100 | 1604 | 0 |
| | | | | | | | |
| **2.90** | **Installation, alignment, & testing** | | **207** | **205** | **17700** | **22644** | **0** |
| 2.91 | HER | 1 | 27 | 17 | 4300 | 4828 | 0 |
| 2.92 | LER | 1 | 29 | 18 | 4590 | 5154 | 0 |
| 2.93 | Interaction region | 1 | 15 | 18 | 790 | 1186 | 0 |
| 2.94 | RF system | 1 | 18 | 15 | 2200 | 2596 | 0 |
| 2.95 | Controls and Diagnostics | 1 | 16 | 19 | 850 | 1270 | 0 |
| 2.96 | Sources and Damping ring | 1 | 36 | 40 | 1360 | 2272 | 0 |
| 2.97 | Linac | 1 | 38 | 47 | 2550 | 3570 | 0 |
| 2.98 | Injection transport | 1 | 20 | 23 | 780 | 1296 | 0 |
| 2.99 | Control room | 1 | 8 | 8 | 280 | 472 | 0 |

Table 26.2: Site and Utilities budget estimate

| **WBS** | **Item** | **Number of units** | **EDIA (mm)** | **Labor (mm)** | **M&S (M€)** | **Total (M€)** | **Repl. Value (M€)o** |
|---|---|---|---|---|---|---|---|
| | | | | | | | |
| **3.00** | **Overall Site and Utility total** | | **0.0** | **0.0** | **157.0** | **157.0** | **0.0** |
| 3.01 | Contingency and VAT | | 0.0 | 0.0 | 26.2 | 26.2 | 0.0 |
| 3.02 | Overall sub-total | | 0.0 | 0.0 | 130.8 | 130.8 | 0.0 |
| 3.10 | Site geological preparation | 1 | 0.0 | 0.0 | 2.5 | 2.5 | 0.0 |
| 3.20 | Tunnel design and documents | 1 | 0.0 | 0.0 | 3.2 | 3.2 | 0.0 |
| 3.30 | Tunnel + surface buildings construction | 1 | 0.0 | 0.0 | 70.1 | 70.1 | 0.0 |
| 3.40 | Utility professional design | 1 | 0.0 | 0.0 | 2.4 | 2.4 | 0.0 |
| 3.50 | Electric substation | 5 | 0.0 | 0.0 | 7.0 | 7.0 | 0.0 |
| 3.60 | Cooling plant | 1 | 0.0 | 0.0 | 40.0 | 40.0 | 0.0 |
| 3.70 | Project management | 1 | 0.0 | 0.0 | 4.9 | 4.9 | 0.0 |
| 3.80 | Acceptance tests | 1 | 0.0 | 0.0 | 0.2 | 0.2 | 0.0 |
| 3.90 | Accessory costs | 1 | 0.0 | 0.0 | 0.5 | 0.5 | 0.0 |

# 27. Operations personnel and costs

The overall operational plan for the SuperB accelerator is to run for particle physics data taking for the detector for about 10 months each year. There will be about a two month down each year for standard routine maintenance, the installation of new accelerator hardware to improve the luminosity performance, and for any realignment of the accelerator components, if necessary. The first month of operation will usually encompass shaking down the accelerator components that have failed during the downtime, testing new hardware, and for trying newly-developed tuning techniques. The detector will also use this month to do similar work for their hardware and software systems. The following nine months will be for a solid straight run, 24hr/7days, to integrate as much data as possible. Every week or so a few shifts will be used for accelerator machine development to study and improve the luminosity towards the ultimate machine performance.

The power needed for SuperB is about 26 MW including about 16 MW for the RF system at full beam capacity and 10 MW for all the other systems including water pumps, lights, controls, safety systems, magnet power supplies, etc. This power will be needed for about 9.5 months per year. The remaining 2.5 months the power will be the maintenance mode power at about 5 to 7 MW.

The staff needed to operate SuperB includes many different skill sets. These skill sets cover all the various occupations needed to run, maintain, improve, and analyze the subcomponents of the accelerator. These skill sets are described in the next few paragraphs.

Accelerator physicists are needed to watch the complex SuperB accelerator, calculate the expected performance, analyze the resulting performance, and predict the improvements by new system upgrades. An estimate of the needed accelerator physicists is about ten to cover all the aspects of the machine.

The accelerator control room operators run the accelerator, manage the safety systems of the accelerator, tune to maximize the luminosity, and tune to minimize the detector backgrounds. For SuperB there needs to be about four operators on every shift: one supervisor, one operator to tune beams in the injector, one operator to tune the beams in the rings, and one operator to tune the interaction region. All this work is done in real time. Given effective availability of the operations staff, about 5 full time people are needed to cover one chair. Thus, a total of about twenty operations staff members are needed to cover control room activities.

There are many other systems to maintain and upgrade through out the SuperB complex. Several of these systems are at a very high technical level and quite involved and they will need constant attention every day. An approximate estimate of the number of staff members, needed for each system, can be made (see Table 27.1). Offices, shop space and storage areas are needed to house these people.

Table 27.1: Operation staff needs (approx. numbers)

| System | People | System | People |
|---|---|---|---|
| RF | 10 | Power supplies | 12 |
| Controls and computing | 20 | Electric power | 8 |
| Cooling systems | 9 | Alignment | 3 |
| Vacuum | 11 | Mechanical design | 6 |
| Building maintenance | 10 | Machinists | 10 |
| HVAC department | 8 | Area technical managers | 6 |
| Project management, human resources, purchasing, safety, and administrative staff | | | 25 |